\makeatletter
\let\my@xfloat\@xfloat
\makeatother

\documentclass{ucalgthes1}  
\makeatletter
\def\@xfloat#1[#2]{
        \my@xfloat#1[#2]%
        \def\baselinestretch{1}%
        \@normalsize \normalsize
}
\makeatother

\usepackage[letterpaper,top=1in, bottom=1.22in, left=1.40in, right=0.850in]{geometry}
\usepackage[width=0.95\textwidth,font=small,format=plain,labelfont=bf,up,textfont=it]{caption}
\usepackage{amssymb,amsmath,amsthm,amsfonts,mathtools,bigints}
\usepackage[utf8]{inputenc}
\usepackage[turkish,english]{babel}
\allowdisplaybreaks 
\usepackage{cancel} 
\usepackage{mathrsfs}
\usepackage{graphicx} 
\usepackage{pdflscape}
\usepackage[inline]{enumitem}
\usepackage{import}
\usepackage[numbers,square,sort&compress]{natbib} 
\usepackage{epstopdf}
\usepackage{aas_macros}

\usepackage[titles]{tocloft}

\usepackage[titletoc]{appendix} 

\usepackage{makeidx} \makeindex
\usepackage{hyperref}
\usepackage{libertine}
\usepackage{float}
\usepackage{fancyvrb} 
\usepackage{listings}
\usepackage{subfig}
\usepackage{array}
\usepackage{multirow} 
\usepackage{tabu} 
\usepackage[section]{placeins} 

\usepackage[T1]{fontenc}

\DeclareSymbolFontAlphabet{\mathrsfs}{rsfs}      
\DeclareMathOperator{\e}{\mathrm{e}}             

\DeclareMathOperator{\Det}{\mathrm{Det}}
\DeclareMathOperator{\acoth}{arcoth}
\DeclareMathOperator{\csch}{csch}
\DeclareMathOperator{\diag}{diag}

\DeclareUnicodeCharacter{00A0}{ }

\newcommand{\cc}{\ensuremath{\mathrm{c}}}     
\newcommand{\G}{\ensuremath{\mathrm{G}}}     
\renewcommand{\d}{\ensuremath{\,\mathrm{d}}}     
\newcommand{\rmd}{\ensuremath{\mathrm{d}}}
\newcommand{\rme}{\ensuremath{\mathrm{e}}}
\renewcommand{\i}{\ensuremath{\mathrm{i}}}       
\newcommand{\R}{\ensuremath{\mathbb{R}}}         

\newcommand{\f}[2]{\ensuremath{\frac{#1}{#2}} }  
\newcommand{\SFRef}[1]{\protect\subref{#1}}
\newcommand{\un}[1]{\ensuremath{\,\mathrm{#1}} }  
\newcommand{\dunits}{\ensuremath{\un{kg}\cdot\un{m^{-3}}}}

\newcommand{\pderiv}[2]{\ensuremath{\frac{\partial #1}{\partial #2}}}
\newcommand{\deriv}[2]{\ensuremath{\frac{\mathrm{d} #1}{\mathrm{d} #2}}}
\newcommand{\sderiv}[2]{\ensuremath{\frac{\mathrm{d}^2 #1 }{\mathrm{d} #2^2 }}}
\newcommand{\spderiv}[2]{\ensuremath{\frac{\partial^2 #1 }{\partial #2^2}}}
\newcommand{\LieD}[1]{\ensuremath{\pounds_{#1}}}
\newcommand{\La}{\ensuremath{\mathcal{L}}}
\newcommand{\I}{\ensuremath{\mathbf{\Pi}}}
\newcommand{\Christoffel}[3]{\genfrac{\{}{\}}{0pt}{}{#1}{#2 #3}}
\DeclarePairedDelimiter\floor{\lfloor}{\rfloor}

\newcommand{\secO}{\ensuremath{2^{\text{nd}} \text{ order}}}

\newcommand{\ppen}{\ensuremath{p_{\perp}}}
\newcommand{\ppeno}{\ensuremath{p_{\perp 0}}}

\let\oldsqrt\sqrt
\def\sqrt{\mathpalette\DHLhksqrt}
\def\DHLhksqrt#1#2{%
\setbox0=\hbox{$#1\oldsqrt{#2\,}$}\dimen0=\ht0
\advance\dimen0-0.2\ht0
\setbox2=\hbox{\vrule height\ht0 depth -\dimen0}%
{\box0\lower0.4pt\box2}}

\newenvironment{myabstract}
{\begin{center} \begin{tabular}{p{0.8\textwidth}} \hline \small} 
{\\ \hline \\ \end{tabular} \end{center}}

\makeatletter
\renewcommand*\env@matrix[1][c]{\hskip -\arraycolsep
  \let\@ifnextchar\new@ifnextchar
  \array{*\c@MaxMatrixCols #1}}
\makeatother

\def\apjl{ApJL }

\def\apjs{ApJS }

\def\aap{A\&A }

\newtheorem{myTheorem}{Theorem}
\title{Exact Solutions for Compact Objects in General Relativity}
\author{Ambrish M. Raghoonundun} \thesisyear{2016}
\thesis{thesis} 
 
\monthname{April} 
\dept{PHYSICS AND ASTRONOMY}
\degree{DOCTOR OF PHILOSOPHY}
\begin{document}
\makethesistitle
\shorthandoff{=} 
\pagenumbering{roman} 
\setcounter{page}{2} \altchapter{Abstract} Seven new solutions to the
interior static and spherically symmetric Einstein's field equations
(EFE) are found and investigated.  These new solutions are a
generalisation of the quadratic density fall-off profile of the
Tolman~VII solution.  The generalisation involves the addition of
anisotropic pressures and electric charge to the density profile.  Of
these new solutions three are found to obey all the necessary
conditions of physical acceptability, including linear stability under
radial perturbations, and causality of the speed of pressure waves
inside the object.  Additionally an equation of state can be found for
all the physically viable solutions.  The generalised pulsation
equation for interior solutions to the EFE that include both electric
charge and pressure anisotropy is derived and used to determine the
stability of the solutions.  However the pulsation equation found is
general and can be used for all new solutions that contain these
ingredients.


\altchapter{Acknowledgements} \label{C:Ack} 

For this thesis in $\floor{100\pi}$ pages, typeset with \LaTeX, corresponding
to several years of work, I would like to thank my supervisor David
Hobill for funding me through my M.Sc. and Ph.D., even when this
was difficult, and for always having his door open for discussions
about physics, teaching, travelling, and mentoring.  We could probably
have done this in less time, and for this we were both at fault.  

Both of my external examiners: Dr. Gene Couch, and Dr. Valeri Frolov
who patiently read documents that unfortunately do not get read as
often as they should.

My teaching mentors while I was a TA and an instructor: Jason Donev
and Mike Wieser, who helped me become a better teacher.

Tracy and Leslie who were always very helpful with everything
administrative, moreso than other people.  

My parents, and sometimes my sister who listened to my frustrations,
with varying degrees of understanding.  

My friends, in no particular order, Ramya\&Preshant, Sandeep, Kady,
Arthur, Ish, James, Brendan, Hannah\&Mark, Arashdeep, Amir, Rahul,
Carrie, Hallah, Monzu, Jim, Julia, Elliot and Mark who listened, and
put up with caustic remarks about academia, life, people, Calgary,
Canada, Canadians, and other frustrations.  We had some good times.

Our Astro group, colleagues and good friends: Mehrnoosh,
Kianoosh, Anna, Wes, Russel and Sujith.

Di who was kind enough to proof-read some of Chapter~\ref{C:Stability}
when I was really struggling with many things.

My roomates, and subsequently brothers: Warren and Christian, who
introduced me to many new things, video games and philosophy being
among the major ones.


\begin{singlespace}
\clearpage
\tableofcontents
\clearpage
\listoftables
\clearpage
\listoffigures
\clearpage 
\end{singlespace}
\pagenumbering{arabic}
\pagestyle{plain}

\chapter{Introduction} \label{C:Introduction}

\begin{myabstract}
  We motivate the work done in this thesis starting from very general
  notions.  We attempt to find {\bf physically relevant} exact
  interior solutions to Einstein's Field equations (EFE) in their
  static and spherically symmetric case.  The energy-momentum is
  assumed to come from a charged fluid having anisotropic pressures.
  New solutions that can be used to model compact objects (Neutron
  stars, Strange stars, etc.\ ) are found.  The equation of state is
  obtained as a result of the solution, and under additional
  assmptions that fix the parameter values to ones close to nuclear
  densities for example, might be useful for nuclear physics
  considerations.
\end{myabstract}

\section{General Relativity}
First put forward in 1915 by Albert Einstein, General Relativity (GR)
introduces the idea that gravitation is not a force, but rather, it is
a consequence of curvature in a four dimensional
\index{space-time}space-time~\cite{Ein15}.  Space-time itself was
introduced earlier (1905) by Einstein as part of the framework of
special relativity (SR) to help elucidate the then pressing problem of
the constancy of the speed of light with respect to aether, viz. the
result of the Michelson-Morley experiment. GR, according to
Wheeler\cite{MTW}, can be summarised by:
\begin{quote}
  Matter tells space how to curve; Space tells matter how to move.
\end{quote}
GR is a physical theory that has produced a number of predictions over
the years.  The Schwarzschild solution~\cite{Sch16} to the full GR
equations predicted the existence black-holes, which have now been
observed indirectly.  The currently accepted cosmological
model~\cite{HawEll73,StoMaaEll95} the
Friedmann--Lema\^itre--Robertson--Walker metric is a solution to the
fundamental equation of GR --the Einstein field equations (EFE)-- and
has been tested by combining observations from the WMAP~\cite{Kom09}
and Planck~\cite{Pla15} satellites.

There have also been several other direct observational test of GR.
The very first one that convinced people to start taking the theory
seriously was the prediction by the theory of the perihelion shift in
the orbit of Mercury. Gravitational waves~\cite{Ein16} which were
predicted barely a year after the main theory was put forward were
detected a few months ago~\cite{Abb16}, confirming GR once again.
Similarly, the gravitational lensing effect predicting that massive
objects bend light rays and measured famously by
Eddington~\cite{DysEdd20} in 1919 also confirmed GR and brought it its
fame initially.  All these attest to the fact that classical GR is a
well established physical theory that predicts measurable quantities.

The areas in which GR still has predictions that have not been tested
completely are the strong regimes where potentials that scale like
\(M/R\) are very large.  Then the non-linearities of the EFE become
important and methods that can take these non-linearities into account
become more valuable.  The detection of gravitational waves tested
some of these strong regime predictions through numerical methods
modelling the space-time around the black-holes producing the
waves~\cite{Abb16}.  Another method of getting predictions for the
strong regime is through perturbation methods that go beyond the
linearised EFE.  These methods have been successful in calculating
Love numbers and subsequent gravitational waves from systems of
compact astrophysical objects~\cite{DamNag09,DamNagVil12}.  However
these are not the methods this thesis investigates.  Instead we
approach the problem of strong gravitational fields by constructing
exact solution to the EFE with a particular application toward
understanding the structure of compact objects.

\section{Exact Solutions}
The reason for looking at exact solutions, aside from the fact that
they are mathematically interesting, is that they also provide a
baseline against which both of the methods mentioned above can be
compared.  Exact solutions like the Schwarzschild Interior
solution~\cite{Sch16} are still being used, despite being non-physical
for this very reason: at best these exact solutions provide a limit on
certain physical parameters , and at worse they guide the
understanding into the behaviour of the gravitational field.  For this
reason the construction of exact solutions can be more rewarding than
just the mathematical exercise.  Some of the solutions found can be
used for physical modelling and thereafter to predict measurable
quantities about physical systems.  Upon measurement, the truthfulness
of the model can then be ascertained, or denied.

The history of the hunt of exact solutions to the interior EFE is a
long and interesting one.  We briefly outline this history in
Section~\ref{pr.sec.IntSol}, and refer the reader to~\cite{SteKra09}
for a more complete list and history.  Of interest for this thesis is
that of the roughly 130 static spherically symmetric perfect fluid
solutions to the EFE that were known in~\citeyear{DelLak98}, only 9
were deemed to be physically viable~\cite{DelLak98, FinSke98}.  The
criteria for physical viability are simple constraints from within the
framework of GR and classical physics, and do not have any
quantum-mechanical component to them.  

Once an interior solution has been deemed to behave physically, it can
be used to predict measurables/observables of the system.  For one of
the known solutions deemed to be physically relevant
in~\cite{DelLak98}, we extract both masses and radii of the compact
object modelled in Chapter~\ref{C:TolmanVII}.  The solution considered
is the Tolman~VII solution~\cite{Tol39}, and we find that it indeed
predicts masses and radii that are in line with current
observations~\cite{RagHob15}.  This not only demonstrates that the
Tolman~VII solution is a viable model for physical systems, it also
shows that GR's strong field solutions are accurate to the limit of
our current observations capabilities in these systems.  Another
prediction that this model produces is an equation of state (EOS) for
the matter inside the star.  This EOS is obtained without any quantum
mechanical assumptions, and while there is no direct way to test the
accuracy of this EOS, that the observed mass and radii of neutron can
be obtained without detailed microphysics, suggests that the
sensitivity of the bulk properties of compact stars to the differences
in many nuclear models is very low.

The above is the primary motivation for finding {\bf new physically
  motivated solutions} to the EFE that can be used to model compact
objects.  Once these solutions have been found, all the predictions
stemming from their use as model for these systems can be
investigated.  Furthermore more elaborate numerical and perturbative
methods can be used and compared with the new exact solutions found.
For example, the calculation of Love numbers in binary systems and the
generation of gravitational waves in these same systems could greatly
benefit from exact interior solutions.

\section{This work}
The solutions presented in~\cite{DelLak98} are all uncharged with a
perfect fluid matter as the source.  In Chapter~\ref{C:NewSolutions},
we generalize the source to include anisotropic pressures in the fluid
in Section~\ref{ns.sec:Ani}, and then additionally include electric
charge in the source in Section~\ref{ns.sec:Cha}.  The physical
reasoning behind these additions is provided in
Chapter~\ref{C:Preliminaries}, with a brief historical overview for
these types of solutions.  Chapter~\ref{C:NewSolutions} is the
mathematical component of this work, and does not investigate the
physical validity of the solutions found.  It is only concerned with
the mathematical consistency of the solutions to the EFE.  During this
process, we find seven new solutions, and generate a few solutions
that had already been found before, when certain of our parameters are
set to zero.  A summary of the solution landscape is given in
Figure~\ref{pr.fig:Flow}.

A question that is often asked is whether static solutions to the EFE
are stable.  The answer is unknown, but~\citeauthor{Cha64L} in a
seminal paper~\cite{Cha64L} attempted to find whether the general
perfect fluid solution was stable under radial perturbations, and came
up with a ``pulsation equation.''  The frequencies of the normal modes
of this equation then determines whether the solutions are stable
under \emph{radial linear} perturbations: real frequencies
corresponding to ``breathing modes'' and imaginary ones to unstable
expanding or contracting ones.  When we began considering this for our
new solutions, we found that the pulsation equation for our case--a
charged fluid with anisotropic pressure--was not available.  This was
the basis of Chapter~\ref{C:Stability}, where we investigate the
stability of the solutions we found in Chapter~\ref{C:NewSolutions},
by first deriving a general pulsation equation.  We then show that for
certain parameter choices, our solutions are stable, and for which
choices the solutions are not.

Chapter~\ref{C:Analysis} instead analyses the new solutions found from
the perspective of physical acceptability.  We list the criteria for
physical acceptability in Chapter~\ref{C:Analysis} and use them in
Section~\ref{an.ssec:PhysRel} onwards on all the new solutions and
conclude that of the seven solutions we found, three have promising
characteristics that make them physically interesting.  We discuss
these solutions extensively.  While we are unable to provide exact cut
off values for some parameters that distinguish between physical and
unphysical solutions we can provide some important relations in terms
of general inequalities.

As a summary, the new aspects of this work are
\begin{enumerate}
\item The construction of seven new solutions to the EFE with various
  combinations of the anisotropic pressures and charge.
\item The derivation of a stability equation for radial perturbations
  that can be applied to all new solutions with charge and/or
  anisotropic pressures.  The pulsation equations for the simpler
  cases of zero charge, or zero pressure anisotropy are recovered when
  those parameters are set to zero, attesting to the accuracyof our
  derivation.  We use the derived equation to prove that the solutions
  we are interested in are indeed stable
\item The analysis of the seven new solutions, and the conclusion that
  three might be physically viable.
\end{enumerate}

The new physical solutions we found could potentially be used to model
astrophysical objects, deduce EOS for these compact stars, calculate
Love numbers in binary systems, and even infer gravitational wave
spectra of radiating neutron stars from the EOS.  These are all avenues
for further work.



\chapter{Preliminaries} \label{C:Preliminaries}
\begin{myabstract}
  We look at the ingredients that make up the EFE.  We then look at
  some known interior solutions to the EFE, and how our work fits in
  the overall picture of finding physical solutions to model stars.
\end{myabstract}

\section{Definition of terms}
This chapter uses the definitions and theorems stated in the
Appendix~\ref{C:AppendixA}.  Basing all work to follow on Einstein's
theory of gravity, Einstein's Field equations(EFE) can be written as
\begin{equation}
  \label{pr.eq:EFE}
  G_{ab} = \kappa T_{ab}.
\end{equation}
The next sections introduces all the elements needed to interpret and
use equation~\eqref{pr.eq:EFE}.  The assumptions and notations that we
will use are the following:
\begin{enumerate}
\item We use geometrical units throughout this thesis, unless
  otherwise stated.  Geometrical units are used to simplify most of
  the equations which would otherwise have the physical constants
  \(G,\) Newton's gravitational constant, and \(c,\) the speed of
  light appear in various factors throughout.  The use of geometrical
  units imply that \(G=c=1,\) so that these constants no longer appear
  in the equations.  In chapters~\ref{C:NewSolutions}
  and~\ref{C:Analysis}, we relax this assumption to calculate values
  in SI units by putting back the value of the constants to
  \(G = 6.67 \times 10^{-11} \un{m^{3} \cdot kg^{-1} \cdot s^{-2}},\)
  and \(c = 3.00 \times 10^{8} \un{m \cdot s^{-1}}.\) Table~\ref{pr.tab:Conv}
  provides conversion factors to and from geometrical units to SI
  units in the main text of this chapter
  \item The matter coupling constant (See Appendix~\ref{C:AppendixA})
    \(\kappa =8\pi.\) 
  \item The metric signature (for a definition of signature see
    Appendix~\ref{C:AppendixA}) we use throughout is \((+,-,-,-)\)
  \item Latin indices are used for space-time tensor indices and take
    the values 0,1,2, and 3.  Greek indices span 1,2,3 only instead
    and are used for spatial tensor indices.
\end{enumerate}
\newcolumntype{C}{>{$}c<{$}}
\begin{table}[ht]
  \centering
  \begin{tabular}[c]{c|c|c|C}
    Physical quantity & SI unit & Geometrical unit & \text{Conversion factor } (\times)\\ 
    \hline\hline
    Length    & \un{m}              & \un{m}      & 1              \\
                      &                     & \un{s}      & c^{-1}    \\
                      &                     & \un{kg}     & c^2 G^{-1}\\
    \hline
    Time         & \un{s}              & \un{s}      & 1              \\
                      &                     & \un{m}      & c         \\
                      &                     & \un{kg}     & c^3 G^{-1}\\
    \hline
    Mass         &\un{kg}              & \un{kg}     & 1              \\
                      &                     & \un{m}      & G c^{-2}   \\
                      &                     & \un{s}      & G c^{-3}   \\  
    \hline
    Energy       &\un{kg \cdot m^2 \cdot s^{-2}}   & \un{kg}     & c^{-2}     \\
                      &                     & \un{m}      & G c^{-4}   \\
                      &                     & \un{s}      & G c^{-5}   \\
    \hline
    Density      &\un{kg \cdot m^{-3}}       & \un{m^{-2}} & G c^{-2}   \\
   
    Pressure     &\un{kg \cdot m^{-1} \cdot s^{-2}}& \un{m^{-2}} & G c^{-4}   \\
   
    Speed        &\un{m \cdot s^{-1}}        & unit-less    & c^{-1}     \\
   
    Mass/Radius  &\un{kg \cdot m^{-1}} & unit-less         &G c^{-2}     \\
   
    Electric charge&\un{A \cdot s}      & m                 &\f{1}{c^2}\sqrt{\f{G}{4\pi\epsilon_0}}\\  
  \end{tabular}
  \caption[Units conversion table]{Conversion from SI units to Geometrical units for relevant physical quantities.} \label{pr.tab:Conv}
\end{table}

\section{Geometry}
In general relativity, space-time is modelled by a (3+1)--dimensional
\textbf{Lorentzian manifold}\index{Manifold!Lorentzian}.  The
space-time is additionally endowed with a symmetric metric \(g_{ab}\)
that is used to measure lengths and angles.  In GR the manifold also
has a symmetric metric connection (and therefore no torsion).  The
definition of a manifold requires some additional ideas from set
theory (notions of sets\index{Set}, subsets, elementary set
operations, the real line, \(\mathbb{R},\) etc.\ ) which we shall
assume and not state explicitly, and other definitions (taken mostly
from references~\cite{Cho82,Cho08,Wal84,MTW}) which we give in
Appendix~\ref{C:AppendixA}.  Next we introduce the geometrical
quantities that are needed in the EFE.

\subsection{The metric}
The metric is the dynamical quantity that specifies the general
relativity component of our models for stars.  We will be considering
static models endowed with spherical symmetry.  As a result the
sixteen components of the general space-time metric will be reduced to
four components, two of which will have arbitrary coefficients, that
depend solely on one spatial coordinate.  As a result, Einstein's
equations of general relativity are greatly simplified into a set of
ordinary differential equations (ODEs), whose solutions and
interpretations will be the main thrust of this thesis.  Depending on
our assumptions about matter, and electromagnetic fields we shall have
either three coupled ODEs (isotropic matter without electric charge),
or four ODEs (anisotropic matter without electric charge), or five
coupled ODEs (anisotropic matter with electric charge).  Each case
will be treated separately and solved to yield viable physical models
that can be used to model compact stars.

As mentioned in appendix~\ref{C:AppendixA}, naively the metric
should have sixteen\((=4 \times 4)\) components.  Symmetry of the
metric, that is \(g_{ab} = g_{ba}\) reduces this to 10 independent
components.  In this section we show how additional constraints will
reduce these ten components to two only.  Doing so will require the
application of symmetry arguments most easily done with Lie
derivatives, the definitions of which are given in
appendix~\ref{C:AppendixA}.

\subsection{Symmetry}
\label{pr.ssec:Sym}
The Lie derivative of the metric along vector field \(\xi\) is 
\[\LieD{\xi} g_{ab} = \xi^{c} g_{ab,c} + \xi^{c}{}_{,b}g_{ac} + \xi^{c}{}_{,a}g_{cb},\] 
so that Killing's equation becomes the above being identically equal
to zero.  For staticity, we want our first killing vector \(T\) to be
\(\partial_{t}\), or in an adapted frame with usual spherical type
coordinates \((x_{0},x_{1},x_{2},x_{3}) \equiv (t,r,\theta,\varphi)\): \(T =
(1,0,0,0)\).  As a result of the constant components of this vector,
we can immediately write \(T^{a}{}_{,b} = 0\) as a result of which the
Lie derivative above reduces to \[\LieD{T}g_{ab} = T^{c} g_{ab,c} =
g_{ab,0} = 0.\] From this we deduce that our metric components can
have no dependencies on the \(t\) coordinate at all, so that
\(g_{ab}(t,r,\theta,\varphi) \equiv g_{ab}(r,\theta, \varphi).\)

For spherical symmetry, we require three additional Killing vectors
which generate \(SO(3)\) and whose Lie brackets are cyclic with each
other as in Appendix~\ref{C:AppendixA}.  As given in the example
involving the unit 2-sphere there, the following three vectors in the
spherical adapted frames satisfy the conditions for spherical
symmetry:
\begin{subequations}
  \begin{equation}
    \label{eq:8}
    P = -\sin\varphi \pderiv{}{\theta} - \cot\theta \cos\varphi \pderiv{}{\varphi} \coloneqq L_{1}
  \end{equation}
  \begin{equation}
    \label{eq:9}
    Q = \cos\varphi \pderiv{}{\theta} - \cot\theta \sin\varphi \pderiv{}{\varphi} \coloneqq L_{2},
  \end{equation}
\begin{equation}
  \label{eq:10}
  R = \pderiv{}{\varphi} \coloneqq L_{3}
\end{equation}
\end{subequations}
As in the definition of spherical symmetry, given completely in
Appendix~\ref{C:AppendixA} the above vectors obey the cyclic Lie
structure \([L_{i},L_{j}] = \epsilon^{ijk} L_{k}.\)

To impose symmetry on the metric, we can now simply use Killing's
equation on any linear combination of the \(L\) vectors given above.
For \(\vec{\xi} = xL_{1} + yL_{2} + z L_{3},\) we require that
\[ \LieD{\xi}g_{ab} = \xi^{c} g_{ab,c} + \xi^{c}{}_{,b}g_{ac} +
  \xi^{c}{}_{,a}g_{cb} = 0,\] and the long calculation~\cite{Sha07}
that results end in the metric being split into two blocks, the first
block containing exclusively the 2-sphere coordinates \(\theta\) and
\(\varphi,\) in the
form\[ \d s^{2}_{\text{2-sphere}} = r^{2} \d \Omega = f^{2}(r,t) (\d
  \theta^{2} + \sin^{2}\theta \d \varphi^{2}),\] where \(f(r,t)\) is
an arbitrary function; and the other part containing the remaining two
coordinates \(t\) and \(r\) through
\[\d s^{2} = g_{AB}\d x^{A} \d x^{B},\] with \(g_{AB}\) independent of
the angles \(\theta\) and \(\varphi.\) Since we also have the Killing
vector \(T\) to contend with, the off diagonal terms of \(g_{AB}\)
have to be zero. Similarly, \(\partial_{t}f(r,t) = 0,\) so that
without loss of generality and through imposition of spherical
symmetry, the metric's 10 components are reduced to only four diagonal ones:
\begin{equation}
  \label{pr.eq:SymMetric}
\d s^{2} = \e^{\nu(r)} \d t^{2} -\e^{\lambda(r)} \d r^{2} - r^{2} \d \Omega^{2}.  
\end{equation}
With the complete set of of Killing vectors, \(T, P, Q\) and \(R\)
that we imposed on the metric, we can also impose the same symmetry
requirements on the other tensors of the theory.  This results in
``collineation theory'' which we consider next when we apply this idea
on the matter and electromagnetic fields.

This section introduced the metric tensor we will be using throughout
this work and reduced its ten components into only four.  From methods
presented in Appendix~\ref{C:AppendixA}, we now have the complete
geometrical description of the system we are considering, since all
the geometrical tensors can now be computed in terms of the metric.

\subsection{The connection coefficients}
As given in the Appendix~\ref{C:AppendixA}, in general relativity the
metric is compatible with the Levi--Civita connection, which means
that they can all be computed once the metric is specified.  From the
form of the metric~\eqref{pr.eq:SymMetric}, a lengthy but
straight-forward calculation yields the following non-zero connection
coefficients,
\begin{subequations}
  \begin{equation}
  \Gamma^{r}_{tt} = \f{1}{2}\e^{\nu-\lambda} \deriv{\nu}{r}, \qquad 
  \Gamma^{t}_{tr} = \f{1}{2}\deriv{\nu}{r}, \qquad
  \Gamma^{r}_{rr} = \f{1}{2}\deriv{\lambda}{r}, \qquad
  \Gamma^{r}_{\theta \theta} = -r\e^{-\lambda}.
  \end{equation}
  \begin{equation}
  \Gamma^{\theta}_{\theta \varphi} = \Gamma^{\varphi}_{r \varphi} = \f{1}{r}, \qquad
  \Gamma^{r}_{\varphi \varphi} = -r \e^{-\lambda} \sin^{2}\theta, \qquad 
  \Gamma^{\theta}_{\varphi \varphi} = -\sin \theta \cos \theta, \qquad
  \Gamma^{\varphi}_{\theta \varphi} = \cot \theta.
  \end{equation}
\end{subequations}
In the above instead of using numbers for the indices, we have used
the names of the four coordinates we used in
metric~\eqref{pr.eq:SymMetric} following the following convention:
\(0 \equiv t, 1 \equiv r, 2 \equiv \theta,\) and \(3 \equiv \varphi.\)

\subsection{The Einstein tensor}
The culmination of the geometrical calculations yields the Einstein
Tensor.  The whole procedure usually involves the computation of the
Riemann tensor, contracting it into the Ricci tensor and scalars, and
then the computation of the Einstein tensor.  We will not show all
these steps, and instead refer the reader to either the numerous tomes
that contain all this information~\cite{Tol66, Iva02, Inv92}, or to
analytical packages like 'GRTensor'~\cite{Grtensor} on
Maple\texttrademark~or 'ctensor' on
Maxima\texttrademark~\cite{maxima}, which allow such calculations.
The Einstein tensor \(G^{a}{}_{b}\) for the metric~\eqref{pr.eq:SymMetric}
is given by
\begin{subequations}
  \begin{equation}
    \label{pr.eq:Ein1}
    G^{t}{}_{t} = \f{\e^{-\lambda}}{r^{2}} \left(-1 + \e^{\lambda} + r\deriv{\lambda}{r} \right) ,
  \end{equation}
  \begin{equation}
    G^{r}{}_{r} = \f{\e^{-\lambda}}{r^{2}} \left(-1 - \e^{\lambda} + r\deriv{\nu}{r} \right),
  \end{equation}
  \begin{equation}
    G^{\theta}{}_{\theta} = \f{\e^{-\lambda}}{4r} \left[2\deriv{\nu}{r} - 2\deriv{\lambda}{r} 
      -r \left(\deriv{\nu}{r}\right)\left(\deriv{\lambda}{r}\right)  + r \left(\deriv{\nu}{r}\right)^{2}
      + 2r\sderiv{\nu}{r} \right],
  \end{equation}
  \begin{equation}
        G^{\varphi}{}_{\varphi} = \f{\e^{-\lambda}}{4r} \left[2\deriv{\nu}{r} - 2\deriv{\lambda}{r} 
      -r \left(\deriv{\nu}{r}\right)\left(\deriv{\lambda}{r}\right)  + r \left(\deriv{\nu}{r}\right)^{2}
      + 2r\sderiv{\nu}{r} \right],
  \end{equation}
\label{pr.eq:Ein}
\end{subequations}

This concludes the geometrical considerations of this Chapter.  The
next section looks instead at the other side of the EFE, concerned
with the source terms of gravitation. Thus we tackle what the models
are made up of: matter and the electromagnetic field.

\section{Source terms}
The other side of the Einstein equations involves sources of curvature.
These sources can be matter, or fields, and we look at each of these
in the next Subsections.
\subsection{Matter}
\label{pr.sec:Matter}
In GR matter is expressed through the stress-energy tensor.  We give a
lengthy and complete derivation from first principles starting from a
Newtonian picture of the form of the stress-energy tensor we
use in this thesis in Appendix~\ref{C:AppendixA}.  The salient points
of this derivation is that in our symmetry case from the previous
Section~\ref{pr.ssec:Sym} the stress-energy tensor for the matter we
are considering can be reduced into the form
\begin{equation*}
  T^{i}{}_{j} =
  \begin{pmatrix}[c]
    \rho    & 0  & 0  & 0  \\
    0     & -p_{r}  & 0  & 0  \\
    0     & 0  & -\ppen & 0  \\
    0 & 0 & 0 & -\ppen
  \end{pmatrix},
\end{equation*}
or in the more compact form~\cite{Mus14, Let80}
\[
  T^{ab} = (\rho + \ppen)u^{a}u^{b} - \ppen g^{ab} +
  ( p_{r} - \ppen)n^{a}n^{b} ,
\]
where \(u^{a}\) is the four-velocity normalized so that
\(u_{a}u^{a} = 1,\) and \(n^{a}\) is a space-like unit vector in the
radial direction, with \(n_{a}n^{a} = -1,\) these being chosen so that
\(n_{a} u^{a} = 0.\) The quantities \(\rho, p_{r},\) and \(\ppen\) are
the energy density, radial pressure, and angular pressure
respectively, and the latter only exists in the case where we consider
anisotropic pressures.  When considering the case with isotropy only,
we have \(\ppen = p_{r},\) with subsequent simplifications of the
above expressions.

The reasoning behind anisotropic pressures will be given in
Section~\ref{ssec:Pr.Anis} and has to do with the use of multiple
species to model the matter component of the star.  An interesting
consequence of spherical symmetry not usually considered is that the
matter collineation induced by the Killing fields impose that the two
pressures \(\ppen\) and \(p_{r}\) be equal at the centre of the
coordinate system, where \(r=0.\) This criterion needs not be
satisfied if the model is not spherically symmetric, but in this work
all our models are strictly spherically symmetric, and we make sure
that \(p_{r}(r=0) = \ppen(r=0).\)

Next we look at the electromagnetic field, and how we include it in
our models.

\subsection{Electromagnetic fields}
\label{pr.sec:EM}
General relativity like special relativity was brought about by
Einstein thinking about the constancy of the speed of light in
different frames.  Light even during his time was modelled as an
electromagnetic wave, underpinned by Maxwell's theory.  Since the
origin of general relativity is so closely related to Maxwell's
equations, it should come as no surprise that these two theories are
completely compatible, and the inclusion of electric charge in
Einstein's theory is not complicated.  In the static case we are
considering we define \(F_{ab}\) as the electromagnetic field strength
tensor (Faraday tensor) introduced in Appendix~\ref{C:AppendixA}, through
\[F_{ab} = \partial_{[a}A_{b]} = \partial_{a} A_{b}
  - \partial_{b}A_{a},\] where \(A_{a}\) is the electromagnetic
4-potential.  In the static case we shall be considering in this
thesis, we impose a gauga where \(A_{a} = (A_{0},A_{\mu})\) with
\(A_{\mu} = 0.\) This is because in the static case magnetic fields do
not exist, and hence the magnetic vector potential vanishes.  The
\(A_{0}\) component encodes the electric field, and in the case of the
charged sphere for example, we get \(A_{0} = -q/r,\) the same as the
classical result.

The energy--momentum associated with the electromagnetic field is then
given through
\(T^{i}{}_{j} = \f{1}{4\pi}\left( F^{i}{}_{c}F^{c}_{j}{} - \f{1}{4}
  g^{i}_{j} F^{cd}F_{cd}\right). \) For the assumed
electric field of \(A = (-q/r, 0, 0, 0),\) this yields
  \begin{equation*}
    \label{pr.eq:TEM}
T^{i}{}_{j}  =
  \begin{pmatrix}[c]
    \f{q^{2}}{\kappa r^{4}}   & 0  & 0  & 0  \\
    0     & \f{q^{2}}{\kappa r^{4}}  & 0  & 0  \\
    0     & 0  & -\f{q^{2}}{\kappa r^{4}} & 0  \\
    0     & 0  & 0  &  -\f{q^{2}}{\kappa r^{4}} 
  \end{pmatrix},
  \end{equation*}
  where \(\kappa = 8\pi\) is the gravitational coupling constant.

  Minimal coupling of matter and fields as we mentioned in
  Appendix~\ref{C:AppendixA} then results in the full
  matter--electromagnetic energy momentum tensor of the form
\begin{equation}
    \label{pr.eq:Ttotal}
T^{i}{}_{j}  =   \begin{pmatrix}[c]
    \rho + \f{q^{2}}{\kappa r^{4}}   & 0  & 0  & 0  \\
    0     & -p_{r}+ \f{q^{2}}{\kappa r^{4}}  & 0  & 0  \\
    0     & 0  & -\ppen- \f{q^{2}}{\kappa r^{4}} & 0  \\
    0     & 0  & 0  & -\ppen - \f{q^{2}}{\kappa r^{4}} 
  \end{pmatrix}.
\end{equation}
and this tensor encodes all the source terms that ``generate''
gravitation.  The interesting conclusion that can be drawn from this
discussion is that there is a simple way to include electromagnetic
fields and matter in the same framework, and once specified, the
solution to the EFE with the source term will yield consistent
solutions incorporating both matter and charge.

\section{Exterior solutions}\label{pr.sec:ExtSoln}
Einstein's field equations are expressed in terms of the Einstein's
tensor \(G_{ab}\) we just defined.  In the static case this tensor is
similar to the Laplacian of the scalar potential \(\nabla^{2} \phi\)
of electrodynamics in that it also consists of at most second
derivatives of the the gravitational ``potential'' \(g_{ab},\) and
that it can be solved for both the vacuum case (Laplace's equation),
and the case where source terms exist (Poisson's equation).  With this
general picture in mind, we discuss the two vacuum solutions we will
pursue as the external solutions we need to match in the different
cases we will find solutions for.  These solutions will assume
\(T^{ab} = 0,\) or \(T^{ab} = T^{ab}_{\text{electromagnetic}}\) in
Subsections~\ref{pr.ssec:SchExt} and~\ref{pr.ssec:ReiNor} respectively

\subsection{The Schwarzschild exterior solution}\label{pr.ssec:SchExt}
This solution was found by \citeauthor{Sch16} in \citeyear{Sch16}.
His derivation of the equation is complicated because he used
Cartesian coordinates instead of the more natural spherical
coordinates for this spherically symmetric solution.  Here we will
just state the main lines of another derivation popularized by
Droste~\cite{Dro17}. 

The assumptions that lead to the Schwarzschild exterior solution are
\begin{enumerate}
\item A spherically symmetric and static space-time.  As we showed in
  the Section~\ref{pr.ssec:Sym}, this implies that the metric is
  reduced to equation~\eqref{pr.eq:SymMetric}.
\item This is a vacuum solution, so there is no source terms,
  therefore Einstein's tensor is annulled, the set of equation to be
  solved thus become \[G_{ab} = 0.\]
\end{enumerate}
As a result of these assumptions, the complete set of linear ODEs to
be solved become the set~\eqref{pr.eq:Ein} equated to zero.
Algebraic manipulation of this set then results into the simple linear
equation for \(\e^{-\lambda}\):
\[ -\e^{-\lambda} \left(1 - r \deriv{\lambda}{r} \right) + 1 = 0 \implies
  \deriv{}{r}\left(\e^{-\lambda}\right) + \f{\e^{-\lambda}}{r} =
  \f{1}{r}.\] Solving the latter differential equation then results in 
\[ \e^{-\lambda} = 1 + \f{A}{r}, \quad\text{and therefore}\quad
  \e^{\nu} = B\left( 1 + \f{A}{r} \right).\] with \(A\) and \(B\)
arbitrary integration constants.  A simple coordinate rescaling can be
used to set the constant \(B = 1.\) Boundary conditions requiring that
this solution be compatible with Newtonian gravity then forces
\(A=-2M,\) where \(M\) is the mass of the object perceived at some
distance.  Then the interpretation of this solution is the external
gravitational field of some mass \(M\) at the symmetry centre of the
solution.  The complete Schwarzschild exterior metric hence becomes
\begin{equation}
  \label{pr.eq:SchExt}
  \d s^{2} = \left( 1 - \f{2M}{r} \right) \d t^{2} - \left( 1 - \f{2M}{r} \right)^{-1} \d r^{2} - r^{2}\left(\d \theta^{2} + \sin^{2}\theta \d \varphi^{2} \right).
\end{equation}

This exterior is usually used to model non-rotating and uncharged
black-holes since the interior can be left unspecified, and is
modelled through the total mass only.  This is one of the consequences
of Birkhoff's theorem.

The Schwarzschild exterior solution will be the solution we want to
match the interior solutions we find in this thesis when we have
uncharged matter.  The physical reason behind this: to first order the
metric~\eqref{pr.eq:SchExt} gives classical Newtonian gravity, and to
second order predicts the perihelion shift of mercury~\cite{Ein16}.
Any object, from a black-hole to the sun, and going through the
``middle'' case of a compact object like a neutron star exerts gravity
in a similar way.  Therefore the compact objects we model should also
behave similarly.

Mathematically the uniqueness of this solution is ensured by
Birkhoff's theorem~\cite{Bir23, Cho08} which we now state without
proof.
\begin{myTheorem}
  Any spherically symmetric solution of the vacuum field equation
  \(G_{ab} = 0,\) must be static and asymptotically flat.  This means
  that the exterior solution must be given by the Schwarzschild
  metric.
  \label{Pr.th.Birkhoff}
\end{myTheorem}
This theorem is physically unexpected since in Newtonian theory,
staticity is unrelated to spherical symmetry.  Thus the Schwarzschild
solution is the only possible solution for a spherically symmetric,
asymptotically flat vacuum space-time.

\subsection{The Reissner--Nordstr\"om solution}\label{pr.ssec:ReiNor}
When the EFE are combined with the Maxwell's equations which we just
investigated in~\ref{pr.sec:EM}, the vacuum metric
solution~\eqref{pr.eq:SchExt} can be generalised to an electro-vacuum
solution: The Reissner--Nordstr\"om solution which includes the
electrical charge.  This solution was found by~\citeauthor{Rei16}
and~\citeauthor{Nor18} after whom it is named~\cite{Rei16,Nor18}.

In this solution the electromagnetic 4-potential is given by
\[A_{\mu} = 0, \qquad A_{0} = -\f{Q}{r},\] since the field is static
(no magnetic potential), and assumed to come from a sphere yielding
the \(1/r\) electric potential.  This is in line with the spherical
symmetry of the solution.

A similar solution method to the one we showed for the Schwarzschild
solution can be used to find the Reissner--Nordstr\"om solution.
However, in this case we will have a source term coming from the
charge that we assume the central body to process.  As a result, the
EFE cannot be equated to zero, and have to be equated to the
energy-momentum tensor components
\( T^{a}{}_{b} = \diag{(-Q^{2}/r^{4},-Q^{2}/r^{4}, Q^{2}/r^{4},
  Q^{2}/r^{4})} \).  The additional terms do not change the solution
procedure drastically and the final Reissner--Nordstr\"om metric then
becomes~\cite{Cho08}
\begin{equation}
  \label{pr.eq:ReiNor}
    \d s^{2} = \left( 1 - \f{2M}{r} + \f{Q^{2}}{2r^{2}}\right) \d t^{2} - \left( 1 - \f{2M}{r} +\f{Q^{2}}{2r^{2}} \right)^{-1} \d r^{2} - r^{2}\left(\d \theta^{2} + \sin^{2}\theta \d \varphi^{2} \right).
\end{equation}
This metric is smooth and Lorentzian with \(t\) a time variable as long as 
\[ \f{2M}{r} - \f{Q^{2}}{2r^{2}} < 1, \] to keep the metric from
changing signature.  For large values of \(r\) the term
\(Q^{2}/r^{2} \) decreases more rapidly than \(2M/r,\) and seems to
suggest that charge effects are more difficult to observe directly in
astrophysical objects.  However as we shall see in our interior
solutions, electric charge also affects \(M\) in non-trivial ways.

This is the metric to which we shall match when we are considering
charged interior solution to the EFE.  The matching of the mass and
charge at the boundary then ensures the consistency of the metric
throughout the space-time.
 
\section{Interior Solutions}\label{pr.sec.IntSol}
Interior solutions refer to equations which solve the EFE while
connecting to some exterior (usually cosmological) solution.  Together
the two provide a complete picture of the local region inside and
outside some matter that is gravitating.  When we look for solutions,
this ``interior'' quality is modelled by two different factors
\begin{enumerate}
  \item The field equations will contain matter, usually in the form of
    fluids, and the exterior solution will either be more fluid, or empty space.
  \item The junction condition must match the interior structure to
    some exterior ones without any singularities occurring at the
    junction.  This junction is usually modelled as a hypersurface
    (called the matching hypersurface), and on this hypersurface, the
    interior and exterior metric must match, and be at least of class
    \(C^{1}.\)
\end{enumerate}
These consistency conditions are important and without their
satisfaction, the models built cannot be deemed to be mathematically
consistent.

We now present a brief history of the process of finding interior
solutions, and how our work fits into the overall solution landscape
already present in the field.

\subsection{Perfect fluid matter}
The first interior solution was found in 1916 by~\citeauthor{Sch16},
and this solution was matched to an exterior solution found by the
same person, now called the Schwarzschild exterior solution.  This
solution modelled the matter content through an incompressible fluid
of constant density \(\rho,\) and for this reason is nowadays deemed
unphysical~\cite{DelLak98, FinSke98}.  The reasoning behind the
unphysicality is that the speed of pressure waves (sound waves) in the
Schwarzschild interior solution is infinite, in contradiction with
relativity.  Following its discovery, a number of studies into its
properties were carried out.  Of note is the re-expression of this
solution's metric in isotropic coordinates~\cite{Wym46}, and the
extension, albeit in a perturbative manner to finite but constant
speeds of sounds in~\cite{BucLan68}.  The latter reference is also
interesting in that it is the first time that an equation of state
(EOS) of the form \(p = \rho - \rho_{b}\) is considered directly.
This form of EOS is important since now the speed of pressure waves is
exactly equal to the speed of light since
\(v_{s} = \d p / d \rho = 1,\) instead of infinite as in the case of
Schwarzschild interior.

Later it was shown that the Schwarzschild interior metric is the
unique spherically symmetric and static metric that was also
conformally flat~\cite{GurGur07} (Admitting the conformal Killing
vector in addition to the one we imposed on our metric.)  The
importance of the Schwarzschild interior solution cannot be denied: It
has been used to model, and limit physical characteristics of compact
stars since its discovery, and the well-known Buchdahl limit of
\(M/R < 4/9,\) is a result that depend crucially on its existence and
properties~\cite{Buc59}.  Recently we showed~\cite{RagHob16a} that one
of the solution this thesis looks into closely: the Tolman~VII
solution, is a possible extension of the Schwarzschild interior
metric.

A different set of interior solutions to the EFE was found in 1939
by~\citeauthor{Tol39}.  All these solutions were matched to the
Schwarzschild exterior solution~\cite{Tol39} and while all the matter
quantities were not computed, some had interesting enough properties
that they were either rediscovered~\cite{DurGeh71,DurRaw79} or used
under different names~\cite{Meh66}.  Some of this set were well known
solutions: The Einstein universe, the Schwarzschild interior we just
talked about, and the Schwarzschild--de Sitter
solution~\cite{RagHob15}.  Others have been generalized in various
ways: an early attempt by~\cite{Wym49}, followed by various others.
An incomplete but comprehensive list is given
in~\cite{DelLak98,FinSke98}.  

From then on there have been many other sets and classes of interior
solutions.  Most of these have numerous problems and are not
interpretable as physical objects.  Usual issues include the
divergence and asymptotic behaviour of the matter variables: the
density being infinite somewhere in the model being a common
occurrence.  In the same year, \citeyear{OppVol39},
\citeauthor{OppVol39} derived an equation that has since been named
the Tolman--Oppenheimer--Volkoff equation of general relativistic
hydrodynamic stability~\cite{OppVol39}.  

This equation revolutionized the task of finding EOSs for compact
object.  This is because for the first time, it allowed a method other
than brute mathematical intuition to be used to model these compact
objects.  If one had some idea of an equation of state, then one could
impose that equation of state onto the geometry and get observable
values like masses and radii of the objects.  This line of approach
was and remains popular in many places, particularly with those who
model neutron star structure.  The hope there is that neutron stars
will give us a glimpse into what neutron-rich matter behaves
like~\cite{LatPra01}.  However certain recent results seem to indicate
that detailed quantum mechanical descriptions are not very important,
since the masses and radii of neutron stars seem to be insensitive to
the differences between most of the nuclear EOS that are
investigated~\cite{YagStiPapYun14}.

\subsection{Anisotropic matter}\label{ssec:Pr.Anis}
All the interior solutions we have mentioned so far are uncharged and
admit only one radial pressure.  The interest in non-radial pressures
began in the 1970's, when anisotropic pressures with
\(\ppen \neq p_{r}\) started to be considered~\cite{BowLia74}.  Later
physical interpretations of these unequal pressures started to
appear~\cite{Let80, Bay82}, and the interpretation that the
anisotropic pressure was just an expression for the existence of
multiple perfect fluids minimally coupled to each other in the
interior gained acceptance.  Many solutions~\cite{CosHer81,Flo74}
containing anisotropic pressures were found and used to model compact
objects, and the analysis led to a number of discoveries.  Of note is
the discovery~\cite{Bon64, BowLia74,Pon88} that by having anisotropic
pressures, one was no longer subject to the Buchdahl limit of
\(M/R < 4/9.\)

At the same time, the generalized TOV equation which includes the
anisotropic pressure was derived, and the realisation that this
equation looked very much like the classical Newtonian hydrostatic
equilibrium equation~\cite{HerBar13} was reached.  This led to a
number of investigations into the structure and stability of Newtonian
stars~\cite{DevGle00,DevGle03} to be done, since the similarity to the
relativistic case meant that result for the Newtonian case could be
extended to the relativistic case easily.  A classic
review~\cite{HerSan97} by~\citeauthor{HerSan97} during the same time
brought anisotropic solutions to the forefront, and interest in such
solutions particularly for gravitational wave generation, and Love
numbers computation became a popular study.

In this thesis, the relationship between the two pressures defines a
measure of anisotropy.  We define \(p_{r}(r) - \ppen(r) = \Delta(r),\)
generally in the beginning, and then continue on to define
\(\Delta = \beta r^{2},\) in certain solutions.  As a result, by
modifying the value of \(\beta,\) a constant parameter, we change the
anisotropic pressure indirectly.  The physical reasoning being the
expression of \(\Delta\) will be explained in the relevant section
which investigate that particular case.

\subsection{Charged matter}
From the charged EFE it is clear that if we have non-zero source terms
(from an electromagnetic energy-momentum tensor,) the exterior
solution to be matched becomes the Reissner--Nordstr\"om exterior
solution. This affects the interior solution to be found as the
electromagnetic repulsion of the charges will contribute to the
overall energy, and thus mass of the object.  The Reissner-Nordstr\"om
solution provides a relationship between the mass and the charge for
the interior solution to be consistent, and either \(M\) or \(Q\)
could be set to zero.  

However this is not always possible in the interior, as was shown by
Bonnor~\cite{Bon60}, since positing zero \(M\) necessitates a negative
mass-energy density in the interior, an unphysical result.  Having an
interior solution as we will shortly demonstrate allows for obtaining
more relationships between the charge and mass, and we shall do so in
the course of this thesis.  Following the discovery of several charged
solutions in the 1950's by~\citeauthor{Pap47}, interest to find even
more of these solutions grew, and many more solutions to match the
exterior Reissner-Nordstr\"om solution have since been found.

Bonnor~\cite{Bon65} found another regular charged solution in 1965.
In this solution the assumption that the mass density be equal to the
charge density was utilised.  Taking inspiration on this, in one of
our solution, we instead assume that the anisotropy measure matches
the charge density.  This solution showed that is was indeed possible
to have a consistent solution in which repelling charges could balance
the attracting gravitational field, in general relativity. 

Many other solutions having singularities were also found during the
same time period however.  Some tried to remove, and/or transform away
these singularities, meeting with meagre success.  Soon after however,
a host of solutions consistently matching to the Reissner--Nordstr\"om
exterior were found~\cite{KroBar75, Flo77, SinYad78, CooCru79}.  Some
of these solutions also tried imposing the strong energy conditions
\(\rho + 3p > 0,\) to various conclusions. Some were not able to
impose it, as some solutions did not allow for this condition to hold
for any choice of \(\rho.\) Of note is the solution
by~\citeauthor{KylMar67} which one of our solutions reduces to when
the anisotropic factor \(\beta = 0.\) \citeauthor{Iva02} summarizes
many of these results and how to obtain them from first principles in
an extensive review in Ref.~\cite{Iva02}, and we use the latter for
many of our derivations and reasoning.

\section{The solution landscape}
\begin{figure}[p]
  \includegraphics[angle=90,
  width=1\textwidth]{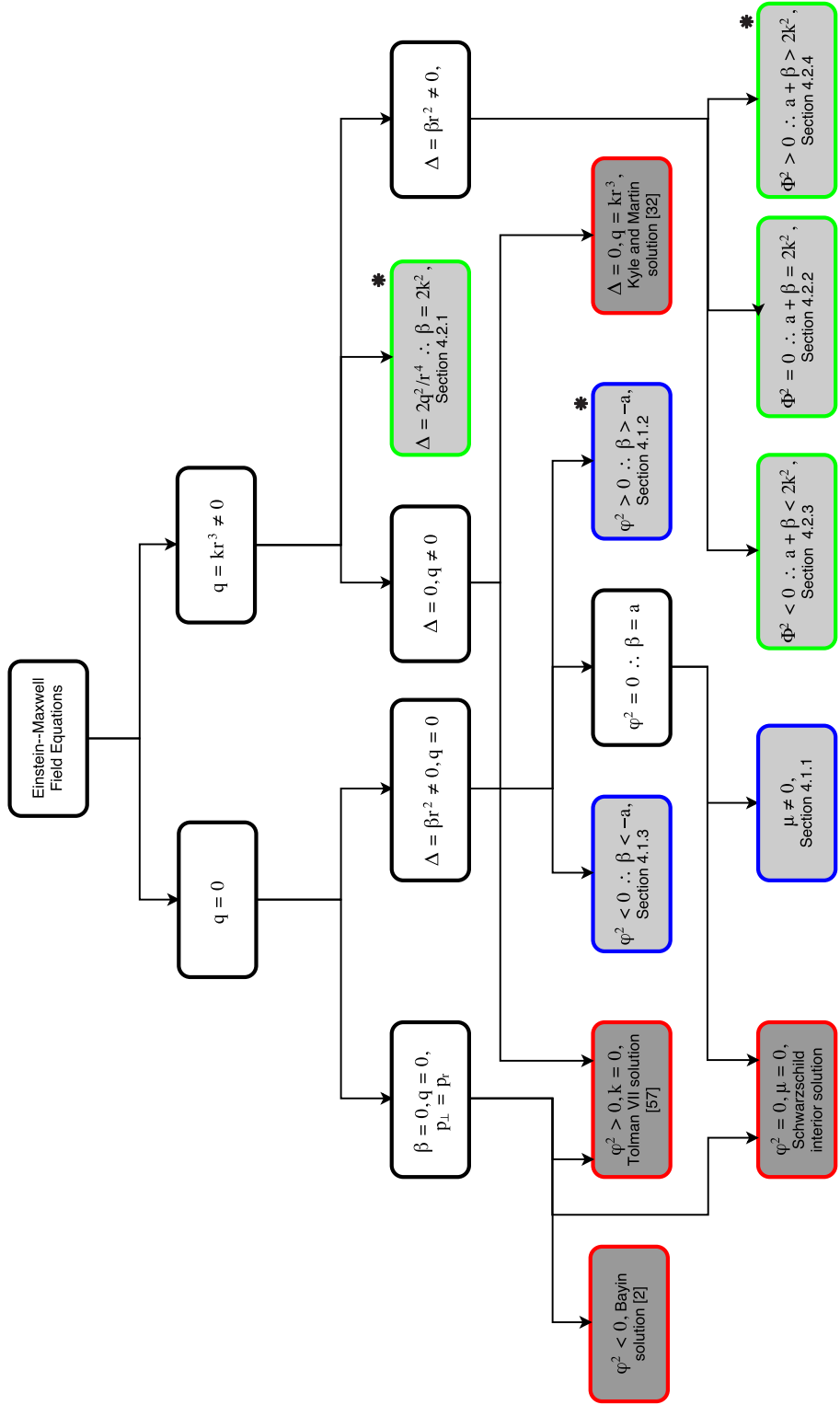}
  \caption[The decision flow leading to solutions]{The solution
    landscape explored in this thesis. Lightly shaded boxes are the
    new solutions described in this work, and darker ones are the
    older known solutions.  In the online coloured version,
    red-bordered boxes are solutions with isotropic pressure,
    blue-bordered ones are uncharged solutions with anisotropic
    pressures only and green-bordered ones are charged with
    anisotropic pressures.}
\label{pr.fig:Flow}
\end{figure}

Having introduced all the main ingredients that are used in this
thesis, we now provide a logical flow for how we approach the finding
of solutions that might be physical.  In Figure~\ref{pr.fig:Flow}
we look at the decision flow in terms of the parameters \(\beta\)
encoding the anisotropy, and \(k\) encoding the electric charge to
generate the new solutions we found.  This diagram summarises the
solution landscape around our solutions as well, showing for example
how through setting certain parameters to zero, we can also generate
already known solutions.

\section{Sturm-Liouville systems}
A detour into the mathematical theory of Sturm--Liouville systems is
required for our section on the stability of our models.  Indeed, the
overall stability of interior solutions, while clear on general
physical intuitions is a lot more complicated to prove mathematically.
At fault is the non-linear structure of the EFE, which causes
perturbations in any matter fields to change every single equation, up
to the metric functions, through a complicated propagation process.
In this thesis we assess the linear stability of our models, and do
not proceed further.  Studies have shown that a range of different
non-linear instabilities are also possible, but their analysis would
require a lot more work.

The TOV equation and the EFE can be reduced in certain cases to a
Sturm--Liouville form. The general form the linear stability equation
for the EFE reduces to:
\begin{equation}
\label{pr.eq:gSL}
  \left\{ \deriv{}{r} \left[ P(r) \deriv{f}{r}\right] + \left( Q(r)
    + \sigma^{2} W(r) \right) f + R \right\} = 0.
\end{equation}
With this form, the most important result of the theory, the ordered
spectrum theorem cannot be used directly, unless the function \(R\)
vanishes.  We show how this is achieved in our derivation and we then
use Theorem~\ref{A.th:S-L} from Appendix~\ref{C:AppendixA} to conclude
that if the fundamental mode of vibration, corresponding to the first
eigenvalue \(\sigma^{2}_{1}\) is positive, then so will all the higher
modes, so that no modes will cause the star to have its radius
perturbed away from equilibrium, proving the radial stability of the
star.  We provide this result in a table that investigates different
parameter values to see which results in stable, and which in unstable
stars.

\section{Astronomical observation of compact objects}
Astronomical observations of compact objects (neutron stars, pulsars,
black holes) have been possible for quite some time
now~\cite{OzeFre16}.  During the last decade both X-ray and
\(\gamma-\)ray telescopes have provided large datasets about precise
timings of pulsars, and from these a number of measurements, most of
them not independent of the underlying model behind these objects, have
been possible.

A recent review by~\citeauthor{OzeFre16} claims that precise masses of
approximately 35 neutron stars are known, and the radii of about 10 of
these are known to varying degrees of precision.  These measurements
already place constraints on the possible EOS of neutron matter, since
the very heavy and small (having large\( M/R\) values) stars eliminate
many of the softer EOS that predict much lower maximum \(M/R\) values
than observed.

Since that initial discovery and observation, a number of theoretical
modelling, usually based on the TOV equation together with some type
of effective field theory or the newer quantum chromodynamics (QCD)
calculations have also been used to understand the structure of these
stars.  However the theory aspect of the problem is fraught with
difficulties.  For example, the composition of matter at high
densities, and in particular the density limit at which matter has to
be modelled as quarks instead of nucleons is not known.  Also unknown
are the relative effects of the presence of boson condensates and
strangeness in such high density matter.  Work in particle
accelerators working with neutron rich nuclei hope to find some of
these parameters, however even those nuclei do not come close to what
a neutron star's composition is thought to be, and their validity to
neutron star matter is contentious.

However this problem is approached, once certain approximations about
the different components of the neutron matter has been made, an EOS
can be deduced, and from utilising this EOS in the TOV equation, a
mass to radius (\(M-R\)) curve corresponding to the EOS can be drawn.
This curve can then be matched to observations of neutron stars, and
appropriate conclusions drawn.  This can be seen as a test of both GR
in the strong field regime (through the use of the TOV equation,) and
as a test of the EOS of cold ultra-dense matter.

\subsection{Mass measurements}
Most of the neutron star mass measurements come from pulsars in binary
system.  These constitute about 10\% of all known pulsars for a total
of 250.  Of these, most are termed as ``recycled'' pulsars because
throughout their lives, they have accrued mass from their companions.
This mass transfer usually increases the mass of the pulsar, but the
clearest signal of this process happening is the spin-up of the the
pulsar~\cite{OzeFre16}.  Accompanying this spin-up is also the
counter-intuitive reduction in the magnetic field associated with the
pulsar, and indeed the mechanism leading to this reduction is very
poorly understood.  

All the pulsar masses that we have come from binary systems, and
method employed usually involved pulsar timings of some sort.  Through
these accurate timings, much about the orbit of the pulsar can be
inferred, and once the Keplerian orbital parameters are obtained,
measuring the mass of the system becomes easy.

The nature of the companion changes the method of measurement of the
pulsar itself (as opposed to the total mass of the whole binary
system,) and some of the numerous methods used to measure their masses
are:
\begin{enumerate}
\item Straight forward pulsar timings for pulsars in binary pulsar
  systems (the companion is a pulsar too).  This has been used for
  example in~\cite{Lyn04}.
\item Shapiro delay which is the delay in the reception of the radio
  pulses associated with the pulsars on Earth due to the propagation
  of the radio signal in the curved space-time near the companion
  star.  This is akin to the ``lensing'' of radio waves.  This is used
  with systems where the pulsar has a white dwarf companion, for
  example in~\cite{RybTay91}
\item Spectroscopic mass measurements, which involves the studying of
  the Balmer lines of hydrogen produced in the companion's atmosphere.
  For this method to be used, the companion has to be optically
  bright.  This has been used for example in~\cite{Ant13}.
\end{enumerate}
These do not provide the complete extent of the methods, but do give
an idea how complicated this field is, and how dependant the methods
are on the companions.  

\begin{figure}[!h]
  \centering
\includegraphics[width=0.9\textwidth]{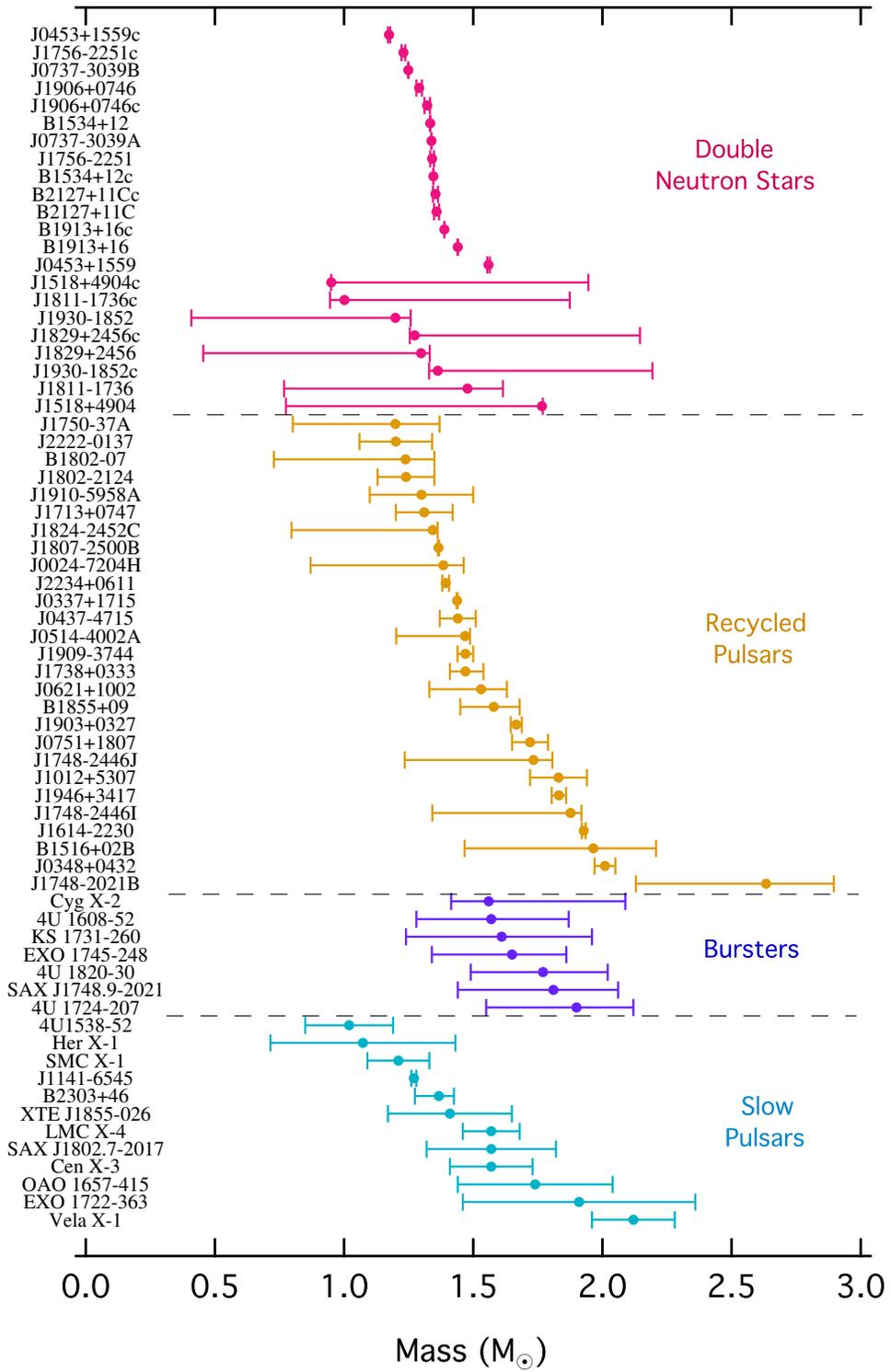}  
  \caption[Pulsar masses]{Figure taken from the review~\cite{OzeFre16} by~\citeauthor{OzeFre16}. The references that give the latest mass measurements are included in the article above, and we will not reproduce it here. }
  \label{pr.fig:PulsMass}
\end{figure}

We provide a list of pulsar masses obtained through various methods in
Figure~\ref{pr.fig:PulsMass}, which is taken from the
review~\cite{OzeFre16}.

Masses of these pulsars are the ``easy'' measurements.  Neutron stars
are incredibly compact, and measuring their radii, which is
typically of the order of a few kilometres is even more challenging.
We look into the methods and observations of these next.

\subsection{Radius measurements}
The field of radii measurements of pulsars has only been active during
the past decade, and the method employed for these measurements is
based on the detection of thermal emission from the surface of the
star either to measure its apparent angular size or to detect the
effects of the neutron-star space-time on this emission to extract the
radius information~\cite{OzeFre16}.  

\begin{figure}[!h]
\subfloat[The mass-radius constraint for qLXMB during quiescence]{\label{pr.fig:PulsRadQ}
  \includegraphics[width=0.50\linewidth]{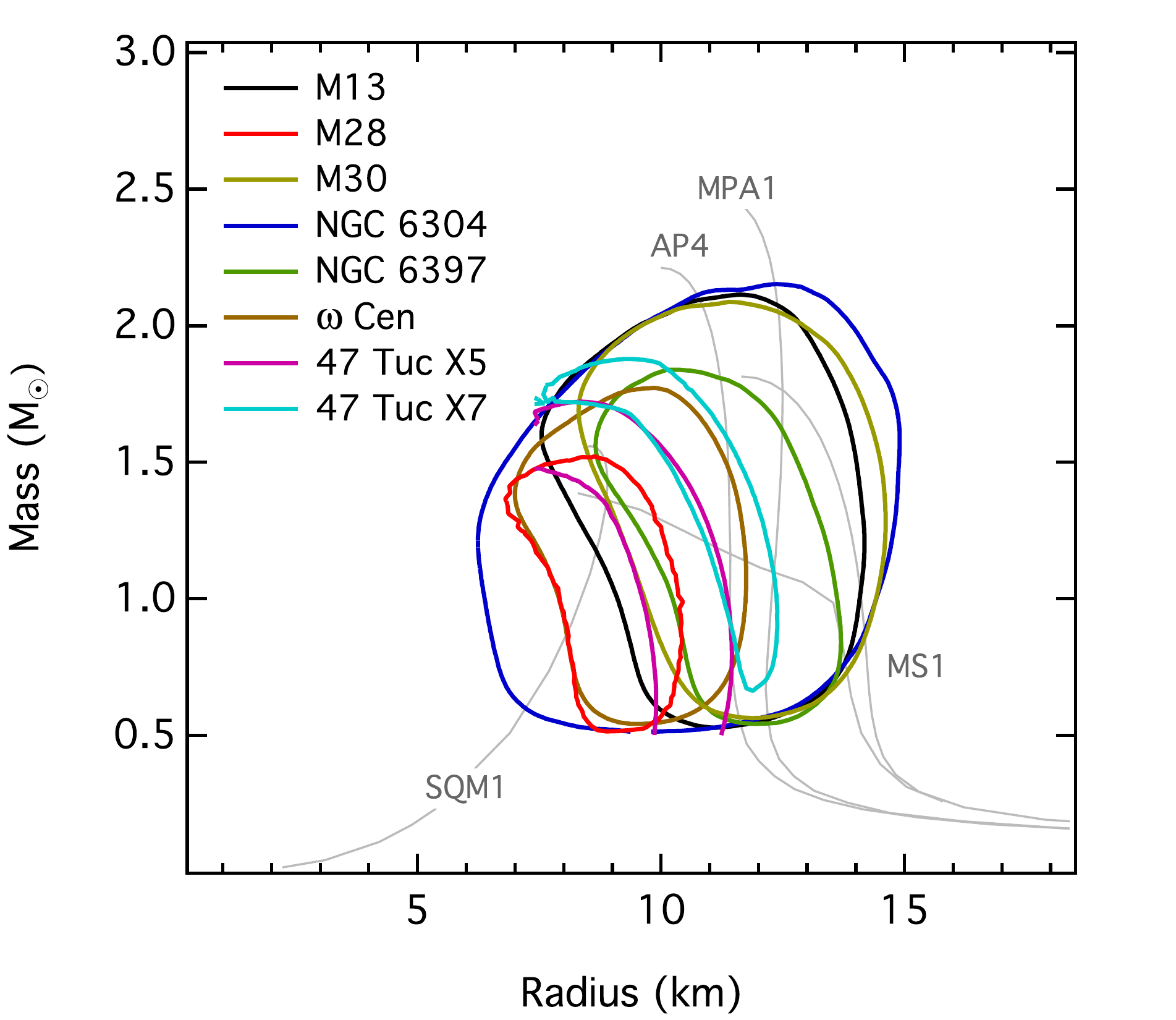}} 
\subfloat[The Mass-Radius Constraint for LXMB during bursts]{\label{pr.fig:PulsRadB}
  \includegraphics[width=0.50\linewidth]{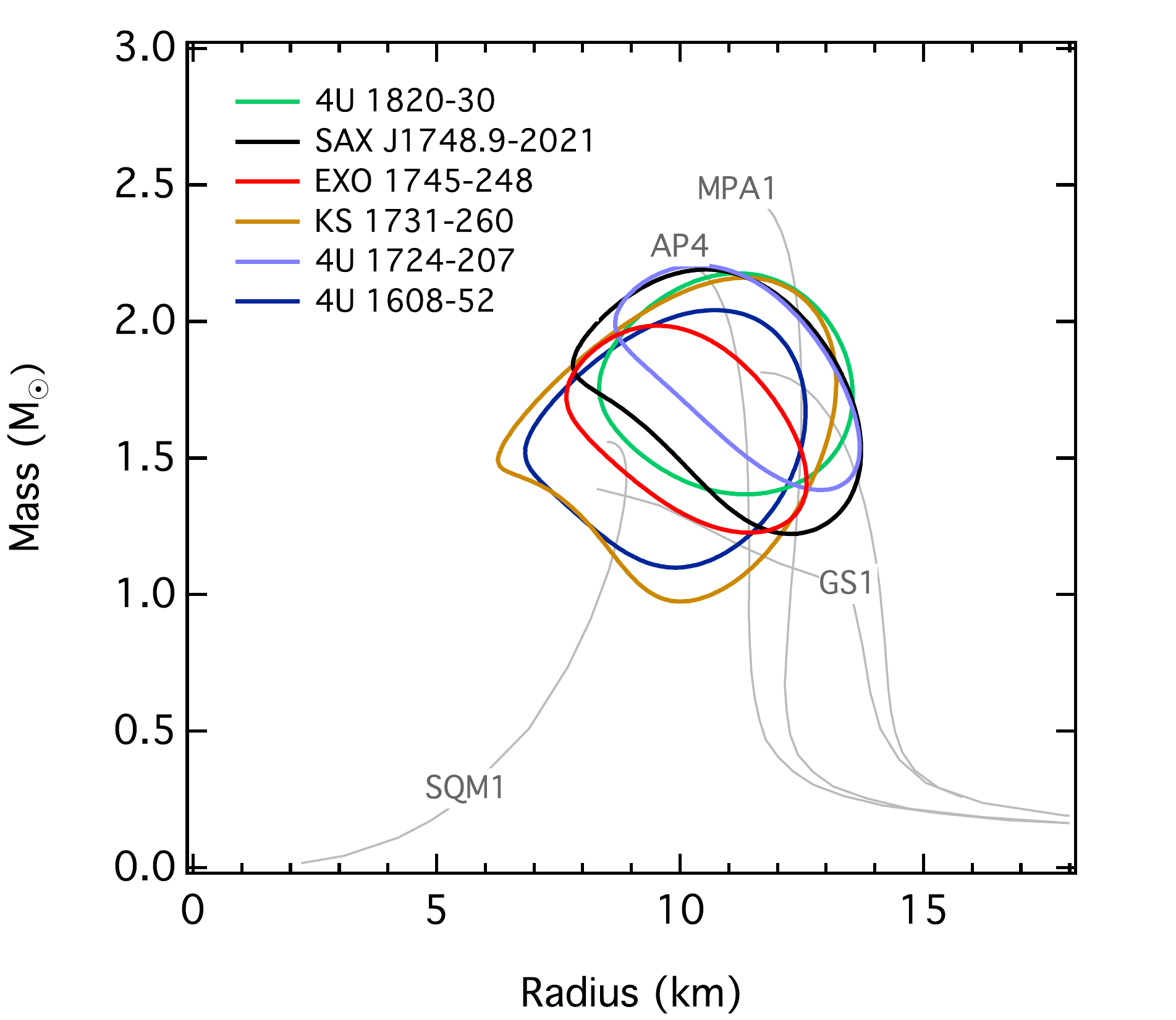}}\\\centering
\subfloat[The Mass-Radius Constraint for millisecond pulsars]{\label{pr.fig:PulsRadM}
  \includegraphics[width=0.55\linewidth]{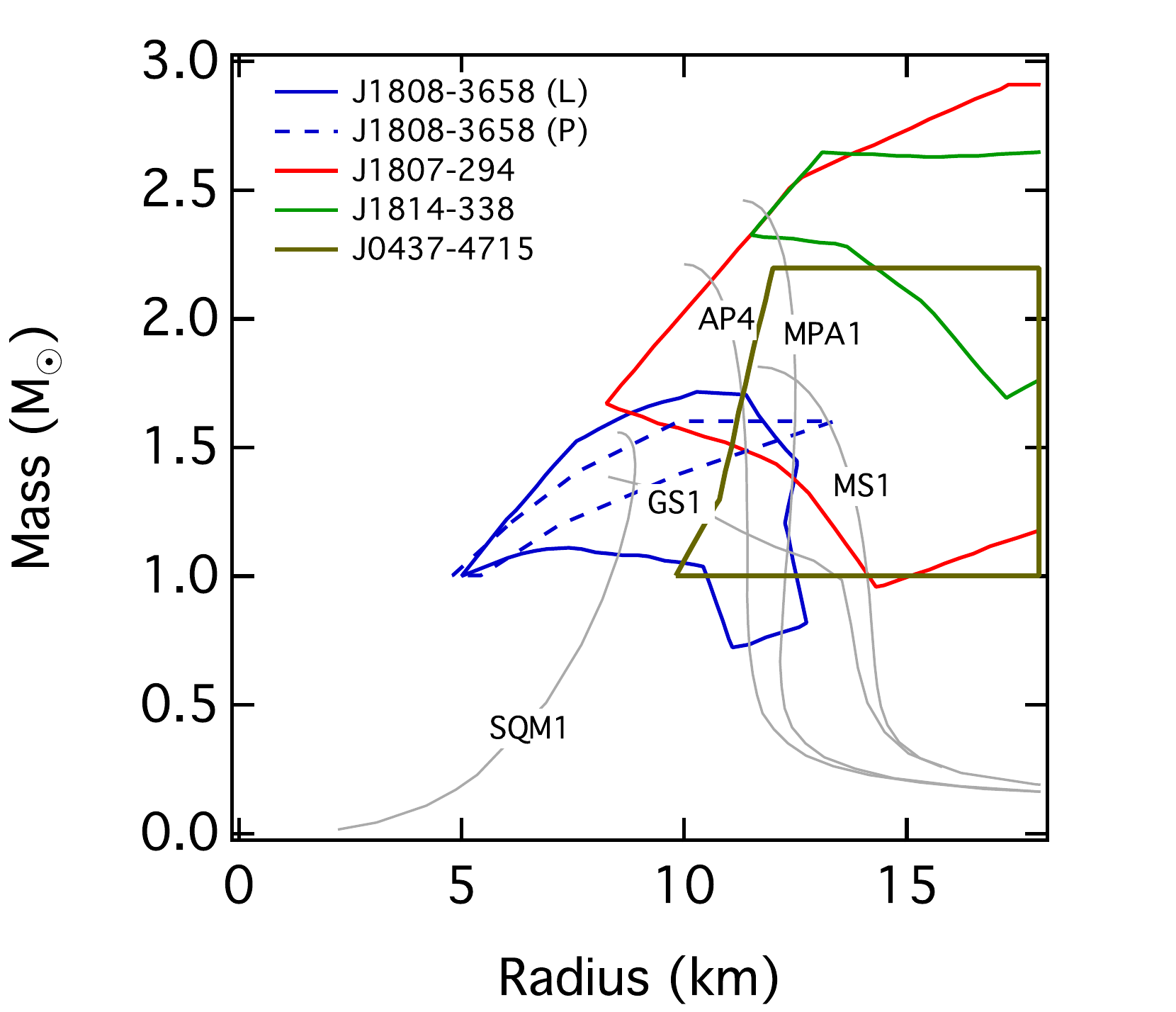}}
\caption[The Mass-Radii constraints from various methods]{The
  mass-radius constraint obtained from the various methods outlined in
  the text. These diagrams were all taken from~\cite{OzeFre16}}
\label{pr.fig:PulsRad}
\end{figure}

The major method that has produced reliable results for the radii
measurements so far is spectroscopic measurements involving the
determination of the angular sizes of the pulsars by measuring the
thermal flux from the pulsar, modelling the spectrum to determine the
effective temperature, and combining this with a distance measurement
to obtain an apparent radii.  The fact that neutron stars
gravitationally lens their own emission make this process
arduous~\cite{PsaOzeDed00}.  Since some pulsars spin very quickly, the
space-time around them can no longer be described by the Schwarzschild
metric (a non-rotating solution), and this introduces other
complications.  The magnetic fields associated with the pulsars stream
the flux from the pulsar, which can then cause a large temperature
difference between different points on the pulsar.  This makes the
inferring of the temperature difficult.  All these complications have
to be either modelled, or sources that exhibit the least of these
chosen for this method to work.  This method was worked reasonably
well in a few quiescent low mass X-ray binaries
(qLXMBs)~\cite{GuiRut14, LatSte14}.  Quiescence refers to LXMBs which
cease to accrete, or accrete at a very low rate. Because of this low
rate, the thermal emission from these stars can observed without too
much interference.

The major source of the radiation coming from LXMBs is a process known
as Type-I X-Ray bursts.  In these, accreted matter to the pulsar
undergoes a helium flash\footnote{A helium flash is a nuclear fusion
  reaction where a large quantity of helium is converted into carbon
  through a triple alpha process.  This reaction is associated with a
  large release of energy leading to thermal runaway (positive
  feed-back) and thus a rise in temperature} that consumes the
accreted matter over the whole surface of the star.  The luminosity of
the star rises rapidly \((\sim 1s)\) when this happens, but the energy
is quickly radiated away \((\sim 50s).\) In some cases the luminosity
reaches the Eddington luminosity limit where the radiation pressure
matches the gravitational force.  When the Eddington limit is exceeded
and the photosphere gets lifted from the surface of the star, the
luminosity then behaves in a typical way that is well understood, and
allows to infer and constrain the \(M-R\) values of the pulsars.

The second method used for radii measurements rely on the periodic
brightness oscillations that spinning neutron stars undergo.  These
oscillations are due to the temperature anisotropies on the surface of
the star.  The amplitude and spectra of the emissions depend on the
neutron star space-time, and on the temperature profile of the stellar
surface.  Theoretical models of this emerging radiation can be used to
constrain the mass and radii of the star.  These theoretical models
started with non-spinning neutron stars~\cite{PecFtaCoh83}, to which
the Doppler shifts and aberration expected from the spinning were
added~\cite{MilLam98, PouBel06}.  Effects like frame dragging and the
oblateness of the star were also subsequently added~\cite{Mor07,
  Cad07}.  Once the models were obtained and deemed accurate,
constraints on mass and radii could be obtained.

These two main methods yielded constraints on both mass and radii,
which we now show on the \(M-R\) diagrams in
Figure~\ref{pr.fig:PulsRad}.  Statistical analyses of the errors stem
from the numerous assumptions and models used, and for this reason
each measurement is represented as a ``patch'' in the \(M-R\) plot.

This preliminary section introduces all the notions that the remaining
chapters will use directly, without introduction.  The first part of
this work will look at a known solution to the interior EFE given
in~\eqref{pr.eq:EFE}.  Then we find new physical solutions to this EFE
in two more general cases in Chapter~\ref{C:NewSolutions}.  Then the
linear stability of these solutions is investigated in
Chapter~\ref{C:Stability}, and a discussion of the possibility to use
the solutions as models in the future investigated in
Chapter~\ref{C:Analysis}.



\chapter{The Tolman VII solution, an example} \label{C:TolmanVII}
\begin{myabstract}
  We apply the Tolman, and/or Ivanov procedure to the spherically
  symmetric and static case, without anisotropic pressure or electric
  charge to give an overview and the simplest example possible for our
  plan of attack before generalizing to the more complicated cases.
  We also present the type of analysis that we hope will be possible
  for the new solutions.  In so doing, the Tolman~VII solution for a
  static perfect fluid sphere to the Einstein equations is re-examined
  and a closed form class of equations of state (EOS) is deduced for
  the first time.  These EOS allow further analysis to be carried out,
  leading to a viable model for compact stars with arbitrary boundary
  mass density to be obtained.  Explicit application of causality
  conditions places further constraints on the model, and recent
  observations of masses and radii of neutron stars prove to be within
  the predictions of the model.  The adiabatic index predicted is
  $\gamma \geq 2,$ but self-bound crust solutions are not excluded if
  we allow for higher polytropic indices in the crustal regions of the
  star.  The solution is also shown to obey known stability criteria
  often used in modelling such stars.  It is argued that this solution
  provides realistic limits on models of compact stars, maybe even
  independently of the type of EOS, since most of EOS usually
  considered do show a quadratic density falloff to first order, and
  this solution is the unique exact solution that has this property.
\end{myabstract}

\section{\label{t7.sec:Intro}Introduction}
The construction of exact analytic solutions to the Einstein equations
has had a long history, nearly one hundred years to be more precise.
However in spite of the fact that the total number of solutions is
large \cite{KraSteMac80} and growing, only a small subset of those
solutions can be thought of as having any physical relevance.  Most
solutions exhibit mathematical pathologies or violate simple
principles of physics (energy conditions, causality, etc.) and are
therefore not viable descriptions of any observable or potentially
observable phenomena.

Indeed works that review exact solutions and their properties
demonstrate the difficulties associated with constructing solutions
that might be relevant to gravitating systems that actually exits in
our Universe.  Even in the simplest case of exact analytic solutions
for static, spherically symmetric fluid spheres, it has been shown
that less than ten percent of the many known solutions can be
considered as describing a realistic, observable object.  For example
\citeauthor*{DelLak98} using computer algebra methods reviewed over
130 solutions and found that only nine could be classified as
physically relevant~\cite{DelLak98}.  A similar study by Finch and
Skea~\cite{FinSke98} arrived at the same conclusion.  The latter
review also introduced a classification that further reduced the
number of physically relevant solutions to those that had exact
analytic equations of state of the form \(p = p(\rho)\)
where \(p\)
is the fluid pressure and \(\rho\)
is the matter density.  This class of solutions was called
``interesting solutions''.

In 1939 \citeauthor{Tol39} introduced a technique for constructing
solutions to the static, spherically symmetric Einstein equations with
material fluid sources~\cite{Tol39}.  That method led to eight exact
analytic expressions for the metric functions, the matter density and
in some cases the fluid pressure.  Beginning with an exact analytic
solution for one of the two metric functions, an expression for the
mass density could be obtained by integration.  With such expressions
for the density and the first metric function in hand, the analytic
expression for the second metric function could be obtained. This
often required an appropriate change of the radial variable to obtain
a simple integral.  All functions could then be written as explicit
functions of the radial coordinate \(r.\)
While the fluid pressure could, in principle, be obtained from the
metric and density functions, Tolman chose not to evaluate the fluid
pressure in some cases due to the fact that to do so would lead to
rather mathematically complicated expressions that might be difficult
to interpret.

Of the eight solutions presented in his paper, three were already
known (the Einstein universe, Schwarzschild--de Sitter solution, and
the Schwarzschild constant density solution), most of the others
``describe situations that are frankly unphysical, and these do have a
tendency to distract attention from the more useful
ones.''~\cite{Kin75}.  One, the so-called Tolman~VII solution appeared
to have some physical relevance but this was one of the solutions for
which no explicit expression for the pressure was given.

The Tolman~VII solution has been rediscovered a number of times and
has appeared under different names, the
Durgapal~\cite{DurGeh71,DurRaw79} and the \citeauthor*{Meh66}
solutions being two examples. That these solutions can be used to
describe realistic physical systems has been noted by many authors
including those of the two review papers mentioned
above~\cite{DelLak98, FinSke98}.  It has been used as an exact
analytic model for spherically symmetric stellar systems and
additional research has investigated its stability
properties~\cite{NeaLak01, NeaIshLak01, BhaMurPan15}.  While these later works were
able to obtain the complicated expressions for the fluid pressure as a
function of the radial coordinate, according to Finch and Skea it
still was not one of the ``interesting solutions'' since it lacked an
explicit expression for the equation of state.  The choice of
parameters that has been taken by different authors in order to
completely specify the solution in many ways prevented the immediate
interpretation of the physical conditions described by the solution.

The reasons mentioned above are not sufficient to use or classify the
Tolman~VII solution as a physically viable one.  Instead we seek
physical motivations for the viability of this solution, and indeed we
find these in many forms:
\begin{enumerate}[leftmargin=\parindent,labelindent=\parindent,label=(\roman*)]
\item \label{lameEmden} From a Newtonian point of view, simple
  thermodynamic arguments yield polytropes of the form
  \(p(\rho) = k \rho^{\gamma},\) (here $\gamma$ is the adiabatic index
  sometimes written in terms of the polytropic index $n$, $\gamma = 1 + 1/n$,
and $k$ is known at the adiabatic constant that can vary from star to star) 
  as viable models for neutron matter.  When coupled with Newtonian
  hydrodynamic stability and gravitation, the result is the Lane-Emden
  differential equation for the density profile, \(\rho(r).\)
  Solutions of the latter, obtained numerically, or in particular
  cases \((\gamma = \infty, 2,\)
  or 1.2) exactly, all have a distinctive density falloff from the
  centre to the edge of the Newtonian star.  This is a feature we wish
  physical solutions to have.  Furthermore this distinctive falloff is
  quadratic in the rescaled radius~\cite{Hor04}, suggesting that even
  in the relativistic case, such a falloff would be a good first
  approximation to model realistic stars, which have a proper
  thermodynamic grounding.

\item \label{schInt} Looking at viable exact relativistic solutions to
  the Einstein equations, the one used extensively before 1939 and
  even much later, was the Schwarzschild interior solution.  This
  solution has the feature that the density is constant throughout the
  sphere, and is not physical: the speed of sound (pressure) waves in
  its interior is infinite.  However this solution provides clear
  predictions about the maximum possible mass of relativistic stars in
  the form of the Buchdahl limit~\cite{Buc59}: \(M \leq 4R/9.\)
  The next best guess in this line of reasoning of finding limiting
  values from exact solutions would be to find an exact solution with
  a density profile that decreases with increasing radius, since a
  stability heuristic for stars demands that
  \(\rmd \rho / \rmd r \leq 0,\)
  as expected from~\ref{lameEmden} in the Newtonian case.  Extension
  to the relativistic Lane-Emden equation also requires~\cite{Hor04}
  that \((\rmd \rho / \rmd r)|_{r=0} = 0,\) a property Tolman~VII has.

\item \label{falloff} Additionally an extensive review of most EOS
  used from nuclear physics to model neutron stars concluded that a
  quadratic falloff in the density is a very close approximation to
  \emph{most} such nuclear models~\cite{LatPra01}: the differences
  between drastically different nuclear models from Tolman~VII being
  only minor if only the density profiles were compared.  Since
  Tolman~VII is precisely the unique exact solution to the full
  Einstein field equations that exhibits a quadratic falloff in the
  density profile, we believe that it captures much of what nuclear
  models have to say about the overall structure of relativistic
  stars.
\end{enumerate}

These three reasons taken together make a strong case for considering
the Tolman~VII solution as the best possible exact solution that is
capable of describing a wide class of EOS for neutron stars.
At the very least it is as good a candidate that captures first order
effects in density of \emph{most} nuclear model EOS, and at best it is
the model that all realistic nuclear models tend to, while including
features like self-boundedness naturally, as we shall show.

The purpose of this Chapter is to re-examine the Tolman~VII solution
by introducing a set of constant parameters that we believe provide a
more intuitive understanding of the physical content of the solution.
In addition the solution now becomes a member of the set of
``interesting solutions'' since we provide an explicit expression for
the EOS.  The EOS will allow for further
exploration of the predictions of the solution as well as a
description of the material that makes up the star. The imposition of
both causality conditions where the speed of sound in the fluid never
exceeds the speed of light, and different boundary conditions will
provide further restrictions on the parameters associated with the
solution.  What this all leads to is a complete analytic model for
compact stars that can be used to compare with recent observations of
neutron star masses and radii.  That the Tolman~VII solution is
consistent with all measurements leads to the conclusion that this
exact solution is not only physically relevant but may be one
physically realized by nature.

This chapter is divided as follows: following the brief introduction
presented in this section, we re-derive the Tolman~VII solution in
section~\ref{t7.sec:pressure}, paying particular attention to the
pressure expression in physically more intuitive variables.  We then
invert the density equation and use the pressure expression just found
to derive an EOS in section~\ref{t7.sec:eos}, where we also carry out
an analysis of the said EOS.  We will then proceed to contrast the two
different types of physical models that the solution admits in
section~\ref{t7.sec:nat+self}, where we will also show how qualitative
differences arise in the stars' structure and quantitative ones appear
in the predicted values of the adiabatic indices of the fluid.  We
shall then provide brief concluding remarks in
section~\ref{t7.sec:conclusion}.

\section{\label{t7.sec:pressure}The Tolman solution}
Beginning with a line element in terms of standard areal
(Schwarzschild) coordinates for a static and spherically symmetric
metric:
\[ \rmd s^{2} = \rme ^{\nu(r)} \rmd t^{2} - \rme ^{\lambda(r)} \rmd
r^{2} - r^{2} \rmd \theta^{2} - r^{2} \sin^{2} \theta \rmd
\varphi^{2}, \]
the Einstein equations for a perfect fluid source lead to three
ordinary differential equations for the two metric variables \(\nu\),
\(\lambda\),
and the two matter variables \(\rho\)
and \(p\).
However these variables will not be the most practical ones to carry
out our analysis.  Instead we introduce two different metric
functions, \(Z(r) = \rme^{-\lambda(r)}\)
and \(Y(r) = \rme^{\nu(r)/2},\)
as derived in Ivanov~\cite{Iva02}.  The reason for these new metric
variables is that with our subsequent density assumption, this will
allow easier linearisation of the differential equations.  The
Einstein equations then reduce to the following set of three coupled
ordinary differential equations (ODEs) for the four variables
\(Z,Y,p,\) and \(\rho:\)
\begin{subequations}
  \label{t7.eq:EinR}
  \begin{alignat}{3}
    \label{t7.eq:EinR1}
    \kappa \rho &= \rme^{-\lambda}\left( \f{\lambda'}{r} -\f{1}{r^2}\right) +\f{1}{r^{2}} 
    &&= \f{1}{r^{2}} - \f{Z}{r^{2}} - \f{1}{r}\deriv{Z}{r}, \\
    \label{t7.eq:EinR2}
    \kappa p &= \rme^{-\lambda} \left( \f{\nu'}{r} + \f{1}{r^2}\right) -\f{1}{r^2} 
    &&= \f{2Z}{rY}\deriv{Y}{r} + \f{Z}{r^{2}} - \f{1}{r^{2}},\\
    \label{t7.eq:EinR3}
    \kappa p &= \rme^{-\lambda} \left( \f{\nu''}{2} - 
      \f{\nu'\lambda'}{4} + \f{(\nu')^2}{4} + \f{\nu'- \lambda'}{2r}\right)  
    &&= \f{Z}{Y}\sderiv{Y}{r} + \f{1}{2Y}\deriv{Y}{r}\deriv{Z}{r} + \f{Z}{rY}\deriv{Y}{r} + \f{1}{2r}\deriv{Z}{r}.
  \end{alignat}
\end{subequations}
Where the primes \((')\) denote differentiation with respect to \(r,\)
and \(\kappa\) is equal to \(8\pi,\) since we use natural units where
\(G = c = 1.\) The first two equations~\eqref{t7.eq:EinR1}
and~\eqref{t7.eq:EinR2} can be added together to generate the simpler
equation
\begin{equation}
  \label{t7.eq:EinR1+2}
  \kappa (p+\rho) = \f{2Z}{rY}\deriv{Y}{r} - \f{1}{r}\deriv{Z}{r},
\end{equation}
which will be useful later on.  To solve this set of ODEs, we shall
assume a specific form for the energy density function that we claim has
physical merit:
\begin{equation}
  \label{t7.eq:Density}
  \rho = \rho_{c} \left[ 1 - \mu \left( \f{r}{r_{b}}\right)^{2} \right],
\end{equation}
where the constants \(r_{b}\) represents the boundary radius as
mentioned previously, \(\rho_{c}\) represents the central density at
\(r=0,\) and \(\mu\) is a ``self-boundness'' dimensionless parameter,
that will span values between zero and one, so that when it is equal
to zero, we have a sphere of constant density.  This form of the
density function for \(\mu > 0\) is physically realistic since it is
monotonically decreasing from the centre to the edge of the sphere, in
contrast to the constant density exact solution (Schwarzschild
interior) frequently used to model such objects.  Additionally we will
need boundary conditions for the system: since we eventually want to
match this interior solution to an external metric.  Since the vacuum
region is spherically symmetric and static, the only candidate by
Birkhoff's theorem is the Schwarzschild exterior solution.  The
Israel-Darmois junction conditions for this system can be shown to be
equivalent to the following two conditions~\cite{Syn60} as is derived
in Appendix~\ref{C:AppendixA}:
\begin{subequations}
  \label{t7.eq:Boundary1+2}
  \begin{align}
    \label{t7.eq:BoundaryP}
    p(r_{b}) &= 0, \quad \text{and,} \\
    \label{t7.eq:BoundaryZ}
    Z(r_{b}) &= 1-\f{2M}{r_{b}} = Y^{2}(r_{b}).
  \end{align}
\end{subequations}
Where \(M = m(r_{b})\) is the total mass of the sphere as seen by an outside
observer, and \(m(r)\) is the mass function defined by
\begin{equation}
  \label{t7.eq:MassFunction}
  m(r) = 4\pi \int_{0}^{r} \rho(\bar{r}) \bar{r}^{2 }\rmd \bar{r}.
\end{equation}
Furthermore we will also require the regularity of the mass function,
that is for mass function to be zero at the \(r=0\) coordinate, from
physical considerations: \(m(r=0) = 0.\) On
imposing~\eqref{t7.eq:BoundaryZ}, we can immediately write \(Z\) in terms
of the parameters appearing in the density assumption:
\begin{equation}
  \label{t7.eq:ZWithParams}
  Z(r) = 1 - \left( \f{\kappa \rho_{c}}{3}\right) r^{2} + 
  \left(\f{\kappa \mu \rho_{c}}{5r_{b}^{2}}\right) r^{4} \eqqcolon 1 - br^2+ ar^4.
\end{equation}
In contrast Tolman assumed the last equation, and then obtained the
density function from equation~\eqref{t7.eq:EinR1}.  The physical
constants \(\mu, \rho_c,\) and \(r_b\) occur frequently enough in the
combinations above that we will also use \(a\) and \(b\) as defined
above when convenient.  The solution method for these ODEs leading to
the Tolman VII solution have been given in multiple
references~\cite{Meh66,Tol39}, and we briefly sketch it.  Essentially
two variable transformations convert the set of ODEs into a simple
harmonic differential equation, and back-substitution to the original
variables solves the system.  The first step in this procedure in the
change of variable from \(r\) to \(x = r^2,\) as a result of which the
derivative terms have to be changed too.

The set of equation~\eqref{t7.eq:EinR} with this variable change then
becomes
\begin{subequations}
  \label{t7.eq:EinX}
  \begin{align}
    \label{t7.eq:EinX1}
    \kappa \rho &= \f{1}{x} - \f{Z}{x} - 2\deriv{Z}{x}, \\
    \label{t7.eq:EinX2}
    \kappa p &= \f{4Z}{Y}\deriv{Y}{x} + \f{Z}{x} - \f{1}{x}\\
    \label{t7.eq:EinX3}
    \kappa p &= \deriv{Z}{x} +\f{1}{Y} \left( 4Z \deriv{Y}{x} + 4xZ\sderiv{Y}{x} +2x \deriv{Y}{x}\right).
  \end{align}
\end{subequations}
We see that by subtracting equation~\eqref{t7.eq:EinX3} from
equation~\eqref{t7.eq:EinX2}, we get the second order differential
equation\[ \f{Z}{x} - \f{1}{x} - \f{4xZ}{Y} \sderiv{Y}{x} - \f{2x}{Y}
\left(\deriv{Y}{x}\right) \left(\deriv{Z}{x} \right) - \deriv{Z}{x} =
0,\] which is further simplified by multiplying by \(-Y/(2x),\)
resulting in\[ 2Z\sderiv{Y}{x} + \left(\deriv{Y}{x}\right)
\left(\deriv{Z}{x} \right) + Y\left( \f{1}{2x^2} -\f{Z}{2x^2}
  +\f{1}{2x}\deriv{Z}{x} \right) = 0.\] Solving the system in this
form is equivalent to solving the whole system, and since we have
already integrated the first Einstein equation~\eqref{t7.eq:EinR1}, and
have an expression for \(Z\) in equation~\eqref{t7.eq:ZWithParams} in
terms of \(r\) and equivalently \(x,\) we substitute those now in the
above differential equation to get
\begin{equation}
  \label{t7.eq:Ydiff1}
  Z \sderiv{Y}{x} + \f{1}{2} \left( \deriv{Z}{x} \right) \deriv{Y}{x} + Y \left[ \f{a}{4} \right] = 0 .
\end{equation}
This last equation has a middle cross term involving a first
derivative of the metric function \(Y\) which we want to find.  From
the form of the equation we see that if we could transform the above
equation with a substitution to get rid of the cross term, we would
end up with a second order equation of the simple harmonic type.  It
turns out that the variable change that permits this is
\begin{equation}
  \label{t7.eq:xiInt}
  \xi = \int_0^x \f{\rmd \bar{x}}{\sqrt{Z(\bar{x})}}  \Rightarrow \deriv{\xi}{x} = \f{1}{\sqrt{Z(x)}},
\end{equation}
Where we retain the integral form of the equation.  Transforming our
dependant variable \(x,\) to \(\xi\) and taking into account that the
derivatives with respect to \(x\) will transform according to
\begin{subequations}
  \begin{align}
    \label{t7.eq:xiDeriv}
    \deriv{A}{x}  &= \deriv{A}{\xi} \deriv{\xi}{x} = \f{1}{\sqrt{Z}} \deriv{A}{\xi}, \\
    \sderiv{A}{x} &= \deriv{}{x}\left( \deriv{A}{\xi} \deriv{\xi}{x} \right) = \sderiv{\xi}{x} \deriv{A}{\xi} 
    +\left(\deriv{\xi}{x} \right)^2 \sderiv{A}{\xi} = \f{1}{Z}\sderiv{A}{\xi} - \f{1}{Z^{3/2}} \deriv{Z}{x} \deriv{A}{\xi},
  \end{align}
\end{subequations}
where \(A\) is a dummy function of both \(x\) and \(\xi,\) we simplify
the differential equation~\eqref{t7.eq:Ydiff1} into\[ Z \left\{ \f{1}{Z}
  \sderiv{Y}{\xi} - \cancel{ \f{1}{2Z\sqrt{Z}}
    \deriv{Z}{x}\deriv{Y}{\xi} }\right\} + \cancel{\f{1}{2}
  \deriv{Z}{x} \left( \f{1}{\sqrt{Z}} \deriv{Y}{\xi}\right)} + \left(
  \f{a}{4} \right) Y = 0,
\] which is indeed in simple harmonic form, as we wanted.  We can as a
result immediately identify three different classes of solutions
depending on the value of the constant \(\phi^2 = a/4,\) summarized
below:
 
\begin{table}[!h]
\centering 
\begin{tabular}{| >{$}c<{$} | >{$}c<{$} | l |}
  \hline
    \phi^2 & Y(\xi) & Solution's name \\
    \hline
    \phi^2 < 0 & c_1 \exp{\left(\sqrt{-\phi^2}\xi\right)} + c_2 \exp{\left(-\sqrt{-\phi^2}\xi\right)} & Bayin~\cite{Bay78} \\
    \phi^2 = 0 & c_1 + c_2 \xi & Schwarzschild interior\\
    \phi^2 > 0 & c_1 \cos{\left( \phi \xi \right)} + c_2 \sin{\left(\phi \xi\right)} & Tolman VII\\
\hline
  \end{tabular}
  \caption{The different solutions that can be generated through different values of the parameter~$\phi.$}
\end{table}

Once we pick a value for \(\phi,\) the solution is completely
specified.  Applying the boundary conditions will then permit us to
find the value of the parameters.  At this stage, we could using the
interpretation scheme given previously for the constants \(\mu,
\rho_c,\) and \(r_b,\) to figure out the sign for \(\phi.\) From
equation~\eqref{t7.eq:ZWithParams}, we find that since all the constants
involved in \(a/4\) are positive definite, \(\phi^2\) must be so too,
and hence we are forced to pick the third solution, i.e.\ Tolman's, if we
want to model physically realistic stars.  We also note that were
\(\mu\) to take the limiting value of zero, we would not only have to
pick the the second solution, which is Schwarzschild's interior
solution, but also assume a constant density, a well known aspect of
this particular solution. 

The complete Tolman VII solution is specified with the two functions
below, together with the previously given density
function~\eqref{t7.eq:Density}, and the metric function \(Z\) in
equation~\eqref{t7.eq:ZWithParams},
\begin{equation}
  \label{t7.eq:Y}
    Y(\xi) = c_{1} \cos (\phi \xi) + c_{2} \sin (\phi \xi) , \quad \text{with } 
    \phi = \sqrt{\f{a}{4}},
\end{equation}
where we have used \(\xi,\) but not actually given an explicit
expression for it in terms of \(r.\) The actual form is found by
performing the integral~\eqref{t7.eq:xiInt}, which can be solved by
consulting an integrals' table~\cite{GraRyz07}, however insight into
the form of the integral is gained through an Euler substitution
of the form \(\sqrt{Z(\bar{x})} = \bar{x}t -1,\) so that the
denominator and Jacobian of the transformation become \[\bar{x}t - 1 =
\f{t^2-bt+a}{t^2-a} \quad \text{and} \quad \deriv{\bar{x}}{t} =
\f{-2(t^2 + a -bt)}{(t^2-a)^2}.\] This allows the integrand to be
expressed as
\[\xi(r) = \int^{\bar x = x}_{\bar x = 0} \f{-2\cancel{(t^2-bt+a)}}{(t^2-a)^{\cancel{2}}} \left( \f{\cancel{t^2-a}}{\cancel{t^2 -bt +a}}\right) \rmd t = \int^{\bar x = x}_{\bar x = 0} \f{-2 \rmd t}{(t^2-a)},\]
which requires another substitution to be solved.  A number of
different substitutions would give different equivalent forms of the
integral, however because of the positive sign of \(a,\) the most
useful substitution is \(t^2 = a \coth^2 \alpha,\) with resulting
Jacobian \(\rmd t = \sqrt{a} (-\csch^2 \alpha) \rmd \alpha.\) These
reduce the denominator of the previous integrand to \(a \csch^2
\alpha,\) so that the final form of \(\xi\) is
\begin{equation}
  \label{t7.eq:xiCoth}
  \xi(r) = \f{2}{\sqrt{a}} \coth^{-1} \left. \left( \f{t}{\sqrt{a}} \right) \right|^{\bar x = x}_{\bar x =0} = \f{2}{\sqrt{a}} \coth^{-1} \left( \f{1+\sqrt{1-br^2+ar^4}}{r^2\sqrt{a}}\right),
\end{equation}
where the last equality comes from back-substituting the multiple
variable changes done before.  Sometimes the equivalent form of the
above equation, in terms of logarithms is more useful, and in this
form the above is expressed as
\begin{equation}
  \label{t7.eq:xiLog}
  \xi(r) = \f{1}{\sqrt{a}} \left[ \log(b+2\sqrt{a}) -  \log(b - 2ar^2 + 2\sqrt{aZ}) \right],
\end{equation}
from the well known hyperbolic identity \(\coth^{-1} x \equiv \f{1}{2}
\left[ \log \left( 1+ \f{1}{x}\right) - \log \left( 1- \f{1}{x}
  \right)\right],\) for \(x \neq 0.\)
Now that we have the full solution of the metric functions, we can
compute the pressure through the relation below, obtained from a
simple rearrangement and variable change of~\eqref{t7.eq:EinR1+2}:
\begin{equation}
    \label{t7.eq:PressureMetric}
    \kappa p(r) =  4 \f{\sqrt{Z}}{Y}\deriv{Y}{\xi} - \f{1}{r}\deriv{Z}{r} - \kappa \rho,
\end{equation}
resulting in the very complicated looking, 
\begin{equation}
  \label{t7.eq:PressureR}
  \kappa p(r) = \f{4\phi [c_{2} \cos{(\phi\xi)} - c_{1} \sin{(\phi\xi)}] \sqrt{1 - br^{2} + ar^{4}}}{c_{1}\cos{(\phi\xi)} + c_{2} \sin{(\phi\xi)}} - 4ar^{2} + 2b - \kappa \rho_{c} \left[ 1 - \mu \left( \f{r}{r_{b}}\right)^{2} \right].
\end{equation}
So far we have not found the expressions of any of the two integration
constants \(c_{1}\) and \(c_{2}.\) To find those, we need to apply the
boundary conditions explicitly and to do so we perform the variable
changes \(r \rightarrow x \rightarrow \xi\) on
equation~\eqref{t7.eq:EinR1+2},
\[ \kappa (p+\rho) = \f{2Z}{rY}\deriv{Y}{r} - \f{1}{r}\deriv{Z}{r} \xrightarrow{r
  \rightarrow x} \f{4Z}{Y}\deriv{Y}{x} - 2\deriv{Z}{x} \xrightarrow{x
  \rightarrow \xi} \f{4\sqrt{Z}}{Y} \deriv{Y}{\xi} - 2 \deriv{Z}{x}.\]
With this equation, together with the boundary
condition~\eqref{t7.eq:BoundaryZ}, we have\[\left. \kappa (p+\rho)
\right|_{x=x_{b}} = \f{4 \cancel{\sqrt{Z(x_{b})}}}{\cancel{Y(x_{b})}}
\left. \deriv{Y}{\xi} \right|_{\xi=\xi_{b}} - 2 \left. \deriv{Z}{x}
\right|_{x=x_{b}},\] where all the \(b\) -subscripted variables are the
values at the boundary.  However since according to the second
boundary condition~\eqref{t7.eq:BoundaryP}, the pressure has to vanish at
the boundary, the latter equation simplifies to\[ \left. \kappa \rho
\right|_{x=x_{b}} = \left. 4 \deriv{Y}{\xi} \right|_{\xi=\xi_{b}} - 2 \left. \deriv{Z}{x}
\right|_{x=x_{b}},\] which can be further simplified and rearranged as
\begin{equation}
  \label{t7.eq:DefAlpha}
  \left. \deriv{Y}{\xi} \right|_{\xi=\xi_{b}} = \f{b-ax_{b}}{4} = \f{\kappa \rho_{c}}{4}\left( \f{1}{3} - \f{\mu}{5}\right) \eqqcolon \alpha.
\end{equation}
Since the ODE for \(Y\) is second order, we also need a further
constraint equation.  This is simply going to be
condition~\eqref{t7.eq:BoundaryZ} restated as
\begin{equation}
  \label{t7.eq:DefGamma}
  Y(x=x_b) = \sqrt{1-bx_{b}+ax_{b}^{2}} = \sqrt{1 - \kappa\rho_{c}r^{2}_{b} \left( \f{1}{3} - \f{\mu}{5}\right)} \eqqcolon \gamma
\end{equation}
These two equations~\eqref{t7.eq:DefAlpha} and~\eqref{t7.eq:DefGamma} constitute the
complete Cauchy's boundary condition on Y.  We now only need to
simplify our integration constants \(c_{1}\) and \(c_{2}\) with
these, to specify the solution completely in terms of the parameters
we chose initially.  To do so we re-express the metric function and
its derivatives in terms of their solutions, yielding two simultaneous
equations:
\begin{align}
  \label{t7.eq:BDYsimul}
  \left. \deriv{Y}{\xi} \right|_{\xi=\xi_{b}} = \phi \left[c_{2} \cos{(\phi \xi_{b})} - c_{1} \sin{(\phi\xi_{b})} \right] = \alpha \Rightarrow &c_{2} \cos{(\phi \xi_{b})} - c_{1} \sin{(\phi\xi_{b})} = \alpha / \phi, \\
\label{t7.eq:BDZsimul}
 Y(\xi=\xi_b) = &c_{2} \sin{(\phi\xi_{b})} + c_{1} \cos{(\phi \xi_{b})} = \gamma.
\end{align}
This system can be solved by first multiplying~\eqref{t7.eq:BDYsimul} by
\(\cos{(\phi \xi_{b})}\), and~\eqref{t7.eq:BDZsimul} by \(\sin{(\phi
  \xi_{b})},\) and adding the equations obtained: yielding \(c_{2}.\)
Similarly switching the multiplicands and performing a subtraction
instead yields \(c_{1},\) both of which we now give.
\begin{align}
  \label{t7.eq:C1solved}
  c_{1} &= \gamma \cos{(\phi\xi_{b})} - \f{\alpha}{\phi} \sin{(\phi \xi_{b})}, \\
  c_{2} &= \gamma \sin{(\phi\xi_{b})} + \f{\alpha}{\phi} \cos{(\phi \xi_{b})}.
\end{align}
We note that all the constants employed in the expressions of the
integration constants are ultimately in terms of the set of parameters
\(\I\) we initially chose, viz.\ \( \I = \{\rho_{c}, r_{b},\mu\}.\)
This completes the specification of the full Tolman~VII solution in
the new constant scheme.

Another quantity we wish to consider is the adiabatic speed of
pressure(sound) waves that the fluid can sustain.  The usual
definition of this quantity in perfect fluids is
\(v^{2} = \rmd p / \rmd \rho.\) However, we will find it convenient to
find an expression of this speed directly from the differential
equations, since the expression and functional form, while completely
equivalent is simpler to work with.  We notice first that from the
expression of the density~\eqref{t7.eq:Density}, we can obtain the
derivative\[\deriv{\rho}{r} = -\f{2 \mu \rho_{c}}{r_{b}^{2}} r,\]
which is zero at \(r = 0,\) or if one of the 2 parameters \(\mu=0\) or
\(\rho_{c}=0.\) For the other equation, we use the conservation of the
energy momentum tensor \(\nabla_{a} T^{a}{}_{b} = 0,\) which as we
have shown before reduces to
\begin{equation}
  \label{t7.eq:TOV}
  \deriv{p}{r} = -\f{\nu'(p+\rho)}{2} = -\f{(p+\rho)}{Y}\deriv{Y}{r},
\end{equation}
in the \(b=0\) case.  These two expressions can be used to find \(\rmd p
/\rmd \rho\) for every value of \(r\) but the centre, so that
\begin{equation}
  \label{t7.eq:SpeedSound}
  v^{2} = \left( \deriv{p}{r} \middle/ \deriv{\rho}{r}\right) = 
  \f{r_{b}^{2}}{4\mu\rho_{c}} \f{\nu'(p+\rho)}{r}= 
  \f{r_{b}^{2}(p+\rho)}{2\mu\rho_{c}rY}\deriv{Y}{r}.
\end{equation}
Since we have expressions for all the terms in this formula, we also
have a closed form for the speed of sound. 

The bulk modulus \(K\) of a fluid is a measure of the resistance of a
fluid to change its volume under an applied pressure.  For perfect
fluids it is related to the speed of sound (pressure waves) in the
media through \(K = \rho v^{2}.\) This is also a quantity which we
calculate for the fluid in the interior, and this calculation show us
that the material we are dealing with has no earthly analogue, since
the order of magnitude of the bulk modulus is much higher than any
currently known substance.

\begin{figure}[h!]
\subfloat[The density]{\label{t7.fig:density}\includegraphics[width=0.5\linewidth]{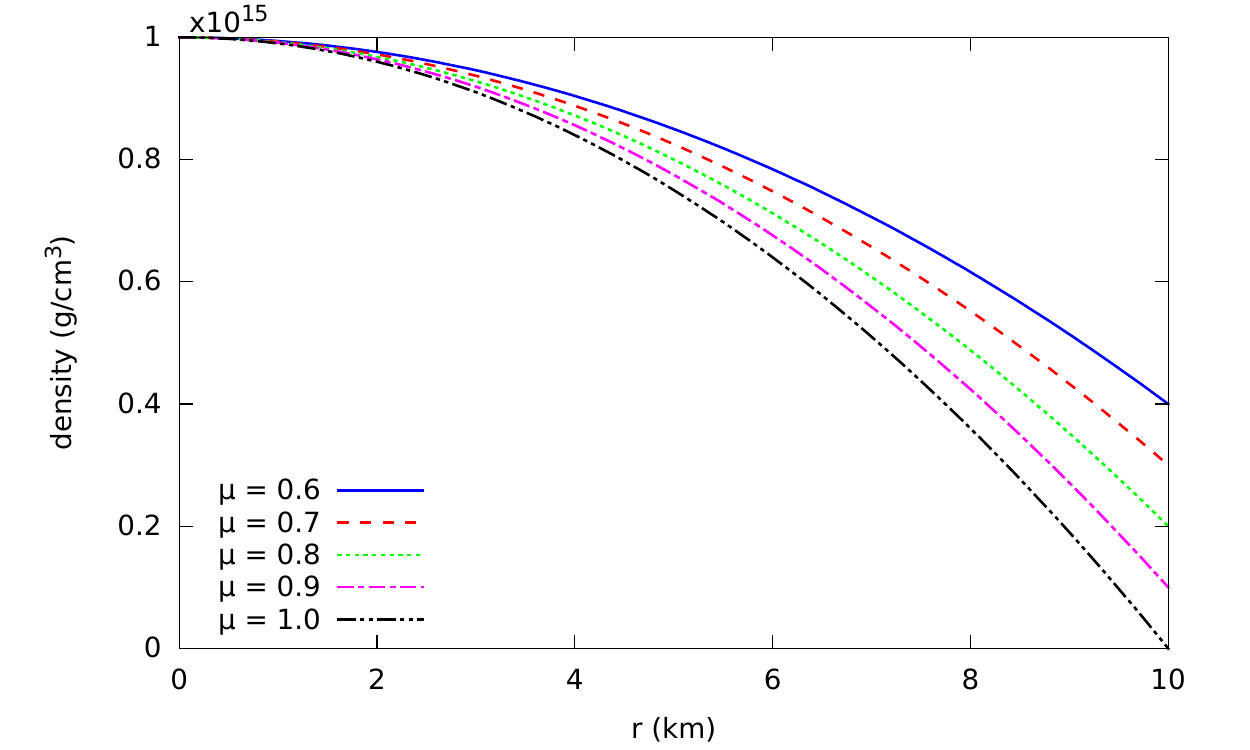}} 
\subfloat[The pressure]{\label{t7.fig:pressure}\includegraphics[width=0.5\linewidth]{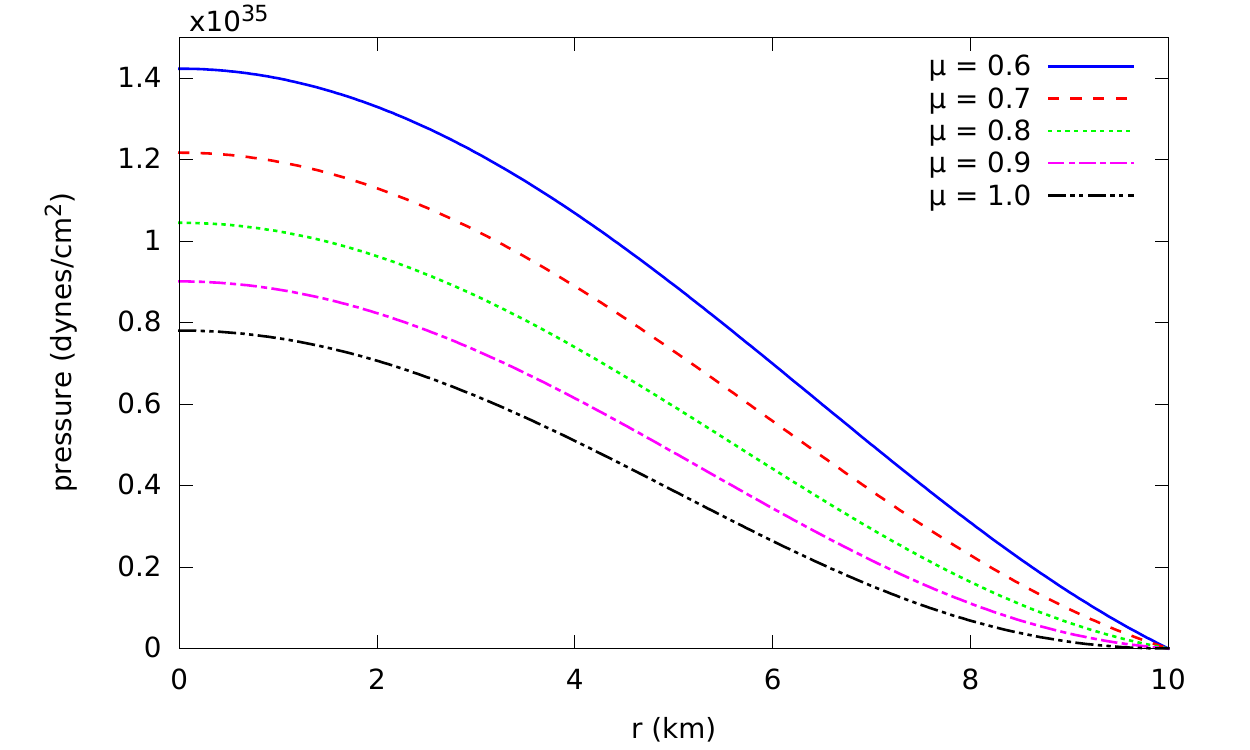}}\\
\subfloat[The speed of sound]{\label{t7.fig:sound}\includegraphics[width=0.5\linewidth]{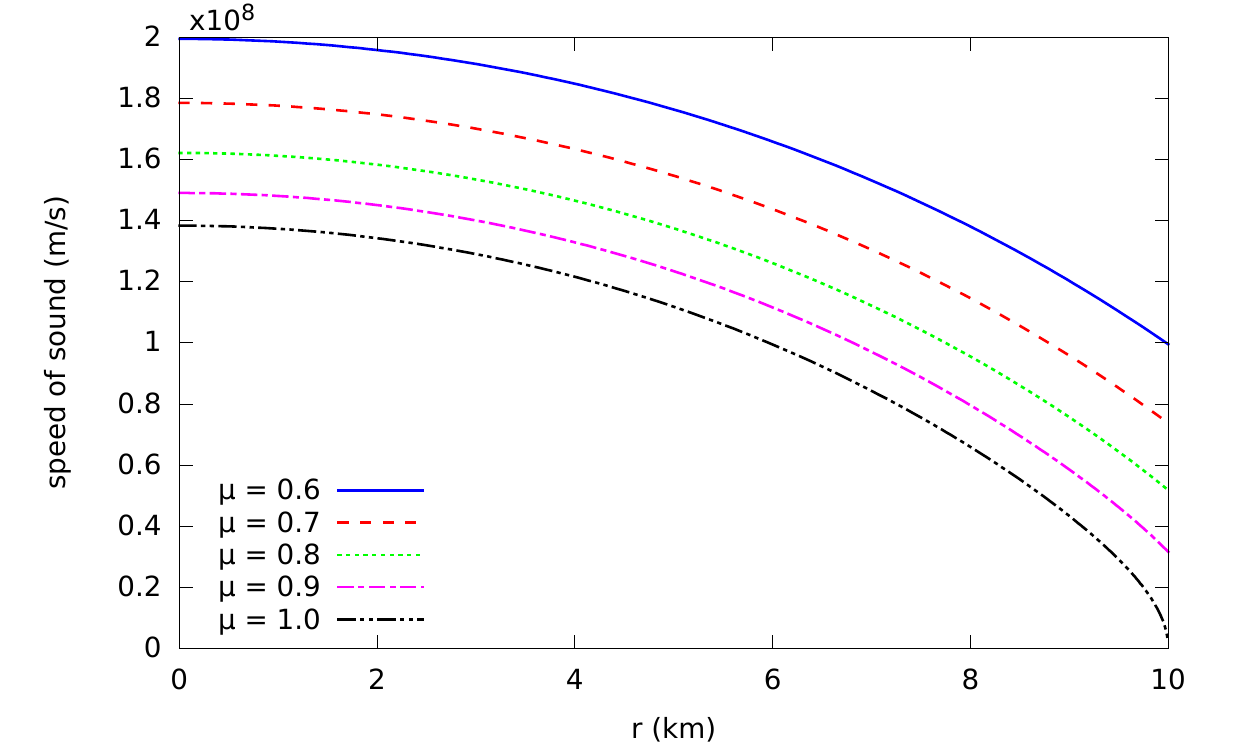}}
\subfloat[The bulk modulus]{\label{t7.fig:bulk}\includegraphics[width=0.5\linewidth]{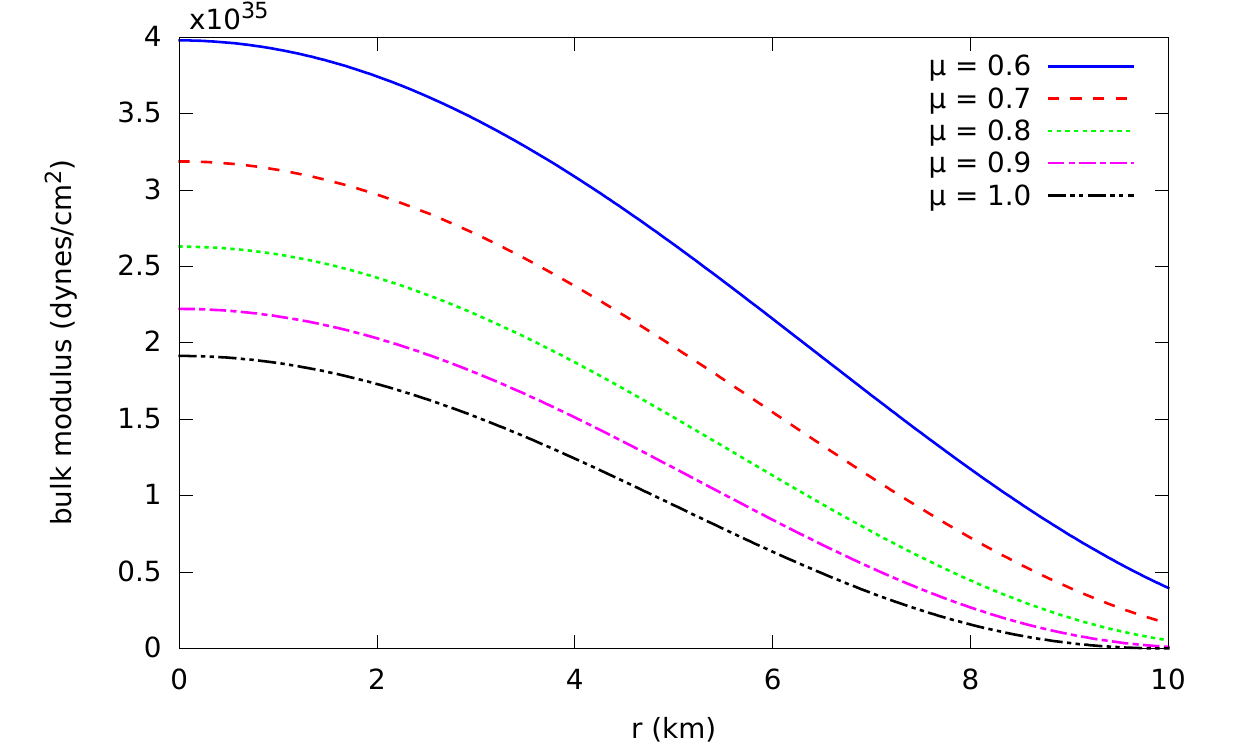}} 
\caption[The matter variables and the speed of sound]{The matter
  variables including the density, pressure, speed of sound, and bulk
  modulus inside the star.  The parameter values are
  $\rho_{c}=\un{1\times 10^{18}\,kg \cdot m^{-3}}, r_b = \un{1 \times
    10^{4}\,m}$
  and $\mu$ taking the various values shown in the legend }
\label{t7.fig:AllMatterCoeff}
\end{figure}

The next step to understanding this solution is to investigate the
behaviour of the different physical variables we have.  However before
we can do that, we have to specify values for our parameters.  We will
use different values of the parameters, and each time we will specify
the values being used.  The primary motivation for the values we will
be using is that we ultimately wish to model compact astrophysical
objects.  As a result central densities \(\rho_{c}\) of \un{1 \times
  10^{15} g\cdot cm^{-3}} will be typical. Similarly radii \(r_{b}\) of
\un{1 \times 10^{6} cm} will often be used for the same reason.  As
can be seen from the density profile~\eqref{t7.eq:Density}, the latter
decreases quadratically with the radial coordinate, as we show in
figure~\ref{t7.fig:density}.

As this point we can interpret the effect of varying the
parameter \(\mu\) on the density profile: It is changing the
surface density from a zero value when \(\mu=1\) to increasingly
higher densities as \(\mu\) is decreasing.  In the
literature~\cite{LatPra01}, models having zero surface densities have
been named ``natural'', and those with non-vanishing surface densities
have been called ``self-bound.''  As a result we will call \(\mu\) the
``self-boundness'' parameter that will allow us to change the surface
density in our models.

Similarly, the complicated expression of the pressure that we have
obtained can also be plotted.  Of importance here is the fact that
while the densities might not vanish at the boundary \(r_{b},\) the
pressure for all parameter values must do so according to our boundary
condition~\eqref{t7.eq:BoundaryP}.  This is eminently clear in
figure~\ref{t7.fig:pressure}, where we see the pressures associated with
the density curves shown in figure~\ref{t7.fig:density}.  Similarly the
speed of sound and bulk modulus, all associated with the matter
content in the star, can be plotted and we show this in
figure~\ref{t7.fig:sound} and~\ref{t7.fig:bulk} respectively.

\begin{figure}[h!]
\subfloat[The $Y(r)$ metric function]{\label{t7.fig:Ymetric}\includegraphics[width=0.5\linewidth]{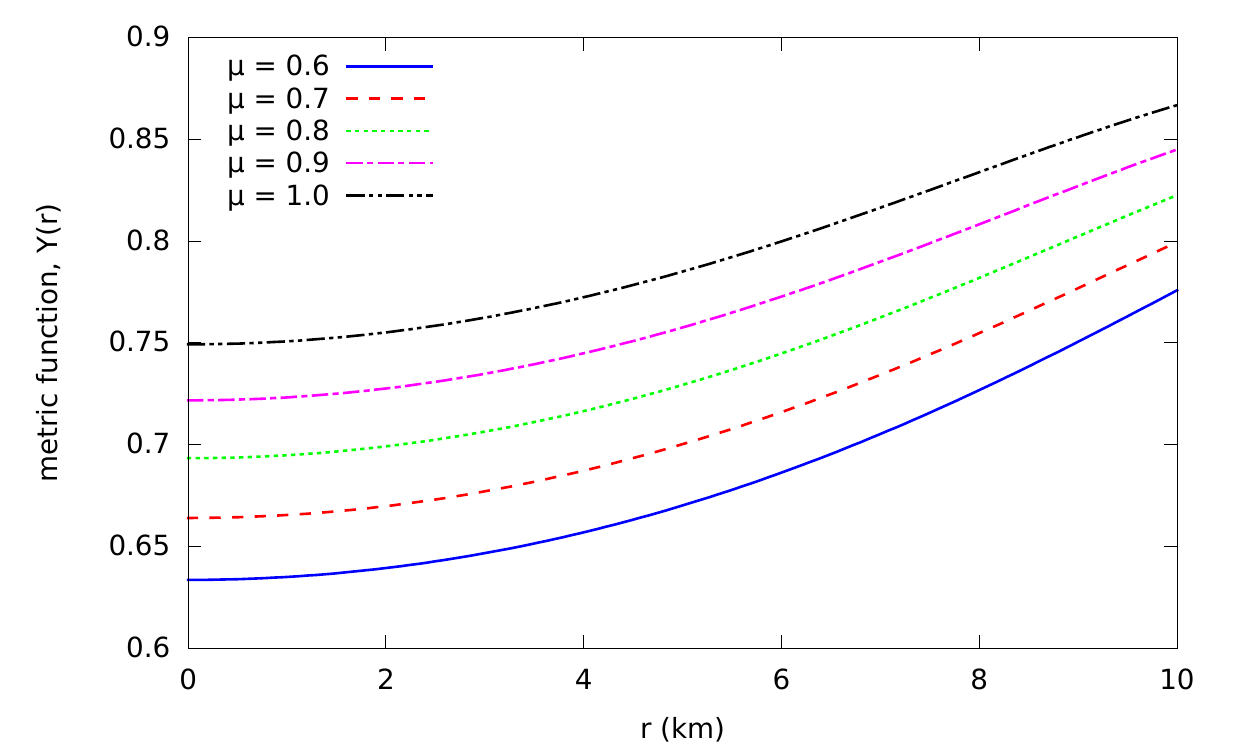}} 
\subfloat[The $Z(r)$ metric function]{\label{t7.fig:Zmetric}\includegraphics[width=0.5\linewidth]{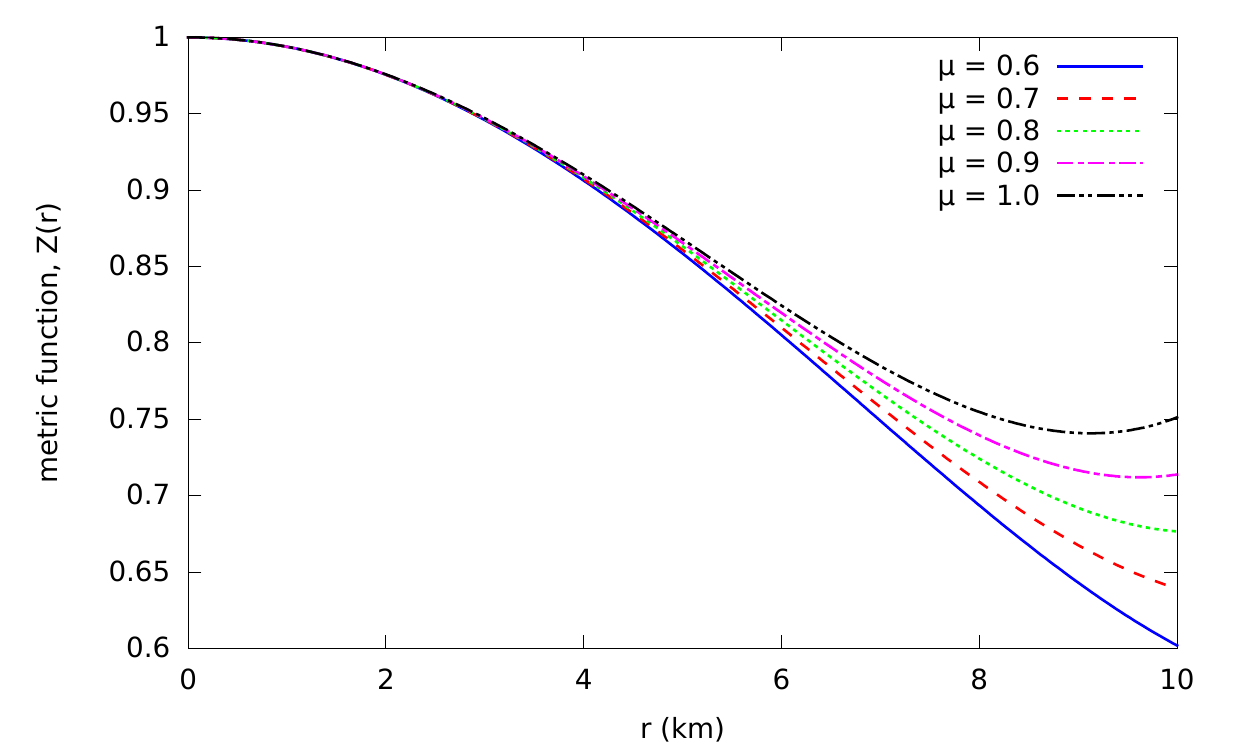}}\\
\subfloat[The $\nu(r)$ metric function]{\label{t7.fig:NuMetric}\includegraphics[width=0.5\linewidth]{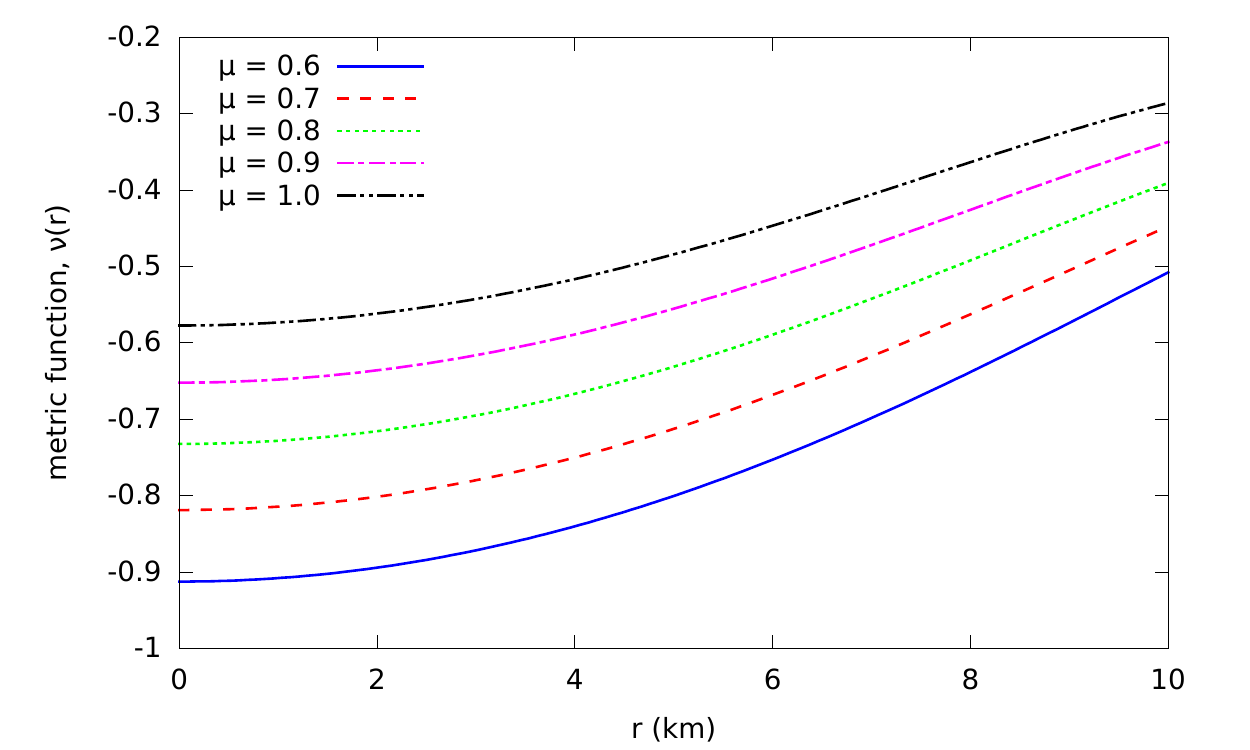}}
\subfloat[The $\lambda(r)$ metric function]{\label{t7.fig:LambdaMetric} \includegraphics[width=0.5\linewidth]{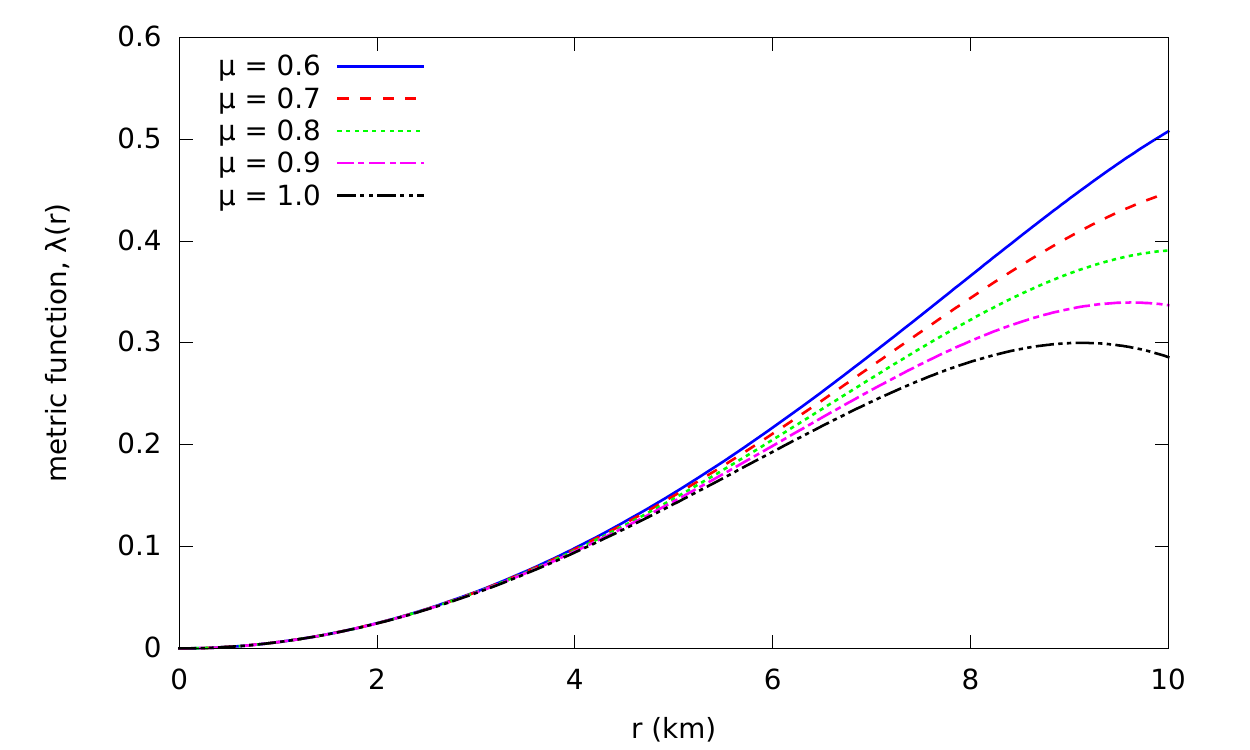}} 
\caption[The metric variables in their two equivalent formulation]{The
  variation of the metric variables with the radial coordinate inside
  the star: we show both~\citeauthor{Iva02}'s $Y(r),$ and $Z(r),$ and
  the more generic $\lambda(r),$ and $\nu(r).$ The parameter values
  are
  $\rho_{c}=\un{1\times 10^{18}\,kg \cdot m^{-3}}, r_b = \un{1 \times
    10^{4}\,m}$
  and $\mu$ taking the various values shown in the legend }
\label{t7.fig:AllMetricCoeff}
\end{figure}

The other variables that solving our differential equations yield are
the metric coefficients \(Z(r)\) and \(Y(r).\) We show both of these
next in figures~\ref{t7.fig:Zmetric} and~\ref{t7.fig:Ymetric} respectively,
again for different values of the self-boundness \(\mu.\) Equivalently
we could give the metric coefficients in Schwarzschild form: the form
most often used in the literature for specifying static spherically
symmetric models.  We do so for now the sake of completeness, giving
\(\lambda(r)\) in figure~\ref{t7.fig:LambdaMetric} and \(\nu(r)\) in
figure~\ref{t7.fig:NuMetric} respectively.

\section{\label{t7.sec:eos}The equation of state}
The nice feature of our density assumption~\eqref{t7.eq:Density} is that
it can be inverted to easily obtain \(r\) as a function of \(\rho,\)
which allows us to generate an equation of state (EOS) for this
solution.  We give the full equation of state below, before starting
to analyse it:
\[ p(\rho) = -\frac{1}{20 \pi h_1 h_2} \left\{ h_1 - h_2 \sqrt{-2 f_1
    \cot^{2} f_2} + 4 \pi h_1 h_2 \rho \right \},\] where \(f_1 (\rho)\) and
\(f_2(\rho)\) are functions of the density:
\[f_1(\rho) = 50 - 3\left(\f{h_{1}}{h_{2}}\right)^{2} -\f{4 \pi
  h_{1}^{2}}{h_{2}} \rho + 32 \pi^{2} h_{1}^{2} \rho^{2}, \]
and\[ f_2(\rho) = \frac{1}{2} \ln \left[ \frac{\sqrt {8 f_1 h_2} +
    h_1 - 16 \pi h_1 h_2 \rho }{20 h_2 C } \right]. \] The constants
\(h_1\) and \(h_2\) are determined by the central density and \(\mu,\)
as follows:
\[h_1 = r_b \sqrt{\frac{5}{2 \pi \rho_c \mu}} \qquad \text{and,}
\qquad h_2 = \frac{3}{8 \pi \rho_c}, \] while the constant \(C\)
is expressible as a complicated function of the parameters only, in
terms of the auxiliary variables \(\sigma\), and \(\chi\),
\[
C = \left( 1 - \f{h_{1}}{4h_{2}}\right) \sqrt{\f{h_{1}(4h_{2}-h_{1})} {8r_{b}^{2} h_{2}-h_{1}^{2} + \chi} }
\exp{\left[\arctan\left(\f{\chi}{\sigma}\right)\right]},
\]
with, 
\begin{align*}
  \chi &= 4\sqrt{h_{2}(4h_{2}r_{b}^{4}-h_{1}^{2}r_{b}^{2} + h_{1}^{2}h_{2})},\\
  \sigma &= 16h_{2}r_{b}^{2}+8\pi \rho_{c} h_{1}^{2}h_{2}(1-\mu)-
  2h_{1}^{2}.
\end{align*}
We note here is that no assumption about the nature of matter, except
for the very general thermodynamic prescription of a perfect fluid has
gone into this solution.  Everything else, and in particular the
equation of state was obtained solely by virtue of the field equations
and the density profile~\eqref{t7.eq:Density}.  With the equation of
state, it is a simple matter to find the derivative \(\rmd p / \rmd
\rho\) for the speed of pressure waves. 

\begin{figure}[h!]
  \centering
  \includegraphics[width=\textwidth]{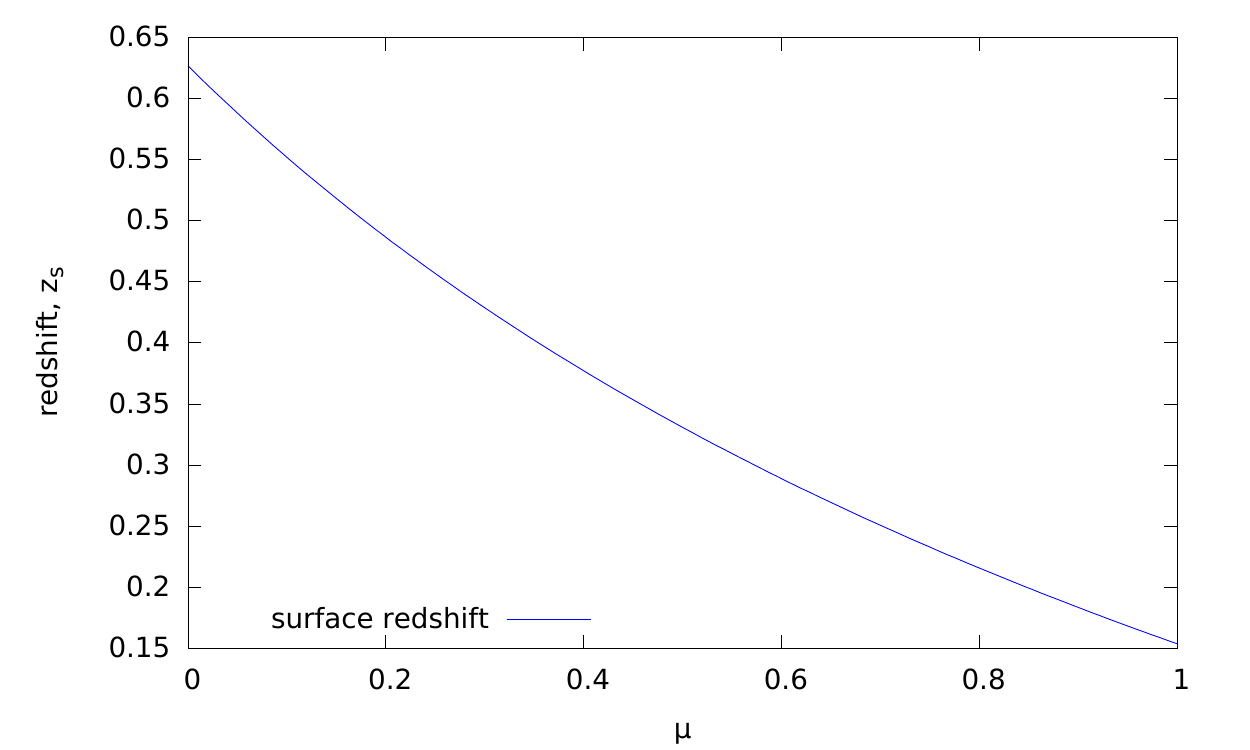}
  \caption[The redshift]{The redshift $z$ at the surface of the sphere for different values of $\mu.$}
  \label{t7.fig:redshift}
\end{figure}

The redshift of light emanating from a star as perceived by distant
observers is another quantity that can potentially be measured.  This
quantity can also be calculated in our model, from the relation \[z_{s} =
\left(1-\f{2m(r_{b})}{r_{b}} \right)^{-\f{1}{2}} - 1.\] We show this value
at the surface of the star for different values of \(\mu\) in
figure~\ref{t7.fig:redshift} next.

\section{\label{t7.sec:nat+self}Physical models} 
The expression for the EOS is somewhat complicated, but it is not
without physical interpretation, contrary to what Tolman~\cite{Tol39}
thought in 1939:
\begin{quote}
  The dependence of \(p\) on \(r\), with \(\rme^{-\lambda/2}\) and
  \(\rme^{-\nu}\) explicitly expressed in terms of \(r\), is so
  complicated that the solution is not a convenient one for physical
  considerations.
\end{quote}
Something that immediately becomes clear is possibility of two
separate interpretations for an EOS.  Both \( \label{t7.eq:EOS1} p(\rho;
\I) \) for \(\rho_{b} = \rho_{c}(1-\mu) \leq \rho \leq \rho_{c},\)
with the values of the elements of \( \I,\) in particular \(\rho_{c}\)
fixed (henceforth called EOS1); and \(\label{t7.eq:EOS2} p(\rho=\rho_{c};
\I), \) with the parameters of \(\I\) varying between limits imposed
by causality (EOS2) could be candidates.  In the literature, both
interpretations have been used, and sometimes even interchanged.
However, each has a completely different content in that the first
interpretation expresses how the pressure of the fluid changes in
moving from the centre of the star \(r=0, \rho=\rho_{c},\) to the
boundary \(r=r_{b}, \rho_{b}=\rho_{c}(1-\mu).\) The second
interpretation by contrast looks closely at the fluid material itself
and how the pressure at a certain point in the star changes as the
density of the fluid at the centre changes.  At this point in our
derivation, we have not yet imposed any causality condition on any
expressions.

\begin{figure}[h!]
\includegraphics[width=\textwidth]{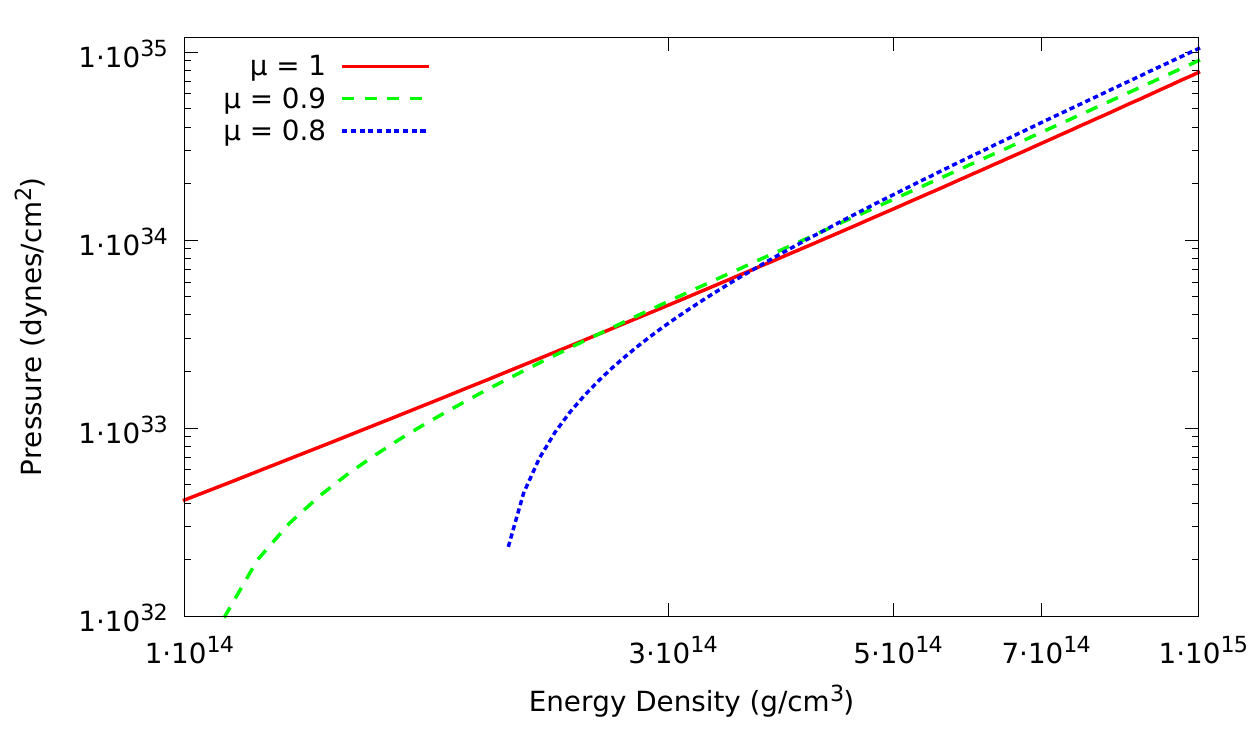}
\caption[Log--log plot of pressure]{\label{t7.fig:loglog} (Colour online) Log--log plot of pressure
  versus density for neutron star models determined by different
  $\mu,$ but same $\rho_{c},$ and $r_{b}.$ The densities and pressures
  are in cgs units, and the $\I$ is fixed by the following: $r_{b} =
  10^{6} \un{cm} , \rho_{c}=10^{15} \un{g \cdot cm^{-3}}.$ Since
  pressure is a decreasing function of distance from the centre, large
  densities indicate points closer to the centre of the star.}
\end{figure}

We first carry out an analysis of EOS1, and find surprisingly that to
a high degree of accuracy, the variation of \(p(\rho; \I),\) with
\(\rho,\) and equivalently \(r,\) is very close to that of a polytrope
of the form \(p = k \rho^{\gamma} - p_{0}.\) This relation is very
obvious from the shape of the curve in the ``natural'' \(\mu=1,\) case
as is seen in figure~\ref{t7.fig:loglog}.

\begin{figure}[h!]
\includegraphics[width=\textwidth]{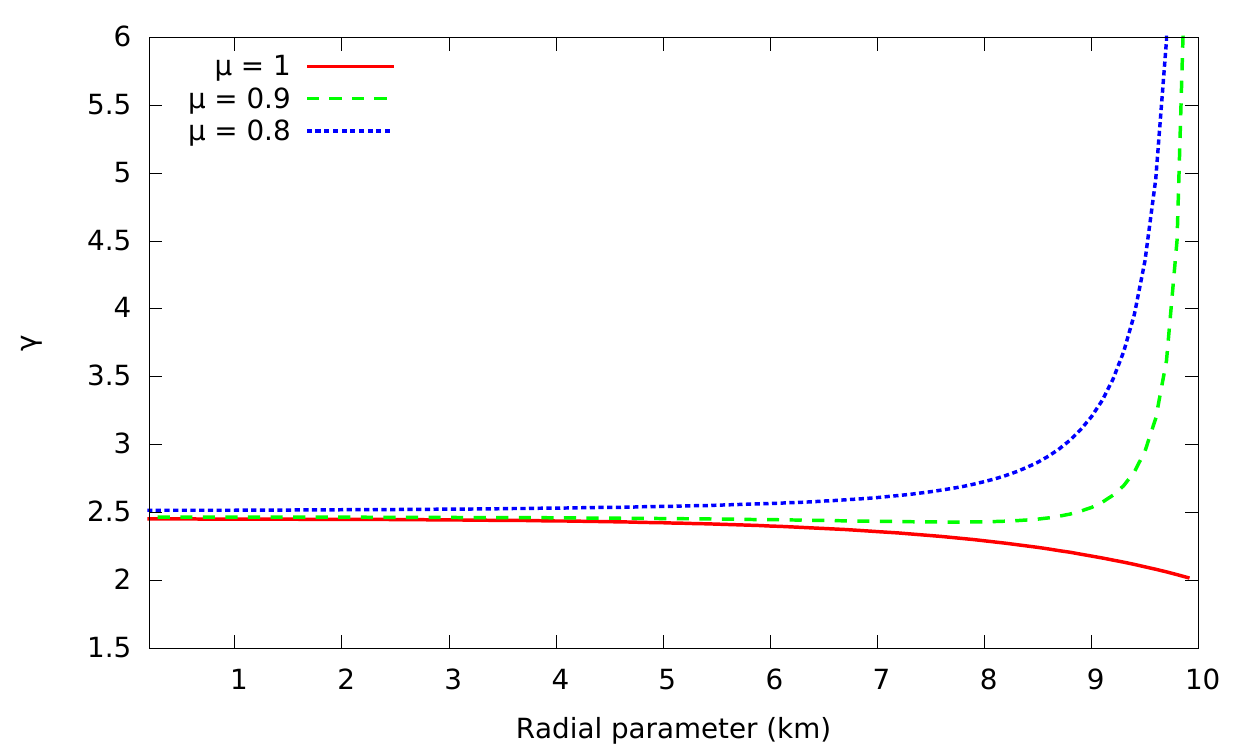}
\caption[The adiabatic index]{\label{t7.fig:gamma} (Colour online) The adiabatic index
  variation from the centre to the boundary of the star for different
  values of the parameter $\mu.$ The other parameter values are the
  same as those in FIG.~\ref{t7.fig:loglog}. }
\end{figure}

Models employing polytropic perfect fluids use similar values for the
adiabatic index \(\gamma,\) as what we find for a range of different
values of parameters \(\I.\) We show this in figure~\ref{t7.fig:gamma}
which treats \(\gamma\) as a continuous variable defined by \(\gamma =
\f{\rmd (\log p)}{\rmd (\log \rho)},\) and can be understood as the slope
of the previous log--log graph.

From this figure it becomes evident how both types of stars have an
interior structure well described by a polytrope with index close to
2.5.  The ``self-bound'' stars exhibit the existence of an envelope
consisting of material that is considerably stiffer than that found in
the interior.  Physically this is intuitive: for fixed \(\rho_{c}\)
and \(r_{b}\) the self bound stars will become more and more massive
as \(\mu\) decreases.  The increasing boundary density discontinuity
requires a stiffer exterior mass distribution to maintain the
equilibrium condition.

Now turning to the second way to characterize the EOS, concentrating
on the behaviour of the fluid material itself, independent of the
geometry of the star, we determine how different physical quantities
depend on the values of the central density \(\rho_{c}.\) The total
mass--energy is defined as,
\begin{equation}
M = 4 \pi \int_{0}^{r_{b}} \bar{r}^{2} \rho(\bar{r}) \rmd \bar{r} = 
\f{4 \pi r_{b}^{3} \rho_{c} \left( 5-3\mu \right)}{15}.
\end{equation}
The mass is important since it is the only directly and reliably
measurable quantity we obtain from neutron star observations.
Lattimer and Prakash~\cite{LatPra01, LatPra07, LatPra05} and
others~\cite{Gle92,Gle96} have ruled out certain EOS based on mass and
spin measurement of neutron stars.  The former have also used Tolman
VII, to constrain other EOS based on nuclear micro-physics, and have
even postulated that Tolman VII could be used as a guideline
discriminating between viable and non-viable EOS~\cite{LatPra05}.  If
this postulate is true, now that we have the complete Tolman VII EOS1,
we can apply the causality condition, independent of measurements
first, and compare with the previous references~\cite{Gle96,LatPra05}.

\begin{figure}[h!]
\includegraphics[width=\textwidth]{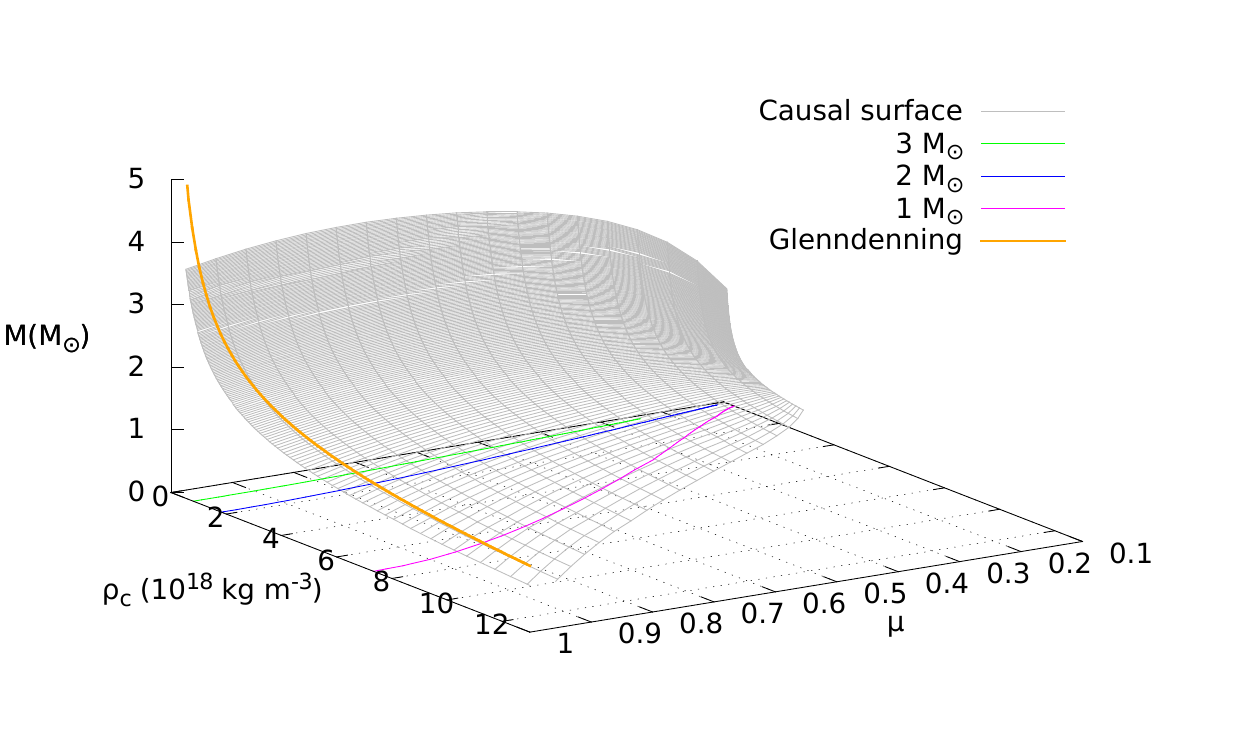}
\caption[The mass of possible stars]{\label{t7.fig:phase} (Colour online) The mass of possible stars
  just obeying causality. The grey surface obeys the equation
  $v_{s}(r=0) =\un{c}.$ Every point below the surface is a possible
  realization of a star, and we can potentially read off the mass,
  central density, and $\mu$ value of that star.  The numbered lines
  represent stars with the same mass that are causal, i.e.\ they are
  projections of the causal surface onto the $\rho_{c}$--$\mu$ plane.
  Glendenning's~\cite{Gle96} curve is shown in orange and represents a
  limit in the natural case only, and according to our results is
  acausal, being above our surface.  The $\mu = 1$ plane's
  intersection with our graph is the graph given in \cite{LatPra05},
  and here too our prediction is more restrictive.}
\end{figure}

We do this in figure~\ref{t7.fig:phase}, where we superimpose the result
of~\cite{Gle96}, on our own analysis of the whole solution space
\(\I.\) The surface shown is that of values at which the speed of
sound \(v_{s} = \left. \left(\sqrt{\rmd p / \rmd \rho} \right)
\right|_{r=0},\) at the centre of the fluid sphere just reaches the
speed of light.  This is a sufficient condition for the solution to be
causal since \(v_{s}\) is a monotonically decreasing function of \(r\)
in the sphere.  Any point located below this surface has coordinate
values for \(M, \rho_{c},\) and \(\mu\) that represents a valid
\textbf{causal} solution to the Tolman~VII differential equations.  The
orange line is the previous result obtained by Glendenning~\cite{Gle96}
from rotational considerations.

Imposing causality to constrain the parameter space \(\I,\) is not a
new idea.  However having an explicit EOS allows one to easily
generate the causal surface shown above in figure~\ref{t7.fig:phase}.

\begin{figure}[h!]
  \includegraphics[width=\textwidth]{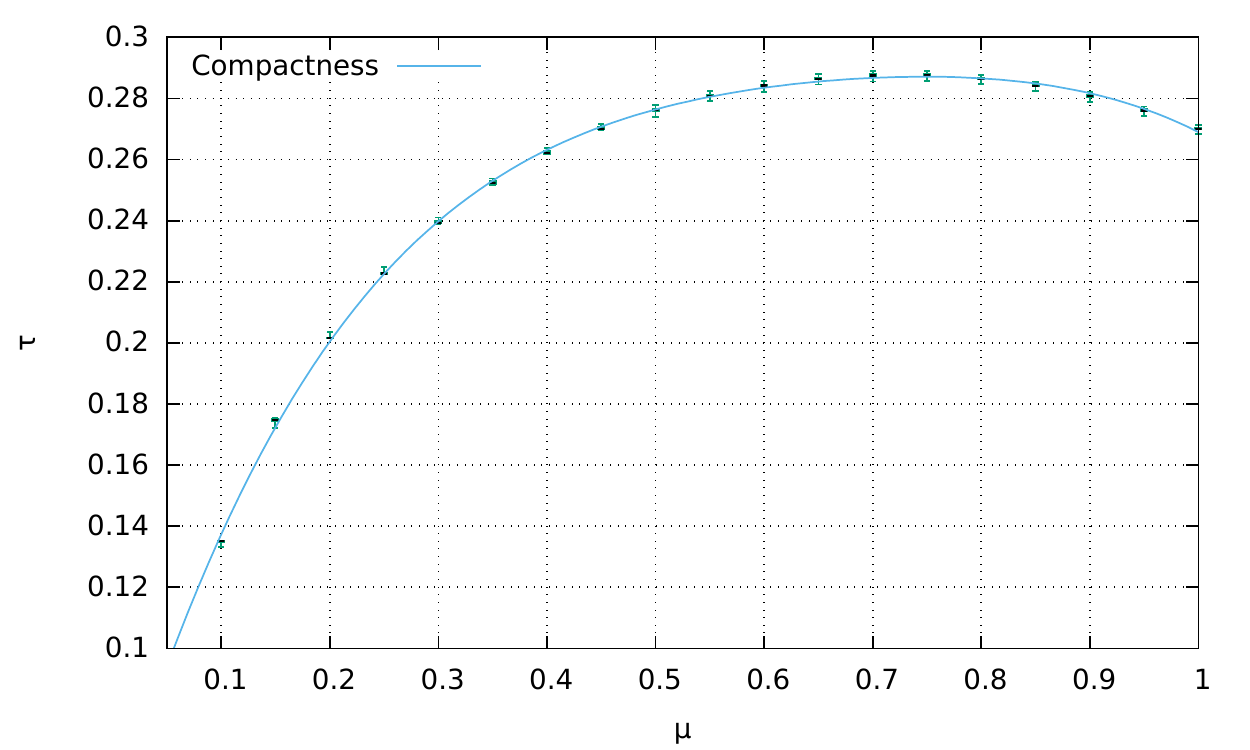}
\caption[The compactness as a function of the
self-boundness]{\label{t7.fig:beta} (Colour online) The compactness as a
  function of the self-boundness parameter $\mu$.  This plot was
  generated by varying $r_{b}$ from 4 km to 20 km for fixed $\mu$ and
  finding $\rho_{c}$ and subsequently the compactness each time, such
  that the sound speed was causal at the centre of the star.  The
  curve shown is a polynomial fit, and the box-and-whisker plots(very
  small in green) show the variation of $\tau$ for fixed $\mu$, but
  different $r_{b}.$ The very small whiskers justify the pertinence of
  $\tau$ as a useful measure in the analysis of the behaviour of the
  model.}
\end{figure}
  
Previously the usual way to denote different EOS2 has been to
calculate the compactness ratio, given by \(\label{t7.eq:compactness}
\tau = \f{\un{G} M}{\un{c^{2}}r_{b}}.\) We found that even in the
case of Tolman VII, this is a stable quantity to characterize a star
since the values of \(\tau\) for large parameter variations \(\I\) is
relatively constant.  This means that even though we might change the
value for \(\I\) of the stars, the ones bordering on causality share
very similar compactness, albeit one that is lower than that
previously thought possible.  We show how this compactness \(\tau,\)
varies with \(\mu,\) in figure~\ref{t7.fig:beta}.  The previous maximal
compactness was about 0.34 from rotational and causality
criteria~\cite{LatPra05}.  Our analysis shows that \(\tau\) should be
below 0.3 for all possible stars, if Tolman~VII is a valid physical
model for stars.

\begin{figure}[h!]
\begin{center}
\includegraphics[width=\textwidth]{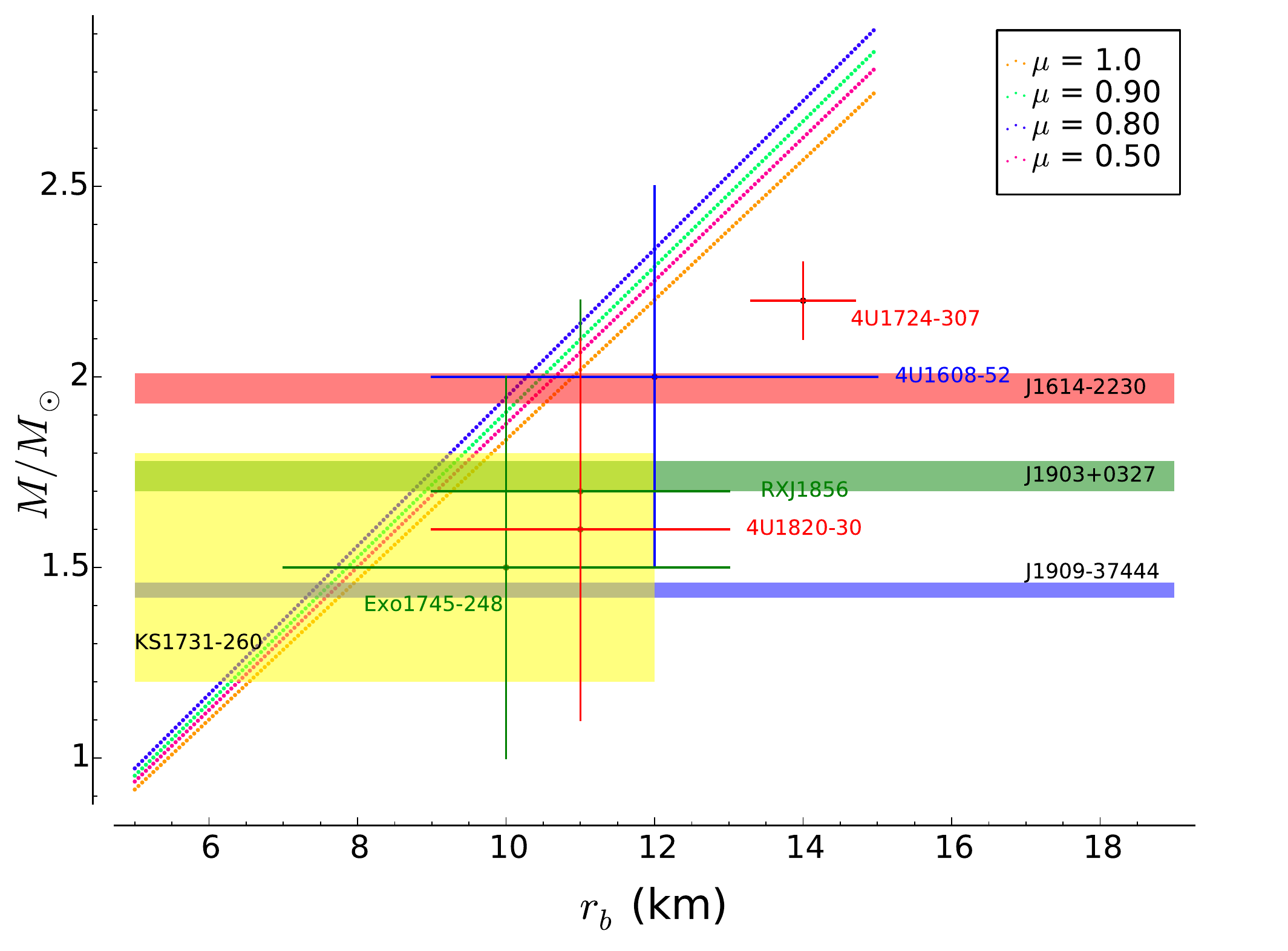}
\caption[The mass $M$ in solar units
  versus radius $r_{b}$ in kilometres]{\label{t7.fig:expt} (Color online) The mass \(M\) in solar units
  versus radius \(r_{b}\) in kilometres of a few stars for which these
  values have been measured. We use error bars to denote observational
  uncertainties, and coloured bands in the case where only the mass is
  known.}
\end{center}
\end{figure}

Recently measurements of the radius of a limited number neutron stars
have been
obtained~\cite{OzeGuvPsa08,OzePsa09,GuvWroCam10,GuvOzeCab10,OzeGouGuv12,SulPou11,SteLatBro10}. These
are shown along with some other stars of known mass in
figure~\ref{t7.fig:expt}. We also superimpose a few of the limiting
causal curves obtained for different values of \(\mu\) from Tolman
VII, to show that Tolman VII is not ruled out by observational
results, even though it predicts lower compactness than most nuclear
models.  However the lines shown are on the edge of causality, that is
they are the counterparts of those on the surface of
figure~\ref{t7.fig:phase}.  Since all observations of compactness are
bounded by the most extreme Tolman VII model we claim that the
solution is actually realized by compact stars in nature.

\section{Conclusion}\label{t7.sec:conclusion}
Thus a complete analysis of the Tolman VII solution was carried out
and it was found that it is a completely valid solution with a huge
potential for modelling physical objects.  The EOS this solution
predicts has been found, and in certain regimes behaves very much like
a polytrope with an adiabatic index of 2.5. Using the EOS, we are able
to compute the speed of pressure waves, and imposing causality on the
latter results in a more restrictive limit on the maximum compactness
of fluid spheres allowable by classical general relativity.  The
solution is also stable under radial perturbations, since the speed of
these pressure waves is finite and monotonically decreasing from the
centre outwards, thus satisfying the stability criterion
in~\cite{AbrHerNun07}.  If we believe as in ref~\cite{LatPra05} that
Tolman VII is an upper limit on the possible energy density
\(\rho_{c},\) for a given mass \(M,\) some known
models~\cite{LatPra07} will have to be reconsidered.





\chapter{New Solutions} \label{C:NewSolutions}
\begin{myabstract}
  We solve our coupled system of differential equations, under two
  different assumptions, and deduce expressions for the metric
  functions, and pressures.  We then apply boundary conditions to
  these solutions and deduce all integration constants in terms of
  parameters that are physically meaningful.  We then look at
  possibilities for using similar methods for finding new solutions.
\end{myabstract}

Following the exposition of the Tolman~VII~\cite{Tol39} solution in
the previous chapter, we now generalize this solution to generate new
exact solutions to the Einstein's interior equations.  We feel that
Tolman~VII is a good candidate for such a generalization procedure
since by itself Tolman VII obeys conditions for physical viability.
Presumably, generalizations of the solution that maintain this
physical viability will be possible, and this is what we attempt to do
in this chapter.

This chapter has two major sections.  In Section~\ref{ns.sec:Ani} we
generalize the field equations to include an anisotropic pressure,
while maintaining spherical symmetry.  This has the advantage of
introducing one additional degree of freedom in the types of functions
we can posit for the matter quantities, thus making the generalization
straightforward.  In Section~\ref{ns.sec:Cha} we solve the
Einstein--Maxwell system by including electric charge in our matter
quantities.  Charged models might seem like a strange concept since it
is expected that astrophysical objects will be charge neutral.
However it is still interesting to see what kind of additional
structure charge introduces in stellar models.  Finally in
section~\ref{ns.sec.Conclusion} we tentatively suggest avenues for
finding new solutions by using similar methods.

\section{Uncharged case with anisotropic pressures}\label{ns.sec:Ani}
In this section, we generalize the Tolman~VII solution by introducing
an anisotropic pressure.  In Appendix~\ref{C:AppendixA} we explain how
the energy-momentum tensor \(T_{ab}\) changes under this new
assumption: the components of the pressure, which we assumed to be the
same in all directions must now be generalized to two different
function, which for intuitive reasons we will call \(p_{r}\) for the
radial pressure component, and \(\ppen\) for the angular pressure
component.  Our starting metric functions do not change from the
original ones, since we are not relaxing our spherical symmetry axiom.
As a result of these, our energy-momentum now becomes
\begin{equation}
  T^{i}{}_{j} =
  \begin{pmatrix}[c]
    \rho   & 0  & 0  & 0  \\
    0     & -p_{r}  & 0  & 0  \\
    0     & 0  & -\ppen & 0  \\
    0     & 0  & 0  & -\ppen 
  \end{pmatrix},
\end{equation}
and the EFE corresponding to the above reduce to the following set
\begin{subequations}
  \label{ns.eq:EinR}
  \begin{alignat}{3}
    \label{ns.eq:EinR1}
    \kappa \rho &= \e^{-\lambda}\left( \f{\lambda'}{r} -\f{1}{r^2}\right) +\f{1}{r^{2}} 
    &&= \f{1}{r^{2}} - \f{Z}{r^{2}} - \f{1}{r}\deriv{Z}{r}, \\
    \label{ns.eq:EinR2}
    \kappa p_{r} &= \e^{-\lambda} \left( \f{\nu'}{r} + \f{1}{r^2}\right) -\f{1}{r^2} 
    &&= \f{2Z}{rY}\deriv{Y}{r} + \f{Z}{r^{2}} - \f{1}{r^{2}},\\
    \label{ns.eq:EinR3}
    \kappa \ppen &= \e^{-\lambda} \left( \f{\nu''}{2} - 
      \f{\nu'\lambda'}{4} + \f{(\nu')^2}{4} + \f{\nu'- \lambda'}{2r}\right)  
    &&= \f{Z}{Y}\sderiv{Y}{r} + \f{1}{2Y}\deriv{Y}{r}\deriv{Z}{r} + \f{Z}{rY}\deriv{Y}{r} + \f{1}{2r}\deriv{Z}{r}.
  \end{alignat}
\end{subequations}
We note that equation~\eqref{ns.eq:EinR3} is different from the
previous~\eqref{t7.eq:EinR3}, since as we now have two pressure
components, this third equation of this set is in terms of the new
pressure.  To find the solution of these ODEs, we will follow a
similar method to the previous chapter, to be able to get a solution
of the same form.  In particular the ansatz for the
density~\eqref{t7.eq:Density} that we used previously will be the
same.  As a result the first ODE is solved in the exact same way as
the previous chapter.  The boundary conditions will be expressed in
the exact same way as in chapter~\ref{C:TolmanVII}, a non-intuitive
result we will show in due course. Schematically, we have:
\begin{equation}
  \label{ns.eq:ZSol}
  \rho = \rho_{c} \left[ 1 - \mu \left( \f{r}{r_{b}}\right)^{2}
\right] \quad \longrightarrow \quad Z(r) = 1 - \left( \f{\kappa
    \rho_{c}}{3}\right) r^{2} + \left(\f{\kappa \mu
    \rho_{c}}{5r_{b}^{2}}\right) r^{4} \eqqcolon 1- br^2+ ar^4.
\end{equation}
The solution to the second and third equation is complicated by the
inequality of the two equations~\eqref{ns.eq:EinR2} and~\eqref{ns.eq:EinR3}.
In Tolman~VII, we equated these two equations:~\eqref{t7.eq:EinR1}
and~\eqref{t7.eq:EinR3}, but here we are forced to take the difference
between the two, and call the new quantity the ``measure of
anisotropy'' \(\Delta:\)
\begin{equation}
  \label{ns.eq:DeltaDef}
  \kappa \Delta = \kappa(p_{r} - \ppen) = \f{Z}{rY}\left( \deriv{Y}{r}\right)
  -\f{Z}{Y}\left( \sderiv{Y}{r} \right) -\f{1}{2Y}\left( \deriv{Z}{r}\right)\left( \deriv{Y}{r}\right) - \f{1}{2r}\left( \deriv{Z}{r}\right) + \f{Z}{r^{2}} - \f{1}{r^{2}}.
\end{equation}
This equation can be rearranged and simplified into a second order ODE
for \(Y,\) which can then be solved with our usual series of variable
transformations:
\begin{equation}
  \label{ns.eq:YDiffR}
2r^{2}Z \left(\sderiv{Y}{r}\right)  + 
\left[ r^{2}\left(\deriv{Z}{r} \right) -2rZ \right] \left( \deriv{Y}{r}\right) +
\left[ 2+2r^{2}\Delta -2Z +r \deriv{Z}{r}\right] Y = 0.   
\end{equation}
The second order ODE will have \(\Delta\) as an undetermined function,
which when set to zero transforms the ODE into the Tolman~VII one for
\(Y\) we had solved previously: in this aspect this is a
generalization of the Tolman~VII solution.  The next step in the
solution is the variable transformation \(x = r^{2},\) where care must
be taken to transform the derivatives to the appropriate form.  A
straight forward derivation yields \(\deriv{}{r} \equiv
2\sqrt{x}\deriv{}{x},\) and similarly \(\sderiv{}{r} \equiv
4x\sderiv{}{x} + 2 \deriv{}{x}.\) Applying these to the above
equation~\ref{ns.eq:YDiffR} results in
\begin{multline*}
  2xZ \left( 4x \sderiv{Y}{x} + 2 \deriv{Y}{x}\right) + \left( 2
    \sqrt{x} \deriv{Y}{x} \right)
  \left( 2 x^{3/2} \deriv{Z}{x} - 2 \sqrt{x}Z\right) \\
  + \left[ 2 + 2x\Delta +\sqrt{x}\left(2\sqrt{x}\deriv{Z}{x}\right) -2Z \right]Y = 0,
\end{multline*}
which can be rearranged into
\[8x^{2}Z \sderiv{Y}{x} + \left( \cancel{4xZ} + 4x^{2} \deriv{Z}{x} \cancel{-4xZ} \right) \deriv{Y}{x} + 2\left( 1+ x\Delta +x\deriv{Z}{x}-Z \right)Y = 0,\]
a clear simplification of some of the cross terms appearing in the
coefficient of the first derivative of \(Y.\) At this stage, dividing
by \(8x^{2}\) will tidy up our equation into
\begin{equation}
  \label{ns.eq:YDiffX}
  Z \sderiv{Y}{x} + \left(\f{1}{2} \deriv{Z}{x} \right) \deriv{Y}{x} + \left( \f{1+ x\Delta +x\deriv{Z}{x}-Z}{4x^{2}} \right) Y = 0.
\end{equation}

The second step of the solution procedure involves another variable
change from \(x\) to \(\xi\) which is defined through
\begin{equation}
  \label{ns.eq:DefXi}
  \xi = \int_{0}^{x} \f{\d \bar{x}}{\sqrt{Z(\bar{x})}} \Rightarrow \deriv{\xi}{x} = \f{1}{\sqrt{Z}}.
\end{equation}
This induces a change in the \(x-\)derivatives, so that we have
\(\deriv{}{x} \equiv \f{1}{\sqrt{Z(x)}}\deriv{}{\xi}\), and
\(\sderiv{}{x} \equiv \f{1}{Z} \sderiv{}{\xi} - \f{1}{2Z^{3/2}}
\deriv{Z}{x} \deriv{}{\xi}.\) The actual expression for \(\xi\) in
terms of \(x\) will be derived later on when it becomes useful.

Applying these changes to our differential equation~\eqref{ns.eq:YDiffX}
results in the elimination of the first derivative term for \(Y,\)
further simplifying the second order ODE:
\[ Z \left\{ \f{1}{Z} \sderiv{Y}{\xi} - \cancel{ \f{1}{2Z^{3/2}} \deriv{Z}{x} \deriv{Y}{\xi}}  \right\} 
+\cancel{ \f{1}{2} \deriv{Z}{x} \left( \f{1}{\sqrt{Z}} \deriv{Y}{\xi}\right) } 
+ \left( \f{1+ x\Delta +x\deriv{Z}{x}-Z}{4x^{2}} \right) Y = 0.\]

At this stage, except for the coefficient of \(Y,\) we have a simple
equation.  However from~\eqref{ns.eq:ZSol} we already have expressions
for both \(Z(x)\) and \(\deriv{Z}{x}\) with which we can reduce that
last coefficient into a simple form consisting of our initial
parameters only, yielding \[ \left( \f{1+ x\Delta +x\deriv{Z}{x}-Z}{4x^{2}} \right) \rightarrow 
\left( \f{1+x\Delta+x(ax-b) - (1-bx+ax^{2})}{4x^{2}} \right) = \left( \f{a}{4} + \f{\Delta}{4x} \right),\]
so that the ODE to be solved for \(Y\) finally becomes
\begin{equation}
  \label{ns.eq:YdiffXi2}
  \sderiv{Y}{\xi} + \left( \f{a}{4} + \f{\Delta}{4x}\right)Y = 0.
\end{equation}
This equation would be very easy to solve if we had a constant term
for the coefficient in brackets.  As mentioned previously, \(\Delta\)
is a function we can pick and is a measure of anisotropy between the
pressures in our model.  From spherical symmetry we must have both the
radial pressure \(p_{r}\) and the tangential pressure \(\ppen\) be
equal at the centre, resulting in \(\Delta\) having to be equal to
zero when \(x=r^{2}=0.\) The energy conditions impose additional
constraints on the absolute value that the pressures can take, and we
will have to ensure compliance with the energy conditions later when
we have the complete expression for both pressures.  However, the
requirement that \(\Delta(r=0)=0\) suggests that setting \(\Delta =
\beta x\) might be a good candidate for a physical solution since one
of the constraints is automatically taken care of, while considerably
simplifying our ODE.  Imposing this results in a simple harmonic ODE:
\[\sderiv{Y}{\xi} + \left( \f{a+\beta}{4}\right)Y = 0,\] whose solutions
we can write immediately in terms of \(\phi^{2} = (a+\beta)/4\) in the
following table, which also redirects us to the relevant section where
the specific solution is looked into in detail.

\begin{table}[!h]
\label{ns.tab:AniSolns}
\centering 
\begin{tabular}{| >{$}c<{$} | >{$}c<{$} | l |}
  \hline
    \phi^2 & Y(\xi) & Solution's analysis \\
    \hline
    \phi^2 < 0 & c_1 \cosh{\left(\sqrt{-\phi^2}\xi\right)} + c_2 \sinh{\left(-\sqrt{-\phi^2}\xi\right)} & section~\ref{ns.ssec:phiNeg} \\
    \phi^2 = 0 & c_1 + c_2 \xi & section~\ref{ns.ssec:phiZero}\\
    \phi^2 > 0 & c_1 \cos{\left( \phi \xi \right)} + c_2 \sin{\left(\phi \xi\right)} & section~\ref{ns.ssec:phiPos}\\
\hline
  \end{tabular}
  \caption[The different solutions with different $\phi^{2}$s]{The different solutions that can be generated through different values of the parameter~$\phi.$ The integration constants $c_{1},$ and $c_{2}$ are determined by our two boundary conditions.}
\end{table}

In the next sections we will analyse the different possibilities
offered by this extension to anisotropic pressures, considering the different ones separately.

\subsection{The $ \phi^{2} = 0 $ case}\label{ns.ssec:phiZero}
\begin{figure}[h!]
  \centering
  \includegraphics[width=\linewidth]{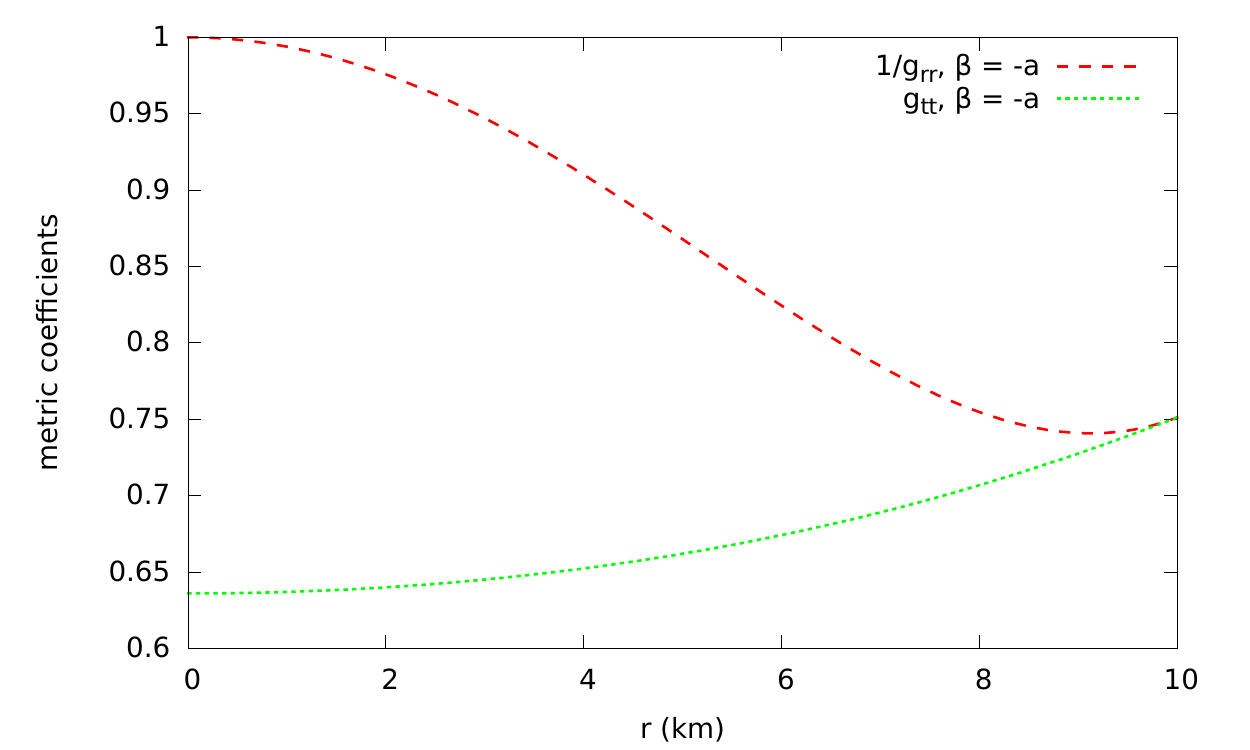}
  \caption[Matching the metric functions for
  $\phi =0$ (Anisotropy only)]{Application of the boundary conditions resulting in the
    value and slope matching of the metric function at $r=r_{b}.$ for
    the $\phi =0$ case. The parameter values are
    $\rho_{c}=\un{1\times 10^{18}\,kg \cdot m^{-3}}, r_b = \un{1
      \times 10^{4}\,m}$ and $\mu = 1$}
\label{ns.fig:MetricPhiZ}
\end{figure}

When \(\phi = 0,\) the only possibility is for
\(\beta = -a = -\f{\kappa\mu\rho_{c}}{5r_{b}^{2}},\) which is either
negative when all the constants in the previous expression are
positive definite: the case we will consider now, or zero when
\(\mu = 0\).  The latter case reduces to the Schwarzschild interior
solution on which there is much historical~\cite{Tol66,Wey52} and
contemporary literature~\cite{Wal84, Inv92}, and so we will not look
at it in detail.  For the \(\beta \neq 0\) case, we have
\(\ppen = p_{r} - \Delta = p_{r} + ax,\) and the angular pressure is
thus larger than the radial pressure everywhere but at the centre.  We
now apply our two boundary conditions to solve for the integration
constants.  From last chapter's arguments, and remembering that the
boundary conditions come from imposing matching conditions on the
interior and exterior metric through the use of the equation relating
pressure and density~\eqref{t7.eq:EinR1+2} which is unchanged even in
the anisotropic case, we have
\begin{itemize}
\item \(\left. \deriv{Y}{\xi} \right|_{\xi=\xi_{b}} = \alpha,\) where
  we can compute the \(\xi-\)derivative for \(Y\) from its expression. This results in \(c_{2} = \alpha\).
\item \(\left. Y \right|_{\xi=\xi_{b}} = \gamma \Rightarrow c_{1} +
  c_{2}\xi_{b} = \gamma,\) as a result of which we have \(c_{1} = \gamma - \f{2 \alpha}{\sqrt{b}} 
\acoth{\left( \f{1+\gamma}{r^{2}_{b} \sqrt{b}} \right). }\)
\end{itemize}
A plot of the metric functions~\ref{ns.fig:MetricPhiZ} will show the
matching of the values and slopes of the metric functions at the
radius \(r_{b},\) as expected from the matching to the Schwarzschild
exterior metric.

We can now give expressions for all quantities, since our system of
equations has been completely solved.  
Starting with the density ansatz,
\begin{equation}
  \label{ns.eq:PhiZDensity}
  \rho(r)  = \rho_{c} \left[ 1 - \mu \left( \f{r}{r_{b}}\right)^{2}\right], 
\end{equation}
which leads to an expression for the first metric function \(Z(r),\)
\begin{equation}
  \label{ns.eq:PhiZZ}
  Z(r) = 1 - \left( \f{\kappa
    \rho_{c}}{3}\right) r^{2} + \left(\f{\kappa \mu
    \rho_{c}}{5r_{b}^{2}}\right) r^{4}.
\end{equation}
The solution of the second metric function after the variable changes
and substitutions give
\begin{equation}
  \label{ns.eq:PhiZY}
  Y(r) = \gamma + \f{2\alpha r_{b}}{\sqrt{\kappa \rho_{c}\mu /5}}\left[ 
  \acoth{\left(\f{1-\sqrt{Z(r)}}{r^{2}\sqrt{\f{\kappa \rho_{c} \mu} {5r_{b}^{2}}}}\right) }  -\acoth{\left( \f{1-\gamma}{r_{b}\sqrt{\kappa \rho_{c} \mu /5}}\right)}  \right],
\end{equation}
where the constants \(\alpha\) and \(\gamma\) are given in terms of
the initial set of parameters \(r_{b}, \rho_{c}\) and \(\mu\) through
\begin{align}
  \label{ns.PhiZalphaGamma}
\alpha &= \f{1}{4} \left( \f{\kappa \rho_{c}}{3} - \f{\kappa\rho_{c}\mu}{5} \right)  &&= \f{\kappa \rho_{c}(5-3\mu)}{60},&\\
\gamma &= \sqrt{1 - \left( \f{\kappa
    \rho_{c}}{3}\right) r_{b}^{2} + \left(\f{\kappa \mu
    \rho_{c}}{5r_{b}^{2}}\right) r_{b}^{4}} &&= \sqrt{1+ \f{\kappa \rho_{c}r_{b}^{2}(3\mu-5)}{15} },&\\
\beta  &=-a &&= - \f{\kappa \rho_{c} \mu}{5r_{b}^{2}}.&
\end{align}
The two pressures can similarly be given in terms of the above
variables.  The radial pressure can be computed from the second
Einstein equation~\ref{ns.eq:EinR2} in a straightforward manner to yield
\begin{multline}
  \label{ns.eq:PhiZpr}
  \kappa p_{r}(r) = \f{2\kappa\rho_{c}}{3} - \f{4\kappa\rho_{c}\mu r^{2}}{5r_{b}^{2}} -\kappa\rho_{c} \left[ 1 - \mu \left( \f{r}{r_{b}}\right)^{2}\right] + \\
+ \left( \f{\kappa\rho_{c}}{3} - \f{\kappa \rho_{c} \mu}{5}\right) \f{\sqrt{1-\f{\kappa\rho_{c}}{3}r^{2} + \f{\kappa\mu\rho_{c}}{5r_{b}^{2}} r^{4}}}{\gamma + \f{2\alpha r_{b}}{\sqrt{\kappa \rho_{c}\mu /5}}\left[ 
  \acoth{\left(\f{1-\sqrt{Z(r)}}{r^{2}\sqrt{\f{\kappa \rho_{c} \mu} {5r_{b}^{2}}}}\right) }  -\acoth{\left( \f{1-\gamma}{r_{b}\sqrt{\kappa \rho_{c} \mu /5}}\right)}  \right]},
\end{multline}
and similarly the tangential pressure is easily written in terms of the above as
\begin{equation}
  \label{ns.eq:PhiZpt}
  \ppen(r) =p_{r} - \beta r^{2} = p_{r} + \f{\kappa\rho_{c}\mu}{5r_{b}^{2}} r^{2}. 
\end{equation}
This completes the solution, since we have given all the functions in
our ODEs in terms of the constants found in our ansatz and our
coordinate variable only.  If an equation of state for this solution
is required, we could invert the density
relation~\eqref{t7.eq:Density}, to get an expression for \(r\) in
terms of \(\rho.\) Simple substitution in the expressions we have for
the pressures~\eqref{ns.eq:PhiZpr} and~\eqref{ns.eq:PhiZpt} will then
give us the equation of state for both pressures \(p_{t}(\rho),\) and
\(\ppen(\rho),\) a process similar to what we did in the previous
chapter.

\subsection{The $ \phi^{2} > 0 $ case}\label{ns.ssec:phiPos}
When \(\phi^{2} > 0,\) we must have that \(a + \beta > 0,\) which can
only mean that \(\beta > -a.\) Since we have an expression for \(a,\)
we get \(\beta > - \f{\kappa \mu \rho_{c}}{5r_{b}^{2}}, \) which
allows \(\beta\) to have negative values, since the fraction in the
last expression is positive definite.  We can also write expressions
for the derivative of \(Y\) by direct computation, which will allow us
to apply boundary conditions to solve for our integration constants as
we show now:
\begin{itemize}
\item \(\left. \deriv{Y}{\xi} \right|_{\xi=\xi_{b}} =  \phi \left[ c_{2} \cos{(\phi \xi_{b})} - c_{1} \sin{(\phi \xi_{b})} \right] =\alpha,\) and solving this results in an equation for \(c_{1}\) and \(c_{2}\) in the form of, 
\( c_{2} \cos{(\phi \xi_{b})} - c_{1} \sin{(\phi \xi_{b})}  = \f{\alpha}{\phi},\) and,
\item \(\left. Y \right|_{\xi=\xi_{b}} = \gamma \Rightarrow c_{2} \sin{(\phi \xi_{b})} +
  c_{1}\cos{(\phi \xi_{b})} = \gamma.\) 
\end{itemize}
We solve this coupled system for \(c_{1}\) and \(c_{2}\) by the usual
process of elimination by multiplication by the appropriate
trigonometric function, and this yields
\begin{align*}
  c_{2} &= \gamma \sin{(\phi\xi_{b})} + \f{\alpha}{\phi} \cos{(\phi\xi_{b})}\\
  c_{1} &= \gamma \cos{(\phi\xi_{b})} - \f{\alpha}{\phi} \sin{(\phi\xi_{b})},
\end{align*}

A plot of the metric functions~\ref{ns.fig:MetricPhiP} at this point
will show the matching of of the values and slopes of the metric
functions at the radius \(r_{b},\) as expected from the Schwarzschild
metric:
\begin{figure}[h!]
  \centering
  \includegraphics[width=\linewidth]{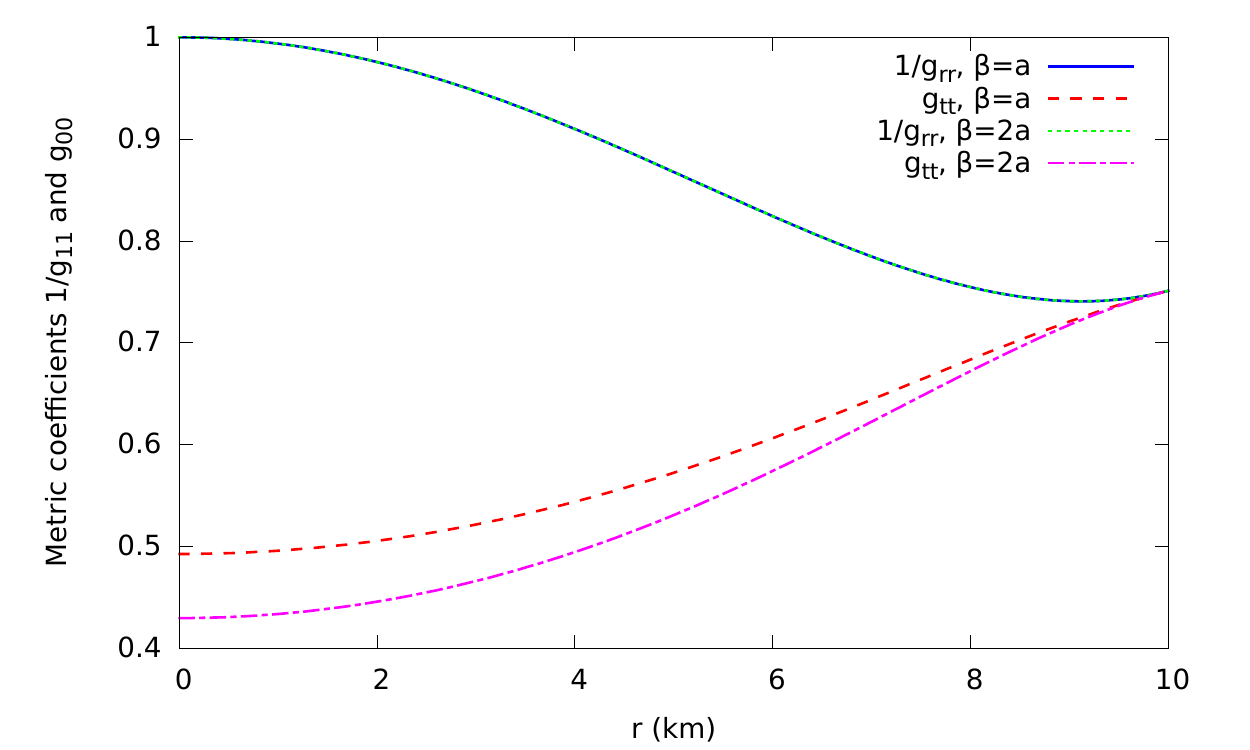}
  \caption[Matching the metric functions for $\phi > 0$ (Anisotropy only)]{Application of the
    boundary conditions resulting in the value and slope matching of
    the metric function at $r=r_{b}.$ for the $\phi > 0$ case.  The parameter values are
    $\rho_{c}=\un{1\times 10^{18}\,kg \cdot m^{-3}}, r_b = \un{1 \times
      10^{4}\,m}$ and $\mu = 1,$ with $\beta$ given in the legend.}
\label{ns.fig:MetricPhiP}
\end{figure}
The complete solution for the \(Y-\)metric function in this case is thus
  \begin{multline}
\label{ns.eq:YphiP}
    Y(r) = \left( \gamma \cos{(\phi\xi_{b})} - \f{\alpha}{\phi} \sin{(\phi\xi_{b})} \right) 
    \cos{\left( \f{2\phi}{\sqrt{a}} \coth^{-1} \left( \f{1+\sqrt{1-br^2+ar^4}}{r^2\sqrt{a}}\right)  \right)} +\\
    + \left(\gamma \sin{(\phi\xi_{b})} + \f{\alpha}{\phi} \cos{(\phi\xi_{b})} \right) 
    \sin{\left(\f{2\phi}{\sqrt{a}} \coth^{-1} \left( \f{1+\sqrt{1-br^2+ar^4}}{r^2\sqrt{a}}\right) \right)},
  \end{multline}
which then allows us to write the matter variables \(\ppen\) and \(p_{r}\) as
\begin{multline}
  \label{ns.eq:PhiPpr}
\kappa p_{r}(r) = \f{2\kappa\rho_{c}}{3} - \f{4\kappa\rho_{c}\mu r^{2}}{5r_{b}^{2}} -\kappa\rho_{c} \left[ 1 - \mu \left( \f{r}{r_{b}}\right)^{2}\right] + 4 \phi\sqrt{1-br^{2}+ar^{4}} \times \\
\times \f{\left[ \gamma \sin{(\phi\xi_{b})} + \f{\alpha}{\phi} \cos{(\phi\xi_{b})} \right] \cos{(\phi \xi)} - \left[ \gamma \cos{(\phi\xi_{b})} - \f{\alpha}{\phi} \sin{(\phi\xi_{b})} \right] \sin{(\phi \xi)} }{\left[ \gamma \sin{(\phi\xi_{b})} + \f{\alpha}{\phi} \cos{(\phi\xi_{b})}\right] \sin{(\phi \xi)} + \left[\gamma \cos{(\phi\xi_{b})} - \f{\alpha}{\phi} \sin{(\phi\xi_{b})} \right]\cos{(\phi \xi)}},  
\end{multline}
and 
\begin{equation}
  \label{ns.eq:PhiPpt}
  \ppen(r) =p_{r} - \beta r^{2}. 
\end{equation}
The variables in the above expressions for this case are given by :
\begin{align*}
  \alpha &= \f{\kappa\rho_{c}\left( 5-3\mu\right)}{60}, &\quad& \beta > -\f{\kappa\mu\rho_{c}}{5r_{b}^{2}}, \\
  \gamma &= \sqrt{1+ \f{\kappa \rho_{c}r_{b}^{2}(3\mu-5)}{15} }, &\quad& \phi^{2} =\f{3\beta + 4\kappa\rho_{c}}{12}, 
\end{align*}
which completes the solution.  As with the previous examples, and in
particular Tolman~VII, we can invert the density relation and generate an equation of state.

\subsection{The $ \phi^{2} < 0 $ case}\label{ns.ssec:phiNeg} 
When \(\phi^{2} < 0,\) we must have that \(a + \beta < 0,\) which can
only mean that \(\beta < -a.\) Since we have an expression for \(a,\)
we get \(\beta < - \f{\kappa \mu \rho_{c}}{5r_{b}^{2}}, \) which
forces \(\beta\) to have negative values only, since the fraction in the
last expression is positive definite.  We can also write expressions
for the derivative of \(Y\) by direct computation, which will allow us
to apply boundary conditions to solve for our integration constants as
we show now:
\begin{itemize}
\item \(\left. \deriv{Y}{\xi} \right|_{\xi=\xi_{b}} =  \phi \left[ c_{2} \cosh{(\phi \xi_{b})} + c_{1} \sinh{(\phi \xi_{b})} \right] =\alpha,\) and solving this results in an equation for \(c_{1}\) and \(c_{2}\) in the form of, 
\( c_{2} \cosh{(\phi \xi_{b})} + c_{1} \sinh{(\phi \xi_{b})}  = \f{\alpha}{\phi},\) and,
\item \(\left. Y \right|_{\xi=\xi_{b}} = \gamma \Rightarrow c_{2} \sinh{(\phi \xi_{b})} +
  c_{1}\cosh{(\phi \xi_{b})} = \gamma.\) 
\end{itemize}
We solve this coupled system for \(c_{1}\) and \(c_{2}\) by the usual
process of elimination by multiplication by the appropriate
trigonometric function, and this yields
\begin{align*}
  c_{2} &= \f{\alpha}{\phi} \cosh{(\phi\xi_{b})} - \gamma \sinh{(\phi\xi_{b})}  \\
  c_{1} &= \gamma \cosh{(\phi\xi_{b})} - \f{\alpha}{\phi} \sinh{(\phi\xi_{b})},
\end{align*}

\begin{figure}[h!]
  \centering
  \includegraphics[width=\linewidth]{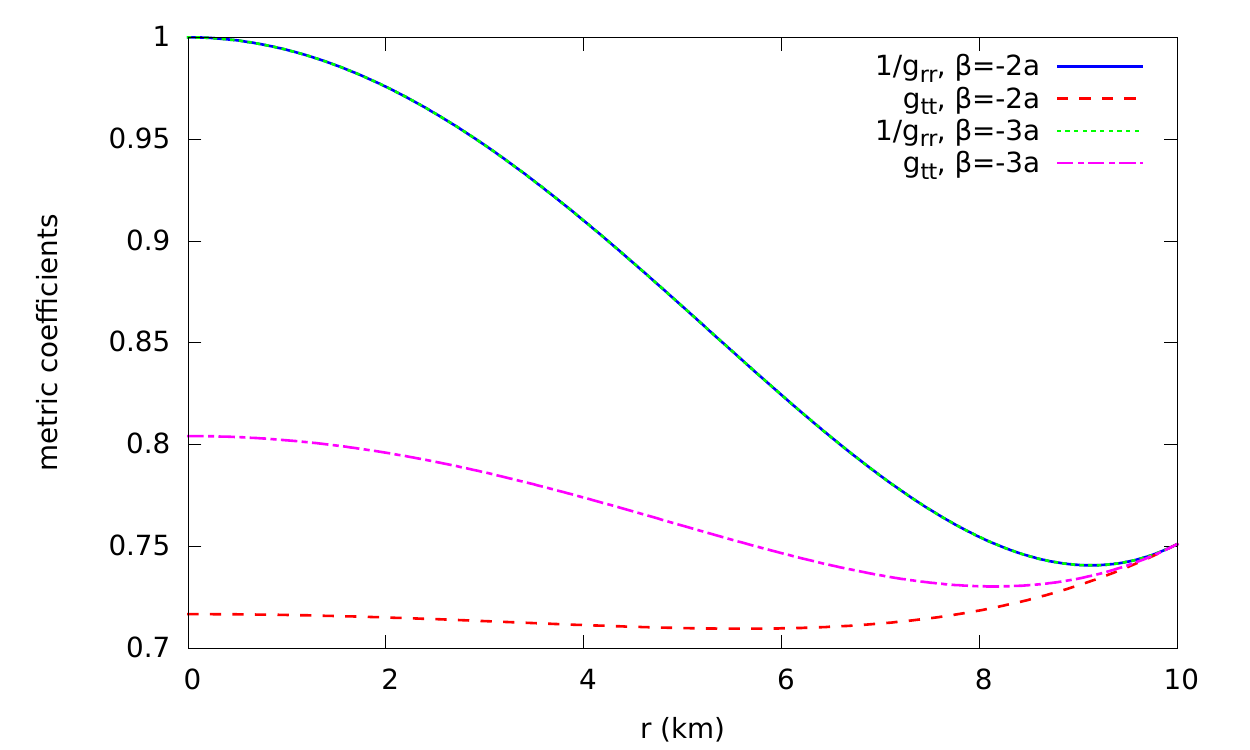}
  \caption[Matching the metric functions for $\phi < 0$ (Anisotropy only)]{Application of the
    boundary conditions resulting in the value and slope matching of
    the metric function at $r=r_{b}.$ for the $\phi < 0$ case.  The parameter values are
    $\rho_{c}=\un{1\times 10^{18}\,kg \cdot m^{-3}}, r_b = \un{1 \times
      10^{4}\,m}$ and $\mu = 1,$ with $\beta$ given in the legend.}
\label{ns.fig:MetricPhiN}
\end{figure}

A plot of the metric functions at this point will show the matching of
of the values and slopes of the metric functions at the radius
\(r_{b},\) as expected from the Schwarzschild metric in
Figure~\ref{ns.fig:MetricPhiN}.
The complete solution for the \(Y-\)metric function in this case is thus
  \begin{multline}
\label{ns.eq:YphiN}
    Y(r) = \left( \gamma \cosh{(\phi\xi_{b})} - \f{\alpha}{\phi} \sinh{(\phi\xi_{b})} \right) 
    \cosh{\left( \f{2\phi}{\sqrt{a}} \coth^{-1} \left( \f{1+\sqrt{1-br^2+ar^4}}{r^2\sqrt{a}}\right)  \right)} +\\
    + \left( \f{\alpha}{\phi} \cosh{(\phi\xi_{b})} - \gamma \sinh{(\phi\xi_{b})} +\right) 
    \sinh{\left(\f{2\phi}{\sqrt{a}} \coth^{-1} \left( \f{1+\sqrt{1-br^2+ar^4}}{r^2\sqrt{a}}\right) \right)},
  \end{multline}
which then allows us to write the matter variable \(p_{r}\) as
\begin{multline}
  \label{ns.eq:PhiNpr}
\kappa p_{r}(r) = \f{2\kappa\rho_{c}}{3} - \f{4\kappa\rho_{c}\mu r^{2}}{5r_{b}^{2}} -\kappa\rho_{c} \left[ 1 - \mu \left( \f{r}{r_{b}}\right)^{2}\right] + 4 \phi\sqrt{1-br^{2}+ar^{4}} \times \\
\times \f{ \left[ \f{\alpha}{\phi} \cosh{(\phi\xi_{b})} -\gamma \sinh{(\phi\xi_{b})} \right] \cosh{(\phi \xi)} + \left[ \gamma \cosh{(\phi\xi_{b})} - \f{\alpha}{\phi} \sinh{(\phi\xi_{b})} \right] \sinh{(\phi \xi)} }{\left[ \gamma \cosh{(\phi\xi_{b})} - \f{\alpha}{\phi} \sinh{(\phi\xi_{b})}\right] \cosh{(\phi \xi)} + \left[\gamma \cosh{(\phi\xi_{b})} - \f{\alpha}{\phi} \sinh{(\phi\xi_{b})} \right]\sinh{(\phi \xi)}},  
\end{multline}
and \(\ppen,\) the tangential pressure through the above as
\begin{equation}
  \label{ns.eq:PhiNpt}
  \ppen(r) =p_{r} - \beta r^{2}. 
\end{equation}
The Greek variables in the above expressions for this case are given by :
\begin{align*}
  \alpha &= \f{\kappa\rho_{c}\left( 5-3\mu\right)}{60}, &\quad& \beta > -\f{\kappa\mu\rho_{c}}{5r_{b}^{2}}, \\
  \gamma &= \sqrt{1+ \f{\kappa \rho_{c}r_{b}^{2}(3\mu-5)}{15} }, &\quad& \phi^{2} =\f{3\beta + 4\kappa\rho_{c}}{12}, 
\end{align*}
which completes the solution.  As with the previous examples, and in
particular Tolman~VII, we can invert the density relation and generate an equation of state.

\section{Charged case with anisotropic pressures}\label{ns.sec:Cha}
In this section we investigate electrically charged solutions.  As has
been noted by numerous authors~\cite{Iva02, KylMar67,VarRahRay10}, in
the static limit, this does not change the difficulty of solving the
EFE, since we add a Maxwell differential equation for the electric
charge that can immediately be integrated and incorporated into a
global charge that is seen from the outside only through the
Reissner-Nordstr\"om external metric.  The EFE do not change
drastically either, and a similar solution procedure to the one
already employed can be used to great effect.  We will give the full
Einstein-Maxwell field equations (EFME) before showing how we solve
then to get new solutions:

The energy-momentum tensor \(T_{ab}\) for the static electromagnetic
field is obtained from the Faraday tensor \(F_{ab}\) through
\[T^{\text{EM}}_{ab} = g_{ac}F^{cd}F_{db} - \f{1}{4} g_{ab}
  F^{cd}F_{cd}.\] As mentioned in Appendix~\ref{C:AppendixA} the
Faraday tensor in our case is
\begin{equation}
  F_{ab} =
  \begin{pmatrix}[c]
    0     & -\f{qY}{r^{2}\sqrt{Z}}  & 0  & 0  \\
    \f{qY}{r^{2}\sqrt{Z}}     & 0  & 0  & 0  \\
    0     & 0  & 0  & 0  \\
    0     & 0  & 0  & 0
  \end{pmatrix},
\end{equation}
and this allows us to write the total stress-energy, \(T^{\text{Total}}\) as
\begin{equation}
\label{ns.eq:energyMomentum}
  T^{i}{}_{j} =
  \begin{pmatrix}[c]
    \rho + \f{q^{2}}{\kappa r^{4}}   & 0  & 0  & 0  \\
    0     & -p_{r}+ \f{q^{2}}{\kappa r^{4}}  & 0  & 0  \\
    0     & 0  & -\ppen- \f{q^{2}}{\kappa r^{4}} & 0  \\
    0     & 0  & 0  & -\ppen - \f{q^{2}}{\kappa r^{4}} 
  \end{pmatrix}.
\end{equation}

As a result the EFME become the set
\begin{subequations}
  \label{ns.eq:EinMR}
  \begin{alignat}{3}
    \label{ns.eq:EinMR1}
    \kappa \rho +\f{q^{2}}{r^{4}} &= \e^{-\lambda}\left( \f{\lambda'}{r} -\f{1}{r^2}\right) +\f{1}{r^{2}} 
    &&= \f{1}{r^{2}} - \f{Z}{r^{2}} - \f{1}{r}\deriv{Z}{r}, \\
    \label{ns.eq:EinMR2}
    \kappa p_{r} - \f{q^{2}}{r^{4}}&= \e^{-\lambda} \left( \f{\nu'}{r} + \f{1}{r^2}\right) -\f{1}{r^2} 
    &&= \f{2Z}{rY}\deriv{Y}{r} + \f{Z}{r^{2}} - \f{1}{r^{2}},\\
    \label{ns.eq:EinMR3}
    \nonumber
    \kappa \ppen + \f{q^{2}}{r^{4}}&= \e^{-\lambda} \left( \f{\nu''}{2} - 
      \f{\nu'\lambda'}{4} + \f{(\nu')^2}{4} + \f{\nu'- \lambda'}{2r}\right)  
    &&= \f{Z}{Y}\sderiv{Y}{r} + \f{1}{2Y}\deriv{Y}{r}\deriv{Z}{r} \\ & && +\f{Z}{rY}\deriv{Y}{r} + \f{1}{2r}\deriv{Z}{r},
  \end{alignat}
\end{subequations}
and as in the Tolman~VII case, adding the first two equations to
each other results in a simpler equation that will make applying
boundary conditions easier:
\begin{equation}
  \label{ns.eq:EinMR1+2}
  \kappa (p_{r}+\rho) = \e^{-\lambda} \left( \f{\nu'}{r} + \f{\lambda'}{r} \right) = \f{2Z}{rY}\deriv{Y}{r} - \f{1}{r}\deriv{Z}{r}.
\end{equation}

We go through the same procedure to simplify these equation, except
for a crucial additional step: instead of using only
equation~\eqref{ns.eq:ZSol} as the initial ansatz, we revert to
Tolman's initial ansatz about the metric function.  He used \(Z(r) = 1
- br^{2} + ar^{4},\) as we shall, the reason being that by not using a
density ansatz right away we do not have to posit a charge ansatz
either, leaving us free until we have an idea about the physics.
However since we already have an interpretation for the density
function we have been using, we also wish to keep this.  To bridge
these concerns we segue into some physical considerations first. 

Considering that we have a spherical object, classical physics
suggests that most of the charge should be lying on the outer surface
of the sphere.  In GR since charge also contributes to the
gravitation, we expect at least something similar to the classical
picture, although we would expect non-zero but lesser charge in the
interior.  A good guess would be to have the charge be a monotonically
increasing function of the radial coordinate, since then most of the
charge is concentrated towards the surface.  Additionally having the
charge be a power of the radial coordinate is extremely convenient in
finding a solution to our differential equation as we will see.
Therefore for the time being, we append to our initial density the
ansatz, \(q(r) = k r^{n},\) with \(n > 0.\)

This initial ansatz for \(Z\) can be fed into the RHS of our first
differential equation~\eqref{ns.eq:EinMR1}, which results in
\[\kappa \rho + \f{q^{2}}{r^{4}} = 3b - 5ar^{2}.\] Consistency, and
the desire to keep the procedure to solving this system of equation
the same as before then demands that the LHS of the differential
equation also be a quadratic function with zero linear term.  This can
be seen as the ``reason'' for postulating the
density~\eqref{ns.eq:ZSol} we did before, which had this same
structure.  Also, due to the structure of this differential equation
we are forced to either pick either \(q(r) = k r^{2},\) in which case
we will have
\[ 3b - 5ar^{2} = (\kappa \rho_{c} + k^{2}) -
  \f{\kappa\rho_{c}\mu}{r_{b}^{2}} r^{2},\] or pick
\(q(r) = k r^{3},\) which results in
\[ 3b - 5ar^{2} = \kappa \rho_{c} - \left(
    \f{\kappa\rho_{c}\mu}{r_{b}^{2}} - k^{2}\right) r^{2}.\] We can
then read off \(a\) and \(b\) in either case, however if we continue
our procedure of defining an anisotropy measure, and performing the
same variable changes shown in the previous sections to simplify the
equation for the \(Y\) metric function, we quickly find out that the
first choice of \( q(r) = k r^{2}\) yields a differential equation for
\(Y\) that is not soluble with elementary functions\footnote{The
  coefficient of \(Y\) in the second order ODE after variable changes
  still contains a $1/r^2$ term, turning the problem into a variable
  coefficient one. Once additional assumptions about $a$ have been
  made, a solution in terms of hypergeometric functions is possible,
  but the assumption about $a$ renders the solution physically
  uninteresting.} contrary to our initial wish.  We therefore discard
this choice and instead restrict ourselves to \(q(r) \propto r^{3}\)
only.  Then the variable changes go through as before and the
differential equation for \(Y\) reduces to:
\begin{equation}
  \label{ns.eq:YdiffXi}
  \sderiv{Y}{\xi} + \f{Y}{4}\left( a + \f{\Delta}{x} - \f{2q^{2}}{x^{3}}\right) = 0.
\end{equation}
From this equation, it is easy to see that setting the value of \(q\)
to zero results in the uncharged anisotropic second order differential
equation we had previously.  Before we attempt to solve this equation
however we have to discuss the boundary and the junction conditions.
As mentioned previously, the correct exterior solution to be matched
in the Einstein-Maxwell case in the external vacuum
Reisner-Nordstr\"om metric.  This metric in Schwarzschild-type coordinates
is given by
\begin{equation}
  \label{ns.eq:Reiss}
  \d s^{2} = \left( 1 - \f{2M}{r} + \f{Q^{2}}{r^{2}}\right) \d t^{2} - 
\left( 1 - \f{2M}{r} + \f{Q^{2}}{r^{2}}\right)^{-1} \d r^{2} - r^{2}\left( \d \theta^{2} + \sin^{2} \theta \d \varphi^{2} \right),
\end{equation}
where \(M\) is the mass function and \(Q\) is the total electric
charge, both enclosed by the interior metric and perceived to external
observers. These quantities (see Appendix~\ref{C:AppendixA},
or~\cite{Iva02} for details) are given by
\begin{equation}
  \label{ns.eq:Mass+Charge}
  M = 4 \pi \int_{0}^{r_{b}} \left(\rho(r) +\f{q^{2}(r)}{8 \pi r^{4}} \right) r^{2} \d r \quad \text{and,} \quad Q =: 4 \pi \int_{0}^{r_{b}} \sigma(r) \sqrt{Z(r)} r^{2} \d r,
\end{equation}
where \(\rho(r)\)
is the mass density associated with the interior solution, and we
similarly define \(\sigma(r)\)
as the charge density associated with the interior solution, and which
is related to the \(q(r)\)
we have in our energy-momentum tensor \(T_{ab}^{\text{EM}}\)
by construction through \(Q^{2}(r_{b}) = q^{2}(r_{b}).\)
This last equation also encodes the charged part of the junction
conditions required of our differential equations.  We note here that
the mass function has been defined differently here than in the
previous cases.  A discussion on why this is the case can be found in
the appendix~\ref{C:AppendixA}.

We are now in a position to be able to solve the differential equation
for \(Y.\) As seen previously the simple-harmonic form
of~\eqref{ns.eq:YdiffXi} under certain conditions allow for
simplifications.  Again we define \(\Phi,\) a temporary variable
different from the previous sections through
\[4\Phi^{2} = a + \f{\Delta}{x} - \f{2q^{2}}{x^{3}},\] requiring that
\(\Phi\) be a number will ensure that the solution of our differential
equation be simple.  Requiring \(\Delta = 0,\) and thus ``switching
off'' anisotropy is the easiest thing to try, and doing so leaves us
with \[4\Phi^{2} = a - \f{2q^{2}}{x^{3}}.\] Here, \(q=0\) gives us
back Tolman's solution as expected and is a solution we already
considered.  However \(2q^{2} = 2 k^{2} x^{3}\) looks promising since
this would allow \(4\Phi^{2} = a-2k^{2},\) a pure number, and be of
the same form \(q \propto r^{3}\) we required before.  However the
solution by Kyle and Martin~\cite{KylMar67} reduces to this same
assumption and is analysed in their article, so that we must look
elsewhere, and consider the case \(\Delta \neq 0.\) This case allows
for two immediate possibilities:
\begin{itemize}
\item requiring \(\f{\Delta}{x} = \f{2q^{2}}{x^{3}}\) effectively
  ``anisotropises'' the electric charge allowing the latter to
  contribute to the anisotropy only, and considerably simplifying the
  solution to \(Y.\) We will look at this solution is
  Section~\ref{ns.ssec:phiA}.
\item If instead we ask that \(\Delta = \beta x,\) and
  \(2q^{2} = 2 k^{2} x^{3},\) we get
  \(4\Phi^{2} = a + \beta -2k^{2},\) which allows an analysis very
  similar to what we did in the previous section since \(\Phi^{2}\)
  can then be of either sign.  We will look at these possibilities in
  Sections~\ref{ns.ssec:phi0}, \ref{ns.ssec:phiN}, and~\ref{ns.ssec:phiP}
\end{itemize}

As with the Tolman~VII case we need boundary conditions to find a
complete closed form solution.  We implement this next, and determine
the integration constants in our solutions.

The boundary conditions here are not very different from the previous
case.  We recall that in Tolman~VII we required that the pressure at
the fluid--vacuum interface vanish and that the metric coefficients be
compatible with the Schwarzschild coefficients through
~\eqref{t7.eq:Boundary1+2}.  Here the first requirement is the same
when applied to the radial pressure only, and the compatibility of
metric coefficients is with Reissner--Nordstr\"om instead:
\begin{subequations}
  \label{ns.eq:Boundary1+2}
  \begin{align}
    \label{ns.eq:BoundaryPca}
    p_{r}(r_{b}) &= 0, \quad \text{and,} \\
    \label{ns.eq:BoundaryZca}
    Z(r_{b}) &= 1-\f{2M}{r_{b}} + \f{Q^{2}}{r^{2}_{b}} = Y^{2}(r_{b}).
  \end{align}
\end{subequations}

Considering~\eqref{ns.eq:EinMR1+2}, we find an expression for the radial
pressure as \[\kappa p_{r} = \f{2Z}{rY}\deriv{Y}{r} - \f{1}{r}\deriv{Z}{r} -\kappa \rho \xrightarrow{r
  \rightarrow x} \f{4Z}{Y}\deriv{Y}{x} - 2\deriv{Z}{x} -\kappa \rho \xrightarrow{x
  \rightarrow \xi} \f{4\sqrt{Z}}{Y} \deriv{Y}{\xi} - 2 \deriv{Z}{x} -\kappa \rho. \]
Applied at the boundary \(r=r_{b}\), conditions~\eqref{ns.eq:Boundary1+2} result in
\[ \kappa p_{r}(r_{b}) = 0 =
\f{4\cancel{\sqrt{Z(r_{b})}}}{\cancel{Y(r_{b})}}
\left. \deriv{Y}{\xi}\right|_{\xi = \xi_{b}} - \left. 2
  \deriv{Z}{x}\right|_{x=x_{b}} -\kappa \rho(r_{b}), \] so that we
have an ``easy-to-use'' equivalent condition on the derivative of
\(Y,\)
\begin{equation}
  \label{ns.eq:BoundaryApp1}
  \kappa\rho(r_{b}) = \left. 4\deriv{Y}{\xi}\right|_{\xi=\xi_{b}} - \left. 2\deriv{Z}{x}\right|_{x=x_{b}} \Rightarrow 
\left. \deriv{Y}{\xi}\right|_{\xi=\xi_{b}} = \f{1}{4} \left[ \f{\kappa \rho_{c}}{3} - \f{\kappa\rho_{c}\mu}{5} -\f{4k^{2}r_{b}^{2}}{5}\right] =: \alpha.
\end{equation}
For the second condition we re-express equation~\eqref{ns.eq:BoundaryZca} in terms of \(Y\) as
\begin{equation}  \label{ns.eq:BoundaryApp2}
Y(r_{b}) = \sqrt{Z(r_{b})} = \sqrt{1 + \f{\kappa \rho_{c}r_{b}^{2}(3\mu-5)}{15} - \f{k^{2}r_{b}^{4}}{5}} =: \gamma
\end{equation}
and subsequent application of the value and slope condition on the
\(Y\)
metric function form a Cauchy boundary pair and results in unique
integration constants for the \(Y\)
metric function in terms of the auxiliary constants \(\alpha\)
and \(\gamma,\) defined through the above equality.

\subsection{Anisotropised charge}
\label{ns.ssec:phiA}
In this section we analyse the solution to the EFME if we require that
the electric charge and anisotropy be related to each other through
the relation \(\Delta = 2(q/x)^{2},\) where we take the functional
form \(q \propto r^{3} = kr^{3}\) as mentioned before.  This
particular choice simplifies the differential equation for our \(Y\)
metric function allowing us to write an expression for the solution
analogous to the Tolman VII solution for \(Y\) directly as
\begin{equation}
  \label{ns.eq:YAnisotropisedCharge}
  Y(\xi) = c_{1} \cos{\left( \Phi \xi \right)} + c_{2} \sin{\left( \Phi \xi \right)}, \quad \text{with } \Phi = \sqrt{\f{a}{4}}.
\end{equation}
However we have to keep in mind that this solution is fundamentally
different from Tolman VII which was a solution to the Einstein's system
of equation and not the Einstein--Maxwell system.  This fact comes in
through three different ways
\begin{enumerate}
\item The charge in this system is non-zero, unlike the Tolman VII
  solution, where \(Q = 0.\)  
\item The presence of anisotropic pressure in the solution means that
  \(\ppen\) is \emph{not} the same as the radial pressure \(p_{r}.\)
  This is clear if we remember that \(\Delta \neq 0\) here.
\item Also, this solution will have to be matched to the
  Reissner-Nordstr\"om metric outside the sphere, as opposed to the
  Schwarzschild solution for Tolman~VII.
\end{enumerate}
If we take care to ensure these conditions, we have a fully-fledged
new solution to the EMFE, onto which we can apply boundary
conditions~\eqref{ns.eq:BoundaryApp1} and~\eqref{ns.eq:BoundaryApp2}.
\begin{itemize}
\item The first condition on the derivative results in  
  \[\left. \deriv{Y}{\xi}\right|_{\xi=\xi_{b}} = \Phi \left[
    c_{2}\cos{(\Phi \xi_{b})} -c_{1}\sin{(\Phi \xi_{b})} \right] =
  \alpha, \]
  which can be rearranged to yield an equation for \(c_{1}\)
  and \(c_{2}\)
  in terms of previously defined constants:
  \(c_{2}\cos(\Phi \xi) -c_{1}\sin(\Phi \xi) = \f{\alpha}{\Phi}.\)
\item The second condition also give us a similar equation:
\[Y(r_{b}) = c_{1}\cos{(\Phi \xi_{b})} + c_{2} \sin{(\Phi \xi_{b})} = \gamma,\]
\end{itemize}
and together this pair of equations can be solved for \(c_1\)
and \(c_2\) through simple algebraic manipulation to give
\begin{align*}
  c_{2} &= \gamma \sin{(\Phi\xi_{b})} + \f{\alpha}{\Phi} \cos{(\Phi\xi_{b})}\\
  c_{1} &= \gamma \cos{(\Phi\xi_{b})} - \f{\alpha}{\Phi} \sin{(\Phi\xi_{b})},
\end{align*}
A plot of the metric functions show us that indeed the conditions stated above are satisfied

\begin{figure}[h!]
  \centering
  \includegraphics[width=\linewidth]{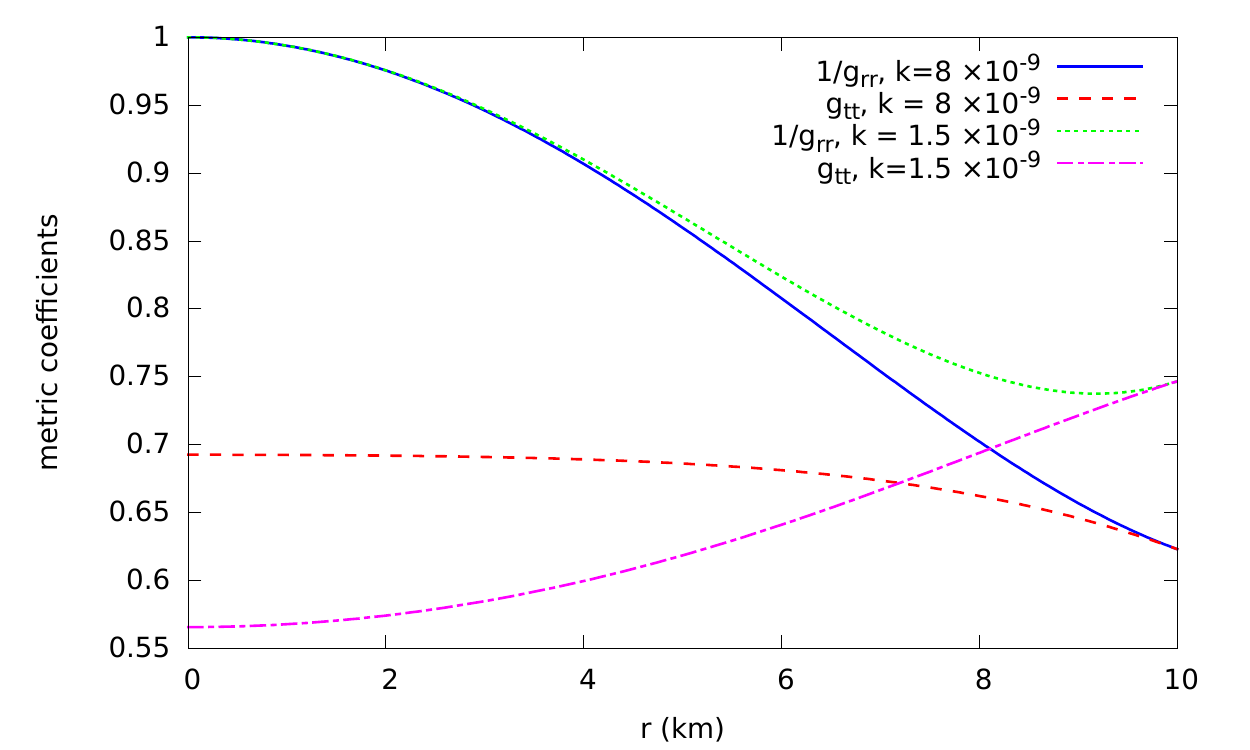}
  \caption[Matching the metric functions, for the anisotropy
  compensating the charge]{Application of the boundary conditions
    resulting in the value and slope matching of the metric function
    at $r=r_{b}.$ for $\Phi \neq 0,$ but where anisotropy compensates
    the charge.  The parameter values are
    $\rho_{c}=\un{1\times 10^{18}\,kg \cdot m^{-3}}, r_b = \un{1 \times
      10^{4}\,m}$ and $\mu = 1$}
\label{ns.fig:MetricAniCha}
\end{figure}

The complete solution where the anisotropy and the charge compensate
for each other thus becomes
\begin{multline}
\label{ns.eq:YphiCCA}
    Y(r) = \left( \gamma \cos{(\Phi\xi_{b})} - \f{\alpha}{\Phi} \sin{(\Phi\xi_{b})} \right) 
    \cos{\left( \f{2\Phi}{\sqrt{a}} \coth^{-1} \left( \f{1+\sqrt{1-br^2+ar^4}}{r^2\sqrt{a}}\right)  \right)} +\\
    + \left(\gamma \sin{(\Phi\xi_{b})} + \f{\alpha}{\Phi} \cos{(\Phi\xi_{b})} \right) 
    \sin{\left(\f{2\Phi}{\sqrt{a}} \coth^{-1} \left( \f{1+\sqrt{1-br^2+ar^4}}{r^2\sqrt{a}}\right) \right)},
  \end{multline}
which then allows us to write the matter variables \(\ppen\) and \(p_{r}\) as
\begin{multline}
  \label{ns.eq:PhiPprCCA}
\kappa p_{r}(r) = \f{2\kappa\rho_{c}}{3} - \f{4}{5}\left( \f{\kappa \rho_{c}\mu}{r_{b}^{2}} - k^{2}\right) r^{2} -\kappa\rho_{c} \left[ 1 - \mu \left( \f{r}{r_{b}}\right)^{2}\right] + 4 \Phi\sqrt{1-br^{2}+ar^{4}} \times \\
\times \left\{ \f{ \left[ \gamma \sin{(\Phi\xi_{b})} + \f{\alpha}{\Phi} \cos{(\Phi\xi_{b})} \right] \cos{(\Phi \xi)} - \left[ \gamma \cos{(\Phi\xi_{b})} - \f{\alpha}{\Phi} \sin{(\Phi\xi_{b})} \right] \sin{(\Phi \xi)} }{\left[ \gamma \sin{(\Phi\xi_{b})} + \f{\alpha}{\Phi} \cos{(\Phi\xi_{b})}\right] \sin{(\Phi \xi)} + \left[\gamma \cos{(\Phi\xi_{b})} - \f{\alpha}{\Phi} \sin{(\Phi\xi_{b})} \right]\cos{(\Phi \xi)}} \right\},  
\end{multline}
and 
\begin{equation}
  \label{ns.eq:PhiPptCCA}
  \ppen(r) =p_{r} - \Delta = p_{r} - 2k^{2}r^{2}   
\end{equation}
The variables in the above expressions for this case are given by :
\begin{align*}
  \alpha &= \f{\left( \kappa\rho_{c}(5 -3\mu) - 12k^{2}r_{b}^{2} \right)}{60}, &\quad& \Delta(r) = 2k^{2} r^{2} = \f{qr}{2k}, \\
  \gamma &= \sqrt{1+ \f{\kappa \rho_{c}r_{b}^{2}(3\mu-5)}{15} -\f{k^{2}r_{b}^{2}}{5}}, &\quad& \Phi^{2} = \f{1}{4}\left(\f{\kappa\rho_{c}\mu}{r_{b}^{2}} - k^{2} \right), 
\end{align*}
which completes the solution.  As can be seen, we could express the
solution in terms of \(q,\)
or \(\Delta\)
exclusively as expected, since these two functions are not independent
in this particular solution.  As with the previous example, we can
invert the density relation and generate an equation of state.  The
total mass and charge of the object modelled by this solution is
obtained through~\eqref{ns.eq:Mass+Charge}, and for this particular
case, these equations simplify to
\begin{equation}
  \label{ns.eq:M+CCCA}
M = 4\pi\rho_{c}r_{b}^{3} \left( \f{1}{3} - \f{\mu}{5} \right) +\f{k^{2}r_{b}^{5}}{10}, \quad \text{and,} \quad  Q = kr_{b}^{3}.  
\end{equation}
The last equation can be used to determine the charge density
\(\sigma(r), \) since from~\eqref{ns.eq:Mass+Charge} we have
\begin{equation*}
  \int_{0}^{r_{b}} \bar{r}^{2} \d r \left[ 4\pi \sigma(\bar r) \sqrt{Z(\bar r)} \right] = Q = kr_{b}^{3} = \int_{0}^{r_{b}} \bar{r}^{2} \d r \left[ 3 k \right].
\end{equation*}
Direct comparisons of terms yield the charge density
\begin{equation}
  \label{ns.eq:sigmaCCA}
  \sigma(r) = \f{3k}{4 \pi \sqrt{Z(r)}}.
\end{equation}
This completes the solution for this case.  We now turn to the case
where we have both charge and anisotropy independently of each other.
As we mentioned previously, this will require a thorough analysis of
the different combinations of charge and anisotropy, and how those
conspire to change the character of the second differential equation
we have.

\subsection*{The full Anisotropy and charged
  solution}\label{ns.ssec:phiABK}
Inspired by the previous sections, and building upon all the
simplifications and discussions so far we look directly at the second
order differential equation for the \(Y\)
metric function in the form of~\eqref{ns.eq:YdiffXi}.  This equation
contains a number of assumptions, all of which we have discussed
before.  Of particular interest in finding a general solution will be
the bracketed terms since different values or functions in the
brackets will lead to fundamentally different solution type for \(Y\)
independently of the form of \(Z.\)
As mentioned earlier also, for simplicity we pick functions for
\(\Delta\)
that give pure numbers for \(\Delta/x,\)
and the form of \(q\)
being determined previously through the choice of \(Z\)
also gives us a pure number for \(q/x^{3}.\)
These choices are reflected in the simplified form of
equation~\eqref{ns.eq:YdiffXi} which becomes
\begin{equation}
  \label{ns.eq.YdiffAC}
\sderiv{Y}{\xi} + \f{Y}{4}\left( a + \beta - 2k^{2} \right) =: \sderiv{Y}{\xi} + \Phi^{2} Y = 0.
\end{equation}

The only choice remaining for the different forms of \(Y\)
thus depends on the overall sign of the term in brackets,
\(\Phi^{2}\).
In this section we will provide conditions and the form of the
complete solutions for the different possibilities in the three sub-sections below.
 
\subsection{The $\Phi^{2} = 0$ case}\label{ns.ssec:phi0}
The fact that the coefficient of \(Y\) in
equation~\eqref{ns.eq.YdiffAC} contains terms of either sign
immediately points us to the possibility of choosing the terms to
annihilate the bracket completely.  For this to happen we have to
choose \(2k^{2} = a + \beta,\) somehow making the charge contribution
to be compensated by the anisotropy (through \(\Delta,\) and hence
\(\beta\)) and density (through \(a,\)) to yield the simplest
anisotropic charged solution of this class.  This choice is the crux
of this special solution, allowing us to express the anisotropy
measure \(\Delta \propto \beta\) in terms of the charge
\(q \propto k.\)

Since the term in brackets vanishes, the solution for \(Y\)
is the simple linear \(Y = c_{1} +c_{2} \xi,\)
with \(c_{1}\)
and \(c_{2}\)
our integration constants.  Applying boundary conditions on this
solution then results in \[\left. \deriv{Y}{\xi} \right|_{\xi=\xi_{b}} = c_{2} = \alpha := 
\f{1}{4} \left( \f{\kappa \rho_{c}}{3} - \f{3\kappa\rho_{c}\mu}{11} - \f{4r_{b}^{2}\beta}{11} \right),\] and 
\[c_{1} + c_{2}\xi_{b} = \gamma :=
\sqrt{1+r_{b}^{2}\kappa\rho_{b}\left( \f{2\mu}{11} - \f{1}{3}\right)
- \f{\beta r_{b}^{4}}{11}},\]
which can be solved together algebraically to give the value of
\(c_{1}.\) This completes the solution for \(Y\) in this particular case.

The \(Z\)
metric function still fixed by the Tolman assumption is
\(Z = 1 -br^{2} + ar^{4},\)
with however different values of \(a,\)
and \(b\)
than previously.  In this particular case these are given by
\[a = \f{2}{11}\left( \f{\kappa \mu \rho_{c}}{r_{b}^{2}} -
  \f{\beta}{2}\right), \quad \text{and} \quad b =
\f{\kappa\rho_{c}}{3}. \]

\begin{figure}[h!]
  \centering
  \includegraphics[width=\linewidth]{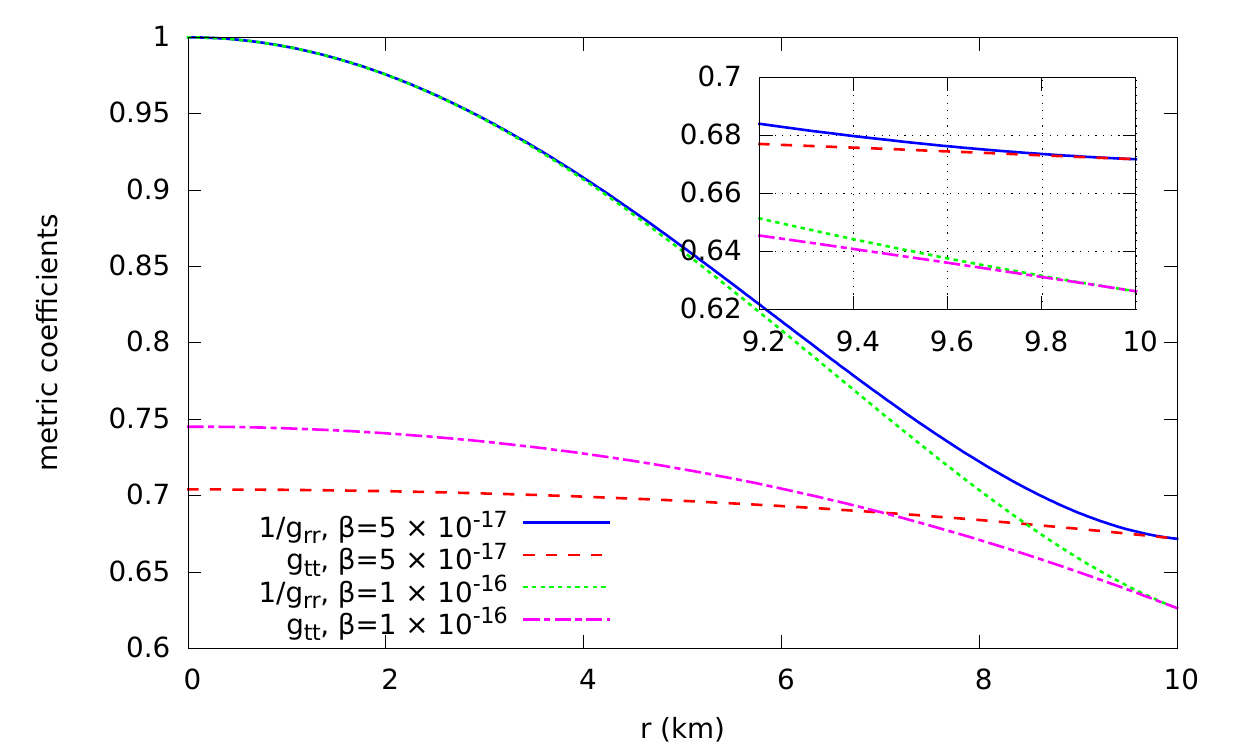}
  \caption[Matching the metric functions for $\Phi = 0$] {Application
    of the boundary conditions resulting in the value and slope
    matching of the metric function at $r=r_{b}.$ for the $\Phi = 0$
    case. The parameter values are
    $\rho_{c}=\un{1\times 10^{18}\, kg \cdot m^{-3}}, r_b = \un{1
      \times 10^{4}\, m}$ and $\mu = 1$}
\label{ns.fig:MetricAniChaPhiZ}
\end{figure}

Clearly in this case, because of the equation connecting \(\beta, a\)
and \(k\),
we can express these constants in terms of each other, and require
only two to completely specify the solution.  We show this
feature, and the consistent matching boundary in
figure~\ref{ns.fig:MetricAniChaPhiZ}.

Once we have the two metric functions, all other quantities are
determined, in particular the radial pressure \(p_{r}\) is given by
\[p_{r} = \f{1}{\kappa} \left[ \f{4c_{2} \sqrt{1 -br^{2} + ar^{4}}
  }{c_{1}+c_{2}\xi} +2b -4ar^{2} \right] - \rho(r),\]
and the tangential pressure \(\ppen,\)
in turn is \(\ppen = p_{r} -\Delta/\kappa,\)
giving\[\ppen = p_{r} - \f{\beta x}{\kappa}.\]
The mass \(M\)
and charge \(Q\)
seen from the exterior, which are still given by~\eqref{ns.eq:M+CCCA}
result in
\[ M = 4\pi \rho_{c}r_{b}^{3} \left( \f{1}{3} - \f{7\mu}{55}\right) +
\f{\beta r_{b}^{5}}{22}, \quad \text{and}, \quad Q = r_{b}^{3}
\sqrt{\left( \f{5\beta}{11} +
    \f{\kappa\mu\rho_{c}}{11r_{b}^{2}}\right)}.\]
This completes the solution for this particular case, and a summary of
all the functions and constants used in this results is given
in Appendix~\ref{C:AppendixB}

\subsection{The $\Phi^{2} < 0$ case}\label{ns.ssec:phiN}
For this to happen we need \((a+\beta-2k^{2})/4 < 0,\) turning our ODE
for \(Y\) into a simple harmonic type equation with the ``wrong''
sign.  As a result we expect a solution in terms of hyperbolic
functions, in this case given
by\[Y = c_{1}\cosh{\left(\Phi \xi \right)} + c_{2}\sinh{\left(\Phi \xi
    \right)}.\] In this particular case, we will not have a
simplification wherein the charge could be compensated completely by
the anisotropy or mass, and we are forced to deal with all three
components.  We however have that the charge contribution will exceed
the mass and anisotropy contribution (since \(2k^{2} > a+\beta,\)) and
this lead us to believe that such a solution has very little chance of
being physical.  We will however reconsider it in detail and come to a
conclusion on its viability as a physical solution later.

\begin{figure}[h!]
  \centering
  \includegraphics[width=\linewidth]{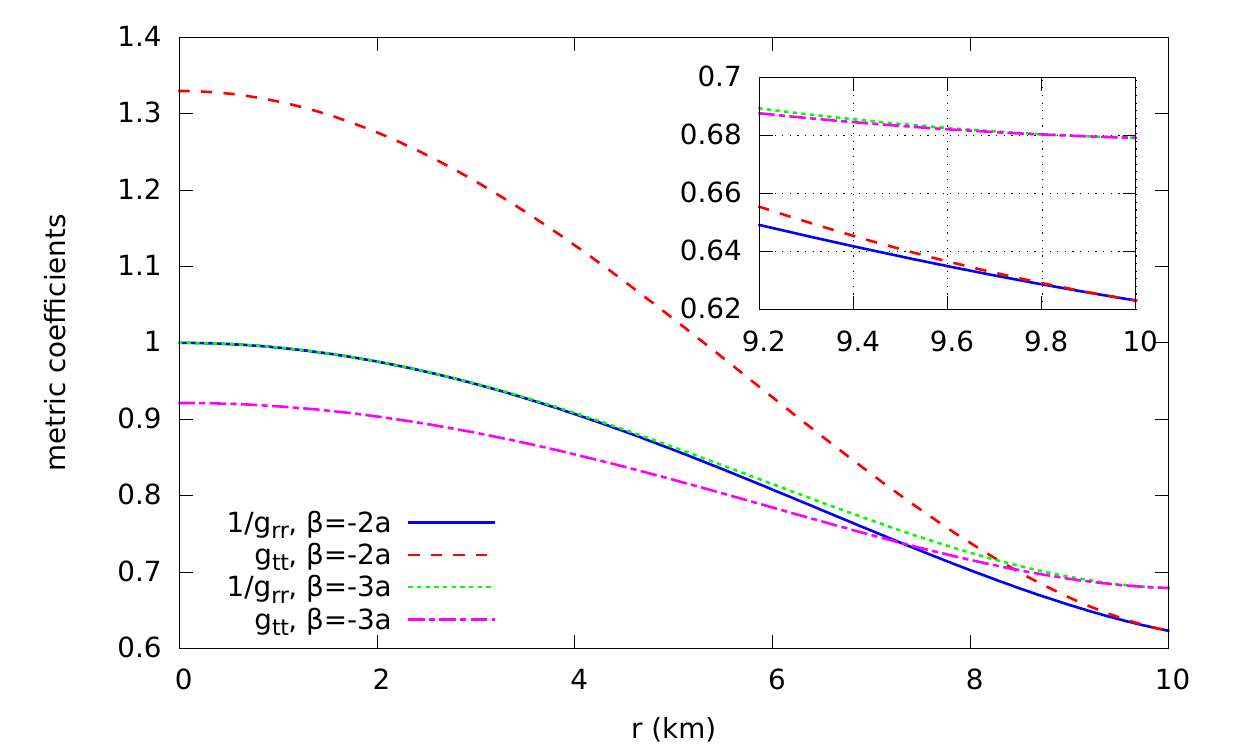}
  \caption[Matching the metric functions for $\Phi < 0$]{Application
    of the boundary conditions resulting in the value and slope
    matching of the metric function at $r=r_{b}$ for $\Phi < 0.$ The
    parameter values are
    $\rho_{c}=\un{1\times 10^{18}\, kg \cdot m^{-3}}, r_b = \un{1 \times
      10^{4} \,m}$ and $\mu = 1$}
\label{ns.fig:MetricAniChaPhiN}
\end{figure}

We apply boundary conditions to this solution to obtain the values of
the constants \(c_{1}\)
and \(c_{2}\)
through 
\begin{enumerate}
\item \(\left. \deriv{Y}{\xi}\right|_{\xi=\xi_{b}} = \Phi \left[ c_{2} \cosh{(\Phi \xi_{b})} + c_{1} \sinh{(\Phi \xi_{b})}\right] = \alpha, \) and 
\item \(Y(\xi_{b}) = c_{2}\sinh{(\Phi \xi_{b})} +c_{1}\cosh{(\Phi \xi_{b})} = \gamma.\)
\end{enumerate}
Then using a procedure very similar to that of previous sections we
obtain for the integration constants
\begin{align}
  c_{2} &= \f{\alpha}{\Phi} \cosh{(\Phi\xi_{b})} - \gamma\sinh{(\Phi\xi_{b})},\\
  c_{1} &= \gamma \cosh{(\Phi\xi_{b})} - \f{\alpha}{\Phi} \sinh{(\Phi\xi_{b})}
\end{align}
We show the matching boundary conditions at the boundary in
figure~\ref{ns.fig:MetricAniChaPhiN}, and note that in this case we
need both \(\beta\) and \(k\) to completely specify one particular solution.

This then completes the solution for the \(Y\) metric coefficient

\subsection{The $\Phi^{2} > 0$ case}\label{ns.ssec:phiP}
For this to happen we need \((a+\beta-2k^{2})/4 > 0,\) turning our ODE
for \(Y\) into a simple harmonic type equation.  As a result we expect
a solution in terms of trigonometric functions, in this case given
by\[Y = c_{1}\cos{\left(\Phi \xi \right)} + c_{2}\sin{\left(\Phi \xi
    \right)}.\] In this particular case, we will not have a
simplification wherein the charge could be compensated completely by
the anisotropy or mass, and we are forced to deal with all three
components.  We however have that the charge contribution will be less
than the mass and anisotropy contribution (since
\(2k^{2} < a+\beta,\)) and this lead us to believe that this will be
the most promising physically acceptable candidate in terms of new
solutions.  We will investigate this solution, and the remaining ones,
in detail and come to a conclusion on their viability as a physical
solution later.

\begin{figure}[h!]
  \centering
  \includegraphics[width=\linewidth]{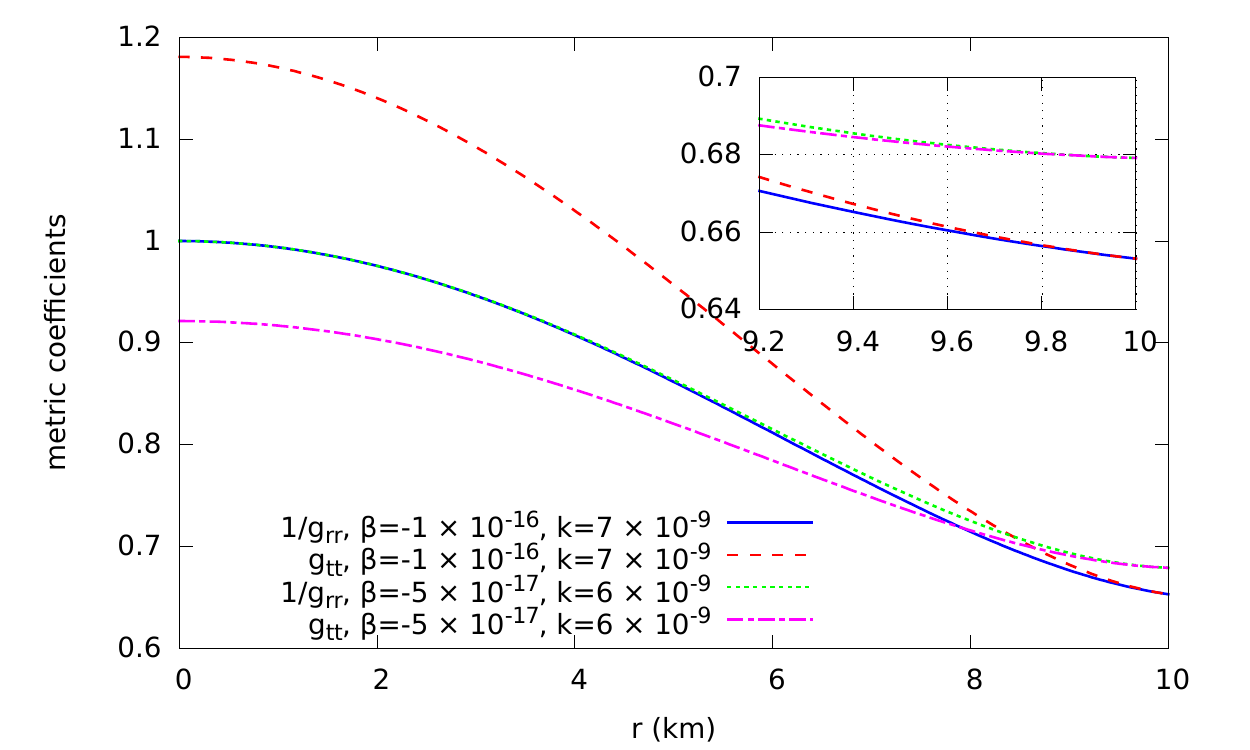}
  \caption[Matching the metric functions for $\Phi < 0$]{Application
    of the boundary conditions resulting in the value and slope
    matching of the metric function at $r=r_{b}$ for $\Phi < 0.$ The
    parameter values are
    $\rho_{c}=\un{1\times 10^{18}\,kg \cdot m^{-3}}, r_b = \un{1 \times
      10^{4}\,m}$ and $\mu = 1$}
\label{ns.fig:MetricAniChaPhiP}
\end{figure}

We apply boundary conditions to this solution to obtain the values of
the constants \(c_{1}\)
and \(c_{2}\)
through 
\begin{enumerate}
\item \(\left. \deriv{Y}{\xi}\right|_{\xi=\xi_{b}} = \Phi \left[ c_{2} \cos{(\Phi \xi_{b})} - c_{1} \sin{(\Phi \xi_{b})}\right] = \alpha, \) and 
\item \(Y(\xi_{b}) = c_{2}\sin{(\Phi \xi_{b})} +c_{1}\cos{(\Phi \xi_{b})} = \gamma.\)
\end{enumerate}
Then using a procedure very similar to that of previous sections we
obtain for the integration constants
\begin{align}
  c_{2} &=  \gamma\sin{(\Phi\xi_{b})} + \f{\alpha}{\Phi} \cos{(\Phi\xi_{b})},\\
  c_{1} &= \gamma \cos{(\Phi\xi_{b})} - \f{\alpha}{\Phi} \sin{(\Phi\xi_{b})}.
\end{align}
We show the matching boundary conditions at the boundary in
figure~\ref{ns.fig:MetricAniChaPhiP}, and note that in this case also
we need both \(\beta\)
and \(k\) to completely specify one particular solution.

\section{Possibilities for other solutions}
While keeping the ansatz for \(Z\)
fixed, but adding both anisotropy and charge to the system of
equations, we managed to tease out new solutions for the EMFE.  From
working with the equations it is clear to us that since the choices
for the charge function are not arbitrary if we want to maintain the
form of \(Z,\)
we can only realistically modify the anisotropy choice.  We note here
that the crux of our solution finding method stems from
equation~\eqref{ns.eq:YdiffXi}, which we then convert by judicious
choices to a simple harmonic equation with no forcing or damping.
Being restricted by the charge \(q\)
which has to be a cubic function not only for the existence of a
simple \(Y\)
solution, but also crucially for \(Z,\)
we can isolate this part of the equation immediately
into\[\sderiv{Y}{\xi} - \f{k^{2}}{2}Y + \f{Y}{4}\left( a +
  \f{\Delta}{x}\right) = 0.\]
Of course picking uncharged solutions does away with both \(k\)
and \(q,\)
and if we want uncharged solutions, this is the way we would proceed.
However, if we want charged solutions, we will have to modify the
terms in brackets in such a way as to keep a simple form for the \(Y\)
ODE. 

The following discussion will to be heavily influenced by choosing
linear ODEs with constant coefficients that are straightforward
generalizations of the harmonic oscillator equation.  We hope that
this will give a simple way of extending this type of work to larger
classes of \emph{physically relevant} solutions to spherical static
stars.

Adding linear first derivative terms in the ODE for \(Y\) we could
presumably posit more complicated forms for \(\Delta,\) for example
\(\Delta = f(\xi,Y,\deriv{Y}{\xi}) x/Y,\) for some particular choice
of the \(f\) function.  However we have to keep in mind the criterion
that \(\Delta\) has to satisfy: it has to vanish at
\(x = r = \xi =0,\) If we manage to pick \(f\) such that this is true,
we will get other new solutions, with possibly new features to be
explored.

As an example for this approach we pick \(f\)
to be \(f = g \deriv{Y}{\xi},\)
for some constant \(g.\)
This converts our undamped ODE into a damped one, whose solutions can
be classified according to the schemes usual to solving second order
ODEs of that form.  Judicious choice of the value of \(g\)
will then ensure that the discriminant of the ODE is in the
appropriate range to admit exponentially decaying envelopes to
sinusoids (the usual characteristic of damped harmonic systems,) for
the metric function \(Y.\)
The difficulty in this type of approach will then be in ensuring that
when the system has been solved that both \(p_{r}\)
and \(\ppen\) behave in physically expected ways.

Picking \(f\)
to be a forcing-type term as is usually encountered in forced
electrical oscillators, will yield another potential class of
solutions.  Furthermore the frequency of the forcing term could be
tuned to different values depending on how we want \(\Delta\)
to behave. As an example of this, we could pick
\(f = g\sin{(\sqrt{\f{a}{4}-k^{2} + \epsilon}\xi)},\)
so that we force our solution for \(Y\)
around its natural frequency, depending on the exact value of
\(\epsilon.\)
Again we would have to ensure that the behaviour of the physical
variables be consistent with our starting assumptions, but hopefully
this should be possible by tuning the frequency of the forcing, or by
restricting the value of key parameters like \(a\)
or \(k.\)
The fact that the latter two parameters have different signs should be
helpful in this endeavour.

We have provided two distinct examples of how new solutions could be
generated from our work, and clear paths to checking consistency of
the new solutions.  We will not look at any of these newer solutions
in detail or check whether they have already been discovered, or even
discuss their stability, but these are things that would have to be
done if they are to be used for modelling physical objects.

\section{Conclusion}\label{ns.sec.Conclusion}
In this chapter we provided the generalization we used to extend the
Tolman~VII ansatz models involving electric charge and anisotropic
pressures.  We computed exact analytic solutions for different cases,
and valid for different values of charge, and anisotropy, recovering
the original solutions in degenerate cases, as expected.  We also
applied boundary conditions on the metric functions \(Y\) and \(Z\) to
complete the closed form solutions in terms of variables that can be
\emph{physically} interpreted.  We obtained the expressions for the
matter variables: the density \(\rho,\) the pressures \(p_{r}\) and
\(\ppen,\) the electric charge \(Q,\) and the mass \(M\) for our
models.  We did not show or mentions possible issues like stability
(see Chapter~\ref{C:Stability},) or divergences and non physicality in
the matter variable (see Chapter~\ref{C:Analysis},) waiting for the
next chapters for these clarifications.  This chapter should be
regarded as the mathematical component of solution finding, and model
building for our solutions.  The physics \emph{per-se} will be in the
analysis Chapter~\ref{C:Analysis} mostly where we will provide
conditions and applicability criteria for each solution in detail, and
predict measurables like masses, radii, and charges for our models.
Comparisons with recent observations will also be done then.



\chapter{Stability analysis} \label{C:Stability}
\begin{myabstract}
  We investigate the general stability\index{Stability} theory of
  spherically static and symmetric space-times, as applied to stellar
  objects.
\end{myabstract}

Spherically static and symmetric objects have been studied, and their
stability analysed for quite some times under different circumstances.
The general theory of stability in relativity is made complicated
since many variables can change at the same time.  Therefore
maintaining consistency can be a difficult task.  The most complete
derivation, that of Chandrasekhar will be extended to include the case
of anisotropic pressures, and electric charges in the following
sections.

\section{Introduction}
The stability analysis of solutions to Einstein's equation has a long
history.  If these solutions are to be used for physical modelling
applications, the need to demonstrate that the solution is indeed
stable becomes even more important.  Global existence and uniqueness
of solutions are usually the other aspects of solutions that are
deemed to be as important as stability, but while the former two have
been shown to be true, global stability of solutions is still an open
question.

In this chapter we aim to show a very restricted version of stability:
we plan to show that the solutions we have presented so far are indeed
stable locally.  Global stability issues are not considered in this
Chapter.  We proceed by first analysing some heuristic methods of
determining stability.  Since these produce contradictions we then
continue with a full-blown linear perturbation analysis of our
solutions.

\section{Heuristics}
Stability analysis based on perturbing the governing equations is
usually a lengthy process, even when the equations we deal with are
simple.  In relativity the differential equations we start with are
not simple, and are also coupled.  Thus perturbing these equations and
finding the linear stability of the system requires lengthy
calculations.  Over the years a number of heuristics have been
developed to determine whether a relativistic system will be stable or
not.  These heuristics work most of the time, however there do exist
cases where they do not hold, and a formal proof of linear stability
is the only sure way of determining stability.

\subsection{The static stability criterion}\index{Harrison stability criterion}
A heuristic based on~\cite{HarThoWak65} and widely used in the
literature~\cite{HaePotYak07} states that for a star to be stable, it
has to satisfy
\begin{equation}
  \label{st.eq:HarrisonStab}
\deriv{M}{\rho_{c}} > 0.
\end{equation}
However, as noted in many
places~\cite{HarThoWak65,HaePotYak07,ZelNov71}, this is only a
necessary condition, and is not sufficient to ensure stability, hence
our classification of it as a heuristic.

In our case, we only ever have one expression for the density, and
hence the mass is always the same function, given by
equations~\eqref{t7.eq:MassFunction},\eqref{ns.eq:Mass+Charge},
and~\eqref{an.eq.Mass}.  The latter two equations include the mass
contribution from the electric charge too, but that does not change
how we implement this condition.  Given these equations, since
\(M=m(r_{b}),\) by taking the derivative of the latter we obtain
\[
\deriv{M}{\rho_{c}} = 4\pi r_{b}^{3} \left( \f{1}{3} - \f{\mu}{5}\right),
\]
onto which we can impose the positivity condition quite easily to yield
\begin{equation}
\label{st.eq:mu}
\mu < \f{5}{3},  
\end{equation}
since \(r_{b} > 0\) for all stars consider.  The above
condition~\eqref{st.eq:mu} is automatically satisfied since our
starting assumption on \(\mu\) was that is was never going to be more
than unity, ensuring that at least this heuristic is always satisfied by
all our solutions, anisotropy or charge notwithstanding.

\subsection{The Abreu--Hern\'andez--N\'u\~nez (AHN)
  criterion}\index{Abreu stability criterion}
This heuristic is based on~\cite{AbrHerNun07}, which analyses the
``cracking'' instability\index{Cracking instability} in anisotropic
pressure models: the precise case we are dealing with in our
solutions.  The method consists in comparing the speed of pressure
waves in the two principal directions of the spherically symmetric
star: the radial sound speed with the tangential sound speed, and then
based on those values at particular points in the sphere, we could
potentially conclude whether the model is stable or unstable under
cracking instability.

Cracking as a concept was introduced previously by~\citeauthor{Her92}
in~\citeyear{Her92}.  It involves the possibility of ``breaking up''
the fluid sphere due to the appearance of total radial forces of
different signs, and hence in different directions, at different
points in the star.  It should be mentioned that this has never been
observed, but that under suitable physical assumptions, it is a likely
scenario, and was investigated as such in both~\cite{Her92}
and~\cite{AbrHerNun07}.  This process is potentially a source of
instability and is characterized most easily through the speed of
pressure waves.

The main message of~\cite{AbrHerNun07} is that if the tangential speed
of pressure wave, \(v_{s\perp}^{2} = \deriv{\ppen}{\rho}\)
is larger that the radial speed of pressure waves,
\(v_{sr}^{2} = \deriv{p_{r}}{\rho},\)
then this could potentially result in cracking instabilities to occur
in the star, rendering the latter unstable.

We shall explicitly check whether this occurs in our models, and hence
classify our solutions according to the AHN scheme.  We would normally
start by testing this condition on the Tolman~VII solution, however
because ``cracking'' only occurs in anisotropic models~\cite{Her92},
Tolman~VII is automatically stable under ``cracking'' instabilities.
In the other solutions that we have constructed, the quantity
that becomes important is \(\Delta,\) since it measures the difference
in the tangential and radial pressures~\eqref{ns.eq:DeltaDef}.  The
AHN condition then reduces to
\begin{equation}
  \label{st.eq:AHNDelta}
  v_{sr}^{2} - v_{s\perp}^{2} < 0 \implies \deriv{\Delta}{\rho} < 0.
\end{equation}
The assumption for \(\Delta\)
in all the new solution we presented has been that
\(\Delta = \beta x,\) so that the above condition simplifies to 
\[
\beta \deriv{x}{\rho} < 0 \implies \f{\beta}{\d \rho/ \d x} < 0.
\]
Since the density expression we use is the same in all the solutions,
we can easily simplify the latter equation
through~\eqref{ns.eq:PhiZDensity}, and remembering that \(x = r^{2},\)
we finally get\[-\f{\beta r_{b}^{2}}{\mu} < 0.\] Since all the
constants, except for \(\beta,\) in the above expression are positive
definite, we have a prescription on the latter from the AHN
prescription: \emph{we will have no cracking instability in our
  solutions if}
\begin{equation}
  \label{st.eq:NoCrackingBeta}
  \beta > 0.
\end{equation}
This concludes the application of this method on our solution.  Once
we start analysing these solutions in Chapter~\ref{C:Analysis}, we can
impose the condition on \(\beta\) to ensure no cracking instabilities.

\subsection{Ponce De Leon's criterion}\index{Ponce de Leon stability criterion}
This method is mostly concerned with the behaviour of the Weyl tensor
for the solution, to conclude whether a certain model is more or less
stable than another comparable model.  It should be noted that this is
a comparative method: nothing is said about the absolute stability:
only the relative stability as compared to another model/solution can
be obtained.  The exact method of performing this comparison was given
in~\cite{Pon88}.

This method starts by calculating a function of the metric variables,
called \(W,\)
in~\cite{Pon88}.  This function, defined in terms of the \(\lambda\)
and \(\nu\) metric variables in the original article, is given by
\begin{equation}
  \label{st.eq:PdLWGeom}
  W(r) \coloneqq \f{r^{2}Z'(Y -rY') + 2rZ ( rY'- Y -r^{2}Y'') + 2rY }{12Y}  
\end{equation}
in our metric variables \(Z\) and \(Y,\) the primes~\((')\) denoting
derivatives with respect to \(r.\) Through the use of Einstein's
equations, this purely geometrical quantity can be rewritten in terms
of the matter variables.  The complete derivation of this equivalence
is in Ref~\cite{Pon88}, and we will not give it here, but the result
reads:
\begin{equation}
  \label{st.eq:PdLWMatter}
  W(r) \equiv m(r) - \f{4 \pi r^{3}}{3} \left( T_{t}{}^{t} + T_{r}{}^{r} - T_{\theta}{}^{\theta}\right).
\end{equation}
In addition to being easier to calculate than~\eqref{st.eq:PdLWGeom},
this expression~\eqref{st.eq:PdLWMatter} can be interpreted quite
simply, particularly in our variables.  Plugging in expressions of the
energy momentum tensor from equation~\eqref{ns.eq:energyMomentum}, and the mass
function from~\eqref{ns.eq:Mass+Charge} for the most general
expressions of these quantities, we obtain
\begin{equation}
  \label{st.eq:PdLWSimp}
  W(r) = \f{4 \pi r^{3}}{3} \left\{ r^{2} 
    \left[ \f{2}{5} \left(\f{\rho_{c} \mu}{r_{b}^{2}} \right)  -\f{3k^{2}}{8 \pi} \right] + \left( \f{k^{2}}{8 \pi} + \Delta \right)\right\},
\end{equation}
where we note that because of the different signs associated with the
terms, \(W(r)\)
can be of either sign.  The stability argument then proceed by
comparing the value of \(W\)
for spheres having the same masses and radii, and then concluding that
the lower the value of \(W,\)
the more stable the corresponding sphere.  The full argument as to why
\(W\)
can be used in such a fashion is very long and given in full
in~\cite{Pon88}, and touched upon in~\cite{RagHob15b}. In both of
these references the relationship of \(W\)
to the Weyl\index{Tensor!Weyl} tensor \(C_{abcd},\)
and to the Newmann-Penrose Weyl scalar \(\Psi_{2}\)
is\index{Newman Penrose Formalism!$\Psi_{2}$} emphasized so that this
stability criterion becomes less strange, but we shall not go into
details here.

Applying this criterion to all our solutions results in a different
expression of \(W\)
in all the sub-classes of solution, and we summarize this in
table~\ref{st.tab:W}
\begin{table}[h]
  \centering
{\tabulinesep=1.2mm
  \begin{tabu}[h]{p{3.3cm}| >{$\displaystyle}c<{$} | >{$\displaystyle}c<{$} }
    \hline
    Solution Name & \text{Specific case} &  W(r) \\ \hline
    Tolman~VII & \Delta = k = 0 & \f{8 \pi \mu \rho_{c}}{15r_{b}^{2}} r^{5} \\ 
    Anisotropic TVII & \Delta =\beta r^2, k = 0 & \f{4 \pi r^{5}}{3} \left[ \f{2}{5} \left(\f{\rho_{c} \mu}{r_{b}^{2}} \right) +\beta \right] \\
    Anisotropic TVII with charge & \Delta = 2k^2 r^2 &  \f{4 \pi r^{3}}{3} \left\{ r^{2} 
    \left[ \f{2}{5} \left(\f{\rho_{c} \mu}{r_{b}^{2}} \right)  + k^2 \left( 2 - \f{3}{8 \pi}  \right) \right] + \f{k^{2}}{8 \pi} \right\} \\
    & \Delta = \beta r^2, k \neq 0 & \f{4 \pi r^{3}}{3} \left\{ r^{2} 
    \left[ \f{2}{5} \left(\f{\rho_{c} \mu}{r_{b}^{2}} \right)  -\f{3k^{2}}{8 \pi} +\beta \right] + \f{k^{2}}{8 \pi} \right\} \\ \hline 
  \end{tabu}}
  \caption{The different expressions of the $W$ function for our different classes of solutions}
\label{st.tab:W}
\end{table}
from which it is immediately clear that adding anisotropy in the
form of \(\Delta \neq 0,\)
changes the value of \(W,\)
and depending on the sign of \(\beta\)
we can get \(W\)
to increase or decrease.  If we admit the ``no cracking'' heuristic
condition~\eqref{st.eq:NoCrackingBeta}, we will have that any addition
of anisotropy will \emph{increase} \(W.\)
By contrast since charge only comes in the form of \(k^{2}\)
in the expression for \(W,\)
all the charge terms contribute positive quantities. However, since in
the general expression of \(W,\)
the charge term occurring with a negative sign is larger in magnitude,
electric charge has a capacity of \emph{reducing} \(W.\)

From this heuristic we therefore conclude that addition of charge
stabilizes the star, and addition of anisotropy destabilizes it.  The
exact effects of both however, and how these two interact with each
other can only be guessed at this point.  We shall elucidate this in
the next section where we perform a full radial perturbation stability
analysis of this system.

\section{Radial Perturbation Analysis}\index{Perturbation}
We will follow~\citeauthor{Cha64L} in the initial phase of our
derivation, with corrections from Chandrasekhar's own erratum,
and~\citeauthor{KnuPed07}\cite{KnuPed07}. However since we will be
considering a more general form of the energy-momentum tensor: one
that admits both electric charge and anisotropic pressures, the later
part of the derivation will be more cumbersome.

The most extensive use of~\citeauthor{Cha64L}'s derivations was
by~\citeauthor{Too65}, who considered a number of models before
integrating the Chandrasekhar's pulsation equations numerically for
polytropes of various orders to obtain the normal mode frequencies in
both general relativity, and in post-Newtonian approximations.  Many
other authors have subsequently investigated stability.  For
example~\citeauthor{Neg04} checks the stability of self-bound Tolman
VII solutions, and determines that this cannot be stable, however with
hand-waving arguments involving the ``'type independance' property of
mass 'M'(sic).'' We prove later in this chapter that the self-bound
Tolman VII solution can be stable.

The extension to anisotropic models was first done
by~\citeauthor*{HilSte76}, who studied the dynamic
stability\index{Stability!Dynamic} of anisotropic models numerically
to find the eigenfrequencies of the normal modes\index{Normal Modes}
for the pulsation\index{Pulsation equation} equation~\cite{HilSte76}.
They found that the same method
(Sturm--Liouville\index{Sturm--Liouville} eigenmode analysis) employed
by~\citeauthor{Cha64} could be extended to anisotropic pressures.
However the work being numerical in nature, the type of anisotropy had
to be specified, and~\citeauthor*{HilSte76} only considered very
specific types of anisotropy.  They also tried to look at more general
non-radial perturbations, but they only did so for the Newtonian case,
citing that ``Since the anisotropic term in the equilibrium equations
is of purely Newtonian origin [...], we will discuss non-radial
pulsations only in the Newtonian approximation.''  The authors
concluded by stating that anisotropic models are as stable as the
isotropic stars, while allowing for a greater concentration of mass in
the star.

Many other authors have worked on the issue of stability, with a
marked preference given to the Newtonian stars which are deemed
complicated enough that most of the physics would be similar when it
comes to stability considerations.  Of note is the article
by~\citeauthor*{LieYau87}, who derive the Newtonian hydrodynamic
stability condition from a quantum mechanical point of view, even
extending the analysis to boson stars~\cite{LieYau87}. By
contrast~\citeauthor*{ShaAza12} look at the proper relativistic
equations throughout up to and including the matching conditions to a
Reissner-Nordstr\"om exterior, but then look at the pulsation and
contractions in a Newtonian and a post-Newtonian limit, concluding
that both anisotropy and electric charge affect the collapse,
pulsations, and hence stability of the star, with charge reducing the
instability of the star~\cite{ShaAza12}.

EOS with quasi-local\index{Quasilocal variables} components (where the
EOS depends on quasi-local variables such as the average density,
total mass or total radius, as well as local ones) have also been
investigated intensively, for example in~\cite{HorIliMar11}.  While
technically more difficult because the boundary conditions on the
pulsation equations are now quasi-local, these EOS allow the density
to increase outwards in violation of \citeauthor{Buc59}'s assumption
of \(\d \rho / \d r \leq 0,\) leading to the maximum
compactness~\cite{Buc59} of \(M/R \leq 4/9,\) while maintaining the
stability of the star with the anisotropy.

The most complete treatment of stability in anisotropic stars was that
of \citeauthor{DevGle03}, where the pulsation equation was obtained
for the general anisotropic case, after a lengthy treatment of both
Newtonian and relativistic stars.  Due the the length of some of the
equations, a number of typographic errors are present, but a very
comprehensive section on different examples of anisotropy concludes
that anisotropy, for the most part stabilizes stars~\cite{DevGle03}.
We wish to extend this result to the electrically charged case.

Charged models have received similar treatment over the years:
\citeauthor{Ste73} studied a constant mass density model with constant
charge density on the surface of the star and concluded from an
analysis which follows \citeauthor{Cha64} closely that in certain
cases when the charge is not too large, the system is stabilized with
inclusion of the electric charge~\cite{Ste73}.  \citeauthor{Gla79} by
contrast looks at a completely general charged isotropic model in the
spirit of \citeauthor{Cha64}, until he obtains the pulsation equation,
but then only applies his result to a dust solution by
\citeauthor{Bon65} to conclude that electric charge decreases the
minimum radius at which dynamical stability is possible.

The only work that tried to extend the pulsation formalism to both
anisotropic and electric charge was that of \citeauthor{EscAlo10},
however we could not use their result since they specialized their
pulsation equation with a restrictive assumption for the type of
anisotropy: one where \(\Delta = C p_{r}\).  Furthermore the interior
solution in their work has to admit
conformal\index{Symmetry!Conformal} symmetry~\cite{EscAlo10}, which we
do not have.  We have tried to use notation consistent with the
mentioned works as much as possible, and thoroughly checked our
expressions, but due to the length of the involved equations,
typographic errors are inevitable.

Following in the footsteps of \citeauthor{Cha64L},
\citeauthor*{DevGle03} and \citeauthor*{EscAlo10}, we now proceed to
perturb the metric with a non-zero radial four velocity \(v,\) whose
time integral will be the perturbation control parameter \(\zeta.\)
This radial perturbation will cause the stress-energy tensor to be
non-diagonal, and as a result all the matter and metric variables will
be perturbed by an amount that can be related to this radial velocity
perturbation.  Each of these perturbations will be consistently
expanded to first order (linear perturbation analysis) in terms of
\(\zeta\) and unperturbed quantities.  Conservation of baryon number
inside the star is then used as a condition on an undetermined
equation of state to close the system into a differential equation of
the form \begin{equation} \label{st.eq:DiffEig} -\spderiv{\zeta}{t} =
  \mathcal{L} \zeta,
\end{equation}
where \(\mathcal{L}\) is some differential operator.  As is usual in
eigenmode analysis, \(\zeta\) is then assumed to have a time
dependence of the form \(\zeta = \e^{\i \sigma t},\) and substitution
in the above differential equation~\eqref{st.eq:DiffEig} results in a
Sturm-Liouville type problem for the frequency \(\sigma.\) The
analysis of the spectrum of \(\sigma,\) for specific test functions of
\(\zeta\) gives a natural ordering of eigenfrequencies, and the sign
of the leading eigenfrequency determines whether the solution is
stable or not.  We will follow this prescription in what follows.

As is usual in deriving the pulsation equation for non-static,
spherically symmetric metric, we will assume the following form for
the line element:
\begin{equation}
  \label{st.eq:metric}
  \d s^{2} = \e^{\nu} \d t^{2} - \e^{\lambda} \d r^{2} - r^{2}\left( \d \theta^{2} + \sin^{2} \theta \d \varphi^{2} \right) .
\end{equation}
From this metric we will immediately be able to write down the
Einstein equations:
\begin{equation}
  \label{eq:Ein}
  \f{8 \pi \G}{\cc^{4}} T^{i}{}_{j} = G^{i}{}_{j},
\end{equation}
which in component form explicitly give:
\begin{subequations} \label{eqs:EinExplicit}
  \begin{align}
    - \f{8 \pi \G}{\cc^{4}} T^{0}{}_{0} &= \e^{-\lambda} \left( \f{\lambda'}{r} -\f{1}{r^{2}} \right) + \f{1}{r^{2}} = -\f{1}{r^{2}} \left( r \e^{-\lambda} \right)' 
+ \f{1}{r^{2}} \label{eq:EEa}, \\
- \f{8 \pi \G}{\cc^{4}} T^{1}{}_{1} &= -\e^{-\lambda} \left( \f{\nu'}{r} +\f{1}{r^{2}} \right) + \f{1}{r^{2}} \label{eq:EEb}, \\
- \f{8 \pi \G}{\cc^{4}} T^{2}{}_{2} &= -\e^{-\lambda} \left( \f{\nu''}{2} - \f{\nu' \lambda'}{4} + \f{\nu'^{2}}{4} + \f{\nu' - \lambda'}{2r} \right) 
+ \e^{-\nu} \left( \f{\dot \lambda \dot \nu}{4} + \f{\ddot \lambda}{2} + \f{\dot \lambda^{2}}{4}\right) \label{eq:EEc}\\
- \f{8 \pi \G}{\cc^{4}} T^{1}{}_{0} &= -\f{\e^{-\lambda}}{r} \dot{\lambda}.\label{eq:EEd}
  \end{align}
\end{subequations}
Here the primes \(({}')\) and dots \((\dot {})\) refer to derivatives
with respect to the radial coordinate \(r\), and time coordinate
\(t\), respectively.  It should also be noted that the coordinates are
denoted in two separate but equivalent ways in this derivation:
\(x^{i} \equiv (x^{0}, x^{1}, x^{2}, x^{3}) \equiv
(t,r,\theta,\varphi)\).  If we assert the same symmetry conditions on
the energy momentum tensor, we have to assume a \(T^{i}{}_{j}\) of the
following form, as discussed in previous chapters:
\begin{equation}
  \label{eq:EnergyMomentum}
  T^{i}{}_{j} = \left( p_{r} + \rho \right) u^{i}u_{j} - \delta^{i}{}_{j} - \left(\ppen - p_{r} \right)n^{i}n_{j}
  + \f{1}{4\pi} (F_{jk}F^{ki} + \f{1}{4} \delta^{i}_{j}F_{ab}F^{ab}), 
\end{equation}
where \(\delta^{i}{}_{j}\) is the Kronecker delta, \(u^{i}\) and
\(u_{j}\) are the contravariant and co-variant space-like
four-velocities, defined through \(u^{i} = \f{\d x^{i}}{\d s}\), so
that \(u_{i} u^{j} = 1,\) and \(n_{i}\) is a time-like four-velocity
so that \(n_{i}n^{j} = -1.\) The last part of the equation
incorporates the electromagnetic part of the energy-momentum tensor
derived from the Faraday tensor \(F_{ab} = A_{a;b} - A_{b;a},\) with
\(A_{a}\) the usual electromagnetic four-potential.  Since we are
considering the static and spherically symmetric case with anisotropic
pressure, we will have two distinct pressures: the radial pressure
\(p_{r},\) and the tangential pressure \(\ppen;\) while \(\rho\)
denotes the energy density, and the only non-zero component of the
vector potential \(A_{a}\) is the time component, so that
\(A_{a} = (A_{0},0,0,0)\).  Further we will also have that the frame
velocities are such that the angular four-velocities, \(u^{2}\) and
\(u^{3}\) vanish.  This results in the energy momentum tensor having
the following form :
\begin{equation}
  T^{i}{}_{j} =
  \begin{pmatrix}[c]
    \rho + \eta  & 0  & 0  & 0  \\
    0     & -p_{r} + \eta & 0  & 0  \\
    0     & 0  & -\ppen-\eta & 0  \\
    0     & 0  & 0  & -\ppen - \eta 
  \end{pmatrix}.
\end{equation}
Here, as derived in Appendix~\ref{C:AppendixA},
\(\eta = \f{\e^{-(\nu+\lambda)}}{8\pi} (F_{01})^{2}\).

We note here that the form of the equations we are using is slightly
more general than the previous versions~\eqref{t7.eq:EinR},
or~\eqref{ns.eq:EinMR} which did not include the time derivative
terms.  The reason for this more general form is that we want to be
able to perturb our solutions in time, and this is impossible to do
with the static set of Einstein's equations we used before.

\subsection{Simplifying the Einstein's equations}
The equations~\eqref{eq:EEa} and \eqref{eq:EEb} can be combined to give the
following more workable form: 
\begin{equation}
  \label{eq:EinCombined}
  \f{\e^{-\lambda}}{r} \f{\partial }{\partial r} \left( \lambda + \nu \right) = 
 \f{8\pi\G}{\un{c}^{4}} \left( T^{1}{}_{1} - T^{0}{}_{0} \right).
\end{equation}

The equations~\eqref{eqs:EinExplicit} and not independent, but rather
are related through the Bianchi identities: \(T^{i}{}_{j;i} = 0\).
Explicitly these give rise to the following two equations:
\begin{subequations}
  \label{eqs:Bianchi}
\begin{equation}
  \label{eq:B1}
  \pderiv{T^{0}{}_{0}}{t} + \pderiv{T^{1}{}_{0}}{r} + 
\f{1}{2} \left( T^{0}{}_{0} - T^{1}{}_{1} \right) \pderiv{\lambda}{t} +
T^{1}{}_{0} \left( \f{1}{2} \pderiv{(\lambda + \nu)}{r} + \f{2}{r}\right) = 0, 
\end{equation}
and, 
\begin{equation}
 \label{eq:B2}
\pderiv{T^{0}{}_{1}}{t} + \pderiv{T^{1}{}_{1}}{r} + 
\f{1}{2} T^{0}{}_{1} \pderiv{(\lambda + \nu)}{t} +
\f{1}{2} \left( T^{1}{}_{1}- T^{0}{}_{0} \right) \pderiv{\nu}{r} + 
\f{1}{r} \left(2T^{1}{}_{1} - T^{2}_{2} - T^{3}_{3} \right) = 0.  
\end{equation}
\end{subequations}
in the \(j=0,1\) cases respectively.  These equations will allow us to
perform a number of simplifications later on.

\subsubsection{The static case}
For fluid balls that are in hydrostatic equilibrium, additionally,
there is no dependence of any of the fields on the time coordinate
\(t\).  All of the above equations then simplify, as a result of the
following zero-subscripted time independent variables replacing the
general time dependant ones:
\begin{equation}
  T^{i}{}_{j} =
  \begin{pmatrix}[c]
    \rho_{0} + \eta_{0}  & 0  & 0  & 0  \\
    0     & -p_{r0} + \eta_{0} & 0  & 0  \\
    0     & 0  & -\ppeno-\eta_{0} & 0  \\
    0     & 0  & 0  & -\ppeno - \eta_{0} 
  \end{pmatrix},   
\label{eq:StaticAssumptions}
\end{equation}
with the metric functions, \(\lambda(r,t) = \lambda_{0}(r),\) and
\(\nu(r,t) = \nu_{0}(r).\) Additionally, all the frame four velocities
also vanish by choice, except for \(u^{0}\).  We will also choose
units so that \(c = G = 1,\) so as to avoid carrying all the \(G\) and
\(c\) terms through this long calculation.  As a result of these
simplifications, the equations given above as~\eqref{eqs:EinExplicit},
\eqref{eq:EnergyMomentum}, \eqref{eq:EinCombined},
and~\eqref{eqs:Bianchi} simplify to the following set:
\begin{subequations}\label{eqs:ECstatic}
  \begin{align}
    \deriv{(r \e^{-\lambda_{0}})}{r} &= 1 - 8\pi r^{2}(\rho_{0} + \eta_{0}),\label{eq:ECa}\\
\f{\e^{-\lambda_{0}}}{r} \deriv{\nu_{0}}{r} &= 
\f{1}{r^{2}} \left( 1-\e^{-\lambda_{0}}\right) + 8\pi (p_{r0} - \eta_{0}), \label{eq:ECb} \\
\deriv{(p_{r0} - \eta_{0})}{r} &= -\f{1}{2} \left( p_{r0} + \rho_{0}\right) \deriv{\nu_{0}}{r} + \f{2}{r}(\ppeno-p_{r0}) + \f{4\eta_{0}}{r}, \label{eq:ECc}\\
\f{\e^{-\lambda_{0}}}{r} \deriv{\left( \nu_{0} + \lambda_{0} \right)}{r} &=
8\pi \left( p_{r0} + \rho_{0}\right). \label{eq:ECd}
  \end{align}
\end{subequations}

If we want to investigate the stability of these equations, we will
have to perform a time dependent perturbation on them, keeping in mind
that the perturbed equations will still obey the full Einstein's
equations~\eqref{eqs:EinExplicit}.  Since we will be introducing time
dependent fields in the pressure \(p\), the density \(\rho\), the
metric coefficients \(\nu\), and \(\lambda\), and the four-velocities
\(u^{i}\), we will need the full set of the time dependant
equations~\eqref{eqs:EinExplicit}.

In linear stability analysis, it is common to expand every perturbed
expression to first order consistently.  This is what we will strive
to do in the following, starting first with expressions for the time
four-velocity:
\[u^{0} = \f{\d t}{\d s} = \sqrt{\f{\d t^{2}}{\d s^{2}}} =
\e^{-\nu/2}.\] The control variable we will be using to do our
perturbation expansion is \(v = \f{\d r}{\d t}.\) Expressing the
radial four-velocity in terms of this variable, we get 
\[u^{1} = \f{\d r}{\d s} = \deriv{r}{t} \cdot \deriv{t}{s} = v
\e^{-\nu/2}, \quad\text{since}\quad \deriv{s}{t} \neq 0.\]  The respective
co-variant versions of the velocities are obtained through the metric
since \(u_{i} = g_{ij} u^{j}.\) Since the metric is diagonal, this
expression simplifies considerably for both radial and time
four-velocities resulting in \[ u_{0} = g_{00}u^{0} = \e^{\nu/2} \quad
\text{and} \quad u_{1} = g_{11}u^{1} = v \e^{\lambda-\nu/2}.\]

\subsection{Perturbing the static case}
We are now ready to perturb the fields, keeping in mind that any terms
that are second order or higher in the perturbation will be discarded.
This process involves the consistent substitution of
\(r \to r+\delta r = r + \zeta,\)
leading to \(\lambda \rightarrow \lambda_{0} + \delta \lambda,\)
and \(\nu \rightarrow \nu_{0}+\delta \nu. \)
As a result we will be getting the following
\begin{subequations}\label{eq:4Vel}
  \begin{align}
  u^{0} &= \e^{(\nu_{0}+\delta \nu)/2} \simeq \e^{-\nu_{0}/2}, \label{eq:t4VelContra} \\  
  u^{1} &= v \e^{-(\nu_{0}+\delta \nu)/2 } \simeq v \e^{-\nu_{0}/2}, \label{eq:r4VelContra}\\
  u_{0} &\simeq \e^{\nu_{0}/2}, \label{eq:t4VelCo}\\
  u_{1} &\simeq v \e^{\lambda_{0}-\nu_{0}/2}. \label{eq:r4VelCo} 
  \end{align}
\end{subequations}
It might seem that terms of the first order are also being culled in
the above, particularly in equation~\eqref{eq:t4VelContra},
and~\eqref{eq:t4VelCo}, but since the four-velocity \(u^{0}\) always
occurs in a product in all the fields we are considering here, instead
of carrying the first order term continuously, and lengthen an already
tedious process, we use only the zeroth order approximation, for these
expressions.

By using the energy-momentum equation~\eqref{eq:EnergyMomentum}, we
can find the corresponding perturbed energy-momentum introduced by the
perturbed fields mentioned above.  A straightforward substitution of
the four-velocities results in perturbed pressures, \(p_{r}
\rightarrow p_{r0}+\delta p_{r}\), and \(\ppen \rightarrow \ppeno +
\delta \ppen\); perturbed energy density fields, \(\rho \rightarrow
\rho_{0} + \delta \rho \); and perturbed electromagnetic fields \(\eta
\rightarrow \eta_{0} + \delta \eta \), resulting in,
\begin{equation} \label{eq:perturbedEM}
  T^{i}{}_{j} = 
  \begin{pmatrix}[c]
    A  & B  & 0  & 0  \\
    -\e^{\lambda_{0} - \nu_{0}} B     & C  & 0  & 0  \\
    0     & 0  & D & 0  \\
    0 & 0 & 0 & D
  \end{pmatrix}, \quad \text{with,} \quad 
  \left\{ \begin{matrix}[l]
  A = \rho_{0}+ \eta_{0} + \delta \rho +\delta \eta, \\  
  B = (p_{r0}+\rho_{0})v, \\
  C = \eta_{0} -p_{r0} + \delta \eta -\delta p_{r} \\
  D = -\eta_{0} - \ppeno - \delta \eta -\delta \ppen
\end{matrix}
\right.
\end{equation}
This expression, through construction, is to first order, as we wanted,
since the pressure and density are the fields we will be concerned
with mostly.  With this new perturbed equation in mind, and the
full Einstein equations, we find that equation~\eqref{eq:ECa} has a
very similar perturbed form, viz
\[
\pderiv{(r \e^{-\lambda_{0}-\delta \lambda})}{r} = 1 -
8\pi r^{2}(\rho_{0}+\delta
\rho + \eta + \delta \eta).
\]
The total derivative is transformed into a partial one since now
\(\delta \lambda\) also depends on time.  Simplifying this expression
to first order results in the following:
\[
\begin{aligned}
  \pderiv{(r \e^{-\lambda_{0}} \e^{-\delta \lambda})}{r} &= 1 -
  8\pi r^{2} (\rho_{0}  + \eta_{0} ) -8\pi r^{2} (\delta \rho +\delta \eta), \\
  \pderiv{(r \e^{-\lambda_{0}}(1-\delta \lambda + \cdots) )}{r} - 1 + 8\pi r^{2}(\rho_{0} + 
\eta_{0}) & = - 8\pi r^{2} (\delta \rho + \delta \eta), \\
\underbrace{\deriv{(r \e^{-\lambda_{0}}) }{r} -1 + 8\pi r^{2}(\rho_{0} + \eta_{0})}_{=0,\text{ from equation~\eqref{eq:ECa}}} - \pderiv{}{r}(r \e^{-\lambda_{0}}\delta \lambda) &=-8\pi r^{2} (\delta \rho + \delta \eta).
\end{aligned}
\]
This simplification give us the relation between the
perturbation in energy density and electromagnetic field, and one of the metric coefficient:
\begin{equation}
  \label{eq:pertLambdaRho}
  \pderiv{}{r}(r \e^{-\lambda_{0}}\delta \lambda) =8\pi r^{2} (\delta \rho + \delta \eta)
\end{equation}
Similarly the perturbed equation relating the other metric
coefficients, and the pressure resulting from
perturbing~\eqref{eq:ECb}, becomes
\[
\f{\e^{-\lambda_{0}} \e^{-\delta \lambda}}{r} \pderiv{}{r}(\nu_{0}+\delta \nu) = 
\f{1}{r^{2}} \left( 1-\e^{-\lambda_{0}} \e^{-\delta \lambda}\right) + 8\pi (p_{r0} + \delta p_{r} -\eta_{0} -\delta \eta)
\]
On expanding and removing the zeroth order terms in accordance to the
static equation~\eqref{eq:ECb} results in the simplification:
\begin{multline*}
  \f{\e^{-\lambda_{0}}}{r} (1-\delta \lambda + \cdots) \left( \deriv{\nu_{0}}{r} + \pderiv{\delta \nu}{r}\right) \simeq \f{1}{r^{2}}[1-\e^{-\lambda_{0}}(1-\delta\lambda + \cdots)] \\
+ 8\pi (p_{r0} - \eta_{0}) +8\pi( \delta p_{r} - \delta \eta ).
\end{multline*}
Rearranging this equation, and collecting terms that constitute the zeroth order form then gives, 
\begin{multline}
\overbrace{\f{\e^{-\lambda_{0}}}{r} \deriv{\nu_{0}}{r} - \f{1}{r^{2}}(1-\e^{-\lambda_{0}}) - 8\pi (p_{r0} - \eta_{0})}^{=0 \text{ from equation~\eqref{eq:ECb}}}  
= \f{1}{r^{2}}(\e^{-\lambda_{0}} \delta \lambda) + 8\pi \delta p -\f{\e^{-\lambda_{0}}}{r} \pderiv{\delta\nu}{r} \\ 
+\f{\e^{-\lambda_{0}}}{r}\delta \lambda \left( \deriv{\nu_{0}}{r} + \pderiv{\delta\nu}{r}\right) + 8\pi (\delta p_{r} - \delta \eta).  
\end{multline}
We now remove all terms that are higher than first order.  Derivatives
of the perturbations are taken to be first order, but products of
perturbations are of second order, and we neglect them.  The previous
equation reorganized thus gives:
\[
\f{\e^{-\lambda_{0}}}{r} \left( \pderiv{\delta \nu}{r} - \delta \lambda \deriv{\nu_{0}}{r} \right) -\f{\e^{-\lambda_{0}}}{r} \underbrace{\delta\lambda \pderiv{\delta\nu}{r}}_{2^{\text{nd}} \text{ order}}= \f{\e^{-\lambda_{0}}}{r^{2}} \delta\lambda + 8\pi (\delta p_{r} - \delta \eta),
\]
which finally results in an equation relating the perturbations only:
\begin{equation}
  \label{eq:pertLambdaNuP}
 \f{\e^{-\lambda_{0}}}{r} \left( \pderiv{\delta \nu}{r}  -\delta \lambda \deriv{\nu_{0}}{r}  \right) = \f{\e^{-\lambda_{0}}}{r^{2}} \delta\lambda + 8\pi (\delta p_{r} - \delta \eta). 
\end{equation}
In the static set of equations~\eqref{eqs:ECstatic}, we notice that
since none of the variables depend explicitly on time, the
off-diagonal terms of the Einstein equations, and energy momentum
tensor vanish.  In the perturbed set, this is unfortunately not the
case, and we have to deal with the additional equation~\eqref{eq:EEd},
contingent upon the static version where
\begin{equation}
  \label{eq:EEdStatic}
  -8\pi T^{0}{}_{1} = \f{\e^{\lambda_{0}}\dot{\lambda_{0}}}{r} = 0.
\end{equation}
If we perturb equation~\eqref{eq:EEd}, use the linearised expression
for \(T^{0}{}_{1}\) from equation~\eqref{eq:perturbedEM}, linearise
the rest, and remove the contribution from the static
part~\eqref{eq:EEdStatic}, we get the following simplifications:
\[
\begin{aligned}
  -8\pi T^{0}{}_{1} &= - \f{\e^{-(\lambda_{0}+ \delta\lambda)}}{r} \pderiv{}{t}(\lambda_{0} + \delta\lambda),\\
  8\pi (p_{r0}+\rho_{0})v &=
  -\f{\e^{\lambda_{0}}[1-\delta\lambda+\cdots]}{r} \pderiv{}{t}(\lambda_{0} +
  \delta\lambda),\\
  &\simeq \f{\e^{-\lambda_{0}}}{r} \underbrace{\deriv{\lambda_{0}}{t}
    \left(\delta\lambda - 1 \right)}_{=0,\text{ no time dependance}}
  -\f{\e^{-\lambda_{0}}}{r}\pderiv{\delta\lambda}{t} +
  \underbrace{\f{\e^{-\lambda_{0}}}{r}\delta\lambda
    \pderiv{\delta\lambda}{t}}_{\secO}.
\end{aligned}
\]
The final equation relating the velocity \(v\) to the metric perturbations is then:
\begin{equation}
  \label{eq:pertvLambda}
  - 8\pi (p_{r0}+\rho_{0}) v =\f{\e^{-\lambda_{0}}}{r}\pderiv{\delta\lambda}{t}. 
\end{equation}
We can also use the Bianchi identities~\eqref{eqs:Bianchi} to get an
equation relating all the different perturbations.  To achieve this,
we first look at the how the static condition simplified this
equation, and then generate the time dependant version with the
perturbation.

The static case with assumptions~\eqref{eq:StaticAssumptions}
transforms the first Bianchi identity~\eqref{eq:B1} into
equation~\eqref{eq:ECc}.  The other Bianchi identity with the perturbed
quantities results in
\begin{multline*}
\pderiv{}{t} \left[ -\e^{\lambda_{0} - \nu_{0}} ( \rho_{0} + p_{r0})v\right]  + \f{2}{r} (\eta_{0} -p_{r0}+\delta\eta-\delta p_{r} + \eta_{0} +\ppeno +\delta \eta +\delta\ppen) \\+ \f{1}{2}\left[ -\e^{\lambda_{0}-\nu_{0}} (p_{r0}+\rho_{0})v\right]\pderiv{}{t} (\nu_{0}+\delta\nu +\lambda_{0}+\delta\lambda) + \pderiv{}{r} (\eta_{0}-p_{r0}+\delta \eta -\delta p_{r}) \\ + \f{1}{2}(\eta_{0}-p_{r0}+\delta\eta -\delta p_{r} - \rho_{0} -\eta_{0}-\delta \rho -\delta\eta) \pderiv{}{r}(\nu_{0}+\delta \nu)  = 0.
\end{multline*}
If we rearrange this equation and cancel the time-derivatives of
static quantities, and additionally realize that the quantity \(v\) is
already first order, we can simplify the
above into:
\begin{multline*}
  -\e^{\lambda_{0}-\nu_{0}} (p_{r0}+\rho_{0}) \pderiv{v}{t} + \pderiv{}{r} (\eta_{0} - p_{r0} + \delta\eta -\delta p_{r}) 
-\f{1}{2} (p_{r0}+\rho_{0}) \deriv{\nu_{0}}{r}  
-\f{1}{2} (p_{r0}+\rho_{0}) \pderiv{\delta\nu}{r} \\
-\f{1}{2}[\e^{\lambda_{0} -\nu_{0}} (p_{r0}+\rho_{0})v ]  \overbrace{ \left[ \deriv{}{t}(\nu_{0}+\lambda_{0}) \right.}^{=0\text{, static}} + \overbrace{\left.\pderiv{}{t} (\delta \nu +\delta \lambda)\right]}^{=0\text{, \secO}} 
-\f{1}{2} (\delta p_{r}+\delta \rho) \pderiv{(\nu_{0}+\cancel{\delta\nu)}}{r} \\+ 
\f{2}{r} (\ppeno + \delta\ppen + 2 \delta\eta + 2\eta_{0} - p_{r0}-\delta p_{r}) = 0.
\end{multline*}
Upon rearrangement, and isolating the parts of the equation that
correspond to the static case to get a further simplification, we
obtain
\begin{multline*}
  -\e^{\lambda_{0}-\nu_{0}} (p_{0}+\rho_{0}) \pderiv{v}{t}  -\overbrace{\f{1}{2} (\rho_{0} + p_{r0}) \pderiv{\nu_{0}}{t} -\pderiv{}{r}(p_{r0}-\eta_{0}) + \f{2}{r}(\ppeno - p_{r0} + 2\eta_{0})}^{=0, \text{ static equation}}  \\- \pderiv{}{r}(\delta p_{r} - \delta \eta) + 
\f{2}{r} (- \delta p_{r} + \delta\ppen ) - \f{1}{2} (p_{r0}+\rho_{0}) \pderiv{\delta\nu}{r}  - \f{1}{2} (\delta p_{r} + \delta \rho) \pderiv{\nu_{0}}{r} = 0.
\end{multline*}
The final constraint equation relating all the perturbations, obtained
from the second Bianchi identity simplifies to,
\begin{equation}
  \label{eq:simpB1}
  \e^{\lambda_{0}-\nu_{0}} (p_{r0}+\rho_{0}) \pderiv{v}{t}  + \pderiv{}{r}(\delta p_{r} - \delta \eta) + \f{1}{2} (\delta p_{r} +\delta\rho)\deriv{\nu_{0}}{r} +\f{1}{2}(p_{r0}+\rho_{0}) \deriv{\delta\nu}{r} + \f{2}{r} (\delta p_{r} - \delta \ppen) = 0.
\end{equation}
The first Bianchi identity imposes no further constraints on the
perturbations, instead it regenerates one of the previous equations.
We will now be using this second equation as our starting point and
find all the terms in it by other methods.  First we will find
expressions for the metric perturbations \(\delta \nu,\) and \(\delta
\lambda.\)

\subsection{Lagrangian description and partial integration}
If we now shift our attention to a Lagrangian description as opposed
to the Eulerian one we have been using thus far, we can define a
Lagrangian displacement in terms of the velocity \(v.\) Let \(\zeta\)
represent such a displacement with respect to the time coordinate,
\(x^{0} = t.\) Then we can define \(v = \pderiv{\zeta}{t},\) and
rewrite equation~\eqref{eq:pertvLambda} in terms of this displacement.
Doing this allows us to immediately integrate the latter equation to:
\begin{align}
- 8\pi (p_{r0}+\rho_{0}) \pderiv{\zeta}{t} &=\f{\e^{-\lambda_{0}}}{r}\pderiv{\delta\lambda}{t}, \notag \\
  - 8\pi (p_{r0}+\rho_{0}) \int \pderiv{\zeta}{t} \d t &=\f{\e^{-\lambda_{0}}}{r} \int \pderiv{\delta\lambda}{t} \d t, \notag\\
- 8\pi (p_{r0}+\rho_{0}) \zeta &= \f{\e^{-\lambda_{0}}}{r} \delta\lambda \label{eq:deltaLambda}.
\end{align}
This equation can be further simplified through the use of another
relation we have already found, viz equation~\eqref{eq:ECd}.
Substituting this equation in the previous~\eqref{eq:deltaLambda}
gives:
\[
\cancel{\f{\e^{-\lambda_{0}}}{r}} \delta\lambda = - \cancel{\f{\e^{-\lambda_{0}}}{r}} \left[\deriv{}{r} \left( \nu_{0} + \lambda_{0} \right)\right] \zeta,
\]
giving us the final form of an equation relating the perturbation
\(\delta\lambda\) to the static variables, and Lagrangian
displacement:
\begin{equation}
  \label{eq:LambdaXi}
  \delta\lambda = -\zeta \left[\deriv{}{r} \left( \nu_{0} + \lambda_{0} \right)\right].
\end{equation}
We now simplify the relationship between the perturbed density
\(\delta\rho\), and the perturbed metric coefficient
\(\delta\lambda\), viz equation~\eqref{eq:pertLambdaRho}, with the
relation just obtained~\eqref{eq:deltaLambda},
\[
  \pderiv{(r \e^{-\lambda_{0}} \delta\lambda)}{r} = 8\pi r^{2} (\delta\rho + \delta \eta), \qquad \text{with} \qquad \delta\lambda = -8\pi (p_{r0}+\rho_{0})\zeta r \e^{\lambda_{0}},
\]
to obtain the following simplification, 
\begin{align*}
  \pderiv{}{r} \left\{ r \bcancel{\e^{-\lambda_{0}}} \left[- \cancel{8\pi} (p_{r0}+\rho_{0})\zeta r \bcancel{\e^{\lambda_{0}}} \right] \right\} &= \cancel{8\pi} r^{2} (\delta\rho + \delta \eta),\\
\pderiv{}{r}\left\{ -r^{2} \zeta (p_{r0}+\rho_{0} )\right\} &= r^{2}(\delta\rho + \delta \eta),
\end{align*}
finally giving us the equation relating the perturbed density in terms of the non perturbed variables:
\begin{equation}
  \label{eq:deltaRho}
  \delta\rho + \delta \eta = - \f{1}{r^{2}} \pderiv{}{r}\left(r^{2}\zeta (p_{r0}+\rho_{0}) \right).
\end{equation}
The above compact form of the equation can be expanded and further
simplified into an alternative version involving the expanded
derivative on the right hand side.  This is done to isolate perturbed
quantities from static ones explicitly, as follows:
\begin{align}
  \delta\rho +\delta \eta &= - \f{1}{r^{2}} \left\{ r^{2} \zeta \pderiv{(p_{r0}+\rho_{0})}{r} + (p_{r0}+\rho_{0}) \pderiv{(r^{2}\zeta)}{r} \right\} \notag,\\
  \delta\rho + \delta \eta &= -\zeta \deriv{p_{r0}}{r} - \zeta \deriv{\rho_{0}}{r} -\left(\f{p_{r0}+\rho_{0}}{r^{2}} \right)\pderiv{(r^{2}\zeta)}{r}.
\end{align}
Recalling that the simplified static Bianchi identity results in an
expression for the static pressure derivative through
equation~\eqref{eq:ECc}, we can substitute the first right hand term
in the above equation, and together with the redefinition of
anisotropy measure, \(\Pi = \ppen - p_{r}\), we have
\[
\delta\rho + \delta \eta = -\zeta \deriv{\rho_{0}}{r} -\zeta\left\{ -\f{1}{2}(p_{r0} + \rho_{0}) \deriv{\nu_{0}}{r} + \f{2}{r}\Pi_{0}  + \f{4\eta_{0}}{r} + \deriv{\eta_{0}}{r} \right\} - \f{p_{r0} + \rho_{0}}{r^{2}} \pderiv{}{r}(r^{2} \zeta),
\]
which is easily rearranged into the more convenient,
\[
\delta\rho +\delta \eta = -\zeta \deriv{\rho_{0}}{r} - \f{p_{r0}+\rho_{0}}{r^{2}} \left\{ \pderiv{(r^{2}\zeta)}{r} - \f{\zeta r^{2}}{2} \deriv{\nu_{0}}{r} \right\} - \zeta \left[ \f{2\Pi_{0}}{r} + \f{4\eta_{0}}{r} + \deriv{\eta_{0}}{r} \right].
\]
We are now in a position to obtain a compact form of the above
equation by multiplying the term in braces by unity:
\(\e^{\nu_{0}/2}\e^{-\nu_{0}/2} = 1\) as shown:
\begin{multline*}
\delta\rho = -\zeta \deriv{\rho_{0}}{r} - \f{(p_{r0}+\rho_{0}) \e^{\nu_{0}/2}}{r^{2}} \left\{ \e^{-\nu_{0}/2}\pderiv{(r^{2}\zeta)}{r} - \f{\zeta r^{2}}{2} \e^{-\nu_{0}/2}\deriv{\nu_{0}}{r} \right\} \\- \zeta \left[ \f{2\Pi_{0}}{r} + \f{4\eta_{0}}{r} + \deriv{\eta_{0}}{r} \right] - \delta \eta.
\end{multline*}
This last non-intuitive step is needed to factorize the derivative on the right hand side into the compact form we are looking for, 
\begin{equation}
  \label{eq:deltaRhof}
\delta\rho = -\zeta \deriv{\rho_{0}}{r} - \f{(p_{r0}+\rho_{0}) \e^{\nu_{0}/2}}{r^{2}} \left\{ \pderiv{}{r}(\e^{-\nu_{0}/2}r^{2}\zeta) \right\} - \zeta \left\{ \f{2\Pi_{0}}{r} + \f{4\eta_{0}}{r} + \deriv{\eta_{0}}{r} \right\} - \delta \eta.  
\end{equation}
We now proceed in a similar fashion to obtain another
expression, this time for the other perturbed metric coefficient
\(\delta\nu\).  To achieve this we first notice that we already have a
promising candidate, viz equation~\eqref{eq:pertLambdaNuP}.  If we
were to substitute for the \(\delta\lambda\) terms in this equation
from the result~\eqref{eq:deltaLambda}, we get,
\[
\f{\e^{-\lambda_{0}}}{r} \pderiv{\delta\nu}{r} = 8\pi \left\{\delta p_{r} -\delta \eta - \f{\zeta(p_{r0}+\rho_{0})}{r} \right\}  + \deriv{\nu_{0}}{r} \left\{ -8\pi (p_{r0}+\rho_{0})\zeta \right\},
\]
upon collecting like terms we get the following more workable form of the equation:
\begin{equation}
  \label{eq:pertNuP}
  \f{\e^{-\lambda_{0}}}{r} \pderiv{\delta\nu}{r} = 8\pi \left[ \delta p_{r} -\delta \eta
 - (p_{r0}+\rho_{0})\zeta \left\{ \deriv{\nu_{0}}{r} + \f{1}{r}\right\} \right].
\end{equation}
Remembering from equation~\eqref{eq:ECd}, that we have an expression
for the first term in the above equation, we can do one more
substitution in this equation to get
\begin{equation}
  \label{eq:pertNuPf}
  (p_{r0}+\rho_{0}) \pderiv{\delta\nu}{r} = \left[ \delta p_{r} - \delta \eta 
 - (p_{r0}+\rho_{0})\zeta \left\{ \deriv{\nu_{0}}{r} + \f{1}{r}\right\} \right] \deriv{(\lambda_{0}+\nu_{0})}{r}.
\end{equation}

Thus far we have obtained the expressions for the perturbations of the
metric functions \(\lambda\) and \(\nu,\) and of the matter density
\(\rho\) when the radius of the star is changed.  Next we find the
perturbation of the electric field \(\delta \eta\) in terms of the
static quantities following the work of \citeauthor{Gla76}\cite{Gla76}
who considers the Maxwell's source equation:
\begin{equation}
  \label{eq:Maxwell}
 \pderiv{}{x^{a}} \left(\sqrt{(-g)} F^{ab}\right) = -\f{4\pi}{c} \sqrt{(-g)} J^{b},
\end{equation}
with \(g = - (r^{2} \sin \theta)^{2} \e^{(\nu+\lambda)} \)
the metric determinant, and \(J^{a} = \epsilon c u^{a}\)
the 4-current density with epsilon being the charge density.  

In the static case the only non-zero components of \(F_{ab}\)
are \(F_{01} = -F_{10} = E,\) the electric field,
so that equation~\eqref{eq:Maxwell} reduces to
\begin{equation}
  \label{eq:MaxwellHere}
\pderiv{}{t} \left(\sqrt{(-g)} F^{10}\right) = -4\pi \sqrt{(-g)}
\epsilon u^{1},
\end{equation}
also since \(F^{ab} = g^{ac}g^{bd}F_{cd},\)
and \(g^{ab}\) is diagonal, 
we are left with \(F^{01} = g^{11}g^{00}F_{01},\)
resulting in \(F^{01} =-\e^{-(\lambda + \nu)} E.\)
When the radial coordinate is perturbed, the electric field changes in
such a way as to cause \(E \to E_{0} + \delta E,\)
and similarly the metric coefficients
\(\lambda \to \lambda_{0} + \delta \lambda,\)
and \(\nu \to \nu_{0} + \delta \nu.\)

Then both sides of equation~\eqref{eq:MaxwellHere} can be simplified
separately: the LHS giving
\begin{align*}
\pderiv{}{t} \left(\sqrt{(-g)} F^{10}\right) &= \pderiv{}{t} \left( - r^{2} \sin{\theta} \e^{(\lambda + \nu)/2} \e^{-(\lambda + \nu)} (E)\right), \\
&=- r^{2} \sin{\theta} \pderiv{}{t} \left[  \e^{-(\lambda_{0} +\delta \lambda + \nu_{0} + \delta \nu )/2}  ( E_{0}+\delta E ) \right], \\
&=- r^{2} \e^{-(\lambda_{0} + \nu_{0})/2} \sin{\theta} \pderiv{}{t} \left[ \left(1 -\f{\delta \lambda}{2} - \f{\delta \nu}{2} + \cdots \right) ( E_{0}+\delta E ) \right], \\
&=- r^{2} \e^{-(\lambda_{0} + \nu_{0})/2} \sin{\theta} \pderiv{}{t} \left(\delta E - E_{0}\f{\delta \lambda}{2} - E_{0}\f{\delta \nu}{2} \right) + O(\delta^{2}).
\end{align*}
Similarly since the velocity \(u^{1},\)
following~\eqref{eq:r4VelContra}, goes to \(v e^{-\nu_{0}/2} = \pderiv{\zeta}{t} e^{-\nu_{0}/2} ,\) under radial perturbations, and the change density \(\epsilon \to \epsilon_{0} + \delta \epsilon,\) 
the RHS of equation~\eqref{eq:MaxwellHere} gives
\begin{align*}
  -4\pi \sqrt{(-g)} \epsilon u^{1} &= -4 \pi r^{2} \sin{\theta} \e^{(\lambda + \nu)/2} \epsilon v e^{-\nu_{0}/2}, \\
  &=-4 \pi r^{2} \sin{\theta} \e^{(\lambda_{0} +\delta \lambda + \nu_{0} + \delta \nu )/2} (\epsilon_{0} + \delta \epsilon) \pderiv{\zeta}{t} e^{-\nu_{0}/2}, \\
&=-4 \pi r^{2} \sin{\theta} \e^{\lambda_{0}/2} (1 + \delta \nu +\delta \lambda+\cdots) (\epsilon_{0} + \delta \epsilon) \pderiv{\zeta}{t} ,\\
&=  -4 \pi r^{2} \sin{\theta} \e^{\lambda_{0}/2}\epsilon_{0}\pderiv{\zeta}{t} + O(\delta^{2}). 
\end{align*}
Identifying both sides of the equation 
\[
- r^{2} \e^{-(\lambda_{0} + \nu_{0})/2} \sin{\theta} \pderiv{}{t} \left(\delta E - E_{0}\f{\delta \lambda}{2} - E_{0}\f{\delta \nu}{2} \right) = -4 \pi r^{2} \sin{\theta} \e^{\lambda_{0}/2}\epsilon_{0}\pderiv{\zeta}{t} \] then allows us to conclude after a time integration that to first order,
\begin{equation}
  \label{eq:deltaE}
2 \f{\delta E}{E_{0}} - \delta \lambda - \delta \nu = \f{8 \pi}{E_{0}} \e^{(\lambda_{0} + \nu_{0}/2)} \epsilon_{0} \zeta,
\end{equation}

As seen in appendix~\ref{C:AppendixA}, the electromagnetic part of the
stress energy tensor can be expressed as
\((T^{0}{}_{0})_{\text{EM}} = \eta = \f{\e^{-(\lambda+\nu)}}{8\pi}
(F_{10})^{2},\) which under a radial perturbation is transformed as
\begin{align*}
  \eta_{0} + \delta \eta &= \f{\e^{-(\lambda_{0}+\delta \lambda +\nu_{0}+ \delta \nu)}}{8\pi} (E_{0} + \delta E )^{2},\\
&= \f{\e^{-(\lambda_{0}+\nu_{0})}}{8\pi} (E_{0})^{2} (1-\delta \nu - \delta \lambda + \cdots) \left(1 + \f{\delta E}{E_{0}}\right)^{2} \\
&=\f{\e^{-(\lambda_{0}+\nu_{0})}}{8\pi} (E_{0})^{2} \left( 1 + 2\f{\delta E}{E_{0}} -\delta \nu -\delta \lambda \right) + O(\delta^{2}).
\end{align*}
The terms in brackets being the same as equation~\eqref{eq:deltaE}, we
can immediately write
\[\eta_{0} + \delta \eta = \f{\e^{-(\lambda_{0}+\nu_{0})}}{8\pi}
(E_{0})^{2} + \e^{-(\lambda_{0}+\nu_{0})} \e^{(\lambda_{0} +
  \nu_{0}/2)} E_{0} \epsilon_{0}\zeta,\]
allowing us to deduce that the perturbation of the stress energy
component of the electromagnetic field is given by
\begin{equation}
  \label{eq:DeltaEta}
\delta \eta = E_{0} \epsilon_{0} \e^{-\nu_{0}/2} \zeta.
\end{equation}

We now have all the perturbations of the field quantities, except for
the pressures.  The latter require a constitutive relation of the
material and the one we will use is baryon conservation, to be able to
close the system of equations and come up with a complete set of
perturbation equations for all the matter and metric fields.

\subsection{Baryon number conservation}
An equation of state will involve a fixed number of baryons, since we
will be considering the static case.  This number will obviously
depend on the other state variables in a non-trivial way.  In the most
general static case we will have
\(\bar{N}(\rho, p_{r}, \ppen, \eta, r).\) However, in all models we
will be analysing we will have a known dependence of the perpendicular
pressure \(\ppen,\) on the radial pressure \(p_{r}\) and the radial
parameter, so that we can simplify the baryon number as
\(\tilde{N}(\rho, p_{r}, \eta, r)\).  Furthermore the charge density
\(\eta,\) will as seen previously, depend on the mass density
\(\rho\), and the radial parameter, so that \(\eta(\rho, r)\), then
without loss of generality we can have the baryon number as
\(N(\rho, p_{r}, r)\).  In whichever way \(N\) is introduced, the
scalar baryon number, \(N\) has to be conserved in any radial
perturbation.  The way this is expressed in general relativity, as
seen previously~\eqref{A.eq:VecCons} is
\begin{equation}
  \label{eq:pertBaryCons}
  \left(N u^{k} \right)_{;k} = 0,
\end{equation}
where \(u^{k}\) is the four-velocity.  Upon expansion this equation results in 
\begin{align*}
 0 &= \pderiv{}{x^{k}} (N u^{k}) + N u^{k} \pderiv{}{x^{k}}(\log\sqrt{-g}), \\
   &= \pderiv{(Nu^{0})}{t} + \pderiv{(Nu^{1})}{r} + Nu^{1} \left( \f{\nu' + \lambda'}{2} + \f{2}{r} \right) + Nu^{0} \left( \f{\dot{\nu}+ \dot{\lambda}}{2}\right),
\end{align*}
  as seen in~\eqref{A.eq:contractedGamma}.  Again considering
the four-velocities introduced previously as a result of radial
perturbation, viz~\eqref{eq:4Vel}, and expanding all the derivative of the products, results in 
\[0 = \e^{-\nu/2} \pderiv{N}{t} + \f{1}{r^{2}} \pderiv{}{r}(N v r^{2}
\e^{-\nu/2}) + \f{N \e^{-\nu/2}}{2} \pderiv{\lambda}{t} + \f{N v
  \e^{-\nu/2}}{2} \pderiv{(\lambda + \nu)}{r}, \] after
simplification.  The next step in obtaining the variation of the
baryon number as a result of metric and field perturbations is to
replace all perturbed variables with their closed form, and expand
consistently to first order.  This has to be carried out for all the
terms in the above equation, and after a tedious but straightforward
process, we obtain the following surviving first-order forms:
\begin{align*}
  \e^{-\nu/2} \pderiv{N}{t} &\rightarrow \e^{-\nu_{0}/2} \pderiv{(\delta N)}{t} \\
  \f{1}{r^{2}} \pderiv{}{r}(N v r^{2} \e^{-\nu/2}) &\rightarrow \f{1}{r^{2}} \pderiv{}{r}\left( N_{0} v r^{2} \e^{-\nu_{0}/2}\right)\\
  \f{N \e^{-\nu/2}}{2} \pderiv{\lambda}{t} &\rightarrow \f{N_{0} \e^{-\nu_{0}/2}}{2} \pderiv{(\delta \lambda)}{t}\\
  \f{N v \e^{-\nu/2}}{2} \pderiv{(\lambda + \nu)}{r} &\rightarrow \f{N_{0} \e^{-\nu/2} v}{2} \pderiv{}{r} (\lambda_{0} + \nu_{0}),
\end{align*}
resulting in the first order perturbed baryon conservation equation to read:
\begin{equation}
  \label{eq:FirstbaryonCons}
   \e^{-\nu_{0}/2} \pderiv{(\delta N)}{t} + \f{1}{r^{2}} \pderiv{}{r}\left( N_{0} v r^{2} \e^{-\nu_{0}/2}\right) + \f{N_{0} \e^{-\nu_{0}/2}}{2} \pderiv{(\delta \lambda)}{t} + \f{N_{0} \e^{-\nu/2} v}{2} \pderiv{}{r} (\lambda_{0} + \nu_{0}) = 0.
\end{equation}
This equation can be readily time-integrated, once the Eulerian
velocity \(v\) is replaced by the corresponding Lagrangian
displacements, \(\deriv{\zeta}{t}\).  Once this substitution is made,
and the equation integrated, we have
\[\delta N + \f{\e^{\nu_{0}/2}}{r^{2}} \pderiv{}{r}(N_{0} r^{2} \zeta \e^{-\nu_{0}/2}) + \f{N_{0}}{2} \overbrace{ \left(\delta \lambda + \zeta \deriv{}{r} (\lambda_{0} + \nu_{0}) \right)}^{=0, \text{from }\eqref{eq:LambdaXi} },  \]
finally resulting in
\begin{equation}
  \label{eq:deltaN}
  \delta N =  - \zeta \deriv{N_{0}}{r} - \f{N_{0}\e^{\nu_{0}/2}}{r^{2}} \pderiv{}{r} (r^{2}\zeta \e^{-\nu_{0}/2}).
\end{equation}

From the functional form of the baryon number, any variation to \(N\)
resulting from metric perturbations, will come from both the radial
pressure \(p_{r}\) and the mass density \(\rho\) allowing us to write
\begin{equation}
  \label{eq:Ndifferential}
  \delta N = \pderiv{N}{\rho} \delta \rho + \pderiv{N}{p_{r}} \delta p_{r}, 
\qquad\text{and}\qquad \d N_{0} = \pderiv{N}{\rho} \d \rho_{0} + \pderiv{N}{p_{r}} \d p_{r0},
\end{equation}
where the first equation involves perturbations, and the second one is
the differential form of the baryon number equation.  We already have
an expression for the density~\eqref{eq:deltaRho} and baryon
number~\eqref{eq:deltaN} perturbation, and using those in the previous
equation, we can find the resulting perturbation in the radial
pressure.  In order to do this, first we
substitute~\eqref{eq:deltaRho} and~\eqref{eq:deltaN}
in~\eqref{eq:Ndifferential}, and solve for \(\delta p_{r}\).  This
results in
\begin{multline*}
\pderiv{N}{p_{r}} \delta p_{r} = -\zeta \deriv{N_{0}}{r} - N_{0} C - 
\pderiv{N}{\rho} \left\{ -\zeta \deriv{\rho_{0}}{r} - (p_{r0}+\rho_{0})C \right. \\ -\left.
\zeta \left( \f{2 \Pi_{0}}{r} + \f{4\eta_{0}}{r} + \deriv{\eta_{0}}{r}\right) - \delta \eta \right\}.
\end{multline*}
Combining all the terms in \(\zeta\), and  replacing the multiply occurring expression given by \( \f{\e^{\nu_{0}/2 }}{r^{2}} \pderiv{}{r}\left( r^{2} \zeta \e^{-\nu_{0}/2}\right) \)
with the place-holder \(B\) to make the equations more concise, we realize that the
first right hand term in the next equation is recognizable as the expression
\(\pderiv{N}{p_{r}} \deriv{p_{r0}}{r}\) from equation~\eqref{eq:Ndifferential}:
\begin{multline*}
\pderiv{N}{p_{r}} \delta p_{r} =-\zeta \left( \deriv{N_{0}}{r} - \pderiv{N}{\rho} \deriv{\rho_{0}}{r} \right) - 
B\left( N_{0} - (p_{r0}+\rho_{0}) \pderiv{N}{\rho} \right)\\ +
\pderiv{N}{\rho} \left[ \zeta \left( \f{2 \Pi_{0}}{r} + \f{4\eta_{0}}{r} + \deriv{\eta_{0}}{r}\right) + \delta \eta \right].
\end{multline*}
Hence dividing this equation throughout by \(\pderiv{N}{p_{r}}\),
since we are  assuming that \(\pderiv{N}{p_{r}}\) is never equal to zero, we can
solve for the perturbation in the radial pressure brought about by
metric perturbations:
\begin{align*}
   \delta p_{r} &= -\zeta \f{\partial N / \partial p_{r}}{\partial N / \partial p_{r}} \deriv{p_{r0}}{r} - 
  \f{B p_{r0}} {p_{r0} \left(\partial N / \partial p_{r}\right) } \left( N_{0} - (p_{r0}+\rho_{0}) \pderiv{N}{\rho} \right) + \\
   & \phantom{= -\zeta} +\f{\partial N / \partial \rho}{\partial N / \partial p_{r}} \left[ \zeta \left( \f{2 \Pi_{0}}{r} + \f{4\eta_{0}}{r} + \deriv{\eta_{0}}{r}\right) + \delta \eta \right],\\
  &= -\zeta \deriv{p_{r0}}{r} - B\gamma p_{r0} + 
  \pderiv{p_{r0}}{\rho_{0}} \left[ \zeta \left( \f{2 \Pi_{0}}{r} + \f{4\eta_{0}}{r} + \deriv{\eta_{0}}{r}\right) + \delta \eta \right].
\end{align*}
Replacing the place-holder \(B\) with its proper expression, we have, 
\begin{equation}
  \label{eq:deltaPr}
  \delta p_{r} = -\zeta \deriv{p_{r0}}{r} - \gamma \f{\e^{\nu_{0}/2} p_{r0} }{r^{2}} \pderiv{}{r}\left( r^{2} \zeta \e^{-\nu_{0}/2}\right)  + 
  \pderiv{p_{r0}}{\rho_{0}} \left[ \zeta \left( \f{2 \Pi_{0}}{r} + \f{4\eta_{0}}{r} + \deriv{\eta_{0}}{r}\right) + \delta \eta \right].
\end{equation}
where we have defined the adiabatic index \(\gamma,\) following Chandrasekhar as
\begin{equation}
  \label{eq:DefGama}
  \gamma = \f{1} {p_{r0} \left(\partial N / \partial p_{r} \right)} \left( N_{0} - (p_{r0}+\rho_{0}) \pderiv{N}{\rho} \right) = \f{1}{p_{r0}} \left(N_{0}\pderiv{p_{r}}{N} - (p_{r0} + \rho_{0}) \pderiv{p_{r}}{\rho} \right),
\end{equation}
  Now
that we have an expression for the perturbed pressure, we can continue
solving for the pulsation equation, which required \(\delta p_{r}\) to be simplified.  
 
\subsection{Separation of variables}
We now have in the form of equations~\eqref{eq:pertNuPf}
and~\eqref{eq:simpB1}, a set of constraints that the perturbations we
are considering must obey.  In order to separate the time dependence
from the spatial dependence in all of these perturbation, we will
assume the usual form of radial oscillations expected in this model,
and postulate that all the perturbed fields can be expressed as the
following:
\begin{equation}
  \label{eq:pertExpXi}
  \zeta(r,t) \rightarrow \zeta(r)\e^{\i \sigma t} \implies v(r,t) = \pderiv{\zeta}{t} = \i \sigma \zeta  \e^{\i \sigma t},
\end{equation}
as a result of which, 
\begin{equation}
  \label{eq:pertExpV}
  \deriv{v}{t} = -\sigma^{2} \zeta \e^{\i \sigma t}.
\end{equation}
The pressures and density are assumed to follow similar time
evolution, with the same frequency \(\sigma\) as above, and this
results in the following postulates:
\begin{equation}
  \label{eq:pertExpPRho}
  \delta p_{r} \rightarrow \delta p_{r} \e^{\i \sigma t}, \qquad \delta \ppen \rightarrow \delta \ppen \e^{\i \sigma t}, 
  \qquad \delta\rho \rightarrow \delta \rho \e^{\i \sigma t}, 
\end{equation}
The metric coefficients will be similarly affected through this time
dependence, and we thus get
\begin{equation}
  \label{eq:pertExpLNu}
  \delta\lambda \rightarrow \delta \lambda \e^{\i \sigma t}, \qquad \delta\nu \rightarrow \delta\nu \e^{\i \sigma t}.
\end{equation}
The electromagnetic field component of the stress-energy will
similarly pulsate through
\begin{equation}
  \label{eq:DeltaEtaSep}
  \delta\eta(r,t) = E_{0} \epsilon_{0} \e^{-\nu_{0}/2}\zeta(r,t) = E_{0} \epsilon_{0} \e^{-\nu_{0}/2}\zeta\e^{\i \sigma t}.
\end{equation}

\subsection{The pulsation equation}
We now have closed forms for all the different perturbations in
equation~\eqref{eq:simpB1}. We want to deduce an equation for how the
spatial dependence of the perturbations are constrained by the
Einstein's equations, to first order.  In what follows the time
dependence of all variables have been eliminated through the
separation of variables method we have used previously, to give:
\begin{multline*}
  \e^{\lambda_{0}-\nu_{0}} \left(p_{r0}+\rho_{0}\right) \left[-\sigma^{2}\zeta \cancel{\e^{\i\sigma t}}\right] + 
  \pderiv{}{r}\left[ (\delta p_{r} -\delta \eta ) \cancel{\e^{\i \sigma t}} \right] +
  \f{1}{2}\deriv{\nu_{0}}{r} (\delta p_{r} + \delta \rho) \cancel{\e^{\i \sigma t}}  \\+
  \f{p_{r0}+\rho_{0}}{2} \deriv{(\delta \nu)}{r} \cancel{\e^{\i \sigma t}} + 
  \f{2}{r} (\delta p_{r} - \delta \ppen) \cancel{\e^{\i \sigma t}} = 0.
\end{multline*}
Since the oscillation frequency \(\sigma\), does not depend on the spatial
variables, it commutes with the derivatives in the above expression to yield:
\begin{equation}
  \label{eq:puls1}
  \sigma^{2} \e^{\lambda_{0}-\nu_{0}} \left(p_{r0}+\rho_{0}\right) \zeta  = 
  \pderiv{}{r}(\delta p_{r} -\delta \eta )  +
  \deriv{\nu_{0}}{r} \f{(\delta p_{r} + \delta \rho)}{2} +
  \f{p_{r0}+\rho_{0}}{2} \deriv{(\delta \nu)}{r}  + 
  \f{2}{r} (\delta p_{r} - \delta \ppen).
\end{equation}
This equation can be further reduced if we recall that
equation~\eqref{eq:pertNuPf} gives us an expression for the partial
derivative of one of the metric perturbations
\(\delta\nu\). Substituting the latter in the above, and rearranging terms results in the equation~ \eqref{eq:puls}
\begin{multline}
  \label{eq:puls}
  \sigma^{2}\e^{\lambda_{0}-\nu_{0}} \left( p_{r0}+\rho_{0}\right)\zeta = 
  \deriv{}{r}(\delta p_{r} - \delta \eta) + 
  \left(\f{2\delta p_{r} + \delta \rho -\delta \eta}{2}\right) \deriv{\nu_{0}}{r} +
  \left(\f{\delta p_{r} - \delta \eta}{2}\right) \deriv{\lambda_{0}}{r} + \\ -
  \f{p_{r0}+\rho_{0}}{2} \zeta \left( \deriv{\nu_{0}}{r} + \f{1}{r} \right)
  \left( \deriv{\lambda_{0}}{r} + \deriv{\nu_{0}}{r} \right) +
  \f{2}{r} (\delta p_{r} - \delta \ppen)
\end{multline}
This equation is the one that we will be using to test the stability
under first order perturbations (linear stability analysis) of any
isotropic new interior solutions we will be using.  Since everything
in the above equation is in terms of the static variables, and all the
terms are separately known in closed form, we can continue simplifying
it into a workable Sturm-Liouville type problem.  

The first person to do this was Chandrasekhar, and we will follow his
method to express the above expression in two different parts.  The
first part will consist of all the parts that Chandrasekhar had to
deal with in his derivation.  Even this part will not be exactly what
Chandrasekhar had, since our expressions for the terms in this
equation, (e.g.\ \(\delta p_{r}\)), have additional contributions from
anisotropy and electric charge not present in the original.  However,
intuitively we can see that the end result should be reducible to
Chandrasekhar's in the limit of zero anisotropic pressure and charge.
With this general guideline in mind, we proceed and systematically
transfer all additional terms not present in Chandrasekhar into an
``extra'' part, like so
\[\sigma^{2}\e^{\lambda_{0}-\nu_{0}} \left( p_{r0}+\rho_{0}\right)\zeta = 
\underbrace{\dotsc}_{\text{Chandrasekhar}} +
\overbrace{\dotsc}^{\text{extra}}.\] We will show this division in the
following equations by boxing the terms present in Chandrasekhar's
derivation, to keep track of how we are advancing in our
simplification.

Equation~\eqref{eq:puls} will first be separated as
\begin{multline}\label{eq:pulstemp}
\sigma^{2}\e^{\lambda_{0}-\nu_{0}} \left( p_{r0}+\rho_{0}\right)\zeta =
 \boxed{\pderiv{\delta p_{r}}{r} + \delta p_{r} \left( \deriv{\nu_{0}}{r} + \f{1}{2} \deriv{\lambda_{0}}{r}\right)
   + \f{\delta \rho}{2} \deriv{\nu_{0}}{r}}  
   + \f{2}{r} (\delta p_{r} - \delta \ppen) \\
 \boxed{- \f{1}{2} (p_{r0} + \rho_{0})\zeta \left( \deriv{\nu_{0}}{r} + 
     \f{1}{r}\right) \left( \deriv{\lambda_{0}}{r} +\deriv{\nu_{0}}{r} \right)} - \f{\delta \eta}{2} \left( \deriv{\nu_{0}}{r} + \deriv{\lambda_{0}}{r} \right)-\pderiv{\delta \eta}{r},
\end{multline}
where the boxed terms are present in Chandrasekhar's derivation.
Simplifying these only, i.e.\ substituting the closed forms for
\(\delta p_{r}\) from equation~\eqref{eq:deltaPr} we have 
\begin{align*}
  \pderiv{}{r} \delta p_{r} &= \pderiv{}{r} \left\{ -\zeta p'_{r0} - \gamma B p_{r0} + \pderiv{p_{r0}}{\rho_{0}}
      \left[ \zeta \left( \f{2\Pi}{r} + \f{4 \eta_{0}}{r} + \deriv{\eta_{0}}{r} \right) + \delta \eta\right] \right\}\\
    &=\boxed{\pderiv{}{r} \left\{ -\zeta p'_{r0} - \gamma B p_{r0}\right\}} + \pderiv{}{r} \left\{ \pderiv{p_{r0}}{\rho_{0}}\left[ \zeta \left( \f{2\Pi}{r} + \f{4 \eta_{0}}{r} + \deriv{\eta_{0}}{r} \right) + \delta \eta\right] \right\},
\end{align*}
and additionally,
\begin{align*}
 \mathclap{ \delta p_{r} \left( \deriv{\nu_{0}}{r} + \f{1}{2} \deriv{\lambda_{0}}{r}\right)} & \\
  &=\left\{ -\zeta p'_{r0} - \gamma B p_{r0} + \pderiv{p_{r0}}{\rho_{0}}
      \left[ \zeta \left( \f{2\Pi}{r} + \f{4 \eta_{0}}{r} + \deriv{\eta_{0}}{r} \right) + \delta \eta\right] \right\}
    \left( \deriv{\nu_{0}}{r} + \f{1}{2} \deriv{\lambda_{0}}{r}\right) \\
   &= \boxed{(-\zeta p'_{r0} - \gamma B p_{r0}) \left\{ \deriv{\nu_{0}}{r} + \f{1}{2} \deriv{\lambda_{0}}{r} \right\}}+\\
     &+\pderiv{p_{r0}}{\rho_{0}}\left[ \zeta \left( \f{2\Pi}{r} + \f{4 \eta_{0}}{r} + \deriv{\eta_{0}}{r} \right) + \delta \eta\right]  \left\{ \deriv{\nu_{0}}{r} + \f{1}{2} \deriv{\lambda_{0}}{r} \right\}.
  \end{align*}
  similarly with \(\delta \rho\) from equation~\eqref{eq:deltaRhof}, we get
\begin{align*}
  \f{\delta \rho}{2} \deriv{\nu_{0}}{r} &= \f{1}{2} \deriv{\nu_{0}}{r} \left[ -\zeta \deriv{\rho_{0}}{r} 
    - \f{\e^{\nu_{0}/2}(p_{r0}+\rho_{0})}{r^{2}} \pderiv{}{r}(\e^{-\nu_{0}/2} r^{2} \zeta)
  - \zeta \left( \f{2\Pi}{r} + \f{4 \eta_{0}}{r} +\deriv{\eta_{0}}{r}\right) -\delta\eta \right] \\
&= \boxed{ \f{1}{2} \deriv{\nu_{0}}{r} \left[ -\zeta \deriv{\rho_{0}}{r} 
    - \f{\e^{\nu_{0}/2}(p_{r0}+\rho_{0})}{r^{2}} \pderiv{}{r}(\e^{-\nu_{0}/2} r^{2} \zeta) \right]}\\
  &\quad - \f{1}{2} \deriv{\nu_{0}}{r} \left[ \zeta \left( \f{2\Pi}{r} + \f{4 \eta_{0}}{r} +\deriv{\eta_{0}}{r} \right) 
+\delta\eta \right].
\end{align*}
This last equation is the final piece needed in the expression we
started with, i.e.\ equation~\eqref{eq:pulstemp}, we have the
intermediate form of the pulsation equation, separated into the boxed
part which Chandrasekhar derived, and the unboxed part resulting from
anisotropic pressure and electric change densities,
\begin{equation*}
  \begin{aligned}
  \sigma^{2}&\e^{\lambda_{0}-\nu_{0}} \left( p_{r0}+\rho_{0}\right)\zeta = 
  \boxed{ \pderiv{}{r} \left( -\zeta p'_{r0} - \gamma B p_{r0} \right) - (\zeta p'_{r0} + \gamma B p_{r0}) \left[ \deriv{\nu_{0}}{r} + \f{1}{2} \deriv{\lambda_{0}}{r} \right]}  \\
&+ \pderiv{}{r} \left\{ \pderiv{p_{r0}}{\rho_{0}}\left[ \zeta \left( \f{2\Pi}{r} + \f{4 \eta_{0}}{r} + \deriv{\eta_{0}}{r} \right) + \delta \eta\right] \right\} - \f{1}{2} \deriv{\nu_{0}}{r} \left[ \zeta \left( \f{2\Pi}{r} + \f{4 \eta_{0}}{r} +\deriv{\eta_{0}}{r} \right) +\delta\eta \right] \\
&\boxed{+\f{1}{2} \deriv{\nu_{0}}{r} \left[ -\zeta \deriv{\rho_{0}}{r} 
    - \f{\e^{\nu_{0}/2}(p_{r0}+\rho_{0})}{r^{2}} \pderiv{}{r}(\e^{-\nu_{0}/2} r^{2} \zeta) \right]} 
-\pderiv{}{r} \delta \eta  -\f{\delta \eta}{2} \left( \deriv{\nu_{0}}{r} + \deriv{\lambda_{0}}{r}\right)  \\
   &\boxed{ - \f{1}{2} (p_{r0} + \rho_{0})\zeta \left( \deriv{\nu_{0}}{r} +  
     \f{1}{r}\right) \left( \deriv{\lambda_{0}}{r} +\deriv{\nu_{0}}{r} \right)} + \\
&+\pderiv{p_{r0}}{\rho_{0}}\left[ \zeta \left( \f{2\Pi}{r} + \f{4 \eta_{0}}{r} + \deriv{\eta_{0}}{r} \right) + \delta \eta\right]  \left\{ \deriv{\nu_{0}}{r} + \f{1}{2} \deriv{\lambda_{0}}{r} \right\} +\f{2}{r} (\delta p_{r} - \delta\ppen) .
\end{aligned}
\end{equation*}
We notice immediately that there are sets of constants that appear
often.  In the interest of economy of equation length, we introduce
two new auxiliary variables,
\( (\delta p_{r} - \delta \ppen) \equiv \delta \Pi, \)
and
\(\zeta \left( \f{2\Pi}{r} + \f{4\eta_{0}}{r} + \deriv{\eta_{0}}{r}
\right) + \delta \eta \equiv A,\)
while at the same time replacing some of the explicit
\(r-\)derivatives, \(\deriv{}{r}\)
with primes \((').\) This reduces the pulsation equation to
\begin{multline}
  \label{eq:pulsAux}
  \sigma^{2}\e^{\lambda_{0}-\nu_{0}} \left( p_{r0}+\rho_{0}\right)\zeta = \boxed{
  \pderiv{}{r}(-\zeta p'_{r0} - \gamma B p_{r0})  - \zeta p'_{r0}\left( \f{\lambda'_{0}}{2} + \nu_{0}\right) -
  \gamma B p_{r0} \left( \nu'_{0} + \f{\lambda'_{0}}{2}\right)} \\
   + \pderiv{}{r} \left( A \pderiv{p_{r0}}{\rho_{0}}\right) 
  + A\pderiv{p_{r0}}{\rho_{0}} \left( \f{\lambda'_{0}}{2} + \nu'_{0}\right)
  - \f{A}{2} \nu'_{0}  + \boxed{ - \f{1}{2} (p_{r0} + \rho_{0})\zeta \left( \nu'_{0} +  
     \f{1}{r}\right) \left( \lambda'_{0} +\nu'_{0} \right)}\\
   \boxed{+\f{\nu'_{0}}{2}  \left[ -\zeta \deriv{\rho_{0}}{r} 
    - \f{\e^{\nu_{0}/2}(p_{r0}+\rho_{0})}{r^{2}} \pderiv{}{r}(\e^{-\nu_{0}/2} r^{2} \zeta) \right]} 
-\f{\delta \eta}{2}(\nu'_{0}+\lambda'_{0}) - \pderiv{\delta \eta }{r} + \f{2}{r} \delta \Pi.
\end{multline}
Here Chandrasekhar factorizes the boxed terms (the only ones he had)
into a form that can be cast into a Sturm-Liouville problem.  We will
proceed similarly, and note in passing that while our boxed
expressions and Chandrasekhar's match, since our metric coefficients
mean different things, (since our energy-momentum tensor is
anisotropic and admits non-zero charge) we will have to be creative in
factorizing these expressions.  Synthesizing all the boxed elements
and reorganizing, we get 
\begin{align*}
\underbrace{\dotsc}_{\text{Chandra}} =& -\pderiv{(\zeta p'_{r0})}{r} 
- \gamma B p_{r0} \left(\nu'_{0} + \f{\lambda'_{0}}{2}\right) -\f{1}{2} (p_{r0} + \rho_{0})\zeta \left( \nu'_{0} +  
     \f{1}{r}\right) \left( \lambda'_{0} +\nu'_{0} \right) + \\
 &-\pderiv{(\gamma B p_{r0})}{r} -\f{\nu'_{0}}{2}  \left[ \zeta \deriv{\rho_{0}}{r} 
    + \f{\e^{\nu_{0}/2}(p_{r0}+\rho_{0})}{r^{2}} \pderiv{}{r}(\e^{-\nu_{0}/2} r^{2} \zeta) \right]
  -\zeta p'_{r0} \left( \f{\lambda'_{0}}{2} + \nu'_{0}\right).
\end{align*}
The first step in the factorization process is to notice that the
second and third terms in the above equation can be written as a total
derivative of a suitably chosen exponential, here
\(\e^{-(\lambda_{0} + 2 \nu_{0})/2} \deriv{}{r} \left[
  \e^{(\lambda_{0} + 2 \nu_{0})/2} \gamma B p_{r0}\right]\), as can be
readily checked by expansion of the latter.  With this factorization
we get
\begin{align*}
\underbrace{\dotsc}_{\text{Chandra}} =& -\pderiv{(\zeta p'_{r0})}{r}  - \e^{-(\lambda_{0} + 2 \nu_{0})/2} \deriv{}{r} \left[
  \e^{(\lambda_{0} + 2 \nu_{0})/2} \gamma B p_{r0}\right] -\f{1}{2} (p_{r0} + \rho_{0})\zeta \left( \nu'_{0} +  
     \f{1}{r}\right) \left( \lambda'_{0} +\nu'_{0} \right) \\
 &-\f{\nu'_{0}}{2}  \left[ \zeta \deriv{\rho_{0}}{r} 
     +\f{\e^{\nu_{0}/2}(p_{r0}+\rho_{0})}{r^{2}} \pderiv{}{r}(\e^{-\nu_{0}/2} r^{2} \zeta) \right]
  -\zeta p'_{r0} \left( \f{\lambda'_{0}}{2} + \nu'_{0}\right).
\end{align*}
The penultimate term is now expanded completely, and we add zero to the
equation in a creative way, as shown:
\begin{align*}
\underbrace{\dotsc}_{\text{Chandra}} =& -\pderiv{(\zeta p'_{r0})}{r}  - \e^{-(\lambda_{0} + 2 \nu_{0})/2} \deriv{}{r} \left[
  \e^{(\lambda_{0} + 2 \nu_{0})/2} \gamma B p_{r0}\right] -\f{1}{2} (p_{r0} + \rho_{0})\zeta \left( \nu'_{0} +  
     \f{1}{r}\right) \left( \lambda'_{0} +\nu'_{0} \right) \\
 &-\f{\nu'_{0}}{2}  \left\{ \zeta \deriv{\rho_{0}}{r} 
    + (p_{r0}+\rho_{0}) \left[ \f{2\zeta}{r} + \deriv{\zeta}{r} -\f{\zeta}{2} \deriv{\nu_{0}}{r}\right] \right\}
  - \zeta p'_{r0} \left( \f{\lambda'_{0}}{2} + \nu'_{0}\right)\\
  &+ \f{\nu'_{0}}{2} \left\{ \cancel{\left[ \zeta p'_{r0} + \zeta \eta'_{0} + \f{2\zeta}{r} \left( \ppeno - p_{r0} + 2 \eta_{0} \right) \right]} + \right. \\
  &- \left. \phantom{\f{\nu'_{0}}{2}\left\{ \right. } \cancel{\left[ \zeta p'_{r0} + \zeta \eta'_{0} + \f{2\zeta}{r} \left( \ppeno - p_{r0} + 2 \eta_{0} \right) \right]}\phantom{+}\right\}.
\end{align*}
We immediately notice that the terms we are adding explicitly contain
quantities that would not have been present in an uncharged and
isotropic set of equations.  However this new form will allow us to
eliminate additional terms and factorize the expression a little bit
more into
\begin{align*}
\underbrace{\dotsc}_{\text{Chandra}} =& -\pderiv{(\zeta p'_{r0})}{r}  - \e^{-(\lambda_{0} + 2 \nu_{0})/2} \deriv{}{r} \left[
  \e^{(\lambda_{0} + 2 \nu_{0})/2} \gamma B p_{r0}\right] -\f{1}{2} (p_{r0} + \rho_{0})\zeta \left( \nu'_{0} +  
     \f{1}{r} \right) \left( \lambda'_{0} +\nu'_{0} \right) \\
 &-\f{\nu'_{0}}{2}  \left\{ \deriv{}{r}[\zeta (\rho_{0} + p_{r0})] + \f{2\zeta}{r}(p_{r0}+\rho_{0}) -\zeta \left[ \cancel{ \
   \deriv{p_{r0}}{r} - \f{2}{r}(\ppeno-p_{r0}+2\eta_{0})-\deriv{\eta_{0}}{r}} + \right. \right. \\
  &\quad \left. \cancel{ \left. +\f{p_{r0}+\rho_{0}}{2} \deriv{\nu_{0}}{r}\right] }  + \zeta \eta'_{0} 
  + \f{2\zeta}{r}(\ppeno-p_{r0} + 2\eta_{0}) \right\} -\zeta p'_{r0} \left( \f{\lambda'_{0}}{2} +\nu'_{0} \right),
\end{align*}
which has the static Einstein expression~\eqref{eq:ECc} spanning the
second and third lines (slashed) and which is equal to zero.  This can
thus be removed, to give
\begin{align*}
\underbrace{\dotsc}_{\text{Chandra}} =& -\pderiv{(\zeta p'_{r0})}{r}  - \e^{-(\lambda_{0} + 2 \nu_{0})/2} \deriv{}{r} \left[
  \e^{(\lambda_{0} + 2 \nu_{0})/2} \gamma B p_{r0}\right] -\f{1}{2} (p_{r0} + \rho_{0})\zeta \left( \nu'_{0} +  
     \f{1}{r}\right) \left( \lambda'_{0} +\nu'_{0} \right) \\
 &-\f{\nu'_{0}}{2}  \left\{[\zeta(\rho_{0} + p_{r0})]' + \f{2\zeta}{r}(p_{r0}+\rho_{0})   + \zeta \eta'_{0} 
  + \f{2\zeta}{r}(\ppeno-p_{r0}+2\eta_{0}) \right\} +\\ 
&-\zeta p'_{r0} \left( \f{\lambda'_{0}}{2} +\nu'_{0}\right).
\end{align*}
We now reorganize the above equation to put the extra terms due to
anisotropy and charge separately, employing the boxed notation to
reference which part was Chandrasekhar's, and which parts got added in
our equations:
\begin{align*}
\underbrace{\dotsc}_{\text{Chandra}} =& \boxed{-\pderiv{(\zeta p'_{r0})}{r}  - \e^{-(\lambda_{0} + 2 \nu_{0})/2} \deriv{}{r} \left[
  \e^{(\lambda_{0} + 2 \nu_{0})/2} \gamma B p_{r0}\right] -\f{1}{2} (p_{r0} + \rho_{0})\zeta \left( \nu'_{0} +  
     \f{1}{r}\right) \left( \lambda'_{0} +\nu'_{0} \right)} \\
 &-\f{\nu'_{0}}{2}  \left\{ \zeta \eta'_{0} 
  + \f{2\zeta}{r}(\ppeno-p_{r0}+2\eta_{0}) \right\} -\boxed{\zeta p'_{r0} \left( \f{\lambda'_{0}}{2} +\nu'_{0}\right)}\\
 &\boxed{-\f{\nu'_{0}}{2}  \left\{ [\zeta(\rho_{0} + p_{r0})]' + \f{2\zeta}{r}(p_{r0}+\rho_{0}) \right\}}.
\end{align*}
Following these simplifications, we are now in a position to recombine
the two parts of the pulsation equation (boxed and unboxed) to give
rise to the full pulsation equation:
\begin{multline*}
   \sigma^{2}\e^{\lambda_{0}-\nu_{0}} \left( p_{r0}+\rho_{0}\right)\zeta = -\pderiv{(\zeta p'_{r0})}{r}  
   - \e^{-(\lambda_{0} + 2 \nu_{0})/2} \left[ \e^{(\lambda_{0} + 2 \nu_{0})/2} \gamma B p_{r0} \right]'
   -\f{\delta \eta}{2} (\nu'_{0}+ \lambda'_{0}) + \f{2}{r} \delta \Pi\\
   -\f{1}{2} (p_{r0} + \rho_{0}) \zeta \left( \nu'_{0} +  \f{1}{r} \right) \left( \lambda'_{0} +\nu'_{0} \right)
   -\f{\nu'_{0}}{2}  \left\{ \zeta \eta'_{0} + \f{2 \zeta}{r}( \ppeno - p_{r0} +2\eta_{0}) \right\} -\zeta p'_{r0} \left( \f{\lambda'_{0}}{2} +\nu'_{0}\right)\\
   -\f{\nu'_{0}}{2}  \left\{ [ \zeta (\rho_{0} + p_{r0} ) ]' + \f{2\zeta}{r}(p_{r0}+\rho_{0}) \right\}
   + \underbrace{ \pderiv{}{r} \left\{ A \pderiv{p_{r0}}{\rho_{0}}\right\} 
   + A\pderiv{p_{r0}}{\rho_{0}} \left\{ \f{\lambda'_{0}}{2} + \nu'_{0}\right\} }
   - \f{A}{2} \nu'_{0} - (\delta \eta)'
\end{multline*}
The same argument used previously involving factorization through an
exponential can be applied to two terms involving the \(A\)s that are
under-braced, to give
\begin{multline}
\label{eq:pulsAux2}
  \sigma^{2}\e^{\lambda_{0}-\nu_{0}} \left( p_{r0}+\rho_{0}\right)\zeta = 
  -\pderiv{(\zeta p'_{r0})}{r}  -\zeta p'_{r0}\left( \f{\lambda'_{0}}{2} + \nu'_{0}\right) 
-\f{1}{2} (p_{r0} + \rho_{0})\zeta \left( \nu'_{0} +  \f{1}{r}\right) \left( \lambda'_{0} +\nu'_{0} \right)\\
  -\f{\nu'_{0}}{2}  \left\{ [ \zeta (\rho_{0} + p_{r0})]'  +\zeta \eta'_{0}+ \f{2\zeta}{r}(\rho_{0} + \ppen0 + 2 \eta_{0}) \right\}  
  -\f{\delta \eta}{2} (\nu'_{0}+\lambda'_{0}) - \f{A}{2} \nu'_{0} - (\delta \eta)' + \f{2}{r} \delta \Pi \\
  +\e^{-(\lambda_{0} + 2 \nu_{0})/2} \left[ \e^{(\lambda_{0} + 2 \nu_{0})/2}  A\pderiv{p_{r0}}{\rho_{0}} \right]'
  -\e^{-(\lambda_{0} + 2 \nu_{0})/2} \left[ \e^{(\lambda_{0} + 2 \nu_{0})/2} \gamma B p_{r0}\right]'.
\end{multline}

The next stage of the simplification is the substitution of the
pressure derivative with the expressions that can be obtained from rearranging~\eqref{eq:ECc} to
\[\f{\nu'_{0}}{2} = \f{1}{p_{r0}+\rho_{0}} \left( \f{A-\delta \eta}{\zeta} - p'_{r0}\right) \iff p'_{r0} = \f{2\Pi_{0}}{r} +
\f{4 \eta_{0}}{r} + \eta'_{0} - \f{\nu'_{0}}{2}(p_{r0}+ \rho_{0}).\]
This allows the first four terms of~\eqref{eq:pulsAux2} to be combined
since many common terms appear after this substitution.  We now show
this step term by term before combining everything: the first
derivative becomes
\begin{align*}
-\pderiv{(\zeta p'_{r0})}{r} &= -\pderiv{}{r} \left[ \zeta \left( \f{2\Pi_{0}}{r} +
\f{4 \eta_{0}}{r} + \eta'_{0} - \f{\nu'_{0}}{2}(p_{r0}+ \rho_{0})\right) \right]\\
  &=-\left[\zeta\left( \f{2\Pi_{0}}{r} +
\f{4 \eta_{0}}{r} + \eta'_{0}\right) \right]' + \f{\zeta}{2} \nu''_{0} (p_{r0}+\rho_{0})+ \cancel{\f{\nu'_{0}}{2} [\zeta(p_{r0}+\rho_{0})]'},
\end{align*}
which we combine with the second term, 
\[-\zeta \left( \f{\lambda'_{0}}{2} + \nu'_{0} \right) p_{r0}=
\f{\zeta}{2} (p_{r0}+\rho_{0}) \left( (\nu'_{0})^{2} + \f{\nu'_{0} \lambda'_{0}}{2} \right) - \zeta \left(\f{\lambda'_{0}}{2} + \nu'_{0} \right) \left( \f{2\Pi_{0}}{r} +
\f{4 \eta_{0}}{r} + \eta'_{0} \right),
 \]
and the third,
\[
  -\f{1}{2} (p_{r0} + \rho_{0})\zeta \left( \nu'_{0} + \f{1}{r}\right)
  \left( \lambda'_{0} +\nu'_{0} \right) = -\f{\zeta}{2}
  (p_{r0}+\rho_{0}) \left[ (\nu'_{0})^{2} + \nu'_{0}\lambda'_{0} + \f{\lambda'_{0}}{r} + \f{\nu'_{0}}{r} \right],
\]
with the forth, 
\begin{multline*}
  -\f{\nu'_{0}}{2}  \left\{ [ \zeta (\rho_{0} + p_{r0})]'  +\zeta \eta'_{0}+ \f{2\zeta}{r}(\rho_{0} + \ppeno + 2 \eta_{0}) \right\} = - \f{\nu'_{0}}{2} [\zeta(p_{r0}+\rho_{0})]' - \f{\zeta \eta'_{0} \nu'_{0}}{2} \\- \f{\nu'_{0}\zeta}{r} (\rho_{0}+p_{r0} - \Pi_{0} + 2\eta_{0}) = \cancel{- \f{\nu'_{0}}{2} [\zeta(p_{r0}+\rho_{0})]'} - \f{\zeta \eta'_{0} \nu'_{0}}{2} -\f{\zeta}{r} \nu'_{0} (p_{r0}+\rho_{0}) + \f{\zeta}{r}\nu'_{0}(\Pi_{0} - 2\eta_{0}), 
\end{multline*}
the resulting equation then becomes
\begin{multline}
\label{eq:pulsAux3}
  \sigma^{2}\e^{\lambda_{0}-\nu_{0}} \left( p_{r0}+\rho_{0}\right)\zeta = 
  \f{\zeta}{2} ( p_{r0}+\rho_{0} )\left( \nu''_{0}  - \f{\nu'_{0} \lambda'_{0}}{2} -\f{\lambda'_{0}}{r}  -\f{3\nu'_{0}}{r} \right) -\f{\zeta\nu'_{0}}{2} 
\left( \f{4\eta_{0}}{r} - \f{2\Pi_{0}}{r} + \eta'_{0} \right)   \\ 
-  \left( \f{\lambda'_{0}}{2} + \nu'_{0} \right) \left[ \zeta \left( \f{2\Pi_{0}}{r} + \f{4\eta_{0}}{r} + \eta'_{0} \right)\right]
-\left[ \zeta \left( \f{2\Pi_{0}}{r} + \f{4\eta_{0}}{r} + \eta'_{0} \right) \right]' -\f{\delta \eta}{2} (\nu'_{0}+\lambda'_{0}) - \f{A}{2} \nu'_{0} \\
  +\e^{-(\lambda_{0} + 2 \nu_{0})/2} \left[ \e^{(\lambda_{0} + 2 \nu_{0})/2}  A\pderiv{p_{r0}}{\rho_{0}} \right]'
  -\e^{-(\lambda_{0} + 2 \nu_{0})/2} \left[ \e^{(\lambda_{0} + 2 \nu_{0})/2} \gamma B p_{r0}\right]' - (\delta \eta)' + \f{2}{r} \delta \Pi.
\end{multline}
By using a rearranged equation~\eqref{eq:EEc} in the static case, we
obtain terms very similar to the first brackets in the RHS
of~\eqref{eq:pulsAux3}, 
\[\f{16\pi G}{c^{4}} (\ppeno + \eta_{0}) \e^{\lambda_{0}} = \left( \nu''_{0} - \f{\nu'_{0}\lambda'_{0}}{2} + \f{(\nu'_{0})^{2}}{2} + \f{\nu'_{0}}{r}  - \f{\lambda'_{0}}{r} \right),\]
and furthermore, the terms on the second line of
equation~\eqref{eq:pulsAux3} can be combined to produce another \(A,\)
which we defined previously.  Together these simplifications can be
substituted into equation~\eqref{eq:pulsAux3} to get a pulsation
equation into the form
\begin{multline}
\label{eq:pulsAux4}
  \sigma^{2}\e^{\lambda_{0}-\nu_{0}} \left( p_{r0}+\rho_{0}\right)\zeta = 
  \f{\zeta}{2} ( p_{r0}+\rho_{0} )\left(  \f{16\pi G}{c^{4}} (\ppeno+\eta_{0}) \e^{\lambda_{0}} -\f{(\nu' _{0})^{2}}{2} -\f{4\nu'_{0}}{r} \right) + 
  \\-\f{\zeta\nu'_{0}}{2}\left( \f{4\eta_{0}}{r} - \f{2\Pi_{0}}{r} + \eta'_{0} \right) +\f{\delta \eta}{2} \nu'_{0} - \f{A}{2} \nu'_{0} + \f{2}{r} \delta \Pi + \\
-  \left( \f{\lambda'_{0}}{2} + \nu'_{0} \right) \underbrace{\left[ \zeta \left( \f{2\Pi_{0}}{r} + \f{4\eta_{0}}{r} + \eta'_{0} \right) + \delta \eta \right]}_{A} 
-\underbrace{\left[ \zeta \left( \f{2\Pi_{0}}{r} + \f{4\eta_{0}}{r} + \eta'_{0} \right) +\delta \eta \right]'}_{A'} + \\
  +\e^{-(\lambda_{0} + 2 \nu_{0})/2} \left[ \e^{(\lambda_{0} + 2 \nu_{0})/2}  A\pderiv{p_{r0}}{\rho_{0}} \right]'
  -\e^{-(\lambda_{0} + 2 \nu_{0})/2} \left[ \e^{(\lambda_{0} + 2 \nu_{0})/2} \gamma B p_{r0}\right]',
\end{multline}
which can be factorised with the exponentials and reorganized into
\begin{multline*}
  \sigma^{2}\e^{\lambda_{0}-\nu_{0}} \left( p_{r0}+\rho_{0}\right)\zeta = 
\f{8 \pi G}{c^{4}} (\ppeno + \eta_{0}) (p_{r0} + \rho_{0})\zeta\e^{\lambda_{0}} - \f{\zeta}{4} (p_{r0} + \rho_{0})\nu'_{0}\left(\nu'_{0}+\f{8}{r} \right) \\
-\f{\nu'_{0}}{2}\left[ \zeta \left( \f{4\eta_{0}}{r} - \f{2\Pi_{0}}{r} + \eta'_{0} \right) + A - \delta \eta \right] + \f{2}{r} \delta \Pi\\
+\e^{-(\lambda_{0} + 2 \nu_{0})/2} \left\{ \e^{(\lambda_{0} + 2 \nu_{0})/2}  A \left[\pderiv{p_{r0}}{\rho_{0}} - 1 \right] \right\}'
  -\e^{-(\lambda_{0} + 2 \nu_{0})/2} \left[ \e^{(\lambda_{0} + 2 \nu_{0})/2} \gamma B p_{r0}\right]'.
\end{multline*}
The term on the second line of the above simplifies too, to convert our pulsation equation into 
\begin{multline}
\label{eq.PulsAux5}
  \sigma^{2}\e^{\lambda_{0}-\nu_{0}} \left( p_{r0}+\rho_{0}\right)\zeta = 
\f{8 \pi G}{c^{4}} (\ppeno+\eta_{0}) (p_{r0} + \rho_{0})\zeta\e^{\lambda_{0}} - \f{\zeta}{4} (p_{r0} + \rho_{0})\nu'_{0}\left(\nu'_{0}+\f{8}{r} \right) + \\
+\e^{-(\lambda_{0} + 2 \nu_{0})/2} \left\{ \e^{(\lambda_{0} + 2 \nu_{0})/2}  A \left[\pderiv{p_{r0}}{\rho_{0}} - 1 \right] \right\}'
  -\e^{-(\lambda_{0} + 2 \nu_{0})/2} \left[ \e^{(\lambda_{0} + 2 \nu_{0})/2} \gamma B p_{r0}\right]' +\\- \nu'_{0}\zeta\left(\f{4\eta_{0}}{r} + \eta'_{0} \right) +  \f{2}{r} \delta \Pi 
\end{multline}
We continue the simplification by substituting equation~\eqref{eq:ECc}
again in the last term of the first line of
equation~\eqref{eq.PulsAux5}, to get in a partial step that
\begin{multline}
  - \f{\zeta}{4} (p_{r0} + \rho_{0})\nu'_{0}\left(\nu'_{0}+\f{8}{r}
  \right) = \f{\zeta p'_{r0} \nu'_{0}}{2} -\f{2\zeta}{r}(p_{r0} + \rho_{0}) \nu'_{0}\\
-\f{\nu'_{0} \zeta}{2} \left( \f{2\Pi_{0}}{r} + \f{4\eta_{0}}{r} +\eta'_{0} \right).
\end{multline}
The above equation is then combined with the last term on the first line
of~\eqref{eq.PulsAux5} and the combination of the last two terms of
the latter results in
\[
\f{p'_{r0}}{p_{r0}+\rho_{0}} (A - \delta\eta) - \boxed{\f{\zeta (p'_{r0})^{2}}{p_{r0}+\rho_{0}}} - \zeta \left(\f{\nu'_{0}}{2}\right) \left( \f{2\Pi_{0}}{r} + \f{12\eta_{0}}{r} + 3\eta'_{0} \right) -\f{4}{r}(A - \delta\eta) + \boxed{\f{4p'_{r0}\zeta}{r}},
\]
after some tedious algebra.  The boxed terms are again the only ones
in Chandrasekhar's analysis.  Again substituting~\eqref{eq:ECc} in
above, and simplifying only the unboxed new terms at this point, we
get
\begin{multline*}
\f{4\zeta p'_{r0}}{p_{r0}+\rho_{0}} \left( \f{\Pi_{0}}{r} + \f{4\eta_{0}}{r} + \eta'_{0} \right) - \f{4\zeta}{r} \left( \f{2\Pi_{0}}{r} + \f{4\eta_{0}}{r} + \eta'_{0} \right) \\
- \f{\zeta}{p_{r0}+\rho_{0}} \left( \f{2\Pi_{0}}{r} + \f{4\eta_{0}}{r} + \eta'_{0} \right) \left( \f{2\Pi_{0}}{r} + \f{12\eta_{0}}{r} + 3\eta'_{0} \right),  
\end{multline*}
allowing us to write down the final form of the pulsation equation
with metric coefficients appearing only in the exponentials, and with
the terms resulting from anisotropy and charge appearing prominently
only from the static contributions,
\begin{multline}
  \label{eq:Pulsation}
  \sigma^{2}\e^{\lambda_{0}-\nu_{0}} \left( p_{r0}+\rho_{0}\right)\zeta
  =
  \f{8 \pi G}{c^{4}} (p_{r0} - \Pi_{0} + \eta_{0}) (p_{r0} + \rho_{0})\zeta \e^{\lambda_{0}} +  \f{2}{r} \delta \Pi -\f{\zeta (p'_{r0})^{2}}{p_{r0}+\rho_{0}} \\
  +\e^{-(\lambda_{0} + 2 \nu_{0})/2} \left\{ \e^{(\lambda_{0} + 2\nu_{0})/2} \left[ \zeta \left( \f{2\Pi_{0}}{r}
  + \f{4\eta_{0}}{r} + \eta'_{0} \right) + \delta \eta \right] \left[\pderiv{p_{r0}}{\rho_{0}} - 1 \right] \right\}' \\
  -\e^{-(\lambda_{0} + 2 \nu_{0})/2} \left[ \e^{(\lambda_{0} + 3 \nu_{0})/2} \gamma \f{p_{r0}}{r^{2}} \left( r^{2} \zeta \e^{-\nu_{0}/2}\right)' \right]' 
+ \f{4p'_{r0}\zeta}{r} 
+\f{4\zeta p'_{r0}}{p_{r0}+\rho_{0}} \left( \f{\Pi_{0}}{r} + \f{4\eta_{0}}{r} + \eta'_{0} \right)\\
  - \f{4\zeta}{r} \left( \f{2\Pi_{0}}{r} + \f{4\eta_{0}}{r} + \eta'_{0} \right) 
  - \f{\zeta}{p_{r0}+\rho_{0}} \left( \f{2\Pi_{0}}{r} +\f{4\eta_{0}}{r} + \eta'_{0} \right) \left( \f{2\Pi_{0}}{r} +\f{12\eta_{0}}{r} + 3\eta'_{0} \right) .
\end{multline}
Since only static zero subscripted values appear in the above, we may
without confusion remove all the zero-subscripts from the values in
all subsequent expressions.  We now check this completely general
equation against special cases to check for consistency with the
literature.

First, if we were to set \(\Pi = 0\)
for isotropy and \(\eta = 0\)
for no charge in the above, we immediately get
\begin{multline*}
   \sigma^{2}\e^{\lambda-\nu} \left( p_{r}+\rho\right)\zeta = 
\f{8 \pi G}{c^{4}} p_{r} (p_{r} + \rho)\zeta \e^{\lambda}  -\f{\zeta (p'_{r})^{2}}{p_{r}+\rho} \\
  -\e^{-(\lambda + 2 \nu)/2} \left[ \e^{(\lambda + 3 \nu)/2} \gamma \f{p_{r}}{r^{2}} \left( r^{2} \zeta \e^{-\nu/2}\right)' \right]' + \f{4p'_{r}\zeta}{r}
\end{multline*}
exactly as expected from~\citeauthor{Cha64}'s result.
Similarly, setting just \(\Pi=0\)
for isotropy, but \(\eta \neq 0\) for electric charge results in
\begin{multline*}
 \sigma^{2}\e^{\lambda-\nu} \left( p_{r}+\rho\right)\zeta = 
\f{8 \pi G}{c^{4}} (p_{r} + \eta) (p_{r} + \rho)\zeta \e^{\lambda}  -\e^{-(\lambda + 2 \nu)/2} \left[ \e^{(\lambda + 3 \nu)/2} \gamma \f{p_{r}}{r^{2}} \left( r^{2} \zeta \e^{-\nu/2}\right)' \right]'  \\
+\e^{-(\lambda + 2 \nu)/2} \left\{ \e^{(\lambda + 2 \nu)/2}  \left[  \zeta \left( \f{4\eta}{r} + \eta' \right) + \delta \eta \right] \left[\pderiv{p_{r}}{\rho} - 1 \right] \right\}' - \f{4\zeta}{r} \left( \f{4\eta}{r} + \eta' \right) + \f{4p'_{r}\zeta}{r}\\
-\f{\zeta (p'_{r})^{2}}{p_{r}+\rho} +\f{4\zeta p'_{r}}{p_{r}+\rho} \left(  \f{4\eta}{r} + \eta' \right)  - \f{\zeta}{p_{r}+\rho} \left( \f{4\eta}{r} + \eta' \right) \left( \f{12\eta}{r} + 3\eta' \right), 
\end{multline*}
which is equivalent to~\citeauthor{Gla76}'s pulsation equation, with a slight notation
change (\(\epsilon \to \rho,\)) and different factorization of terms.  In contrast setting
\(\eta = 0\) for no electric charge but \(\Pi \neq 0\) for anisotropic pressures results in 
\begin{multline*}
 \sigma^{2}\e^{\lambda-\nu} \left( p_{r}+\rho\right)\zeta = 
\f{8 \pi G}{c^{4}} (p_{r} - \Pi) (p_{r} + \rho)\zeta \e^{\lambda} +  \f{2}{r} \delta \Pi -\f{\zeta (p'_{r})^{2}}{p_{r}+\rho} \\
+\e^{-(\lambda + 2 \nu)/2} \left\{ \e^{(\lambda + 2 \nu)/2}  \zeta \left(\f{2\Pi}{r} \right) \left[\pderiv{p_{r}}{\rho} - 1 \right] \right\}'
  -\e^{-(\lambda + 2 \nu)/2} \left[ \e^{(\lambda + 3 \nu)/2} \gamma \f{p_{r}}{r^{2}} \left( r^{2} \zeta \e^{-\nu/2}\right)' \right]' \\
+ \f{4p'_{r}\zeta}{r} +\f{4\zeta p'_{r}}{p_{r}+\rho} \left( \f{\Pi}{r} \right) - \f{4\zeta}{r} \left( \f{2\Pi}{r}  \right) - \f{\zeta}{p_{r}+\rho} \left( \f{4\Pi^{2}}{r^{2}}  \right).
\end{multline*}
as given in~\citeauthor{DevGle03}, with the caveat that the latter use
a slightly different definition of the anisotropy so that \(\Pi \to -\Pi,\)
make a few sign mistakes along the way, and use the \(G=c=1\)
normalization.

With the final form of the pulsation equation~\eqref{eq:Pulsation},
and the boundary conditions ensuring that the radial pulsations are
such that
\begin{equation}
\label{Stab.eq:B1}
\zeta = 0 \quad \text{at} \quad r = 0 \qquad \text{or equivalently} \qquad \zeta \sim r \quad\text{as}\quad r \to 0, 
\end{equation}
so that the fluid incurs no radial motion when at the centre,
and
\begin{equation}
\label{Stab.eq:B2}
\delta p_{r} = 0 \quad \text{at} \quad r = r_{b}  
\end{equation}
in accordance with the definition of the boundary of the star, and the
Israel-Darmois condition, the radial stability of the star reduces to
an eigenvalue problem for the pulsation frequency \(\sigma\) with
amplitude \(\zeta.\) In~\citeauthor{Cha64}, this equation is
integrated, after multiplication by the integrating factor
\(r^{2} \zeta \e^{(\lambda + \nu)/2}, \) and then over the whole range
of \(r,\) giving
\begin{multline}
  \label{eq:PulsIntegrated}
\sigma^{2} \int_{0}^{r_{b}} r^{2} \e^{(3\lambda-\nu)/2}(\rho + p_{r}) \zeta^{2} \d r = \f{8\pi G}{c^{4}} \int_{0}^{r_{b}} r^{2} (p_{r} -\Pi +\eta)(p_{r}+\rho) \e^{(3\lambda + \nu)/2}\zeta^{2} \d r \\
+2\int_{0}^{r_{b}} r \e^{(\lambda + \nu)/2} (\delta \Pi) \zeta \d r - \int_{0}^{r_{b}} \f{r^{2} \zeta^{2}}{p_{r}+\rho} \e^{(\lambda + \nu)/2} (p'_{r})^{2} \d r + 4\int_{0}^{r_{b}} r p'_{r} \e^{(\lambda + \nu)/2} \zeta^{2} \d r\\
+\int_{0}^{r_{b}} r^{2}\e^{-\nu/2} \left\{ \e^{(\lambda + 2 \nu)/2}  \left[ \zeta \left( \f{2\Pi}{r} + \f{4\eta}{r} + \eta' \right) + \delta \eta \right] \left[\pderiv{p_{r}}{\rho} - 1 \right] \right\}' \zeta \d r \\- \cancel{\left[\e^{(\lambda + 2\nu)/2} \zeta \gamma p_{r} (r^{2} \zeta \e^{-\nu/2})' \right]_{0}^{r_{b}}} + \int_{0}^{r_{b}} \f{\gamma p_{r}}{r^{2}} [(r^{2} \zeta \e^{-\nu/2})']^{2} \e^{(\lambda+3\nu)/2} \d r \\ 
+ \int_{0}^{r_{b}} \f{4r^{2} \zeta^{2} p'_{r}}{p_{r}+\rho} \left( \f{\Pi}{r} + \f{4\eta}{r} + \eta' \right) \e^{(\lambda+\nu)/2} \d r 
- 4 \int_{0}^{r_{b}} r \e^{(\lambda+\nu)/2}\left( \f{2\Pi}{r} + \f{4\eta}{r} + \eta' \right) \zeta^{2} \d r\\
- \int_{0}^{r_{b}} \f{r^{2}\zeta^{2}}{p_{r}+\rho} \e^{(\lambda+\nu)/2} \left( \f{2\Pi}{r} +\f{4\eta}{r} + \eta' \right) \left( \f{2\Pi}{r} +\f{12\eta}{r} + 3\eta' \right) \d r.
\end{multline}
Integration by parts generates the struck out term, while the
boundary conditions cause the former to vanish. The other integrals
that have not been simplified have to be integrated for each specific
solution, once all the different metric functions \(\lambda\)
and \(\nu,\)
pressure \(p_{r},\)
density \(\rho,\)
anisotropy \(\Pi,\)
and charge \(\eta\)
have been specified. The amplitude of the radial oscillation \(\zeta,\)
also needs to be specified, and in the literature, different trial
functions such that the boundary conditions are satisfied are picked,
while making sure that the \(\zeta\)s
simplify the integrals at the same time.

However, \citeauthor{BarThoMel66} rewrite the pulsation equation in a
canonical Sturm-Liouville form first, and for completeness, we obtain
this form too.  The differential equation~\eqref{eq:Pulsation} can be
multiplied by the integrating factor
\(r^{2} \zeta \e^{(\lambda + \nu)/2}, \)
explicitly shown in the following after having additionally imposed
geometrical units such that \(G=c=1,\) to get
\begin{multline}
  \label{eq:PulsMultiplied}
\sigma^{2} r^{2} \e^{(3\lambda-\nu)/2}(\rho + p_{r}) \zeta^{2}  - 8\pi r^{2} (p_{r} -\Pi +\eta)(p_{r}+\rho) \e^{(3\lambda + \nu)/2}\zeta^{2} 
-2 r \e^{(\lambda + \nu)/2} (\delta \Pi) \zeta \\ +  \f{r^{2} \zeta^{2}}{p_{r}+\rho} \e^{(\lambda + \nu)/2} (p'_{r})^{2}  - 4 r p'_{r} \e^{(\lambda + \nu)/2} \zeta^{2} +\e^{-\nu/2}\left[ \e^{(\lambda_{0} + 3 \nu_{0})/2} \gamma \f{p_{r0}}{r^{2}} \left( r^{2} \zeta \e^{-\nu_{0}/2}\right)' \right]' r^{2}\zeta \\ 
- r^{2}\e^{-\nu/2} \left\{ \e^{(\lambda + 2 \nu)/2}  \left[ \zeta \left( \f{2\Pi}{r} + \f{4\eta}{r} + \eta' \right) + \delta \eta \right] \left[\pderiv{p_{r}}{\rho} - 1 \right] \right\}' \zeta  \\  
-  \f{4r^{2} \zeta^{2} p'_{r}}{p_{r}+\rho} \left( \f{\Pi}{r} + \f{4\eta}{r} + \eta' \right) \e^{(\lambda+\nu)/2} 
+ 4  r \e^{(\lambda+\nu)/2}\left( \f{2\Pi}{r} + \f{4\eta}{r} + \eta' \right) \zeta^{2} \\
+  \f{r^{2}\zeta^{2}}{p_{r}+\rho} \e^{(\lambda+\nu)/2} \left( \f{2\Pi}{r} +\f{4\eta}{r} + \eta' \right) \left( \f{2\Pi}{r} +\f{12\eta}{r} + 3\eta' \right) =0.
\end{multline}

From the above equation we can deduce a generalized Sturm-Liouville
form, i.e.\ equation~\eqref{eq:PulsMultiplied} can be put in the form
\begin{equation}
\label{Stab.eq:gSL}
  f \left\{ \deriv{}{r} \left[ P(r) \deriv{f}{r}\right] + \left( Q(r)
    + \sigma^{2} W(r) \right) f + R \right\} = 0,
\end{equation}
for the function \(f \neq 0,\)
since it encodes the radial perturbations which cannot vanish in a
perturbation calculations, and which is defined through
\(\zeta \eqqcolon  \f{\e^{\nu/2} f(r)}{r^{2}}.\) This substitution then gives
\begin{equation}
\label{Prel.eq:STCoeff}
  \begin{aligned}
    P =& \f{\gamma p_{r} \e^{(\lambda + 3\nu)/2}}{r^{2}},\\
    Q =& - \f{8\pi (p_{r} -\Pi +\eta)(p_{r}+\rho) \e^{3(\lambda + \nu)/2}}{r^{2}}  + \f{\e^{(\lambda+3\nu)/2} (p'_{r})^{2}}{r^{2}(p_{r}+\rho)}  -\f{4p'_{r}\e^{(\lambda+3\nu)/2}}{r^{3}}\\
       & - \f{4\e^{(\lambda+3\nu)/2} p'_{r}}{r^{2}(p_{r}+\rho)} \left( \f{\Pi}{r} + \f{4\eta}{r} + \eta' \right) - \f{4\e^{(\lambda+3\nu)/2}}{r^{3}}\left( \f{2\Pi}{r} + \f{4\eta}{r} + \eta' \right) \\
       &+\f{\e^{(\lambda+3\nu)/2}}{r^{2}(p_{r}+\rho)}  \left( \f{2\Pi}{r} +\f{4\eta}{r} + \eta' \right) \left( \f{2\Pi}{r} +\f{12\eta}{r} + 3\eta' \right),\\
    R =& -\f{2 (\delta \Pi) \e^{(\lambda + 2\nu)/2}}{r} - \left\{ \left[ \f{\e^{(\lambda+3\nu)/2}}{r^{2}} \left(\f{2\Pi}{r} + \f{4\eta}{r}+\eta'\right) f + \e^{(\lambda+2\nu)/2}\delta \eta \right]\left[\deriv{p_{r}}{\rho} - 1\right] \right\}',\\
    W =& \f{[\rho + p_{r}] \e^{(3\lambda+\nu)/2}}{r^{2}}.
  \end{aligned}
\end{equation}
This would be a simple Sturm-Liouville problem if and only if \(R\)
could be absorbed in \(Q,\) and we will see that with our assumptions
about \(\delta \Pi\) and \(\delta \eta\) this is indeed the case.
Then, since \(f\neq 0,\) the parenthesized part of
equation~\eqref{Stab.eq:gSL} vanishes, and we retrieve a
Sturm-Liouville problem, with weight \(W\) so that associated with
this problem are the orthogonal eigenfunctions corresponding to the
different eigenfrequencies.  The orthogonality relation obeyed by this
equation is then
\begin{equation}
  \label{eq:orthoStrum}
\int_{0}^{r_{b}} \e^{(3\lambda-\nu)/2} (\rho +p_{r})r^{2}\zeta^{i}\zeta^{j} \d r = \delta^{ij} = \int_{0}^{r_{b}} \e^{(3\lambda+\nu)/2} \f{(\rho +p_{r})}{r^{2}}f^{i}f^{j} \d r = \delta^{ij},
\end{equation}
with \(\delta^{ij}\)
being the Kronecker delta and \(\zeta^{i}\)
or equivalently \(f^{i}\)
the eigenfunctions associated with eigenfrequency \(\sigma^{i}.\)
Similarly the boundary conditions that need to be satisfied by the
functions \(f\)
stemming from the BCs on \(\zeta\)
from~\eqref{Stab.eq:B1} and~\eqref{Stab.eq:B2} now become 
\begin{equation}
  \label{Stab.eq:BCf}
f(r=0) \sim r^{3}, \qquad \text{and} \qquad \delta p_{r}(r=r_{b}) = 0. 
\end{equation}
This weight function, and the BCs will be useful when we start
computing the integrals in the next section.

Before ending this section, we recap what has been achieved so far.
We provide in the above equations the complete first order radial
pulsation equation valid for all solutions admitting both electric
charge and pressure anisotropy, in all their generality.  If one were
to find new spherically symmetric and static solution through various
means~\cite{Lak03,BooVisWei03}, and even include electric
fields, and/or anisotropic pressures in those solutions, then one could use
equation~\eqref{eq:PulsIntegrated} to investigate its stability
right-away, without going through the lengthy derivation we just
presented.  The result is general enough to be used in spherically
symmetric and static cases where
\begin{enumerate}
\item The baryon number \(N\) and the electric charge density \(\eta\)
  are both  functions of the mass density \(\rho,\)
  the radial pressure \(p_{r},\) and the radial coordinate \(r\) only; 
\item the anisotropic pressure \(\ppen\)
  is a function of the radial pressure \(p_{r}\) and the radius \(r\) only
\end{enumerate}
If these are satisfied, then this first order pulsation
equation~\eqref{eq:PulsIntegrated}, and the associated eigenfunction
orthogonality relation~\eqref{eq:orthoStrum} can be used to test the
model against radial perturbations.

\section{Applying the stability criterion on our solutions}
In our solutions, we have expressions for the metric functions
\(\nu, \lambda,\)
and matter functions \(\rho, p_{r}.\)
However due to the complicated and lengthy expressions involved, it
will be more convenient to pick test functions \(\zeta\)
that simplify the integrals of~\eqref{eq:PulsIntegrated} without
requiring explicit expansion of those functions.  Furthermore the
expressions for \(\Pi\)
and \(\eta\)
are simple enough in our solutions that their computation will not be
overly taxing, and we will simplify those.  In particular
\(\Pi = -\Delta\)
in our solutions, with \(\Delta = \beta r^{2},\)
that is always some polynomial of \(r\)
allows the computation of derivatives such as \(\Pi'\)
easily.  As a result,
\begin{equation}
  \label{eq:anisVariation}
\delta \Pi = \Pi' \delta r = - 2\beta r  \zeta.
\end{equation}
We also calculated the perturbation equation for the electromagnetic
component of the stress tensor, \(\eta\)
in equation~\eqref{eq:DeltaEta}. In our solutions we assume that
\(\eta = q^{2}/(\kappa r^{4}),\)
with \(q^{2} = k^{2} r^{6}.\)
As a result, we have \(\eta = (kr)^{2}/\kappa,\)
allowing us to find the expression for \(\eta' = 2k^{2}r/\kappa.\)
To calculate the perturbation in \(\eta\),
we proceed through equation~\eqref{eq:DeltaEta} which states that
\(\delta \eta = \zeta E_{0} \epsilon_{0} \e^{-\nu/2} .\)
Substituting all the relevant quantities results in
\begin{equation}
  \label{eq:etaVariation}
\delta \eta = \f{6k^{2}}{\kappa} r \zeta,
\end{equation}
and using the fact that the static electric field is
\(E_{0} = (q \e^{(\lambda_{0} + \nu_{0})/2})/r^{2},\) while the static
charge density is \(\epsilon_{0} = [6k/\kappa]\e^{-\lambda_{0}/2},\)
from equation~\eqref{ns.eq:sigmaCCA}, we can finally proceed to
simplify the integral equation~\eqref{eq:PulsIntegrated}.

Indeed, with these two ingredients, we can calculate most of the terms
in the integral equation~\eqref{eq:PulsIntegrated}, which simplifies
to
\begin{multline}
  \label{eq:PulsSpec}
\sigma^{2} \int_{0}^{r_{b}} r^{2} \e^{(3\lambda-\nu)/2}(\rho + p_{r}) \zeta^{2} \d r = \kappa \int_{0}^{r_{b}} r^{2} p_{r} (p_{r}+\rho) \e^{(3\lambda + \nu)/2}\zeta^{2} \d r \\
+\kappa \int_{0}^{r_{b}} r^{4} \left(\beta +\f{k^{2}}{\kappa}\right) (p_{r}+\rho) \e^{(3\lambda + \nu)/2}\zeta^{2} \d r 
 +  \int_{0}^{r_{b}} \f{\gamma p_{r}}{r^{2}} [(r^{2} \zeta \e^{-\nu/2})']^{2} \e^{(\lambda+3\nu)/2} \d r \\
- \int_{0}^{r_{b}} \f{r^{2} \zeta^{2}}{p_{r}+\rho} \e^{(\lambda + \nu)/2} (p'_{r})^{2} \d r + \int_{0}^{r_{b}} r^{2}\e^{-\nu/2} \left\{ \e^{(\lambda + 2 \nu)/2}  \zeta r \left[ \f{12k^{2}}{\kappa} - 2\beta \right] \left[\pderiv{p_{r}}{\rho} - 1 \right] \right\}' \zeta \d r  \\ 
+ \int_{0}^{r_{b}} \f{4r^{3} \zeta^{2} p'_{r}}{p_{r}+\rho} \left(\f{6k^{3}}{\kappa} - \beta \right) \e^{(\lambda+\nu)/2} \d r 
- 8 \int_{0}^{r_{b}} r^{2} \zeta^{2} \e^{(\lambda+\nu)/2}\left( \f{3k^{2}}{\kappa} - \beta\right)  \d r\\
+ 4\int_{0}^{r_{b}} r p'_{r} \e^{(\lambda + \nu)/2} \zeta^{2} \d r - \int_{0}^{r_{b}} \f{4r^{4}\zeta^{2}}{p_{r}+\rho} \e^{(\lambda+\nu)/2} \left( \f{3k^{2}}{\kappa} - \beta \right) \left( \f{9k^{2}}{\kappa} - \beta \right) \d r.
\end{multline}
In this form, finding the frequency \(\sigma\) depends on guessing a
correct test function that will allow the computation of the integrals
of the above equation.  \citeauthor{Cha64} could do this with much
less effort since he did not have as many terms to satisfy at the same
time.  If we try to copy and adapt the method
of~\citeauthor{EscAlo10}\cite{EscAlo10}, we find that the same test
functions do not yield analytic closed form integrals for our case.
Since we only wish to find the frequency and do not require the
eigenfunctions for some \(\zeta,\) we turn to numerical integration.
However, since this is the case, we decided to
use~\citeauthor*{BarThoMel66}'s formulation, since theirs is a clearer
formulation for numerical work~\cite{BarThoMel66}.  We do this next.

\subsection{Numerical integration to obtain the fundamental frequency}
To calculate the the fundamental mode we follow~\cite{BarThoMel66}
instead, since in their formulation, simple numerical integration
obviates the need to guess an accurate test function for \(\zeta:\)
usually a hard procedure.  Before being able to use their result
however, we have to absorb the \(R\) term in
equation~\eqref{Prel.eq:STCoeff} into \(Q\) or \(P.\) since we have
expressions for the perturbations of the electric field and anisotropy
now in the form of~\eqref{eq:etaVariation}
and~\eqref{eq:anisVariation} respectively, we proceed to simplify
\(R.\) The first term containing the perturbed anisotropy simplifies
as
\[-\f{2(\delta \Pi) \e^{(\lambda+2\nu)/2}}{r} f = \f{4\beta
  \e^{(\lambda+3\nu)/2}}{r^{2}} f^{2},\]
which can then be absorbed into \(Q,\)
since it contains the \(f^{2}\)
term.  The second term in \(R\)
can similarly be simplified as
\begin{multline*}
-f \left\{ \left[ \f{\e^{(\lambda+3\nu)/2}}{r^{2}} \left(\f{2\Pi}{r} + \f{4\eta}{r}+\eta'\right) f + \e^{(\lambda+2\nu)/2}\delta \eta \right]\left[\deriv{p_{r}}{\rho} - 1\right] \right\}' =\\* f \left(2\beta - \f{12k^{2}}{\kappa} \right) \left\{ \left[ \left( \deriv{p_{r}}{\rho} - 1 \right) \f{\e^{(\lambda+3\nu)/2}}{r}\right] f' +  \left[ \f{\e^{(\lambda+3\nu)/2}}{r} \left(\deriv{p_{r}}{\rho} - 1 \right) \right]' f \right\},
\end{multline*}
so that the second term can be absorbed in \(Q.\) However, since the
first term of the above equation cannot be easily eliminated in this
general form, we will have to find an additional integrating factor
before being able to reduce it to a simple Sturm-Liouville form.  We
therefore use ideas from~\citeauthor*{HorIliMar11}, who look at a
slightly different problem involving quasi-local
quantities~\cite{HorIliMar11} to be able to simplify this more
generalized problem\footnote{As was pointed out by Dr. Gene Couch
  through private communication, it is the form of \(R\) after
  substituting the expressions of \(\delta \Pi\) and \(\delta \eta\)
  from our solutions that allow for the existence of this integrating
  factor, Were we to have had the general expression for \(R,\) this
  step whould not have been as straight forward.  Indeed we are not
  claiming that a general nonlinear differential equation with any
  \(R\) is liable to this same simplification method.}.  We first
expand out the pulsation equation again and express it as a canonical
second order PDE of the form
\begin{equation}
  \label{eq:canonic2PDE}
  C_{2}f'' + C_{1}f' + C_{0}f  = -\sigma^{2} f,
\end{equation}
with 
\begin{align*}
  C_{0} =& -8\pi\left[ p_{r} + \left( \beta + \f{k^{2}}{\kappa} \right)r^{2} \right] \e^{\nu} 
           -\f{24\e^{\nu-\lambda}} {p_{r} + \rho}  \left( \f{k^{2}}{\kappa} - \f{\beta}{2} \right)  + \f{(p'_{r})^{2} \e^{\nu-\lambda}}{(p_{r}+\rho)^{2}} - \f{4p'_{r} \e^{\nu-\lambda}}{r(p_{r}+\rho)} + \\ 
         &  -\f{4rp'_{r} \e^{\nu-\lambda}}{(p_{r}+\rho)^{2}} \left( \f{6k^{2}}{\kappa} - \beta\right) + 
\f{4 r^{2} \e^{\nu-\lambda}}{(p_{r}+\rho)^{2}} \left( \f{3k^{2}}{\kappa} - \beta \right) \left(\f{9k^{2}}{\kappa} -\beta \right) + \\
         & + \f{2r^{2}}{(p_{r}+\rho) \e^{(3\lambda + \nu)/2}} \left( \beta - \f{6k^{2}}{\kappa} \right) \left[ \f{\e^{(\lambda+3\nu)/2}}{r}\left( \pderiv{p_{r}}{\rho} - 1 \right)\right]', \\ 
  C_{1} =& \left( \f{\gamma p_{r} \e^{(\lambda+3\nu)/2}}{r^{2}}\right)' \f{r^{2}\e^{-(3\lambda+\nu)/2}}{p_{r}+ \rho} 
           + \f{2r\e^{\nu-\lambda}}{p_{r}+\rho} \left( \beta - \f{6k^{2}}{\kappa}\right) \left( \deriv{p_{r}}{\rho} - 1 \right) ,\\
  C_{2} =&\f{P}{W} = \f{\gamma p_{r} \e^{\nu-\lambda}} {p_{r}+\rho} .
\end{align*}
Then by multiplying equation~\eqref{eq:canonic2PDE} by another integrating factor 
\[F(r) = \exp{\left( \int_{0}^{r} \f{C_{1}(\bar{r}) - C'_{2}(\bar{r})} {C_{2}(\bar{r})} \d \bar{r} \right)} 
\implies F'(r) = \f{C_{1}(r) - C'_{2}(r)}{C_{2}(r)} F(r), \]  
a factor that depends crucially on both the anisotropy \(\beta\) and charge \(k.\)  
We see that~\eqref{eq:canonic2PDE} then becomes
\begin{equation}
\label{Stab.eq:IntFact}
(C_{2}F)f'' + (C_{1}F) f' + (C_{0} F) f = -\sigma^{2} F f,  
\end{equation}
which is factorizable into  a true Sturm-Liouville equation~\eqref{Stab.eq:gSL} with the following coefficients:
\begin{equation}
\label{Prel.eq:STCoeff}
  \begin{aligned}
    P = FC_{2} = & \f{\gamma p_{r} \e^{(\lambda + 3\nu)/2}}{r^{2}} \left\{ \exp{\left[ \left(\beta - \f{6k^{2}}{\kappa}\right) \int^{r} \f{2\bar{r}}{\gamma p_{r}} \left( \pderiv{p_{r}}{\rho} - 1 \right) \d \bar{r}\right]}\right\} ,\\
    Q = FC_{0} = & \left\{ \exp{\left[\left( \beta - \f{6k^{2}}{\kappa}\right) \int^{r} \f{2\bar{r}}{\gamma p_{r}} \left( \pderiv{p_{r}}{\rho} -1 \right) \d \bar{r}\right]}\right\} \times \left\{ -\f{4p'_{r}\e^{(3\nu+\lambda)/2}}{r^{3}}  + \right.\\
    & - \f{8 \pi \e^{3(\lambda+\nu)/2}}{r^{2}}\left[p_{r} + \left(\beta + \f{k^{2}}{\kappa} \right)r^{2}\right] \left( p_{r} + \rho\right) -\f{24\e^{(3\nu+\lambda )/2}}{r^{2}} \left( \f{k^{2}}{\kappa} - \f{\beta}{2} \right) + \\
    & +\f{4\e^{(3\nu+\lambda)/2}}{p_{r}+\rho} \left(\f{3k^{2}}{\kappa} - \beta \right) \left(\f{9k^{2}}{\kappa} - \beta\right) - \f{4p'_{r} \e^{(3\nu+\lambda)/2}}{r(p_{r}+\rho)}\left( \f{6k^{2}}{\kappa} -\beta\right)\\
    &\left. + 2\left( \beta - \f{6k^{2}}{\kappa}\right) \left[ \f{\e^{(\lambda+3\nu)/2}}{r} \left( \pderiv{p_{r}}{\rho} - 1 \right)\right]'  + \f{\e^{(3\nu+\lambda)/2} (p'_{r})^{2}}{r^{2}(p_{r}+\rho)}\right\}, \\
    W = F =& \f{(p_{r}+\rho) \e^{(3\lambda+\nu)/2}}{r^{2}} \left\{ \exp{\left[\left( \beta - \f{6k^{2}}{\kappa}\right) \int^{r} \f{2\bar{r}}{\gamma p_{r}} \left( \pderiv{p_{r}}{\rho} -1 \right) \d \bar{r}\right]}\right\},\\ 
    R = 0  \qquad &\text{in true Sturm-Liouville canonical form,}
  \end{aligned}
\end{equation}
since \( [(FC_{2}) f']' + (FC_{0}) f = -\sigma^{2} F f,\)
yields~\eqref{Stab.eq:IntFact} after simplification, as can be checked
explicitly by expansion.

We can therefore and finally use the main result of Sturm-Liouville
theory (refer to appendix~\ref{C:AppendixA}) which directly gives the fundamental
frequency of normal modes of our model.  If this fundamental mode is
positive, then all the other modes are too, and the model is deemed
stable.  This fundamental frequency is given~\cite{BarThoMel66} by the maximal value of 
\begin{equation}
  \label{Stab.eq:fundamentalF}
  \sigma^{2}_{0} = \int_{0}^{r_{b}} \left[P\left( \deriv{f}{r} \right)^{2}  - Q f^{2} \right] \d r,
\end{equation}
where \(f\)
is taken to be a function that obeys the normalization
condition 
\begin{equation}
\label{Stab:eq:Normf}
\int_{0}^{r_{b}} W f^{2} \d r = 1,  
\end{equation}
and the BCs given in~\eqref{Stab.eq:BCf}.  Finding a function \(f\)
that obeys the normalization condition~\eqref{Stab:eq:Normf} is
\emph{a priori} difficult.  So if we wish to eschew this additional
condition, we could instead modify~\eqref{Stab.eq:fundamentalF} to
\begin{equation}
  \label{Stab.eq:fundamentalFg}
  \sigma^{2}_{0} = \f{\int_{0}^{r_{b}} \left[P\left( \deriv{f}{r} \right)^{2}  - Q f^{2} \right] \d r}{\int_{0}^{r_{b}} W f^{2} \d r},
\end{equation}
but now require that the normalization on different eigenfunctions
\(f_{i}\)
to the Sturm-Liouville system~\eqref{Prel.eq:STCoeff} be such that the weight \(W\) annihilates the integral:
\[\int_{0}^{r_{b}} W f_{i}f_{j} \d r = 0, \qquad \text{whenever} \quad i \neq j. \]
We are free to choose different functions \(f,\) however the true
value of the fundamental frequency will only be obtained with a proper
eigenfunction of the above equation.  Since we only want to test for
stability, and do not want the actual frequency of the fundamental
mode, we will choose the simplest \(f\) possible, e.g.\ \(f=r^{3},\)
in accordance with~\eqref{Stab.eq:BCf}, and then find the value of
\(\sigma_{0}\) by numerically computing the integrals involved in
equation~\eqref{Stab.eq:fundamentalFg}, with the coefficients \(P, Q\)
and \(W\) given by~\eqref{Prel.eq:STCoeff}.
\begin{table}
\begin{minipage}{\textwidth}
  \centering
  \begin{tabular}{c|>{$}l<{$}| >{$}c<{$}| c}
    \hline\hline
    Solution   & \text{Parameters} & \sigma_0^{2} \un{(Hz)} & Stable? \\ \hline
               & \mu = 1 & &  \\
       Natural & \rho_c = 7.43 \times 10^{-10} \mathrm{m}^{-2}& &  \\
    Tolman~VII & r_b = 1\times 10^4 \mathrm{m} & 54.8 & Y \\
               & k = 0 & &  \\
               & \beta = 0 & & \\
\hline
               & \mu = 0.7 & &  \\
    Self-bound & \rho_c = 7.43 \times 10^{-10} \mathrm{m}^{-2}& &  \\
    Tolman~VII & r_b = 1\times 10^4 \mathrm{m} & 77.6& Y \\
               & k = 0 & &  \\
               & \beta = 0 & & \\
\hline
               & \mu = 1 & &  \\
    Tolman~VII & \rho_c = 7.43 \times 10^{-10} \mathrm{m}^{-2}& &  \\
       with    & r_b = 1\times 10^4 \mathrm{m} & 91.4& Y \\
    anisotropy & k = 0 & &  \\
               & \beta \sim a/3 =1\times 10^{-17} \mathrm{m}^{-4}& & \\
\hline
               & \mu = 1 & &  \\
    Tolman~VII & \rho_c = 7.43 \times 10^{-10} \mathrm{m}^{-2}& &  \\
       with    & r_b = 1\times 10^4 \mathrm{m} & 821& Y \\
    anisotropy & k = 0 & &  \\
               & \beta \sim 10a/3 =1\times 10^{-16} \mathrm{m}^{-4}& & \\
\hline
    Self-bound & \mu = 0.7 & &  \\
    Tolman~VII & \rho_c = 7.43 \times 10^{-10} \mathrm{m}^{-2}& &  \\
       with    & r_b = 1\times 10^4 \mathrm{m} & 139& Y \\
    anisotropy & k = 0 & &  \\
               & \beta \sim a/3 =1\times 10^{-17} \mathrm{m}^{-4}& & \\
\hline
    Tolman~VII & \mu = 1 & &  \\
    with       & \rho_c = 7.43 \times 10^{-10} \mathrm{m}^{-2}& &  \\
    anisotropy & r_b = 1 \times 10^4 \mathrm{m} & 91.1& Y \\
    and        & k = 1\times 10^{-10} \mathrm{m}^{-2} & &  \\
    charge & \beta \sim a/3 =1\times 10^{-17} \mathrm{m}^{-4}& & \\
\hline
    Tolman~VII & \mu = 1 & &  \\
    with       & \rho_c = 7.43 \times 10^{-10} \mathrm{m}^{-2}& &  \\
    anisotropy & r_b = 1 \times 10^4 \mathrm{m} & 215& Y \\
    and        & k = 9\times 10^{-10} \mathrm{m}^{-2} & &  \\
    charge\footnotemark & \beta \sim a/3 =1\times 10^{-17} \mathrm{m}^{-4}& & \\
\hline
    Tolman~VII & \mu = 1 & &  \\
    with       & \rho_c = 7.43 \times 10^{-10} \mathrm{m}^{-2}& &  \\
    anisotropy & r_b = 1 \times 10^4 \mathrm{m} & -1360& N \\
    and        & k = 5\times 10^{-9} \mathrm{m}^{-2} & &  \\
    charge\footnotemark[1] & \beta \sim a/3 =1\times 10^{-17} \mathrm{m}^{-4}& & \\
  \end{tabular}
  \caption{Eigenfrequency of the fundamental mode of various solutions with different parameter values, and their stability}
\label{Stab.tab:StabilityModels}
\end{minipage}
\end{table}
We did not write our own integral methods, instead replying on the
proved and tested QUADPACK suite of integration routines available in
MAXIMA.  These can deal with all type of integrals, even oscillating
ones through a number of methods designed to robustly integrate and
constrain the numerical errors within tight bounds.  We provide the
MAXIMA routine we used in appendix~\ref{C:AppendixC}, and here provide
a table~\ref{Stab.tab:StabilityModels} with different parameter
values, and an approximate fundamental mode frequency obtained for
those parameter values.  We note that the positivity of this
fundamental frequency is the only criterion needed to prove the
stability of the solution for those parameter values, and that the
value of \(\sigma\)
is only approximately equal to the true eigenfrequency, since our
calculations depend on the test function \(f\)
we chose.  However the
sign of the fundamental frequency is correct (i.e.\ positive), since
if even one test function gives a positive value, we can be certain
that some others will too since the true eigenfrequency is the maximum
value of all possible \(\sigma\)
when we span the \(L^{2}\)
function (Lebesgue) space, according to theorem~\eqref{A.th:S-L}
given in appendix~\ref{C:AppendixA}.
\footnotetext[1]{Using large
  $k \sim 1 \times 10^{-9} m^{-2} $ resulted in the $Q$ integral not
  converging suggesting that even the QUADPACK routines have trouble
  dealing with the integrals when $\beta \sim 6k^{2}/ \kappa,$ since
  then integrating factor, and hence the integrand accumulate many
  numerical errors.  Furthermore, it also seems that any charge at
  all, without anisotropy distabilises the star, suggesting that
  purely charged solutions are unstable.  Since we did the
  calculations numerically, we can only guess the maximum $k$ value
  allowed, but further work should be able to find it.}
\section{Conclusion}
In this chapter we have shown that all of our new solutions, together
with the original Tolman~VII solution are stable under first order
radial perturbations, for certain values of the parameters.  Of
course, for ``excessive'' charge or anisotropy, the stability is
compromised, as expected.  This results adds to the heuristics we
discussed in the beginning, and complements the conclusions reached
that our solutions can indeed be stable.  If any instability is to
occur in these solutions, while the charge and anisotropy are within
``normal'' ranges, it will occur from second order effects, from
non-linear perturbations, or from non-radial pulsations: all of which
are beyond the scope of this work.

Our final calculations were done numerically, and a better way to
approach this would have been to find a suitable test function
\(\zeta\)
that would have allowed an expression of the fundamental frequency to
be obtained as a function of both \(\Pi\)
and \(k.\)
This would have placed constraints on the maximum allowed anisotropy
and charge of our model.  We did not proceed in this direction because
such an endeavour would have taken more time, and we were only
interested in \emph{showing} the viability of our model for modelling
stars.  This approach should lead to interesting results in charge and
anisotropic bounds in the future.

The proof of stability relied on finding the normal mode frequency of
linear radial oscillations which were obtained in a generalized way
from the non-static Einstein equations.  Our general expressions are
valid for charged and anisotropic solutions too: to our knowledge,
this generalized formulation, in both~\citeauthor{BarThoMel66}'s
formulation in the form of equations~\eqref{Prel.eq:STCoeff}
and~\eqref{Stab.eq:gSL}; and~\citeauthor{Cha64}'s formulation in the
form of equation~\eqref{eq:PulsMultiplied} is new, and will be useful
to prove the stability for all static solutions, new and old that
contain electric charge and/or anisotropic pressures.

Recently there has been a renewed interest in the stability of neutron
stars, particularly because pulsation in such stars could potentially
produce measurable gravitation waves.  \citeauthor{KruHoAnd15} for
example~\cite{KruHoAnd15} look at polar modes of perturbations in the
stars, a considerably more difficult problem, to see if gravitational
modes would be generated.  The pulsation equation we generated is only
for radial modes, but could presumably be extended to study the
seismology of stars in a similar fashion.  In a similar vein,
\citeauthor{ChiSouKas15} look at \(f-\)mode (fundamental modes)
oscillation as we do, but pay more attention to the damping of the
mode, thus requiring more than just a linear approximation for the
perturbation calculations, in the search of universal relations over
many EOS, for gravitational waves again.  This could also be a way to
continue this work~\cite{ChiSouKas15}.

On this note, we move on to look at the predictions of the new
solutions in the next chapter.



\chapter{Analysis of new solutions} \label{C:Analysis}
\begin{myabstract}
  We investigate the solutions we found previously in
  chapter~\ref{C:NewSolutions} and deduce the behaviour of the matter
  variables, and metric functions.  We interpret these in view of
  using these solutions to model compact stars, and to achieve this
  goal, we test a number of criteria that is believed to be necessary
  for these mathematical solutions to be viable as models of actual
  astrophysical objects.  We also compute observables such as the masses
  and radii of the models and compare them with observed values of neutron
  star masses and radii, showing that some of our models might indeed
  describe actual stars.
\end{myabstract}

\section{Properties of the fluid as described by the new solutions}
We will now analyse at the solutions found previously, starting
with those that include anisotropic pressures, and moving to the
charged ones with anisotropic pressures later.  Along the way we will
provide conditions that the parameters must satisfy to abide
by the constraints of physicality we impose.  Other constraints
stemming from the definiteness of the metric functions will also be
used to further restrict the values of our parameters.  Just as a
reminder, and as introduced in chapter~\ref{C:TolmanVII}, a
brief interpretation of our ans\"atze in terms of physically
meaningful concepts will be discussed.

\subsection{The free parameters}
The fundamental ansatz we have used consistently is the Tolman~VII
ansatz that assumes a specific form of the metric variable
\(Z = 1 - br^{2} + ar^{4}.\) In Tolman's solution, the coefficients
\(a\) and \(b\) of this quartic function correspond to well defined
combinations of three different physical values: the central density
\(\rho_{c},\) the coordinate radius of the boundary \(r_{b},\) and the
self-boundness parameter \(\mu.\) In the more general charged case
however, the charge density \(k\) also determines the \(a\)
coefficient, and this changes the simple and straight-forward
interpretation we had in Tolman~VII.

The other parameter that can be interpreted is \(\beta,\) which
governs the difference between the tangential pressure \(\ppen\) and
the radial pressure \(p_{r}.\) In the isotropic cases for example,
\(\beta = 0,\) and in one specific case which we called
``anisotropised charge,'' \(\beta\) can be used to express the charge
parameter \(k,\) so that only one of these last two variables is
independant, and thus enough to specify that particular solution.

This list of parameters, \(\{\rho_{c}, r_{b}, \mu, \beta \}\)
for the uncharged case, and \(\{\rho_{c}, r_{b}, \mu, \beta, k \}\)
for the charged case are free by construction.  They correspond to the
exact number of parameters expected for the system of differential
equations associated with each case and therefore all the integration
constants used in the solutions can be expressed uniquely in terms of
the respective set of parameters only.  As a result of this
construction, and immediate interpretation of the constants, we can
already impose naive restrictions on the values of these parameters.

In particular, to model a realistic star, the central density has to
be positive definite so that \(\rho_{c} > 0,\) or else we would be
talking about matter having undefined characteristics.  Similarly the
boundary radius of stars has to be positive definite too, and we
immediately get \(r_{b} > 0.\) To get similar bounds on \(\mu,\) we
investigate the expression for mass density that is also common to all
our solutions, in the form of equation~\eqref{t7.eq:Density}, which we
rewrite
here\[ \rho = \rho_{c} \left[ 1 - \mu \left( \f{r}{r_{b}}\right)^{2}
  \right].\] Clearly if \(\mu\) is negative, we will have increasing
mass density with increasing \(r.\) This is not what we would like for
a stable configuration of matter\footnote{It is \citeauthor{Wei72} who
  famously said ``It is difficult to imagine that a fluid sphere with
  a larger density near the surface than near the centre could be
  stable'' \cite{Wei72}.  He was proved wrong with certain anisotropic
  models~\cite{HorIliMar11}, but the statement remains a good rule of
  thumb.}, so we restrict \(\mu\) to values that are greater than
zero.  Similarly, by modelling a star's interior, we will naturally
restrict the coordinate \(r\) to values less than the boundary, so
that \(r \leq r_{b}.\) As a result, \(r/r_{b} \leq 1,\) implying that
having \(\mu > 1\) will again result in negative densities, which we
want to avoid.  As a result, we will also restrict
\(0 \leq \mu \leq 1.\)

Naively this is unfortunately as far as we can go.  To restrict the
values of these parameters, and the other constants further we will need
additional constraints.  

\subsection{Constraints for physical relevance}\label{an.ssec:PhysRel}
We will list a series of constraints that have been discussed in the
literature~\cite{DelLak98,DurPanPhu84,BurHob09, FinSke98}. These are
simple and ``obvious'' criteria that gravitationally stable spherical
balls of matter should have.  These in one form or another have been
used to restrict parameter values allowed by interior solutions to
Einstein's equations.  We shall use this set of criteria in the
following sections to understand and interpret the solutions we have
found.  The list we will use is:
\begin{enumerate}[label=(\roman*)]
\item \label{an.it.RegularMetric} The metric coefficients must be
  regular (not be singular, and be at least differentiable) everywhere
  including at the centre of the star when \(r=0.\)
\item \label{an.it.ExteriorMetric}The metric functions must match an
  exterior solution (the Schwarzschild exterior or the Reissner-Nordstr\"om)
  to the Einstein equations at the boundary where \(r=r_{b}.\)
\item \label{an.it.ExteriorObservables}The integrated ``observables''
  including total charge, total mass, and proper radius must
  correspond to those parameters in the exterior vacuum solution
  matched at the boundary.
\item \label{an.it.DefiniteRPressure}The radial pressure \(p_{r}\)
  must be positive and finite everywhere inside the fluid, including
  the centre \(r=0.\)
\item \label{an.it.BoundaryRPressure}The radial pressure must vanish
  at the boundary, \(p_{r}(r_{b}) = 0.\)
\item \label{an.it.Delta}The tangential pressure must be equal to the
  radial pressure at the centre of the star,
  \(p_{r}(r = 0) = \ppen(r = 0) = p_{c} \implies \Delta( r = 0) = 0.\)
\item \label{an.it.MonotonicMatter}All three, the pressures \(p_{r}\)
  and \(\ppen,\)
  and the density \(\rho\)
  must be decreasing functions of \(r\)
  so that their first derivatives with respect to \(r\)
  is negative everywhere, except possibly at \(r=0,\)
  and at \(r=r_{b},\) where it could be zero.
\item \label{an.it.EnergyConditions} The energy strong condition, the
  most restrictive one of the energy conditions~\cite{HawEll73} states
  that for realistic matter, with our type of energy-momentum, we must
  have that \(\sum_\alpha{p_{\alpha}} + \rho \geq 0,\)
\item \label{an.it.CausalSpeed}The speed of pressure (sound) waves
  \(v_{s} = \sqrt{\left( \deriv{p_{r}}{\rho} \right)}\)
  is causal in the interior, so that in geometrical units,
  \(0 \leq v_{s} \leq 1.\)
\item \label{an.it.DecreasingSpeed}The speed of sound decreases
  monotonically with increasing coordinate \(r\)\cite{AbrHerNun07}.
\end{enumerate}

We shall now look at these conditions in detail, and determine which
ones can be implemented without an explicit solution.  These we will
apply directly, and then in specialized sections we will look at
those conditions that require the full solutions, and constrain the
latter further in their respective sections.

\subsection{Implementing the constraints}\label{an.ssec:impConstraints}
The first condition~\ref{an.it.RegularMetric} requires the complete
analytic form of both metric functions and as a result we have to wait
before we can implement it completely.  Of interest in implementing
this condition is the capacity to express the integration constants in
terms of the elements of our parameter list.  In all the solutions we
consider, the \(Z\) metric coefficient is expressible in the form
\(1 - br^{2} + ar^{4},\) where \(a\) and \(b\) are slightly different
functions of the parameter list depending on the solution we look at.
However for all the solutions this metric function is equal to unity
at \(r=0,\) and as a result this condition is automatically satisfied.
In view of this, we can modify this particular constraint to read
\begin{enumerate}[label=(\roman*)]
\item The \(Y\)
  metric coefficient must be regular (not be singular, and be at least
  differentiable) everywhere including at the centre of the star when
  \(r=0.\)
\end{enumerate}

Considering the next two constraints~\ref{an.it.ExteriorMetric}
and~\ref{an.it.ExteriorObservables}, these were imposed as one of the
boundary conditions used to generate our solutions: indeed, our
computation of the integration constants \(c_{1}\) and \(c_{2}\)
crucially depended on these particular assumptions, and all the
solutions we have proposed so far automatically obey these
constraints.  As a result we do not impose these constraints
explicitly again, using them instead to check the consistency of the
final solutions at the last stage in the form of
``inside-and-outside'' metric plots for different parameter values.
Of physical importance however is the value of the external
observables mentioned in condition~\ref{an.it.ExteriorObservables}.
In our cases these correspond to the mass \(M = m(r_{b}),\) electric
charge \(Q = q(r_{b}),\) and proper radius
\(R = \int_{0}^{r_{b}} \f{\bar{r} \d \bar{r}}{\sqrt{Z(\bar{r})}}.\)
Since these quantities depend only on the \(Z\) metric function (and the
\(Z\) metric function does not change drastically from solution to
solution), we can compute these quantities right away to get
\begin{equation}
  \label{an.eq.Mass}
  M = m(r_{b}) = 4 \pi \rho_{c} r_{b}^{3} \left( \f{1}{3} - \f{\mu}{5} \right) + \f{k^{2}r_{b}^{5}}{10}, 
\end{equation}
for the mass.  The charge is simply given by
\begin{equation}
  \label{an.eq.Charge}
  Q = kr_{b}^{3},
\end{equation}
since we defined the charge density indirectly through an integral
incorporating the metric function instead in
equation~\eqref{ns.eq:sigmaCCA}.  As a result the difficulty arises in
calculating the charge density of the fluid sphere instead of the
total charge.  For the proper radius of the sphere we have instead
\begin{equation}
  \label{an.eq.Radius}
  R = \int_{0}^{r_{b}} \f{\bar{r}\d \bar{r}}{\sqrt{1 - b\bar{r}^{2} + a\bar{r}^{4}}} = 
  \f{1}{2\sqrt{a}} \left[ \log \left( \f{2\sqrt{a(1 - br^{2}_{b} + a r^{4}_{b})} + 2 a r^{2}_{b}- b}{2\sqrt{a} - b }\right)\right],
\end{equation}
where we have used integration tables from Reference~\cite{GraRyz07}
to compute the final form of the integrand.  It is immediately clear
that since \(R\) must be well-defined, we must have that
\(2\sqrt{a} > b,\) which translates into a well defined constraint on
our physically interpretable parameters.  However as we mentioned
before, the exact values of \(a\) and \(b\) depend on which solutions
we are considering.  As a result the
conditions~\ref{an.it.RegularMetric}-~\ref{an.it.ExteriorObservables}
reduce to \(2\sqrt{a} > b.\)

We now consider condition~\ref{an.it.DefiniteRPressure}, and to
implement it we look at the expression of the pressure in our
solutions.  Without specifying a solution, consider
equation~\eqref{ns.eq:EinMR1+2}, which gives us
\[ \kappa p_{r} = \f{2(1-br^{2}+ar^{4})}{r} \left( \f{1}{Y}
    \deriv{Y}{r}\right) + 2b -4ar^{2} - \kappa\rho_{c}\left[ 1 - \mu
    \left(\f{r}{r_{b}}\right)^{2}\right] > 0.\] However even this form
of the equation proves insufficiently simple to be able to deduce any
constraints directly from it.  We therefore leave
condition~\ref{an.it.DefiniteRPressure} for consideration later when
we have a more definite form of the pressure.

Condition~\ref{an.it.BoundaryRPressure} is already implemented as the
second boundary condition in chapter~\ref{C:NewSolutions}, and all our
solution obey it by construction.  This condition is technically
equivalent to the Israel-Darmois junction condition on the metric and
derivatives as has been shown for example in~\cite{Poi04, MisSha64},
and as we discussed in Appendix~\ref{C:AppendixA}.

The next condition~\ref{an.it.Delta} concerns the tangential pressure,
and is due to spherical symmetry.  The only way to admit an
anisotropic pressure, while still having spherical symmetry is to
ensure that the pressures \(p_{r}\) and \(\ppen\) are equal at the
centre of the star.  This forces our anisotropy measure \(\Delta\) to
vanish at \(r=0.\) By construction, in all solutions, we posited
\(\Delta = \beta r^{2},\) which satisfies this condition, since
\(\Delta = 0\) at \(r=0.\)

The strong energy condition~\ref{an.it.EnergyConditions} has to be
used to provide a constraint on the type of matter we can have in our
solutions.  This is easily implemented once we have expressions of the
density and pressures, and we cannot really implement it until we have
specified parameter values for those quantities.  We will show how we
test this in each solution's section, and just add here that in both
the anisotropic case and the charged anisotropic cases this condition
reduces to \(p_{r} +2\ppen + \rho \geq 0, \) so that this condition
unfortunately tell us nothing about the electric charge, or charge
densities involved.

Causality is one of the important conditions, and we implement this
through condition~\ref{an.it.CausalSpeed} by enforcing that the speed
of pressure (sound) waves in the fluid not propagate at arbitrary
speeds.  The speed of light,\(\un{c}\) in vacuum is taken to be unity
in all our calculations, and to impose this condition we require that
the speed of sound waves be less than one.  To help us impose this, we
first have to find an expression for the speed of these waves.

\subsubsection{The speed of sound}
As we saw in appendix~\ref{C:AppendixA}, the speed of pressure waves
is given by \(v^{2} = \deriv{p_{r}}{\rho}.\) We can either invert the
density relation we have and derive an equation of state once we have
obtained the pressure from the \(Y\) metric coefficient, or we could
simply compute \(v^{2} = \deriv{p_{r}}{r} \Big/ \deriv{\rho}{r}.\)
However in doing the latter we have to be careful, since we are mostly
interested in the behaviour of the speed at the centre where \(r=0,\)
and the density relation has a turning point there\footnote{The reason
  for being interested in the value at $r=0$ is simply because the
  additional constraints we have make it so that the speed of sound is
  monotonically decreasing with increasing $r.$ Since the maximum
  value of the speed of sound occurs at the centre (an expected
  result, from an intuitive Newtonian picture), we should compute it
  there to impose causality more efficiently.}, so that
\(\deriv{\rho}{r} = 0\) when \(r=0.\) In view of this, we carefully
proceed by taking the limit
\(\lim_{r \to 0} \left(\deriv{p_{r}}{r} \Big/
  \deriv{\rho}{r}\right),\) which we then check against the actual
expression obtained for the speed of sound from the equation of state:
This we did in the Tolman~VII case, where luckily both methods give the
same result showing that the ``short-cut'' evaluation is actually
valid even at \(r=0.\)

We assume for the time being that a similar result holds for the more
general cases we will be considering in this chapter: the reason being
that the mathematical structure of the new solutions is not very
different from Tolman~VII.  However we keep in mind that this must be
checked later on when we have a full expression for the pressures.
The reason for wanting to evaluate the speed of sound without
specifying a solution for the metric functions in detail is two--fold:
\begin{enumerate*}[label=\itshape\alph*\upshape)]
\item doing this now ensures a certain independence from the more
  inconvenient details (values ranges of constants) we will have to
  deal with later on,
\item were we to get a finite speed that is unconditionally larger
  than the speed of light, we could reject a solution class right
  here, without going through a complicated calculation involving a
  solution that is obviously unphysical.
\end{enumerate*}

We provide such a derivation now, starting from the definition of the
``measure of anisotropy,'' \(\Delta = \kappa(p_{r} - \ppen)\) in
chapter~\ref{C:NewSolutions}.  However, we note that we are dealing
with the charged case in the most general formulation of the problem,
ending up with more terms in this equation than in the previous
chapter. Also this derivation is more transparent in the original
metric variables \(\lambda\) and \(\nu,\) so we re-express everything
in this equivalent set:
\begin{equation}
  \label{an.eq.Delta}
  \Delta = \e^{-\lambda} \left( \f{\nu'}{r} + \f{1}{r^2}\right) -\f{1}{r^2} - \e^{-\lambda} \left( \f{\nu''}{2} - 
      \f{\nu'\lambda'}{4} + \f{(\nu')^2}{4} + \f{\nu'- \lambda'}{2r}\right) +\f{2q^{2}}{r^{4}}.
\end{equation}
We need to simplify, rearrange and factorise this equation first, and
recognise that some parts can be expressed as the derivative of the
pressure variable.  This derivation yielding the famous
Tolman--Oppenheimer--Volkoff (TOV) equation in the uncharged and
isotropic case was obtained by Oppenheimer and
Volkoff~\cite{OppVol39}, and we follow in their footsteps here.

We multiply equation~\eqref{an.eq.Delta} by \(-2/r,\)
and move the charge term on the left hand side to give
\[ \f{4q^{2}}{r^{5}} - \f{2\Delta}{r} = \f{2}{r} \left[ \e^{-\lambda}
    \left( \f{\nu''}{2} - \f{\nu'\lambda'}{4} + \f{(\nu')^2}{4} +
      \f{\nu'- \lambda'}{2r}\right) - \e^{-\lambda} \left( \f{\nu'}{r}
      + \f{1}{r^2}\right) + \f{1}{r^2} \right].\] Next we want to
group expressions that could be factorised as the derivative of a
product, and choose \(\e^{-\lambda}\) as a common factor to give
\[ \f{4q^{2}}{r^{5}} - \f{2\Delta}{r} = \e^{-\lambda} \left(
  \f{\nu''}{r} - \f{\nu'}{r^{2}} -\f{2}{r}\right) -\lambda'
\e^{-\lambda} \left( \f{1}{r^{2}} + \f{\nu'}{2r}\right) +\e^{-\lambda}
\left(\f{\nu'}{2}\right)\left(\f{\nu'}{r}\right) + \f{2}{r^{3}},\]
we then add zero to the second bracket in the above equation in the
form shown to be able to isolate product derivatives in the next step as shown,
\begin{align*}
  \f{4q^{2}}{r^{5}} - \f{2\Delta}{r} &= \e^{-\lambda} \left(
  \f{\nu''}{r} - \f{\nu'}{r^{2}} -\f{2}{r}\right) -\lambda'
\e^{-\lambda} \left( \f{1}{r^{2}} \right. \underbrace{+\f{\nu'}{r} - \f{\nu'}{r}}_{=0} + \left. \f{\nu'}{2r}\right) +\e^{-\lambda}
\left(\f{\nu'}{2}\right)\left(\f{\nu'}{r}\right) + \f{2}{r^{3}},\\
&= \e^{-\lambda}\deriv{}{r} \left( \f{\nu'}{r} + \f{1}{r^{2}}\right) + \left( \f{\nu'}{r} + \f{1}{r^{2}} \right) \deriv{}{r}\left( \e^{-\lambda}\right) - \deriv{}{r}\left(\f{1}{r^{2}}\right) + \f{\e^{-\lambda}\nu'}{2} \left( \f{\nu'}{r} + \f{\lambda'}{r}\right).
\end{align*}
This last step allows us to factorise the derivative terms into an
expression resembling the second Einstein's
equation~\eqref{ns.eq:EinMR2} with the other term being related to
equation~\eqref{ns.eq:EinMR1+2}:  
\[\f{4q^{2}}{r^{5}} - \f{2\Delta}{r} = \deriv{}{r} \left[ \e^{-\lambda} \left( \f{\nu'}{r} +\f{1}{r^{2}} \right) -\f{1}{r^{2}} \right] + \f{\nu'}{r} \left[ \e^{-\lambda} \left( \f{\nu' + \lambda'}{r} \right) \right].
\]
Substituting the latter two in the
above equation thus results in a generalised TOV equation having both
charge and anisotropic pressures (through \(\Delta,\))
\[\f{4q^{2}}{r^{5}} - \f{2\Delta}{r} = \deriv{}{r} \left( \kappa p_{r}
  - \f{q^{2}}{r^{4}}\right) + \f{\nu' \kappa(p_{r} + \rho)}{2},
\]
upon rearranging we have the final form of the generalised TOV equation
which has the coveted radial pressure derivative in terms of other variables:
\begin{equation}
  \label{an.eq.gTOV}
\deriv{p_{r}}{r} = \f{2 q q'}{\kappa r^{4}} - \f{2\Delta}{\kappa r} - \f{\nu'}{2} \left( p_{r} + \rho \right).
\end{equation}

The density \(\rho\) does not change from Tolman's in our solutions
previously, so that we can calculate its \(r\)--derivative to get
\begin{equation}
  \label{an.eq.densityDeriv}
\deriv{\rho}{r} = -\f{2\rho_{c}\mu r}{r_{b}^{2}}.
\end{equation}
From these two, we can compute the formal speed of sound, technically
valid only for \(r \neq 0,\) but extensible to that case too,
\begin{equation}
  \label{an.eq.soundSpeed}
v_{s}^{2} = \left( \deriv{p_{r}}{r} \middle/ \deriv{\rho}{r} \right) = 
\left( \f{r_{b}^{2}}{\kappa \rho_{c} \mu}\right)\left[ \f{\nu' \kappa \left( p_{r} + \rho \right)}{4 r} -\f{qq'}{r^{5}} + \f{\Delta}{r^{2}} \right].   
\end{equation}
This equation allows us to understand how adding different parameters
and assumptions like charge and anisotropic pressures modify the speed
of sound in a very intuitive manner.  For zero charge \(q,\) and zero
anisotropy \(\Delta,\) we get back the same speed of sound as in the
Tolman~VII case from equation~\eqref{t7.eq:SpeedSound} as expected.
For models with charge only, we expect the speed of sound to be lower
than the same model with no charge, suggesting a ``softening'' of the
equation of state in the presence of electric charge.  Similarly
addition of some positive (negative) anisotropy ``stiffens''
(``softens'') the equation of state, making the speed of pressure
waves larger (smaller) than the model would have had in the isotropic
case.  Finally with both charge and anisotropic pressures, the
contributions of each could somehow conspire to not change the overall
``stiffness'' of the EOS, with the ``stiffening'' effect of anisotropy
contributing more (since it scales \(\sim 1/r^{2}\)) than the
softening of the electric charge (which scales \( \sim 1/r^{5}\)) if
the anisotropy is positive.

After this derivation of an important quantity that will be useful in
the analysis done in this chapter, we can continue looking at the
criteria of physical applicability.
Condition~\ref{an.it.DecreasingSpeed} is easily implemented from our
newly crafted speed of sound expression.  This condition demands that
\(\deriv{v_{s}}{r} < 0\)
so that the speed of sound is maximum at the centre of the star.  From
equation~\eqref{an.eq.soundSpeed}, it is clear that without prior
knowledge of \(\nu'\)
or \(p_{r},\)
this is difficult to implement, and hence we wait until we have full
solutions in order to use this condition to extract restrictions on
our parameters.

The next condition~\ref{an.it.MonotonicMatter} is easier to implement
and check since we already have derivatives of all the relevant
variables.  The density derivative is always given
by~\eqref{an.eq.densityDeriv}, and is obviously always negative since
the the other parameters in the equation are positive.  The pressure
derivative given by~\eqref{an.eq.gTOV} is more complicated, however we
know from the Tolman~VII solution that the last term in that equation
is always negative.  Anisotropic contributions through positive
\(\Delta\)
will only make this derivative more negative, so that should not cause
any problems.  However picking negative \(\Delta\)'s
might offset the last Tolman term, and this particular condition gives
us that in the uncharged case,
\[\Delta < \f{\kappa \nu'}{4} \left( p_{r}+ \rho \right) r \implies 
  \beta < \f{\kappa \nu'}{4r} \left( p_{r}+ \rho \right) \] which
gives us an idea as to the range \(\beta\) can take, if it is
negative.  A similar argument applies for the charge, and the
inequality above becomes more complicated in the most general case
where we could presumably have negative \(\beta\) and large charge
that could potentially force the pressure derivative to become
positive in the star.  These are cases we have to ensure against when
we get our solutions.

The final part of this condition concerns the tangential pressure
\(\ppen.\)
Its \(r\)--derivative
has to be negative too, and since \(\ppen = p_{r} - \Delta,\)
we only have to check that
\[ \deriv{\ppen}{r} = \deriv{p_{r}}{r} - \deriv{\Delta}{r} < 0
  \implies \deriv{p_{r}}{r} < \deriv{\Delta}{r} .\] Since
\(\Delta = \beta r^{2}\) in our solutions, we want
\(\beta > \f{1}{2r} \deriv{p_{r}}{r},\) a condition we can combine
with the previous one in the right circumstances.

This concludes our discussion of the conditions for physical
relevance.  We should note that most of these depend on the final form
of the metric functions \(Y\) and equivalently \(\nu\) to be
implemented, but promise to restrict our parameter space depending on
the type of solution.  We should also keep in mind that this list is
not exhaustive: other more stringent criteria might become important,
for example from stability analysis, or from more accurate
thermodynamics.

In the next section, we go into the details of each solution in full,
check the behaviour of the matter and metric functions, and by
ensuring that the conditions above hold, restrict the range of
applicability of the solutions we found to interesting physical cases
when possible.

\section{The solutions}\label{an.sec:sol}
We will look at each solution in turn, and determine what parameter
values are valid for each.  For reference of which solution is being
discussed, please refer to the relevant section in
chapter~\ref{C:NewSolutions}, and for a quick list of the various
functions and expressions please refer to appendix~\ref{C:AppendixB}.
We start with the anisotropic uncharged generalisations to Tolman~VII.

\subsection{The $\phi^{2}= 0$ case from~\ref{ns.ssec:phiZero}}\label{an.ssec:phiZero}
This is a special case of the general anisotropic pressure case, where
the pressure anisotropy \(\beta\) is fixed to the value of \(-a.\) To
help us in evaluating all the different conditions, it will be helpful
to compute first the expressions for quantities at both the centre of
the star when \(r=0\) and at the matter--vacuum boundary, when
\(r=r_{b}.\) The \(Z\) metric for example gives
\begin{equation}
  \label{an.eq:Z0+Zr_b}
Z(0) = 1, \qquad\text{and}\qquad Z(r_{b}) = 1 - \kappa \rho_{c}r_{b}^{2} \left( \f{1}{3} - \f{\mu}{5} \right).
\end{equation}
Since \(Z\)
is the metric function as it appears in the line element, it cannot
change sign, so that \(Z(r_{b}) > 0.\)
This allows us to use the second equation above to conclude that
\begin{equation}
  \label{an.eq:rho_cFirst}
  \rho_{c} < \f{15}{(5-3\mu)\kappa r_{b}^{2}},
\end{equation}
a relation that will be needed later.

To check our first condition~\ref{an.it.RegularMetric} we have to
ensure that the metric function \(Y\)
is regular in the interior of the solution.  The expression for \(Y\)
is given in equation~\eqref{ns.eq:PhiZY} as 
\[ Y(r) = \gamma + \f{2\alpha r_{b}}{\sqrt{\kappa \rho_{c}\mu
      /5}}\left[ \acoth{\left(\f{1-\sqrt{Z(r)}}{r^{2}\sqrt{\f{\kappa
              \rho_{c} \mu} {5r_{b}^{2}}}}\right) } -\acoth{\left(
        \f{1-\gamma}{r_{b}\sqrt{\kappa \rho_{c} \mu /5}}\right)}
  \right]. \] Evaluating this metric function at \(r=r_{b}\)
annihilates the square bracket, which makes \(\gamma,\) a finite
number as the answer for \(Y(r_{b}).\) For the value of \(Y(0),\) we
have to take a formal limit of the function in the first \(\acoth,\)
since substitution evaluation results in an undefined \(0/0.\) Using
l'H\^opital's rule on this function results in
\[\lim_{r \to 0} \left[ \acoth{\left(\f{1-\sqrt{Z(r)}}{r^{2}\sqrt{\f{\kappa \rho_{c}
        \mu} {5r_{b}^{2}}}}\right)} \right] = \lim_{r \to 0} \left[ \acoth{\left( \f{-\f{1}{2} Z^{-1/2}}{2r\sqrt{\f{\kappa \rho_{c}
        \mu} {5r_{b}^{2}}} }\right)} \right]  = \acoth(\infty) = 0,
\]
the second limit depending on \(Z(0) = 1.\)
As a result of this \(Y(0)\) reduces to the finite value of
\begin{equation}
  \label{an.eq.Y0}
Y(0) = \gamma - \f{2\alpha r_{b}}{\sqrt{\kappa \rho_{c}\mu
    /5}}\left[ \acoth{\left( \f{1-\gamma}{r_{b}\sqrt{\kappa \rho_{c}
          \mu /5}}\right)} \right],
\end{equation}
so that we are reasonably sure that since \(Y\)
is regular at both \(r=0\)
and \(r=r_{b},\)
the extreme values of \(r,\)
it remains so for every value of \( 0 \leq r \leq r_{b},\)
proving that condition~\ref{an.it.RegularMetric} is satisfied by this
solution.

\begin{figure}[!htb]
\subfloat[The $Y(r)$ metric function]{\label{an.fig:PhiZ,Ymetric}
  \includegraphics[width=0.5\linewidth]{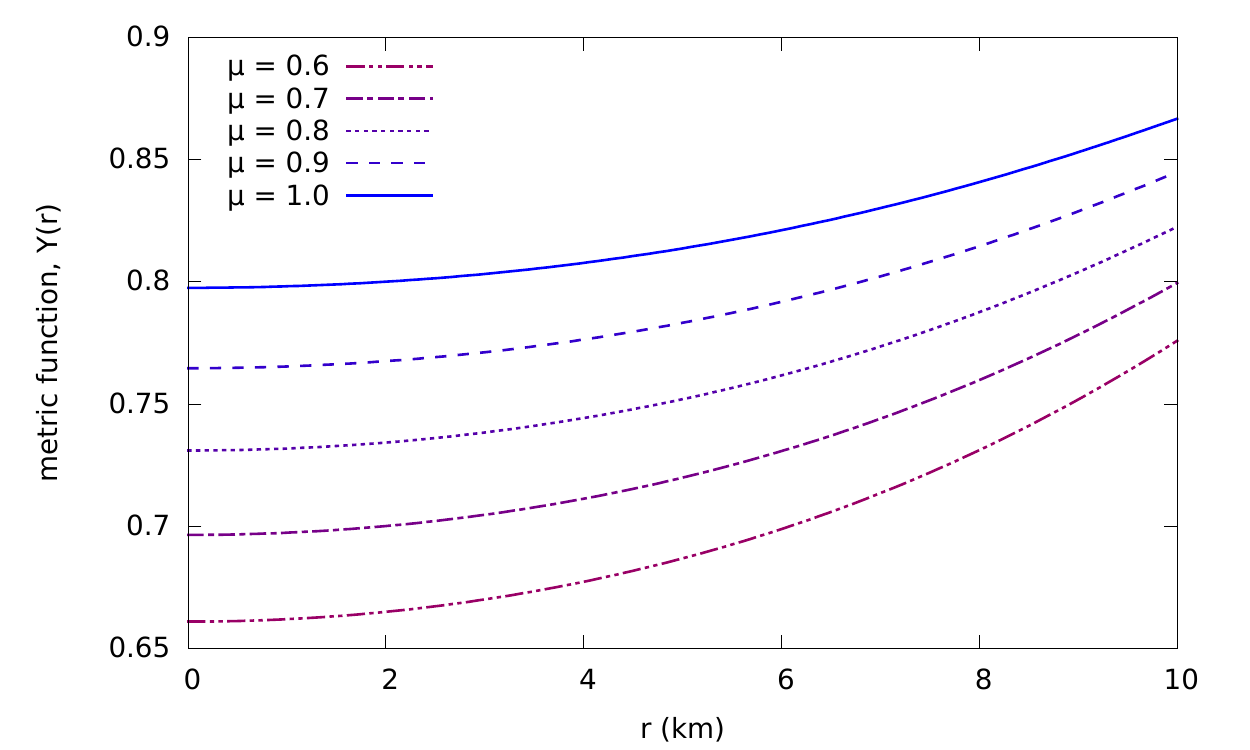}} 
\subfloat[The $Z(r)$ metric function]{\label{an.fig:PhiZ,Zmetric}
  \includegraphics[width=0.5\linewidth]{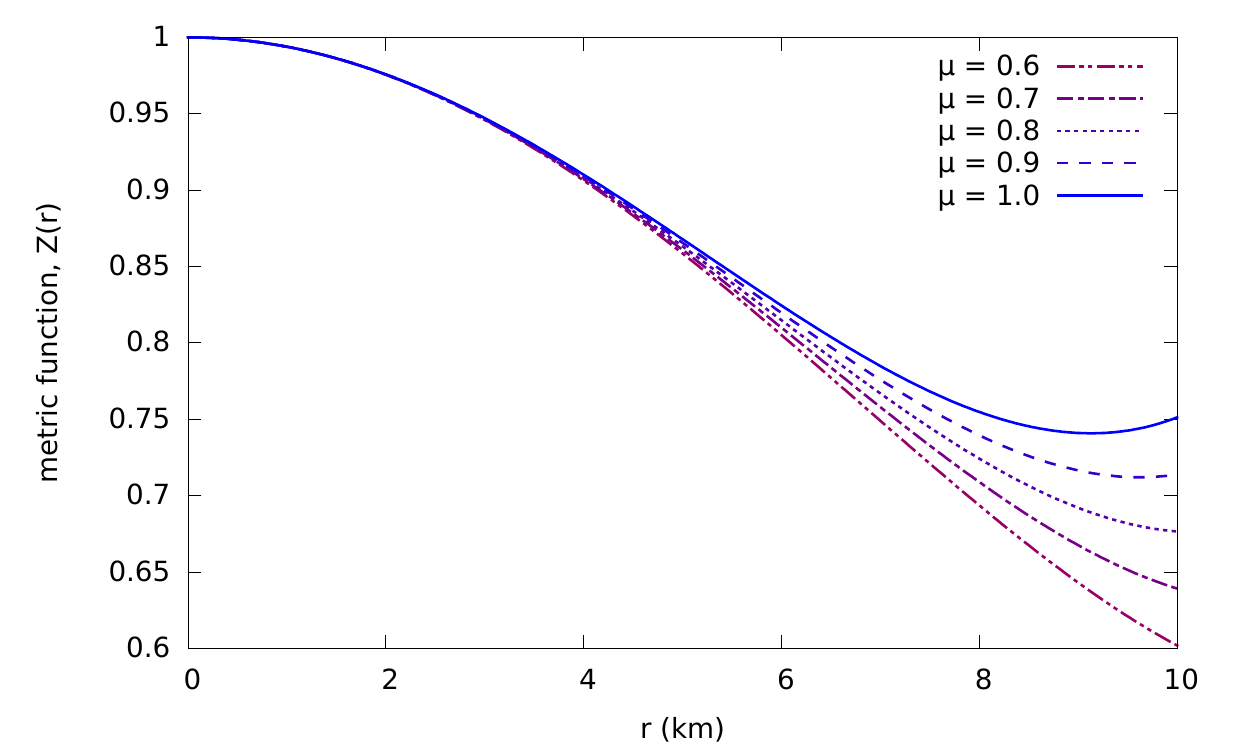}}\\
\subfloat[The $\lambda(r)$ metric function]{\label{an.fig:PhiZ,LambdaMetric} 
  \includegraphics[width=0.5\linewidth]{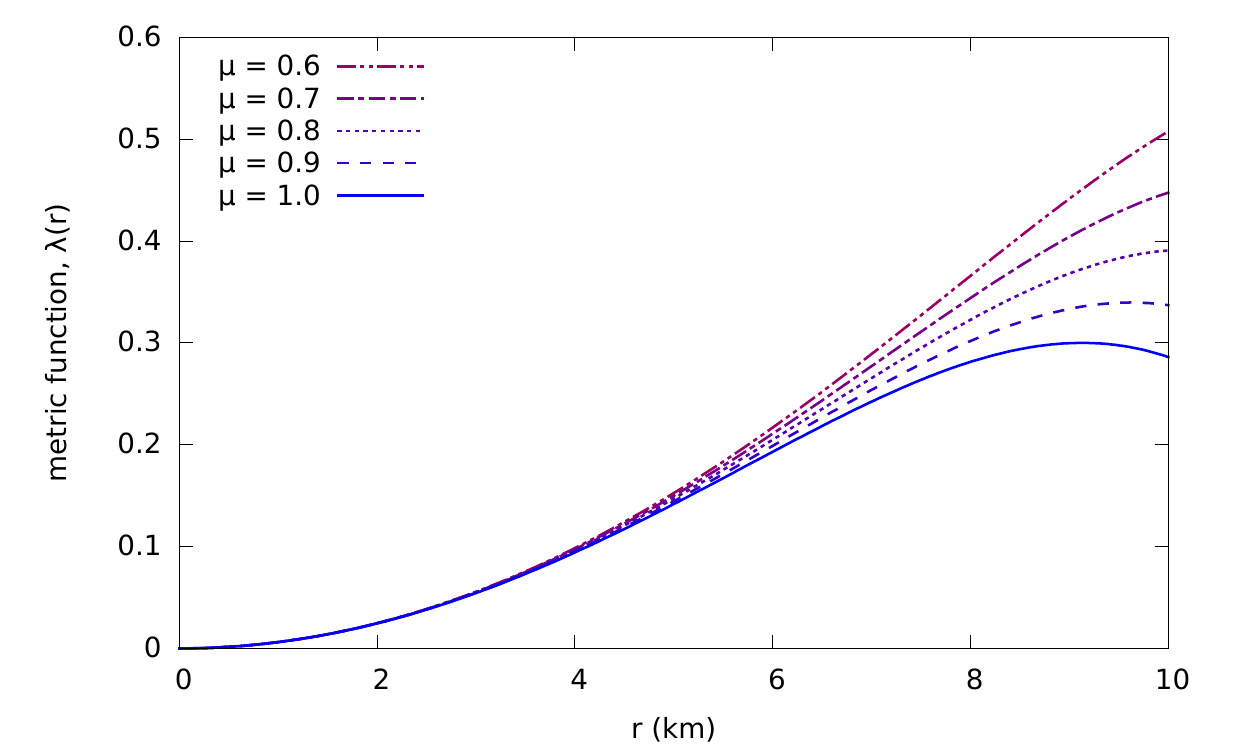}}
\subfloat[The $\nu(r)$ metric function]{\label{an.fig:PhiZ,NuMetric}
  \includegraphics[width=0.5\linewidth]{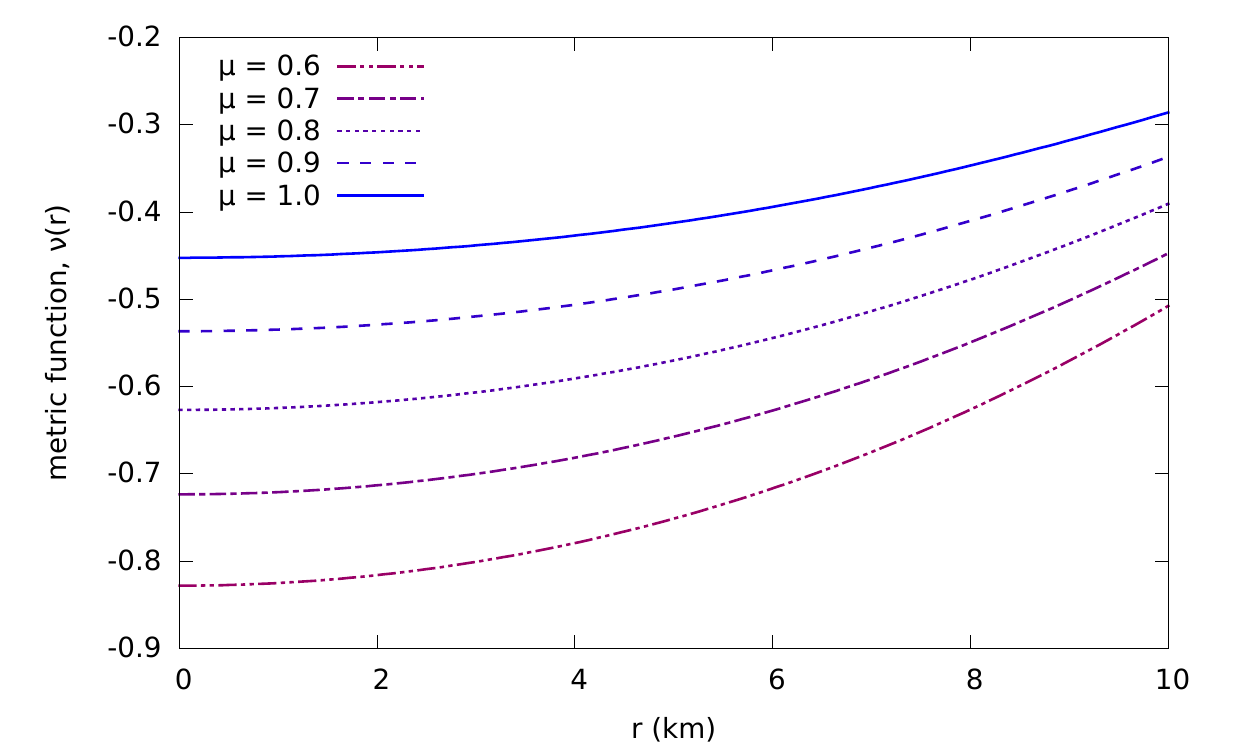}} 
\caption[The metric variables in their two equivalent
formulation]{Variation of metric variables $Y(r),$ and $Z(r)$
  in~\citeauthor{Iva02}'s formulation and $\lambda(r)$ and $\nu(r)$ in
  the usual formulation, with the radial coordinate inside the
  star. The parameter values are
  $\rho_{c}=\un{1\times 10^{18}} \,\dunits, r_b = \un{1 \times 10^{4}
    \,m}$ and $\mu$ taking the various values shown in the legend }
\label{an.fig:PhiZ,MetricCoeff}
\end{figure}

We now show a plot of these metric functions in their two different
forms for specific parameter values in
figure~\ref{an.fig:PhiZ,MetricCoeff}.  Turning to
conditions~\ref{an.it.ExteriorMetric}--~\ref{an.it.ExteriorObservables},
which reduce to \(2\sqrt{a} > b,\)
we have for this particular case that
\begin{equation}
\label{an.eq:rho_cSecond}
  2\sqrt{\left(\f{\kappa\mu\rho_{c}}{5 r_{b}^{2}}\right)} >
\f{\kappa\rho_{c}}{3} \implies \rho_{c} < \f{36 \mu}{5 \kappa
  r_{b}^{2}},
\end{equation}
a limit on the maximum value of the central density for a given type
of star (spanning from natural with \(\mu = 1\)
to various other ones with different ``self-boundness'') at a given
radius \(r_{b}.\)
The only additional assumption that went into this relation is the
positive values of all parameters in the above equation.  We keep this
in mind to restrict our parameter space later.

The next condition~\ref{an.it.DefiniteRPressure} concerning the
positive definiteness of the pressure is harder to implement.  To test
this, we first write down the pressure in terms of known variables as
\begin{multline}
\label{an.eq:pr}
  \begin{aligned} p_{r}(r) = &\f{2\kappa\rho_{c}}{3} - \f{4\kappa\rho_{c}\mu r^{2}}{5r_{b}^{2}} -\kappa\rho_{c} \left[ 1 - \mu \left( \f{r}{r_{b}}\right)^{2}\right] + \\
&+ \left( \f{\kappa\rho_{c}}{3} - \f{\kappa \rho_{c} \mu}{5}\right) \f{\sqrt{1-\f{\kappa\rho_{c}}{3}r^{2} + \f{\kappa\mu\rho_{c}}{5r_{b}^{2}} r^{4}}}{\gamma + \f{2\alpha r_{b}}{\sqrt{\kappa \rho_{c}\mu /5}}\left[ 
  \acoth{\left(\f{1-\sqrt{Z(r)}}{r^{2}\sqrt{\f{\kappa \rho_{c} \mu} {5r_{b}^{2}}}}\right) }  -\acoth{\left( \f{1-\gamma}{r_{b}\sqrt{\kappa \rho_{c} \mu /5}}\right)}  \right]} \end{aligned}
\end{multline}

First we remember that the pressure \(p_{r}\) is zero at the boundary,
and the expression we have should give this same result.  We can
consider this a consistency check on our arithmetic, and indeed, since
\(\gamma = \sqrt{Z(r_{b})}, \) and the vanishing of the square
brackets in the denominator of~\eqref{an.eq:pr}, we are left with the
simple
\[p_{r}(r_{b}) = \f{2\kappa\rho_{c}}{3} - \f{4\kappa\rho_{c}\mu }{5}
-\kappa\rho_{c} \left[ 1 - \mu \right] + \left( \f{\kappa\rho_{c}}{3}
  - \f{\kappa \rho_{c} \mu}{5}\right) = 0, \]
confirming that at least the expression of the pressure is consistent.

Evaluating this expression at \(r=0,\)
results in having to evaluate \(Y(0),\)
which we already found before in equation~\eqref{an.eq.Y0}, and
\(Z(0)\) which is just one.  With these, we get that
\[p_{r}(0) = \kappa \rho_{c} \left( \f{5-3\mu}{ 15 \left\{ \gamma - \f{2\alpha r_{b}}{\sqrt{\kappa \rho_{c}\mu/5}}\left[ \acoth{\left( \f{1-\gamma}{r_{b}\sqrt{\kappa \rho_{c}
          \mu /5}}\right)} \right] \right\} } - \f{1}{3} \right),
\]
an expression that is complicated enough that we cannot immediately
infer its behaviour.  We will see this happen many times in the course
of this chapter, and as a result we will depend heavily on graphing
these functions to determine their behaviour.  Since our parameter
space is four dimensional in the uncharged chase, we will be forced to
assume specific values for certain parameters, and we will of course
use the constraints we already have to pick these values consistently.

Since we are interested in modelling neutron stars, whose typical
radii \(r_b\) is of the order of tens of kilometres, and that have
typical central densities of the order of nuclear densities, we use
these as baseline values and compute the pressures for different
values of \(\mu.\) We check first that our previous inequalities are
satisfied: inequality~\eqref{an.eq:rho_cFirst} gives that
\(\rho_{c} < 1.6 \times 10^{18} \dunits\) when \(\mu=0,\) and
\(\rho_{c} < 4.0 \times 10^{18} \dunits\) when \(\mu=1.\) The second
inequality~\eqref{an.eq:rho_cSecond} instead gives, \(\rho_{c} < 0\)
when \(\mu=0,\) scaling linearly with it, and
\(\rho_{c} < 3.9 \times 10^{18} \dunits\) when \(\mu=1.\) Thus taking
the more restrictive inequality to dictate the value of \(\rho_{c},\)
we pick \(\rho_{c} = 1 \times 10^{18} \dunits\) and
\(r_{b} = 10 \un{km},\) as they satisfy both inequalities even up to
\(\mu=0.6,\) and plot the radial pressure for these choices in
figure~\ref{ns.fig:phiZ,Pr}

\begin{figure}[!htb]
  \centering
  \includegraphics[width=\linewidth]{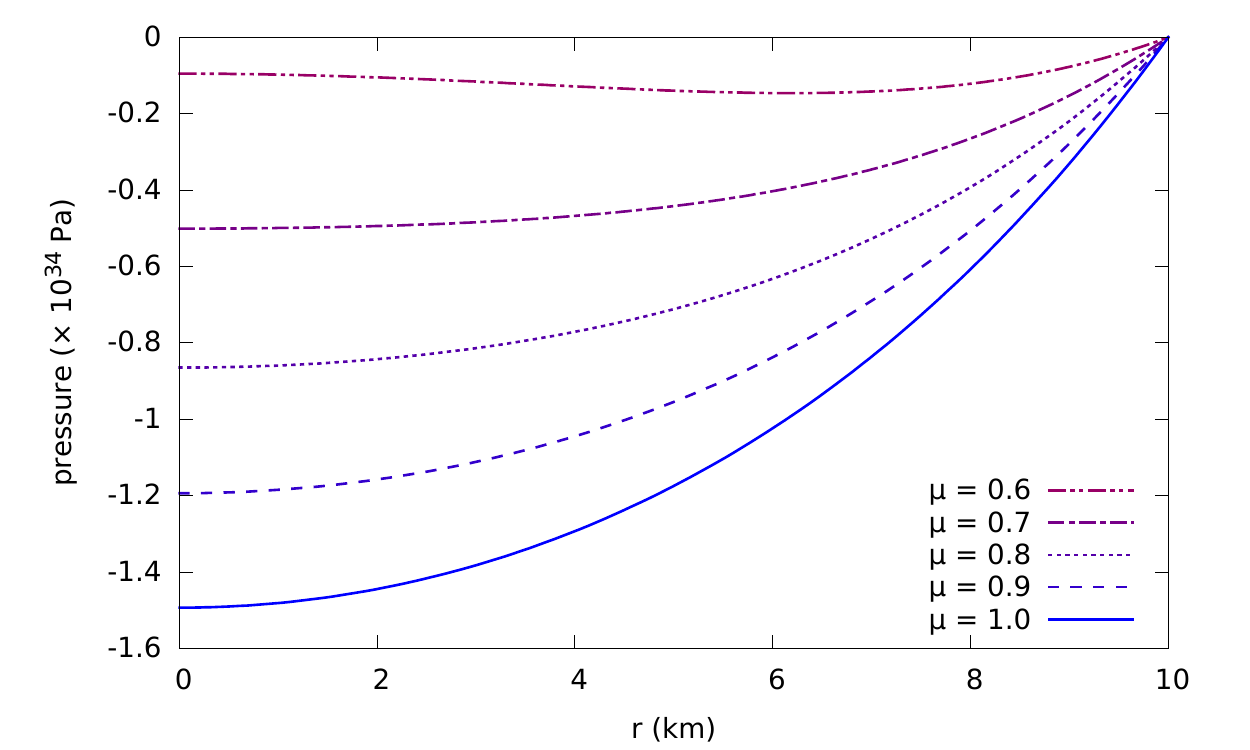}
  \caption[The radial pressure for $\phi =0$]{ The radial pressure in the
    interior of the star for this solution. The parameter values are
    $\rho_{c}=1 \times 10^{18} \,\dunits , r_b = 1 \times 10^{4} \un{m}$
    and $\mu$ varies between $1$ and $0.6$}
\label{ns.fig:phiZ,Pr}
\end{figure}

It is immediately clear that the pressures are negative, something no
normal fluid should have, even though every precaution in choosing
values for our parameters was taken.  However this is a well known
issue with interior solutions: more often than not, and very clearly
here, the solutions found behave non-physically.  Declaring this solution unphysical, 
having failed criterion~\ref{an.it.DefiniteRPressure}, no further analysis of this solution will be carried out.

\FloatBarrier\subsection{The $\phi^{2}<0$ case from~\ref{ns.ssec:phiNeg}}\label{an.ssec:phiNeg}
In this case, \(\beta\)
is no longer fixed to one value as previously: instead it takes on a
range of possible values and as long as the inequality
\(\beta < -\f{\kappa\mu\rho_{c}}{5r_{b}^{2}}\)
is satisfied, the value of \(\phi\)
will be appropriate for this solution.  We first consider the metric
functions in their two forms for different value of \(\mu\)
through plots in figure~\ref{an.fig:PhiN,MetricCoeff}.

\begin{figure}[!h]
\subfloat[The $Y(r)$ metric function]{\label{an.fig:phiN,Ymetric}
  \includegraphics[width=0.5\linewidth]{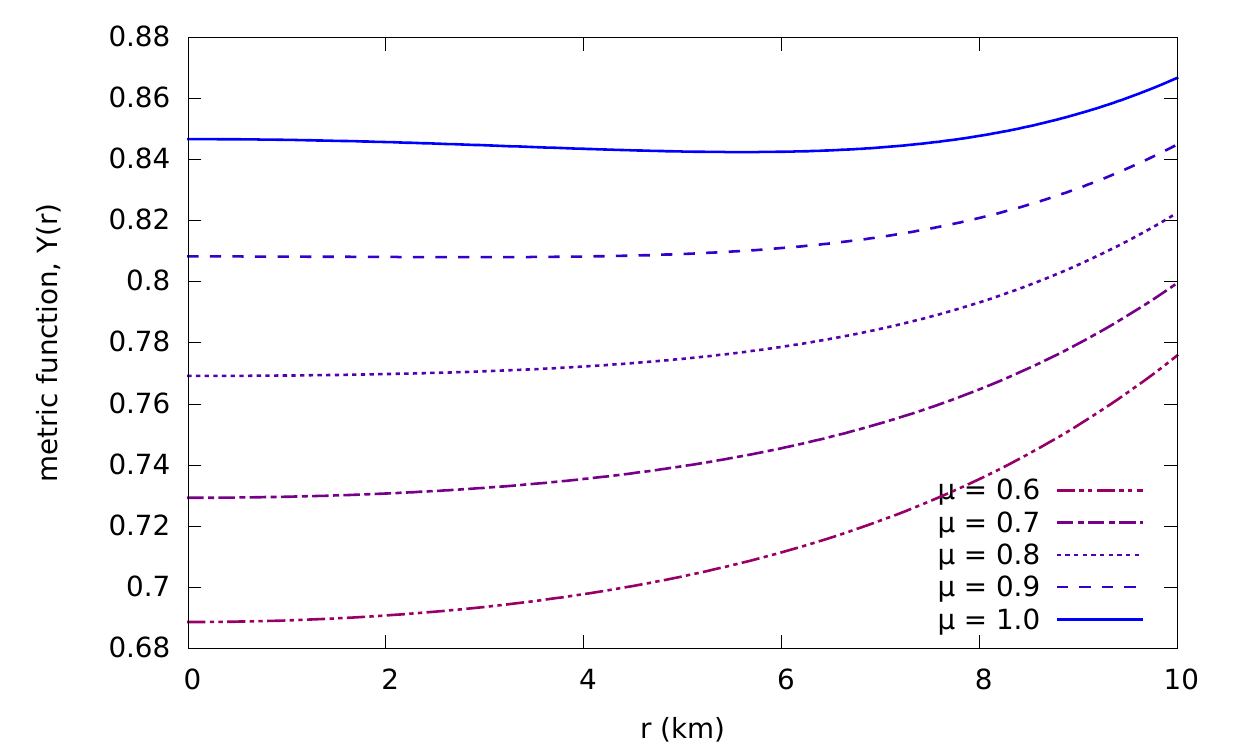}} 
\subfloat[The $Z(r)$ metric function]{\label{an.fig:phiN,Zmetric}
  \includegraphics[width=0.5\linewidth]{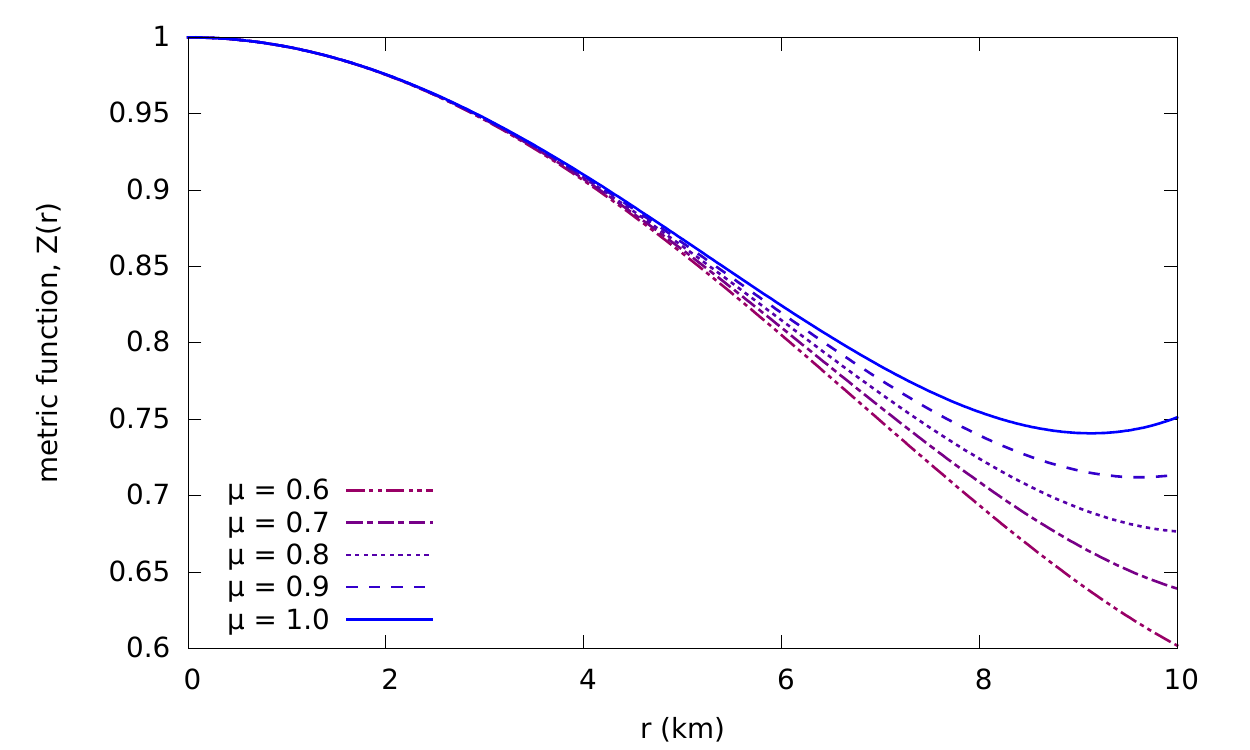}}\\
\subfloat[The $\lambda(r)$ metric function]{\label{an.fig:phiN,LambdaMetric} 
  \includegraphics[width=0.5\linewidth]{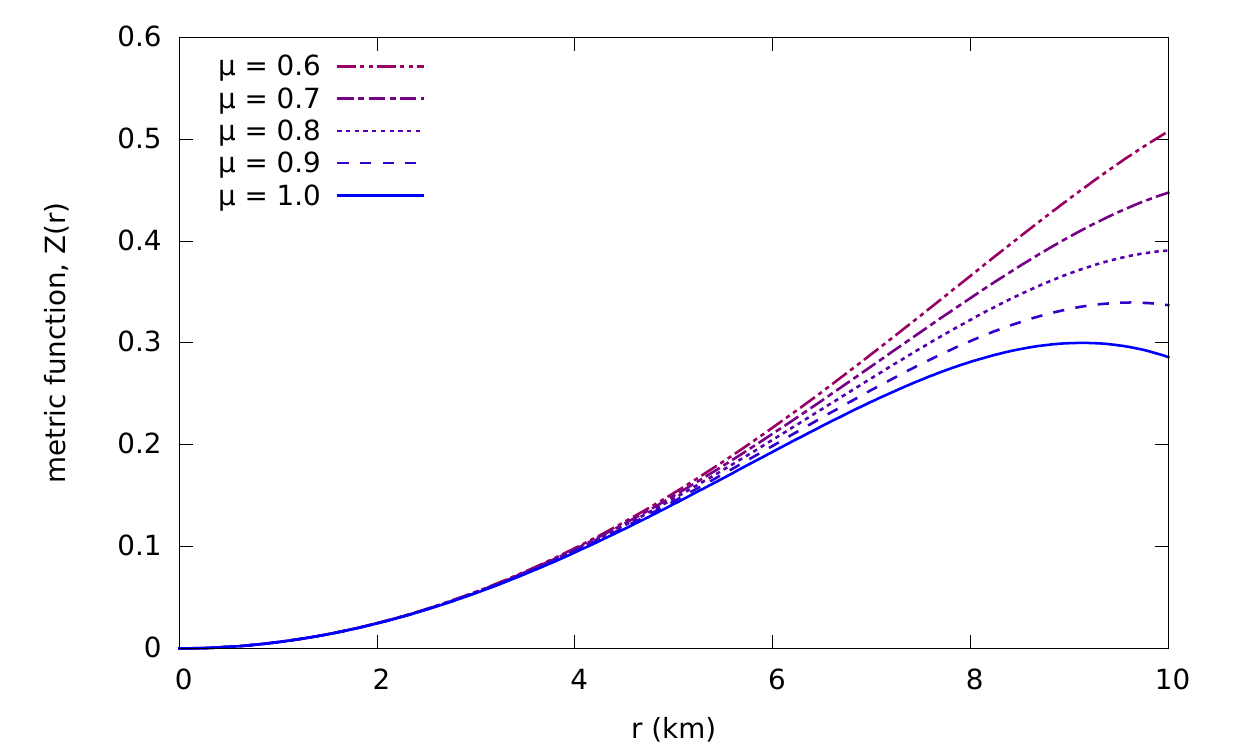}}
\subfloat[The $\nu(r)$ metric function]{\label{an.fig:phiN,NuMetric}
  \includegraphics[width=0.5\linewidth]{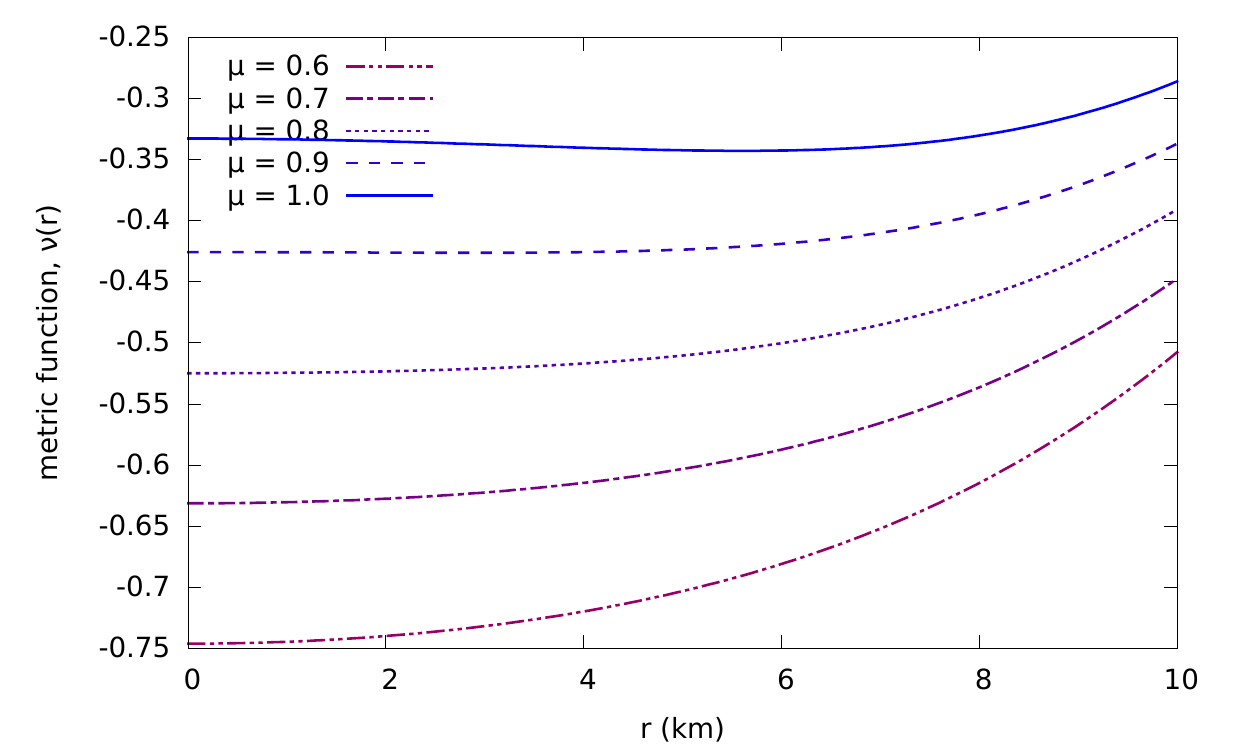}} 
\caption[The metric variables in their two equivalent formulations]
{Variation of metric variables $Y(r),$ and $Z(r)$ in
  \citeauthor{Iva02}'s formulation, and $\lambda(r),$ with $\nu(r)$ in
  the usual formulation, with the radial coordinate inside the
  star. The parameter values are
  $\rho_{c}=\un{1\times 10^{18}}\,\dunits, r_b = \un{1 \times 10^{4}
    \,m}, \mu$
  taking the various values shown in the legend, and $\beta$ being set
  to $-2a$ }
\label{an.fig:PhiN,MetricCoeff}
\end{figure}

Considering the form of \(Z,\) it is clear that changing the value of
\(\beta\) will not affect it.  The metric function \(Y\) however is a
different story and we show in the next
figure~\ref{an.fig:PhiN,MetricCoeffVaryingbeta} how it changes for
different values of the anisotropy factor.  We forgo a direct analysis
of the limiting behaviour of the algebraic expressions (which can be
found in Appendix~\ref{C:AppendixB}) since the latter are quite
complicated, consisting of products of hyperbolic functions.

As it stands we see that
both metric functions are well-behaved in the interior for a range of
parameter values, so that we can be reasonably sure that
condition~\ref{an.it.RegularMetric} is satisfied.

\begin{figure}[!h]
\subfloat[The $Y(r)$ metric function, $\mu=1$]{\label{an.fig:phiN,Ymetric}
  \includegraphics[width=0.5\linewidth]{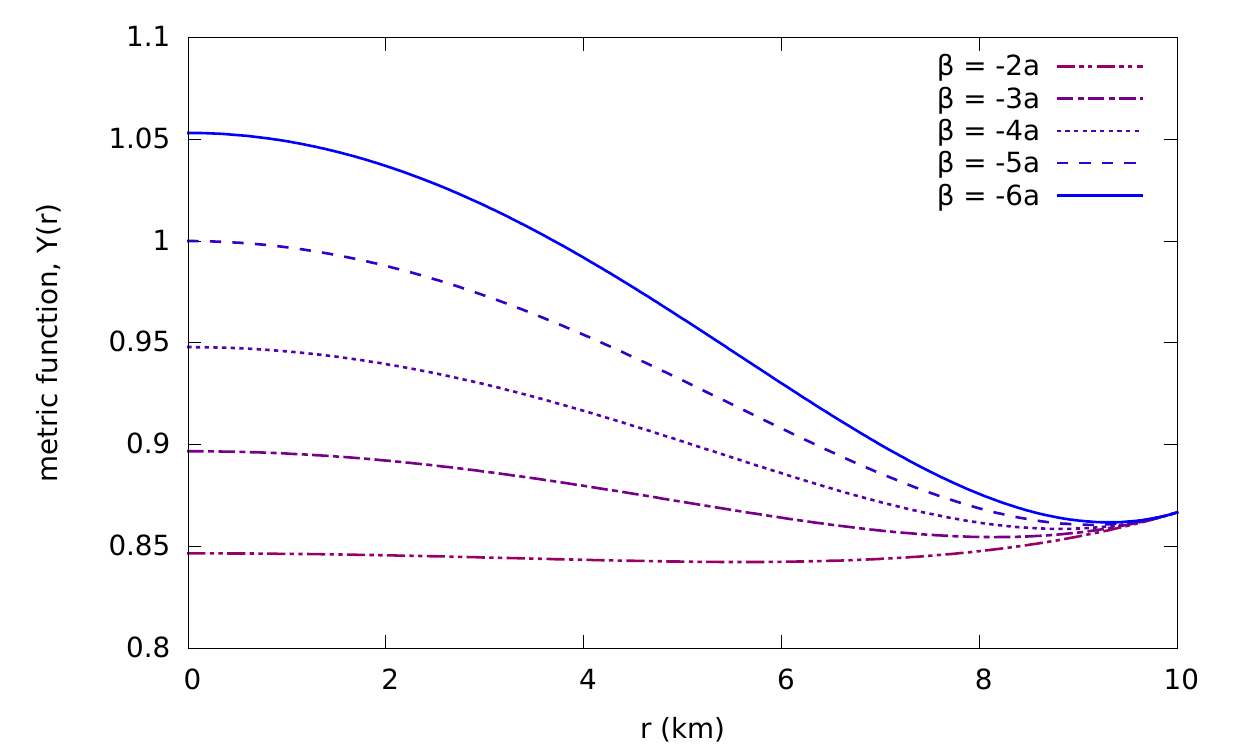}} 
\subfloat[The $Y(r)$ metric function, $\mu=0.6$]{\label{an.fig:phiN,Zmetric}
  \includegraphics[width=0.5\linewidth]{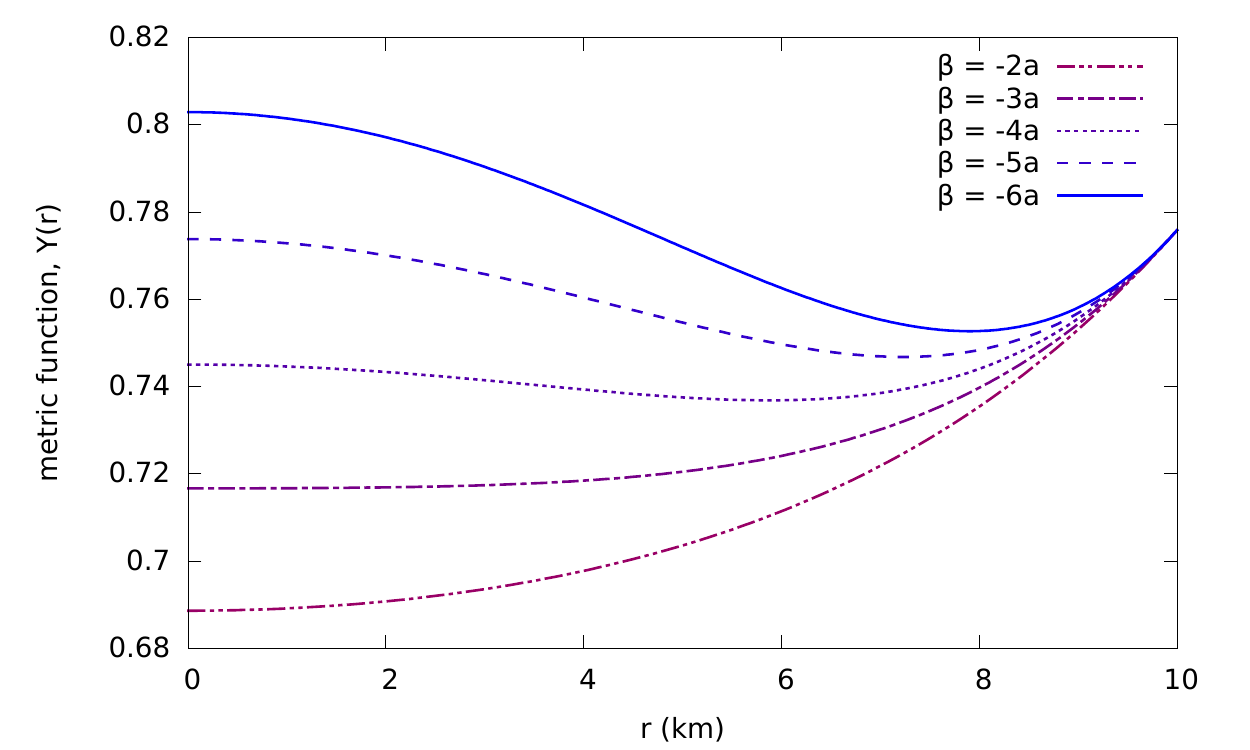}}
\caption[Variation of $Y$ metric variable ]{Variation of the $Y$ metric variables
  with the radial coordinate inside the star. The parameter values are
  $\rho_{c}=\un{1\times 10^{18}} \dunits, r_b = \un{1 \times 10^{4}
    m},\, \mu$
  is set to 1 on the left, and 0.6 on the right, and $\beta$ is set
  to the various values shown in the legend }
\label{an.fig:PhiN,MetricCoeffVaryingbeta}
\end{figure}

The discussion at the beginning of the previous section is still
valid, and in particular the inequalities~\eqref{an.eq:rho_cFirst}
and~\eqref{an.eq:rho_cSecond} must still hold in view
of~\ref{an.it.ExteriorMetric}--~\ref{an.it.ExteriorObservables}, which
yield the exact same results as the \(\phi = 0\) case.  To ensure that
both these conditions remain satisfied, we next look at the pressure
in view of condition~\ref{an.it.DefiniteRPressure}.  Here too the
algebraic expression is very complicated, and instead we provide
graphs of the behaviour for varying parameters \(\mu\) in
figure~\ref{an.fig:PhiN,radialPressureVaryingMu} and \(\beta\) in
figure~\ref{an.fig:PhiN,radialPressureVaryingBeta}.

\begin{figure}[!h]
\subfloat[The radial pressure , $\mu=1$]{\label{an.fig:phiN,PrMu1}
  \includegraphics[width=0.5\linewidth]{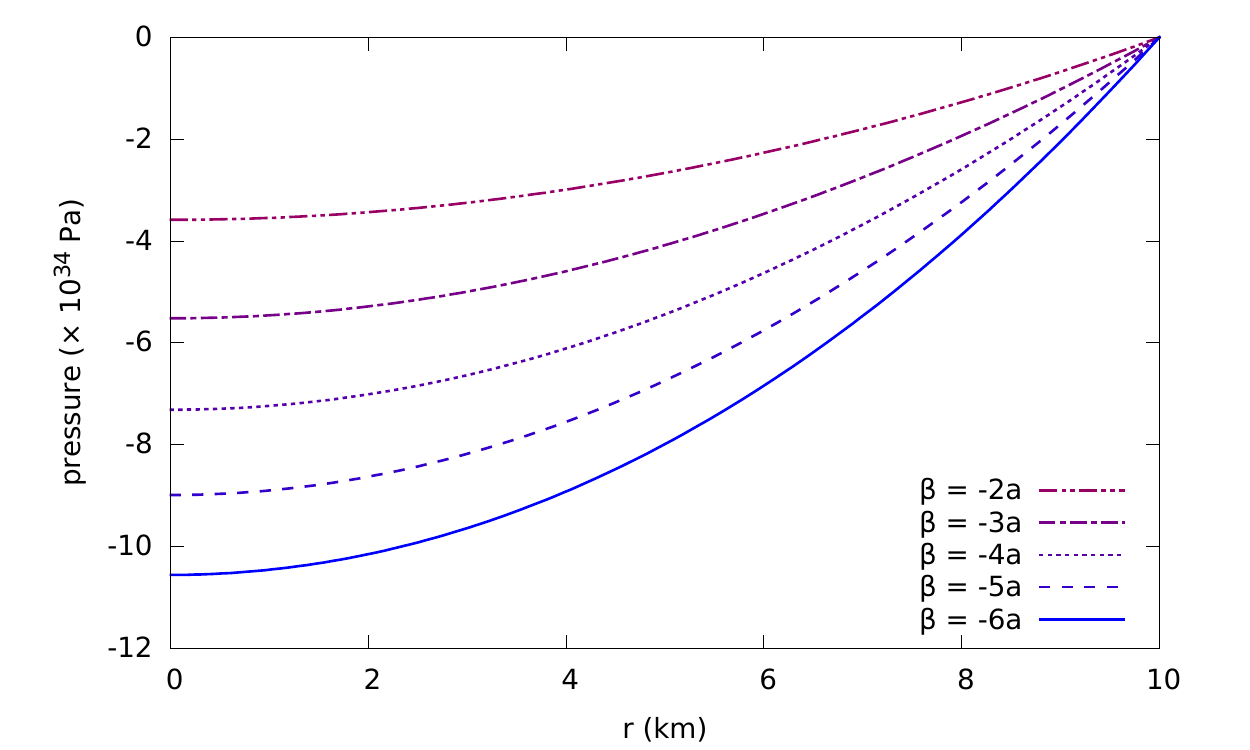}} 
\subfloat[The radial pressure, $\mu=0.6$]{\label{an.fig:phiN,PrMu6}
  \includegraphics[width=0.5\linewidth]{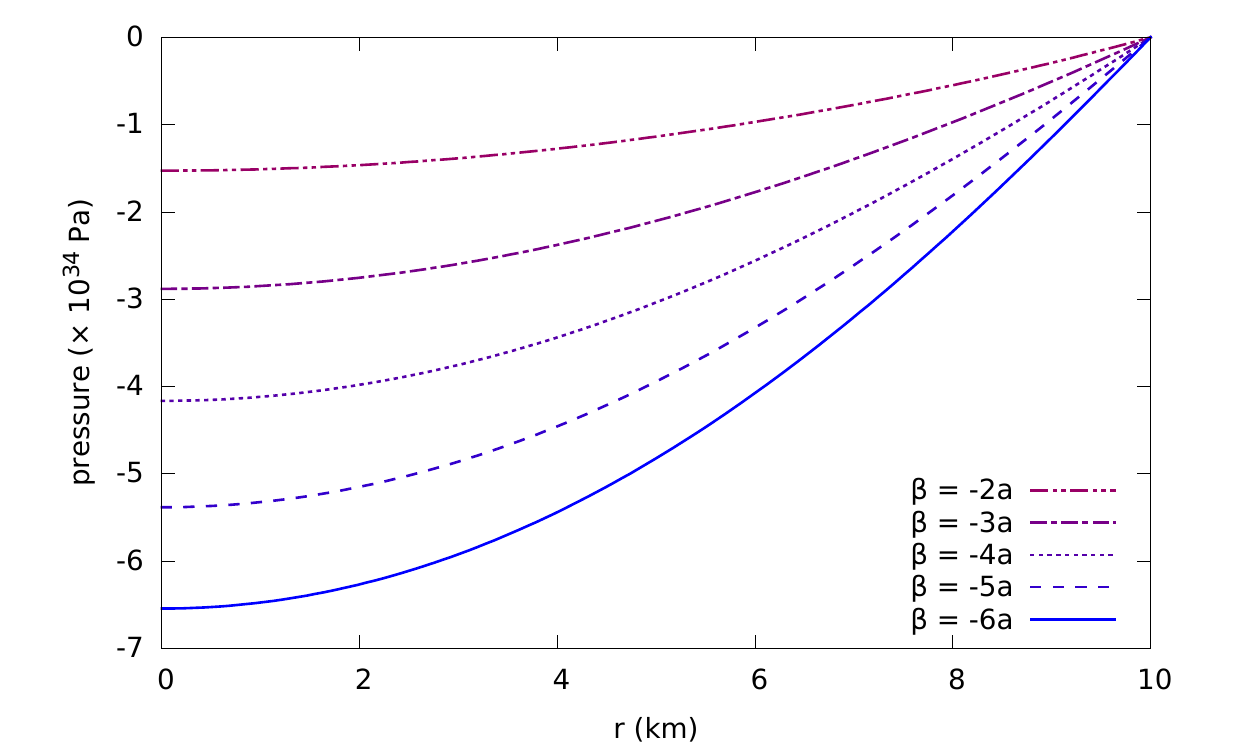}}
\caption[The radial pressure for $\phi^{2} < 0$]{Variation of the radial pressure
  with the radial coordinate inside the star. The parameter values are
  $\rho_{c}=\un{1\times 10^{18}} \dunits, r_b = \un{1 \times 10^{4}
    m},\, \mu$
  is set to 1 on the left, and 0.6 on the right, and $\beta$ is set
  to the various values shown in the legend }
\label{an.fig:PhiN,radialPressureVaryingBeta}
\end{figure}
It is clear that our boundary condition requiring that the pressure
vanish at the boundary is working, as is evident in all the pressure
graphs we are showing.  It should be noted that even though \(\beta\)
is extremely important in the tangential pressure \(\ppen\) component,
its contribution to the radial pressure \(p_{r}\) is not zero: an
unintuitive result stemming from the non-linearity of the EFE.
\begin{figure}[!htb]
\subfloat[The radial pressure , $\beta=-2a$]{\label{an.fig:phiN,PrBeta2a}
  \includegraphics[width=0.5\linewidth]{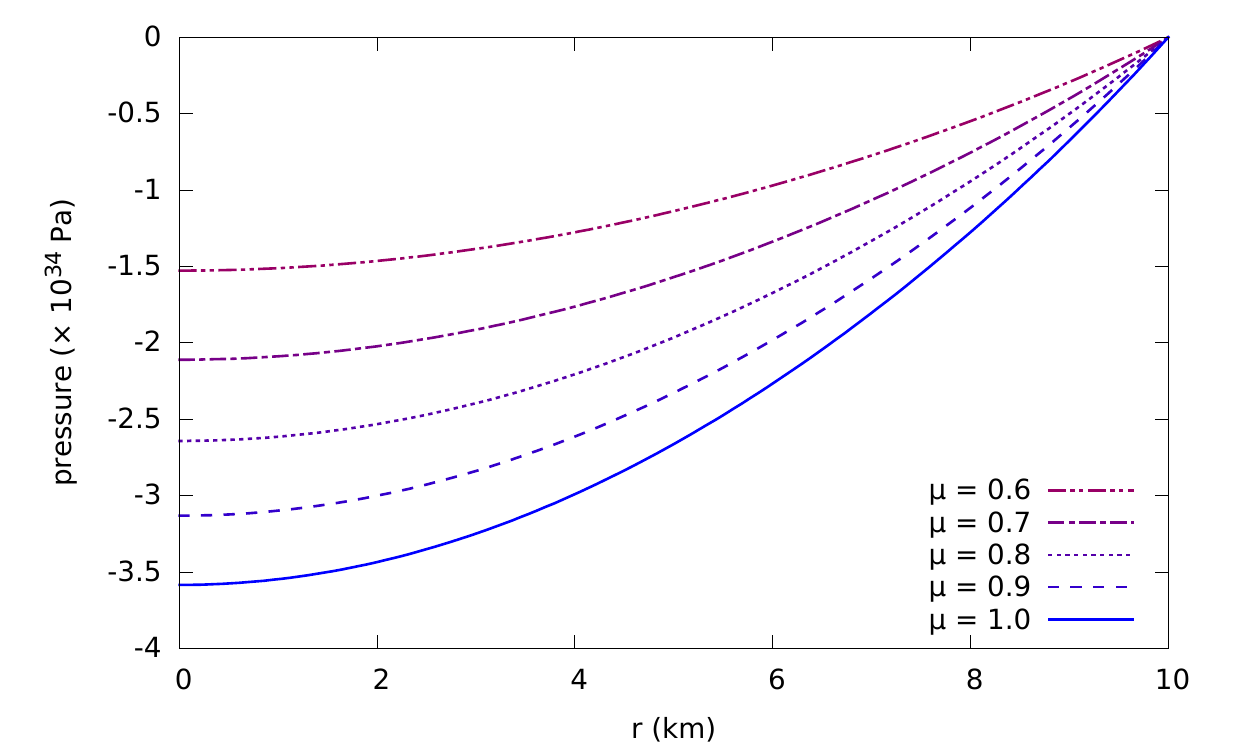}} 
\subfloat[The radial pressure, $\beta=-6a$]{\label{an.fig:phiN,PrBeta6a}
  \includegraphics[width=0.5\linewidth]{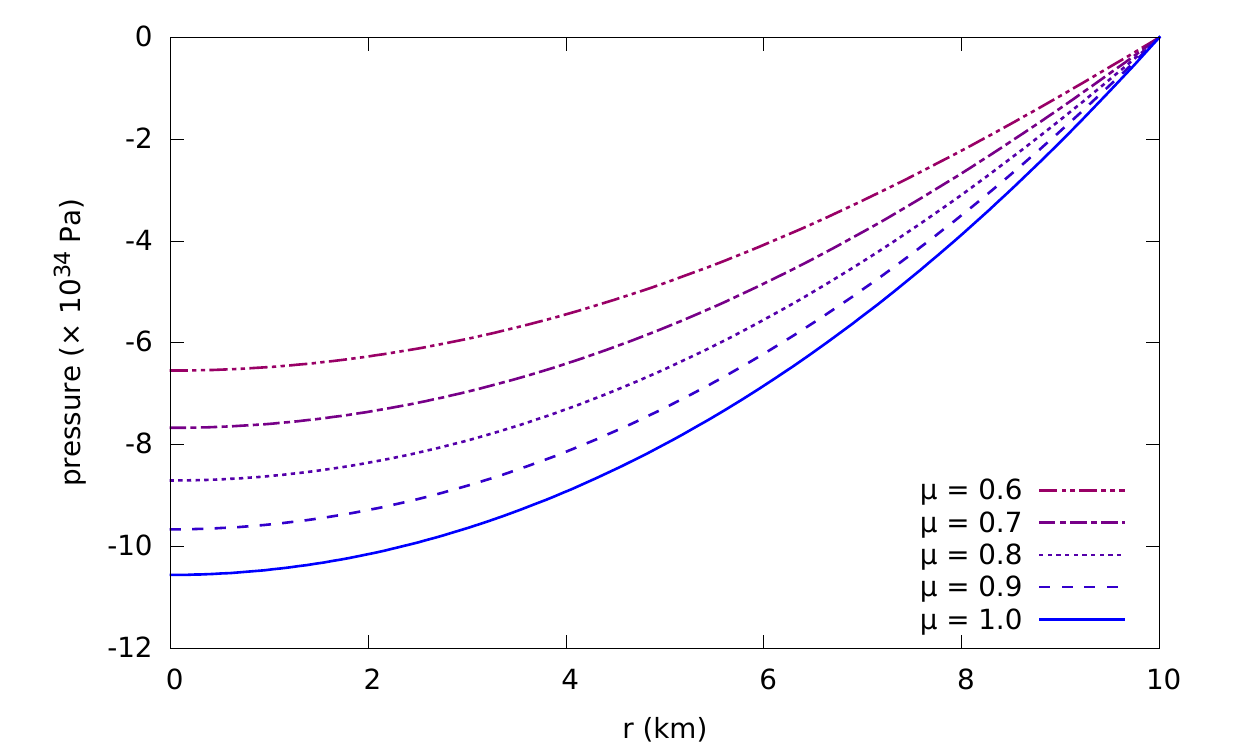}}
\caption[The radial pressure for $\phi^{2} < 0$]{Variation of the radial pressure
  with the radial coordinate inside the star. The parameter values are
  $\rho_{c}=\un{1\times 10^{18}} \dunits, r_b = \un{1 \times 10^{4}
    m},\, \beta$
  is set to $-2a$ on the left, and $-6a$ on the right, and $\mu$ is set
  to the various values shown in the legend }
\label{an.fig:PhiN,radialPressureVaryingMu}
\end{figure}
From these four figures it is immediately clear that for sensible
values of the parameters the radial pressures are all negative. This
result destroys the viability of this solution as a whole.  We
therefore look further for the other case of this solution.

\subsection{The $\phi^{2}>0$ case
  from~\ref{ns.ssec:phiPos}}\label{an.ssec:phiPos} In this case too,
\(\beta\) is no longer fixed to one value: instead it takes on a range
of possible values and as long as the inequality
\(\beta > -\f{\kappa\mu\rho_{c}}{5r_{b}^{2}}\) is satisfied, the value
of \(\phi\) will be appropriate for this solution.  Because of the
above inequality, this solution offers us the possibility of having
different signs for \(\beta.\) Since the latter could be negative
while still having a positive \(\phi^{2}.\) As a result, we need to
investigate the effect of the sign of \(\beta\) on our solutions too.

We first consider the metric
functions in their two forms for different value of \(\mu\)
through plots in figure~\ref{an.fig:PhiP,MetricCoeff}.
\begin{figure}[!htb]
\subfloat[The $Y(r)$ metric function]{\label{an.fig:phiP,Ymetric}
  \includegraphics[width=0.5\linewidth]{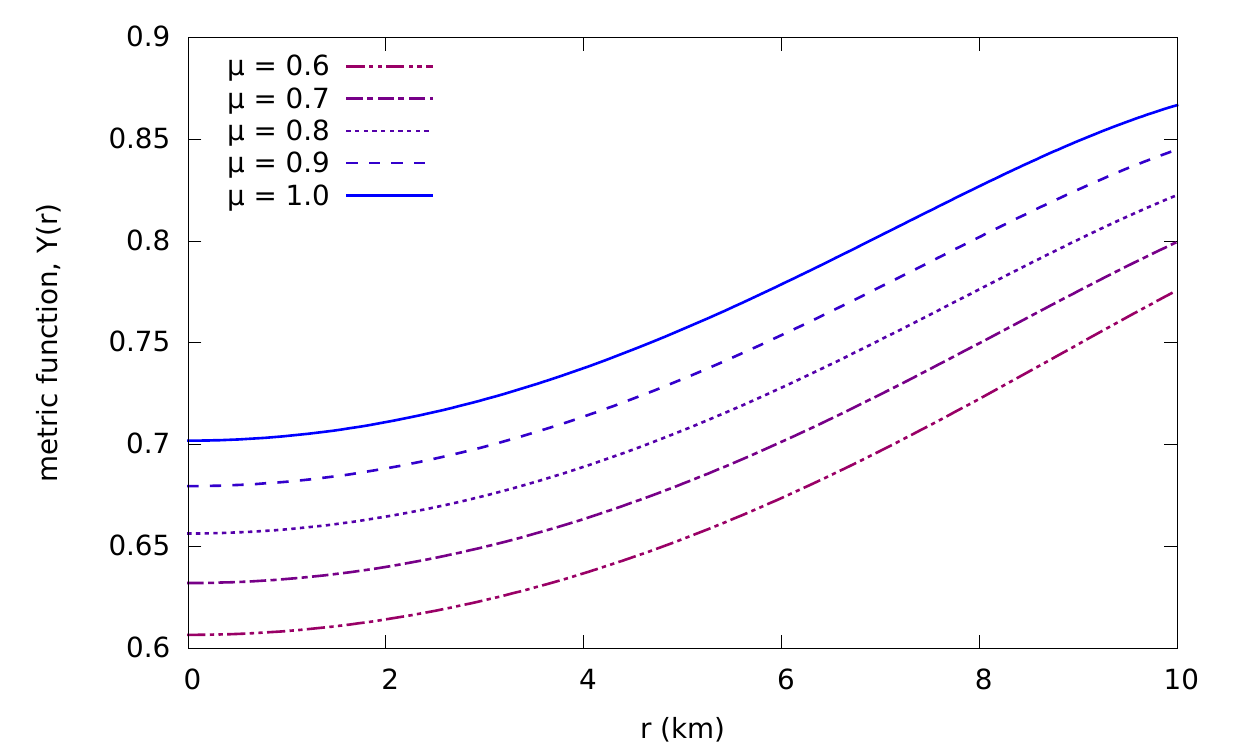}} 
\subfloat[The $Z(r)$ metric function]{\label{an.fig:phiP,Zmetric}
  \includegraphics[width=0.5\linewidth]{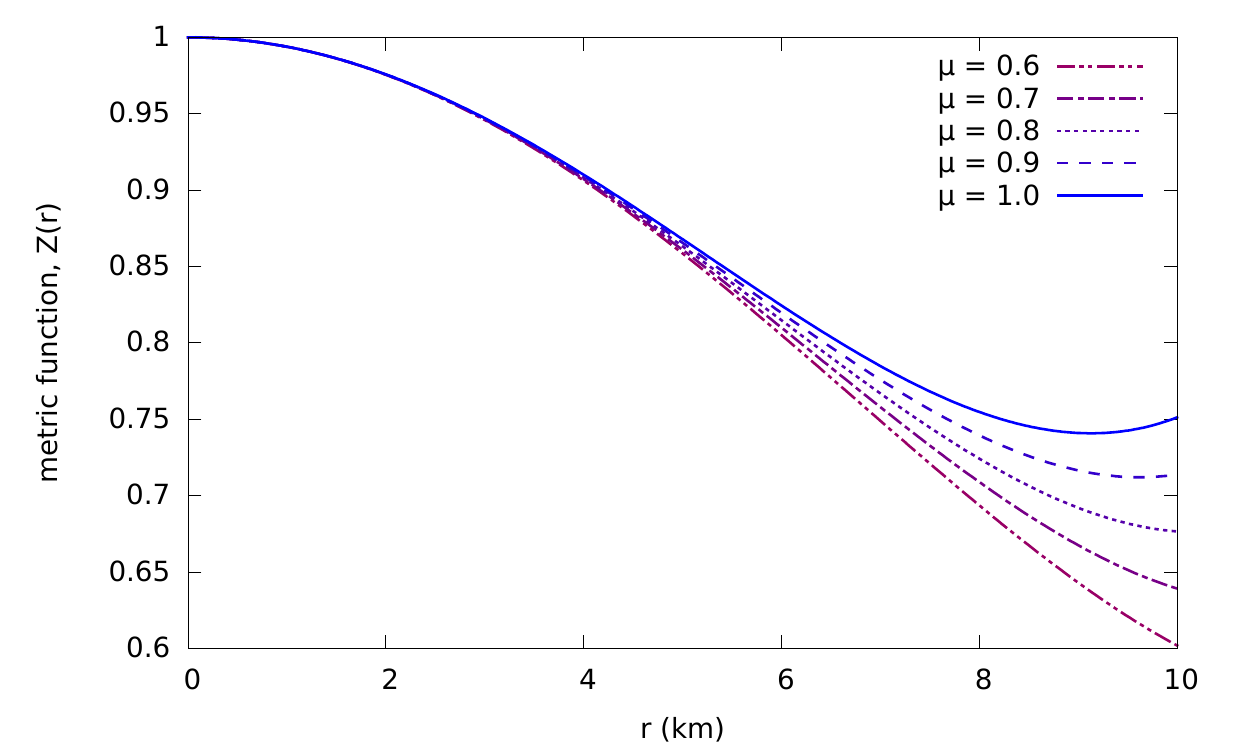}}\\
\subfloat[The $\lambda(r)$ metric function]{\label{an.fig:phiP,LambdaMetric} 
  \includegraphics[width=0.5\linewidth]{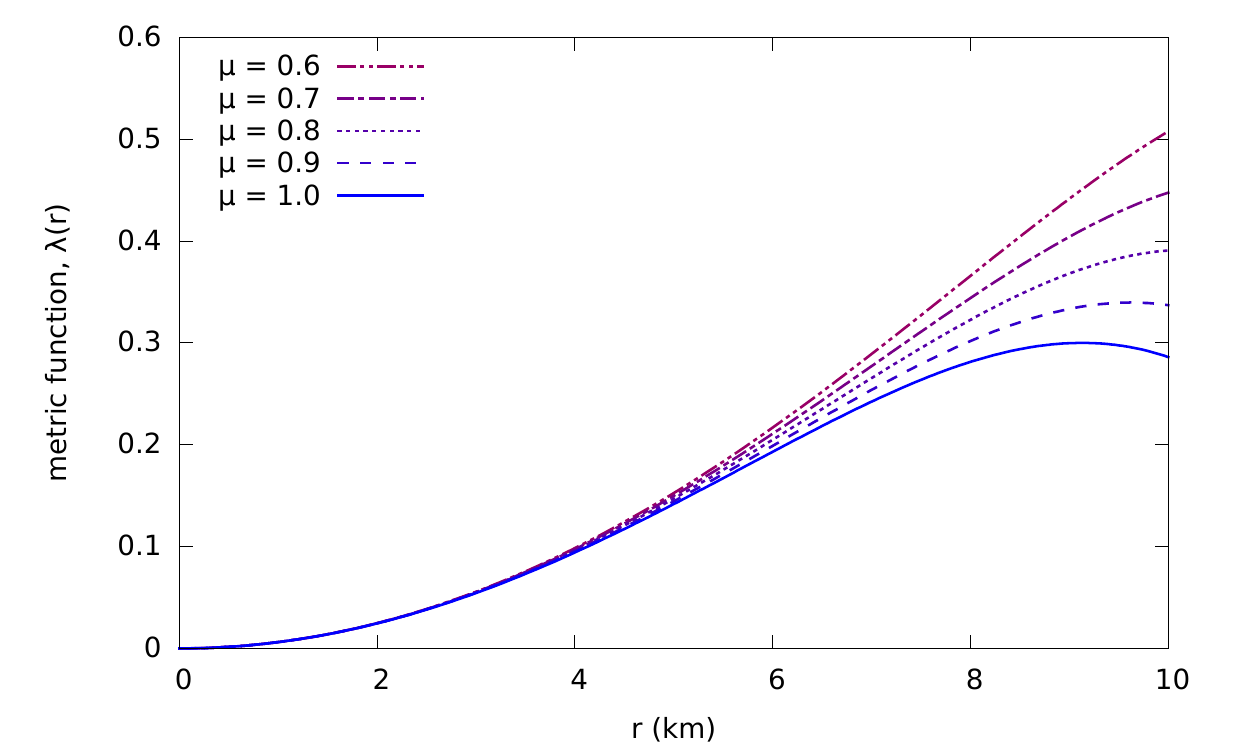}}
\subfloat[The $\nu(r)$ metric function]{\label{an.fig:phiP,NuMetric}
  \includegraphics[width=0.5\linewidth]{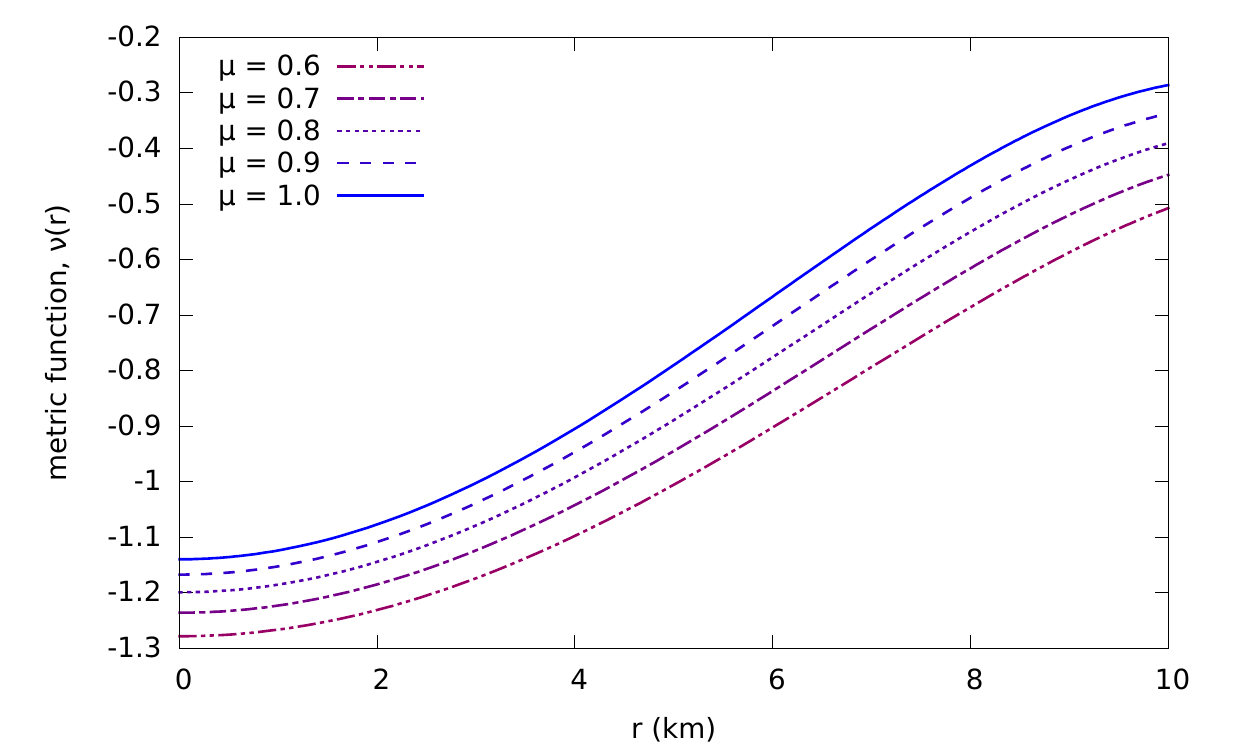}} 
\caption[Variation of metric variables for $\phi^{2} > 0$]{Variation of metric variables with the radial coordinate
  inside the star. The parameter values are $\rho_{c}=\un{1\times
    10^{18}} \dunits, r_b = \un{1 \times 10^{4} m}, \mu$
  taking the various values shown in the legend, and  $\beta$ being set to $-2a$ }
\label{an.fig:PhiP,MetricCoeff}
\end{figure}
These are all well behaved for various \(\mu\)s, but in this case also
we need to check if a similar behaviour holds for various anisotropy
factor \(\beta,\) and indeed we see that this is so in
figure~\ref{an.fig:PhiP,MetricCoeffVaryingbeta}.  Since both metric
functions are continuous in the regions we want, we conclude that
condition~\ref{an.it.RegularMetric} is satisfied for this solution.
\begin{figure}[!htb]
\subfloat[The $Y(r)$ metric function, $\mu=1$]{\label{an.fig:phiP,Ymetric}
  \includegraphics[width=0.5\linewidth]{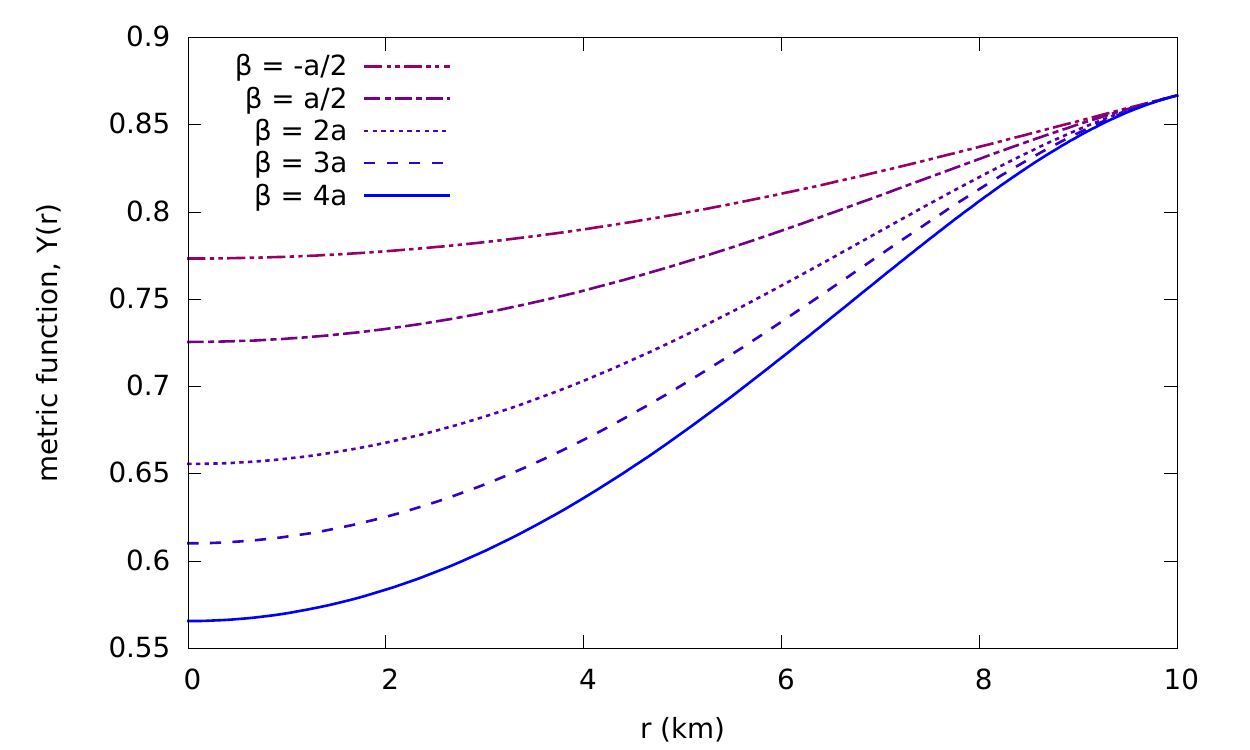}} 
\subfloat[The $Y(r)$ metric function, $\mu=0.6$]{\label{an.fig:phiP,Zmetric}
  \includegraphics[width=0.5\linewidth]{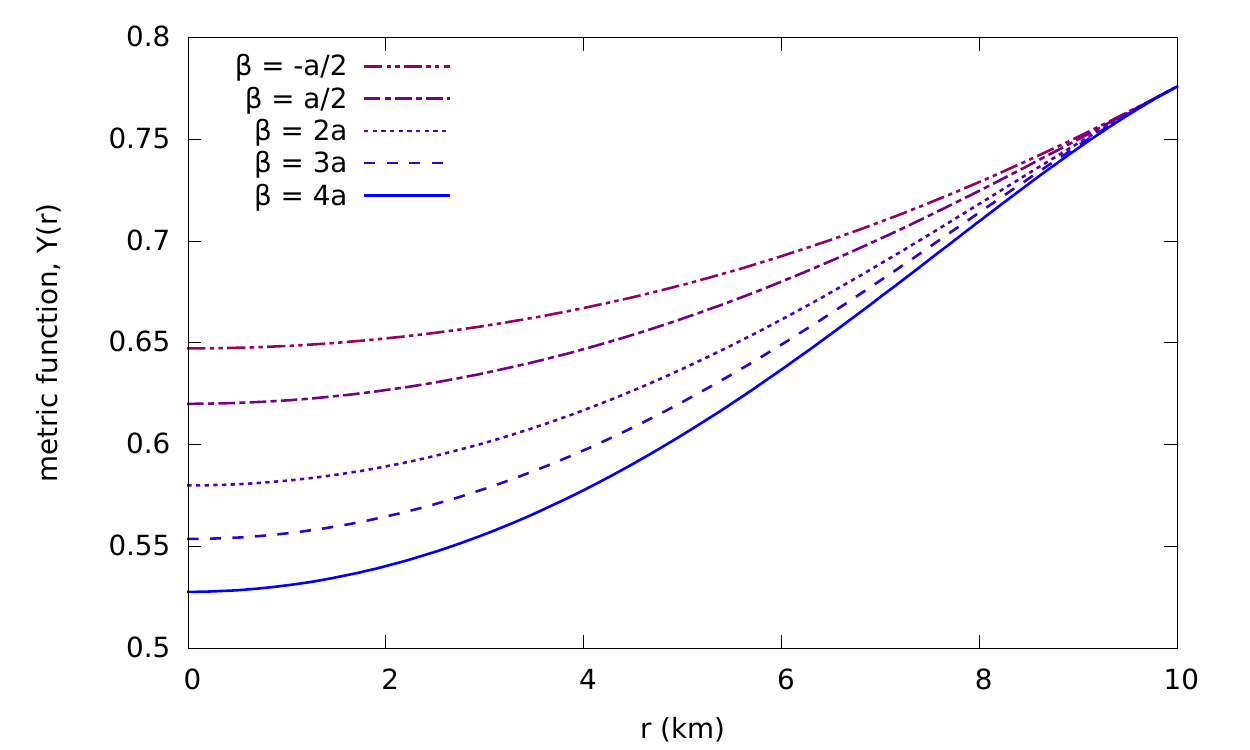}}
\caption[Variation of $Y$ metric variables for $\phi^{2} > 0$]{Variation of the $Y$
  metric variables with the radial coordinate inside the star. The
  parameter values are
  $\rho_{c}=\un{1\times 10^{18}} \dunits, r_b = \un{1 \times 10^{4}
    m},\, \mu$ is set to 1 on the left, and 0.6 on the right, and
  $\beta$ is set to the various values shown in the legend }
\label{an.fig:PhiP,MetricCoeffVaryingbeta}
\end{figure}

The discussion leading to inequalities~\eqref{an.eq:rho_cFirst}
and~\eqref{an.eq:rho_cSecond} must still hold in view of conditions
~\ref{an.it.ExteriorMetric} and~\ref{an.it.ExteriorObservables}, which
yield the exact same results as the \(\phi \leq 0\) cases.  Ensuring
that both these conditions remain satisfied, we next look at the
pressure in view of condition~\ref{an.it.DefiniteRPressure}.

\begin{figure}[!h]
\subfloat[The radial pressure , $\mu=1$]{\label{an.fig:phiP,PrMu1}
  \includegraphics[width=0.5\linewidth]{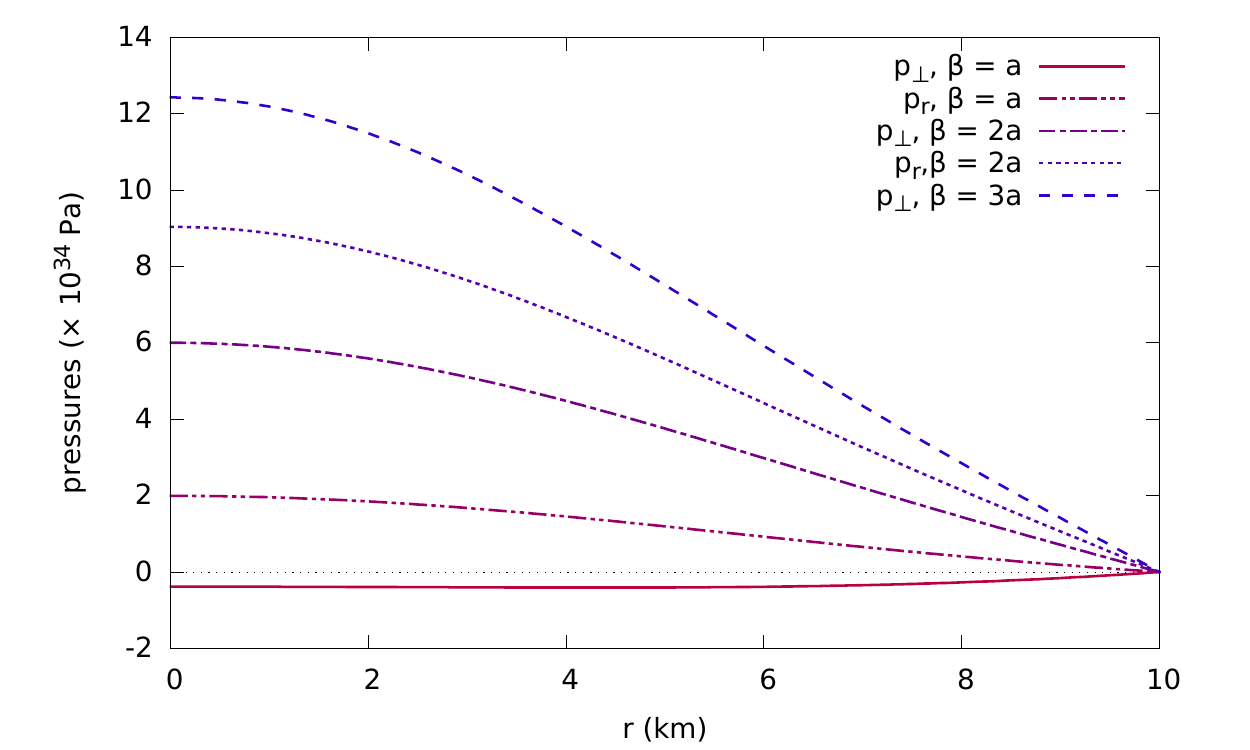}} 
\subfloat[The radial pressure, $\mu=0.6$]{\label{an.fig:phiP,PrMu6}
  \includegraphics[width=0.5\linewidth]{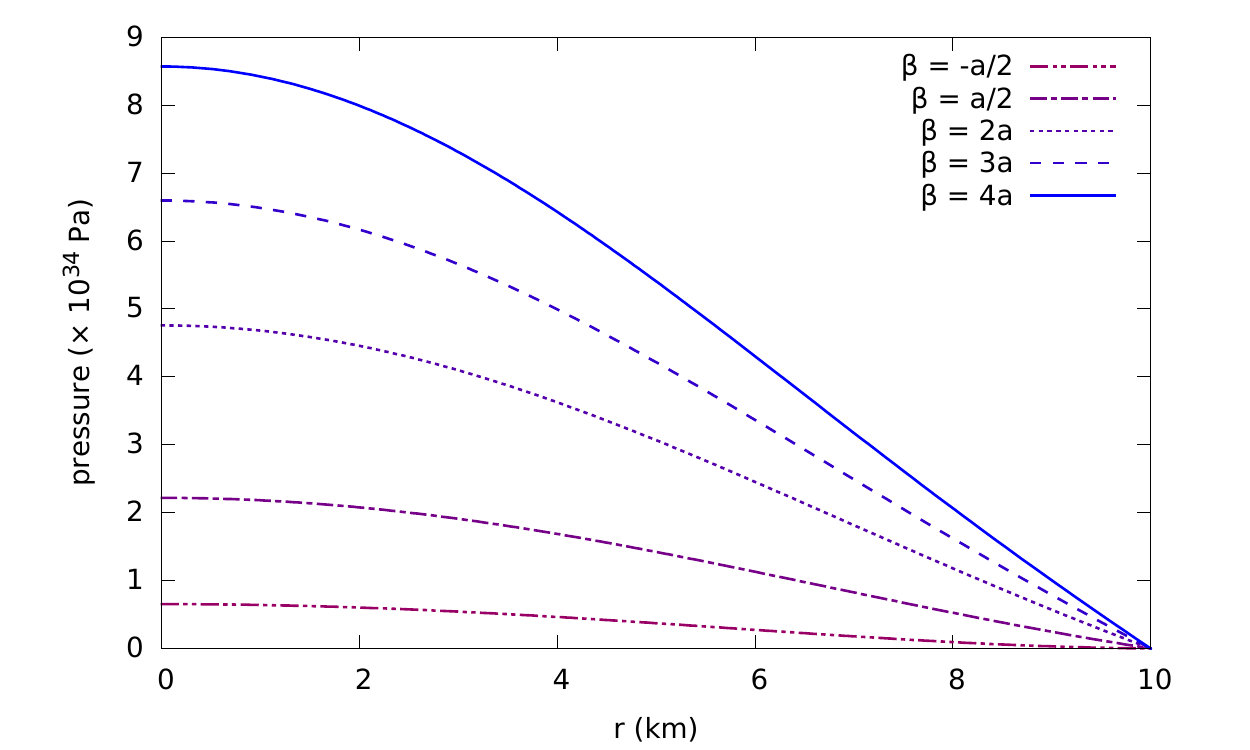}}
\caption[The radial pressure for $\phi^{2}>0$]{Variation of the radial pressure
  with the radial coordinate inside the star. The parameter values are
  $\rho_{c}=\un{1\times 10^{18}} \dunits, r_b = \un{1 \times 10^{4}
    m},\, \mu$
  is set to 1 on the left, and 0.6 on the right, and $\beta$ is set
  to the various values shown in the legend }
\label{an.fig:PhiP,radialPressureVaryingBeta}
\end{figure}

We see in figure~\ref{an.fig:PhiP,radialPressureVaryingBeta} that even
for widely differing \emph{positive} values of \(\beta,\) we still see
\emph{positive} pressures.  However even for the small negative value
of \(\beta = -a/2,\) we get a negative radial pressure, suggesting
that even though the solution is valid for those negative values of
\(\beta,\) any \(\beta < 0\) will yield negative pressures in the
natural case where \(\mu=1\).  Furthermore, since that same value of
\(\beta\) produces positive radial pressures when \(\mu=0.6,\) it is
probable that the extreme value of \(\beta\) is \(\mu\) dependant.

As a side note here, we can try to consider how \(\beta\) is dictating
the type of solution we have.  When \(\beta\) is zero, this solution
reduces to Tolman VII which has perfectly physical pressures and
densities.  When \(\beta \leq -a,\) as in the previous two
sections~\ref{an.ssec:phiNeg} and~\ref{an.ssec:phiZero} all the radial
pressures were negative.  Therefore the region
\(-a \leq \beta \leq 0, \) is the problematic one which we should
investigate in more detail.

In the next figure~\ref{an.fig:PhiP,radialPressureVaryingMu} we check
if the same is true for different values of \(\mu,\) and find that
this is mostly so.  These two sets of figures thus confirms that there
are parameter values for which we can get both positive pressures and
densities, very much like normal matter in this particular solution,
while emphasizing that even smaller values of \(|\beta|\) than
previously yield negative pressures in the natural
case\footnote{Technically we could, as have for
  example~\cite{BauRen93} and others, use the first zero of the
  pressure to fix the radius of the star, i.e.\ define $r_{b}$ such
  that $r_{b} = \min{\{r \in \R^{+} | p_{r}(r) = 0\}}$, and thus not
  deal with the part where the radial pressure gets to negative
  values.  However this will not work in this case because of our
  imposed boundary matching conditions which has to occur at the
  specific $r_{b}$ that has been \emph{already} specified.  The other
  conditions about the derivative of the pressure also start failing
  if we were to admit this here.}, if negative.  This confirms that we
will have to find a way to constrain \(\beta\) if we were to try to
use this solution as a model for stars.

This is however the first time we obtain positive pressures with the
new solutions, and the similarity to Tolman~VII, which we use to get
this solution is finally paying off.  We now need to proceed onto
checking the other conditions for physical viability.  This will take
the form of finding a range of values for \(\beta,\)

 \begin{figure}[!htb]
\subfloat[The radial pressure , $\beta=-a/5$]{\label{an.fig:phiP,PrBetaa}
  \includegraphics[width=0.5\linewidth]{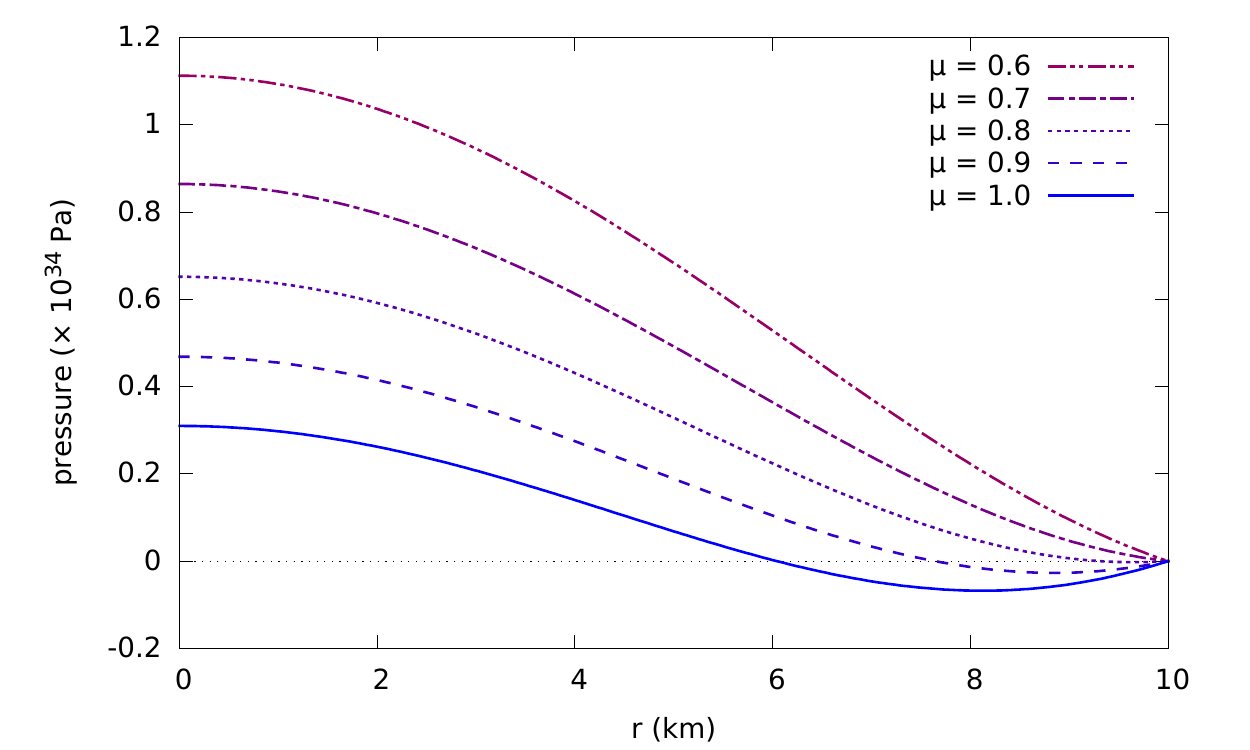}} 
\subfloat[The radial pressure, $\beta=5a$]{\label{an.fig:phiP,PrBeta6a}
  \includegraphics[width=0.5\linewidth]{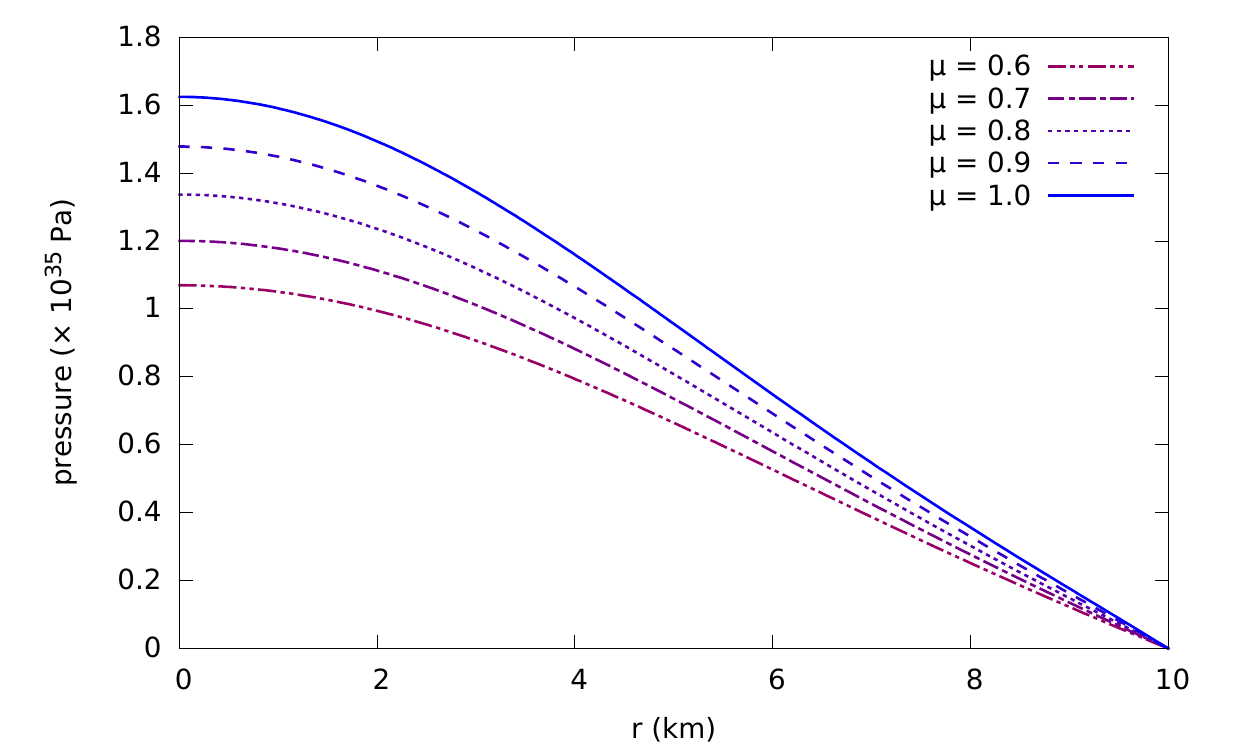}}
\caption[The radial pressure for $\phi^{2}>0$]{Variation of the radial pressure
  with the radial coordinate inside the star. The parameter values are
  $\rho_{c}=\un{1\times 10^{18}} \dunits, r_b = \un{1 \times 10^{4}
    m},\, \beta$
  is set to $-a/5$ on the left, and $5a$ on the right, and $\mu$ is set
  to the various values shown in the legend }
\label{an.fig:PhiP,radialPressureVaryingMu}
\end{figure}

The next conditions~\ref{an.it.BoundaryRPressure}
and~\ref{an.it.Delta} are true as can be seen both by construction in
the boundary conditions, and through the radial pressure pressure
graphs we have shown so far in
figures~\ref{an.fig:PhiP,radialPressureVaryingMu}
and~\ref{an.fig:PhiP,radialPressureVaryingBeta}.  To show that the
construction of \(\ppen,\) in the form of
\(p_{r} - \ppen = \beta r^{2},\) also leads to
condition~\ref{an.it.Delta}, we provide a
plot~\ref{an.fig:PhiP,radialTangentialPressures} of the tangential
pressure and the radial pressure for a few stars, noting the equal
values of the former at the centre of the star where \(r = 0.\) This
plot~\ref{an.fig:PhiP,radialTangentialPressures} also shows that while
the radial pressure always vanishes at the boundary radius
\(r=r_{b},\) the tangential pressure can take on negative values,
which might sound problematic for ordinary matter, and we investigate
this aspect by invoking the next
condition~\ref{an.it.EnergyConditions} involving the energy
conditions.  Before however, we check the
condition~\ref{an.it.MonotonicMatter}, which involves the derivative
of the matter variables.

\begin{figure}[!htb]
\subfloat[The radial and tangential pressures , $\beta=2a$]{\label{an.fig:phiP,PrPTBeta2a}
  \includegraphics[width=0.5\linewidth]{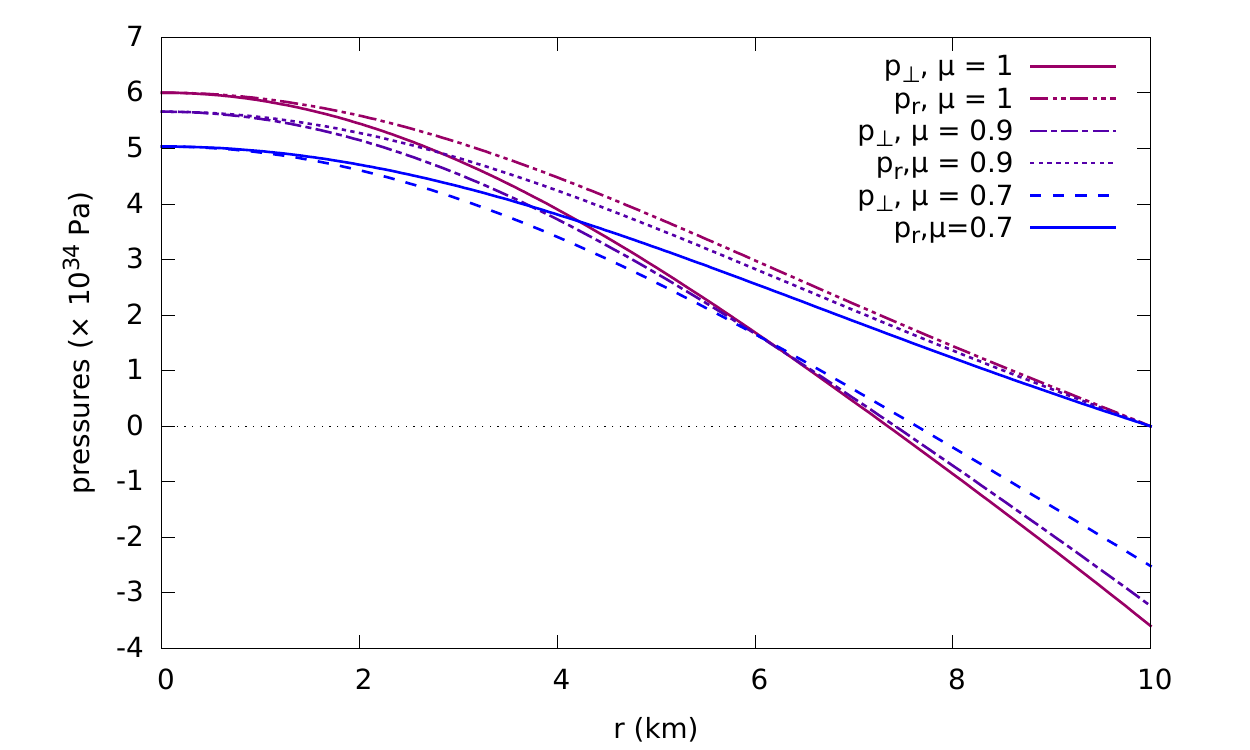}} 
\subfloat[The radial and tangential pressure, $\mu=1$]{\label{an.fig:phiP,PrPTMu1}
  \includegraphics[width=0.5\linewidth]{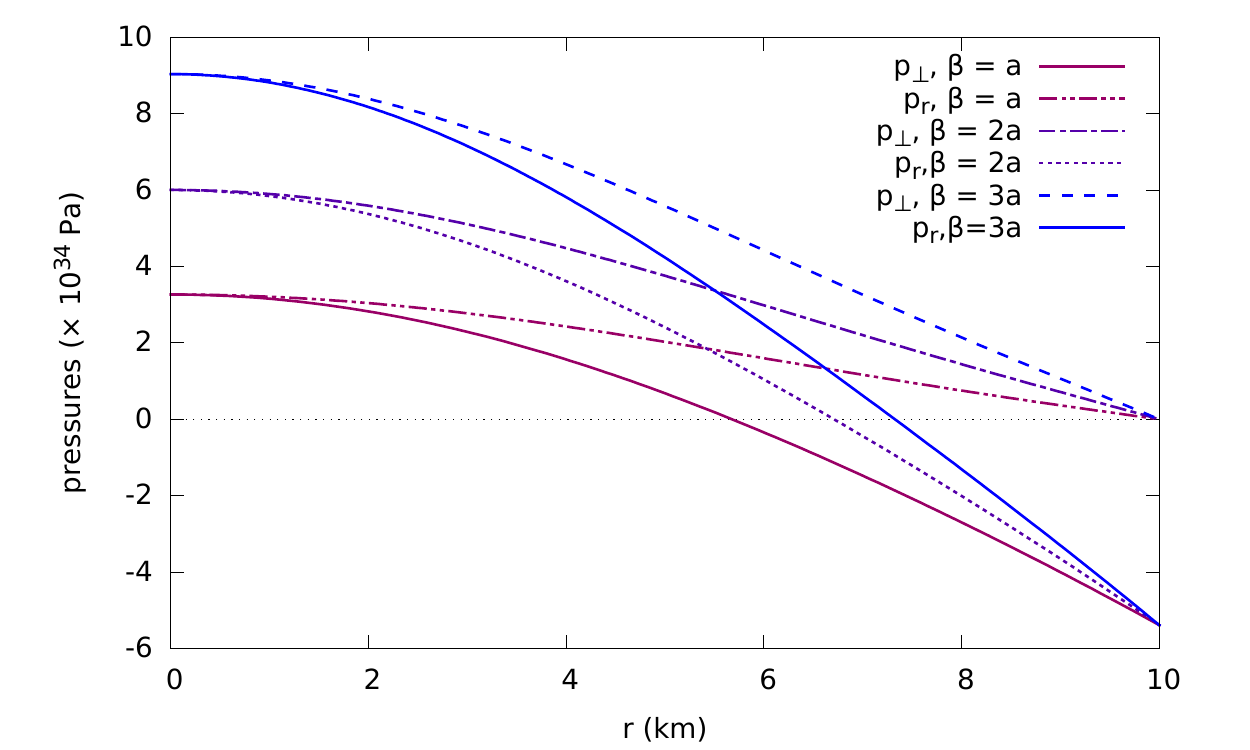}}
\caption[The radial pressure for $\phi^{2} >0$]{Variation of the radial pressure
  with the radial coordinate inside the star. The parameter values are
  $\rho_{c}=\un{1\times 10^{18}} \dunits, r_b = \un{1 \times 10^{4}
    m},\, \beta$
  is set to $2a$ on the left and $\mu$ is given in the legend, while $mu = 1$ on the right, and $\beta$ is set
  to the various values shown in the legend}
\label{an.fig:PhiP,radialTangentialPressures}
\end{figure}

The strong energy condition~\ref{an.it.EnergyConditions} states that
as long as the sum of all the pressures and the energy densities is
positive, we can be certain that the energy condition is satisfied and
that such matter might plausibly exist.  In the anisotropic case, this
condition reduces to \(2\ppen + p_{r}+\rho \geq 0,\)
and using the definition of \(\ppen\)
in terms of \(p_{r}\)
and \(\beta\)
from equation~\eqref{ns.eq:PhiPpt}, this relation imposes a constraint
on \(\beta\) in the form of
\begin{equation}
  \label{an.eq:energyCondition}
  2\beta r^{2} \leq 3 p_{r} + \rho \implies \beta \leq \f{3p_{r}+\rho}{2r^{2}}.
\end{equation}

As a result, we are now forced to pick values of \(\beta\)
within this range if we want to talk about physical stars, and we will
endeavour to ensure that this holds true in the final solutions we
use.  To clarify what this range implies, we should attempt to find an
extremum for \(\beta\)
with the expressions we have for the densities and pressures, however
the above equation has \(\beta\)
on both sides since \(p_{r}\)
depends on trigonometric functions of \(\phi,\)
which in turn contains \(\beta.\)
As a result we are left with a transcendental equation to solve, and no
closed-form solution is possible.  We should also note that in the
form in which we have given the energy condition, both the pressure
and density have the same geometrical units of \([L]^{-2},\)
which is a relief since we are adding them, and not the SI--units we
have been using on our plots.  In those units, \(\rho\)
and \(p_{r}\)
have values closer to each other, making the evaluation of the
inequalities even more important.
\begin{figure}[!htb]
\subfloat[The speed of sound, $v_{s},$ with $\beta=a$]{\label{an.fig:phiP,SpeedBeta=a}
  \includegraphics[width=0.5\linewidth]{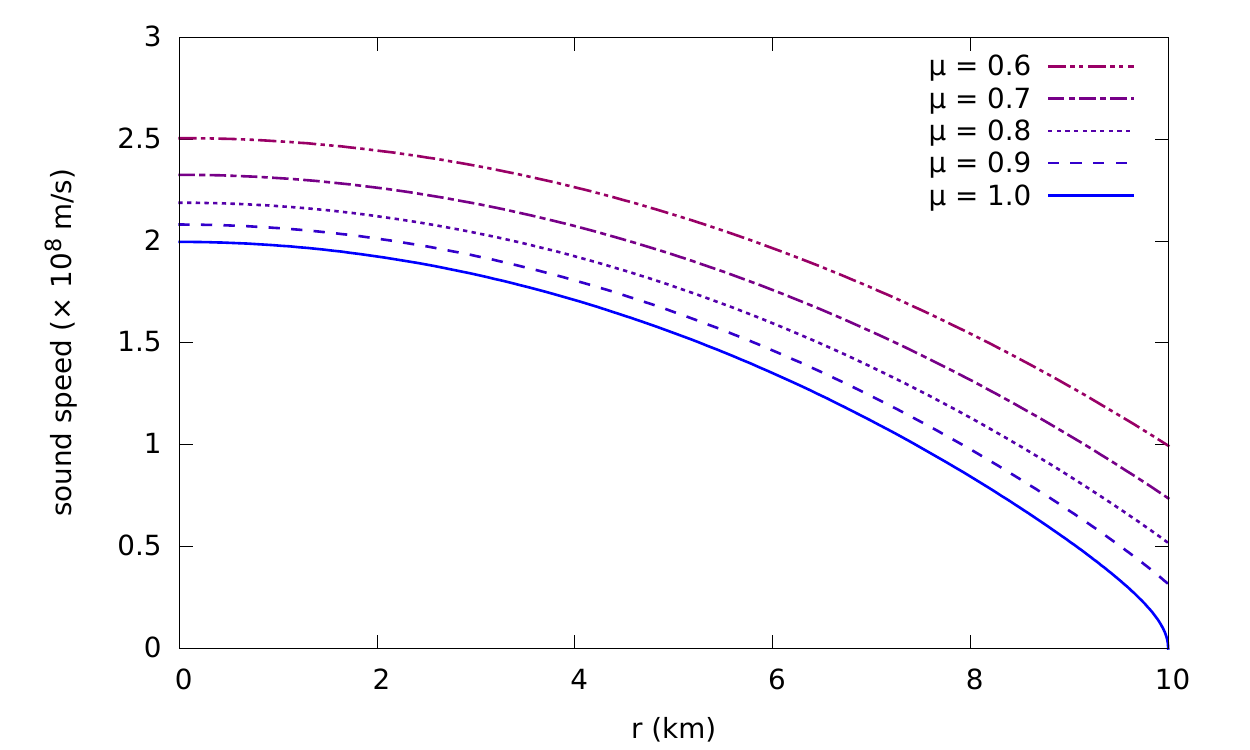}} 
\subfloat[The speed of sound, $v_{s},$ with $\beta=2a$ ]{\label{an.fig:phiP,SpeedBeta=2a}
  \includegraphics[width=0.5\linewidth]{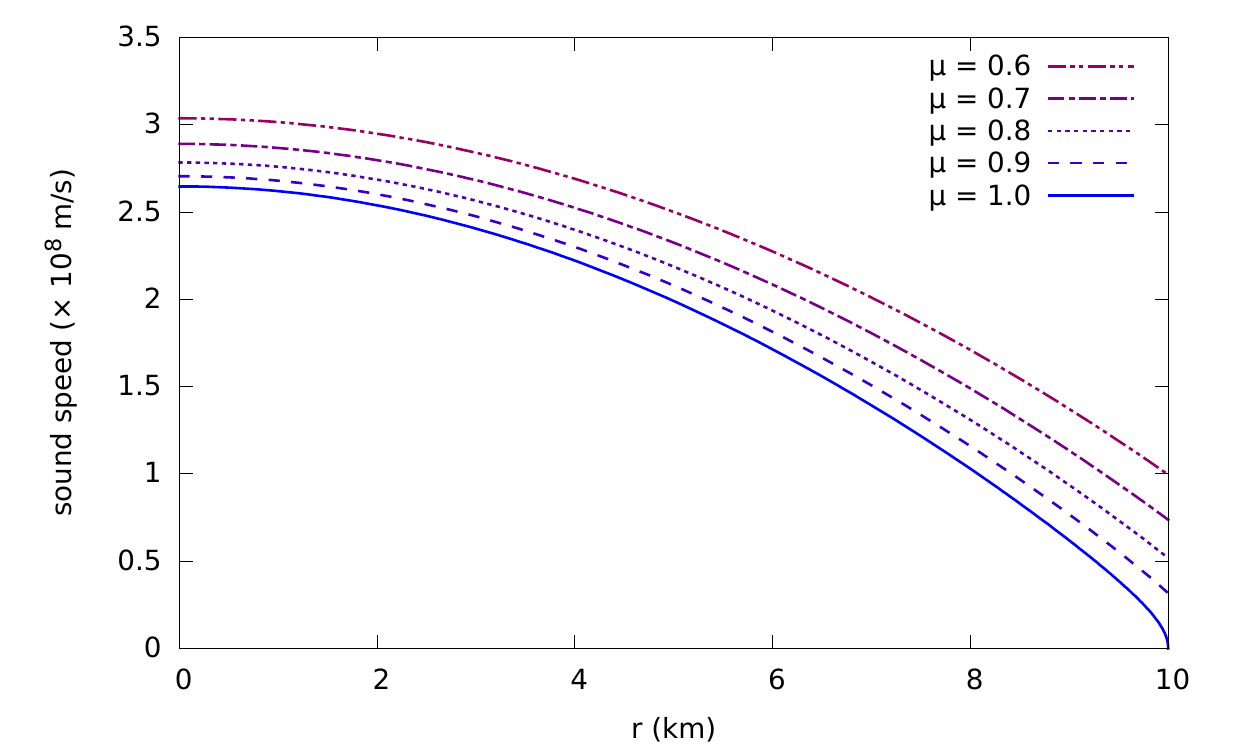}}\\
\subfloat[The speed of sound, $v_{s},$ with $\mu=1$ ]{\label{an.fig:phiP,speedMu=1} 
  \includegraphics[width=0.5\linewidth]{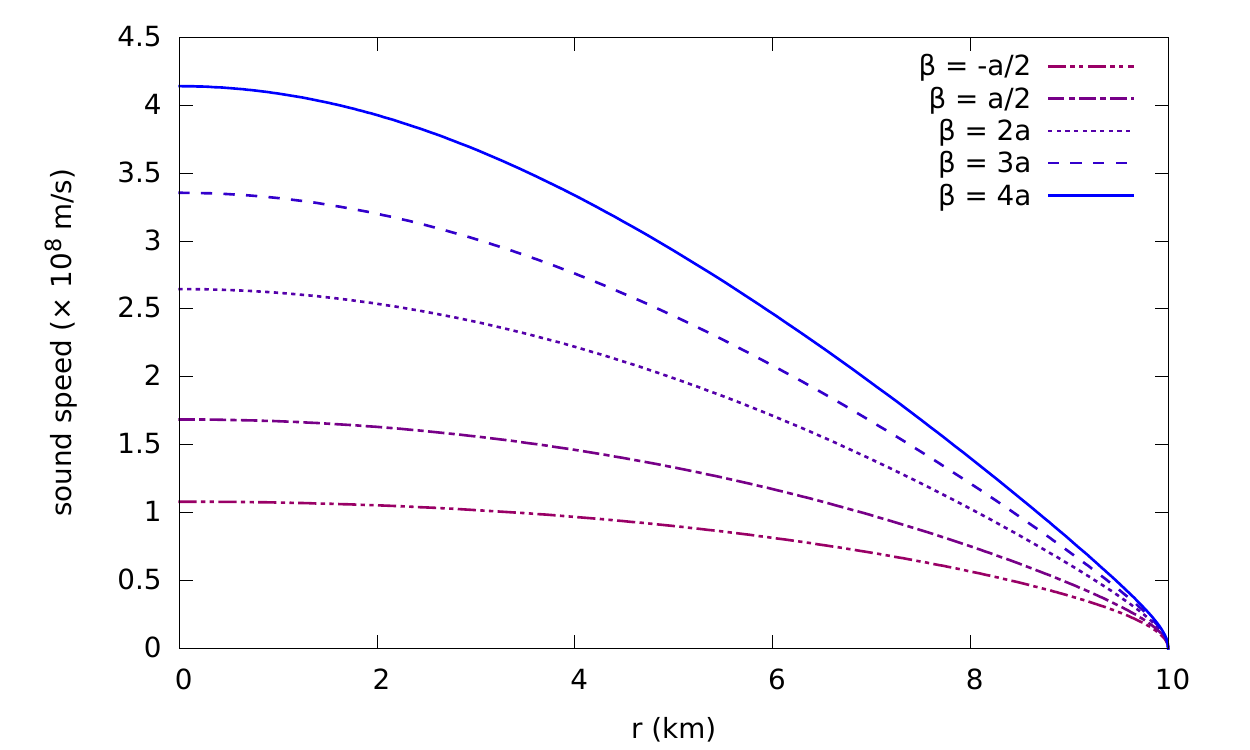}}
\subfloat[The speed of sound, $v_{s},$ with $\mu=0.6$ ]{\label{an.fig:phiP,speedRMu=0.6}
  \includegraphics[width=0.5\linewidth]{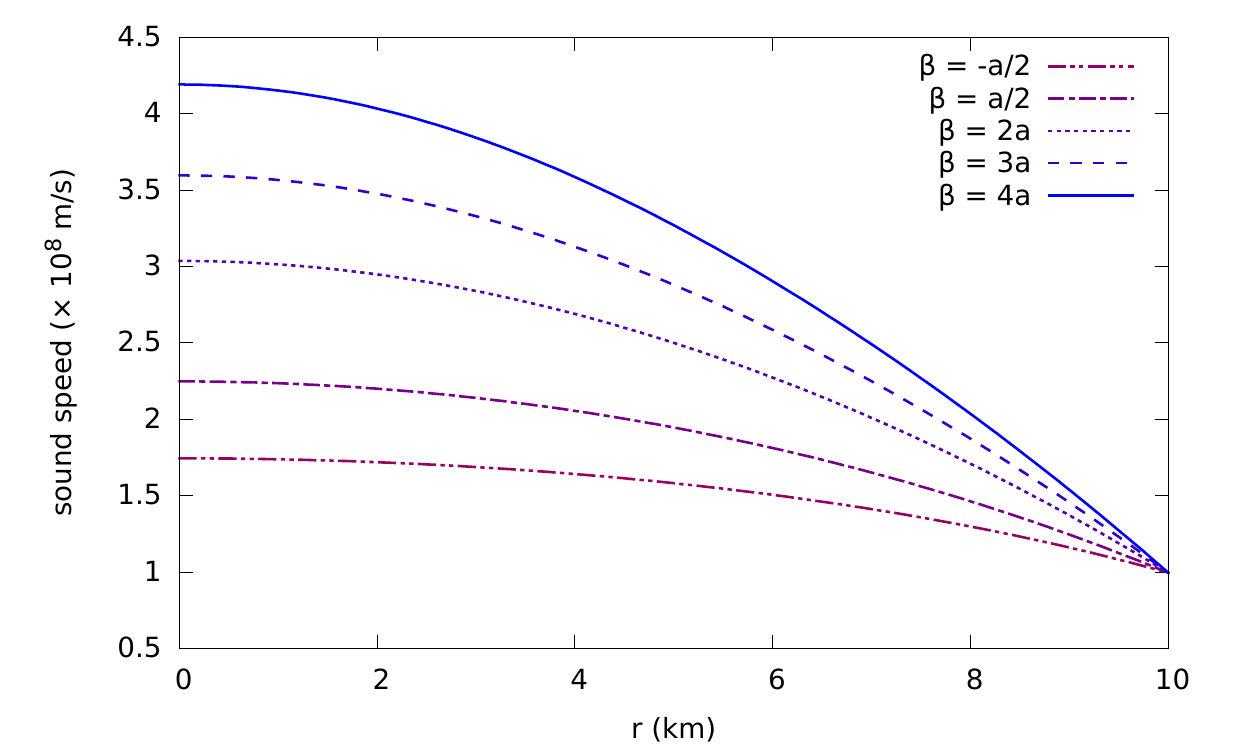}} 
\caption[The speed of sound with $r$ ]{Variation of speed of sound
  with the radial coordinate inside the star. The parameter values are
  $\rho_{c}=\un{1\times 10^{18}} \dunits, r_b = \un{1 \times 10^{4}
    m}, \mu$
  is fixed in the bottom two plots at one on the left, and 0.6 on the
  right, but takes on the values in the legend for the top
  graphs. $\beta$ is set respectively to $a$ and $2a$ in the left and
  right top plots, but varies in the bottom ones.}
\label{an.fig:PhiP,SoundSpeed}
\end{figure}

If we were to use equation~\eqref{ns.eq:EinMR1+2}, which gives us a
ready expression of both density and pressure, we can compute the
inequality~\eqref{an.eq:energyCondition} above as 
\[\beta \leq \f{3}{2 \kappa r^{2}} \left( \f{2Z}{rY}\deriv{Y}{r} - \f{1}{r} \deriv{Z}{r} \right) -\f{\rho}{r^{2}},\]
whose right-hand-side can be evaluated.  The logical places to
evaluate it are where \(\ppen\) is
\begin{enumerate}[label=(\alph{*}), ref=\theenumi.(\alph{*})]
\item the largest, that is  
 where \(r=r_{b},\)
and when we do so, we are left with
\begin{equation}\label{an.eq:v@r_b} \beta \leq \f{3}{\kappa r_{b}^{2}} \left[
  2(1-br_{b}^{2}+ar_{b}^{4})
  \left(\left. \f{1}{rY}\deriv{Y}{r}\right)\right|_{r=r_{b}}
  +2a - 4 b r_{b}^{2} \right] - \f{\rho_{c}(1-\mu)}{r^{2}_{b}}
\end{equation}
\item or equal to the radial pressure, i.e.\ where \(r=0\)
  and when we evaluate this, we get
\begin{equation}\label{an.eq:v@0} \beta \leq \f{3}{\kappa} \left[
  \f{2}{r^{2}}\left(\left. \f{1}{rY}\deriv{Y}{r}\right)\right|_{r=0}
  -\f{2a}{r^{2}} \right] -\f{\rho_{c}}{r^{2}}
\end{equation}
\end{enumerate}
expressions which depend crucially on the value of the term in round
parentheses, and hence the complicated \(Y\) metric functions that
include \(\beta.\) The second expression~\eqref{an.eq:v@0} is
useless\footnote{Instead of an evaluation, the limit of this
  expression could be taken as \(r \to 0.\) This limit does not
  necessarily exist, but once a solution with parameters is specified,
  this equation could be used to check the parameters' validity.  I
  only noticed this after an examiner pointed it out, and did not use
  this condition in the subsequent analysis.} because the \(r\)s in
the denominator causes the expression to diverge even if the term is
round brackets were finite.

However from the form of the first equation~\eqref{an.eq:v@r_b}, if we
were to find a way to evaluate some approximation to
\((1/rY) \deriv{Y}{r}\)
we could easily find a suitable range for \(\beta.\)
We keep this in mind since for the time being, we have no other
constraints that we can use.

To continue with our program of applying the constraints to our
equations, we turn to condition~\ref{an.it.CausalSpeed}, which
requires our speed of pressure waves in the interior to be less than
the speed of light.  First we show, to give an idea about how the
speed of sound changes with the different parameters \(\beta\)
and \(\mu,\)
plots of this speed for differing parameter values in
figure~\ref{an.fig:PhiP,SoundSpeed}

We see clearly
that for certain parameter values the speed at the centre is larger
than the speed of light, telling us that causality is being violated.
A formal analysis should allow us to cull parameter values that get us
such violations.

To implement this, we can use equation~\eqref{an.eq.soundSpeed}
directly.  In this particular case, the equation has no electric
charge, so that \(q = q' = 0,\)
and \(\Delta = \beta r^{2}.\)
Therefore the speed of sound condition reduces to
\begin{equation}
  \label{an.eq.soundSpeedPhiP}
\left( \f{r_{b}^{2}}{\kappa \rho_{c} \mu}\right)\left[ \f{\nu' \kappa \left( p_{r} + \rho \right)}{4 r} + \beta \right] \leq 1,
\end{equation}
in geometrical units.  Re-expressing
equation~\eqref{an.eq.soundSpeedPhiP} in terms of the metric variable
\(Y\) instead through the use of~\eqref{ns.eq:EinMR1+2}, we have
\begin{equation*}
  \left( \f{r_{b}^{2}}{\kappa \rho_{c} \mu}\right)
  \left\{ \f{1}{2} \left( \f{1}{rY} \deriv{Y}{r} \right)  \left[\f{2Z}{rY}\deriv{Y}{r} - \f{1}{r} \deriv{Z}{r} \right]+ \beta \right\} \leq 1.
\end{equation*}
Upon rearranging, and considering that the factor
\(1/(rY)\deriv{Y}{r}\)
occurs very frequently, we
define\[\psi := \f{1}{rY} \deriv{Y}{r}, \qquad \text{with} \qquad
\psi_{0} := \left( \left. \f{1}{rY} \deriv{Y}{r}\right) \right|_{r=0},
\qquad \text{and} \qquad \psi_{b} := \left( \left. \f{1}{rY}
    \deriv{Y}{r}\right)\right|_{r=r_{b}},\]
where the second expression is to be understood as a formal limit,
and we get the quadratic inequality
\begin{equation}
  \label{an.eq.soundSpeedPhiPinY}
\left( \f{r_{b}^{2}}{\kappa \rho_{c} \mu}\right)
  \left\{ \beta + Z\psi^{2} - (2ar^{2}-b) \psi \right\} \leq 1.
\end{equation}
 The highest value of the speed of
sound should occur at the centre of the star since both the density
and pressure are maximal there, so we evaluate the
above~\eqref{an.eq.soundSpeedPhiPinY} at \(r=0,\) such that \(Z = 1,\) to get
\[
\left( \f{r_{b}^{2}}{\kappa \rho_{c} \mu}\right) \left( \beta + \psi_{0}^{2} + b\psi_{0} \right) \leq 1 \implies
\psi_{0}^{2} + b\psi_{0} + \left( \beta -  \f{\kappa \rho_{c} \mu}{r_{b}^{2}}\right) \leq 0
\]

This inequality can be solved by first finding the roots of the quadratic above, 
\[ \psi_{0\pm} = -\f{b}{2} \pm \sqrt{\f{b^{2}}{4} - \beta +
  \f{\kappa\rho_{c}\mu}{r_{b}^{2}}},\] to finally get
\begin{equation}
  \label{an.eq:psiRange}
\psi_{0-} \leq  \psi_{0} \leq \psi_{0+}.
\end{equation}
We finally have an equation for the range of \(\psi_{0},\)
and recall that we needed this information before in
equation~\eqref{an.eq:v@0}. Proceeding similarly, but evaluating the
speed of sound at the boundary instead, we get the analogous
inequality \(\psi_{b-} \leq \psi_{b} \leq \psi_{b+},\)
with
\[\psi_{b\pm} = \f{(2ar_{b}^{2} - b )}{2Z_{b}} \pm
\f{\sqrt{(2ar_{b}^{2} - b)^{2} - 4Z_{b}\left(\beta -
      \f{r_{b}^{2}}{\kappa\rho_{c}\mu} \right)}}{2Z_{b}},\]
where \(Z_{b} = 1 - br_{b}^{2} + ar_{b}^{4}.\)
This is a considerably more complicated expression, but we can
consider these two conditions to be our causality criteria on all
variables involved, however we will only test them once we start
specifying parameters for the star.  Furthermore, considering
inequality~\eqref{an.eq:v@r_b}, which has \(\psi_{b}\)
as one if its terms, we finally have a complete prescription to figure
out if the value of \(\beta\)
we are using is within the range of physical acceptability.  The
square-root including \(\beta\)
in the expression of \(\psi_{b\pm}\)
makes computation unwieldy, but once one specifies the set
\(\{\rho_{c},\mu,r_{b}\}, \beta\) is completely constrained. 

Continuing on with the next constraint~\ref{an.it.DecreasingSpeed}, we
need the speed of sound derived earlier to be decreasing with
increasing radius, i.e.\ \(\deriv{v_{s}}{r} < 0.\)
We can easily write down this condition, but from
figure~\ref{an.fig:PhiP,SoundSpeed}, it is clear that for whatever
parameter we might choose, even when these parameters do not obey the
causality criterion, the speed of sound is a decreasing function.

This concludes the implementation of the constraint onto this
particular solution.  We should note here that this particular
solution \emph{can potentially} satisfy all the criteria we have.
Additional restrictions on \(\beta\)
were also found, in particular \(\beta >0,\)
to prevent negative radial pressures is absolutely necessary.
Additional restrictions depending on the value of the other parameter
in the set \(\{\mu, r_{b}, \rho_{c}\}\)
were also found, and will be checked when value for those are picked.
Having spelled out all the conditions in this section, we move onto
the next solutions: the charged ones.

\subsection{The anisotropised charge case, $\Phi^{2} \neq 0$}
In this special case of the more general charged anisotropic case, the
charge is chosen so that it annihilates the contribution of the
anisotropy so that the matter density is the only term in the
differential equation for \(Y.\)

We start by ensuring that the metric functions are well behaved in the
star, as per condition~\ref{an.it.RegularMetric}.  All the charged
solutions have a slightly different definition of the \(Z\)
metric function, as compared to the uncharged solutions.  In this
particular case, as we saw in chapter~\ref{C:NewSolutions}, it is
given by
\begin{equation}
  \label{an.eq:ZchargedAnisotropised}
Z(r) = 1 - \left(\f{\kappa \rho_{c}}{3} \right)r^{2} +
\f{1}{5}\left( \f{\kappa \rho_{c}\mu}{r_{b}^{2}} - k^{2}\right) r^{4}.
\end{equation}
Considering the value of this metric function at \(r=0,\)
and \(r=r_{b},\) we get
\[Z(0) = 1, \qquad \text{and} \qquad Z_{b} = Z(r_{b}) = 1 -
\kappa\rho_{c}r^{2}_{b} \left( \f{1}{3} - \f{\mu}{5}\right)-
\f{k^{2}r^{4}_{b}}{5},\]
a result different from the uncharged case due to the presence of
electric charge through \(k.\)
Since this metric function cannot change sign, we can use the second
equation to obtain a constraint on the maximum charge:
\begin{equation}
  \label{an.eq:ZconstraintCA}
  k^{2} < \f{1}{r^{2}_{b}} \left[ \f{5}{r^{2}_{b}} - \kappa\rho_{c}\left(\f{5}{3} - \mu \right) \right].
\end{equation}
Plugging in the typical values we usually use for the central density,
boundary radius and self-boundness, viz.\
\(\{\mu=1, \rho_{c}=1 \times 10^{18}\un{kg\cdot m^{-3}}, r_{b} =
1\times 10^{4} \un{m} \},\) we find that the above
inequality~\eqref{an.eq:ZconstraintCA} gives us
\(|k| < 2.2 \times 10^{-8} \un{m^{-2}}.\) The charge density \(k\) is
in geometrical units, having been normalised by a factor of
\(\sqrt{G/(4\pi \cc^{4}\epsilon_{0})},\) Einstein's constant
multiplied by Coulomb's.  When we will go back to physical units to
make connections to observed values, this is the factor we will use to
convert to say S.I.\ charge units.  In particular the above limit
corresponds to a charge density of
\(|k| < 2.6 \times 10^{9} \un{C \cdot m^{-3}}.\) As a result, for a
model having a radius of \(r_{b} = 10 \un{km},\) the total maximal
electric charge \(Q = kr_{b}^{3} = 2.6 \times 10^{21} \un{C}. \) This
limit varies non-linearly with \(\mu,\) but the maximum value of \(k\)
is not very sensitive to the change in \(\mu.\)

The next condition we check comes from the necessity for the proper
radius \(R\)
to exist.  As we derived earlier in the section
following~\eqref{an.eq.Radius}, this means that \(2\sqrt{a} > b,\)
and using the expressions for these constants for this solution, we
get straight-forwardly that
\begin{equation}
\label{an.eq:RconstraintCA}
k^{2} < \kappa \rho_{c}\left(\f{\mu}{r_{b}^{2}} -
  \f{5\kappa\rho_{c}}{36} \right).  
\end{equation}
The value of \(k^{2}\)
must clearly be positive, so that the term in brackets must be
positive.  This immediately yields
\(\rho_{c} < \f{36\mu}{5\kappa r_{b}^{2}},\)
the same constraint on density as for the uncharged case.  The
inequality on \(k\)
can then be supplemented with the usual values of \(\rho_{c},\)
and \(r_{b},\)
to yield a function of \(\mu,\)
which we plot next in figure~\ref{an.fig:AniCha,kLim}.
\begin{figure}[!htb]
  \includegraphics{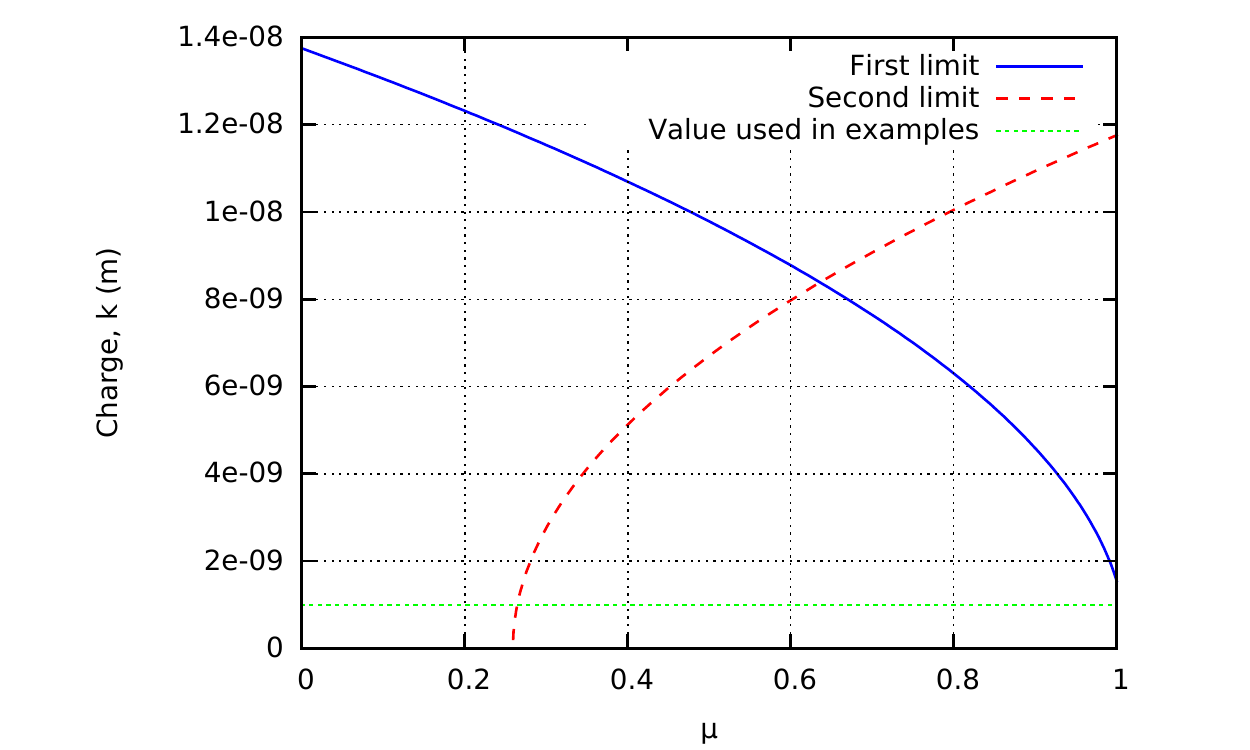}
\caption[The limiting value of $k$ for different $\mu$]{The limiting value of $k$ for different $\mu$: the blue line corresponds to equation~\eqref{an.eq:ZconstraintCA}, the red line to equation~\eqref{an.eq:RconstraintCA}, where values of the central density, $\rho_{c} = 1\times 10^{18} \un{kg \cdot m^{-3},}$ and the boundary radius $r_{b}=10000\un{m}$ have been supplied to produce the lines.}
\label{an.fig:AniCha,kLim}
\end{figure}
As we can see from the figure and inequalities only values of \(k\)
below \emph{both} the red and blue lines can be used as a valid charge
for models having densities and radii specified in the legend.  In
this case, there can be no charge for models having low \(\mu.\)
This trend is seen for all parameter values.

We next check the second
metric function's behaviour: \(Y\)
is given by equation~\eqref{ns.eq:YphiCCA}, which is complicated.
Instead we give plots of the two metric function.
\begin{figure}[!htb]
\subfloat[The $Y(r)$ metric function]{\label{an.fig:YmetricCAa}
  \includegraphics[width=0.5\linewidth]{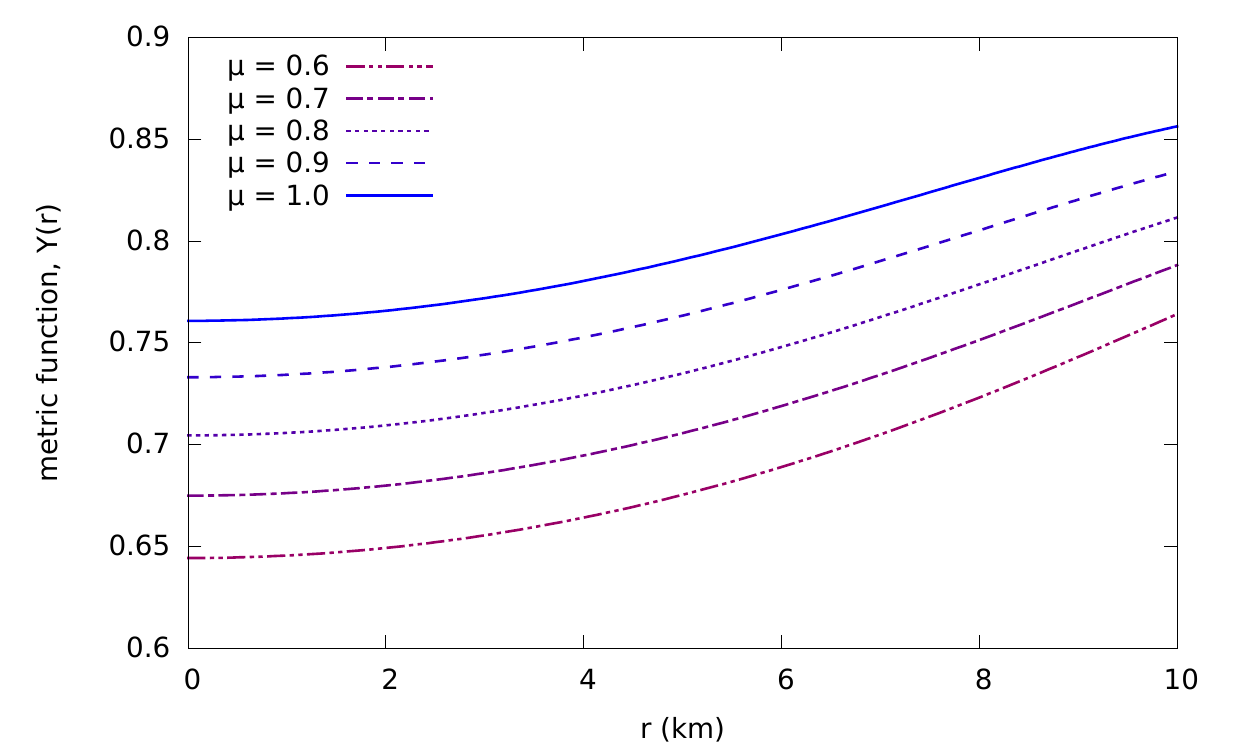}} 
\subfloat[The $Z(r)$ metric function]{\label{an.fig:ZmetricCAa}
  \includegraphics[width=0.5\linewidth]{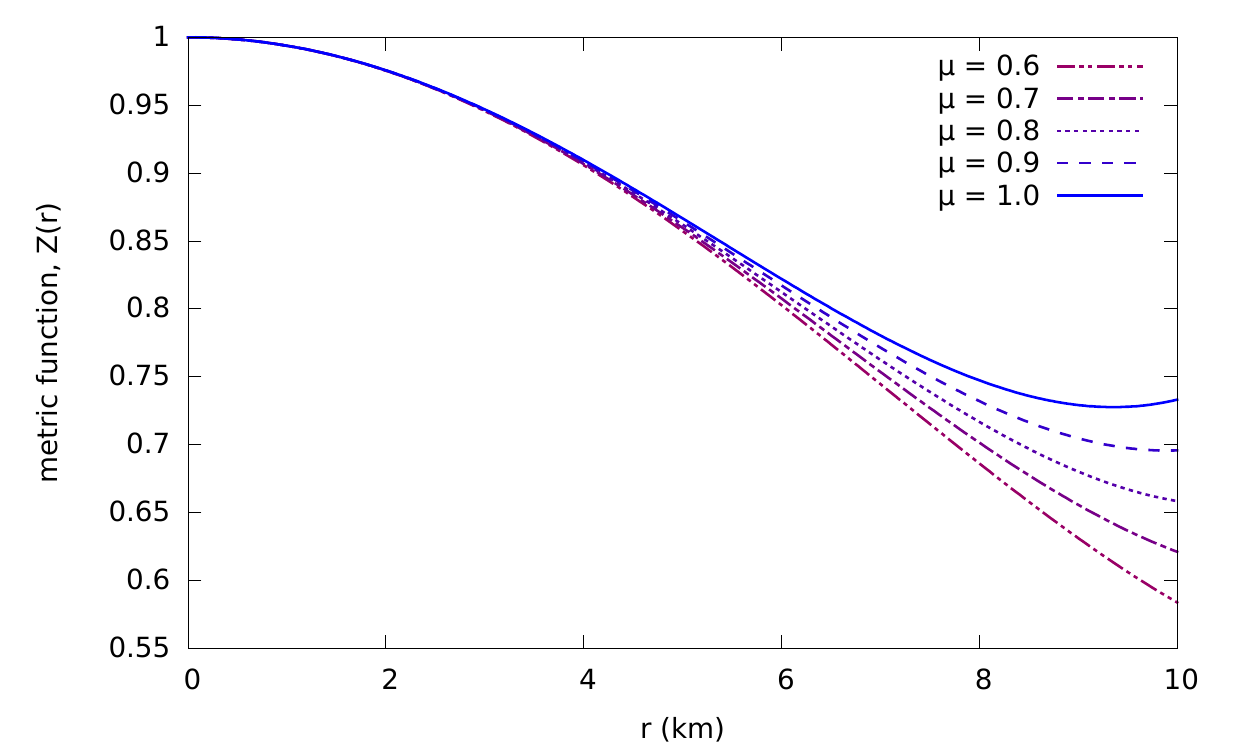}}\\
\subfloat[The $\lambda(r)$ metric function]{\label{an.fig:LambdaMetricCAa} 
  \includegraphics[width=0.5\linewidth]{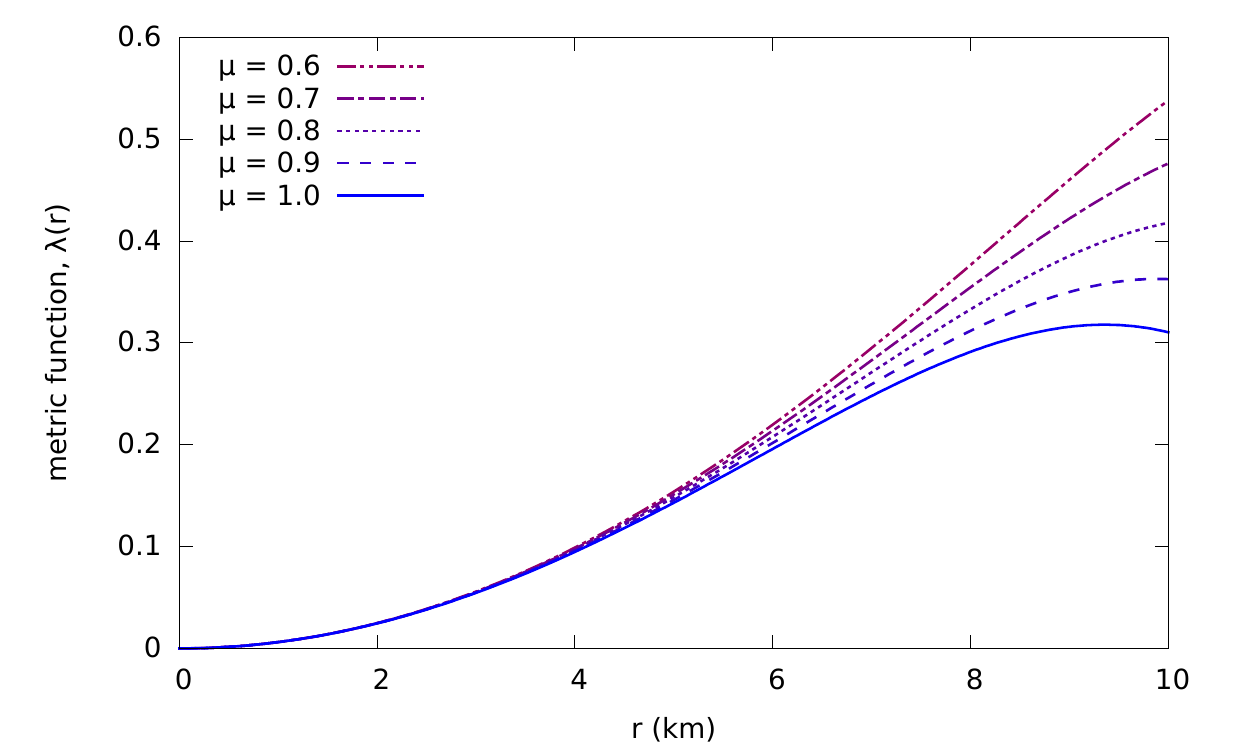}}
\subfloat[The $\nu(r)$ metric function]{\label{an.fig:NuMetricCAa}
  \includegraphics[width=0.5\linewidth]{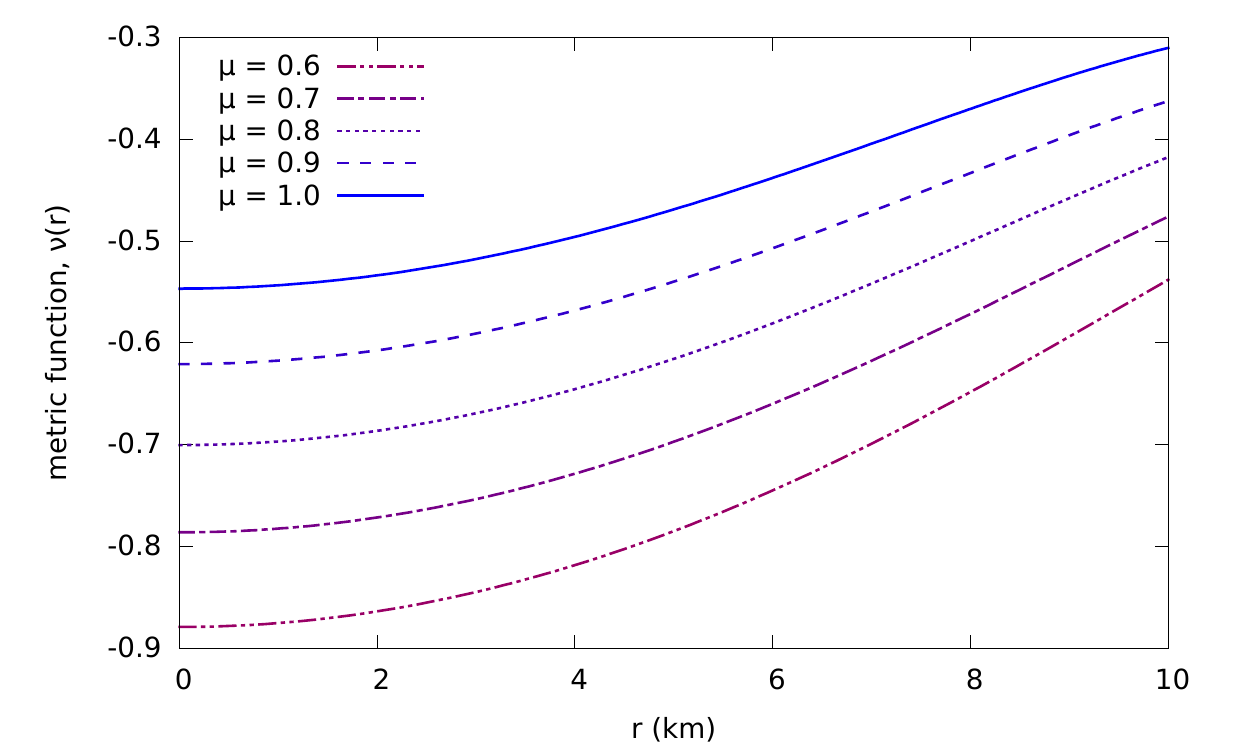}} 
\caption[Variation of metric variables]{Variation of metric variables with the radial coordinate
  inside the star. The parameter values are $\rho_{c}=\un{1\times
    10^{18}} \dunits, r_b = \un{1 \times 10^{4} m}, k = \un{3 \times 10^{-9} m^{-2}} $ and $\mu$
  taking the various values shown in the legend }
\label{an.fig:AniCha,MetricCoeff}
\end{figure}
As we see in figure~\ref{an.fig:AniCha,MetricCoeff}, all the metric
functions are well behaved for the range of parameters we picked.  

Next we look at the behaviour of the radial pressure \(p_{r}\)
in the star.  The expression of the former was previously given
in~\eqref{ns.eq:PhiPprCCA}, and here we only provide a plot of the
radial pressure for differing parameter values instead in
figure~\ref{an.fig:AniCha,radialPressures},
\begin{figure}[!htb]
\subfloat[The radial pressure , $k=1\times10^{-9}$]{\label{an.fig:AniCha,Pk=1e-9}
  \includegraphics[width=0.5\linewidth]{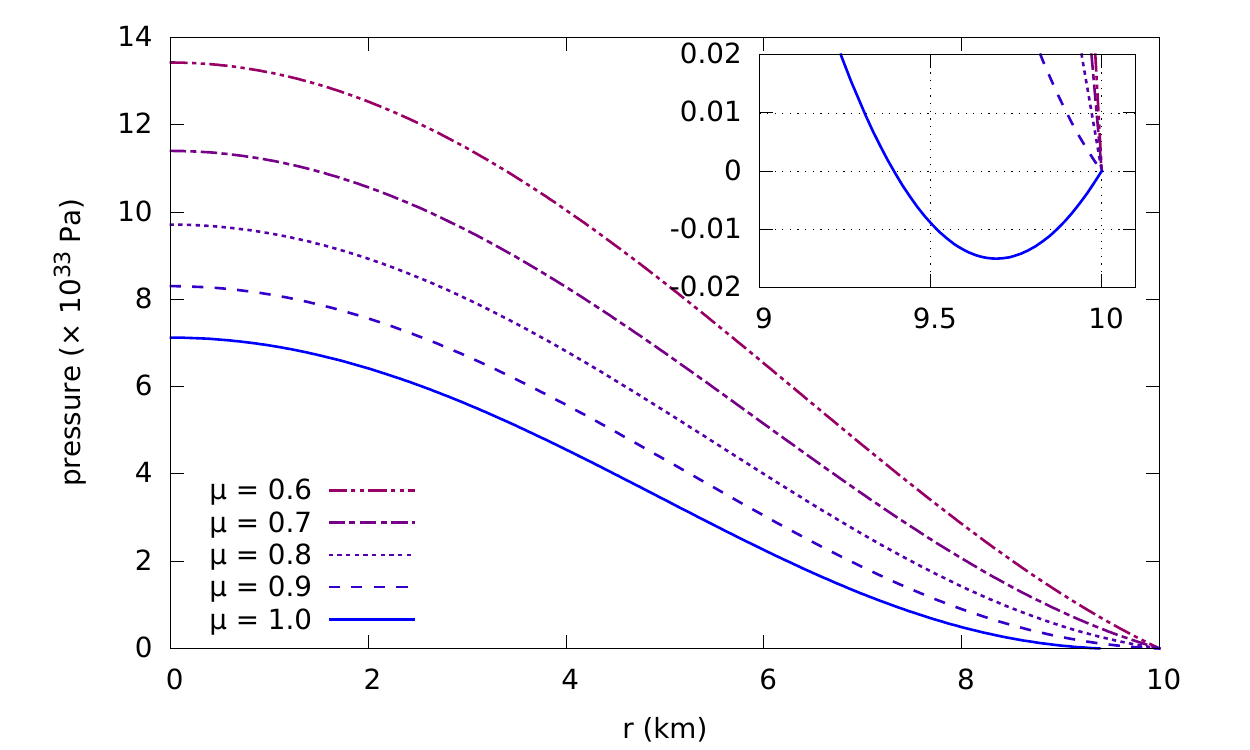}} 
\subfloat[The radial pressure, $k=3\times10^{-9}$]{\label{an.fig:AniCha,Pk=3e-9}
  \includegraphics[width=0.5\linewidth]{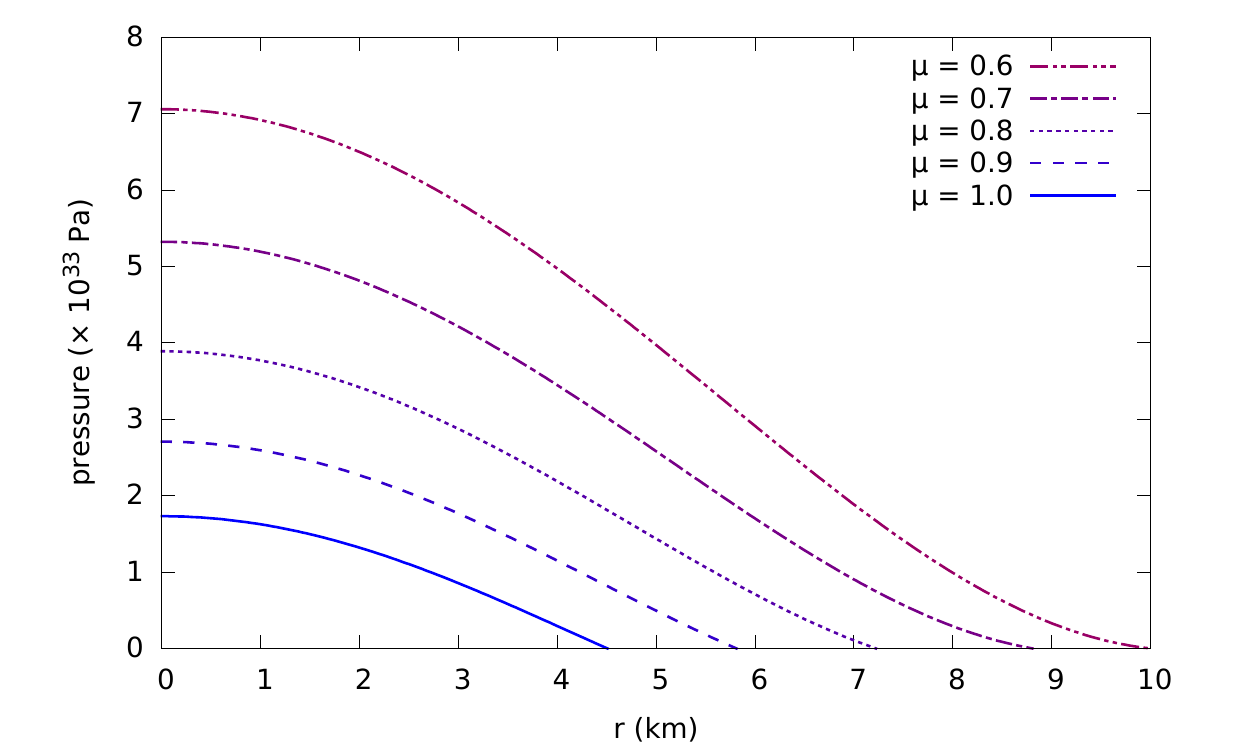}} \\
\subfloat[The radial pressure , $\mu=1$]{\label{an.fig:AniCha,Pm=1}
  \includegraphics[width=0.5\linewidth]{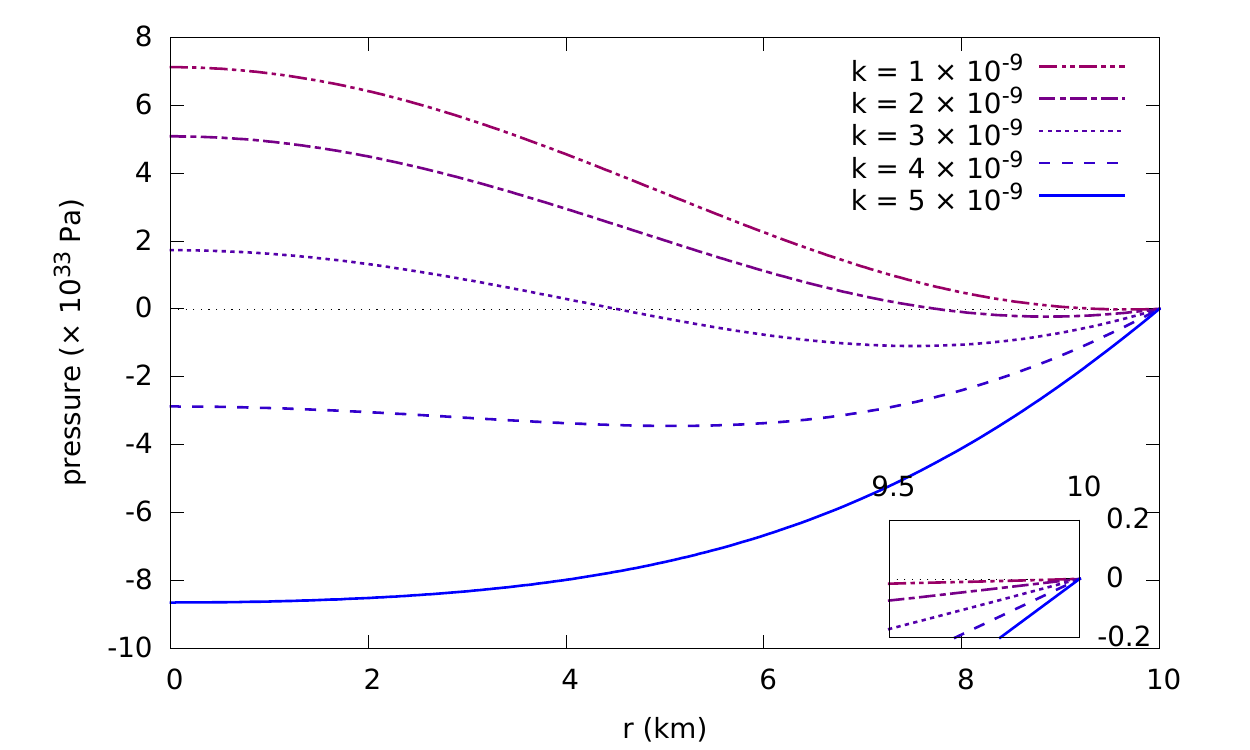}} 
\subfloat[The radial pressure , $\mu=0.6$]{\label{an.fig:AniCha,Pm=0.6}
  \includegraphics[width=0.5\linewidth]{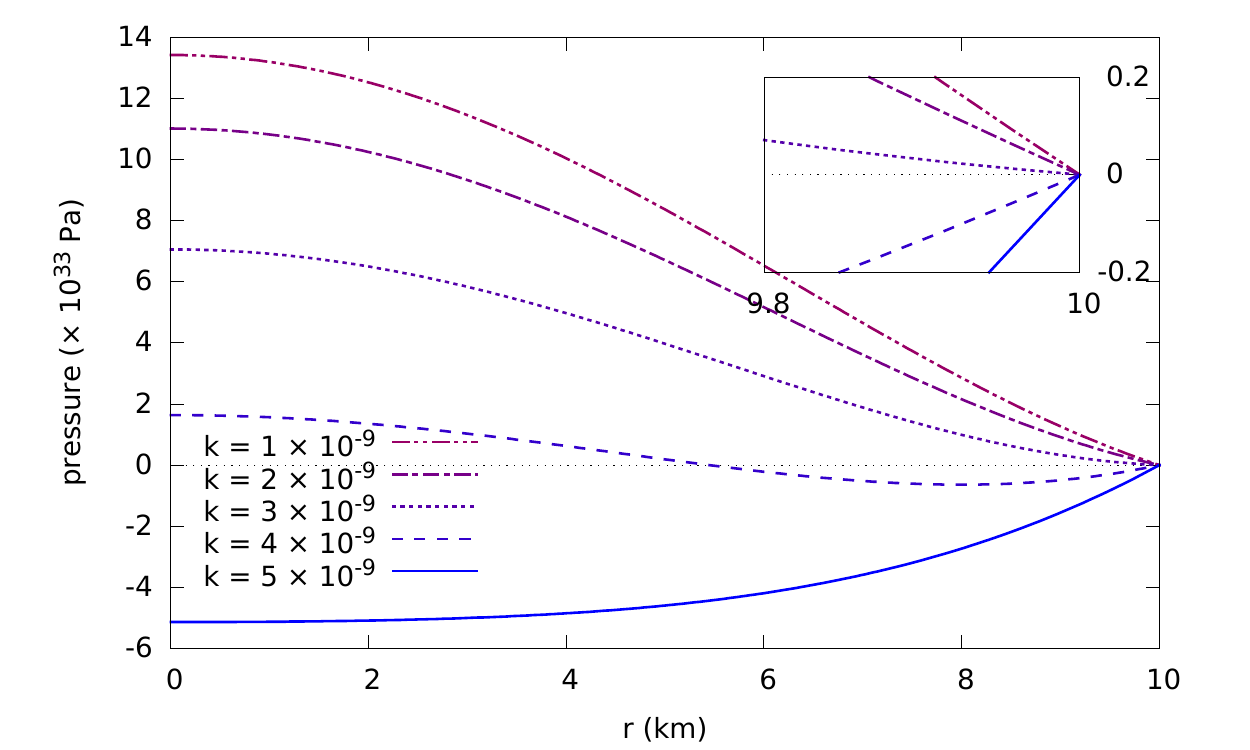}} 
\caption[The radial pressure for anisotropised charge]{Variation of the radial
  pressure with the radial coordinate inside the star. The parameter
  values are
  $\rho_{c}=\un{1\times 10^{18}} \dunits, r_b = \un{1 \times 10^{4}
    m},\, \beta$
  is fixed by $k$ in this solution and $\mu$ is given in the legend
  for the top plots, and $k$ is given in the bottom ones.}
\label{an.fig:AniCha,radialPressures}
\end{figure}

As we can see immediately, the radial pressures get to negative values
in the star for low values of \(\mu,\) and the ``natural'' case is
also plagued by this feature.  The effect is only enhanced with higher
charge values \(k,\) as is obvious in the top right
pane~\ref{an.fig:AniCha,Pk=3e-9}.  The bottom panes
in~\ref{an.fig:AniCha,radialPressures} vary \(k\) instead for fixed
\(\mu,\) and shows that there must exist some critical relation
between \(k\) and \(\mu\) in this model that will allow the pressure
inside the star to be always positive.  We now try to find this
critical value.  Considering the shape of the radial pressure graphs,
we will have negative pressures if when we solve \(p_{r}(r) = 0,\) for
\(r,\) the solution be less than \(r_{b}.\) However solving the
complicated equation~\eqref{ns.eq:PhiPprCCA} is impossible
analytically.  The alternative way we could try finding the critical
value of \(k\) is through the derivative of the radial pressure, whose
expression we already have.  This is possible because from physical
consideration, we need \(p_{r}(r=0) > 0,\) and from the boundary
conditions also that \(p_{r}(r=r_{b}) = 0.\) The only way to have
negative pressure is therefore to have a turning point at \(r=r_{t},\)
where \(0 < r_{t} < r_{b}.\) If \(r_{t}\) then exists, we are assured
that the pressure has become negative somewhere.
Equation~\eqref{an.eq.gTOV} gives us the expression of the pressure
derivative.  We can simplify this equation in our particular case,
from \(q = kr^{3},\) and \(\Delta = 2k^{2}r^{2}\) to give the
condition for a turning point in the pressure if
\begin{equation}
  \label{an.eq:pressureTurningPt}
  0 = \deriv{p_{r}}{r} = \f{2k^{2}r}{\kappa} - \f{\nu'}{2} (p_{r}+\rho).
\end{equation}
This equation is only true for some particular \(r\) and if and only
if the pressure becomes negative somewhere in the region we are
interested in.  We will use it to check for valid values of \(k.\)

Next we turn to the tangential pressure in
figure~\ref{an.fig:AniCha,tangentialPressures}
\begin{figure}[!htb]
\subfloat[The tangential pressure , $k=1\times10^{-9}$]
  {\includegraphics[width=0.5\linewidth]{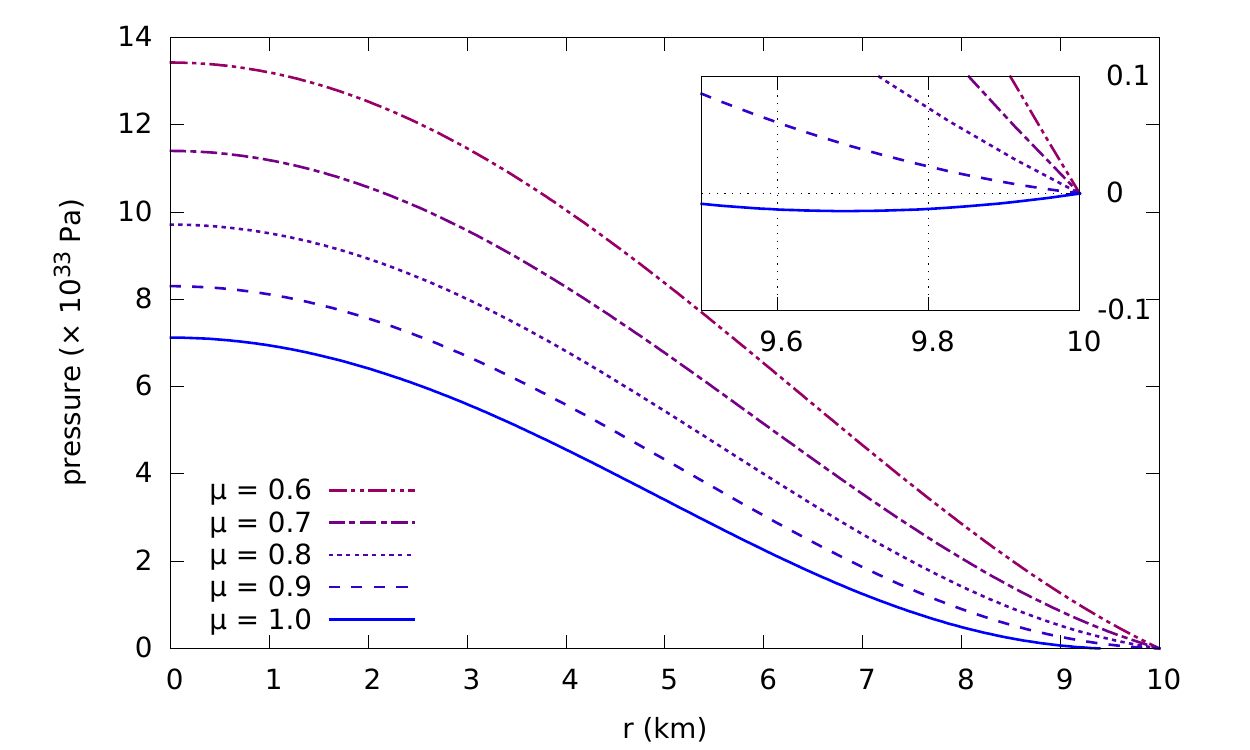}}
\subfloat[The tangential pressure, $k=3\times10^{-9}$]
 {\includegraphics[width=0.5\linewidth]{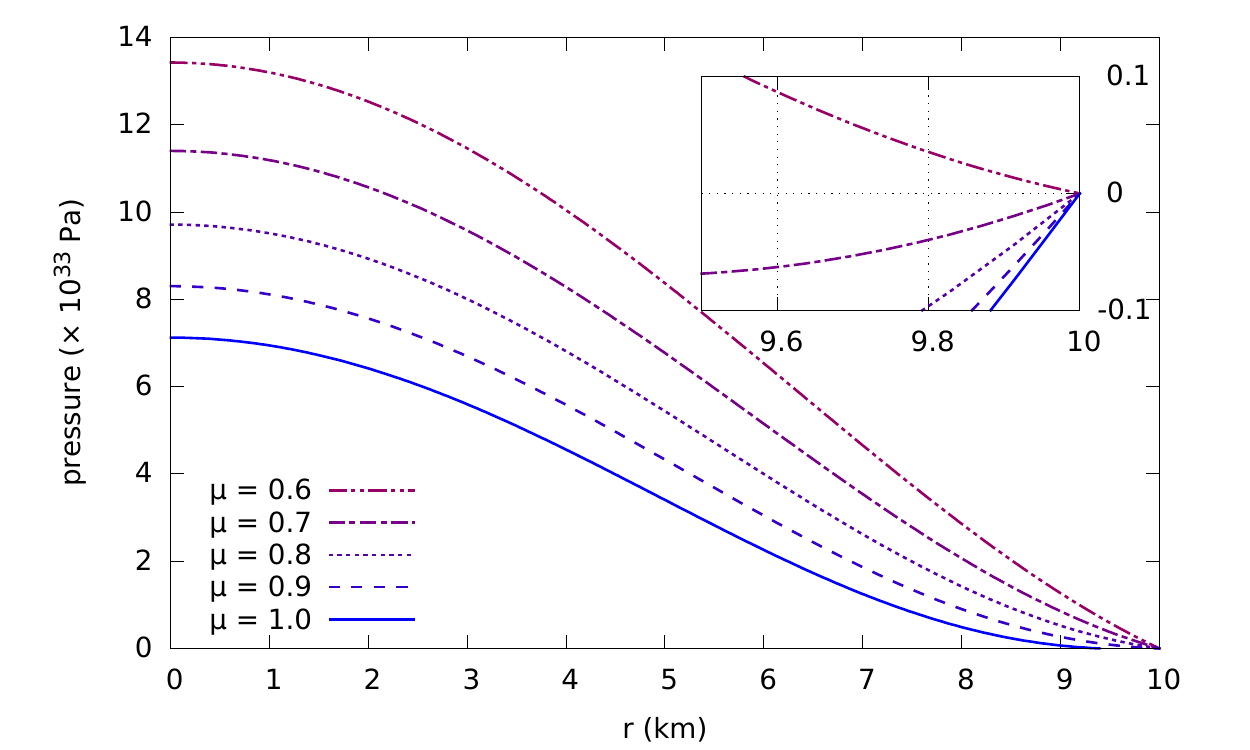}}
\caption[The tangential pressure for anisotropised charge ]{Variation of the tangential
  pressure with the radial coordinate inside the star. The parameter
  values are
  $\rho_{c}=\un{1\times 10^{18}} \dunits, r_b = \un{1 \times 10^{4}
    m},\, \beta$
  is fixed by $k$ in this solution and $\mu$ is given in the legend.}
\label{an.fig:AniCha,tangentialPressures}
\end{figure}
where we see a similar trend.  The tangential pressure however can be
negative, so we cannot further constrain our parameters just yet.  To
do that we turn to our next constraint~\ref{an.it.MonotonicMatter},
which tells us that all our matter fields must be decreasing function
of the radial coordinate.  We only need concern ourselves with the
pressures, and indeed the previous condition we derived had this same
flavour, and came about from the non monotonicity of the pressure.  We
can hence be confident that condition~\eqref{an.eq:pressureTurningPt}
above is exactly the constraint we require for this physical condition
to hold.

The next constraint~\ref{an.it.EnergyConditions} about the energy
conditions is more interesting.  We want the dominant energy condition
to hold, and in this particular case, this translates to
\begin{equation}
  \label{an.eq:AniChaEnergyCondition}
  3p_{r} + \rho -4k^{2}r^{2} \geq 0.
\end{equation}
This will also have to be tested when we are ready to try to model
stars, and will hopefully provide valuable insight into the possible
values of \(k.\)

If we now consider the speed of pressure waves according
to~\ref{an.it.CausalSpeed}, we find that in this case because
\(q = kr^{3},\)
and \(\Delta = 2k^{2} r^{2},\)
the expression reduces to
\begin{equation}
  \label{an.eq:AniChaSpeed}
v_{s}^{2} = 
\left( \f{r_{b}^{2}}{\kappa \rho_{c} \mu}\right)\left[ \f{\nu' \kappa \left( p_{r} + \rho \right)}{4 r} +k^{2} \right],
\end{equation}
a speed that is larger that the Tolman~VII case by precisely the
charge factor \(k^{2}.\)
As a result we expect the parameter value that we can use in this case
to have to be slightly lower than previously, since we still want to
maintain causality: \(v_{s} = \un{c},\)
at the centre of the star.  We show the behaviour of the speed of
sound in figure~\ref{an.fig:AniCha,SoundSpeed}.
\begin{figure}[!htb]
\subfloat[The speed of sound, $v_{s},$ with $k=1 \times 10^{-9} m$]{\label{an.fig:AniCha,Speedk=1e-9}
  \includegraphics[width=0.5\linewidth]{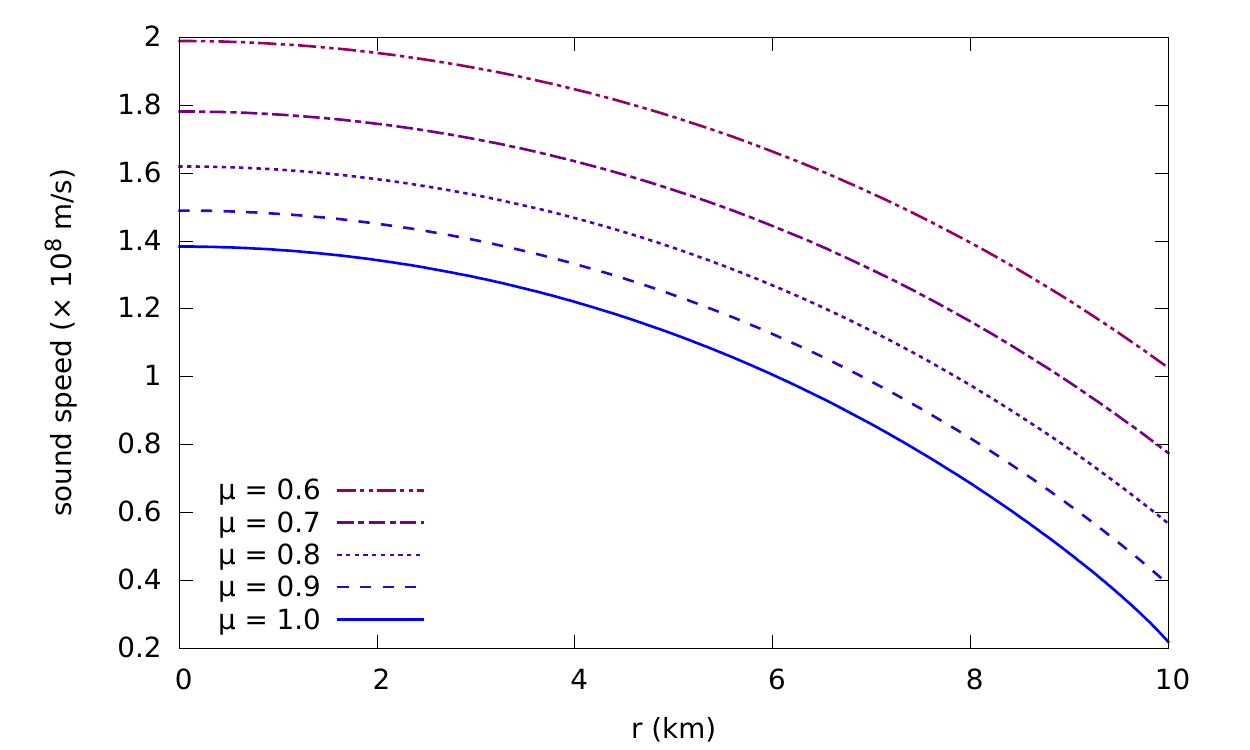}} 
\subfloat[The speed of sound, $v_{s},$ with $k=3 \times 10^{-9} m$ ]{\label{an.fig:AniCha,Speedk=3e-9}
  \includegraphics[width=0.5\linewidth]{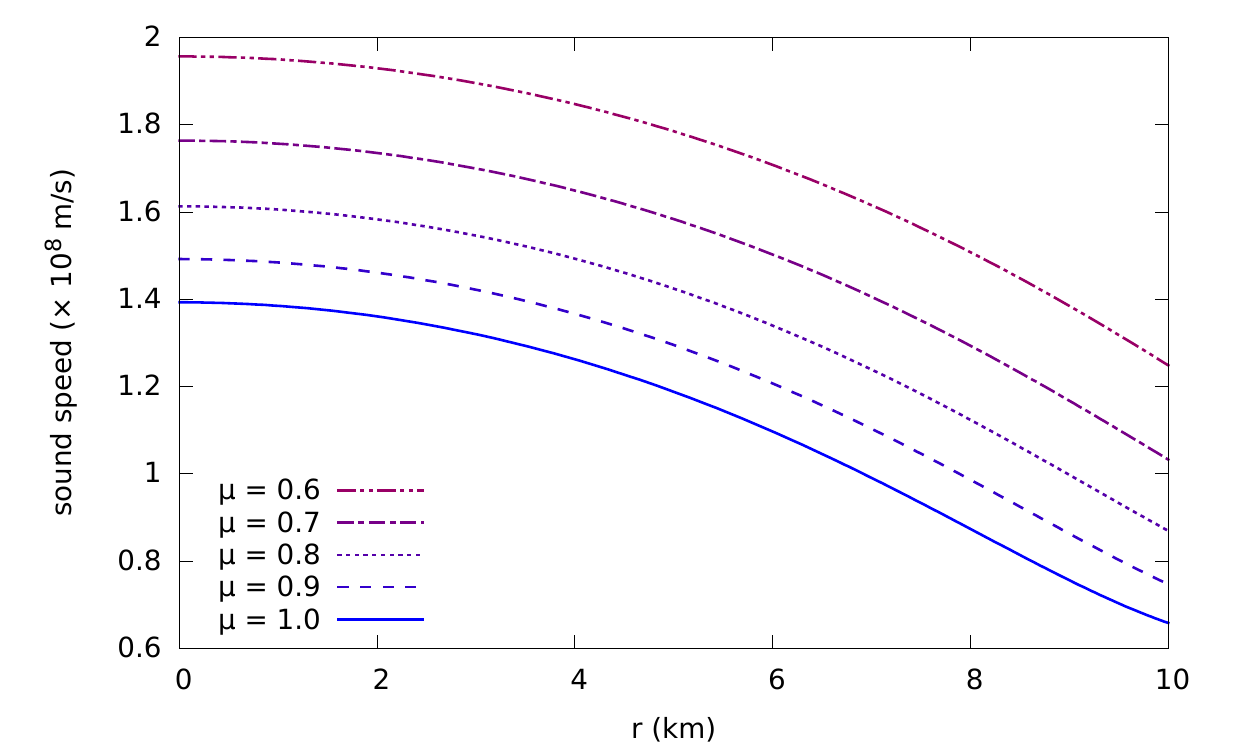}}\\
\subfloat[The speed of sound, $v_{s},$ with $\mu=1$ ]{\label{an.fig:AniCha,speedMu=1} 
  \includegraphics[width=0.5\linewidth]{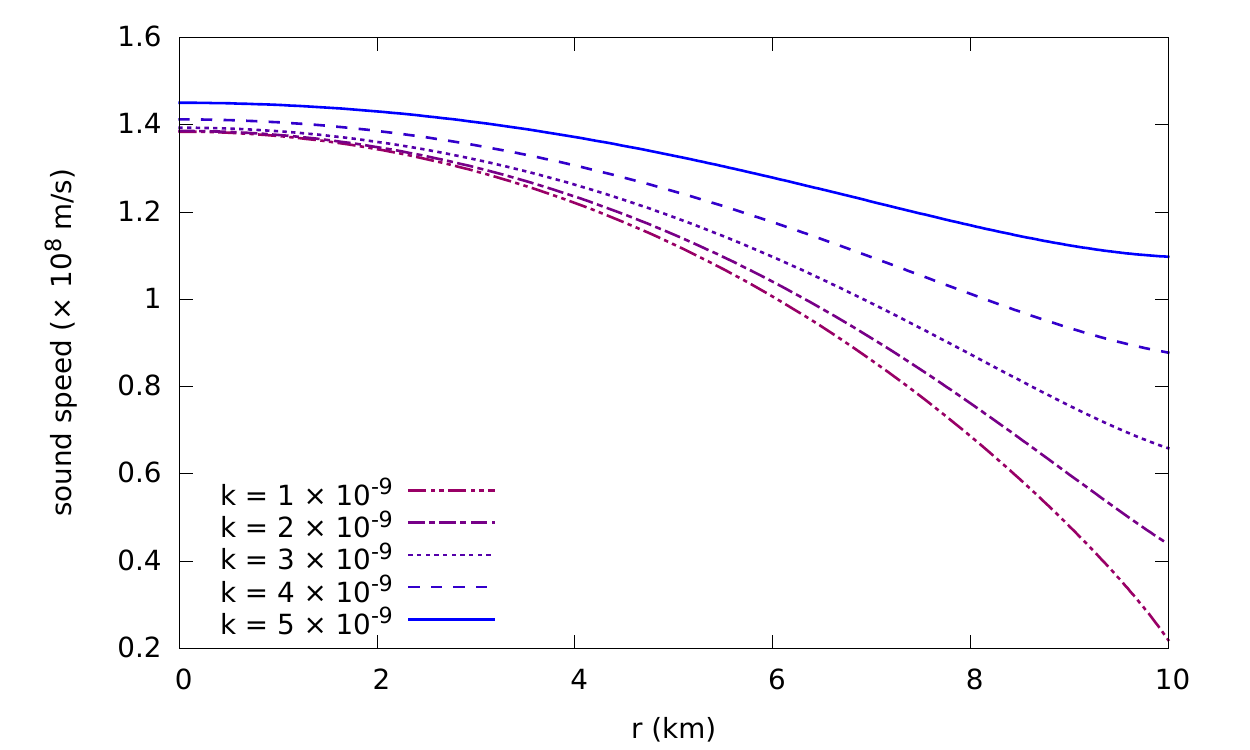}}
\subfloat[The speed of sound, $v_{s},$ with $\mu=0.6$ ]{\label{an.fig:AniCha,speedRMu=0.6}
  \includegraphics[width=0.5\linewidth]{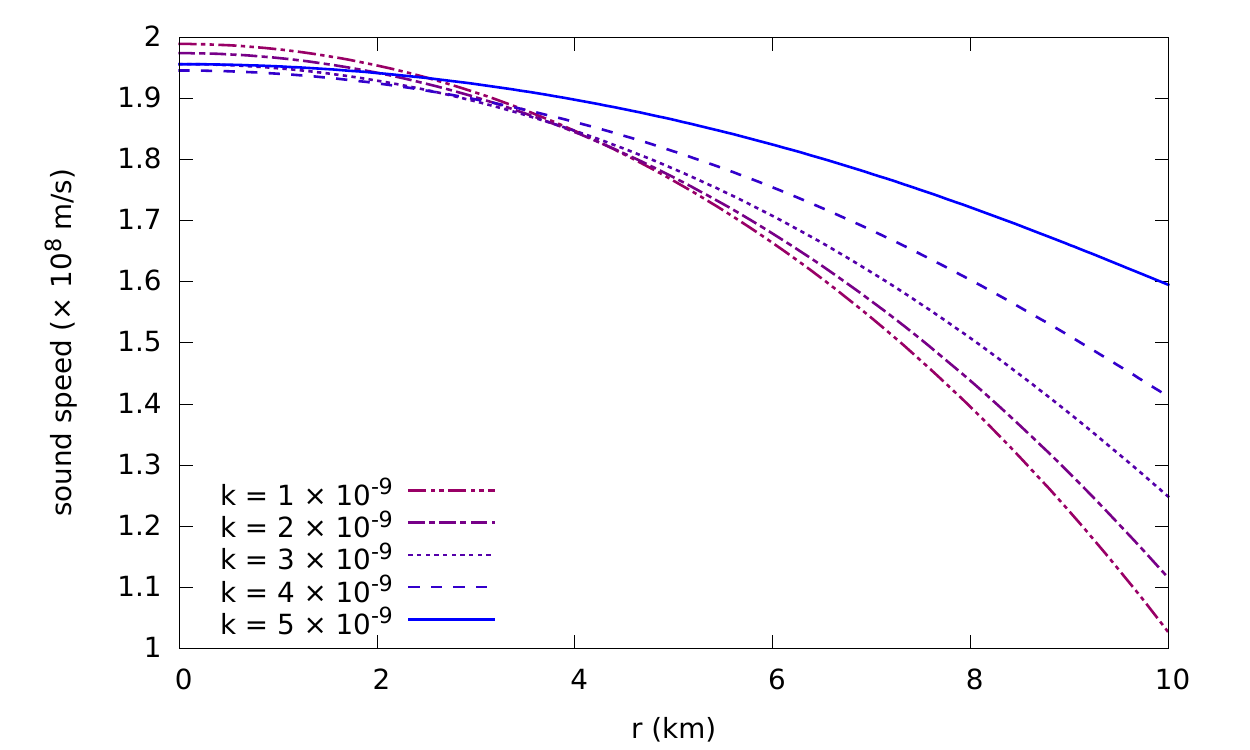}} 
\caption[The speed of sound with $r$ ]{Variation of speed of sound
  with the radial coordinate inside the star. The parameter values are
  $\rho_{c}=\un{1\times 10^{18}} \dunits, r_b = \un{1 \times 10^{4}
    m}, \mu$ is fixed in the bottom two plots at one on the left, and
  0.6 on the right, but takes on the values in the legend for the top
  graphs. $k$ is set respectively to $1 \times 10^{-9} \un{m} $ and
  $3 \times 10^{-9} \un{m}$ in the left and right top plots, but
  varies in the bottom ones.}
\label{an.fig:AniCha,SoundSpeed}
\end{figure}
The other conditions seen previously will be useful in determining
constraints on \(k,\)
and hopefully this one can be used for the other parameters, as we did
in Tolman~VII.  Similarly the next
condition~\ref{an.it.DecreasingSpeed} on the derivative of the speed
of sound will also be useful to restrict the other parameters.  We
can see from the form of~\eqref{an.eq:AniChaSpeed} that the derivative
with respect to \(r\)
is not going to be affected by the additional \(k^{2}\)
factor in that expression.  However this is incorrect since \(\nu'\)
is \(k\)
dependent too in this solution, but since we can see the behaviour of
the speed of sound in figure~\ref{an.fig:AniCha,SoundSpeed}, we see
that this condition is mostly satisfied, except
in~\ref{an.fig:AniCha,speedMu=1}, for large values of charge which we
had already determined to be too large.

This completes the constraints section of this solution.  Like the
previous solution, this solution looks to be viable too as a model for
neutron stars, if we consider the ``self-bound'' ones with
\(\mu \neq 1,\)
since not abiding by this constraint gives us negative radial
pressures.  We will look at this solution in detail too in the next section.

\FloatBarrier\subsection{The charged case, $\Phi^{2} = 0$, derived in \ref{ns.ssec:phi0}}
In this charged and anisotropic case, we annihilate the coefficient of
\(Y\)
in our differential equation to get the simplest solution of the
charged case.  By doing so, we express the measures of charge \(k,\)
anisotropy \(\beta,\)
and Tolman~VII parameters \(\{\rho_{c}, \mu, r_{b}\}\)
in terms of each other.  The metric function for \(Y\)
then becomes a simple linear function of the radial-like coordinate
\(\xi.\)  We now look at the behaviour of this solution, while applying all the
constraints, up to the point at which we run into unphysical
behaviour.

The first constraint we look at is the regularity of the metric \(Z,\)
and in particular its unchanging sign from the centre of the star
where its value is \(Z(0) = 1.\)
We require that even at the boundary radius, this metric function
remain positive, and this
yields\[Z(r_{b}) = 1 - \left( \f{\kappa \rho_{c} r_{b}^{2}}{3} \right)+ \f{2}{11} \left( \kappa\mu\rho_{c}r_{b}^{2} -\f{\beta r_{b}^{4}}{2} \right) > 0,\] which on simplification results in an immediate limit on \(\beta:\)
\begin{equation}
  \label{an.eq:betaLim1}
\beta < \f{11}{2r_{b}^{2}} \left[ 1 + \kappa \rho_{c} r_{b}^{2} \left( \f{2\mu}{11} - \f{1}{3} \right) \right].
\end{equation}
With our typical values for the constants above,
\(\{\rho_{c}=1\times10^{18} \dunits, r_{b}=10^{4} \un{m}\}\)
we get a function of \(\mu,\) graphed in figure~\ref{an.fig:AniCha,betaLim}.
\begin{figure}[!htb]
  \includegraphics{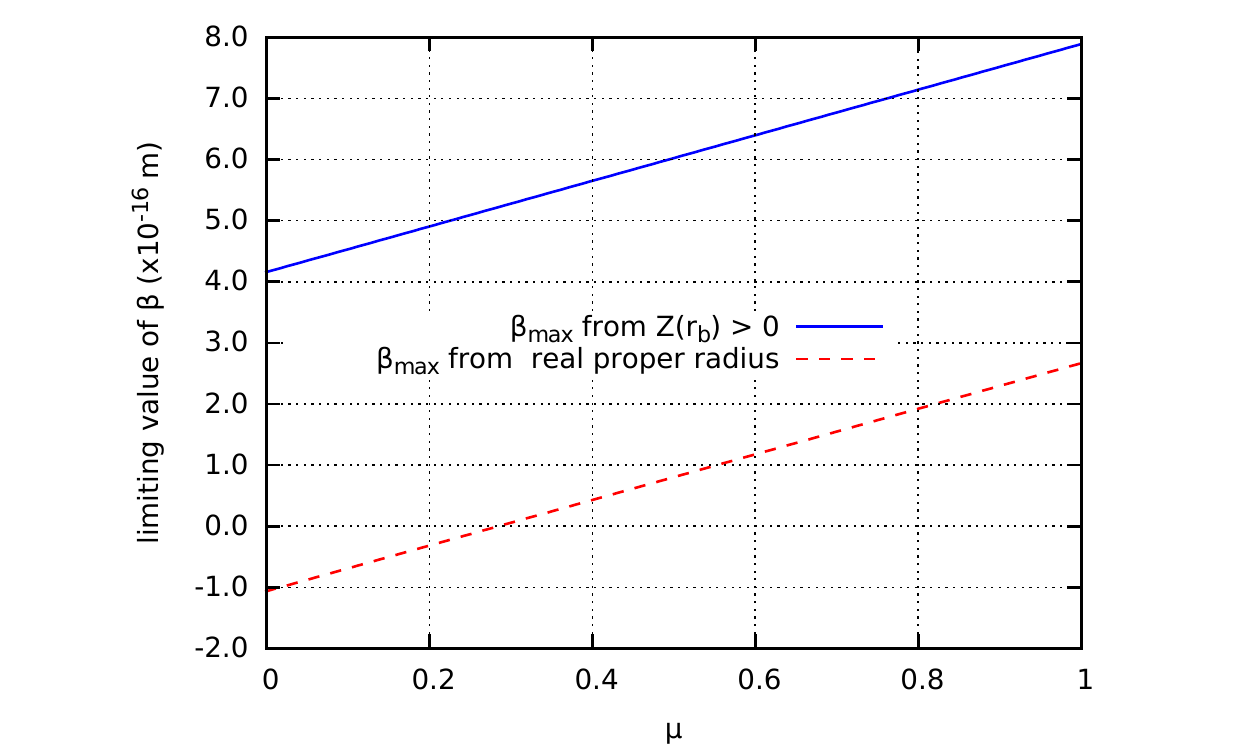}
\caption[The limiting value of $\beta$ for different $\mu$]{The limiting value of $\beta$ for different $\mu.$ Here the parameters are taken to be  $\rho_{c} = 1\times 10^{18} \un{kg \cdot m^{-3},}$ and the boundary radius $r_{b}=10000\un{m},$ and show the two different constraints we are looking at.}
\label{an.fig:AniCha,betaLim}
\end{figure}
For the next constraint we turn to the fact that the value of the
proper radius has to exist, and as discussed previously, this means
that \(2\sqrt{a} > b.\)
In this particular case, this relation gives us that
\begin{equation}
  \label{an.eq:betaLim2}
\beta < \kappa\rho_{c}\left( \f{2\mu}{r_{b}^{2}} - \f{11\kappa\rho_{c}}{36},\right),
\end{equation}
another more restrictive inequality on \(\beta.\)
From this one it is clear that for some values of \(\mu\)
we will need zero or even negative \(\beta.\)
We keep this in mind as we proceed, and for the time being restrict
\(\mu > 0.4,\) so as to have positive \(\beta\) only.

Next we look at the metric functions, and their general behaviour, for
the restricted values of \(\beta\)
we just found.  This is shown in
figure~\ref{an.fig:CAPhiZ,MetricCoeff}
\begin{figure}[!htb]
\subfloat[The $Y(r)$ metric function]{\label{an.fig:YmetricCAPhiZ}
  \includegraphics[width=0.5\linewidth]{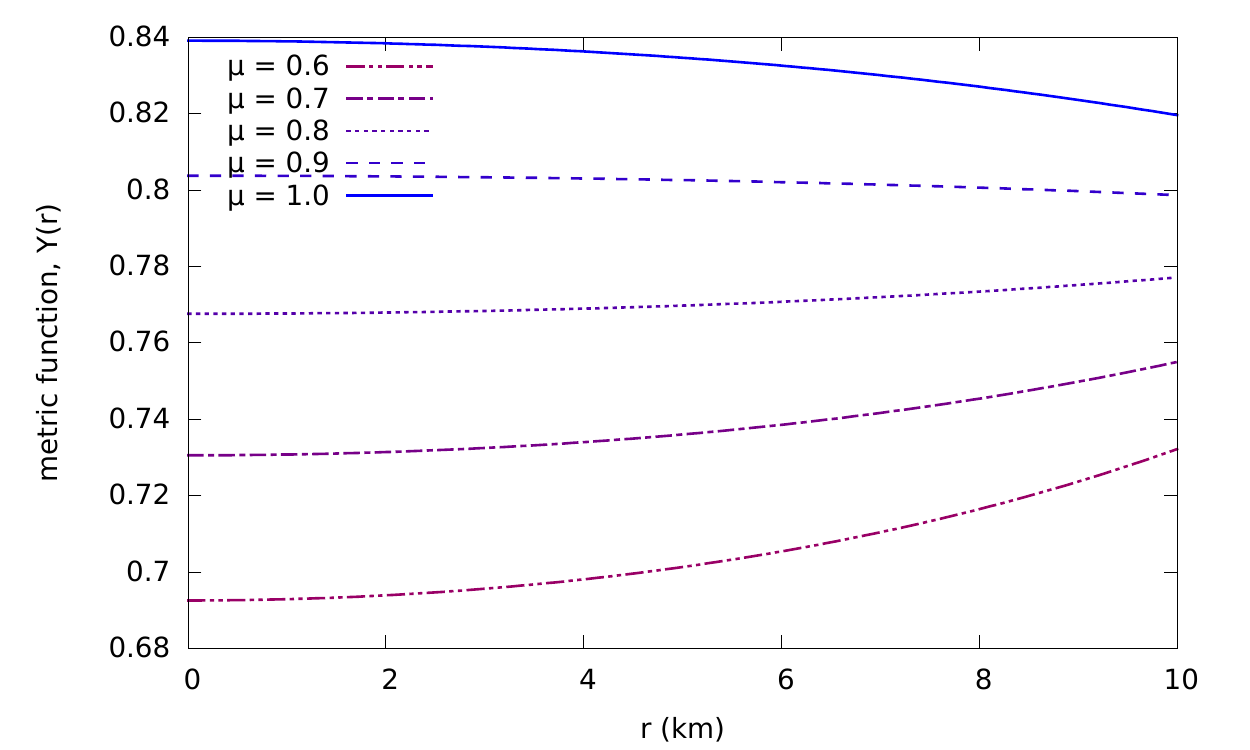}} 
\subfloat[The $Z(r)$ metric function]{\label{an.fig:ZmetricCAPhiZ}
  \includegraphics[width=0.5\linewidth]{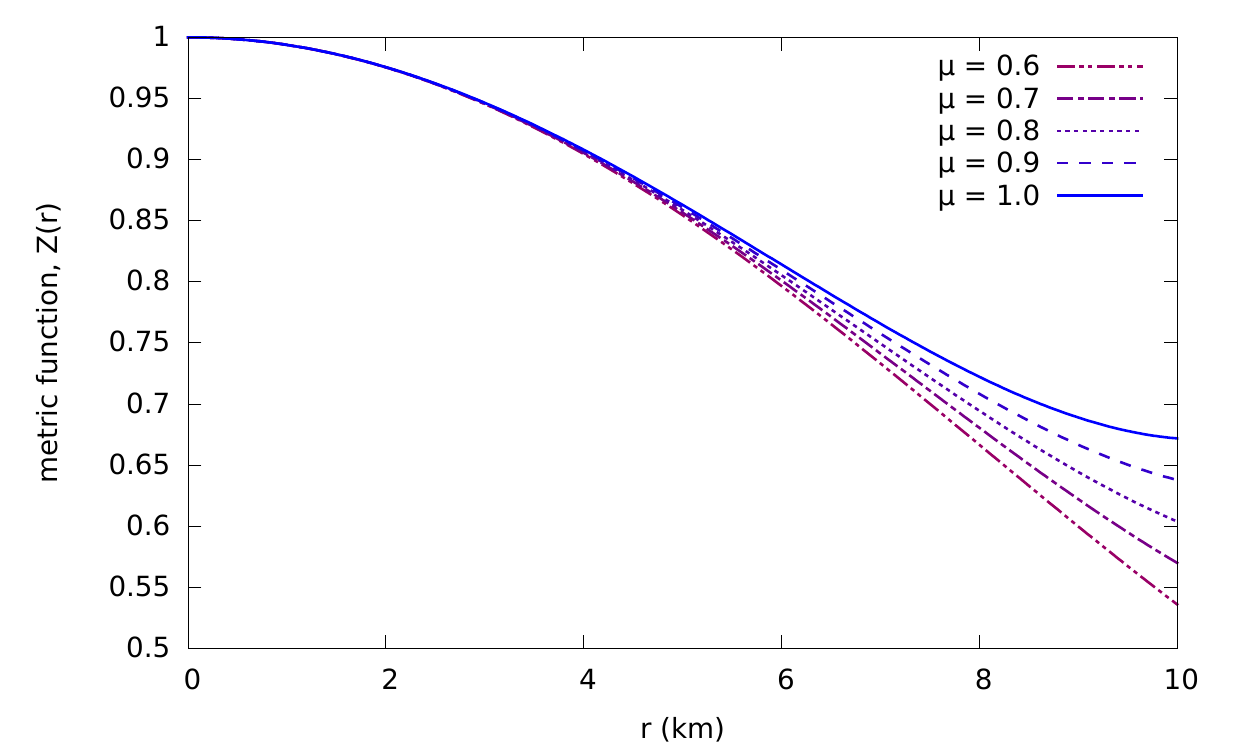}}\\
\subfloat[The $\lambda(r)$ metric function]{\label{an.fig:LambdaMetricCAPhiZ} 
  \includegraphics[width=0.5\linewidth]{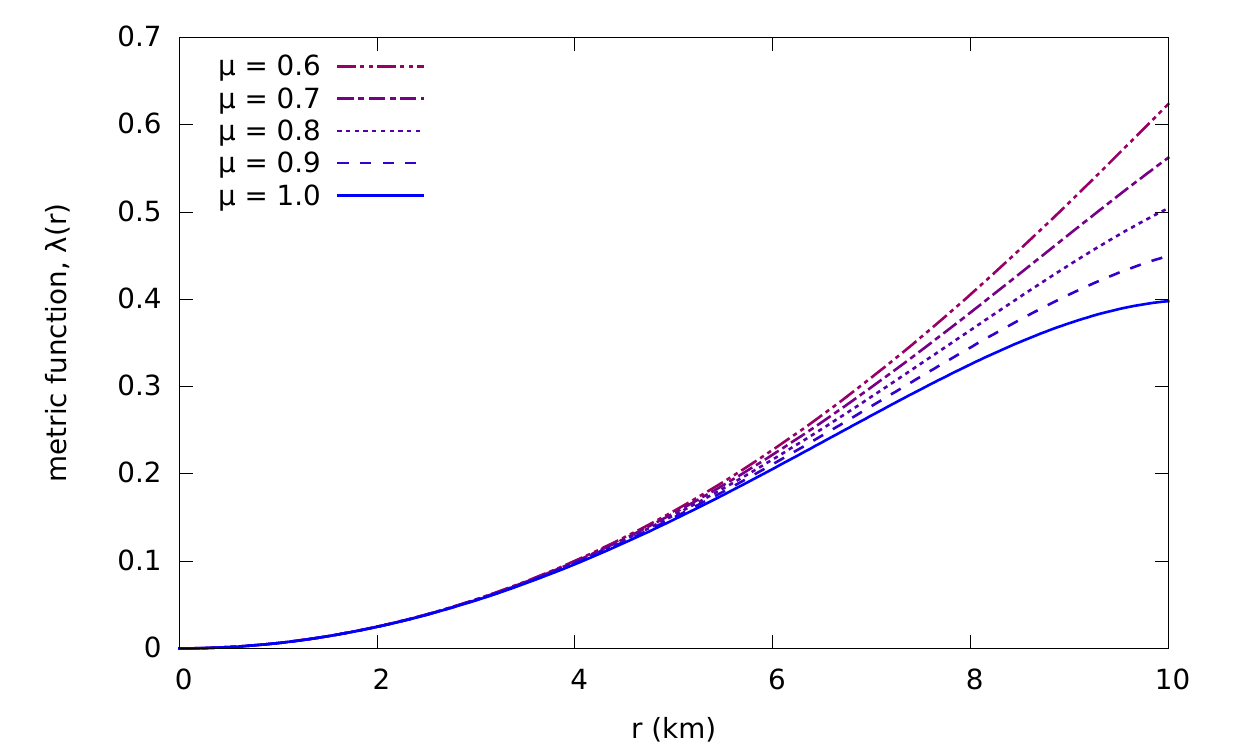}}
\subfloat[The $\nu(r)$ metric function]{\label{an.fig:NuMetricCAPhiZ}
  \includegraphics[width=0.5\linewidth]{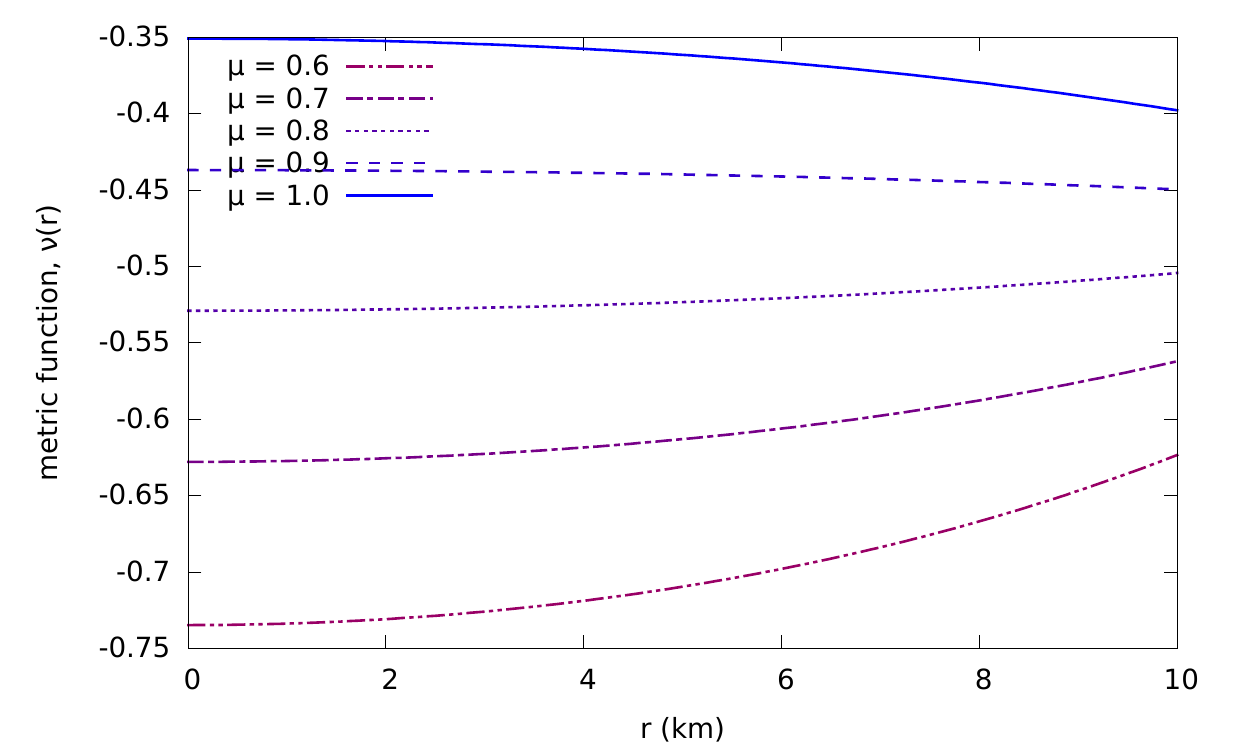}} 
\caption[Variation of metric variables for $\Phi^{2} = 0$]{Variation
  of metric variables with the radial coordinate inside the star. The
  parameter values are
  $\rho_{c}=\un{1\times 10^{18}} \dunits, r_b = \un{1 \times 10^{4}
    m}, \beta = \un{5 \times 10^{-17} m} $ and $\mu$ taking the
  various values shown in the legend }
\label{an.fig:CAPhiZ,MetricCoeff}
\end{figure}
The only striking feature here is the strange behaviour of the \(Y\)
and hence \(\nu\)
metric function, whose derivatives seem to be of either sign.  This is
a sign of trouble, since the metric derivatives have to behave
smoothly, and here it seems that for some parameter values, the metric
could be constant.  Suspecting unphysicality, we turn to the next
condition~\ref{an.it.DefiniteRPressure} which requires positive
pressures.  
\begin{figure}[!htb]
\subfloat[The radial pressure , $k=1\times10^{-9}$]{\label{an.fig:CAPhiZ,Pk=1e-9}
  \includegraphics[width=0.5\linewidth]{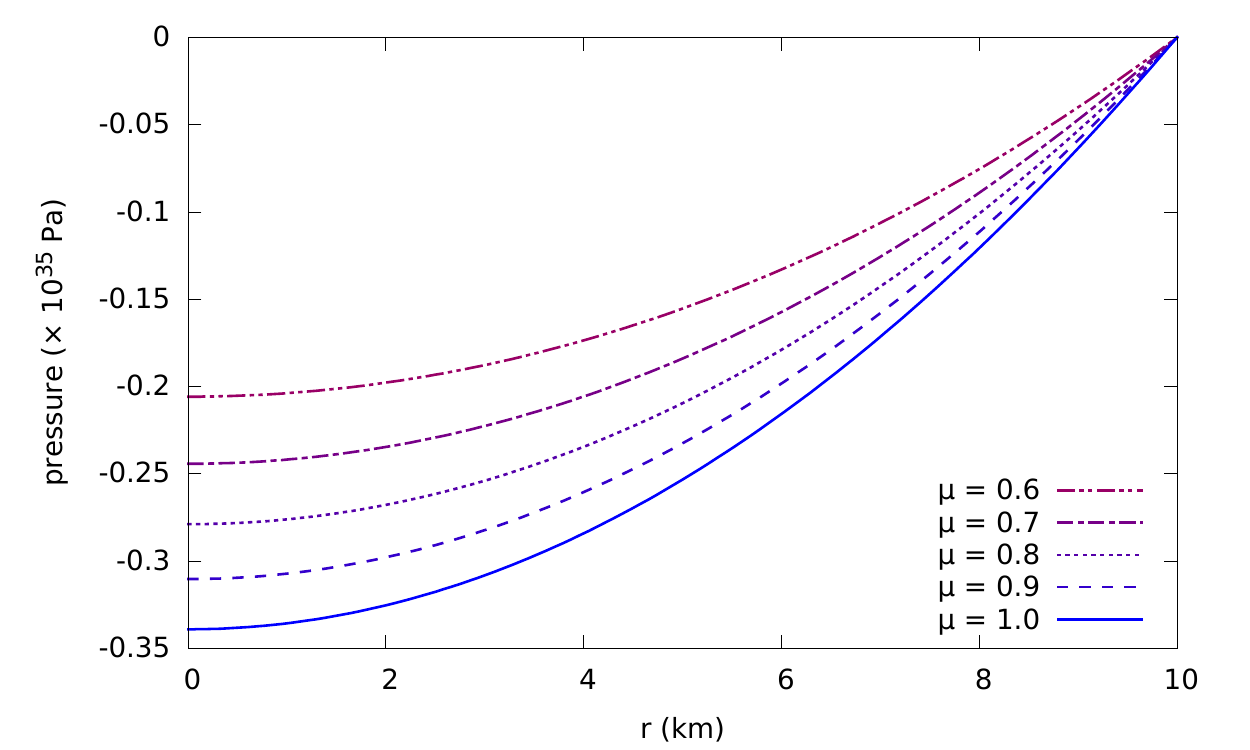}} 
\subfloat[The radial pressure, $k=3\times10^{-9}$]{\label{an.fig:CAPhiZ,Pk=3e-9}
  \includegraphics[width=0.5\linewidth]{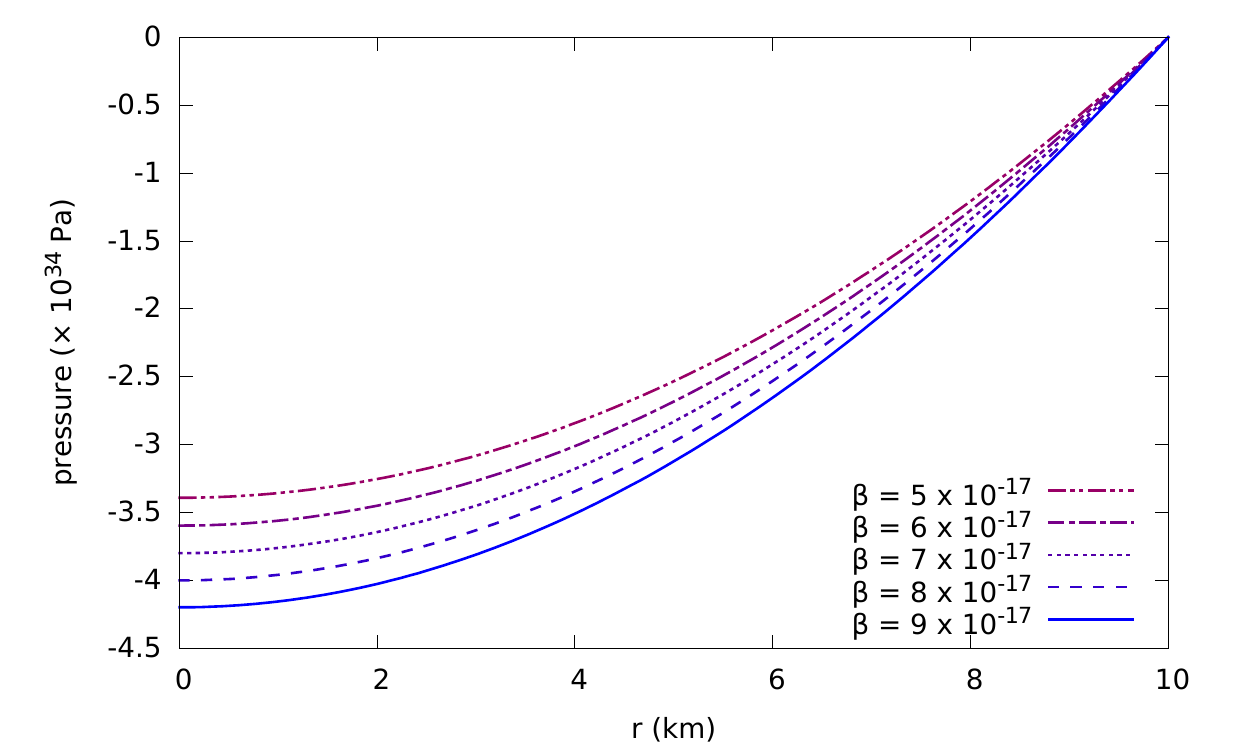}} 
\caption[Variation of radial pressure ]{Variation of the radial
  pressure with the radial coordinate inside the star. The parameter
  values are
  $\rho_{c}=\un{1\times 10^{18}} \dunits, r_b = \un{1 \times 10^{4}
    m},\, \beta$
  is fixed to $5 \times 10^{-17}$ in left panel and $\mu$ is given in the legend, while
  for the right panel  $\mu$ is fixed to 1, and $k$ varies as shown in the legend}
\label{an.fig:CAPhiZ,radialPressures}
\end{figure}
We show in figure~\ref{an.fig:CAPhiZ,radialPressures} how the pressure
changes while varying both the value of the anisotropy \(\beta,\)
and self-boundness \(\mu,\)
but unfortunately find that in both cases, we only get negative
pressures.  As a result, we forgo this particular solution as
unphysical, and do not waste time ensuring any of the other conditions
hold.

\FloatBarrier\subsection{The charged case, $\Phi^{2} < 0$, derived in
  \ref{ns.ssec:phiN}} We now turn to this more general case where
none of the constants are specified or fixed at the beginning.  To get
to this class of solutions, we have to ensure that
\(a + \beta -2k^{2} < 0,\)
as mentioned previously.  The numerical values, and ranges surmised
for these constants from the previous sections will provide a
guideline for what value we pick initially for our plots, but we will
go through a formal derivation from the conditions here too, to check
if any of the conditions yield different constraints.

The definiteness of the metric functions and observables, or
conditions~\ref{an.it.RegularMetric}--~\ref{an.it.ExteriorObservables}
yield the same constraints on \(k\)
as as~\eqref{an.eq:ZconstraintCA} and~\eqref{an.eq:RconstraintCA},
since \(Z\)
does not depend explicitly on \(\beta.\)
As a result we maintain a charge density limit of
\(|k| < 1.2 \times 10^{8} \un{C \cdot m^{-3}},\)
for the usual parameter values of
\(\rho_{c} = 1 \times 10^{18} \un{kg \cdot m^{-3}},\)
and \(r_{b} = 10^{4} \un{m}.\)
We now give plots of the metric functions for this particular case in
figure~\ref{an.fig:CAPhiN,MetricCoeff}, where we notice that they are
all well behaved.  The surprising change in sign for the \(\nu\)
metric function is of no concern, since it is \(\e^{\nu}\)
that appears in the line element, and that function does not change
sign.
\begin{figure}[!htb]
\subfloat[The $Y(r)$ metric function]{\label{an.fig:YmetricCAPhiZ}
  \includegraphics[width=0.5\linewidth]{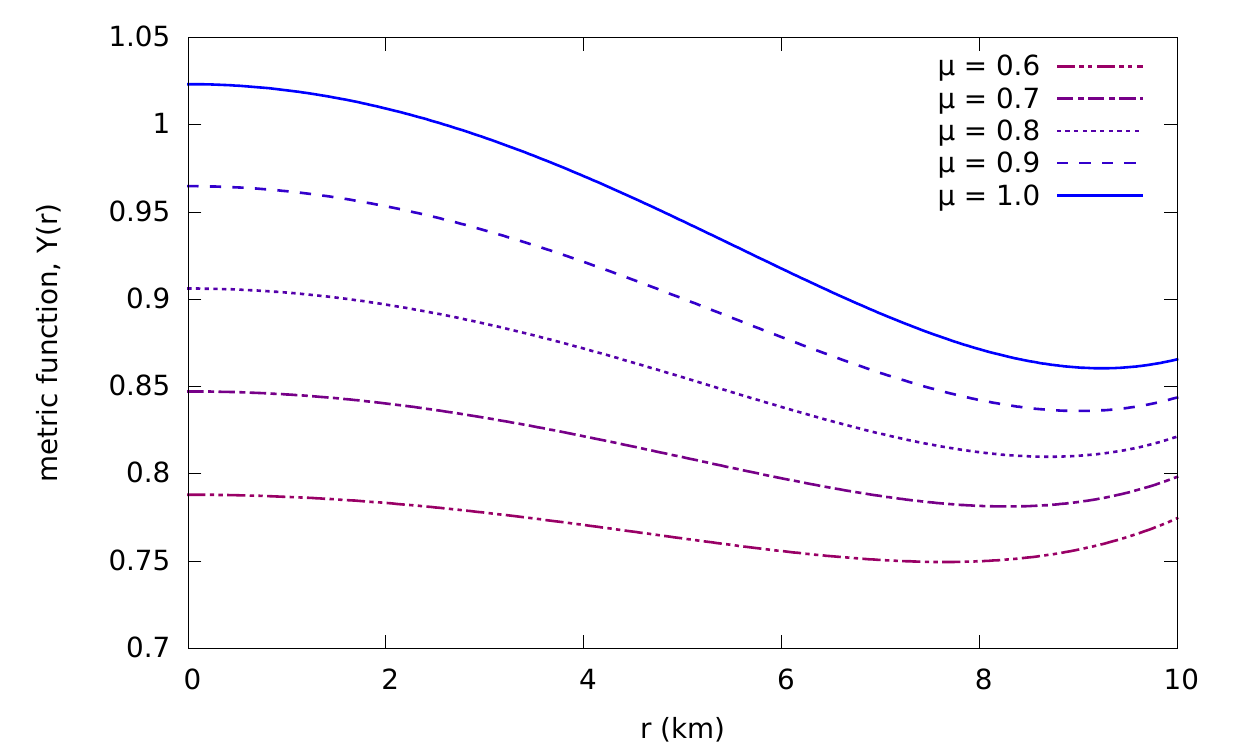}} 
\subfloat[The $Z(r)$ metric function]{\label{an.fig:ZmetricCAPhiZ}
  \includegraphics[width=0.5\linewidth]{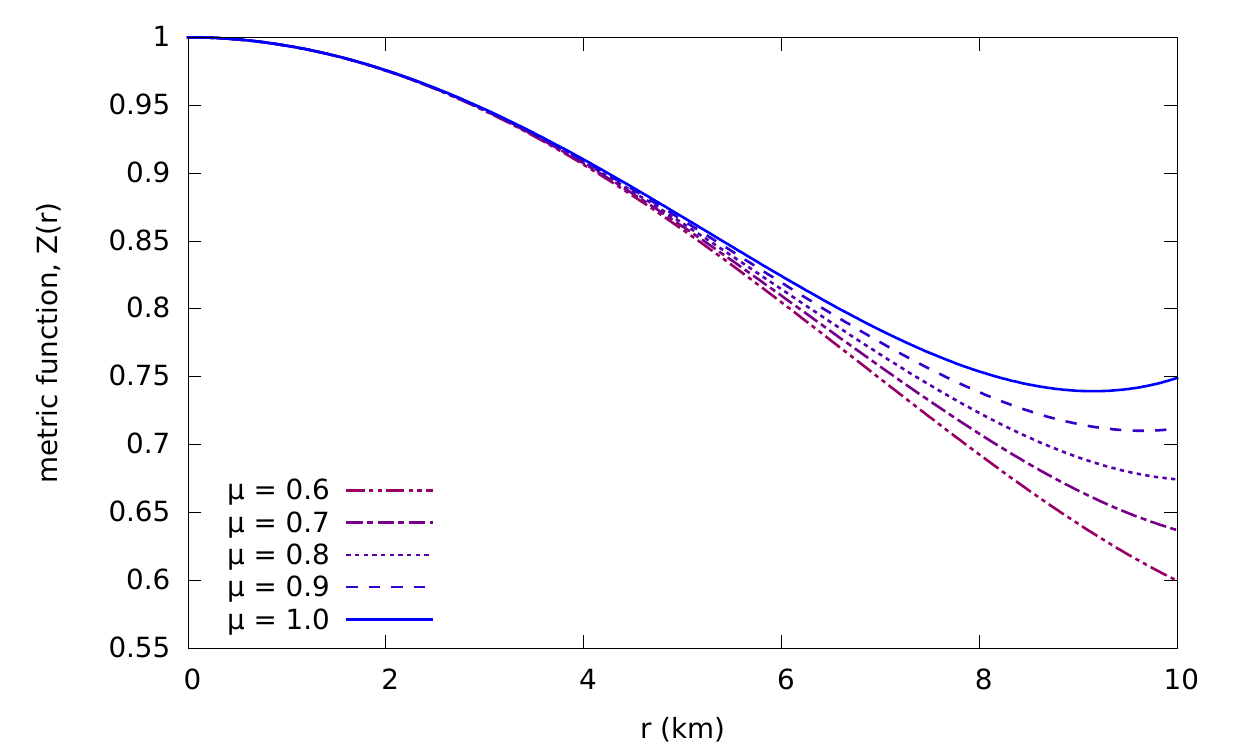}}\\
\subfloat[The $\lambda(r)$ metric function]{\label{an.fig:LambdaMetricCAPhiZ} 
  \includegraphics[width=0.5\linewidth]{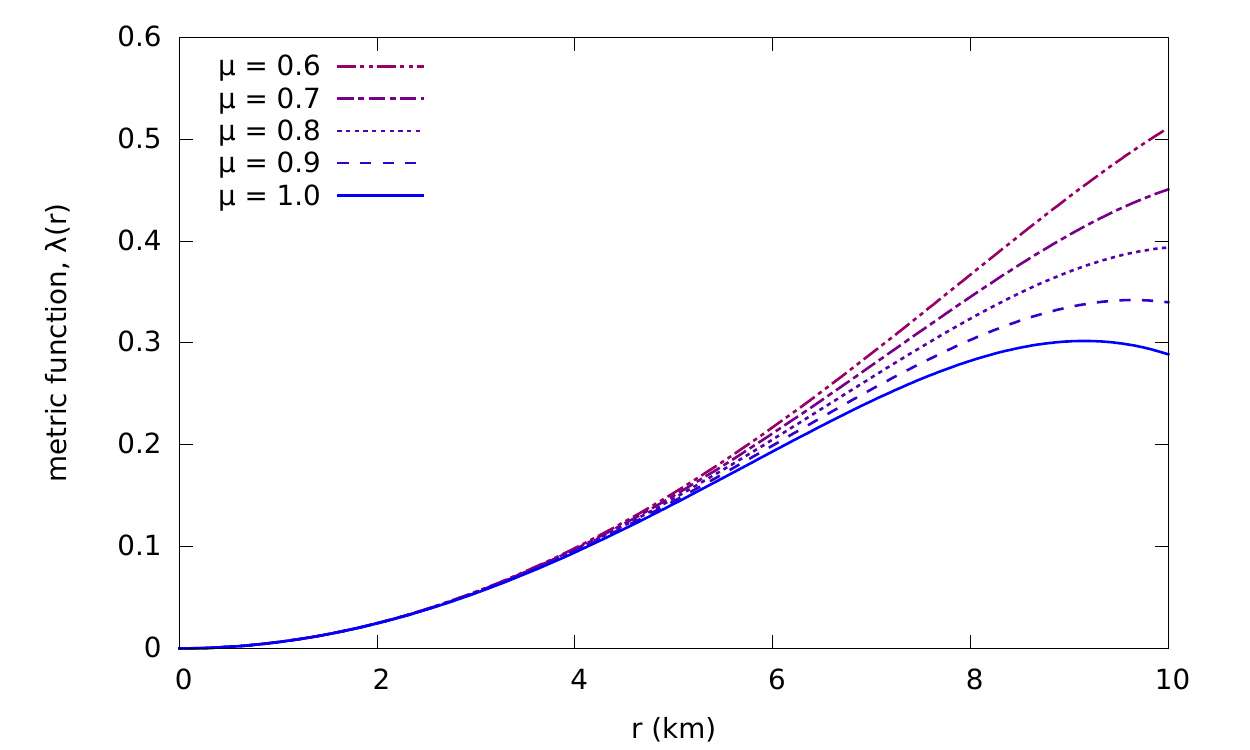}}
\subfloat[The $\nu(r)$ metric function]{\label{an.fig:NuMetricCAPhiZ}
  \includegraphics[width=0.5\linewidth]{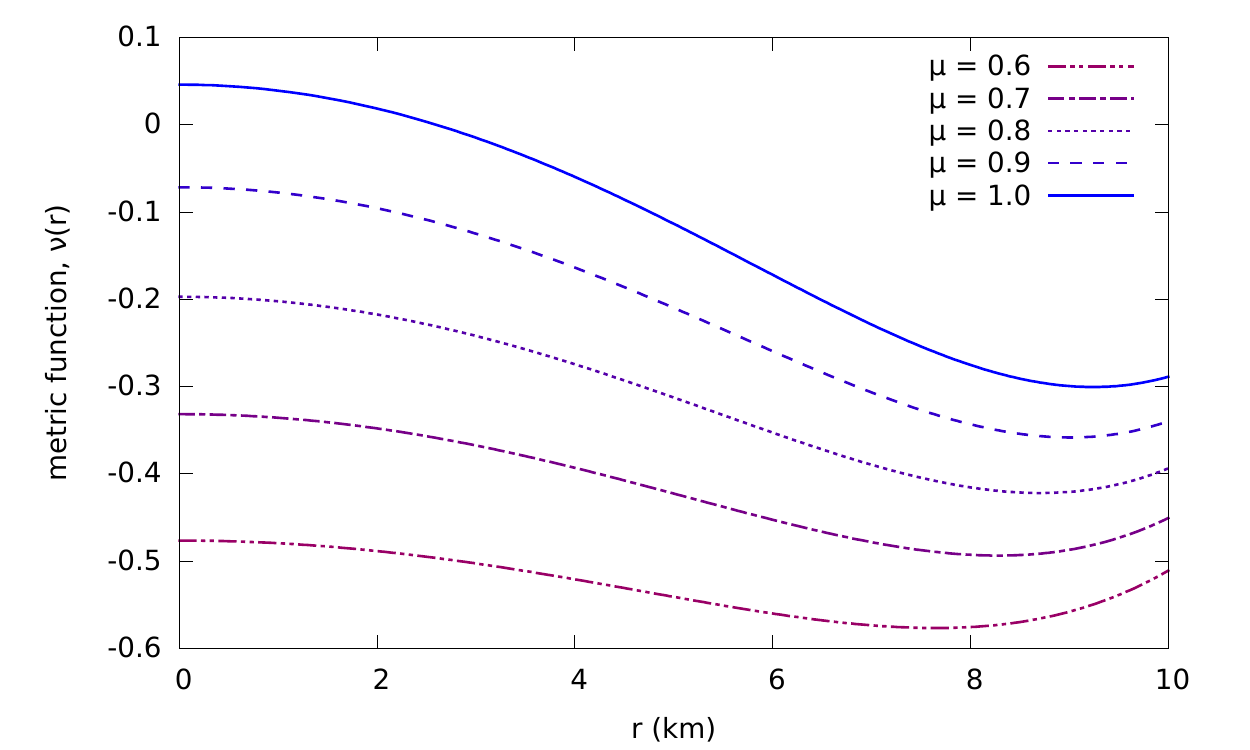}} 
\caption[Variation of metric variables for $\Phi^{2} < 0$]{Variation of metric variables with the radial coordinate
  inside the star. The parameter values are $\rho_{c}=\un{1\times
    10^{18}} \dunits, r_b = \un{1 \times 10^{4} m}, \beta = \un{-2 \times 10^{-16} m}, k = 1 \times 10^{-9} \un{m^{-2}} $ and $\mu$ taking the various values shown in the legend }
\label{an.fig:CAPhiN,MetricCoeff}
\end{figure}
Next we look at condition~\ref{an.it.DefiniteRPressure} on the
pressure.  From the general trend of how these solutions have worked
so far, we suspect that we will get negative pressures, and indeed
this is exactly what we find, as exemplified in figure~\ref{an.fig:CAPhiN,radialPressures}
\begin{figure}[!htb]
\centering
\subfloat[Varying $\mu$]
{\includegraphics[width=0.5\textwidth]{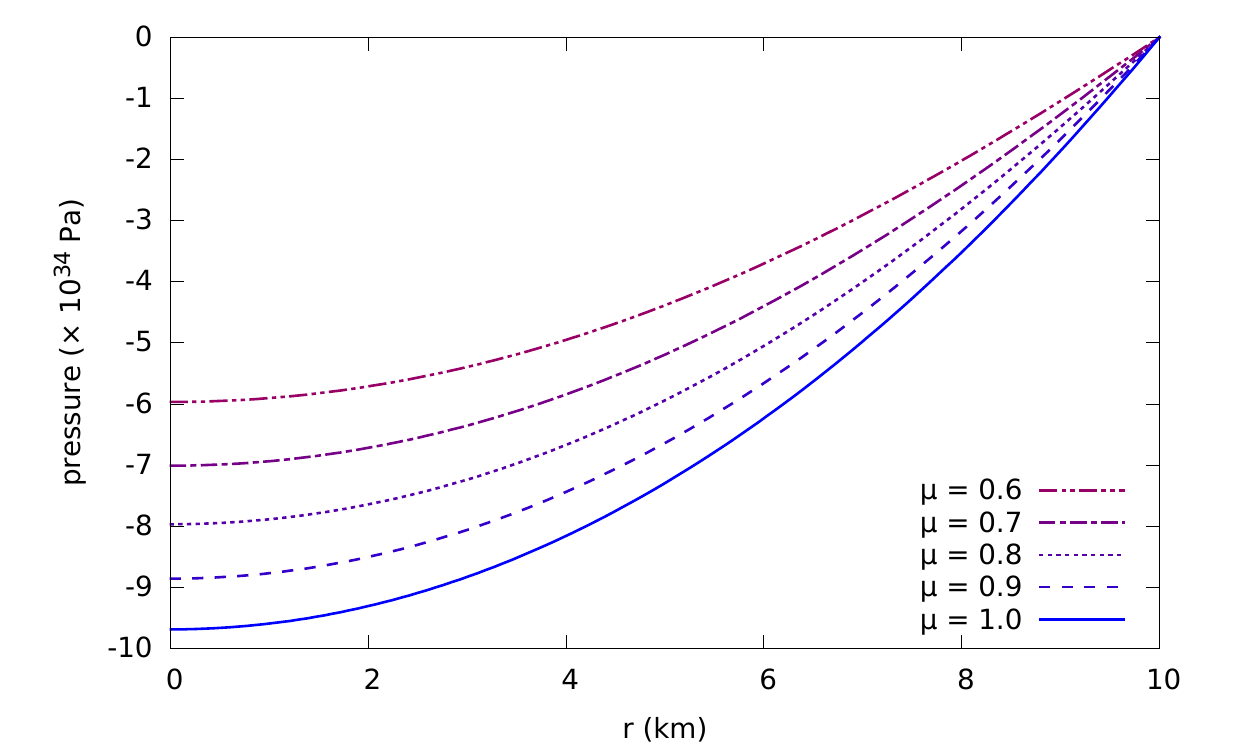}}
\subfloat[Varying $k$]
{\includegraphics[width=0.5\textwidth]{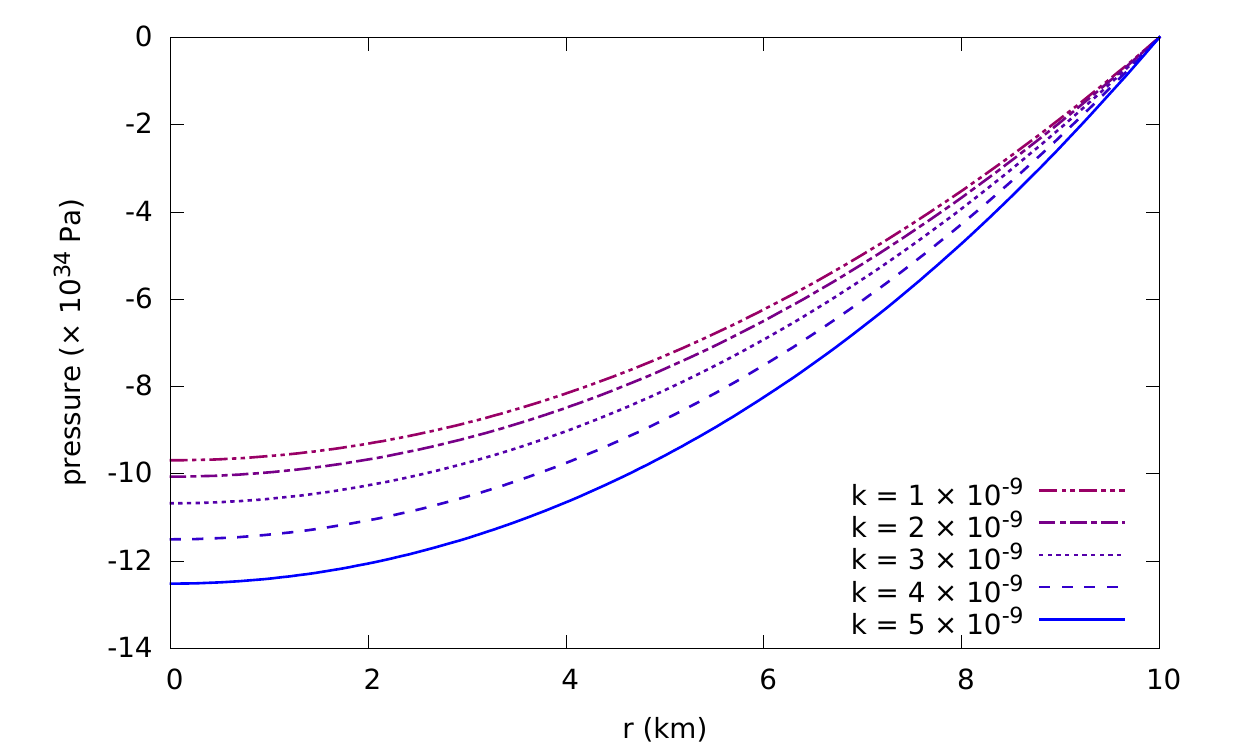}}\\
\subfloat[Varying $\beta$]
{\includegraphics[width=0.5\textwidth]{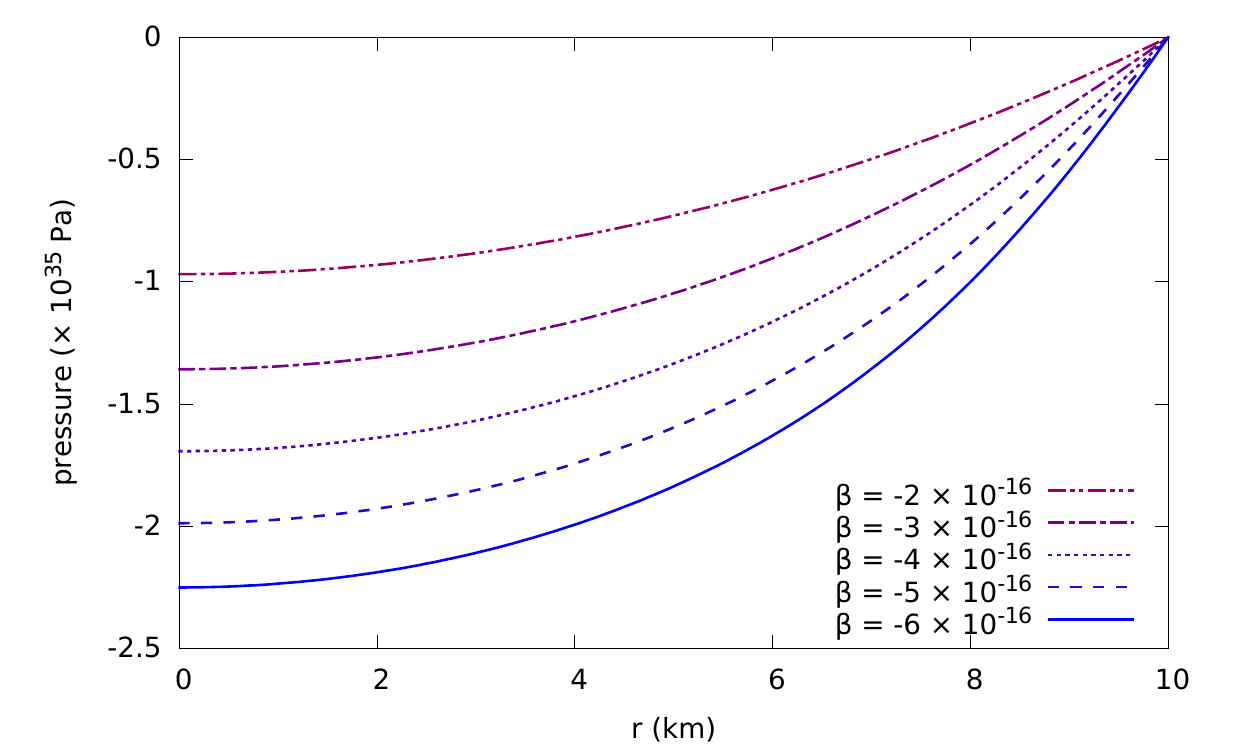}} 
\caption[Variation of radial pressure for $\Phi^{2} < 0$]{Variation of the
  radial pressure with the radial coordinate inside the star. The
  parameter values are
  $\rho_{c}=\un{1\times 10^{18}} \dunits, r_b = \un{1 \times 10^{4}
    m},\, \beta$
  is fixed to $(-2 \times 10^{-16})\times \mu$ in left and middle
  panel, while it is fixed to $(-2 \times 10^{-16})$ in the right
  one.$k$ is fixed to $1 \times 10^{-9} \un{m^{-2}}$ in the right and left
  plots, and changes according to the legend in the middle one. $\mu$
  is fixed to 1 in the middle and right panels, and varies as shown in
  the legend in the left panel}
\label{an.fig:CAPhiN,radialPressures}
\end{figure}
As a result we leave off this solution as being unphysical, and turn
to the next one that looks more promising since it is a very
straight-forward generalisation of Tolman~VII with charge and
anisotropic pressures as the ``bells-and-whistle.''

\FloatBarrier\subsection{The charged anisotropic case with $\Phi^{2} >0$, derived in
  \ref{ns.ssec:phiP}} In this case too, \(\beta\)
is no longer fixed to one value: instead it takes on a range of
possible values and as long as the inequality
\(\beta > \f{1}{5} \left( 11k^2 - \f{\kappa \mu
    \rho_c}{r_b^2}\right)\)
is satisfied, the value of \(\Phi\)
will be appropriate for this solution.  Because of the above
inequality, this solution offers us the possibility of having
different signs for \(\beta: \)
since the latter could be negative and we would still have a positive
\(\Phi^{2}.\)
However we remember the previous case where we only had anisotropy,
and negative \(\beta\)s
only gave us negative pressures.  Here the situation seems even worse
because the charge \(k\)
already comes with a negative sign, suggesting we might run into
trouble right at the beginning.

We start as we have been doing by looking at the behaviour of the
metric functions in figure~\ref{an.fig:CAPhiP,MetricCoeff},
\begin{figure}[!htb]
\subfloat[The $Y(r)$ metric function]{\label{an.fig:YmetricCAPhiP}
  \includegraphics[width=0.5\linewidth]{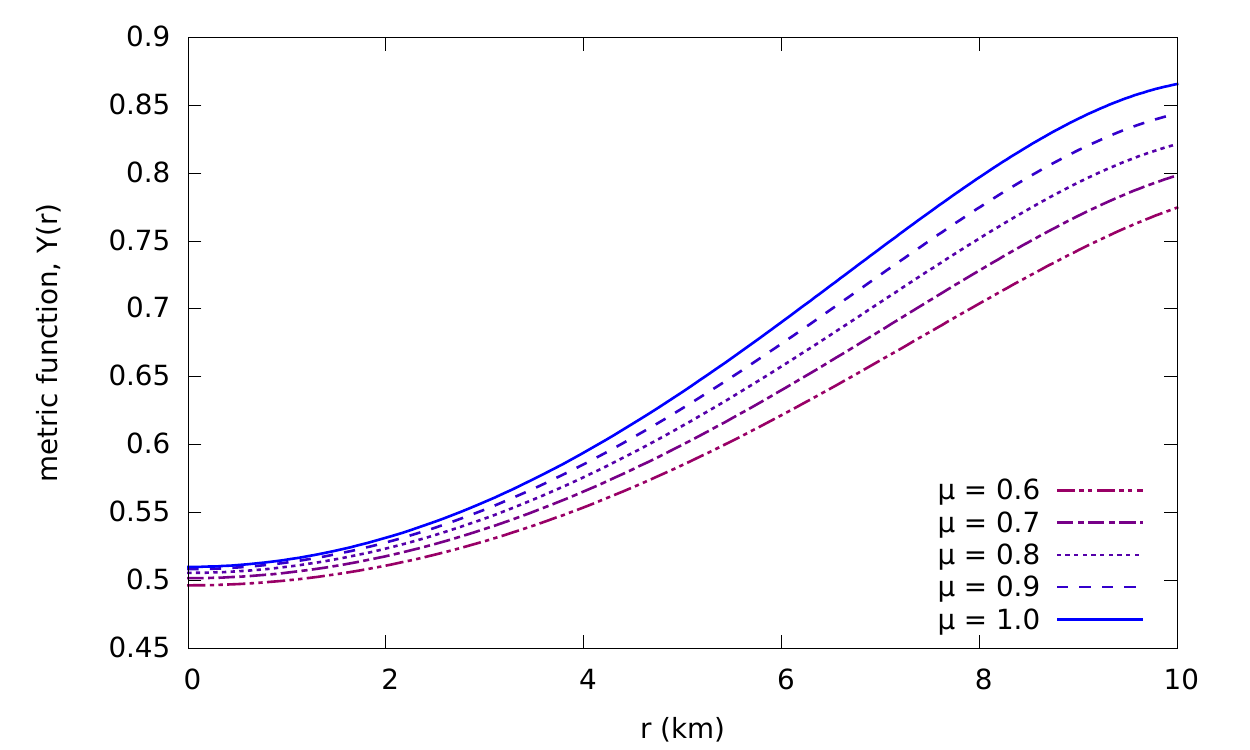}} 
\subfloat[The $Z(r)$ metric function]{\label{an.fig:ZmetricCAPhiP}
  \includegraphics[width=0.5\linewidth]{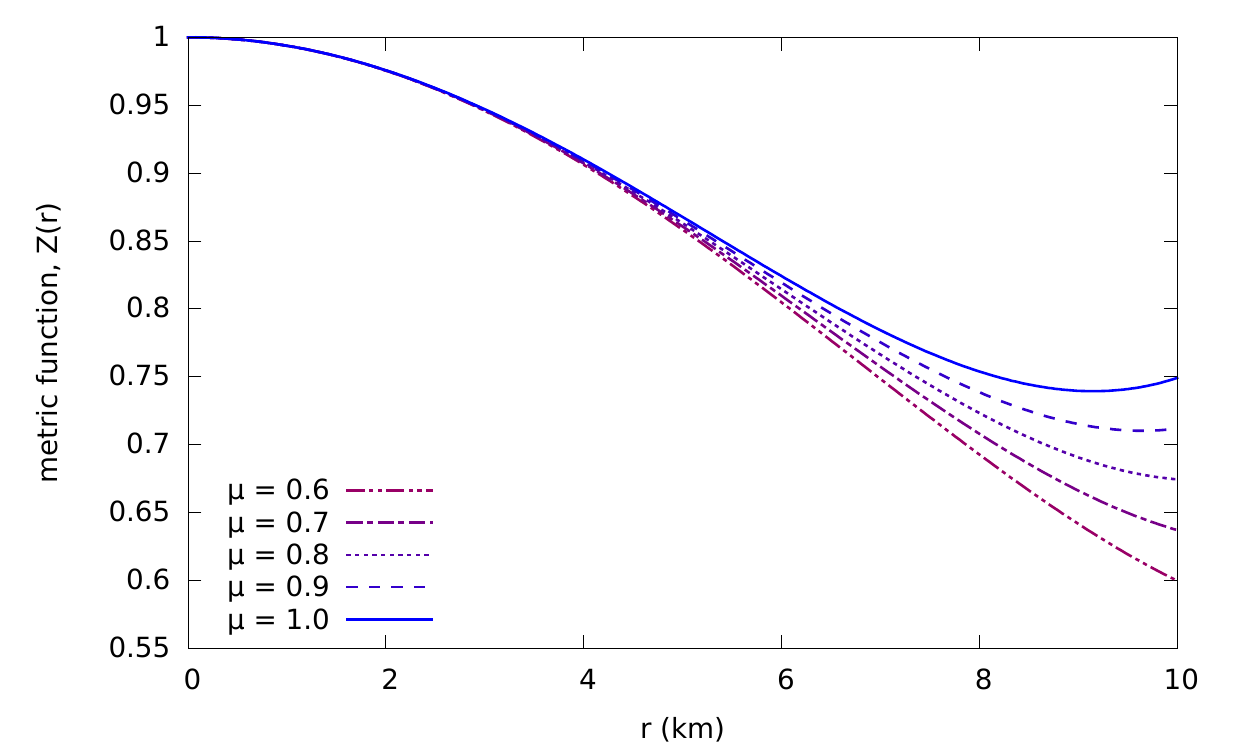}}\\
\subfloat[The $\lambda(r)$ metric function]{\label{an.fig:LambdaMetricCAPhiP} 
  \includegraphics[width=0.5\linewidth]{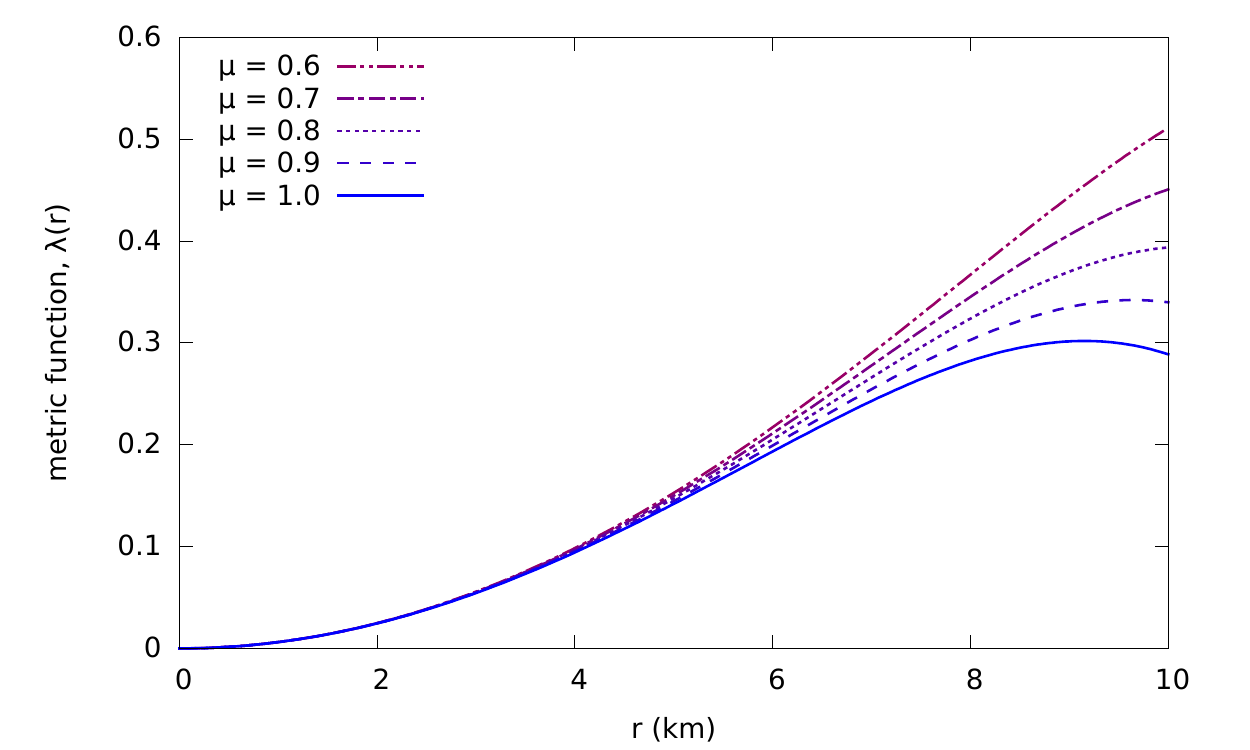}}
\subfloat[The $\nu(r)$ metric function]{\label{an.fig:NuMetricCAPhiP}
  \includegraphics[width=0.5\linewidth]{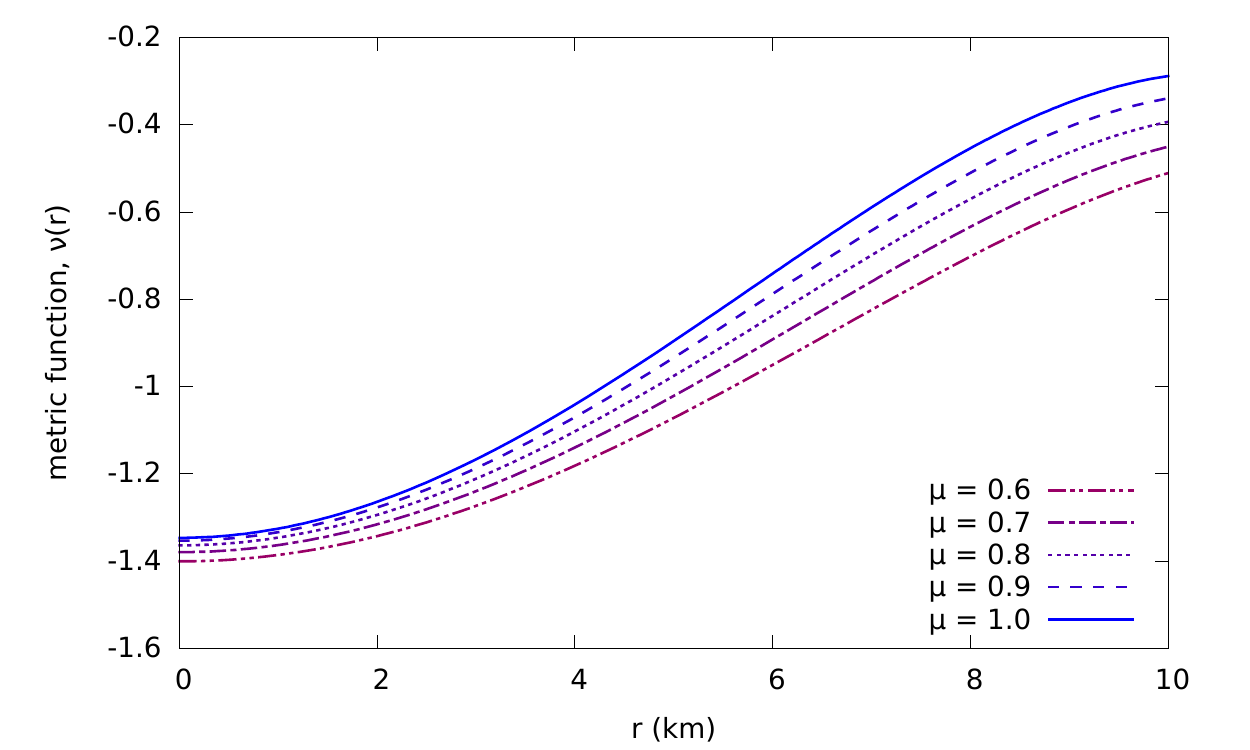}} 
\caption[Variation of metric variables for $\Phi^{2} > 0$]{Variation of metric variables with the radial coordinate
  inside the star. The parameter values are $\rho_{c}=\un{1\times
    10^{18}} \dunits, r_b = \un{1 \times 10^{4} m}, \beta = \un{2 \times 10^{-16} m}, k = 1 \times 10^{-9} \un{m^{-2}} $ and $\mu$ taking the various values shown in the legend }
\label{an.fig:CAPhiP,MetricCoeff}
\end{figure}
and find that they are all well behaved, because they do not show sign
changes for example.  The condition on the existence of the proper
radius reduces to the same as previously, viz.\
equation~\eqref{an.eq:RconstraintCA}, and we keep this in mind as we
proceed here too.

Next we look at the radial pressures. Since in this case we have all
of anisotropic pressures, electric charge, and self-boundness, we show
how the pressure varies with all these parameters in
figure~\ref{an.fig:CAPhiP,radialPressures}.  We expect from the
structure of the equations that at some critical values of each of
\(k\),
\(\mu\)
and \(\beta,\)
none independent of each other, for the pressure to become negative.
However we notice that for the range of parameters we chose in the
plots in~\ref{an.fig:CAPhiP,radialPressures}, the pressure is
surprisingly, but advantageously never negative.  This further
strengthens our perception that this solution will be well suited as
the model for an actual physical object.

\begin{figure}[!htb]
\centering
\subfloat[Varying $\mu$ ] {\label{an.fig:CAPhiP,radialPressuresA}
\includegraphics[width=0.5\textwidth]{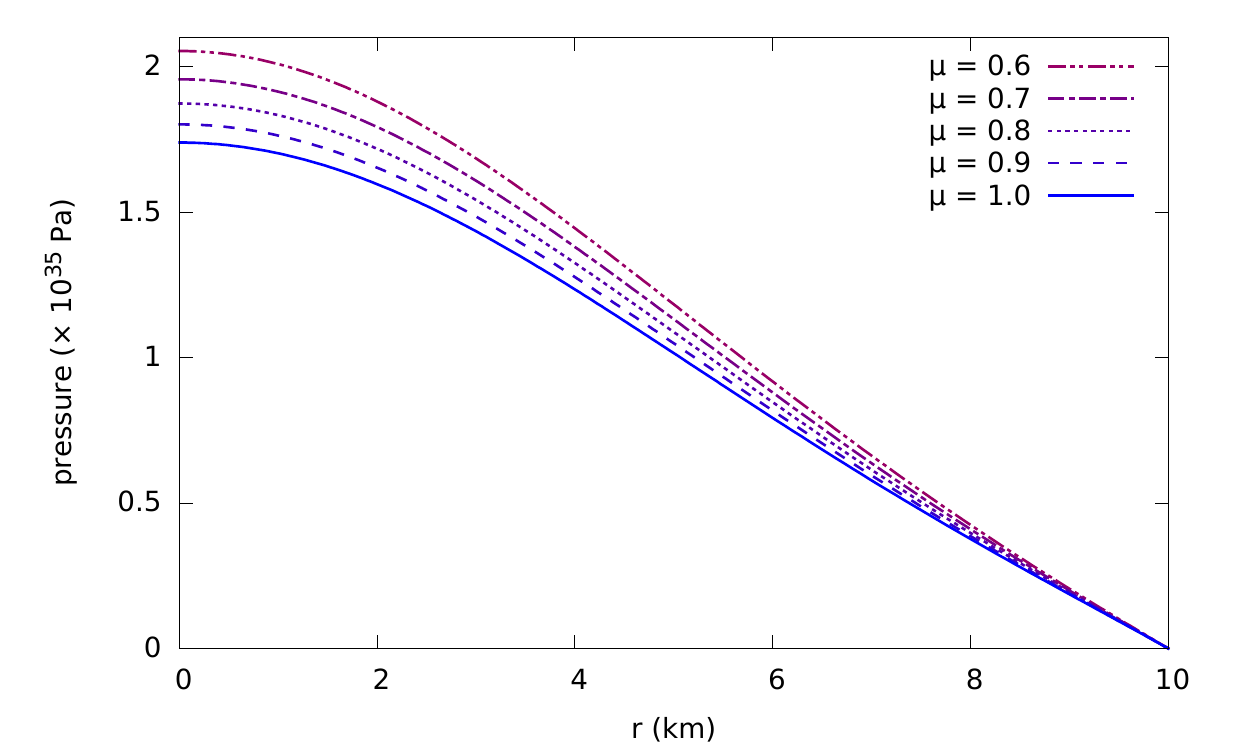}}
\subfloat[Varying $\beta$ such that $\beta \propto \mu$]{\label{an.fig:CAPhiP,radialPressuresB}
\includegraphics[width=0.5\textwidth]{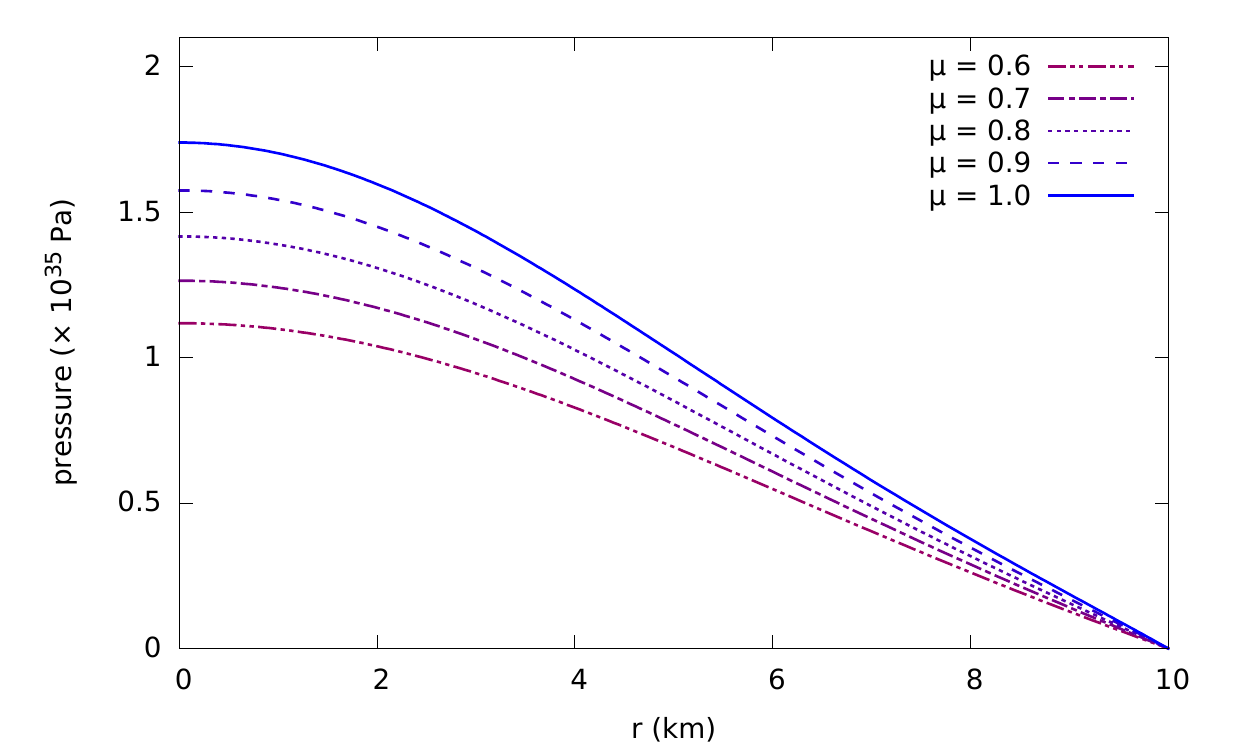}}\\
\subfloat[Varying $k$]{\label{an.fig:CAPhiP,radialPressuresC}
\includegraphics[width=0.5\textwidth]{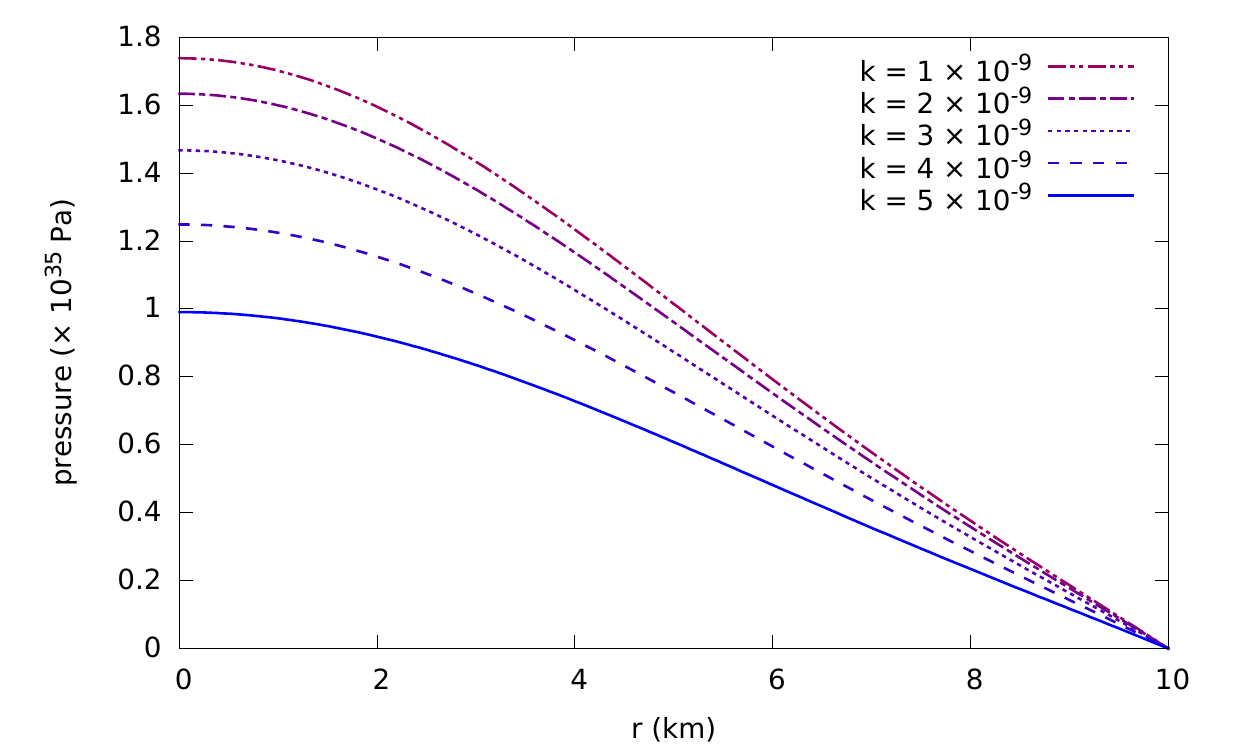}}
\subfloat[Varying $\beta$]{\label{an.fig:CAPhiP,radialPressuresD}
\includegraphics[width=0.5\textwidth]{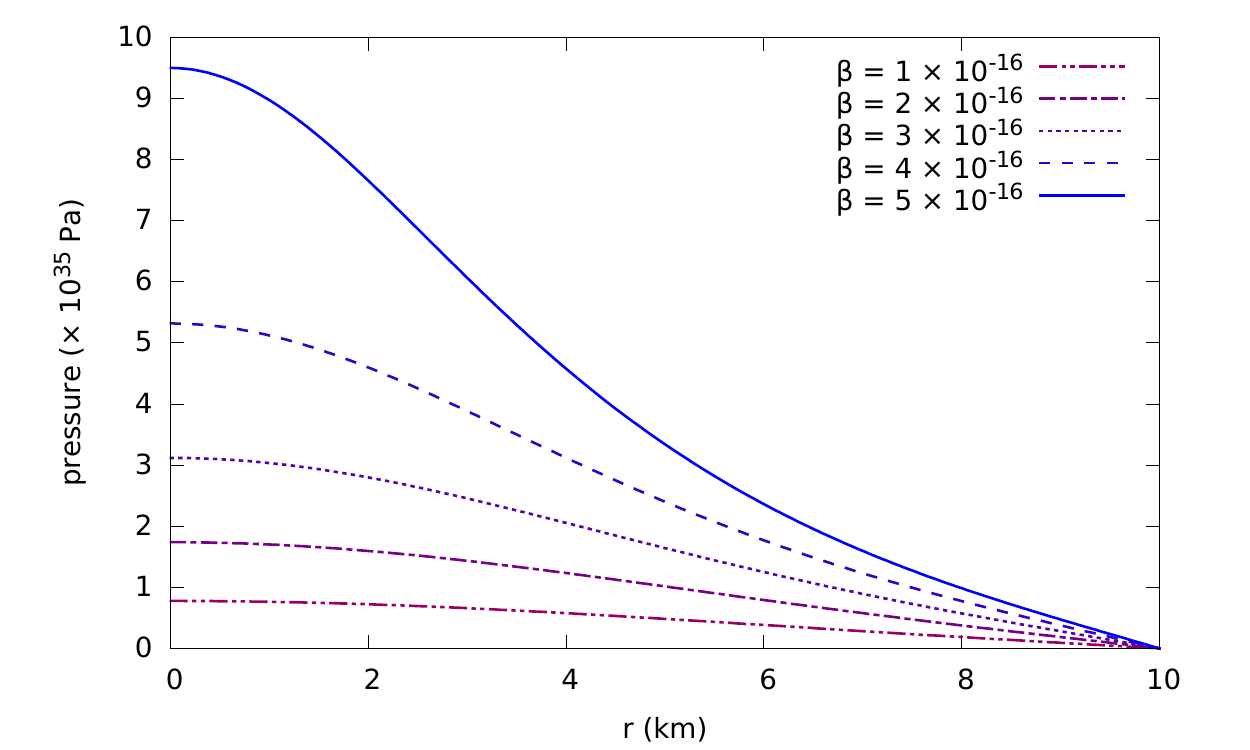}} 
\caption[Variation of radial pressure  for $\Phi^{2} > 0$]{Variation of the
  radial pressure with the radial coordinate inside the star. The
  parameter values are
  $\rho_{c}=\un{1\times 10^{18}} \dunits, r_b = \un{1 \times 10^{4}
    m},\, \beta$
  is fixed to $(2 \times 10^{-16})$ in the two left plots, while it is
  varied in two different ways in the right ones. $k$ is fixed to
  $1 \times 10^{-9} \un{m^{-2}}$ in~\SFRef{an.fig:CAPhiP,radialPressuresA},
  \SFRef{an.fig:CAPhiP,radialPressuresB}
  and~\SFRef{an.fig:CAPhiP,radialPressuresD}, but changes according to
  the legend in~\SFRef{an.fig:CAPhiP,radialPressuresC}. $\mu$ is fixed
  to 1 in the bottom two panels, and varies as shown in the legend in
  the top panels}
\label{an.fig:CAPhiP,radialPressures}
\end{figure}

Even though we do not see negative radial pressures, this is dependant
on some particular choices, and we will derive conditions for this not
to hold momentarily.  Before doing so, we take a look at the
tangential pressures in~\ref{an.fig:CAPhiP,tangentialPressures} ,
which can be negative without implying unphysicality,
\begin{figure}[!htb]
\centering
\subfloat[Varying $\mu$ ] {\label{an.fig:CAPhiP,tangentialPressuresA}
\includegraphics[width=0.5\textwidth]{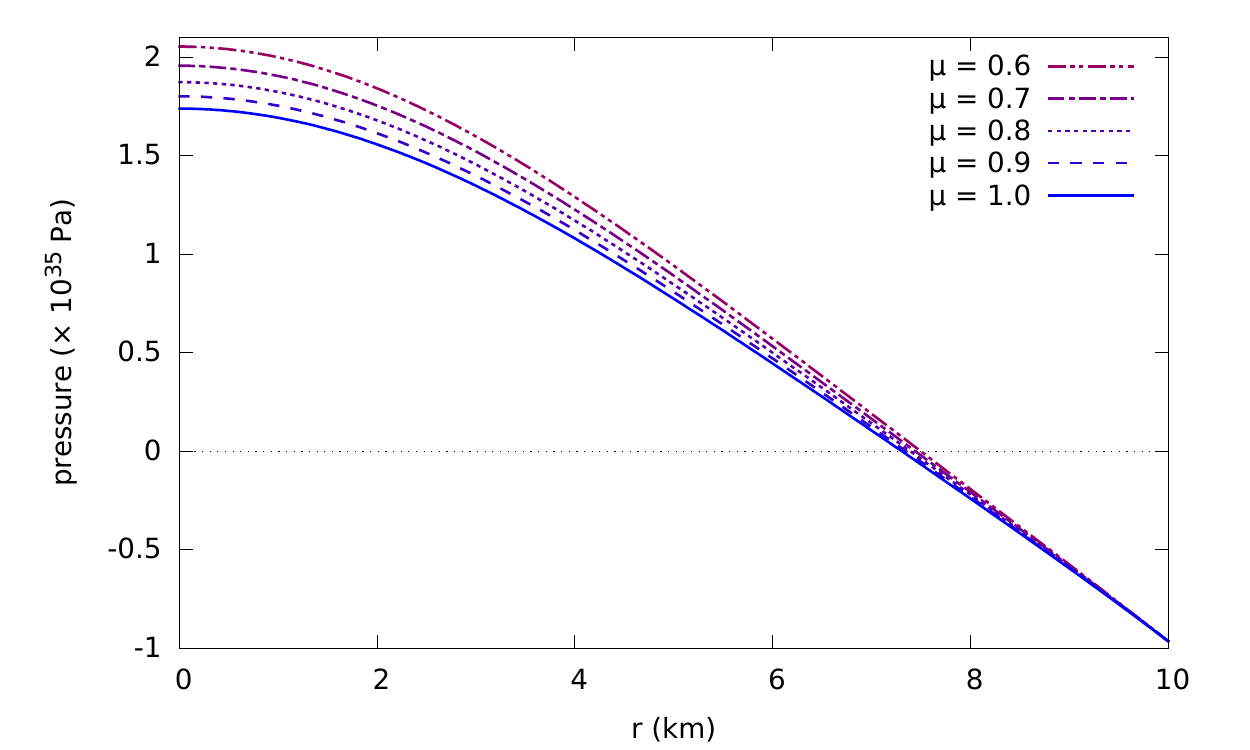}}
\subfloat[Varying $\beta$ such that $\beta \propto \mu$]{\label{an.fig:CAPhiP,tangentialPressuresB}
\includegraphics[width=0.5\textwidth]{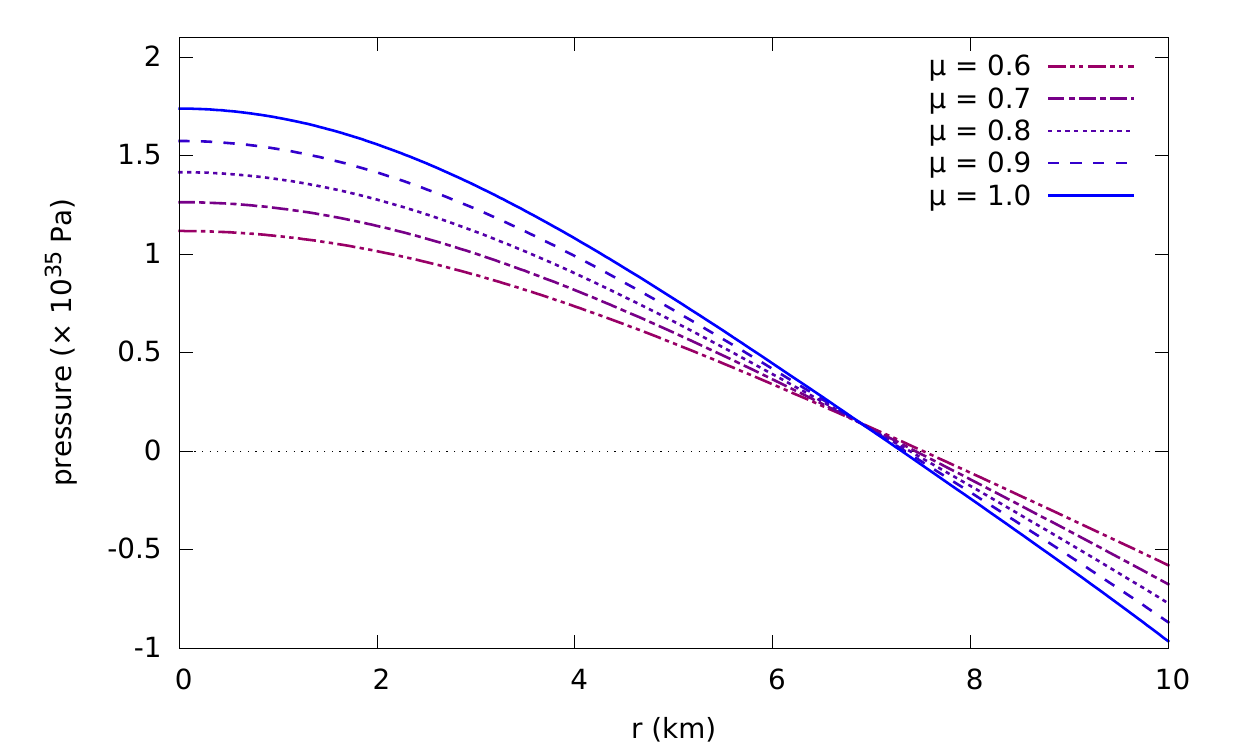}}\\
\subfloat[Varying $k$]{\label{an.fig:CAPhiP,tangentialPressuresC}
\includegraphics[width=0.5\textwidth]{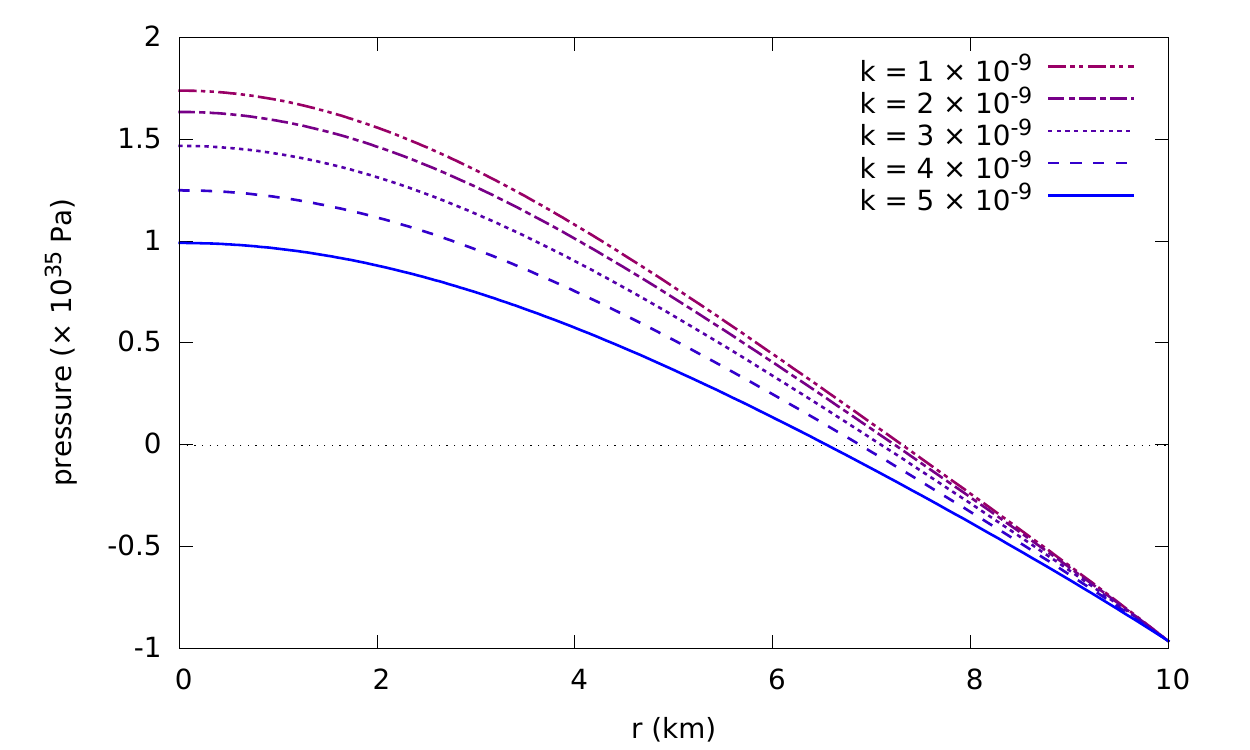}}
\subfloat[Varying $\beta$]{\label{an.fig:CAPhiP,tangentialPressuresD}
\includegraphics[width=0.5\textwidth]{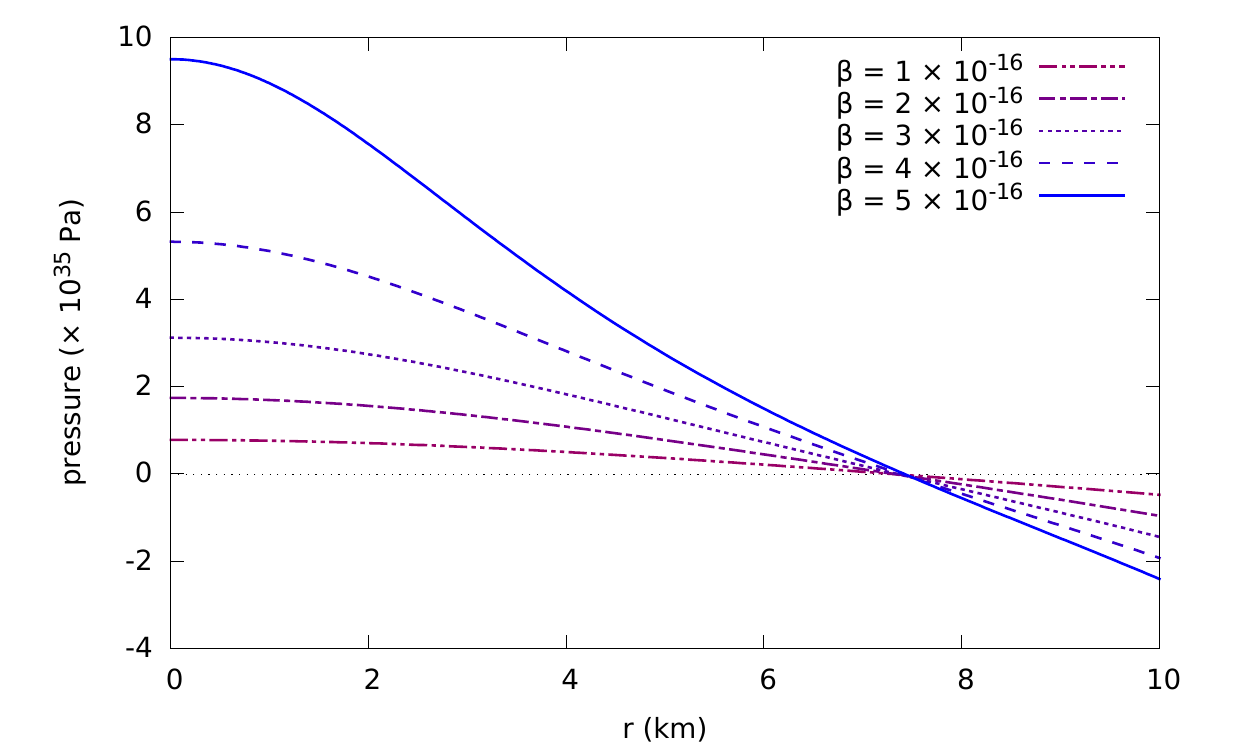}} 
\caption[Variation of radial pressure variables]{Variation of the
  radial pressure with the radial coordinate inside the star. The
  parameter values are
  $\rho_{c}=\un{1\times 10^{18}} \dunits, r_b = \un{1 \times 10^{4}
    m},\, \beta$
  is fixed to $(2 \times 10^{-16})$ in the two left plots, while it is
  varied in two different ways in the right ones. $k$ is fixed to
  $1 \times 10^{-9} \un{m^{-2}}$ in~\SFRef{an.fig:CAPhiP,radialPressuresA},
  \SFRef{an.fig:CAPhiP,radialPressuresB}
  and~\SFRef{an.fig:CAPhiP,radialPressuresD}, but changes according to
  the legend in~\SFRef{an.fig:CAPhiP,radialPressuresC}. $\mu$ is fixed
  to 1 in the bottom two panels, and varies as shown in the legend in
  the top panels}
\label{an.fig:CAPhiP,tangentialPressures}
\end{figure}
and as expected we do see negative tangential pressures.  We first
figure out the limiting values of \(k,\)
since this comes directly from the \(Z\)
metric function, through constraint~\ref{an.it.RegularMetric}.  This
results in the same values as before, i.e.\
\(|k| < 1.55 \times 10^{-9} \un{m^{-2}}.\)
The next two constraints~\ref{an.it.ExteriorObservables}
and~\ref{an.it.ExteriorMetric} were also already implemented and
results in the same parameter ranges as before.  From the graphs of
the radial pressures we show in~\ref{an.fig:CAPhiP,radialPressures},
it is also clear that the constraints concerning the positivity of the
radial pressures hold.  However since we still have no constraints on
the parameter values like \(\beta,\)
we proceed as before and invoke the energy
condition~\ref{an.it.EnergyConditions}, since all the previous ones
have clearly been satisfied by some suitable choice of parameters.  In
this case, we get the constraint equation on \(\beta,\) and \(k\)
to be
\[
\beta r^{2} + \f{q^{2}}{r^{4}} \leq \f{\rho+3p_{r}}{2}.
\]
Since we have been positing \(q = kr^{3}\) throughout this solution, this can be further simplified into
\begin{equation}
  \label{an.eq:betaLim,CAPhiP}
\beta + k^{2} \leq \f{\rho+3p_{r}}{2r^{2}}.
\end{equation}
We are in a position to evaluate this inequality in full, since we
have the full solution now, knowing all the metric components.  The
procedure we employed before for the anisotropised charged works here
too, and we are left instead with a relation on both \(k\)
and \(\beta,\)
instead of just \(\beta\).  In this case these relations give instead
\[
\beta + k^{2} \leq \f{3}{2 \kappa r^{2}} \left( \f{2Z}{rY}\deriv{Y}{r} - \f{1}{r} \deriv{Z}{r} \right) -\f{\rho}{r^{2}}.\]
Again evaluating this expression both at the boundary and at the centre we get the following:
\begin{enumerate}[label=(\alph{*}), ref=\theenumi.(\alph{*})]
\item  where \(r=r_{b},\)
\begin{equation}\label{an.eq:Enegy@r_b,CAPhiP} \beta +k^{2} \leq \f{3}{\kappa r_{b}^{2}} \left[
  2(1-br_{b}^{2}+ar_{b}^{4})
  \left(\left. \f{1}{rY}\deriv{Y}{r}\right)\right|_{r=r_{b}}
  +2a - 4 b r_{b}^{2} \right] - \f{\rho_{c}(1-\mu)}{r^{2}_{b}}
\end{equation}
\item where \(r=0,\) we get
\begin{equation}\label{an.eq:Energy@0,CAPhiP} \beta + k^{2} \leq \f{3}{\kappa} \left[
  \f{2}{r^{2}}\left(\left. \f{1}{rY}\deriv{Y}{r}\right)\right|_{r=0}
  -\f{2a}{r^{2}} \right] -\f{\rho_{c}}{r^{2}}
\end{equation}
\end{enumerate}
expressions which depend crucially on the value of the term in round
parentheses, and hence the complicated \(Y\)
metric functions that include both \(\beta,\) and \(k^{2}.\) 

As in the previous section, we use the criteria on the speed of sound
to constrain the terms in brackets.  We show plots of the speed of
sound in figure~\ref{an.fig:CAPhiP,radialSpeed},
\begin{figure}[!htb]
  \subfloat[$k=2 \times 10^{-9} \un{m^{-2}},$ and $\beta=5 \times 10^{-17}$]{\label{an.fig:CAPhiP,rSpeedk=1e-9}
  \includegraphics[width=0.5\linewidth]{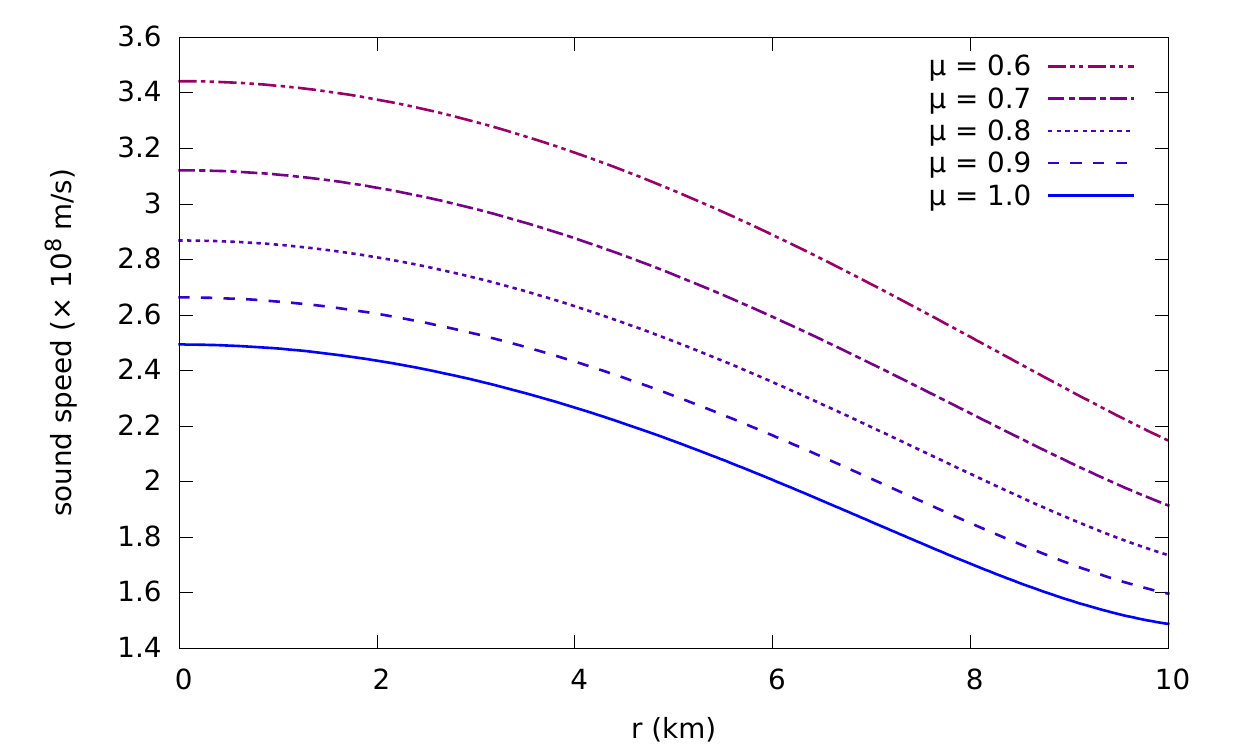}} 
\subfloat[$k=1 \times 10^{-9} m$ and $\beta=5 \times 10^{-17} \times \mu$]{\label{an.fig:CAPhiP,rSpeedk=3e-9}
  \includegraphics[width=0.5\linewidth]{Analysis/Figures/chargedAnisotropicPhiP/speedRk9b2}}\\
\subfloat[$\mu=1, \beta=7 \times 10^{-17}$ ]{\label{an.fig:CAPhiP,rSpeedMu=1} 
  \includegraphics[width=0.5\linewidth]{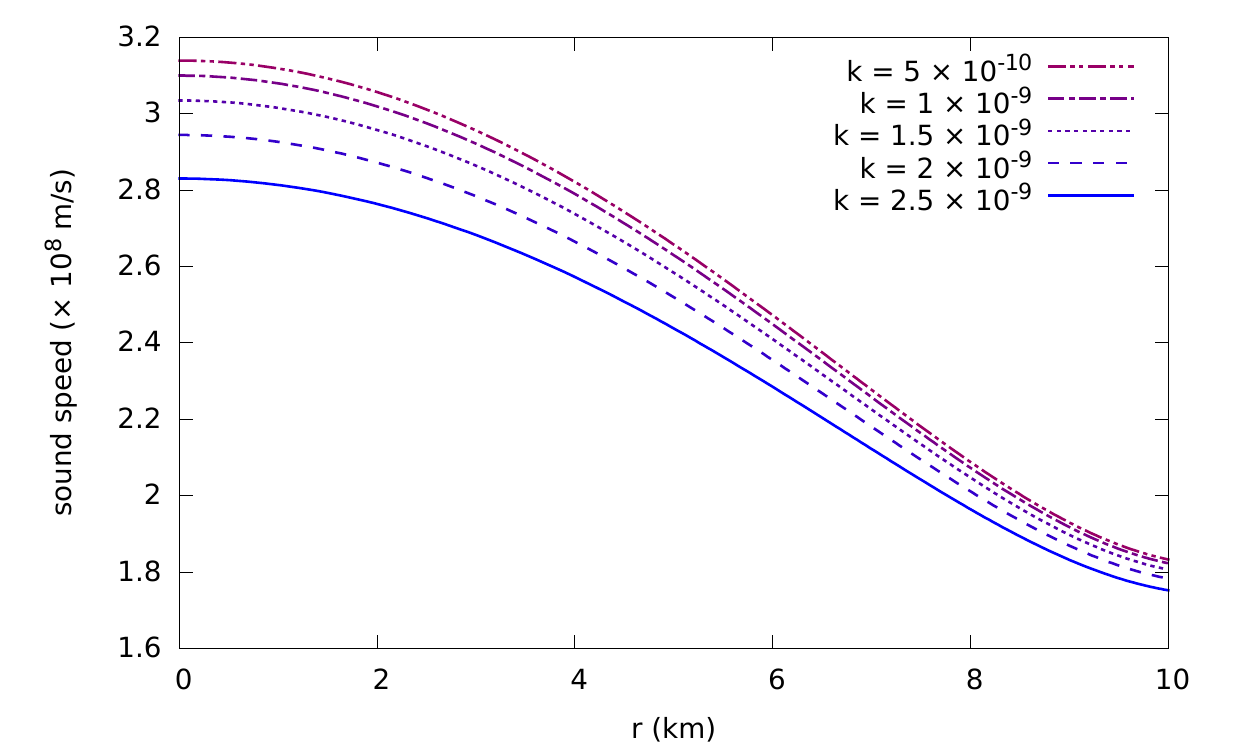}}
\subfloat[$\mu=1, k=1 \times 10^{-9}$  ]{\label{an.fig:CAPhiP,rSpeedRMu=0.6}
  \includegraphics[width=0.5\linewidth]{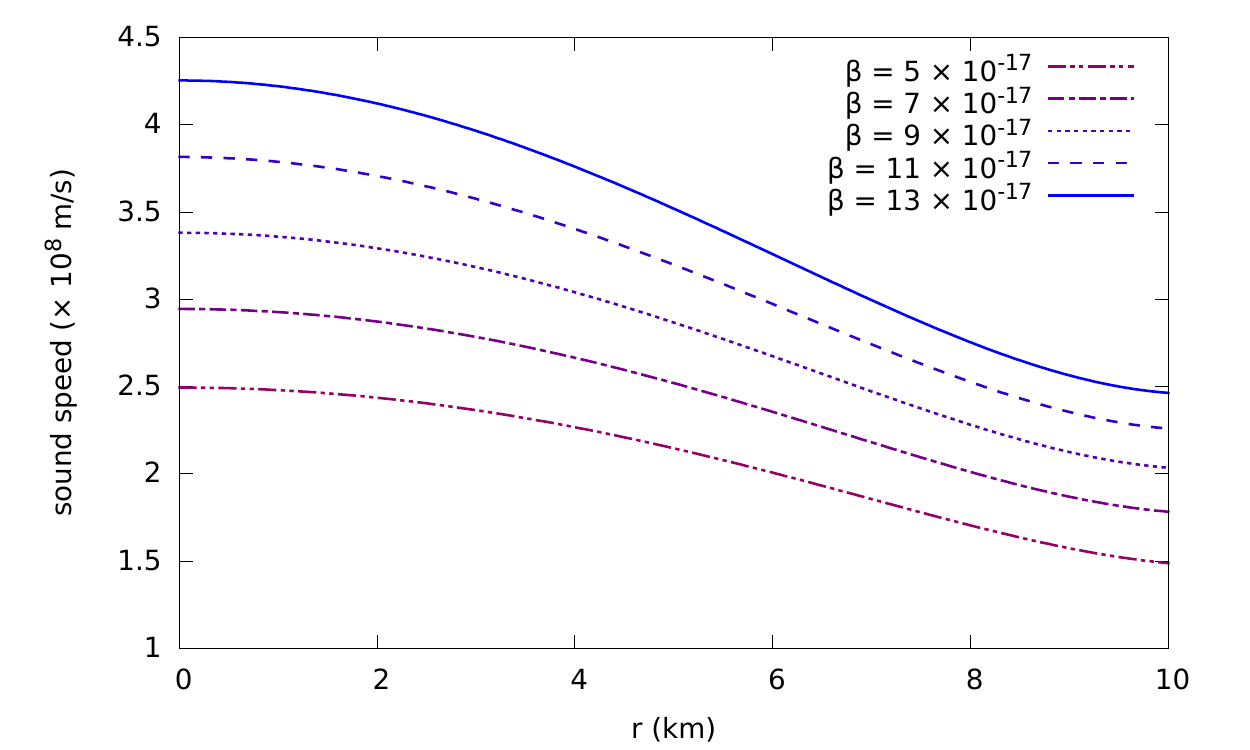}} 
\caption[The speed of sound with $r$]{Variation of speed of sound
  with the radial coordinate inside the star. The parameter values are
  $\rho_{c}=\un{1\times 10^{18}} \dunits, r_b = \un{1 \times 10^{4}
    m}, \mu$
  is fixed in the bottom two plots at one, but takes on the values in
  the legend for the top graphs. $k$ is set to
  $1 \times 10^{-9} \un{m} $ top plots, but varies in the bottom right
  one. $\beta$ is fixed to $5 \times 10^{-17}$ in the top left, and is
  proportional to $mu$ with the same proportionality factor in the
  right one.}
\label{an.fig:CAPhiP,radialSpeed}
\end{figure} 
and if we consider the expressions of this same speed, in this general
case, we have from equation~\eqref{an.eq.soundSpeed}, since
\(\Delta = \beta r^{2},\) and \(q = kr^{3} \implies q' = 3kr^{2},\)
\begin{equation}
\label{an.eq:vCond,CAPhiP}
\left( \f{r_{b}^{2}}{\kappa \rho_{c} \mu}\right)\left[ \f{\nu' \kappa \left( p_{r} + \rho \right)}{4 r} -3k^{2} + \beta \right] \leq 1.
\end{equation}
We re-express this in terms of the previously defined variable
\(\psi,\)
in a similar vein as previously after substituting for \(\nu'\)
to get an inequality on \(\psi:\)
\[
\left( \f{r_{b}^{2}}{\kappa \rho_{c} \mu}\right) \left\{ \beta - 3k^{2} + Z\psi^{2} - (2ar^{2}-b)\psi \right\} \leq 1.
\]
We expect from the shape of the velocity plots to have the maximum
speed at the centre of the star, so we evaluate the above equation at
\(r=0 \implies Z = 1,\) to get
\[
\left( \f{r_{b}^{2}}{\kappa \rho_{c} \mu}\right) \left[\beta -3k^{2} +
  \psi_{0}^{2} + b\psi_{0}\right] \leq 1 \implies \psi_{0}^{2} + b
\psi_{0} + \left( \beta - 3k^{2} - \f{\kappa\rho_{c}\mu}{r_{b}^{2}}\right) \leq 0.
\]

We solve this inequality for \(\psi_{0},\) and get that \( \psi_{0-}\leq \psi_{0} \leq \psi_{0+}\) with \[ \psi_{0\pm} = -\f{b}{2} \pm \sqrt{3k^{2}-\beta + \f{b^{2}}{4}  + \f{\kappa\rho_{c}\mu}{r_{b}^{2}}}. \]

For the usual values we use in our
plots for example:
\(\rho_{c} = 1 \times 10^{18} \dunits, r_{b} = 1 \times 10^{4} \un{m},
\mu=1,\)
and \(k = 1 \times 10^{9} \un{m^{-2}},\)
the inequality results in \(\beta < 3.65 \times 10^{-6},\)
a less restrictive constraint on \(\beta,\)
than the previous ones we had in other solutions: as a result we can
``crank-up'' the anisotropy in this particular solution to larger
values, while still maintaining the energy conditions. 

\section{Application of the models to physical objects}
In the last section, we investigated all the classes of the new
solutions we derived previously in detail, and found out that only
three specific ones give us sensible values for the physical matter
variables.  In this section we look at these three viable solutions in
greater detail by
\begin{enumerate}[label=(\alph*)]
\item deriving an equation of state for each solution.  This equation
  of state comes directly from our assumptions, and general
  relativity: no matter interactions being assumed.
\item We find the masses, radii, and total electric charge each
  solution admits, and how changing parameters change these observables.
\item We restrict the parameters for each solutions, so that within
  the restricted parameter ranges, these solutions behave physically
  with no (unphysical characteristics)
\item We compare our solutions with known observed values of radii and
  masses of visible astrophysical objects.
\end{enumerate}

\subsection{The equation of state}
We derived the equation of state of the Tolman VII solution and
presented it in~\ref{t7.sec:eos}.  This was possible because of the
simple nature of the density relation which is easily invertible.  As
a result, all instances of \(r\)
in the expressions of the pressures can be converted to some function
of \(\rho\), through the inverted equation~\eqref{t7.eq:Density}:
\begin{equation}
  \label{an.eq:InvDensity}
 r(\rho) = r_{b}\sqrt{\f{1}{\mu} \left( 1 - \f{\rho}{\rho_{c}}\right)}.
\end{equation}
With this prescription applied to the pressures we already found in
the previous section for the three specifically physical solutions, we
arrive at the equation of state for each of these solutions.  A note
of caution: these equations of states were obtained from a purely
general relativistic method, with no nuclear physics assumptions,
however as we will show shortly, they still predict values for
observables that are in line with \emph{all} current measurements from
neutron stars, were the TOV method applied to them.

\FloatBarrier\subsubsection{The solution with anisotropy only}
The first equation of state we look at is for the anisotropic
uncharged case, whose pressure is given by~\eqref{ns.eq:PhiPpr}.  With
the inverted density relation the equation of state is obtained by
replacing each occurrence of \(r\)
through~\eqref{an.eq:InvDensity}.  Since the expression is the exact
same one as before, we will here give the expressions of all the
components of the \(p_{r}\)
function in terms of \(\rho\)
instead of rewriting the full pressure again.  The \(Z\)
metric function in this solution is given by
\begin{equation}
  \label{an.eq:EOS:Z, PhiP}
Z(\rho) = 1 + \f{\kappa r_{b}^{2}}{5\mu}\left( \f{\rho^{2}}{\rho_{c}} -\f{\rho}{3} - \f{2\rho_{c}}{3}\right).
\end{equation}
All the evaluated constants like \(a, b, \alpha, \beta, \gamma\)
and \(\Phi\)
remain the same, but the expression of \(\xi\) does change into 
\begin{equation}
  \label{an.eq:EOS:xi, PhiP}
\xi(\rho) = 2r_{b}\sqrt{\f{5}{\kappa\mu\rho_{c}}} 
\acoth{\left( \f{\sqrt{15\mu\rho_{c} - \kappa r_{b}^{2}(2\rho_{c}^{2} + \rho_{c}\rho - 3\rho^{2})} + \sqrt{15\mu\rho_{c}}}{r_{b}\sqrt{3\kappa} (\rho_{c}-\rho)}\right)}
\end{equation}
when put in terms of \(\rho.\)
With these two functions in terms of \(\rho,\)
the equation of state is easily written from
equation~\eqref{ns.eq:PhiPpr} by straight forward substitution.  The
resulting expression as can be guessed from the length of the previous
two equations is very long, and we will not attempt to write it down
in full here.  We however give plots of what the equation of state
look like in figure~\ref{an.fig:PhiP,EOS}, where we also notice some
interesting features that merit discussion.
\begin{figure}[!htb]
\subfloat[The radial pressure , $\beta=1 \times 10^{-16}$]{\label{an.fig:phiP,EOSbeta1}
  \includegraphics[width=0.5\linewidth]{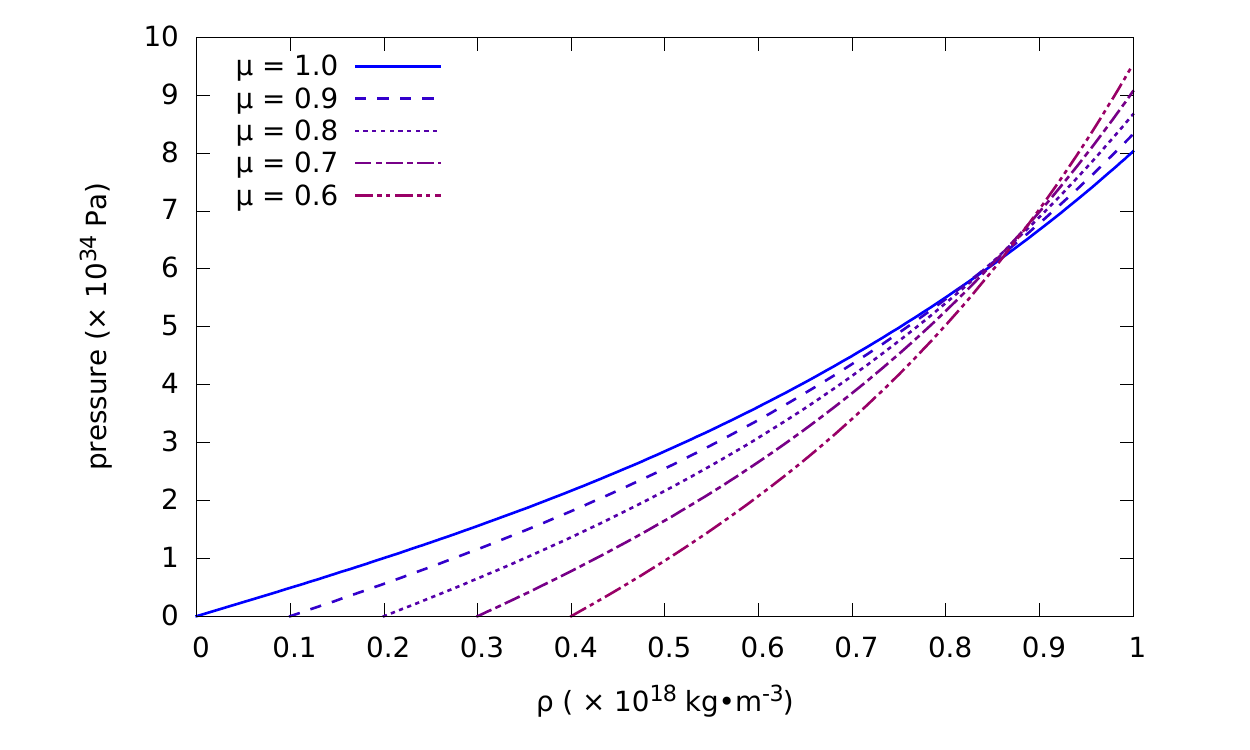}} 
\subfloat[The radial pressure, $\mu=1$]{\label{an.fig:phiP,EOSmu1}
  \includegraphics[width=0.5\linewidth]{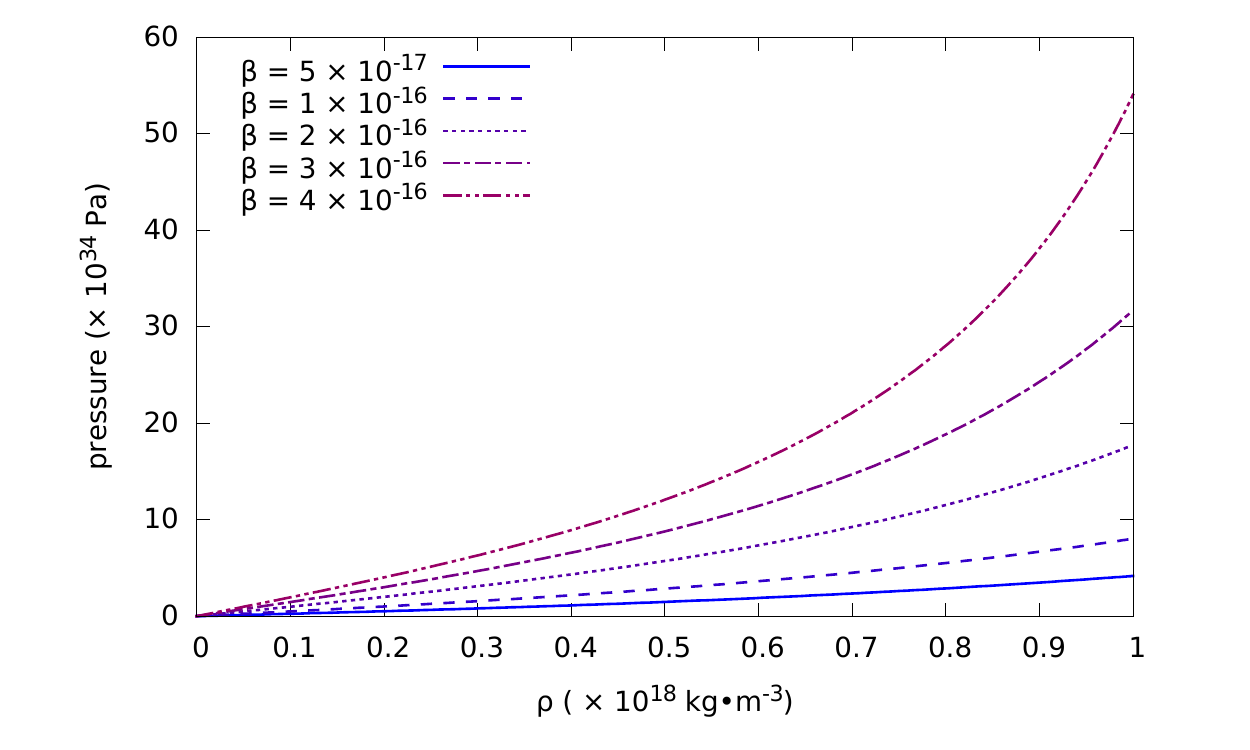}}
\caption[The radial pressure with $r$ for various
parameters]{Variation of the radial pressure with the density inside
  the star. The parameter values are
  $\rho_{c}=\un{1\times 10^{18}} \dunits, r_b = \un{1 \times 10^{4}
    m},\, \beta$ is set to $1 \times 10^{-16}$ on the left, and varies
  on the right, and $\mu$ is set to one on the right but takes the
  various values shown in the legend on the left}
\label{an.fig:PhiP,EOS}
\end{figure}

The first feature that immediately jumps at us is that
the~\ref{an.fig:phiP,EOSbeta1} plot has all the EOS curves with
\(\mu \neq 1\)
starting at non-zero densities.  This is easily understood since self
bound stars with \(\mu \neq 1\)
do not have boundaries at zero density.  This also means that the full
functional form of the EOS (whose plot we have culled at zero
pressure) extends to pressures for densities lower than the boundary
density for \emph{that} star, and hence to negative pressures.  Of
course, such regions do not exist in our models, since we match our
solution to an exterior metric before that happens.  However it
appears in the plots shown because we are plotting the pressures at
densities that these particular stars do not have ``access'' to.

Another feature that is obvious is that this negative pressure does
not occur in~\ref{an.fig:phiP,EOSmu1}.  This is simply because all the
EOS shown there are ``natural'' with \(\mu=1,\)
and hence are valid up to zero densities.  The other characteristic of
the second figure is how drastically the magnitude of the pressure
function changes by changing the value of \(\beta,\)
something that suggest that for high enough anisotropies, the energy
conditions might be violated: a conclusion we arrived at previously,
through a more pedestrian approach.

\FloatBarrier\subsubsection{The solution where charge compensates
  anisotropy} In this solution, \(\beta\)
is compensated by \(k,\)
so that the expressions of the functions and constants, in particular
\(a,\)
and hence \(Z\)
do change a bit.  In this particular case, the metric function \(Z\)
is thus given by 
\begin{equation}
  \label{an.eq:EOS:Z, AniCha}
Z(\rho) = 1 -\f{k^{2}r_{b}^{4}}{5\mu^{2}\rho_{c}^{2}} (\rho_{c}-\rho)^{2} + \f{\kappa r_{b}^{2}}{5\mu}\left( \f{\rho^{2}}{\rho_{c}} -\f{\rho}{3} - \f{2\rho_{c}}{3}\right).
\end{equation}
which has the same components as the previous~\eqref{an.eq:EOS:Z,
  PhiP}, with an additional piece containing the charge \(k.\)
Similarly, the proper radius \(\xi\) is also changed, and its expression is given by
\begin{multline}
  \label{an.eq:EOS:xi, Anicha}
\xi(\rho) = 2r_{b}\sqrt{\f{5}{\kappa\mu\rho_{c} - k^{2}r_{b}^{2}}} \times \\
\times \acoth{\left[ \f{\sqrt{3 k^{2} r_{b}^{4} (\rho_{c}-\rho)^{2} + \kappa \mu r_{b}^{2}\rho_{c} (3\rho^{2} - \rho\rho_{c}-2\rho_{c}^{2}) + 15 \mu^{2} \rho_{c}^{2} } + \sqrt{15} \mu\rho_{c} } { r_{b} (\rho_{c}-\rho) \sqrt{3 \kappa \mu \rho_{c} - 3k^{2} r_{b}^{2}} }\right]}
\end{multline}
which is also similar to the previous~\ref{an.eq:EOS:xi, PhiP}, except
for the additional \(k\)
factors.  Putting these two expressions in the
pressure~\eqref{ns.eq:PhiPprCCA}, we get the equation of state of this
solution whose plots we show next in~\ref{an.fig:AniCha,EOS}.
\begin{figure}[!htb]
\subfloat[The radial pressure , $k=1 \times 10^{-9}$]{\label{an.fig:AniCha,EOSbeta1}
  \includegraphics[width=0.5\linewidth]{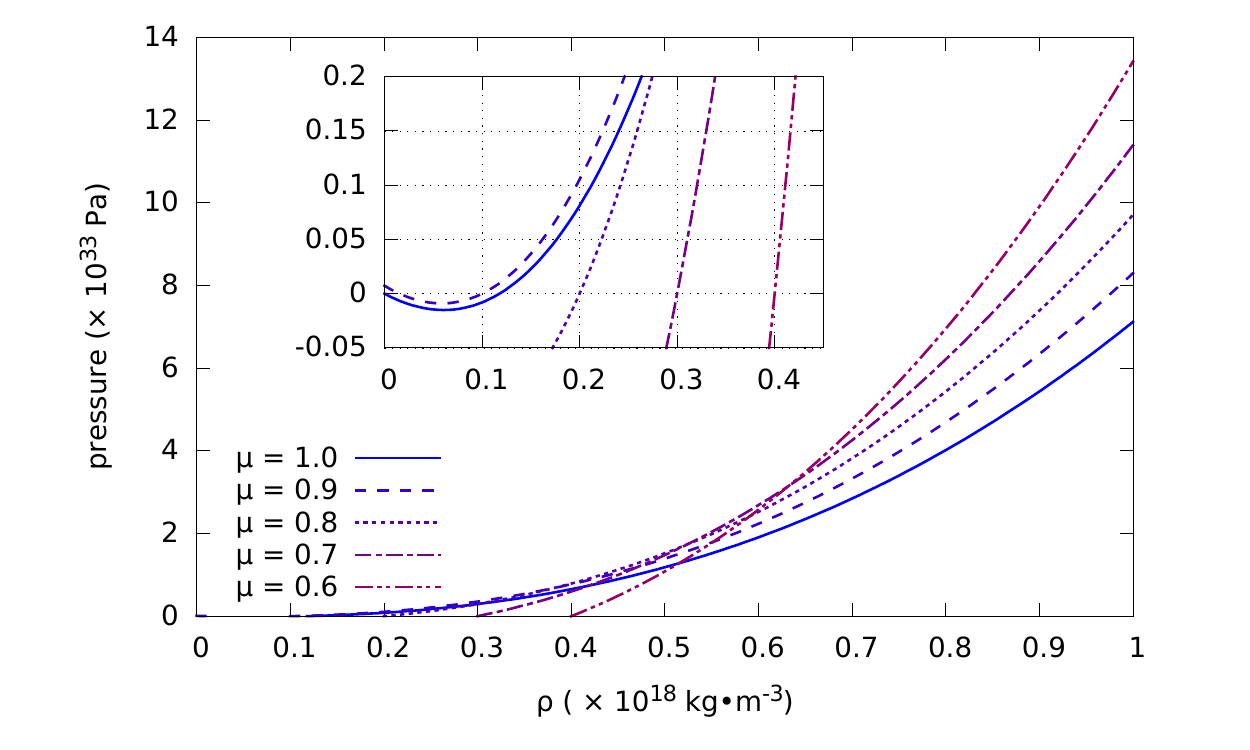}} 
\subfloat[The radial pressure, $\mu=1$]{\label{an.fig:Anicha,EOSmu1}
  \includegraphics[width=0.5\linewidth]{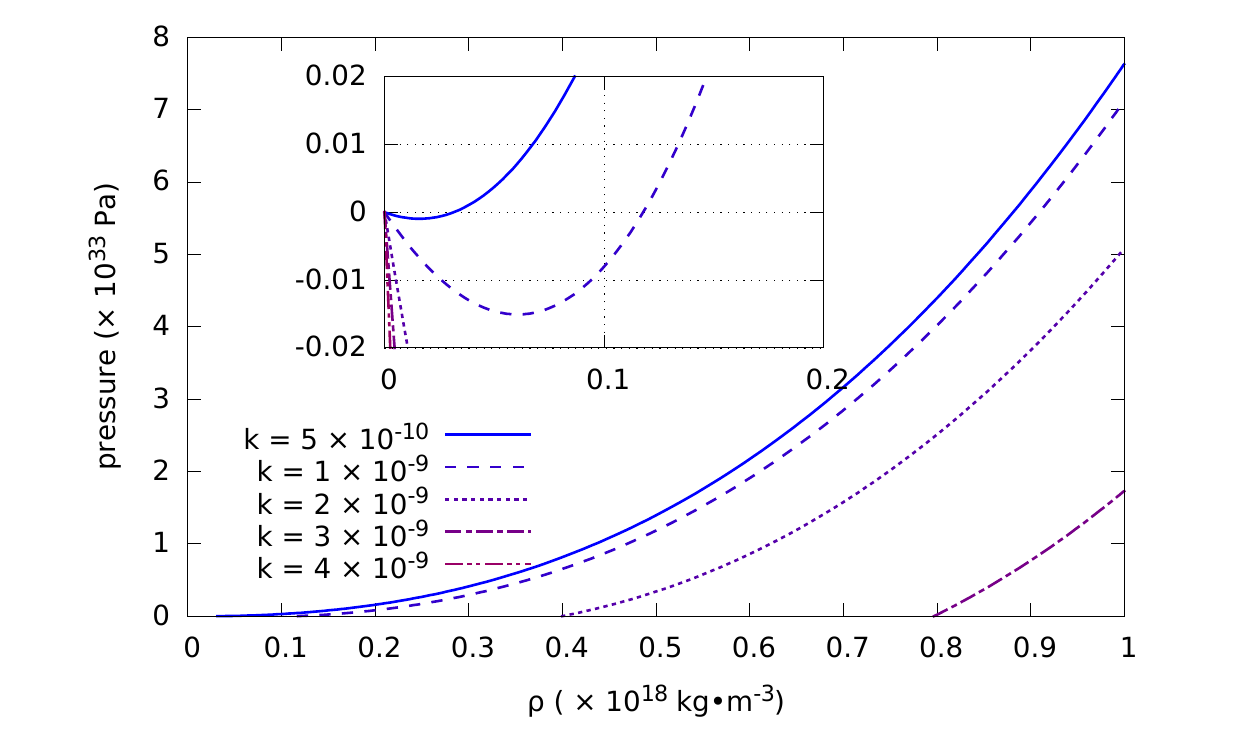}}
\caption[The radial pressure with $r$ for various
parameters]{Variation of the radial pressure with the density inside
  the star. The parameter values are
  $\rho_{c}=\un{1\times 10^{18}} \dunits, r_b = \un{1 \times 10^{4}
    m},\, k$ is set to $1 \times 10^{-9}$ on the left, and varies on
  the right, and $\mu$ is set to one on the right but takes the
  various values shown in the legend on the left}
\label{an.fig:AniCha,EOS}
\end{figure}
Here also we notice particular trends in the two plots, very similar
to the previous solution: changing \(\mu\)
prevents all the densities from being ``accessible'' to the solution
as before.  The trend on the right panel~\ref{an.fig:Anicha,EOSmu1} is
striking in its regularity: the effect of charge \(k\)
on the pressure is made clear, increasing the charge has a similar
effect as increasing the self-boundness inasmuch as certain densities
become inaccessible, but it does so in such a way that the shape of the
EOS does not change.

\FloatBarrier\subsubsection{The solution where both charge and
  anisotropy exists} Since the metric function \(Z\)
does not depend on the anisotropy, having both anisotropy and charge
does not change the functional form of \(Z\)
from the previous case, and we still have~\eqref{an.eq:EOS:Z, AniCha}
as the expression of \(Z(\rho).\)
The same argument applied to \(\xi(\rho),\)
since the difference in the pressure expressions come from the other
constants that have \(\beta\) in them along with \(k\) in this particular case.
\begin{figure}[!htb]
\centering
\subfloat[The radial pressure , $\beta=1 \times 10^{-16}, k=1 \times 10^{-9}$]{\label{an.fig:cAphiP,EOSbeta1}
  \includegraphics[width=0.5\linewidth]{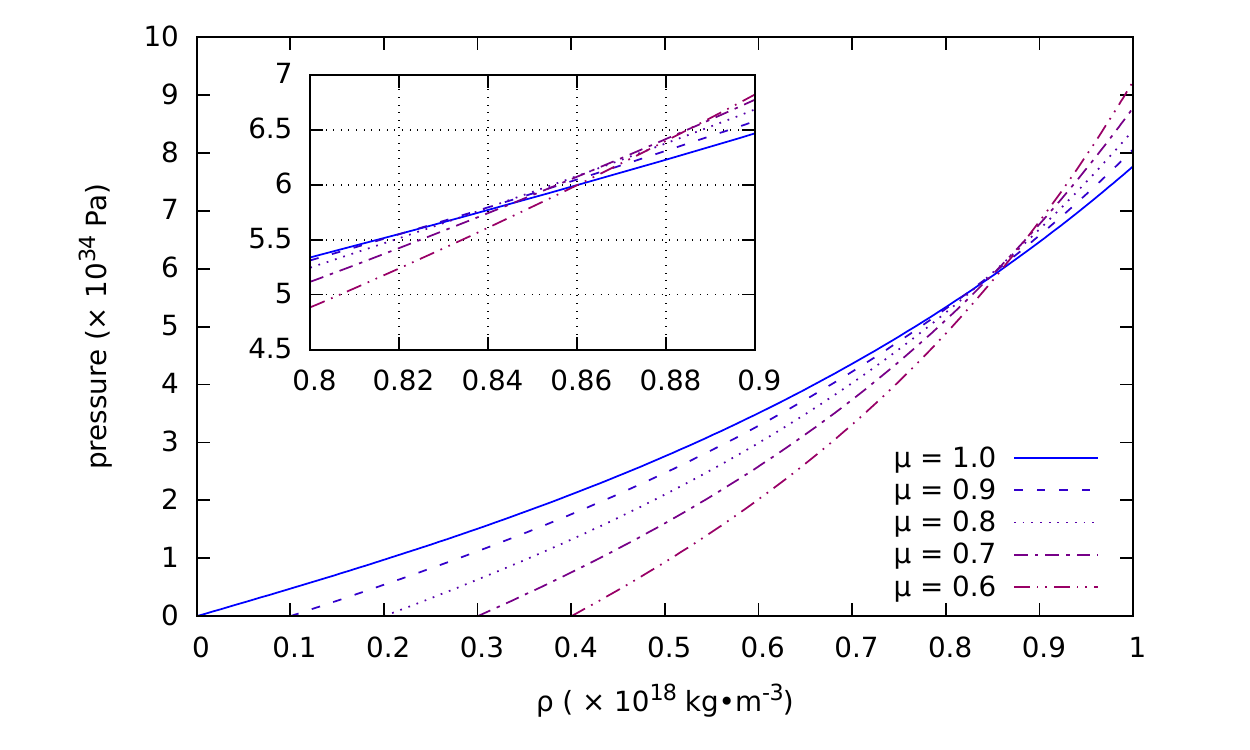}} 
\subfloat[The radial pressure, $\mu=1, \beta = 1 \times 10^{-16}$]{\label{an.fig:cAphiP,EOSmu1}
  \includegraphics[width=0.5\linewidth]{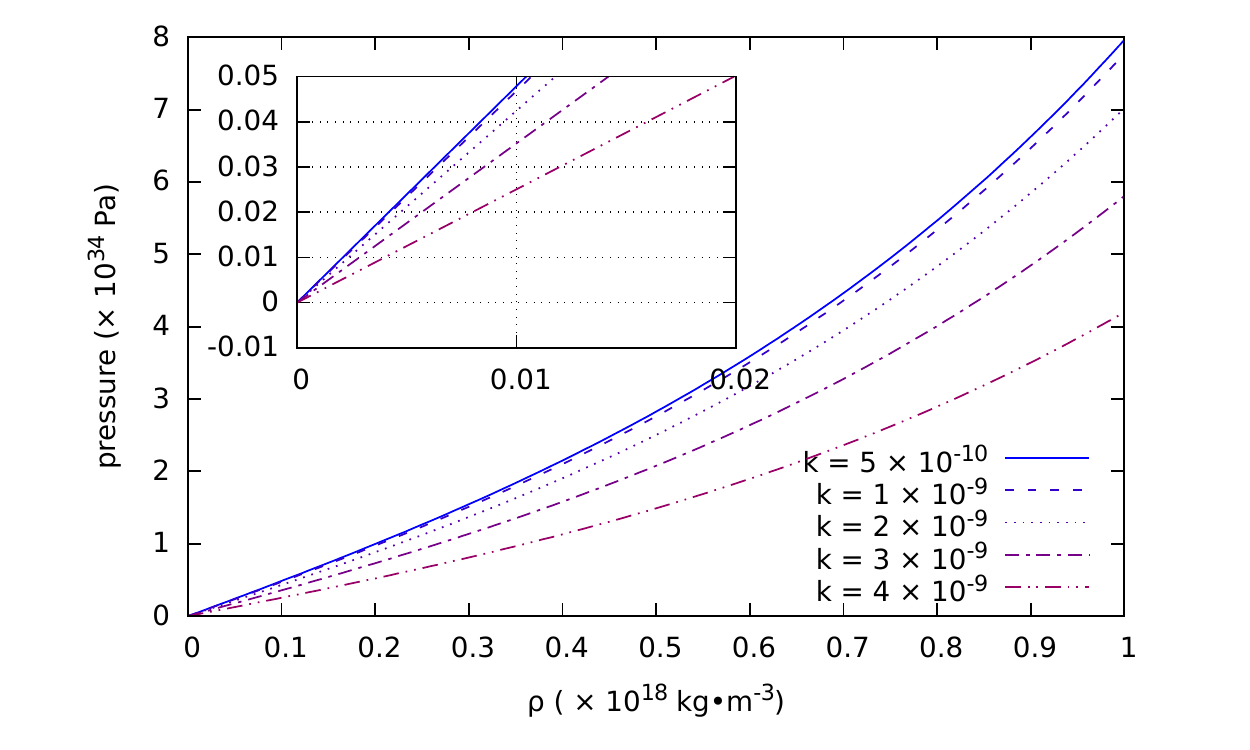}}\\
\subfloat[The radial pressure, $\mu=1, k=1 \times 10^{-9}$]{\label{an.fig:cAphiP,EOSmu1}
  \includegraphics[width=0.5\linewidth]{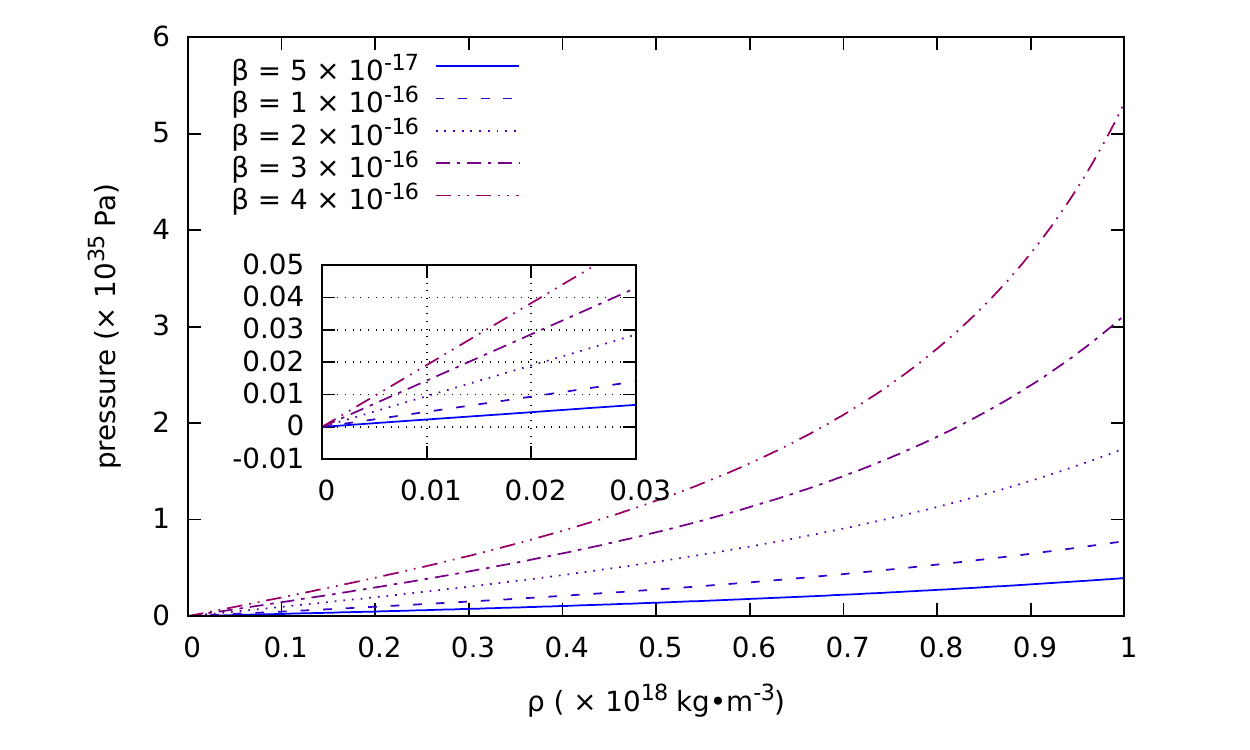}}
\caption[The radial pressure with $r$ for various
parameters]{Variation of the radial pressure with the density inside
  the star. The parameter values are
  $\rho_{c}=\un{1\times 10^{18}} \dunits, r_b = \un{1 \times 10^{4}
    m},\, \beta$ is set to $1 \times 10^{-16}$ on the left, and varies
  on the right, and $\mu$ is set to one on the right but takes the
  various values shown in the legend on the left}
\label{an.fig:cAPhiP,EOS}
\end{figure}

\subsection{Observables: masses, and radii}\label{An.ssec:Obser}
As in Tolman~VII, in this section we use causality, i.e.\ the
criterion that the speed of sound inside the star not exceed the speed
of light to limit the maximum possible masses, radii, and electric
charge that the models we are pursuing can admit.  

To implement this constraint, we find  the expression for the speed of
sound  in the  star, and  from  the shape  of the  equation of  state,
\(p(\rho),\) we know that the speed of sound is positive definite, and
a maximum  at the  centre of the  star.  We find  this expression  as a
function of the free parameters \emph{at  the centre of the star,} for
all the models we have.

Once this expression was found, we varied the parameters \(\mu\) and
\(r_{b}\) to find the maximum possible value for the central density
\(\rho_{c}\) for which the speed of sound \(v_{s}\) is just causal at
the centre, in Tolman~VII.  This allowed us then to calculate the
resulting mass of the star, since all the three parameters were known.
This method works in Tolman~VII where those are the only parameters
completely determining the solution.  In the new models however, this
is no longer the case.

We now specialize to the different solutions we have, and consider the
application of stability to each separately, and in doing so, encounter
a number of complications.  Indeed, the very fact that made the
finding of solutions easier: the greater number of parameters that
could be freely given, now hinders a straight forward physical
interpretation, since the speed of pressure waves is now dependant on
all of the new parameters too.  As a reminder,
equation~\eqref{an.eq.soundSpeed} when taken to the limit of \(r=0,\)
the centre reduces to
\[v_{s}(r=0,\mu, \rho_{c},r_{b},k,\beta) = \left( \f{r_{b}^{2}}{\kappa
      \rho_{c} \mu}\right) [\beta - 3k^{2} + \psi^{2}_{0} +
  \f{\kappa\rho_{c}}{3} \psi_{0}],\] where
\[\psi_{0} = \lim_{r \to 0} \left( \f{1}{rY} \deriv{Y}{r}\right).\]
\(\psi\) being the complicated part of these expressions we only show
plots of how one might go about restricting these parameter ranges.

\subsubsection{The Anisotropic case only}
In the anisotropic new solution we found, we have the anisotropy
factor \(\beta,\) but no charge \(k.\) To present how this additional
parameter changes the observables, we first investigate how the speed
of sound \(v_{s}\) changes with the different parameters and
\(\beta.\) Since we are concerned mostly with causality, and because
we are using natural units, the zeros of the function
\(v^{2}_{s} - 1\) give the parameter value we want to get the
coordinates of the causality surface.  This is more clearly shown in
the Figures~\ref{an.fig:PhiP,CausalityVs} where we show how
\(v^{2}_{s} - 1\) behaves for certain fixed parameters chosen to be
within the range of realistic stars, while another parameter varies on
the \(x-\)axis.
\begin{figure}[!htb]
\subfloat[Varying $\rho_{c}$]{\label{an.fig:PhiP.VsRhowdBeta}
  \includegraphics[width=0.5\linewidth]{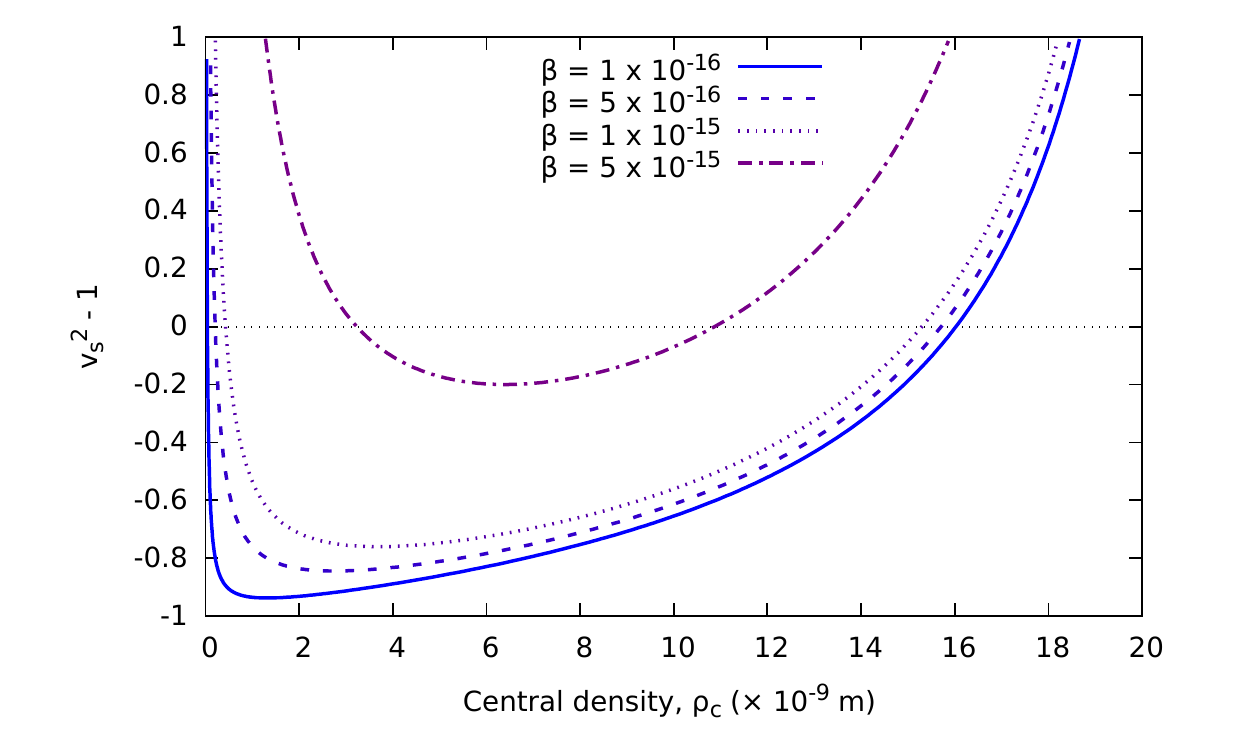}} 
\subfloat[Varying $\beta$]{\label{an.fig:PhiP.VsBetawdRho}
  \includegraphics[width=0.5\linewidth]{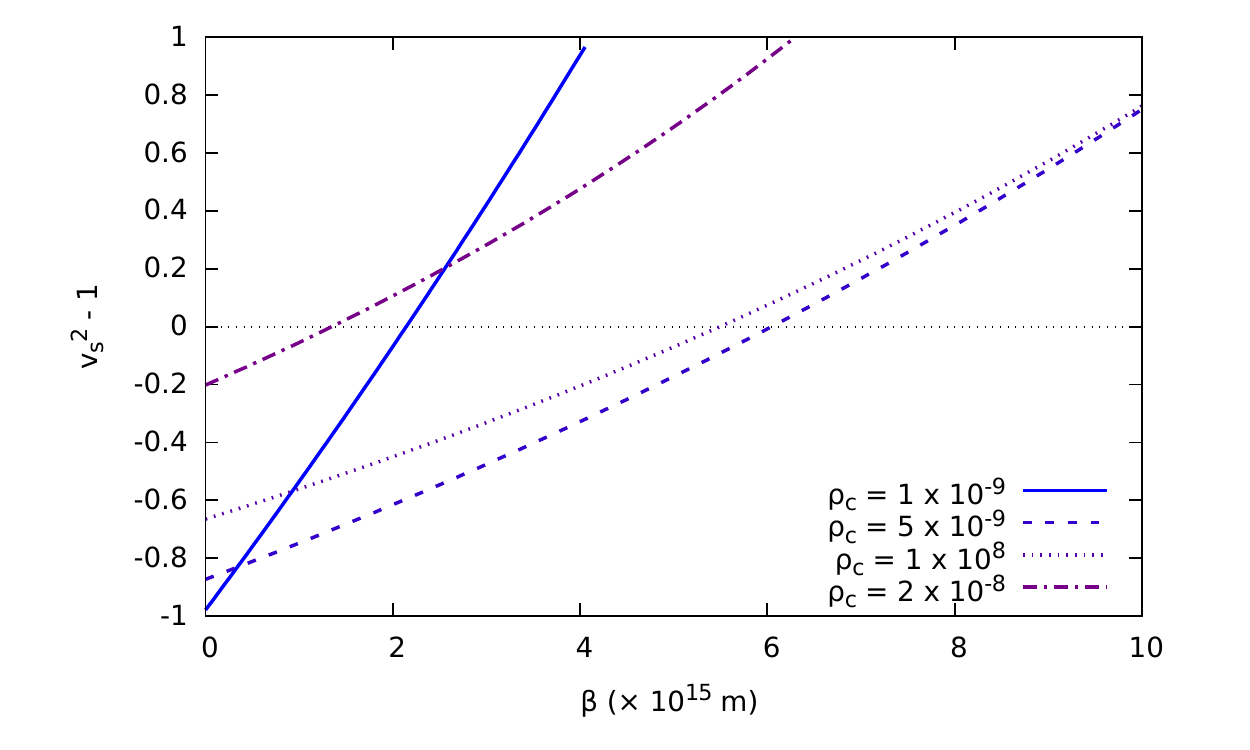}} \\
\subfloat[Varying $r_{b}$]{\label{an.fig:PhiP.VsRbwdBeta}
  \includegraphics[width=0.5\linewidth]{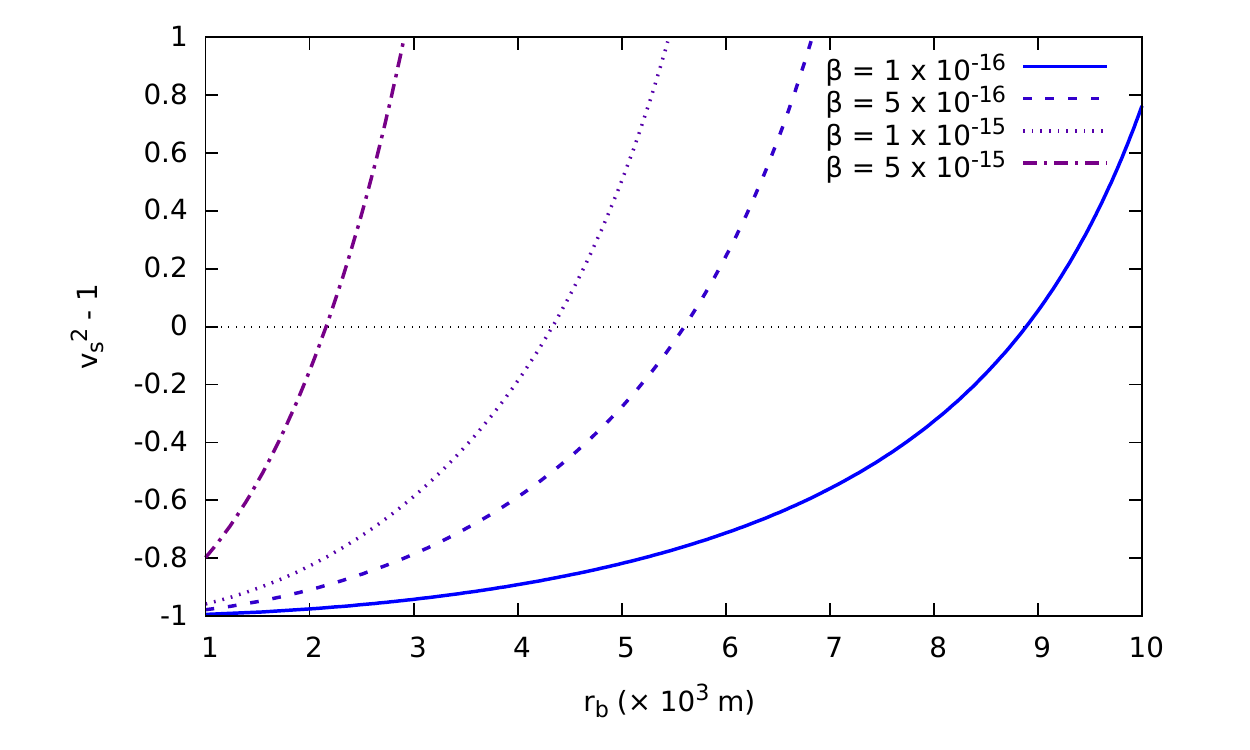}} 
\subfloat[Varying $\mu$]{\label{an.fig:PhiP.VsMuwdBeta}
  \includegraphics[width=0.5\linewidth]{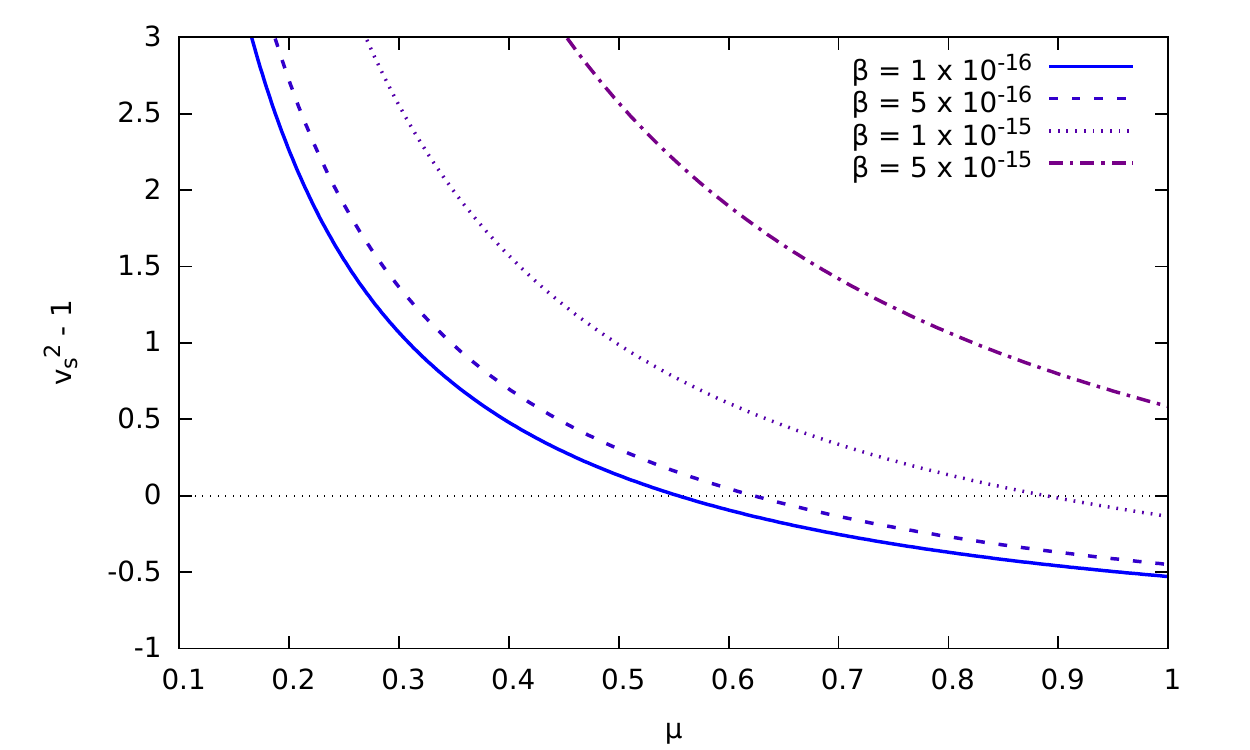}} 
\caption[The causality function with different parameters]{Variation
  of the function $v^{2}_{s} - 1$ at the centre of the star with
  different parameters being varied. In
  panel~\SFRef{an.fig:PhiP.VsRhowdBeta} the parameters are chosen as
  $r_{b} = 3 \times 10^{3}m$ and $\mu =1$;
  In~\SFRef{an.fig:PhiP.VsRbwdBeta}, they are
  $\rho_{c}=1 \times 10^{-9} m^{-2}, \mu = 1$;
  In~\SFRef{an.fig:PhiP.VsMuwdBeta}, they are
  $\rho_{c}=4 \times 10^{-9} m^{-2}, r_{b} = 3 \times 10^{3} m$;
  In~\SFRef{an.fig:PhiP.VsBetawdRho}, they are
  $r_{b}= 3 \times 10^{3}, \mu=1.$ These values were chosen after
  analysis of the different shapes of the curves in terms of the
  different parameters.}
\label{an.fig:PhiP,CausalityVs}
\end{figure}
Figure~\ref{an.fig:PhiP,CausalityVs} shows different values of the
parameters \(\rho_{c}, \beta, r_{b},\) and \(\mu\) where the causality
function \(v^{2}_{s} - 1\) crosses the horizontal axis.  These
solutions to the function give parameter ranges for which the speed of
sound is causal at the centre, and hence everywhere in the star.  Any
value of the parameters that allow for \emph{negative} values of the
function are causal parameter choices, and can potentially be used to
model a compact object.

We notice that the parameter change induce non-trivial parameter range
changes.  The graphs shown have been chosen to have parameters in
specially picked ranges to emphasize the issues that are involved in
finding appropriate ranges for the models.  In particular, notice that
in Figure~\ref{an.fig:PhiP.VsBetawdRho}, where the central density
\(\rho_{c}\) is varied for different values of the anisotropic
parameter \(\beta,\) or vice-versa in~\ref{an.fig:PhiP.VsRhowdBeta}, result
in the solution for the causality function to range through many
different values.

The interpretation that can be afforded to the strange shape
of~\ref{an.fig:PhiP.VsRhowdBeta} is that for low central densities, the
stiffness of the star has to be huge resulting in huge pressure wave
velocities, violating causality.  These models are also unstable,
since there is not enough mass to hold the star together.  In the
middle range between the two solutions -- where the curves intersect
the horizontal axis -- we have central densities that are big enough
to hold the star together, \emph{and} small enough to maintain the
stiffness low so that the speed of pressure wave is not too high.
This is the range of \(\rho_{c}\) we are interested in to model
physical stars.

In Figure~\ref{an.fig:PhiP.VsBetawdRho} we see how \(\beta\) changes both
the shape/slope of the velocity profile.  In this diagram by contrast,
any value of \(\beta\) corresponding to the causality function being
positive is rejected, and only lower values of \(\beta\) are then used.

When considering variations of the boundary radius \(r_{b},\) in
Figure~\ref{an.fig:PhiP.VsMuwdBeta}, we notice that stars with larger
radii, for fixed central densities and anisotropies are closer to the
causality limit.  Furthermore, higher anisotropies in larger values of
\(\beta\) ensures that the maximum radius \(r_{b}\) that can be used
is smaller than without anisotropy.

If variations in the self-boundedness \(\mu\) is sought instead, we
see that the natural case is almost always causal (at least for the
chosen parameter range,) but the closer one gets to the Schwarzschild
interior solution with \(\mu \to 0,\) the less causal the same models
become.

An alternative way to look at these parameter spaces is through
three-dimensional plots.  Next we plot the same type of surface as
Figure~\ref{t7.fig:phase}, but with different values of the
anisotropies.  Additionally we also show another set of
three-dimensional plots, for the natural \(\mu=1\) case but with
varying values of anisotropy beta.  These plots give an idea as to how
the anisotropy parameter alone changes the masses and maximum central
densities for causal stars.

In the first series of three dimensional plots, we show the mass in
solar units, the central mass density and the self-boundness parameter
\(\mu\) on the \(z, x\) and \(y\) axes respectively.  The surface
shows the triplets that make the star just causal at the centre of the
star, and since the speed of sound is a monotonic function of the
radial parameter, this implies that the speed of sound is always
causal in the star.
\begin{figure}[!h]
  \subfloat[Natural and two anisotropic]{\label{an.fig:phiP,CausBeta=0+5+10}
  \includegraphics[width=0.5\linewidth]{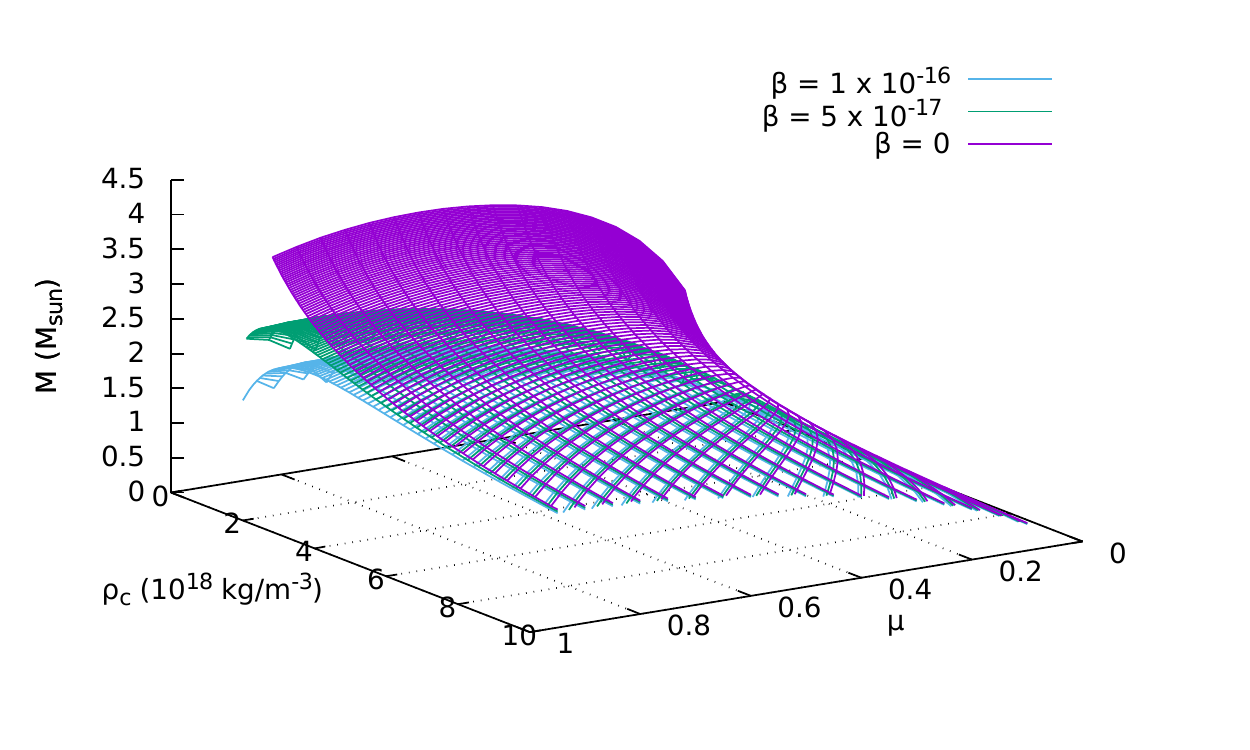}} 
\subfloat[three anisotropic]{\label{an.fig:phiP,CausBeta=0+5+10}
  \includegraphics[width=0.5\linewidth]{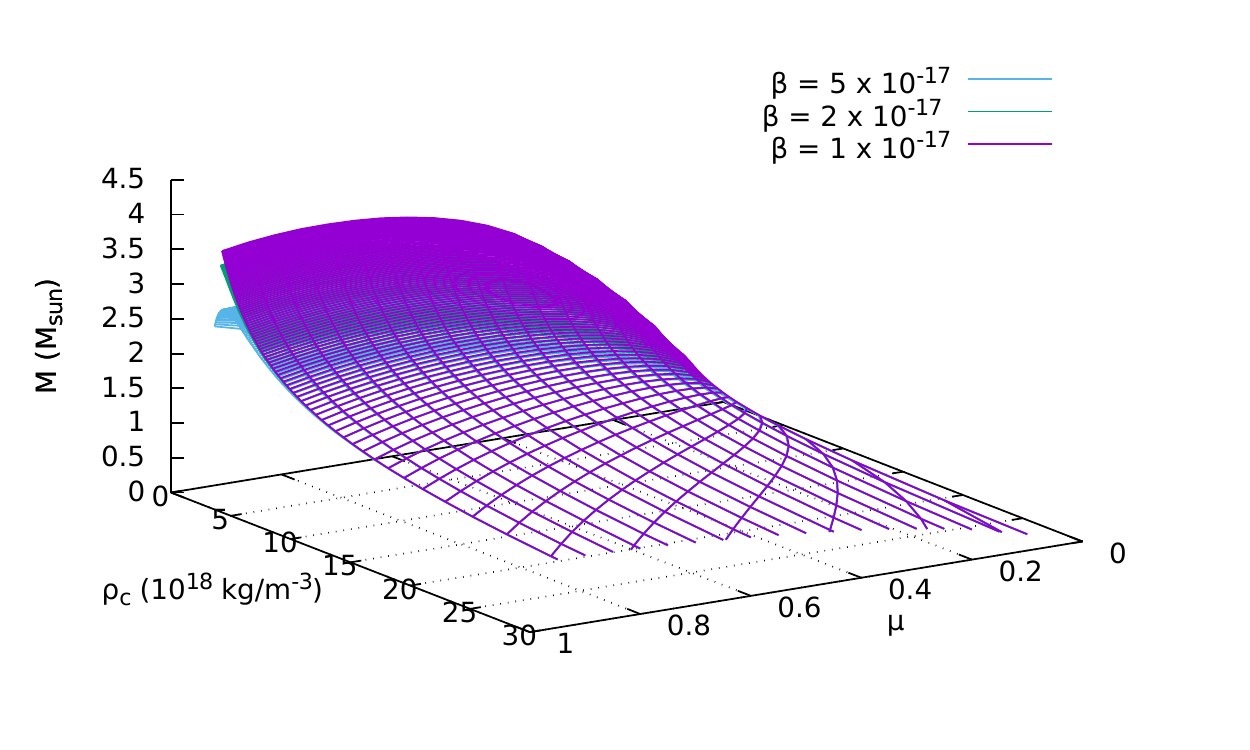}}
\caption[The causality surfaces for different anisotropies]{The causality surfaces for a
  variety of values of $\beta$ }
\label{an.fig:PhiP,CausalityBetaOnly}
\end{figure}

In Figure~\ref{an.fig:PhiP,CausalityBetaOnly}, we first plot the
isotropic Tolman~VII solution's causality surface.  Note that any
triplet underneath that causality curve represents a viable physical
star having parameters that form a causal star.  The other two
surfaces are for two different anisotropic parameters \(\beta,\) and
we see that the higher the anisotropy, the lower the causality
surface, implying that those stars have lower maximum masses typically
than the Tolman~VII solution, for the same parameters.  However since
we are usually mostly concerned with stars that are not on the edge of
causality, this should not be a problem to model more complex
anisotropic stars.

\begin{figure}[!h]
  \subfloat[Natural and two anisotropic]{\label{an.fig:phiP,CausMu=10+4}
  \includegraphics[width=0.5\linewidth]{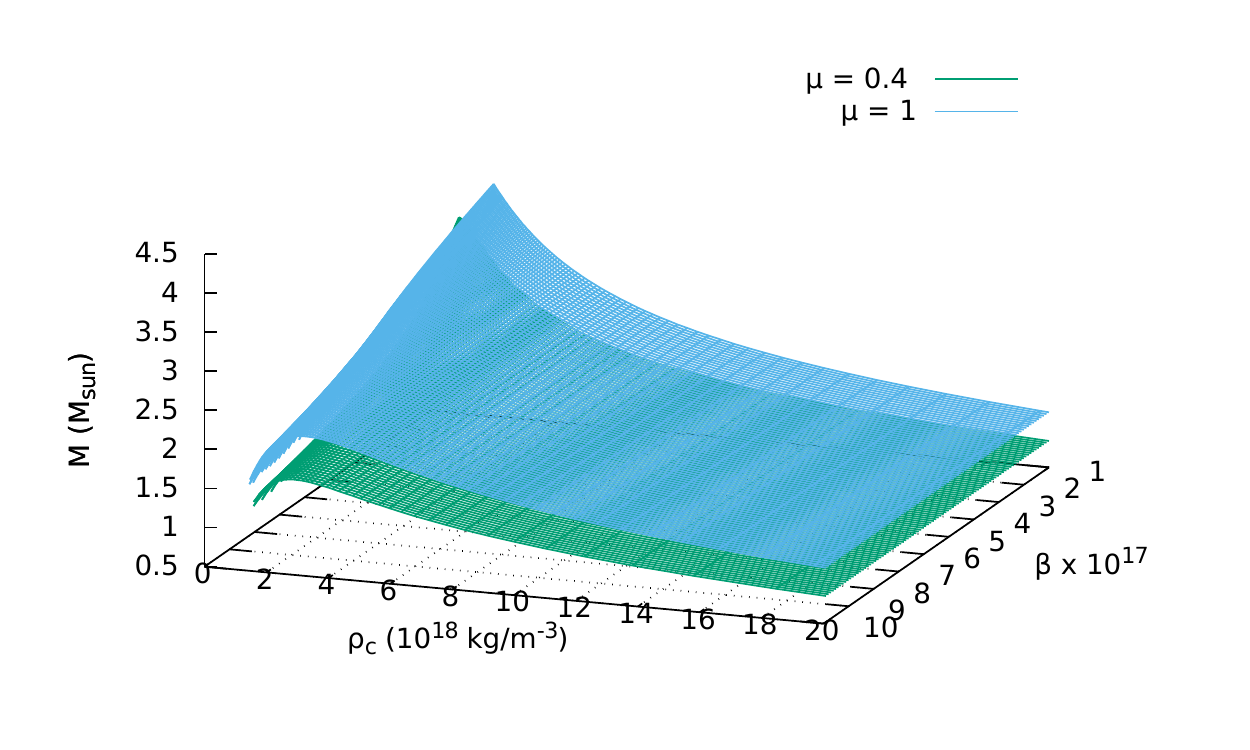}} 
\subfloat[three anisotropic]{\label{an.fig:phiP,CausMu=10+8+6}
  \includegraphics[width=0.5\linewidth]{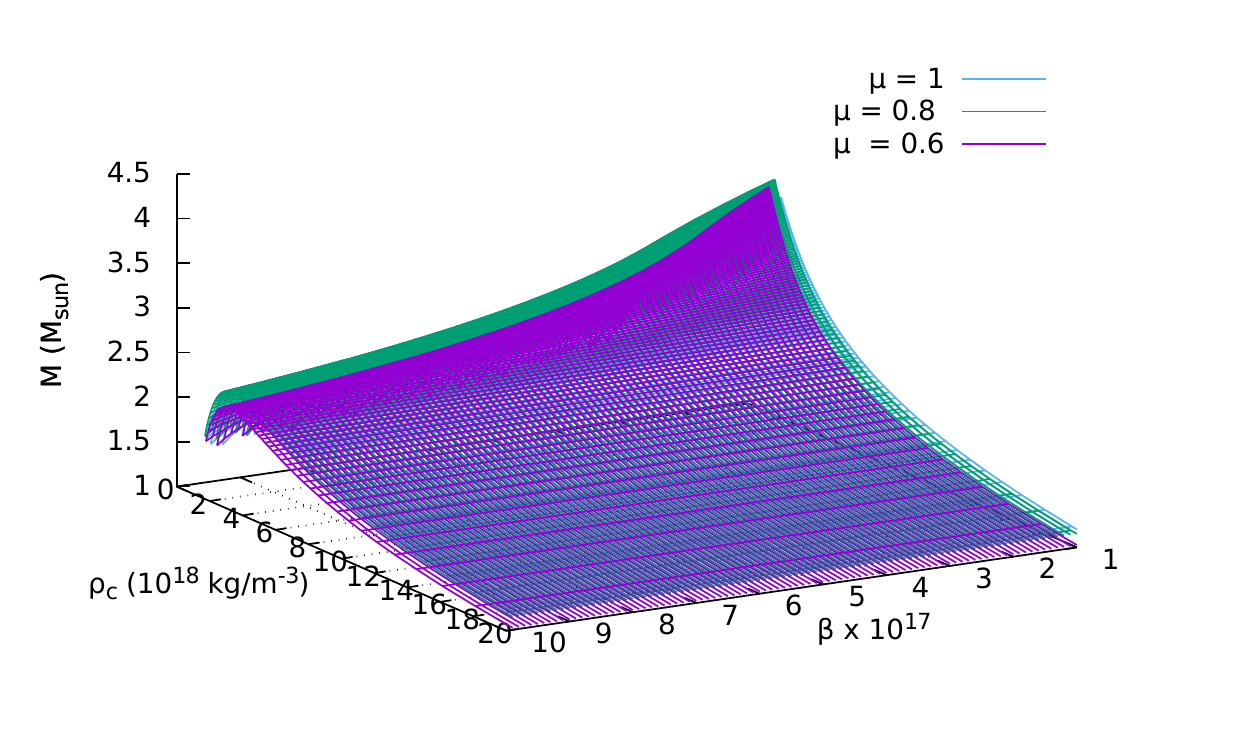}}
\caption[The causality surfaces for different $\mu$]{The causality surfaces for a
  variety of values of $\mu$ }
\label{an.fig:PhiP,CausalityMuOnly}
\end{figure}

\subsubsection{The ``anisotropised charge'' case}
Here since only one of the parameters of either \(k\) or \(\beta\)
remain, we choose one, and express the other one in terms of it.  Here
for convenience we chose \(k = \sqrt{\beta/2},\) and vary \(\beta\)
according to the legend.  This particular value of charge ensures that
all the anisotropic pressure is accounted for by the charge as
explained in detail in Section~\ref{ns.ssec:phiA}.  

\begin{figure}[!htb]
\subfloat[Varying $\rho_{c}$]{\label{an.fig:AC.VsRhowdBeta}
  \includegraphics[width=0.5\linewidth]{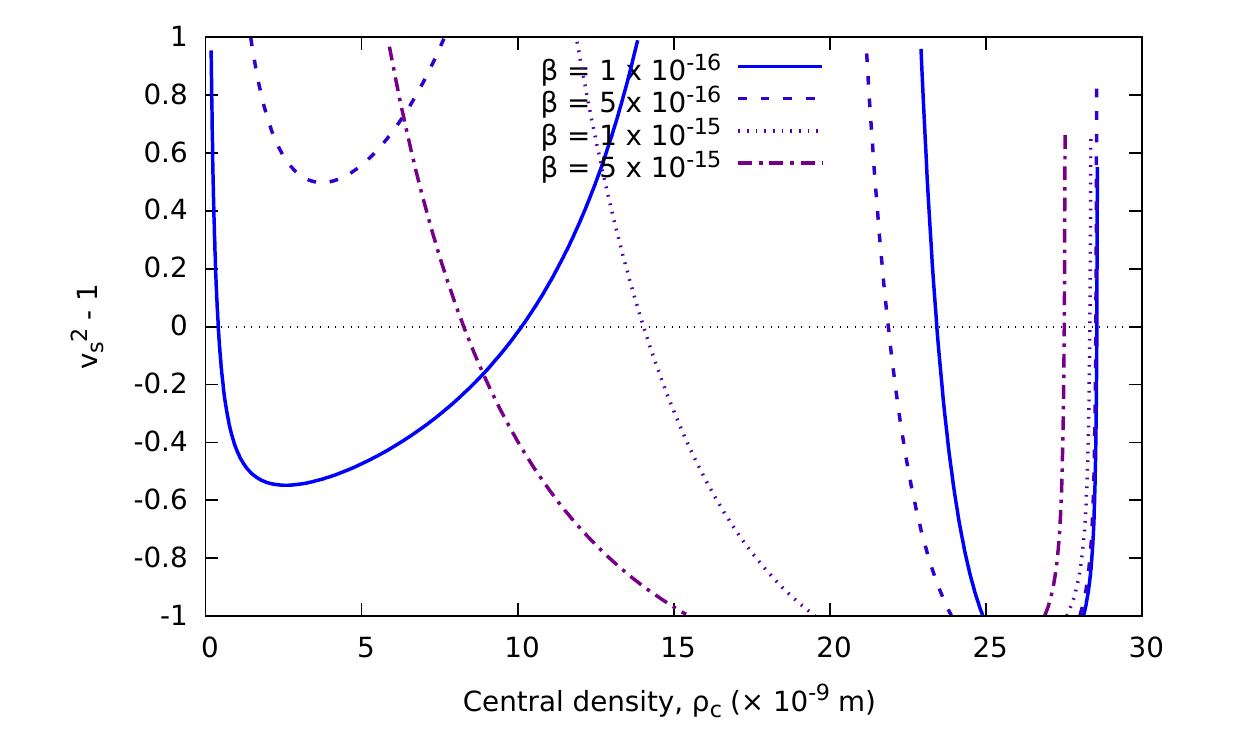}} 
\subfloat[Varying $\beta$]{\label{an.fig:AC.VsBetawdRho}
  \includegraphics[width=0.5\linewidth]{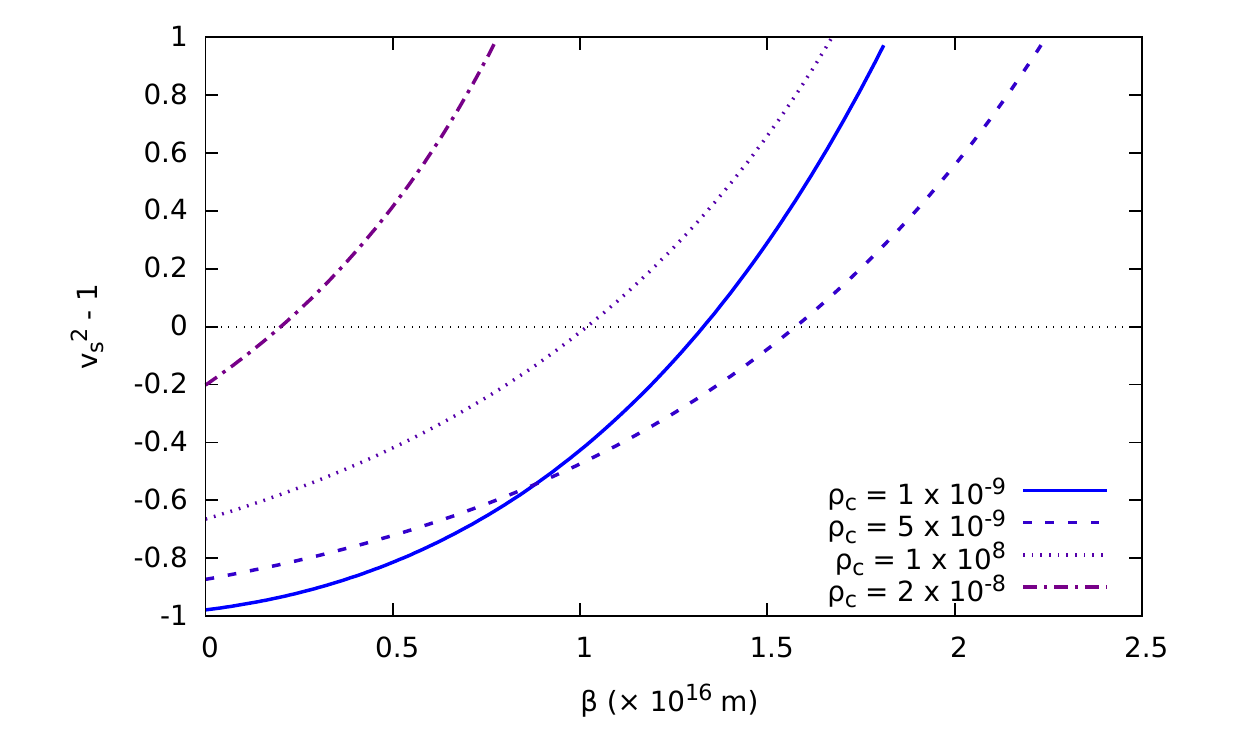}} \\
\subfloat[Varying $r_{b}$]{\label{an.fig:AC.VsRbwdBeta}
  \includegraphics[width=0.5\linewidth]{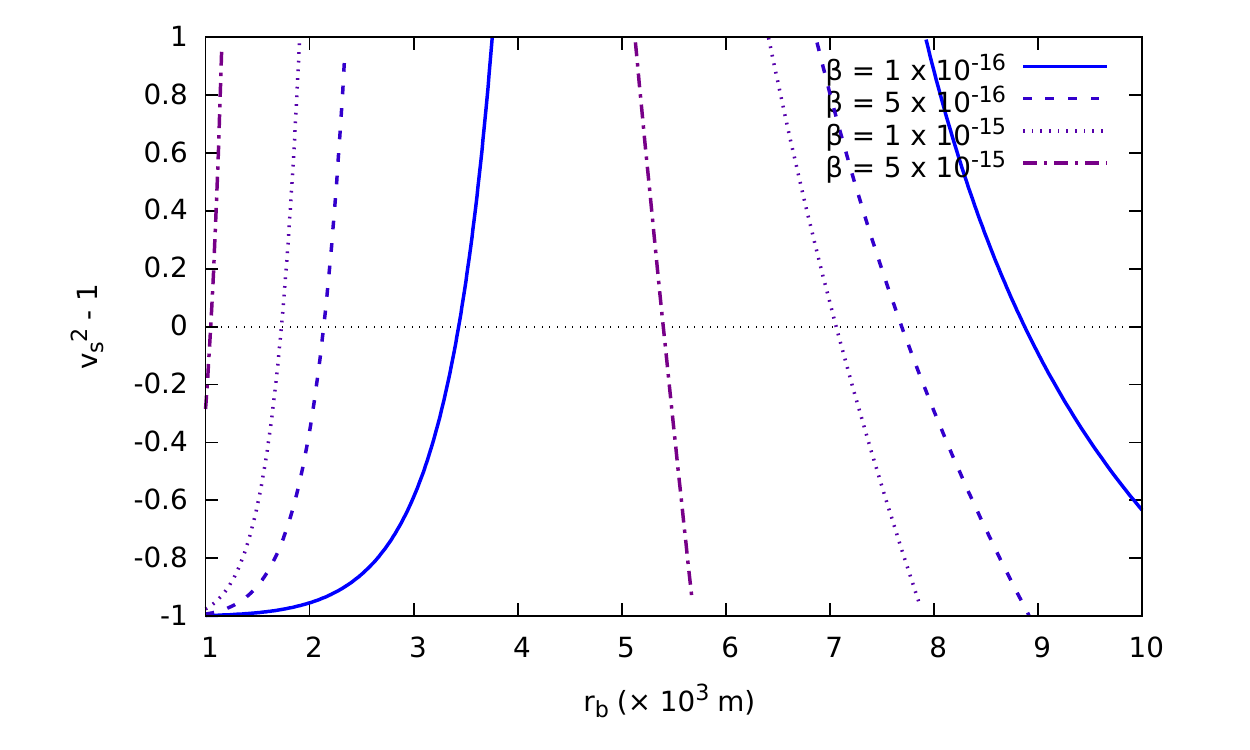}} 
\subfloat[Varying $\mu$]{\label{an.fig:AC.VsMuwdBeta}
  \includegraphics[width=0.5\linewidth]{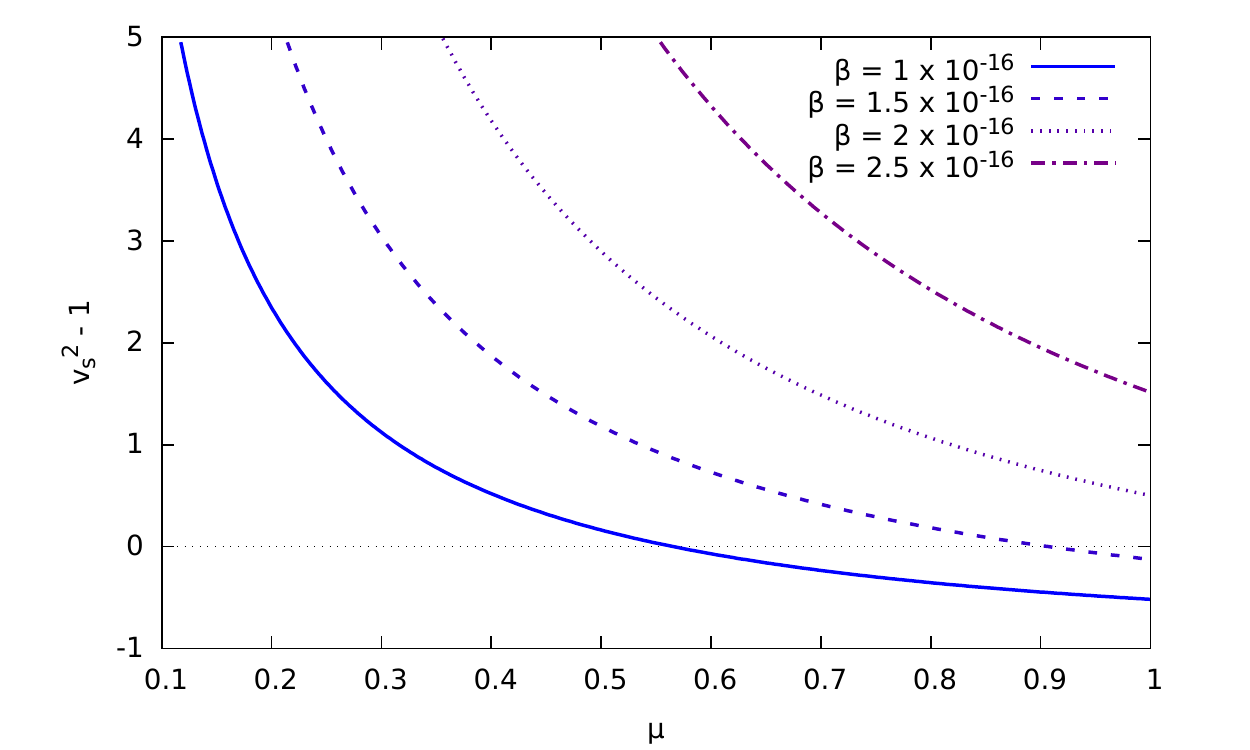}} 
\caption[The causality function with various parameters]{Variation of the function
  $v^{2}_{s} - 1$ at the centre of the star with different parameters
  being varied. In all the plots, $k=\sqrt{\beta/2},$ as expected in
  this model.  In panel~\SFRef{an.fig:AC.VsRhowdBeta} the parameters are
  chosen as $r_{b} = 3 \times 10^{3}m$ and $\mu =1$;
  In~\SFRef{an.fig:AC.VsRbwdBeta}, they are
  $\rho_{c}=1 \times 10^{-9} m^{-2}, \mu = 1$;
  In~\SFRef{an.fig:AC.VsMuwdBeta}, they are
  $\rho_{c}=4 \times 10^{-9} m^{-2}, r_{b} = 3 \times 10^{3} m$;
  In~\SFRef{an.fig:AC.VsBetawdRho}, they are
  $r_{b}= 3 \times 10^{3}, \mu=1.$ These values were chosen after
  analysis of the different shapes of the curves in terms of the
  different parameters.}
\label{an.fig:AC.CausalityVs}
\end{figure}

In Figure~\ref{an.fig:AC.CausalityVs} we plot the same causality
function as in the above section with respect to the different
parameters, with the charge always fixed to match the anisotropic
pressure.  As a result of varying \(\rho_{c},\) in
Figure~\ref{an.fig:AC.VsRhowdBeta} we see how the different values of
the causality function can be.  Again we are only interested in the
range \(|v^{2}_{s}| < 1|,\) and more specifically the solution of
\(v^{2}_{s} - 1 = 0\) for the limiting value of \(\rho_{c}\) for
causality.  The initial parts of the plot mimic the
plot~\ref{an.fig:PhiP.VsRhowdBeta} we had previously when charge was not
important.  We see that for some anisotropies/charge the initial part
of the function--for lower value of \(rho_{c}\)--is always above the
zero--axis, so that no plausible value of low central density is
admissible.  This is physically intuitive since for large charge, we
would require the mass to be large enough to gravitationally
compensate for electromagnetic repulsion.  For higher densities
however, we run into a different problem, where \(v_{s}\) is no longer
real, i.e.\ \(v^{2}_{s} < 0\) sometimes.  These are obviously
unphysical and cannot be used.  However certain values of the central
density in the higher ranges are admissible for causal solutions, and
these are the ones we would choose to model physical objects.  
By contrast the behaviour of causality is monotonic
in~\ref{an.fig:AC.VsBetawdRho} where \(\beta\) is varied on the
horizontal axis.  This makes it clear that only lower values of the
anisotropic parameter \(\beta\) are useful and definite limits on
their maximum value for a given \(\rho_{c}\) can be chosen.

Another complicated solution space structure is seen
in~\ref{an.fig:AC.VsRbwdBeta}, where asymptotes to the causality
function exist in the middle of the diagram.  However this structure
means that two distinct set of radii \(r_{b}\) are possible for a given
choice of the other parameters. This very unintuitive result is one
which makes having figures involving masses, radii and central
densities as in Figure~\ref{t7.fig:phase} difficult for this
particular case, since discrete ``island--''like regions of parameter
values where the star is causal is expected instead of the ``sheet''
type surface we saw previously.  

The monotonicity of the last~\ref{an.fig:AC.VsMuwdBeta} suggests that
once again most natural cases are causal, and as the anisotropic
parameter is increased, the possibility of a causal star decreases,
even if it were natural with \(\mu = 1.\) This is to be expected since
self-bound stars, with charge are even less physically plausible than
just self-bound ones in quark stars.

This concludes the analysis of this particular solution.  We see that
parameter ranges for a causal, ``anisotropised'' charged model are
possible, since the pressures can be chosen to be positive everywhere
in the star as shown in Section~\ref{ns.ssec:phiA}, and these models
can be causal as we just showed.  These being the more stringent
criteria that physical stars have to obey, we can conclude that these
model can be viable for modelling physical stars.

\subsubsection{The general case with both charge and anisotropy}
When both the charge and anisotropic parameter can be varied, there is
more possibilities for unwanted behaviour in the speed of sound, and
causality to occur, as we now see.  In the set of plots shown in
Figure~\ref{an.fig:CP,CausalityVs}, we find how the causality depends
on the initial parameters of this solution.  

\begin{figure}[!htb]
\subfloat[Varying $\rho_{c}$]{\label{an.fig:CP.VsRhowdBeta}
  \includegraphics[width=0.5\linewidth]{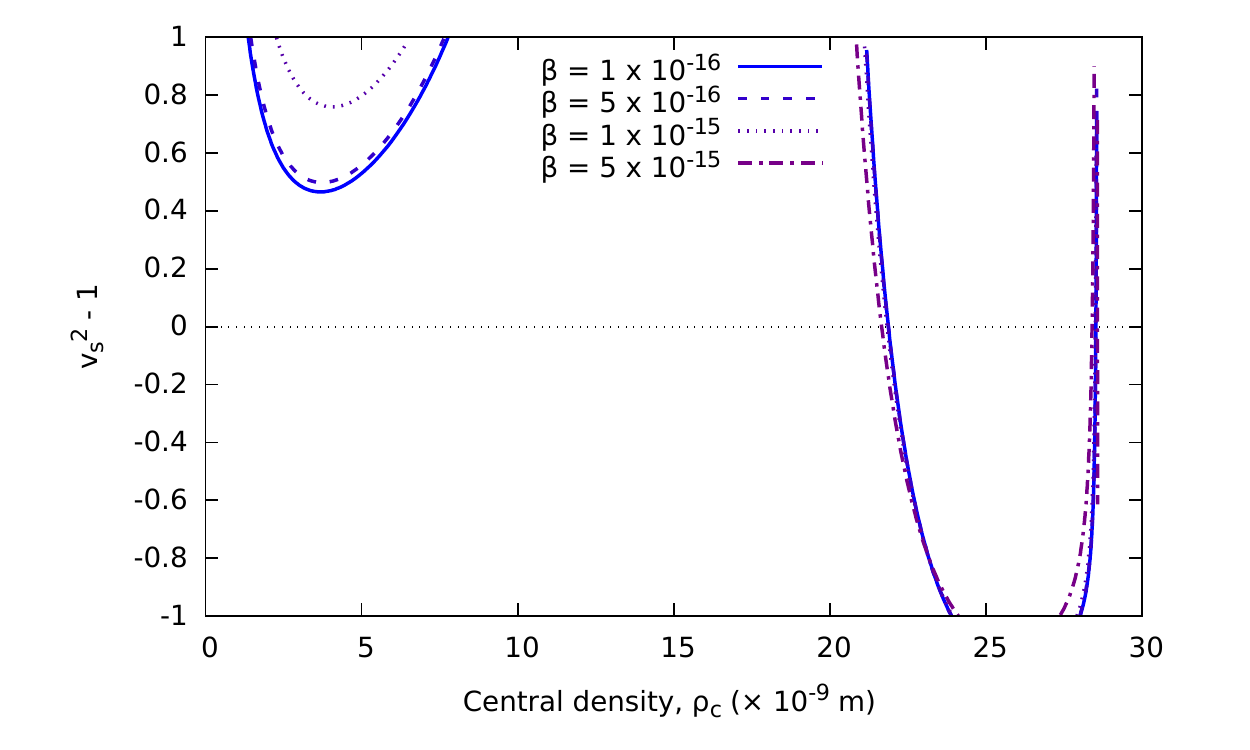}} 
\subfloat[Varying $\beta$]{\label{an.fig:CP.VsBetawdRho}
  \includegraphics[width=0.5\linewidth]{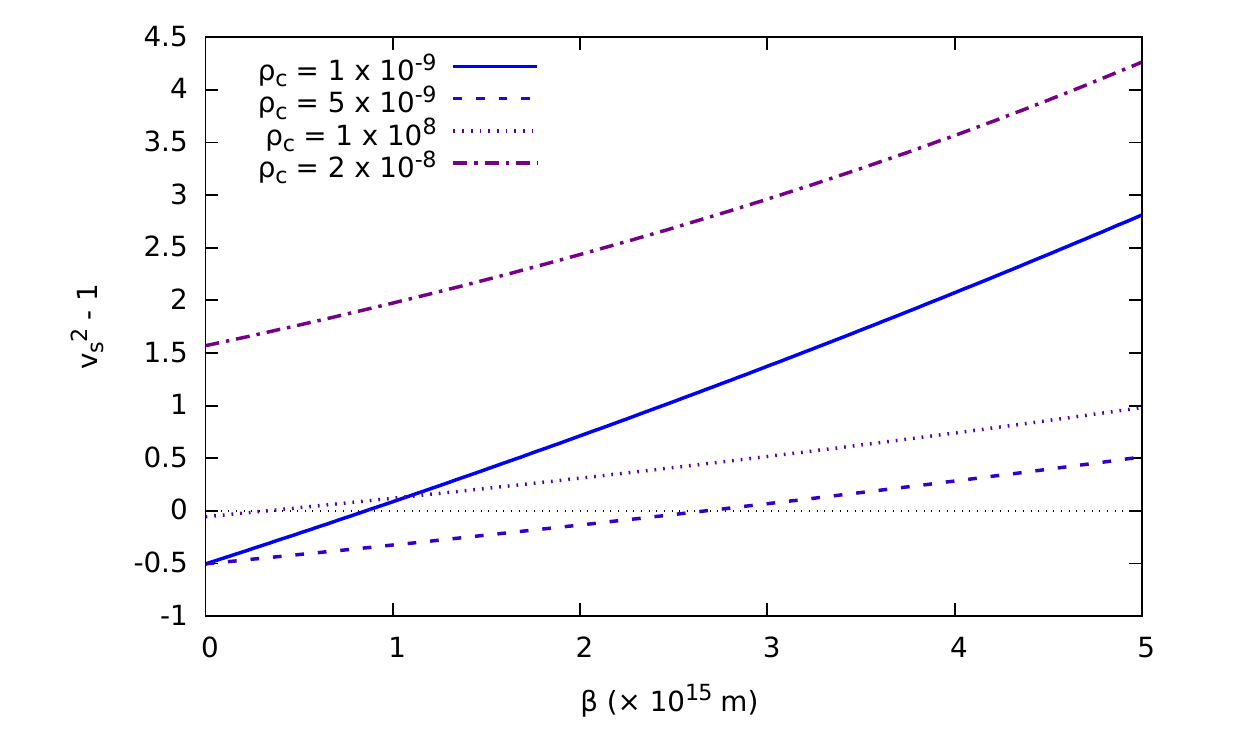}} \\
\subfloat[Varying $r_{b}$]{\label{an.fig:CP.VsRbwdBeta}
  \includegraphics[width=0.5\linewidth]{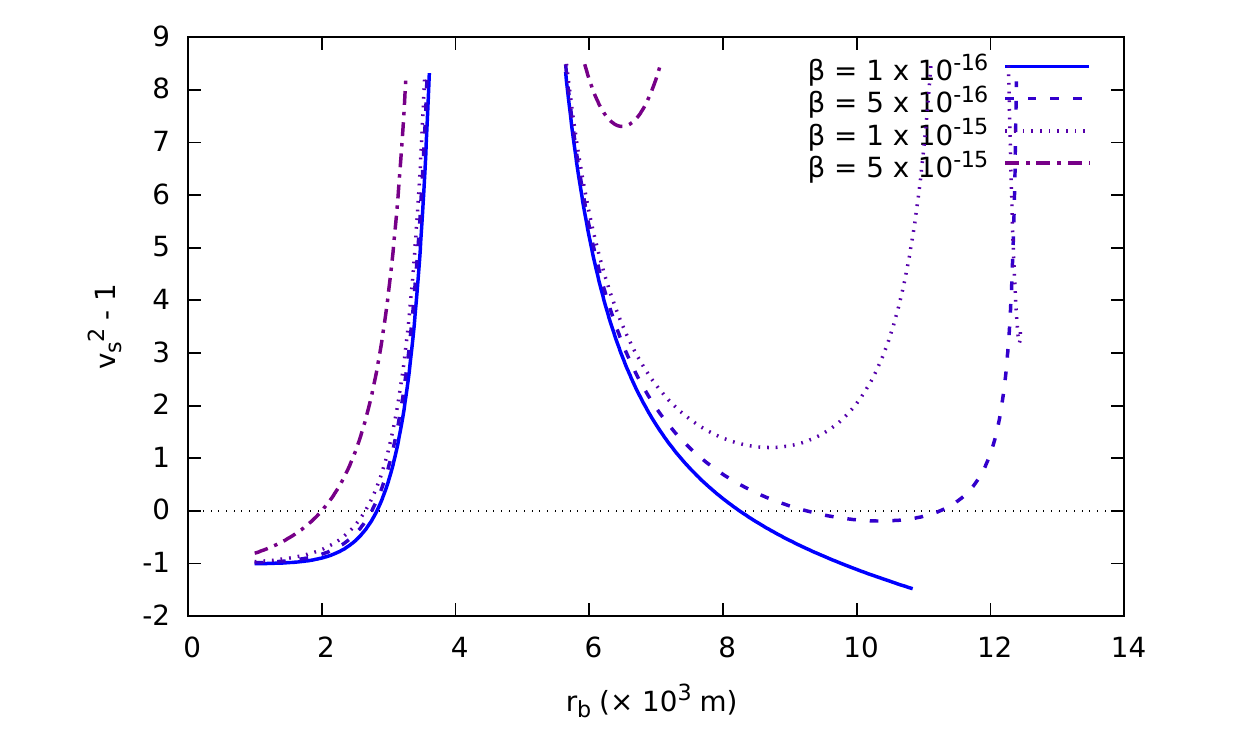}} 
\subfloat[Varying $\mu$]{\label{an.fig:CP.VsMuwdBeta}
  \includegraphics[width=0.5\linewidth]{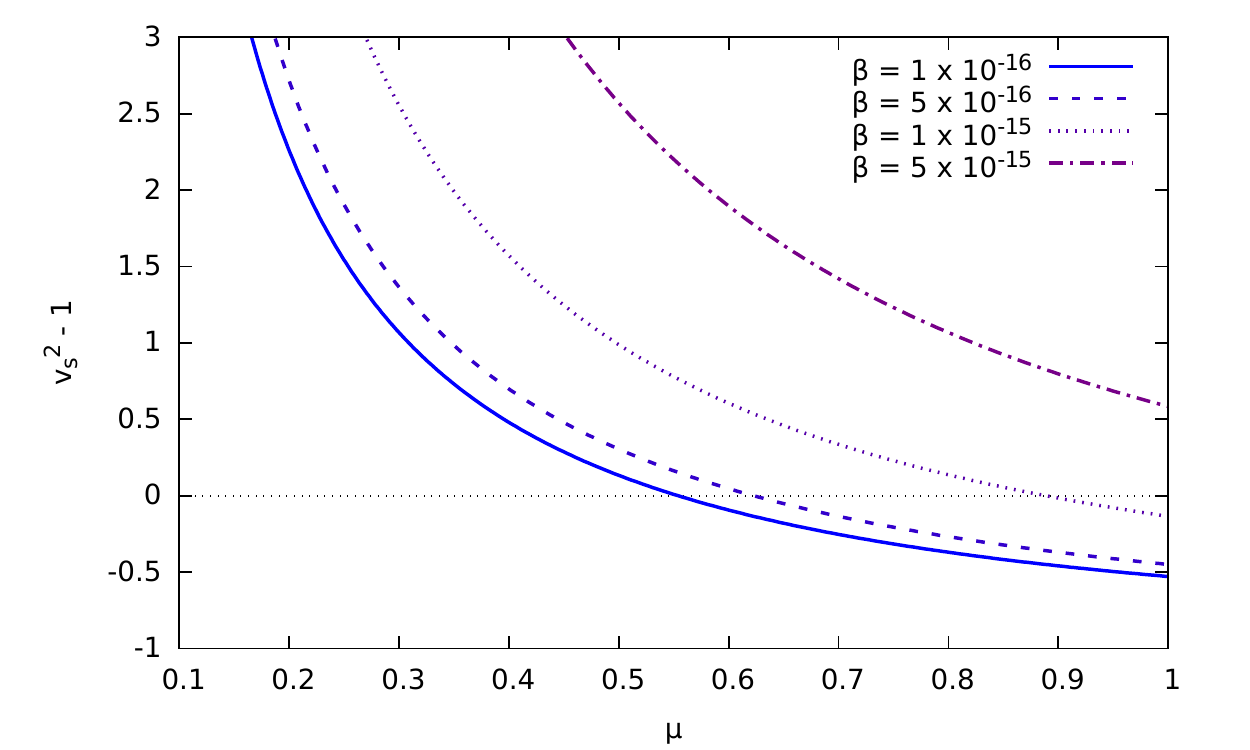}} \\
\subfloat[Varying $k$]{\label{an.fig:CP.VsKwdBeta}
  \includegraphics[width=0.5\linewidth]{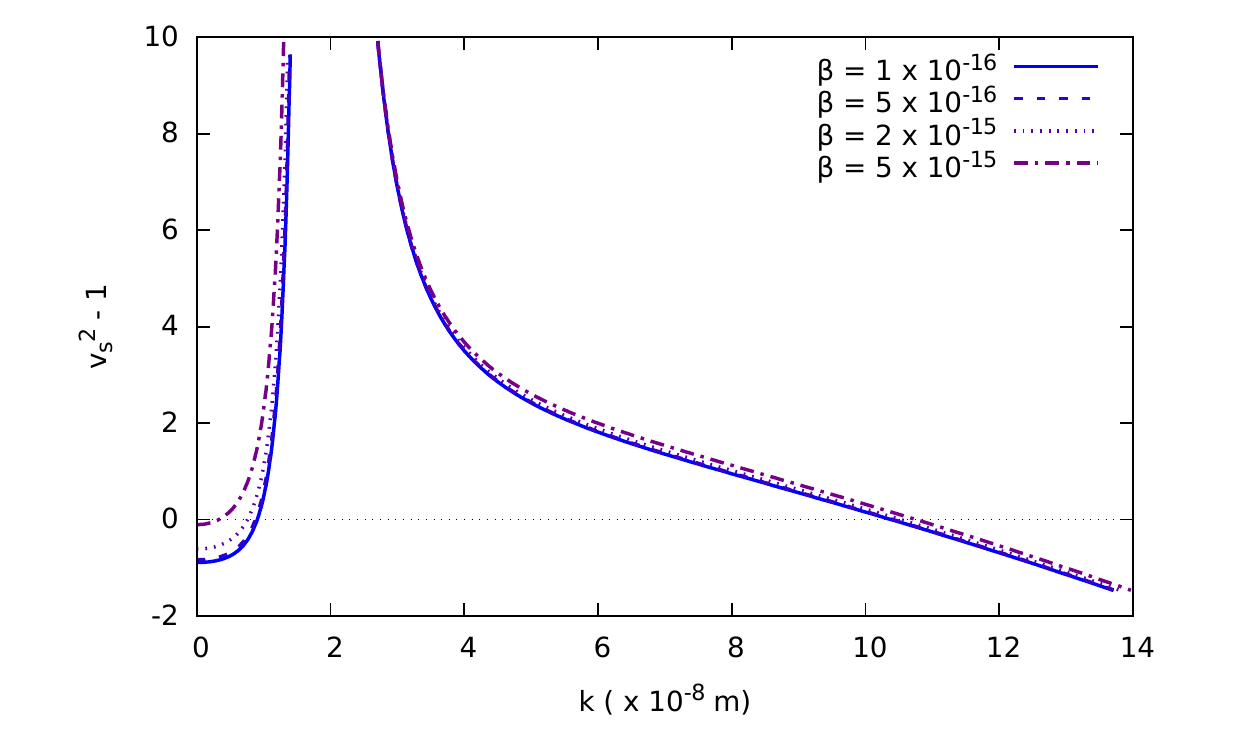}} 
\subfloat[Varying $k$]{\label{an.fig:CP.VsKwdRho}
  \includegraphics[width=0.5\linewidth]{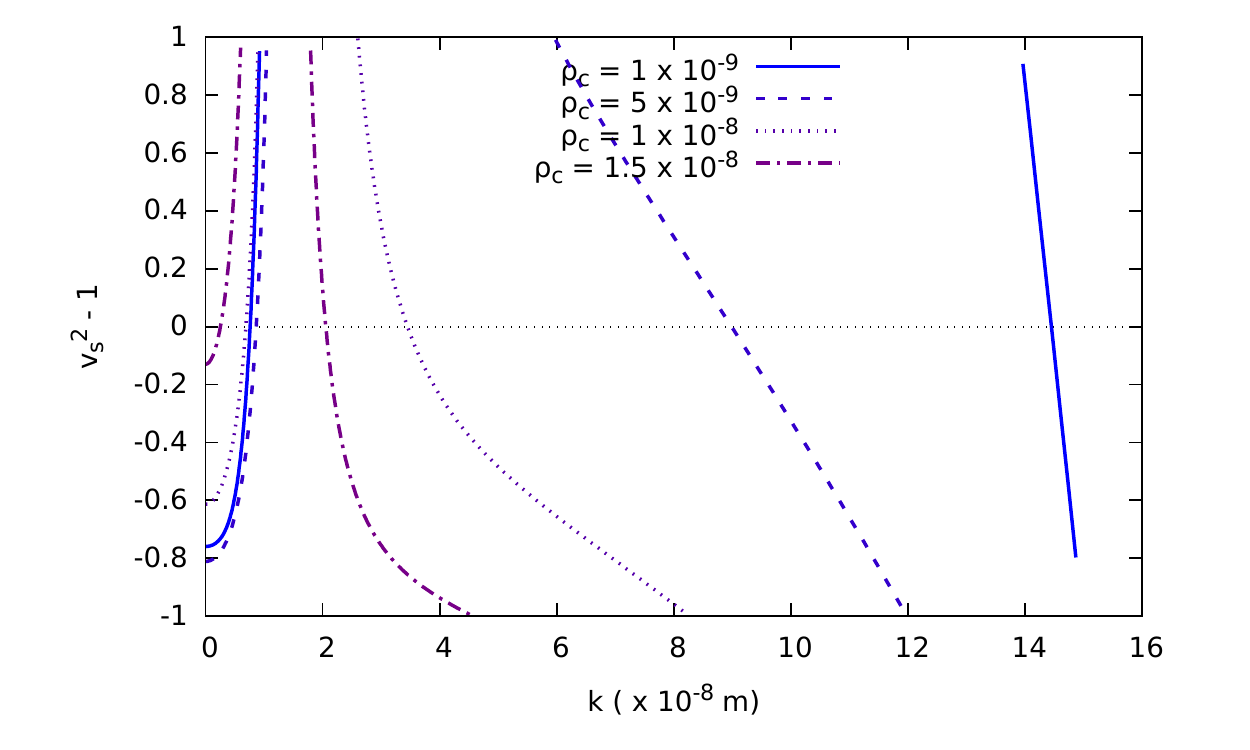}}
\caption[The causality function with various parameters]{Variation of the function
  $v^{2}_{s} - 1$ at the centre of the star with different parameters
  being varied. In all the plots, $k$ is specified separately from
  $\beta$ as expected in this model.  In
  panel~\SFRef{an.fig:CP.VsRhowdBeta} the parameters are chosen as
  $r_{b} = 3 \times 10^{3} \un{m}, k= 1 \times 10^{-8} m,$ and
  $\mu =1$; In~\SFRef{an.fig:CP.VsRbwdBeta}, they are
  $\rho_{c}=1 \times 10^{-9} \un{m}^{-2}, \mu = 1$ and
  $k = 1 \times 10^{-8} \un{m^{-2}}$; In~\SFRef{an.fig:CP.VsMuwdBeta}, they
  are $\rho_{c}=4 \times 10^{-9} m^{-2}, r_{b} = 3 \times 10^{3} m$
  and $k = 7 \times 10^{-9} \un{m^{-2}}$; In~\SFRef{an.fig:CP.VsBetawdRho},
  they are $r_{b}= 3 \times 10^{3}, \mu=1,$ and
  $k = 7 \times 10^{-9} \un{m^{-2}}$; In~\SFRef{an.fig:CP.VsKwdRho},
  they are $r_{b}= 3 \times 10^{3}, \mu=1.$ and
  $\beta = 5 \times 10^{-16} \un{m}$; In~\SFRef{an.fig:CP.VsKwdRho},
  they are $r_{b}= 3 \times 10^{3}, \mu=1,$ and
  $\beta = 5 \times 10^{-16} \un{m}$; In~\SFRef{an.fig:CP.VsKwdBeta},
  they are $r_{b}= 3 \times 10^{3}, \mu=1$ and
  $\rho_{c} = 4 \times 10^{-9} \un{m}.$
These values were chosen after
  analysis of the different shapes of the curves in terms of the
  different parameters.}
\label{an.fig:CP,CausalityVs}
\end{figure}

In~\ref{an.fig:CP.VsRhowdBeta}, for example, we find that the trend we
noticed in~\ref{an.fig:AC.VsRhowdBeta} is only accentuated in that the
initial part of the curves do not even cross the horizontal axis, thus
reducing the range of applicable central densities that could
potentially be used for modelling purposes.  The other striking
feature is that for \emph{all} values of \(\beta\) there exists
certain densities that have imaginary speeds of sound.  This makes the
choice of the parameter set that can be used tricky to specify exactly.

In the next plot~\ref{an.fig:CP.VsBetawdRho}, we see that as with the
previous case~\ref{an.fig:AC.VsBetawdRho}, we have a monotonic
dependence of the causality with respect to \(\beta,\) although
different \(\rho_{c}\) do change the shape of the curves.  Of note
here is that for high values of \(\rho_{c},\) as suggested by
Figure~\ref{an.fig:CP.VsRhowdBeta}, the speed of sound is non-causal
for all values of \(\beta,\) rendering that particular set of
parameter values impossible to use for modelling causal stars.

In Figure~\ref{an.fig:CP.VsRbwdBeta}, we see that indeed for certain
values of \(r_{b},\) an asymptote exists in the causality function.
As a result, only the values of \(r_{b}\) that are very small (smaller
than actual neutron stars) are admissible, or, for low enough
\(\beta,\) some large radii are still valid, however at even larger
radii, these same models start having imaginary speeds of sounds, so
that a very small set of \(r_{b}\) is actually admissible in the end.
This plot, together with Figure~\ref{an.fig:CP.VsKwdBeta}
and~\ref{an.fig:CP.VsBetawdRho}, accentuate the difficulty in
specifying a definite range of values where the model is causal.

The monotonicity of the speed with \(\mu\) in
Figure~\ref{an.fig:CP.VsMuwdBeta} makes interpretation easy, and as
previously, we see that natural models have a better chance of being
causal.  However high values of \(\beta\) ensures that no value of
\(\mu\) can be used for any models we want.  

The next two plots~\ref{an.fig:CP.VsKwdBeta}
and~\ref{an.fig:CP.VsKwdRho} finally show how by changing the electric
charge the model can go from causal initially, to non-causal through
an asymptote, but come back to causality with higher values of charge,
independently of the anisotropy factor \(\beta.\) This is very
counter-intuitive, as one would expect that higher charge would cause
the stiffness of the star to increase considerably.  However since in
general relativity, charge also contributed to the energy density,
this stiffening is not permanent and the star starts becoming causal
again for higher values of charge.

This concludes this section, which looked at the difficulties in
giving strict values for ranges of the parameters we are looking at.
Even in these simplistic models that we produced, the behaviour of the
causality with respect to any of the the parameters is complicated and
has to be approached with care, since certain sets of parameter values
can push the star into non-causality, and therefore probably instabilities.

\section{Solution validity: parameter value ranges}
Using both stability and causality, it should be possible to restrict
the values of the parameter set \(\rho_{c}, r_{b}, \mu, k,\) and
\(\beta.\) In the previous sections we have already given a number of
inequalities that restrict the values of these parameters.  However in
both Chapter~\ref{C:Stability} and Section~\ref{An.ssec:Obser} we
found out that except for the inequalities mentioned, no fixed value
for either the charge \(k\) or anisotropy parameter \(\beta\) can be
specified.  As a result, even specifying a range of values for each
parameter is impossible, however a well established algorithm in the
form of the list of inequalities given previously can be used to get
bounds on each parameter, once others are stated.  These bounds are
also parameter dependent as we showed in the form of plot in
figures~\ref{an.fig:PhiP,CausalityVs},
\ref{an.fig:AC.CausalityVs} and~\ref{an.fig:CP,CausalityVs}.

From the surface plots we showed that the range of values of masses
and radii produced by the stars.  The maximum masses for the just
causal stars is around four solar masses.  This is higher than the
measured mass of compact objects for the same radii, implying that any
triplet of parameter values below the surface, generating a mass
smaller than the highest one possible can be modelled through these
models and equations of state.

We can therefore conclude that all the 3 models, we have given EOS
figures for: the anisotropic model with \(\Phi^{2} > 0,\) the
``anisotropised charged'' model, and the charged anisotropic model
with \(\Phi^{2} > 0\) are viable models that can be used to model
compact objects.  The values for \(\beta\) that can be used in the
first model is typically around \(1 \times 10^{-17} \un{m^{-4}}\) in
geometrical units. This corresponds to about
\(1 \times 10^{27} \un{Pa \cdot m^{-2}}.\) Since we picked a function
from a mathematical point of view, there is no fundamental quantity we
associate with such a unit, except that \(\beta r^{2}\) is a pressure.

The typical values of \(k\) that we have been using are
\(k = 3 \times 10^{-9} \un{m^{-2}}.\) This correspond to total charges
\(q = kr^{3}_{b} = 3 \times 10^{3} \un{m}\) in geometrical units.
Converting to SI units we get that the typical charges that can be
associated with the model is around \(q = 3 \times 10^{20} C,\) for
causal stars. 

\section{Comparison with actual observation}
The table in Figure~\ref{pr.fig:PulsMass} give masses of neutron
stars, with the maximum mass being around 2.5 solar masses.  This
means that any or our models, including Tolman~VII can provide
reasonable models for them.  What more the models we provide are
causal, and exact.  If gravitational wave calculations, or Love number
calculations in neutron star binaries have to be carried out, then our
models which from a relativistic perspective have all the attributes
of physical relevance should be considered, since many of the
calculations will be greatly simplified, with an exact solution in hand.  

The same arguments apply to the radii measurements.  All of our models
have used radii of about \( 10 \un{km},\) but the parameters can be
changed down so that we have even more compact star, with
\(r_{b} \sim 5 \un{km}.\) While the latter are close to just being
non-causal, they are still not ruled out by the measurements in
Figure~\ref{pr.fig:PulsRad}.  If these objects need to be modelled,
one of our solutions can be used there too.

The prediction that we do get from our models are the typical values
of the anisotropy parameter, and the total charge that a compact
object can admit.  While direct measurement of these quantities is not
currently feasible, these are definite predictions that could be used
to infer the viability of our models.


\clearpage
\addcontentsline{toc}{chapter}{References}
\bibliography{bibliographyT}

\begin{thebibliography}{135}
\providecommand{\natexlab}[1]{#1}
\providecommand{\url}[1]{\texttt{#1}}
\expandafter\ifx\csname urlstyle\endcsname\relax
  \providecommand{\doi}[1]{doi: #1}\else
  \providecommand{\doi}{doi: \begingroup \urlstyle{rm}\Url}\fi

\bibitem[{Abreu} et~al.(2007){Abreu}, {Hern{\'a}ndez}, and
  {N{\'u}{\~n}ez}]{AbrHerNun07}
H.~{Abreu}, H.~{Hern{\'a}ndez}, and L.~A. {N{\'u}{\~n}ez}.
\newblock {Sound speeds, cracking and the stability of self-gravitating
  anisotropic compact objects}.
\newblock \emph{Classical and Quantum Gravity}, 24:\penalty0 4631--4645,
  September 2007.
\newblock \doi{10.1088/0264-9381/24/18/005}.

\bibitem[{Antoniadis} et~al.(2013){Antoniadis}, {Freire}, {Wex}, {Tauris},
  {Lynch}, {van Kerkwijk}, {Kramer}, {Bassa}, {Dhillon}, {Driebe}, {Hessels},
  {Kaspi}, {Kondratiev}, {Langer}, {Marsh}, {McLaughlin}, {Pennucci}, {Ransom},
  {Stairs}, {van Leeuwen}, {Verbiest}, and {Whelan}]{Ant13}
J.~{Antoniadis}, P.~C.~C. {Freire}, N.~{Wex}, T.~M. {Tauris}, R.~S. {Lynch},
  M.~H. {van Kerkwijk}, M.~{Kramer}, C.~{Bassa}, V.~S. {Dhillon}, T.~{Driebe},
  J.~W.~T. {Hessels}, V.~M. {Kaspi}, V.~I. {Kondratiev}, N.~{Langer}, T.~R.
  {Marsh}, M.~A. {McLaughlin}, T.~T. {Pennucci}, S.~M. {Ransom}, I.~H.
  {Stairs}, J.~{van Leeuwen}, J.~P.~W. {Verbiest}, and D.~G. {Whelan}.
\newblock {A Massive Pulsar in a Compact Relativistic Binary}.
\newblock \emph{Science}, 340:\penalty0 448, April 2013.
\newblock \doi{10.1126/science.1233232}.

\bibitem[{Bardeen} et~al.(1966){Bardeen}, {Thorne}, and {Meltzer}]{BarThoMel66}
J.~M. {Bardeen}, K.~S. {Thorne}, and D.~W. {Meltzer}.
\newblock {A Catalogue of Methods for Studying the Normal Modes of Radial
  Pulsation of General-Relativistic Stellar Models}.
\newblock \emph{\apj}, 145:\penalty0 505, August 1966.
\newblock \doi{10.1086/148791}.

\bibitem[{Baumgarte} and {Rendall}(1993)]{BauRen93}
T.~W. {Baumgarte} and A.~D. {Rendall}.
\newblock {Regularity of spherically symmetric static solutions of the Einstein
  equations}.
\newblock \emph{Classical and Quantum Gravity}, 10:\penalty0 327--332, February
  1993.
\newblock \doi{10.1088/0264-9381/10/2/014}.

\bibitem[{Bayin}(1978)]{Bay78}
S.~S. {Bayin}.
\newblock {Solutions of Einstein's field equations for static fluid spheres}.
\newblock \emph{\prd}, 18:\penalty0 2745--2751, October 1978.
\newblock \doi{10.1103/PhysRevD.18.2745}.

\bibitem[{Bayin}(1984)]{Bay84}
S.~{\c S}. {Bayin}.
\newblock {Relativistic stars with anisotropic pressure.}
\newblock \emph{Annals of the New York Academy of Sciences}, 422:\penalty0
  330--330, 1984.
\newblock \doi{10.1111/j.1749-6632.1984.tb23365.x}.

\bibitem[Bayin(1982)]{Bay82}
Sel{\c c}uk~{\c S}. Bayin.
\newblock Anisotropic fluid spheres in general relativity.
\newblock \emph{Phys. Rev. D}, 26:\penalty0 1262--1274, Sep 1982.
\newblock \doi{10.1103/PhysRevD.26.1262}.
\newblock URL \url{http://link.aps.org/doi/10.1103/PhysRevD.26.1262}.

\bibitem[Bergmann(1976)]{Ber76}
P.G. Bergmann.
\newblock \emph{Introduction to the Theory of Relativity}.
\newblock Dover Books on Physics Series. Dover Publications, 1976.
\newblock ISBN 9780486632827.

\bibitem[{Bhar} et~al.(2015){Bhar}, {Murad}, and {Pant}]{BhaMurPan15}
P.~{Bhar}, M.~H. {Murad}, and N.~{Pant}.
\newblock {Relativistic anisotropic stellar models with Tolman VII spacetime}.
\newblock \emph{\apss}, 359:\penalty0 13, September 2015.
\newblock \doi{10.1007/s10509-015-2462-9}.

\bibitem[{Birkhoff} and {Langer}(1923)]{Bir23}
G.~D. {Birkhoff} and R.~E. {Langer}.
\newblock \emph{{Relativity and modern physics}}.
\newblock 1923.

\bibitem[{B{\"o}hmer} and {Mussa}(2011)]{BohMus10}
C.~G. {B{\"o}hmer} and A.~{Mussa}.
\newblock {Charged perfect fluids in the presence of a cosmological constant}.
\newblock \emph{General Relativity and Gravitation}, 43:\penalty0 3033--3046,
  November 2011.
\newblock \doi{10.1007/s10714-011-1223-5}.

\bibitem[{Bondi}(1964)]{Bon64}
H.~{Bondi}.
\newblock {Massive Spheres in General Relativity}.
\newblock \emph{Royal Society of London Proceedings Series A}, 282:\penalty0
  303--317, November 1964.
\newblock \doi{10.1098/rspa.1964.0234}.

\bibitem[{Bonnor}(1960)]{Bon60}
W.~B. {Bonnor}.
\newblock {The mass of a static charged sphere}.
\newblock \emph{Zeitschrift fur Physik}, 160:\penalty0 59--65, February 1960.
\newblock \doi{10.1007/BF01337478}.

\bibitem[{Bonnor}(1965)]{Bon65}
W.~B. {Bonnor}.
\newblock {The equilibrium of a charged sphere}.
\newblock \emph{\mnras}, 129:\penalty0 443, 1965.

\bibitem[{Boonserm} et~al.(2005){Boonserm}, {Visser}, and
  {Weinfurtner}]{BooVisWei03}
P.~{Boonserm}, M.~{Visser}, and S.~{Weinfurtner}.
\newblock {Generating perfect fluid spheres in general relativity}.
\newblock \emph{\prd}, 71\penalty0 (12):\penalty0 124037, June 2005.
\newblock \doi{10.1103/PhysRevD.71.124037}.

\bibitem[{Boonserm} et~al.(2015){Boonserm}, {Ngampitipan}, and
  {Visser}]{BooNgaVis15}
P.~{Boonserm}, T.~{Ngampitipan}, and M.~{Visser}.
\newblock {Modelling anisotropic fluid spheres in general relativity}.
\newblock \emph{ArXiv e-prints}, January 2015.

\bibitem[{Bowers} and {Liang}(1974)]{BowLia74}
R.~L. {Bowers} and E.~P.~T. {Liang}.
\newblock {Anisotropic Spheres in General Relativity}.
\newblock \emph{ApJ}, 188:\penalty0 657, March 1974.
\newblock \doi{10.1086/152760}.

\bibitem[Buchdahl(1959)]{Buc59}
H.~A. Buchdahl.
\newblock General relativistic fluid spheres.
\newblock \emph{Phys. Rev.}, 116:\penalty0 1027--1034, Nov 1959.
\newblock \doi{10.1103/PhysRev.116.1027}.

\bibitem[Buchdahl and Land(1968)]{BucLan68}
H.~A. Buchdahl and W.~J. Land.
\newblock The relativistic incompressible sphere.
\newblock \emph{Journal of the Australian Mathematical Society}, 8:\penalty0
  6--16, 2 1968.
\newblock ISSN 1446-8107.
\newblock \doi{10.1017/S1446788700004559}.

\bibitem[{Burke} and {Hobill}(2009)]{BurHob09}
J.~{Burke} and D.~{Hobill}.
\newblock {New Physically Realistic Solutions for Charged Fluid Spheres}.
\newblock \emph{ArXiv e-prints}, October 2009.

\bibitem[{Cadeau} et~al.(2007){Cadeau}, {Morsink}, {Leahy}, and
  {Campbell}]{Cad07}
C.~{Cadeau}, S.~M. {Morsink}, D.~{Leahy}, and S.~S. {Campbell}.
\newblock {Light Curves for Rapidly Rotating Neutron Stars}.
\newblock \emph{\apj}, 654:\penalty0 458--469, January 2007.
\newblock \doi{10.1086/509103}.

\bibitem[{Chandrasekhar}(1964)]{Cha64}
S.~{Chandrasekhar}.
\newblock {The Dynamical Instability of Gaseous Masses Approaching the
  Schwarzschild Limit in General Relativity.}
\newblock \emph{\apj}, 140:\penalty0 417, August 1964.
\newblock \doi{10.1086/147938}.

\bibitem[Chandrasekhar(1964)]{Cha64L}
S.~Chandrasekhar.
\newblock Dynamical instability of gaseous masses approaching the
  {S}chwarzschild limit in general relativity.
\newblock \emph{Phys. Rev. Lett.}, 12:\penalty0 437--438, Apr 1964.
\newblock \doi{10.1103/PhysRevLett.12.437.2}.

\bibitem[{Chirenti} et~al.(2015){Chirenti}, {de Souza}, and
  {Kastaun}]{ChiSouKas15}
C.~{Chirenti}, G.~H. {de Souza}, and W.~{Kastaun}.
\newblock {Fundamental oscillation modes of neutron stars: Validity of
  universal relations}.
\newblock \emph{\prd}, 91\penalty0 (4):\penalty0 044034, February 2015.
\newblock \doi{10.1103/PhysRevD.91.044034}.

\bibitem[Choquet-Bruhat(2008)]{Cho08}
Y.~Choquet-Bruhat.
\newblock \emph{General Relativity and the Einstein Equations}.
\newblock Oxford Mathematical Monographs. OUP Oxford, 2008.
\newblock ISBN 9780199230723.

\bibitem[Choquet-Bruhat et~al.(1982)Choquet-Bruhat, DeWitt-Morette, and
  Dillard-Bleick]{Cho82}
Y.~Choquet-Bruhat, C.~DeWitt-Morette, and M.~Dillard-Bleick.
\newblock \emph{Analysis, Manifolds, and Physics}.
\newblock Number pt. 1 in Analysis, Manifolds, and Physics. North-Holland
  Publishing Company, 1982.
\newblock ISBN 9780444860170.

\bibitem[{Cooperstock} and {de La Cruz}(1979)]{CooCru79}
F.~I. {Cooperstock} and V.~{de La Cruz}.
\newblock {Static and stationary solutions of the Einstein-Maxwell equations}.
\newblock \emph{General Relativity and Gravitation}, 10:\penalty0 681--697,
  June 1979.
\newblock \doi{10.1007/BF00756904}.

\bibitem[Cosenza et~al.(1981)Cosenza, Herrera, Esculpi, and Witten]{CosHer81}
M.~Cosenza, L.~Herrera, M.~Esculpi, and L.~Witten.
\newblock Some models of anisotropic spheres in general relativity.
\newblock \emph{Journal of Mathematical Physics}, 22\penalty0 (1):\penalty0
  118--125, 1981.
\newblock \doi{http://dx.doi.org/10.1063/1.524742}.

\bibitem[{Damour} and {Nagar}(2009)]{DamNag09}
T.~{Damour} and A.~{Nagar}.
\newblock {Relativistic tidal properties of neutron stars}.
\newblock \emph{\prd}, 80\penalty0 (8):\penalty0 084035, October 2009.
\newblock \doi{10.1103/PhysRevD.80.084035}.

\bibitem[{Damour} et~al.(2012){Damour}, {Nagar}, and {Villain}]{DamNagVil12}
T.~{Damour}, A.~{Nagar}, and L.~{Villain}.
\newblock {Measurability of the tidal polarizability of neutron stars in
  late-inspiral gravitational-wave signals}.
\newblock \emph{\prd}, 85\penalty0 (12):\penalty0 123007, June 2012.
\newblock \doi{10.1103/PhysRevD.85.123007}.

\bibitem[{Delgaty} and {Lake}(1998)]{DelLak98}
M.~S.~R. {Delgaty} and K.~{Lake}.
\newblock {Physical acceptability of isolated, static, spherically symmetric,
  perfect fluid solutions of Einstein's equations}.
\newblock \emph{Computer Physics Communications}, 115:\penalty0 395--415,
  December 1998.
\newblock \doi{10.1016/S0010-4655(98)00130-1}.

\bibitem[{Dev} and {Gleiser}(2000)]{DevGle00}
K.~{Dev} and M.~{Gleiser}.
\newblock {Anisotropic Stars: Exact Solutions}.
\newblock \emph{ArXiv Astrophysics e-prints}, December 2000.

\bibitem[{Dev} and {Gleiser}(2003)]{DevGle03}
K.~{Dev} and M.~{Gleiser}.
\newblock {Anisotropic Stars II: Stability}.
\newblock \emph{General Relativity and Gravitation}, 35:\penalty0 1435--1457,
  August 2003.
\newblock \doi{10.1023/A:1024534702166}.

\bibitem[D'Inverno(1992)]{Inv92}
R.~D'Inverno.
\newblock \emph{Introducing Einstein's Relativity}.
\newblock Oxford University Press, 1992.

\bibitem[{Droste}(1917)]{Dro17}
J.~{Droste}.
\newblock {The field of a single centre in Einstein's theory of gravitation,
  and the motion of a particle in that field}.
\newblock \emph{Koninklijke Nederlandse Akademie van Wetenschappen Proceedings
  Series B Physical Sciences}, 19:\penalty0 197--215, 1917.

\bibitem[Durgapal and Gehlot(1971)]{DurGeh71}
M.~C. Durgapal and G.~L. Gehlot.
\newblock Spheres with varying density in general relativity.
\newblock \emph{Journal of Physics A: General Physics}, 4\penalty0
  (6):\penalty0 749--755, 1971.

\bibitem[{Durgapal} and {Rawat}(1980)]{DurRaw79}
M.~C. {Durgapal} and P.~S. {Rawat}.
\newblock {Non-rigid massive spheres in general relativity}.
\newblock \emph{Mon. Not. R. Astron. Soc.}, 192:\penalty0 659--662, September
  1980.

\bibitem[{Durgapal} et~al.(1984){Durgapal}, {Pande}, and
  {Phuloria}]{DurPanPhu84}
M.~C. {Durgapal}, A.~K. {Pande}, and R.~S. {Phuloria}.
\newblock {Physically realizable relativistic stellar structures}.
\newblock \emph{\apss}, 102:\penalty0 49--66, July 1984.
\newblock \doi{10.1007/BF00651061}.

\bibitem[{Dyson} et~al.(1920){Dyson}, {Eddington}, and {Davidson}]{DysEdd20}
F.~W. {Dyson}, A.~S. {Eddington}, and C.~{Davidson}.
\newblock {A Determination of the Deflection of Light by the Sun's
  Gravitational Field, from Observations Made at the Total Eclipse of May 29,
  1919}.
\newblock \emph{Philosophical Transactions of the Royal Society of London
  Series A}, 220:\penalty0 291--333, 1920.
\newblock \doi{10.1098/rsta.1920.0009}.

\bibitem[{Einstein}(1915)]{Ein15}
A.~{Einstein}.
\newblock {Die Feldgleichungen der Gravitation}.
\newblock \emph{Sitzungsberichte der K{\"o}niglich Preu{\ss}ischen Akademie der
  Wissenschaften (Berlin), Seite 844-847.}, 1915.

\bibitem[{Einstein}(1916)]{Ein16}
A.~{Einstein}.
\newblock {N{\"a}herungsweise Integration der Feldgleichungen der Gravitation}.
\newblock \emph{Sitzungsberichte der K{\"o}niglich Preu{\ss}ischen Akademie der
  Wissenschaften (Berlin), Seite 688-696.}, 1916.

\bibitem[{Esculpi} and {Alom{\'a}}(2010)]{EscAlo10}
M.~{Esculpi} and E.~{Alom{\'a}}.
\newblock {Conformal anisotropic relativistic charged fluid spheres with a
  linear equation of state}.
\newblock \emph{European Physical Journal C}, 67:\penalty0 521--532, June 2010.
\newblock \doi{10.1140/epjc/s10052-010-1273-y}.

\bibitem[{Finch} and {Skea}(1998)]{FinSke98}
M.~R. {Finch} and J.~E.~F {Skea}.
\newblock A review of the relativistic static sphere.
\newblock Unpublished, available at
  \url{www.dft.if.uerj.br/usuarios/JimSkea/papers/pfrev.ps}, 1998.

\bibitem[Florides(1974)]{Flo74}
P.~S. Florides.
\newblock A new interior schwarzschild solution.
\newblock \emph{Proceedings of the Royal Society of London. Series A,
  Mathematical and Physical Sciences}, 337\penalty0 (1611):\penalty0 529--535,
  1974.
\newblock ISSN 00804630.

\bibitem[Florides(1977)]{Flo77}
P.~S. Florides.
\newblock {The Complete Field of a General Static Spherically Symmetric
  Distribution of Charge}.
\newblock \emph{Nuovo Cim.}, A42:\penalty0 343--359, 1977.
\newblock \doi{10.1007/BF02862400}.

\bibitem[{Glazer}(1976)]{Gla76}
I.~{Glazer}.
\newblock {General relativistic pulsation equation for charged fluids}.
\newblock \emph{Annals of Physics}, 101:\penalty0 594--600, October 1976.
\newblock \doi{10.1016/0003-4916(76)90024-5}.

\bibitem[{Glazer}(1979)]{Gla79}
I.~{Glazer}.
\newblock {Stability analysis of the charged homogeneous model}.
\newblock \emph{\apj}, 230:\penalty0 899--904, June 1979.
\newblock \doi{10.1086/157149}.

\bibitem[{Glendenning}(1996)]{Gle96}
N.~{Glendenning}.
\newblock \emph{{Compact Stars. Nuclear Physics, Particle Physics and General
  Relativity.}}
\newblock Springer-Verlag New York, 1996.

\bibitem[Glendenning(1992)]{Gle92}
Norman~K. Glendenning.
\newblock First-order phase transitions with more than one conserved charge:
  Consequences for neutron stars.
\newblock \emph{Phys. Rev. D}, 46\penalty0 (4):\penalty0 1274--1287, Aug 1992.
\newblock \doi{10.1103/PhysRevD.46.1274}.

\bibitem[Gradshteyn and Ryzhik(2007)]{GraRyz07}
I.~S. Gradshteyn and I.~M. Ryzhik.
\newblock \emph{Table of integrals, series, and products}.
\newblock Elsevier/Academic Press, Amsterdam, seventh edition, 2007.
\newblock ISBN 978-0-12-373637-6; 0-12-373637-4.
\newblock Translated from the Russian, Translation edited and with a preface by
  Alan Jeffrey and Daniel Zwillinger, With one CD-ROM (Windows, Macintosh and
  UNIX).

\bibitem[{Guillot} and {Rutledge}(2014)]{GuiRut14}
S.~{Guillot} and R.~E. {Rutledge}.
\newblock {Rejecting Proposed Dense Matter Equations of State with Quiescent
  Low-mass X-Ray Binaries}.
\newblock \emph{\apjl}, 796:\penalty0 L3, November 2014.
\newblock \doi{10.1088/2041-8205/796/1/L3}.

\bibitem[G{\"u}rses and G{\"u}rsey(2007)]{GurGur07}
M.~G{\"u}rses and Y.~G{\"u}rsey.
\newblock Conformal uniqueness and various forms of the schwarzschild interior
  metric.
\newblock \emph{Il Nuovo Cimento B (1971-1996)}, 25\penalty0 (2):\penalty0
  786--794, 2007.
\newblock ISSN 1826-9877.
\newblock \doi{10.1007/BF02724751}.

\bibitem[{G{\"u}ver} et~al.(2010){G{\"u}ver}, {{\"O}zel}, {Cabrera-Lavers}, and
  {Wroblewski}]{GuvOzeCab10}
T.~{G{\"u}ver}, F.~{{\"O}zel}, A.~{Cabrera-Lavers}, and P.~{Wroblewski}.
\newblock {The Distance, Mass, and Radius of the Neutron Star in 4U 1608-52}.
\newblock \emph{\apj}, 712:\penalty0 964--973, April 2010.
\newblock \doi{10.1088/0004-637X/712/2/964}.

\bibitem[G\"uver et~al.(2010)G\"uver, Wroblewski, Camarota, and
  \"Ozel]{GuvWroCam10}
Tolga G\"uver, Patricia Wroblewski, Larry Camarota, and Feryal \"Ozel.
\newblock The mass and radius of the neutron star in 4u 1820–30.
\newblock \emph{The Astrophysical Journal}, 719\penalty0 (2):\penalty0 1807,
  2010.

\bibitem[{Haensel} et~al.(2007){Haensel}, {Potekin}, and
  {Yakovlev}]{HaePotYak07}
P.~{Haensel}, A.Y {Potekin}, and D.G {Yakovlev}.
\newblock \emph{Neutron~Stars~1 : Equation of State and Structure}, volume~1.
\newblock Springer, 2007.

\bibitem[{Harrison} et~al.(1965){Harrison}, {Thorne}, {Wakano}, and
  {Wheeler}]{HarThoWak65}
B.~K. {Harrison}, K.~S. {Thorne}, M.~{Wakano}, and J.~A. {Wheeler}.
\newblock \emph{{Gravitation Theory and Gravitational Collapse}}.
\newblock 1965.

\bibitem[Hawking and Ellis(1973)]{HawEll73}
S.W Hawking and G.F.R Ellis.
\newblock \emph{The large scale structure of space-time.}
\newblock Cambridge University Press, 1973.
\newblock ISBN 0521099064.

\bibitem[Herrera(1992)]{Her92}
L.~Herrera.
\newblock Cracking of self-gravitating compact objects.
\newblock \emph{Physics Letters A}, 165\penalty0 (3):\penalty0 206 -- 210,
  1992.
\newblock ISSN 0375-9601.
\newblock \doi{http://dx.doi.org/10.1016/0375-9601(92)90036-L}.

\bibitem[{Herrera} and {Barreto}(2013)]{HerBar13}
L.~{Herrera} and W.~{Barreto}.
\newblock {Newtonian polytropes for anisotropic matter: General framework and
  applications}.
\newblock \emph{\prd}, 87\penalty0 (8):\penalty0 087303, April 2013.
\newblock \doi{10.1103/PhysRevD.87.087303}.

\bibitem[Herrera and Santos(1997)]{HerSan97}
L.~Herrera and N.O. Santos.
\newblock Local anisotropy in self-gravitating systems.
\newblock \emph{Physics Reports}, 286\penalty0 (2):\penalty0 53 -- 130, 1997.
\newblock \doi{10.1016/S0370-1573(96)00042-7}.

\bibitem[{Hillebrandt} and {Steinmetz}(1976)]{HilSte76}
W.~{Hillebrandt} and K.~O. {Steinmetz}.
\newblock {Anisotropic neutron star models - Stability against radial and
  nonradial pulsations}.
\newblock \emph{\aap}, 53:\penalty0 283--287, December 1976.

\bibitem[Horedt(2004)]{Hor04}
G.P. Horedt.
\newblock \emph{Polytropes: Applications in Astrophysics and Related Fields}.
\newblock Astrophysics and Space Science Library. Springer Netherlands, 2004.
\newblock ISBN 9781402023507.

\bibitem[{Horvat} et~al.(2011){Horvat}, {Iliji{\'c}}, and
  {Marunovi{\'c}}]{HorIliMar11}
D.~{Horvat}, S.~{Iliji{\'c}}, and A.~{Marunovi{\'c}}.
\newblock {Radial pulsations and stability of anisotropic stars with a
  quasi-local equation of state}.
\newblock \emph{Classical and Quantum Gravity}, 28\penalty0 (2):\penalty0
  025009, January 2011.
\newblock \doi{10.1088/0264-9381/28/2/025009}.

\bibitem[{Israel}(1966)]{Isr66}
W.~{Israel}.
\newblock {Singular hypersurfaces and thin shells in general relativity}.
\newblock \emph{Nuovo Cimento B Serie}, 44:\penalty0 1--14, July 1966.
\newblock \doi{10.1007/BF02710419}.

\bibitem[Ivanov(2002)]{Iva02}
B.~V. Ivanov.
\newblock Static charged perfect fluid spheres in general relativity.
\newblock \emph{Physical Review D}, 65:\penalty0 104001, 2002.

\bibitem[{Kinnersley}(1975)]{Kin75}
W.~{Kinnersley}.
\newblock {Recent Progress in Exact Solutions}.
\newblock In G.~{Shaviv} and J.~{Rosen}, editors, \emph{Relativity and
  Gravitation}, page 109, 1975.

\bibitem[{Knutsen} and {Pedersen}(2007)]{KnuPed07}
H.~{Knutsen} and J.~{Pedersen}.
\newblock {A remark concerning Chandrasekhar's derivation of the pulsation
  equation for relativistic stars}.
\newblock \emph{\physscr}, 75:\penalty0 87--89, January 2007.
\newblock \doi{10.1088/0031-8949/75/1/014}.

\bibitem[{Komatsu} et~al.(2009){Komatsu}, {Dunkley}, {Nolta}, {Bennett},
  {Gold}, {Hinshaw}, {Jarosik}, {Larson}, {Limon}, {Page}, {Spergel},
  {Halpern}, {Hill}, {Kogut}, {Meyer}, {Tucker}, {Weiland}, {Wollack}, and
  {Wright}]{Kom09}
E.~{Komatsu}, J.~{Dunkley}, M.~R. {Nolta}, C.~L. {Bennett}, B.~{Gold},
  G.~{Hinshaw}, N.~{Jarosik}, D.~{Larson}, M.~{Limon}, L.~{Page}, D.~N.
  {Spergel}, M.~{Halpern}, R.~S. {Hill}, A.~{Kogut}, S.~S. {Meyer}, G.~S.
  {Tucker}, J.~L. {Weiland}, E.~{Wollack}, and E.~L. {Wright}.
\newblock {Five-Year Wilkinson Microwave Anisotropy Probe Observations:
  Cosmological Interpretation}.
\newblock \emph{\apjs}, 180:\penalty0 330--376, February 2009.
\newblock \doi{10.1088/0067-0049/180/2/330}.

\bibitem[Kramer et~al.(1980)Kramer, Stephani, MacCallum, and
  Herlt]{KraSteMac80}
D.~Kramer, H.~Stephani, M.~A.~H. MacCallum, and E.~Herlt.
\newblock \emph{Exact solutions of {Einstein's} field equations}.
\newblock Deutscher Verlag der Wissenschaften, Berlin, and Cambridge University
  Press, Cambridge, 1980.
\newblock 1.2.

\bibitem[{Krori} and {Barua}(1975)]{KroBar75}
K.~D. {Krori} and J.~{Barua}.
\newblock {A singularity-free solution for a charged fluid sphere in general
  relativity}.
\newblock \emph{Journal of Physics A Mathematical General}, 8:\penalty0
  508--511, April 1975.
\newblock \doi{10.1088/0305-4470/8/4/012}.

\bibitem[{Kr{\"u}ger} et~al.(2015){Kr{\"u}ger}, {Ho}, and
  {Andersson}]{KruHoAnd15}
C.~J. {Kr{\"u}ger}, W.~C.~G. {Ho}, and N.~{Andersson}.
\newblock {Seismology of adolescent neutron stars: Accounting for thermal
  effects and crust elasticity}.
\newblock \emph{\prd}, 92\penalty0 (6):\penalty0 063009, September 2015.
\newblock \doi{10.1103/PhysRevD.92.063009}.

\bibitem[Kyle and Martin(1967)]{KylMar67}
C.F. Kyle and A.W. Martin.
\newblock Self-energy considerations in general relativity and the exact fields
  of charge and mass distributions.
\newblock \emph{Il Nuovo Cimento A}, 50\penalty0 (3):\penalty0 583--604, 1967.
\newblock ISSN 0369-3546.
\newblock \doi{10.1007/BF02823540}.

\bibitem[Lake(2003)]{Lak03}
Kayll Lake.
\newblock All static spherically symmetric perfect-fluid solutions of
  einstein's equations.
\newblock \emph{Phys. Rev. D}, 67\penalty0 (10):\penalty0 104015, May 2003.
\newblock \doi{10.1103/PhysRevD.67.104015}.

\bibitem[{Lattimer} and {Prakash}(2001)]{LatPra01}
J.~M. {Lattimer} and M.~{Prakash}.
\newblock {Neutron Star Structure and the Equation of State}.
\newblock \emph{The Astrophysical Journal}, 550:\penalty0 426--442, March 2001.
\newblock \doi{10.1086/319702}.

\bibitem[{Lattimer} and {Prakash}(2007)]{LatPra07}
J.~M. {Lattimer} and M.~{Prakash}.
\newblock {Neutron star observations: Prognosis for equation of state
  constraints}.
\newblock \emph{Phys. Rep.}, 442:\penalty0 109--165, April 2007.
\newblock \doi{10.1016/j.physrep.2007.02.003}.

\bibitem[{Lattimer} and {Steiner}(2014)]{LatSte14}
J.~M. {Lattimer} and A.~W. {Steiner}.
\newblock {Neutron Star Masses and Radii from Quiescent Low-mass X-Ray
  Binaries}.
\newblock \emph{\apj}, 784:\penalty0 123, April 2014.
\newblock \doi{10.1088/0004-637X/784/2/123}.

\bibitem[Lattimer and Prakash(2005)]{LatPra05}
James~M. Lattimer and Madappa Prakash.
\newblock Ultimate energy density of observable cold baryonic matter.
\newblock \emph{Phys. Rev. Lett.}, 94\penalty0 (11):\penalty0 111101, Mar 2005.
\newblock \doi{10.1103/PhysRevLett.94.111101}.

\bibitem[Letelier(1980)]{Let80}
Patricio~S. Letelier.
\newblock Anisotropic fluids with two-perfect-fluid components.
\newblock \emph{Phys. Rev. D}, 22:\penalty0 807--813, Aug 1980.
\newblock \doi{10.1103/PhysRevD.22.807}.

\bibitem[{Lieb} and {Yau}(1987)]{LieYau87}
E.~H. {Lieb} and H.-T. {Yau}.
\newblock {A rigorous examination of the Chandrasekhar theory of stellar
  collapse}.
\newblock \emph{\apj}, 323:\penalty0 140--144, December 1987.
\newblock \doi{10.1086/165813}.

\bibitem[{LIGO Scientific Collaboration and Virgo Collaboration}
  et~al.(2016){LIGO Scientific Collaboration and Virgo Collaboration}, Abbott,
  Abbott, Abbott, Abernathy, and et~al.]{Abb16}
{LIGO Scientific Collaboration and Virgo Collaboration}, B.~P. Abbott,
  R.~Abbott, T.~D. Abbott, Abernathy, and et~al.
\newblock Observation of gravitational waves from a binary black hole merger.
\newblock \emph{Phys. Rev. Lett.}, 116:\penalty0 061102, Feb 2016.
\newblock \doi{10.1103/PhysRevLett.116.061102}.

\bibitem[{Lyne} et~al.(2004){Lyne}, {Burgay}, {Kramer}, {Possenti},
  {Manchester}, {Camilo}, {McLaughlin}, {Lorimer}, {D'Amico}, {Joshi},
  {Reynolds}, and {Freire}]{Lyn04}
A.~G. {Lyne}, M.~{Burgay}, M.~{Kramer}, A.~{Possenti}, R.~N. {Manchester},
  F.~{Camilo}, M.~A. {McLaughlin}, D.~R. {Lorimer}, N.~{D'Amico}, B.~C.
  {Joshi}, J.~{Reynolds}, and P.~C.~C. {Freire}.
\newblock {A Double-Pulsar System: A Rare Laboratory for Relativistic Gravity
  and Plasma Physics}.
\newblock \emph{Science}, 303:\penalty0 1153--1157, February 2004.
\newblock \doi{10.1126/science.1094645}.

\bibitem[Maxima(2014)]{maxima}
Maxima.
\newblock Maxima, a computer algebra system. version 5.34.1, 2014.
\newblock URL \url{http://maxima.sourceforge.net/}.

\bibitem[Mehra(1966)]{Meh66}
A.~L. Mehra.
\newblock Radially symmetric distribution of matter.
\newblock \emph{Journal of the Australian Mathematical Society}, 6:\penalty0
  153--156, 5 1966.
\newblock ISSN 1446-8107.
\newblock \doi{10.1017/S1446788700004730}.

\bibitem[{Miller} et~al.(1998){Miller}, {Lamb}, and {Psaltis}]{MilLam98}
M.~C. {Miller}, F.~K. {Lamb}, and D.~{Psaltis}.
\newblock {Sonic-Point Model of Kilohertz Quasi-periodic Brightness
  Oscillations in Low-Mass X-Ray Binaries}.
\newblock \emph{\apj}, 508:\penalty0 791--830, December 1998.
\newblock \doi{10.1086/306408}.

\bibitem[{Misner} and {Sharp}(1964)]{MisSha64}
C.~W. {Misner} and D.~H. {Sharp}.
\newblock {Relativistic Equations for Adiabatic, Spherically Symmetric
  Gravitational Collapse}.
\newblock \emph{Physical Review}, 136:\penalty0 571--576, October 1964.
\newblock \doi{10.1103/PhysRev.136.B571}.

\bibitem[Misner et~al.(1973)Misner, Thorne, and Wheeler]{MTW}
Charles~W Misner, Kip~S Thorne, and John~A Wheeler.
\newblock \emph{Gravitation}.
\newblock W. H. Freeman, 2 edition, 1973.

\bibitem[{Morsink} et~al.(2007){Morsink}, {Leahy}, {Cadeau}, and
  {Braga}]{Mor07}
S.~M. {Morsink}, D.~A. {Leahy}, C.~{Cadeau}, and J.~{Braga}.
\newblock {The Oblate Schwarzschild Approximation for Light Curves of Rapidly
  Rotating Neutron Stars}.
\newblock \emph{\apj}, 663:\penalty0 1244--1251, July 2007.
\newblock \doi{10.1086/518648}.

\bibitem[Mussa(2014)]{Mus14}
A.~Mussa.
\newblock \emph{Spherical symmetry and hydrostatic equilibrium in theories of
  gravity}.
\newblock PhD thesis, University College London, 2014.

\bibitem[{Neary} and {Lake}(2001)]{NeaLak01}
N.~{Neary} and K.~{Lake}.
\newblock {r-modes in the Tolman VII solution}.
\newblock \emph{ArXiv General Relativity and Quantum Cosmology e-prints}, June
  2001.

\bibitem[Neary et~al.(2001)Neary, Ishak, and Lake]{NeaIshLak01}
Nicholas Neary, Mustapha Ishak, and Kayll Lake.
\newblock The {Tolman VII} solution, trapped null orbits and w - modes.
\newblock \emph{Physical Review D}, 64:\penalty0 084001, 2001.

\bibitem[{Negi}(2004)]{Neg04}
P.~S. {Negi}.
\newblock {Hydrostatic Equilibrium of Insular, Static, Spherically Symmetric,
  Perfect Fluid Solutions in General Relativity}.
\newblock \emph{Modern Physics Letters A}, 19:\penalty0 2941--2956, 2004.
\newblock \doi{10.1142/S0217732304015063}.

\bibitem[{Nordstr{\"o}m}(1918)]{Nor18}
G.~{Nordstr{\"o}m}.
\newblock {On the Energy of the Gravitation field in Einstein's Theory}.
\newblock \emph{Koninklijke Nederlandse Akademie van Wetenschappen Proceedings
  Series B Physical Sciences}, 20:\penalty0 1238--1245, 1918.

\bibitem[Oppenheimer and Volkoff(1939)]{OppVol39}
J.~R. Oppenheimer and G.~M. Volkoff.
\newblock On massive neutron cores.
\newblock \emph{Phys. Rev.}, 55\penalty0 (4):\penalty0 374--381, Feb 1939.
\newblock \doi{10.1103/PhysRev.55.374}.

\bibitem[{Ozel} and {Freire}(2016)]{OzeFre16}
F.~{Ozel} and P.~{Freire}.
\newblock {Masses, Radii, and Equation of State of Neutron Stars}.
\newblock \emph{ArXiv e-prints}, March 2016.

\bibitem[{{\"O}zel} et~al.(2012){{\"O}zel}, {Gould}, and
  {G{\"u}ver}]{OzeGouGuv12}
F.~{{\"O}zel}, A.~{Gould}, and T.~{G{\"u}ver}.
\newblock {The Mass and Radius of the Neutron Star in the Bulge Low-mass X-Ray
  Binary KS 1731-260}.
\newblock \emph{\apj}, 748:\penalty0 5, March 2012.
\newblock \doi{10.1088/0004-637X/748/1/5}.

\bibitem[\"Ozel and Psaltis(2009)]{OzePsa09}
Feryal \"Ozel and Dimitrios Psaltis.
\newblock Reconstructing the neutron-star equation of state from astrophysical
  measurements.
\newblock \emph{Phys. Rev. D}, 80\penalty0 (10):\penalty0 103003, Nov 2009.
\newblock \doi{10.1103/PhysRevD.80.103003}.

\bibitem[\"Ozel et~al.(2009)\"Ozel, G\"uver, and Psaltis]{OzeGuvPsa08}
Feryal \"Ozel, Tolga G\"uver, and Dimitrios Psaltis.
\newblock The mass and radius of the neutron star in exo 1745–248.
\newblock \emph{The Astrophysical Journal}, 693\penalty0 (2):\penalty0 1775,
  2009.

\bibitem[Papapetrou(1947)]{Pap47}
A.~Papapetrou.
\newblock {A Static solution of the equations of the gravitational field for an
  arbitrary charge distribution}.
\newblock \emph{Proc. Roy. Irish Acad.(Sect. A)}, A51:\penalty0 191--204, 1947.

\bibitem[Pauli(1958)]{Pau58}
W.~Pauli.
\newblock \emph{Theory of Relativity}.
\newblock Dover Books on Physics. Dover Publications, 1958.
\newblock ISBN 9780486641522.

\bibitem[{Pechenick} et~al.(1983){Pechenick}, {Ftaclas}, and
  {Cohen}]{PecFtaCoh83}
K.~R. {Pechenick}, C.~{Ftaclas}, and J.~M. {Cohen}.
\newblock {Hot spots on neutron stars - The near-field gravitational lens}.
\newblock \emph{\apj}, 274:\penalty0 846--857, November 1983.
\newblock \doi{10.1086/161498}.

\bibitem[Penrose and Rindler(1987)]{Pen87}
R.~Penrose and W.~Rindler.
\newblock \emph{Spinors and Space-Time: Volume 1, Two-Spinor Calculus and
  Relativistic Fields}.
\newblock Cambridge Monographs on Mathematical Physics. Cambridge University
  Press, 1987.
\newblock ISBN 9780521337076.

\bibitem[Peter et~al.(2001)Peter, Denis, and Kayll]{Grtensor}
Musgrave Peter, Pollney Denis, and Lake Kayll.
\newblock {GRTensorII} [computer program], February 2001.
\newblock URL \url{http://grtensor.phy.queensu.ca/}.
\newblock Version 1.79 (R4).

\bibitem[{Piessens} et~al.(1983){Piessens}, {de Doncker-Kapenga}, and
  {Ueberhuber}]{PieDeDUeb83}
R.~{Piessens}, E.~{de Doncker-Kapenga}, and C.~W. {Ueberhuber}.
\newblock \emph{{Quadpack. A subroutine package for automatic integration}}.
\newblock 1983.

\bibitem[{Planck Collaboration} et~al.(2015){Planck Collaboration}, {Ade},
  {Aghanim}, {Arnaud}, {Ashdown}, {Aumont}, {Baccigalupi}, {Banday},
  {Barreiro}, {Bartlett}, and et~al.]{Pla15}
{Planck Collaboration}, P.~A.~R. {Ade}, N.~{Aghanim}, M.~{Arnaud},
  M.~{Ashdown}, J.~{Aumont}, C.~{Baccigalupi}, A.~J. {Banday}, R.~B.
  {Barreiro}, J.~G. {Bartlett}, and et~al.
\newblock {Planck 2015 results. XIII. Cosmological parameters}.
\newblock \emph{ArXiv e-prints}, February 2015.

\bibitem[Poisson(2004)]{Poi04}
E.~Poisson.
\newblock \emph{A Relativist's Toolkit: The Mathematics of Black-Hole
  Mechanics}.
\newblock Cambridge University Press, 2004.
\newblock ISBN 9781139451994.

\bibitem[Ponce~de Leon(1988)]{Pon88}
J.~Ponce~de Leon.
\newblock Weyl curvature tensor in static spherical sources.
\newblock \emph{Phys. Rev. D}, 37\penalty0 (2):\penalty0 309--317, Jan 1988.
\newblock \doi{10.1103/PhysRevD.37.309}.

\bibitem[{Poutanen} and {Beloborodov}(2006)]{PouBel06}
J.~{Poutanen} and A.~M. {Beloborodov}.
\newblock {Pulse profiles of millisecond pulsars and their Fourier amplitudes}.
\newblock \emph{\mnras}, 373:\penalty0 836--844, December 2006.
\newblock \doi{10.1111/j.1365-2966.2006.11088.x}.

\bibitem[{Psaltis} et~al.(2000){Psaltis}, {{\"O}zel}, and {DeDeo}]{PsaOzeDed00}
D.~{Psaltis}, F.~{{\"O}zel}, and S.~{DeDeo}.
\newblock {Photon Propagation around Compact Objects and the Inferred
  Properties of Thermally Emitting Neutron Stars}.
\newblock \emph{\apj}, 544:\penalty0 390--396, November 2000.
\newblock \doi{10.1086/317208}.

\bibitem[Raghoonundun and Hobill(2015{\natexlab{a}})]{RagHob15b}
Ambrish Raghoonundun and David Hobill.
\newblock The geometrical structure of the {T}olman {VII} solution.
\newblock \emph{Classical and Quantum Gravity. (in preparation)},
  2015{\natexlab{a}}.

\bibitem[Raghoonundun and Hobill(2016)]{RagHob16a}
Ambrish Raghoonundun and David Hobill.
\newblock The geometrical structure of the {T}olman {VII} solution.
\newblock \emph{Classical and Quantum Gravity. (in preparation)}, 2016.

\bibitem[Raghoonundun and Hobill(2015{\natexlab{b}})]{RagHob15}
Ambrish~M. Raghoonundun and David~W. Hobill.
\newblock Possible physical realizations of the tolman vii solution.
\newblock \emph{Phys. Rev. D}, 92:\penalty0 124005, Dec 2015{\natexlab{b}}.
\newblock \doi{10.1103/PhysRevD.92.124005}.

\bibitem[Reissner(1916)]{Rei16}
H.~Reissner.
\newblock Über die eigengravitation des elektrischen feldes nach der
  einsteinschen theorie.
\newblock \emph{Annalen der Physik}, 355\penalty0 (9):\penalty0 106--120, 1916.
\newblock ISSN 1521-3889.
\newblock \doi{10.1002/andp.19163550905}.

\bibitem[{Ryba} and {Taylor}(1991)]{RybTay91}
M.~F. {Ryba} and J.~H. {Taylor}.
\newblock {High-precision timing of millisecond pulsars. I - Astrometry and
  masses of the PSR 1855 + 09 system}.
\newblock \emph{\apj}, 371:\penalty0 739--748, April 1991.
\newblock \doi{10.1086/169938}.

\bibitem[Schutz(2009)]{sch09}
B.~Schutz.
\newblock \emph{A First Course in General Relativity}.
\newblock Cambridge University Press, 2009.
\newblock ISBN 9780521887052.

\bibitem[{Schwarzschild}(1916)]{Sch16}
K.~{Schwarzschild}.
\newblock {On the Gravitational Field of a Mass Point According to Einstein's
  Theory}.
\newblock \emph{Abh.~Konigl.~Preuss.~Akad.~Wissenschaften Jahre 1906,92,
  Berlin,1907}, 1916, 1916.

\bibitem[{Sharif}(2007)]{Sha07}
M.~{Sharif}.
\newblock {Matter Collineations of Static Spacetimes with Maximal Symmetric
  Transverse Spaces}.
\newblock \emph{Acta Physica Polonica B}, 38:\penalty0 2003, June 2007.

\bibitem[{Sharif} and {Azam}(2012)]{ShaAza12}
M.~{Sharif} and M.~{Azam}.
\newblock {Effects of electromagnetic field on the dynamical instability of
  expansionfree gravitational collapse}.
\newblock \emph{General Relativity and Gravitation}, 44:\penalty0 1181--1197,
  May 2012.
\newblock \doi{10.1007/s10714-012-1333-8}.

\bibitem[{Singh} and {Yadav}(1978)]{SinYad78}
T~{Singh} and R.~B. {Yadav}.
\newblock {Some exact solutions of charged fluid spheres in Einstein-Cartan
  theory}.
\newblock \emph{Acta Physica Polonica Series B}, 9(10):\penalty0 831--836,
  1978.

\bibitem[{Steiner} et~al.(2010){Steiner}, {Lattimer}, and {Brown}]{SteLatBro10}
A.~W. {Steiner}, J.~M. {Lattimer}, and E.~F. {Brown}.
\newblock {The Equation of State from Observed Masses and Radii of Neutron
  Stars}.
\newblock \emph{\apj}, 722:\penalty0 33--54, October 2010.
\newblock \doi{10.1088/0004-637X/722/1/33}.

\bibitem[Stephani et~al.(2009)Stephani, Kramer, MacCallum, Hoenselaers, and
  Herlt]{SteKra09}
H.~Stephani, D.~Kramer, M.~MacCallum, C.~Hoenselaers, and E.~Herlt.
\newblock \emph{Exact Solutions of Einstein's Field Equations}.
\newblock Cambridge Monographs on Mathematical Physics. Cambridge University
  Press, 2009.
\newblock ISBN 9781139435024.

\bibitem[{Stettner}(1973)]{Ste73}
R.~{Stettner}.
\newblock {On the stability of homogeneous, spherically symmetric, charged
  fluids in relativity}.
\newblock \emph{Annals of Physics}, 80:\penalty0 212--227, September 1973.
\newblock \doi{10.1016/0003-4916(73)90325-4}.

\bibitem[{Stoeger} et~al.(1995){Stoeger}, {Maartens}, and {Ellis}]{StoMaaEll95}
W.~R. {Stoeger}, R.~{Maartens}, and G.~F.~R. {Ellis}.
\newblock {Proving almost-homogeneity of the universe: an almost
  Ehlers-Geren-Sachs theorem}.
\newblock \emph{\apj}, 443:\penalty0 1--5, April 1995.
\newblock \doi{10.1086/175496}.

\bibitem[{Suleimanov} et~al.(2011){Suleimanov}, {Poutanen}, {Revnivtsev}, and
  {Werner}]{SulPou11}
V.~{Suleimanov}, J.~{Poutanen}, M.~{Revnivtsev}, and K.~{Werner}.
\newblock {A Neutron Star Stiff Equation of State Derived from Cooling Phases
  of the X-Ray Burster 4U 1724-307}.
\newblock \emph{\apj}, 742:\penalty0 122, December 2011.
\newblock \doi{10.1088/0004-637X/742/2/122}.

\bibitem[Synge(1960)]{Syn60}
J.L. Synge.
\newblock \emph{Relativity: the general theory}.
\newblock Series in physics. North-Holland Pub. Co., 1960.

\bibitem[{Tolman}(1939)]{Tol39}
R.~C. {Tolman}.
\newblock {Static Solutions of Einstein's Field Equations for Spheres of
  Fluid}.
\newblock \emph{Physical Review}, 55:\penalty0 364--373, February 1939.
\newblock \doi{10.1103/PhysRev.55.364}.

\bibitem[Tolman(1966)]{Tol66}
R.C. Tolman.
\newblock \emph{Relativity, Thermodynamics and Cosmology}.
\newblock The international series of monographs on physics. Clarendon Press,
  1966.

\bibitem[{Tooper}(1965)]{Too65}
R.~F. {Tooper}.
\newblock {Adiabatic Fluid Spheres in General Relativity.}
\newblock \emph{\apj}, 142:\penalty0 1541, November 1965.
\newblock \doi{10.1086/148435}.

\bibitem[{Varela} et~al.(2010){Varela}, {Rahaman}, {Ray}, {Chakraborty}, and
  {Kalam}]{VarRahRay10}
V.~{Varela}, F.~{Rahaman}, S.~{Ray}, K.~{Chakraborty}, and M.~{Kalam}.
\newblock {Charged anisotropic matter with linear or nonlinear equation of
  state}.
\newblock \emph{\prd}, 82\penalty0 (4):\penalty0 044052, August 2010.
\newblock \doi{10.1103/PhysRevD.82.044052}.

\bibitem[Wald(1984)]{Wal84}
Robert~M. Wald.
\newblock \emph{General Relativity}.
\newblock {University Of Chicago Press}, June 1984.
\newblock ISBN 0226870332.

\bibitem[{Weinberg}(1972)]{Wei72}
S.~{Weinberg}.
\newblock \emph{{Gravitation and Cosmology: Principles and Applications of the
  General Theory of Relativity}}.
\newblock July 1972.

\bibitem[Weyl(1952)]{Wey52}
H.~Weyl.
\newblock \emph{Space, Time, Matter}.
\newblock Dover Books on Advanced Mathematics. Dover Publications, 1952.
\newblock ISBN 9780486602677.

\bibitem[Wyman(1946)]{Wym46}
Max Wyman.
\newblock Schwarzschild interior solution in an isotropic coordinate system.
\newblock \emph{Phys. Rev.}, 70:\penalty0 74--76, Jul 1946.
\newblock \doi{10.1103/PhysRev.70.74}.

\bibitem[Wyman(1949)]{Wym49}
Max Wyman.
\newblock Radially symmetric distributions of matter.
\newblock \emph{Phys. Rev.}, 75:\penalty0 1930--1936, Jun 1949.
\newblock \doi{10.1103/PhysRev.75.1930}.

\bibitem[{Yagi} et~al.(2014){Yagi}, {Stein}, {Pappas}, {Yunes}, and
  {Apostolatos}]{YagStiPapYun14}
K.~{Yagi}, L.~C. {Stein}, G.~{Pappas}, N.~{Yunes}, and T.~A. {Apostolatos}.
\newblock {Why I-Love-Q: Explaining why universality emerges in compact
  objects}.
\newblock \emph{\prd}, 90\penalty0 (6):\penalty0 063010, September 2014.
\newblock \doi{10.1103/PhysRevD.90.063010}.

\bibitem[Zeldovich and Novikov(1971)]{ZelNov71}
Y.B. Zeldovich and I.D. Novikov.
\newblock \emph{Relativistic Astrophysics. Vol. 1: Stars and Ralativity}.
\newblock Chicago, 1971.

\end{thebibliography}
\bibliographystyle{plainnat}

\begin{appendices} 
\chapter{}
\label{C:AppendixA}
\begin{myabstract}
  We state the definitions and mathematical machinery necessary for
  general relativity, and the tools we eventually use in this thesis. 
\end{myabstract}

\section{Geometry}
We introduce all the geometry needed for this work in detail starting from set theory. 
\subsection{Topology}
We will not delve into a full topological machinery.  This would take
us too far into fundamental complexities of topological spaces that we
will not encounter in our application.  Instead, we give the bare
minimum required to understand the definitions used later.

A collection \(U\) of subsets of a set \(X\) defines a
\textbf{topology} \index{Topology} on \(X\) if \(U\) contains
\begin{itemize}
\item the empty set and the set \(X\) itself,
\item the union of every one of its sub-collections,
\item the intersection of every one of its finite sub-collections.
\end{itemize}
The sets in \(U\) are then called the \textbf{open sets}
\index{Set!Open} of the topological space \((X,U)\).

A \textbf{neighbourhood} \index{Neighborhood} of a point \(x\) in \(X\)
is a set \(N(x)\) containing an open set which contains the point
\(x\).  A family of neighbourhoods of \(x\) introduces a notion of
``nearness to \(x\).''  A topological space is \textbf{Hausdorff}
\index{Hausdorff} if any two distinct points possess disjoint
neighbourhoods.  All spaces we will use in this thesis will be
Hausdorff, and this definition is meant to discriminate against
certain topological spaces that would not have properties useful for
our purposes.

A collection \(\{U_{j}\}\) of open subsets of \(X\) is a
\textbf{covering}\index{Cover} if each element in \(X\) belongs to at
least one \(U_{i}.\) This means that \(\cup_{i} U_{i} = X.\) If the
system \(\{U_{i}\}\) has a finite number of elements, the covering is
said to be finite.  A \textbf{subcovering}\index{Subcover} of the
covering \(\mathcal{U}\) is a subset of \(\mathcal{U}\) which is
itself a covering.

The covering \(\mathcal{V} = \{V_{i}\}\) is a
\textbf{refinement}\index{Refinement} of the covering \(\mathcal{U} =
\{U_{i}\}\) if for every \(V_{i}\) there exists a \(U_{j}\) such that
\(V_i \subset U_{j}.\) Thus this new finer cover is in some sense
smaller that the original cover.  A covering \(\mathcal{U}\) is
\textbf{locally finite}\index{Locally finite} if for every point \(x\)
there exists a neighbourhood \(N(x),\) which has a non-empty
intersection with only a finite number of members of \(\mathcal{U}.\) 

A subset \(A \subset X\) is \textbf{compact}\index{Compact} if it is
Hausdorff and if every covering of \(A\) has a finite subcovering.

If \((X,U)\) and \((Y,V)\) are two topological spaces, we can build a
\textbf{product space}\index{Product Space}, denoted by \(X \times
Y,\) such that elements of \(X \times Y\) come from both \(X\) and
\(Y\) in the following way: \(X \times Y \equiv \{(x,y) : x \in X, y
\in Y \}.\) We also need a collection of subsets to define this
\textbf{product space topology}\index{Product Space Topology}, and for
those we pick all subsets of \(X \times Y\) which can be expressed as
unions of the sets of the form \(U_{1} \times V_{1}\) with \(U_{1} \in
U\) and \(V_{1} \in V.\)

We now state a theorem, due to Tychonoff that will allow us to
eventually define tensors as objects on these topological spaces.
\begin{myTheorem}
  \label{th:Prel.Tyc}
  Let \((X,U)\) and \((Y,V)\) be compact topological spaces. Then the
  product space \(X \times Y\) is compact in the product topology.
  This result holds even if we take the product of infinitely many
  compact topological spaces.
\end{myTheorem}
We shall not prove this theorem, but will make use of its implications
very commonly.

\subsection{Mappings}
A \textbf{mapping} \index{Mapping} \(f\) from a set \(X\) to a set
\(Y\) associates every \(x\) in \(X\) to a uniquely determined element
\(y = f(x) \in Y.\) Different notations depending on whether the sets,
or the elements are the purpose of the discussion exist in literature.
If the sets themselves are being considered, then it is usual to see
\(f:X \rightarrow Y\).  If it is the elements that are being
considered, \(f:x \mapsto y = f(x)\) is more usual.  Mappings are also
called functions\index{Function}, however we will differentiate
between the two terms, reserving function for a more restricted form
of mapping.

A \textbf{composite mapping} of two the mappings \(f:X \rightarrow Y\)
and \(g:Y \rightarrow Z\) is the mapping \(g \circ f: X \rightarrow
Z\) such that \(x \mapsto g(f(x)).\)

For some \(M\) a subset of \(X\), The symbol \(f(M)\) denotes the
subset \(\{f(x): x \in M\}\) of \(Y\); \(f(M)\) is the \textbf{image}
of \(M\) under the mapping \(f\).  Similarly for some \(N \subset Y,\)
the symbol \(f^{-1} (N)\) denotes the subset \(\{x: f(x) \in N\}\);
\(f^{-1}(N)\) is the \textbf{inverse image} of \(N\) under the mapping
\(f\).

Now we turn to a classification of functions that is going to be
important subsequently: 

If for every \(y \in f(X)\) there is \emph{only} one \(x \in X\) such
that \(f(x) = y \) then \(f\) has an \textbf{inverse mapping},
\(f^{-1}\) and is said to be \textbf{one-one} or
\textbf{injective}\index{Function!injective}.  This is usually notated
as \(f^{-1} : f(X) \rightarrow X\) or \(f^{-1}:y \mapsto x = f^{-1}(y).\) 

The mapping \(f\) is said to map \(X\) \textbf{onto} \(Y\) if \(f(X) =
Y.\) Then \(f\) is also called \textbf{surjective} \index{Function!Surjective}.

The mapping \(f\) is a \textbf{bijection} \index{Function!Bijective}
if it is both one-one and onto.

A mapping \(f\) from a topological space \(X\) to a topological space
\(Y\) is \textbf{continuous} at \(x \in X\) if given any neighbourhood
\(N \subset Y\) of \(f(x)\) there exists a neighbourhood \(M\) of \(x
\in X\) such that \(f(M)\subset N.\) \(f\) is continuous on \(X\) if
it is continuous at all points \(x\) on \(X.\)

A \textbf{homeomorphism}\index{Homeomorphism} \(f:X \rightarrow Y\) is
a bijection \(f\) which is bicontinuous, i.e.\ both \(f\) and
\(f^{-1}\) are continuous.
 
\subsection{Manifolds}
We can finally define a manifolds using these previous ideas.

An n-dimensional topological \textbf{manifold}\index{Manifold} is a
Hausdorff topological space such that every point has a neighbourhood
homeomorphic to \(\mathbb{R}^{n}.\) As such this definition is terse
and not immediately useful.  The Hausdorff property is necessary to
restrict pathological topologies from our models, homeomorphism of
local neighbourhoods to a euclidean space \(\mathbb{R}^{n}\) ensures
the existence of local coordinates, and a topological space is a more
general structure than the pre-relativistic Euclidean space.

A \textbf{chart}\index{Chart} \((U, \varphi)\) of a manifold \(M\) is
an open set \(U\) of \(M,\) called the domain of the chart, together
with a homeomorphism \(\varphi: U \rightarrow V\) of \(U\) onto an
open set \(V\) in \(\mathbb{R}^{n}.\) The coordinates \((x^{1},
x^{2},..., x^{n})\) of the image \(\varphi(x) \in \mathbb{R}^{n}\) of
the point \(x \in U \subset X\) are called the \textbf{coordinates} of
\(x\) in the chart \((U, \varphi)\).  A chart \((U, \varphi)\) is also
called a \textbf{local coordinate system.}  This takes care of local
coordinates, but not of coordinate transformations.  The next
definition, by introducing compatibility conditions, allows coordinate
transformations.

An \textbf{atlas}\index{Atlas} of class \(C^{k}\) on a manifold \(X\)
is a set \(\{(U_{a},\varphi_{a})\}\) of charts of \(X\) such
that the domains \(\{U_{a}\}\) cover \(X\) and the homeomorphisms
satisfy the following compatibility condition.

\begin{figure}[!h] \centering
  \includegraphics[width=9cm]{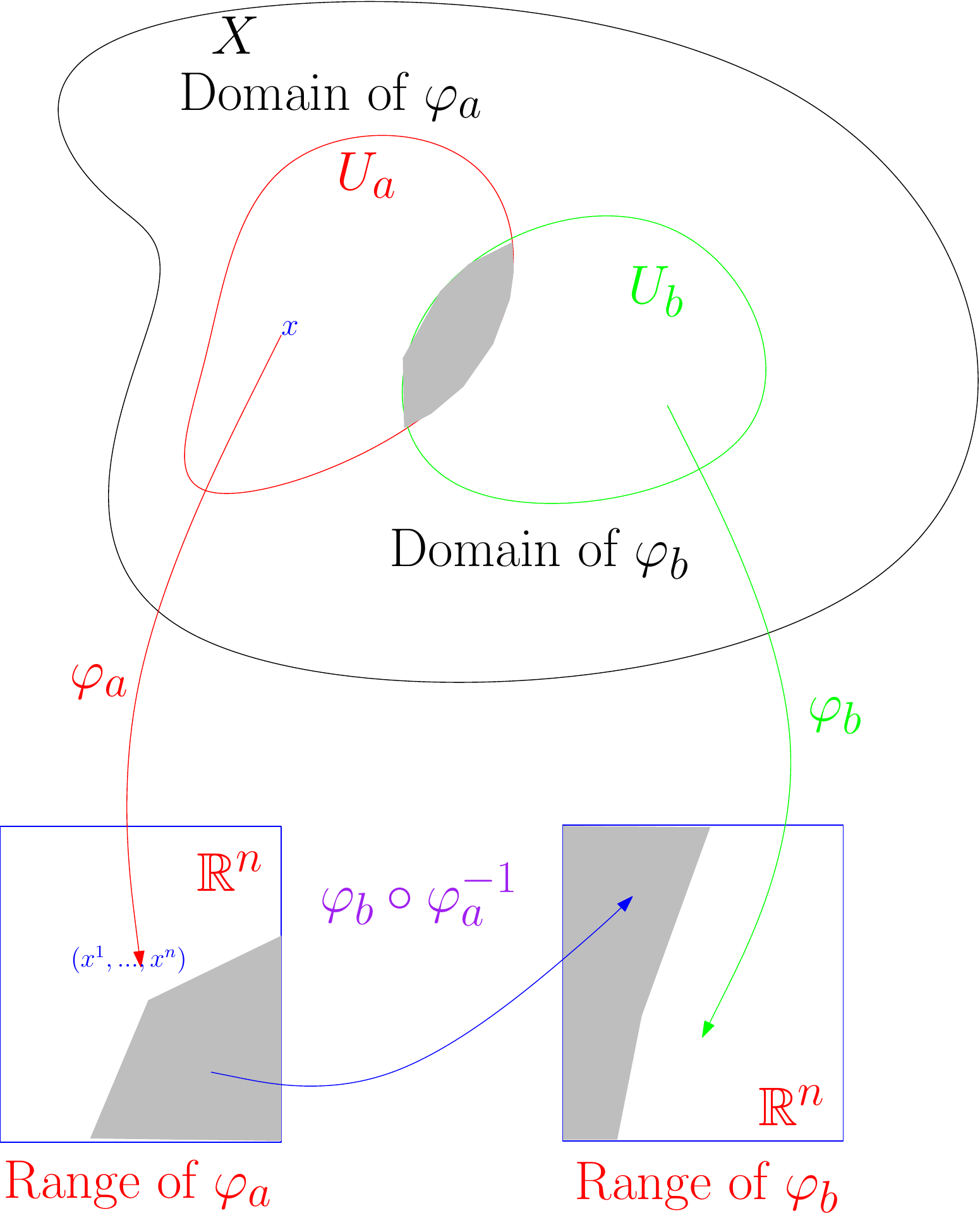}
  \caption[Overlapping patches of a manifold]{The compatibility
criterion given in the text that allows charts to overlap, and so
allow coordinate transformations to be possible generically}
  \label{fig.Prel.2patches}
\end{figure}

The maps \(\varphi_{b} \circ \varphi_{a}^{-1} : \varphi_{a}\left(
  U_{a} \cap U_{b} \right) \rightarrow \varphi_{b}\left( U_{a} \cap
  U_{b} \right)\) are maps of open sets of \(\mathbb{R}^{n}\) into
\(\mathbb{R}^{n}\) of class \(C^{k}\).  In other terms, when
\((x^{i})\) and \((y^{i})\) are the coordinates of \(x\) in the charts
\((U_{a},\varphi_{a})\) and \((U_{b},\varphi_{b})\) respectively, the
mapping \(\varphi_{b} \circ \varphi_{a}^{-1} \) is given in
\(\varphi_{a}\left( U_{a} \cap U_{b} \right)\) by \(n\) real valued
\(C^{k}\) functions of \(n\) variables, \((x^{i}) \mapsto y^{j} =
f^{j}(x^{i}).\) This property is easier to visualize than read, and
figure~\ref{fig.Prel.2patches} makes it clear what is happening, and
how the overlapping of two coordinate systems is dealt with in the
theory.
With these definitions, we can now talk of ``interesting'' manifolds
that we use in general relativity.  A topological manifold \(X\)
together with an equivalence class of \(C^{k}\) atlases is a \(C^{k}\)
structure on \(X\), and we call \(X\) a \(\mathbf{C^k
}\)\textbf{manifold}.  A \textbf{differential manifold} is a manifold
such that the maps \(\varphi_{b} \circ \varphi_{a}^{-1}\) of open sets
\(\mathbb{R}^{n}\) into \(\mathbb{R}^{n}\) are differentiable, but the
expressions differential manifold and \textbf{smooth manifold} are
often used to mean a \(C^{k}\) manifold where \(k\) is large enough
for the given context.  It is usual to use infinitely differentiable
functions, belonging to the \(C^{\infty}\) class in physics.  However
sometimes we are more interested in analytic functions, of class
\(C^{\omega}.\) An \textbf{analytic}\index{Function!Analytic} function
\(f\) can be expanded in a Taylor's series about the point of
analyticity \(x_{0}\), and converges to the function value,
\(f(x_{0})\) in some neighbourhood of \(x_{0}.\)

A \textbf{diffeomorphism}\index{Diffeomorphism} \(f:X \rightarrow Y\)
is a bijection \(f\) which is continuously differentiable (of class
\(C^{1}\)). Nevertheless homeomorphisms of class \(C^{1}\) are not
necessarily diffeomorphisms, as the simple counterexample of
\(f:\mathbb{R} \rightarrow \mathbb{R}: x \mapsto y=x^{3}\), which is a
class \(C^{1}\) homeomorphism shows.  In the latter example,
\(f^{-1}:y \mapsto x^{1/3}\) is continuous but not differentiable at
\(x=0.\) 

\subsection{Calculus}
Calculus on manifolds is complicated because the implicit function
theorem does not hold in these spaces without more
assumptions~\cite{Cho82}.  Here we will assume that our spaces have
the required properties ensuring the existence of differentiable
functions.

A function \(f\) on an \(n\) dimensional manifold \(X\) is a mapping
\(f: X \rightarrow \mathbb{R},\) specified by \(f:x \mapsto f(x).\) Its
representative in local coordinates of the chart \((U,\varphi)\) is a
function on an open set of \(\mathbb{R}^{n},\) defined through \(
f_{\varphi} := f \circ \varphi^{-1} : (x^{i}) \mapsto f(\varphi^{-1}(x^{i})).\)

The function \(f\) is \textbf{differentiable
}\index{Function!Differentiable} at \(x\) if \(f_{\varphi}\) is
differentiable at \(\varphi^{-1}(x)\).  This definition is chart
independent if \(X\) is a differential manifold.  The
\textbf{gradient, }\index{Gradient}also called \textbf{differential,}
of \(f\) is represented in a chart by the partial derivatives of
\(f_{\varphi}.\) If \((U_a,\varphi_a)\) and \((U_{b}, \varphi_{b})\)
are two charts containing point \(x,\) it holds that at \(x\)
\begin{equation}
  \label{eq:Prel.Differential}
  \pderiv{f_{\varphi_{a}}}{x^{i}_{a}} = \pderiv{f_{\varphi_{b}}}{x^{j}_{b}} \pderiv{x^{j}_{b}}{x^{i}_{a}}.
\end{equation}
This equivalence relation allows us to call the differential of \(f\)
a covariant vector, which we will define formally later.

A differential mapping \(f\) between differentiable manifolds, the
source \(X\) of dimension \(n\) and the target \(Y\) of dimension
\(m\), that is \(f:X \rightarrow Y,\) is defined analogously.  The
differential at \(x \in X\) is represented in a chart at \(x \in X\)
and a chart at \(f(x) \in Y\) by a linear map from \(\mathbb{R}^{n}\)
to \(\mathbb{R}^{m}.\)

\subsection{Vectors}
Once we have a differential manifold, we can define vectors and
tensors to characterise differential properties of objects.  However
we have to be careful because we do not want our definitions to be
reliant on the local coordinates: we want generic vectors and tensors
as geometrical objects intrinsically tied to the manifold.  This is
done by defining a tangent vector space \(T_{x} (X)\) at each \(x \in
X\), such that the tangent vector space ``linearise'' the manifold
locally around \(x.\) Many equivalent definitions of tangent vectors,
and their spaces exist, but we will use the one most convenient for
our purposes.  We will assume the basic axioms of a vector space,
although that these are satisfied can be proved formally from our
definitions.

A \textbf{tangent vector }\index{Vector!Tangent}\(v_{x}\) to a
differential manifold at a point \(x\) is an equivalence class of
triplets \((U_{a}, \varphi_{a}, v_{\varphi_{a}})\) where \((U_{a},
\varphi_{a})\) are charts containing \(x,\) while \(v_{\varphi_{a}} =
(v^{i}_{\varphi_{a}}), i = 1,...,n,\) are vectors in
\(\mathbb{R}^{n}.\) The equivalence relation is given by\[
v^{i}_{\varphi_{a}} = v^{j}_{\varphi_{b}}
\pderiv{x^{i}_{a}}{x^{j}_{b}},\] where \(x^{i}_{a}\) and \(x^{i}_{b}\)
are local coordinates in the charts \((U_{a}, \varphi_{a})\) and
\((U_{b}, \varphi_{b})\) respectively.  The vector \(v_{\varphi} \in
\mathbb{R}^{n}\) is the
\textbf{representative}\index{Vector!Representation} of the vector
\(v\) in the chart \((U,\varphi)\).  

The vector \(v\) is attached to
the manifold by the assumption that the numbers \(v^i_{\varphi}\) are
the components of \(v_{\varphi}\) in the frame of \(\mathbb{R}^{n}\)
defined by the tangent to the coordinate curves, where \emph{only one}
coordinate varies.  This definition is compatible to the equivalence
relation given above, and is usually expressed as the short expression
\begin{equation}
  \label{eq:Prel.ConVector}
  v := v^{i}\pderiv{}{\mathbf{x}^{i}} = v^{i}\partial_{i}.
\end{equation}

Tangent vectors at \(x\) make up a vector space, the \textbf{tangent
  space }\index{Tangent space}of \(X\) at \(x\) denoted by \(T_{x}X.\)
An arbitrary set of \(n\) linearly independent tangent vectors
constitute a \textbf{frame }\index{Frame} at \(x.\) The
\textbf{natural frame }\index{Frame!Natural} associated to a
chart\((U, \varphi)\) is the set of \(n\) vectors \(e_{(i)},
i=1,...,n,\) such that \(e^{j}_{(i), \varphi} = \delta^{j}_{i}.\)
These \(n\) vectors are the tangent vectors to the images in \(X\) of
the coordinate curves of the chart.  The numbers \(v^{i}_{\varphi a}\)
are then the \textbf{components }\index{Vector!Component} of the
vector \(v\) in the natural frame.

With this definition of a vector, we can generalise the notion to
vector fields.  The general idea is to associate a vector at each
point \(x\) of the manifold \(X.\)

Formally, A \textbf{vector field}\index{Vector!Field} on \(X\) assigns
a tangent vector at \(x \in X\) to each point \(x,\) the tangent
vector belonging to the tangent spaces \(T_{x} X.\) Since each point
\(x\) has its own tangent space, vector operations on manifolds are
not trivial, as the field has to pick a vector from a different
tangent space for each different point.  A complete definition would
however require additional concepts, and so we will not introduce
them, and instead refer the reader to~\cite{Cho82}

The relations governing vector~\eqref{eq:Prel.ConVector} show that
given a differentiable function \(f\) on \(X,\) the quantity \(v(f)\)
defined for points \(x\) in the domain \(U\) of chart \((U,\varphi)\)
by
\begin{equation}
  \label{eq:Prel.FunctionVector}
  v(f) := v_{\varphi}^{i} \pderiv{f_{\varphi}}{x_{\varphi}^{i},}
\end{equation}
is chart independent.  \(v\) defines a mapping between differentiable
functions.  From this definition, the linearity of \(v\), expressed as
\( v(f + g) = v(f) + v(g),\) can be proved.  \(v\) is also a
\textbf{derivation}\index{Derivation} since in addition to linearity
it satisfies the \textbf{Leibniz law}\index{Liebniz law}, expressed as
\(v(fg) = f v(g) + v(f) g.\) In the same vein, \(v(f)\) is called the
derivation of \(f\) along vector \(v.\) If we take for \(v\) a vector
of a natural frame \(e_{\varphi,(i)},\) so that
\(v^{j} = \delta^{j}_{i}\) then the quantity \(v(f)\) reduces to the
simple expression
\[v(f) \equiv e_{\varphi,(i)}(f) =
\pderiv{f(\varphi^{-1})}{x^{i}_{\varphi}}.\] In general the
differential operator associated to the vector \(e_{(i)}\) of an
arbitrary frame is called a \textbf{Pfaff
  derivative}\index{Derivative!Pfaff}, and is denoted by
\(\partial_{i}.\) In the natural frame, this corresponds to the
partial derivative.

Given a \(C^{\infty}\) map \(h:X \rightarrow Y,\) we define a map
\(h_{*} : T_{x}X \rightarrow T_{h(x)}Y,\) called the \textbf{induced
  linear map,} or \textbf{push-forward}\index{Push-forward} which maps
tangent vectors of a curve \(\gamma\) at \(x \in X\) to the tangent
vector to the curve \(h(\gamma)\) at \(h(x) \in Y.\) 

\subsection{Curves}
Here the notion of a curve will become useful because identifying
vectors with tangent vectors to curves give us back the intuitive
understanding we have for vectors.  A \textbf{parametrized
  curve}\index{Curve!Parametrized} \(\gamma\) on a manifold \(X\) is a
continuous mapping from an open interval \(I \subset \mathbb{R}\) into
\(X\) specified by \(\gamma: I \rightarrow X\) such that \(\gamma:
\lambda \mapsto \gamma(\lambda).\) The curve is oriented in the
direction of increasing \(\lambda.\) A provable, but not quite obvious
consequence of this definition is that a curve is invariant under
reparametrizations which preserve orientation.  As a result
continuous, smooth and monotonously increasing mappings \(I
\rightarrow I' : \lambda \mapsto \lambda'\) preserve curves.

Once we have both the notions of curves, and vector fields, we turn to
an \textbf{integral curve}\index{Curve!Integral} of \(v\) in \(X\),
which is a curve \(\gamma\) in \(X\) such that at each point \(x\) on
\(\gamma,\) the tangent vector is \(v_{x}.\) This integral curve is
\textbf{complete}\index{Curve!Complete} if it is defined for all
values of \(\lambda \in I \subset \mathbb{R}.\) A set of complete
integral curves of a vector field is a
\textbf{congruence.}\index{Curve!Congruence} The concept of an
integral curve in this context comes about because in the curve
\(\gamma\) defined above, it holds
that \[\deriv{\gamma(\lambda)}{\lambda} = v(\gamma (\lambda )) \qquad
\lambda \in I \subset \mathbb{R},\] which is an ordinary differential
equation whose solution is the integral curve.  If we were using a
dynamical systems' terminology, the same integral curve would be
called a \textbf{trajectory.}  By using the following theorem which
can be proved through existence and uniqueness of solutions to exact
differential equations, we find that trajectories are unique.
Formally,
\begin{myTheorem}
  Suppose \(v\) is a \(C^{r}\) vector field on the manifold \(X,\)
  then for every \(x \in X,\) there exists an integral curve of \(v,\)
  given by \(\lambda \mapsto \gamma(x,\lambda),\) such that
  \begin{enumerate}
  \item \(\gamma(\lambda,x)\) is defined for some \(\lambda\)
    belonging to some interval \(I(x) \subset \mathbb{R},\) containing
    \(\lambda=0,\) and is of class \(C^{r+1}\) there.
  \item \(\gamma(0,x) = x\) for every \(x \in X.\)
  \item This curve is unique: Given \(x \in X\) there is no \(C^{1}\)
    integral curve of \(v\) defined on an interval strictly greater
    than \(I(x),\) \emph{and} passing through \(x.\)
  \end{enumerate} 
\end{myTheorem}

The same uniqueness theorem also ensures that in a given congruence,
no curves will intersect, since intersection would require more than
one possible curve through a given point.  The fundamental reason we
introduced curves is that curves provide a natural way to map a
manifold onto itself.  To see how this happens, consider \(\lambda\)
and \(\mu,\) two parameters belonging to the same \(I \subset
\mathbb{R}\), such that the sum \(\lambda + \mu \in I.\) Then since
\(\gamma(\mu,x)\) is a point on one trajectory, we can consider the
point \(\gamma(\lambda, \gamma(\mu,x)),\) which must clearly be
another point further ``down'' the same trajectory.  We can thus
identify \(\gamma(\lambda, \gamma(\mu, x)) = \gamma(\lambda+\mu, x).\)

Since each curve of a congruence is a one--dimensional set of points
(parametrised by \(\lambda,\) say,) the set of all curves of a
congruence to an \(n\)--dimensional manifold is an
\((n+1)\)--dimensional smooth manifold, \(\Sigma_{v}.\) 

\begin{figure}[!h] \centering
  \includegraphics[width=11cm]{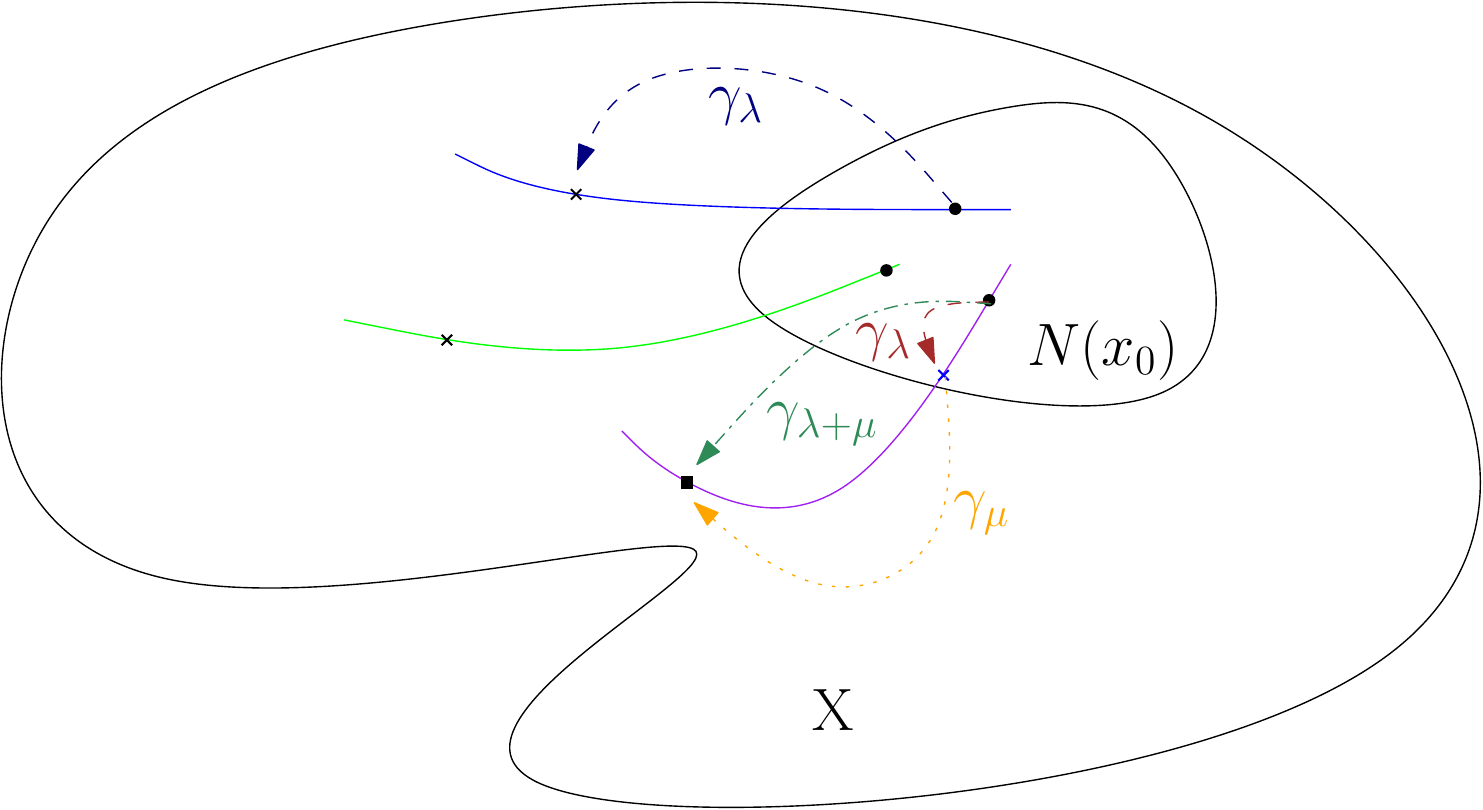}
  \caption[Vector flow on a manifold]{$\gamma_{\lambda}$ maps each dot
    into a cross on the same integral curve, as does $\gamma_{\mu}.$
    We also show how compositions of flows work with  $\gamma_{\lambda+\mu}.$}
  \label{fig:Prel.vectorFlow}
\end{figure}

The mapping \(\gamma: \Sigma_{v} \rightarrow X : (x,\lambda) \mapsto
\gamma(\lambda,x)\) is called the \textbf{flow}\index{Vector!Flow} of
the vector field \(v.\) If both \(X\) and \(v\) are of the same
differentiable class, then so is the flow.  Now, because the
constituents of \(\Sigma_{v}\) are both defined on open
neighbourhoods, and because of the form of the map involved, for every
\(x_{0} \in X,\) there must be a neighbourhood \(N(x_{0}) \subset X,\)
and also an interval \(I(x_{0}) \subset \mathbb{R}\) on the product of
which \(\gamma\) is defined.  Since products of smooth open
neighbourhoods are also smooth and open, if both \(X\) and \(v\) are
smooth, the domain of \(\Sigma_{v},\) denoted by \(N(x_{0}) \times
I(x_{0})\) must be smooth and open, by theorem~\ref{th:Prel.Tyc}. With
this flow map, we can define a \textbf{local transformation} of \(X\)
generated by the vector field \(v\) through \(\gamma(\lambda,.) \equiv
\gamma_{\lambda} : x \mapsto \gamma(\lambda,x)\) defined on
\(N(x_{0})\subset X \) for \(\lambda \in I(x_{0}).\) Under this
mapping, a point \(x \in N(x_{0})\) goes to a point
\(\gamma_{\lambda}(x) \in X\) along the integral curve of \(v\) at
\(x\), the location of \(\gamma_{\lambda}(x)\) along the curve being
determined by the curve parameter \(\lambda.\) Pictorially, the
situation look like figure~\ref{fig:Prel.vectorFlow}, where the
different flows are shown.
We are now in a position to define global transformations on
manifolds.  We do this by extending the domain of our curve parameter
\(\lambda\) to the whole real line.  However we take note that the
interval \(I(x_{0}) \subset \mathbb{R}\) depends on \(N(x_{0})\) in
general.  The intersection \(I\) of all the intervals \(I(x_{0})\)
corresponding to a set of neighbourhoods \(\{N(x_{0})\}\) covering
\(X\) may be empty and this is a case we want to avoid.  However, if
\(X\) is compact\index{Compact}, \(I\) is never empty, since it is
then given by a finite intersection, by the definition of compactness.
Therefore when \(I\) is not empty, then \(\gamma_{\lambda}\) with
\(\lambda \in I\) defines a \textbf{global
  transformation}\index{Global Transformation} of \(X.\) Moreover, we
can now extend \(\gamma_{\lambda}\) for all \(\lambda \in \mathbb{R}\)
through the relation \(\gamma_{\lambda + \mu} = \gamma_{\lambda} \circ
\gamma_{\mu},\) as we show pictorially in
figure~\ref{fig:Prel.vectorFlow}.  We can also define an inverse
transformation for each \(\gamma_{\lambda},\) denoted by
\(\gamma_{-\lambda},\) which undoes the flow \(\gamma_{\lambda}.\) 

As a result of the existence of such a global transformation on
compact manifolds, we have
\begin{myTheorem}
  A smooth vector field on a manifold \(X,\) which vanishes outside a
  compact set \(K \subset X,\) generates a one parameter group of
  diffeomorphisms of \(X.\)
\end{myTheorem}
This theorem allows for points to be ``dragged'' along congruences
globally on the manifold, and this process is sometimes called
\textbf{Lie dragging.}\index{Lie dragging}

Points are not the only things that can be dragged with a congruence
of curves.  If a function \(f\) is defined on a manifold, then the
group of diffeomorphisms generated by a vector field defined on the
manifold defines a new function \(f^{*}_{\lambda}.\) This works by
carrying \(f\) along the congruence: if a point \(x\) on a certain
integral curve is mapped to a point \(y,\) a parameter value
\(\lambda\) away on the same curve, (as in
figure~\ref{fig:Prel.vectorFlow},) then the new function
\(f^{*}_{\lambda}\) has the same value at \(y\) as \(f\) had at \(x,\)
that is \(f(x) = f^{*}_{\lambda}(y).\) If the value of
\(f^{*}_{\lambda}(y)\) takes the same value as \(f(y),\) so that the
dragged function and the original function have the same value at the
same point, the function is said to be
\textbf{invariant}\index{Function!Invariant} under the mapping. If
additionally this condition holds for all values of \(\lambda,\) the
function \(f\) is said to be \textbf{Lie dragged,}\index{Function!Lie
  dragged} and then since \(f\) does not seem to depend on any motion
along the congruence, we see that \(\d f / \d \lambda = 0.\)

Vector fields can also be dragged along congruences, and this notion
can be used to define the \textbf{Lie
  derivative}\index{Derivative!Lie} of a vector field.  

\subsection{Forms}
Before considering tensors, it is convenient to define a dual vector
space to the tangent space.  The \textbf{cotangent space
}\index{Cotangent space} \(T^{*}_{x}\) to \(X\) is the dual of
\(T_{x},\) that is the space of 1-forms on \(T_{x}.\) These form a
vector space of \textbf{covariant vectors.}\index{Vector!Covariant}
Covariant vectors are geometrical objects independent of the choice of
coordinates.  The components of a covariant vector at \(x \in X,\) in
a chart \((U,\varphi)\) containing \(x,\) is a set of \(n\) numbers
\(\omega_{i},i=1,...,n.\) Under a change of chart from \((U,
\varphi_{a})\) to \((U,\varphi_{b})\) the numbers \(\omega_{i}\)
transform through \begin{equation}
  \label{eq:Prel.CoVector}
  \omega_{i} = \omega_{j}\pderiv{x^{j}_{a}}{x^{i}_{b}}.
\end{equation}
Covariant vectors can also be defined through the equivalence
relation~\eqref{eq:Prel.CoVector}, analogous to the definition of
contravariant vectors given earlier.  A
\textbf{coframe}\index{Coframe} is a set of \(n\) linearly independent
covariant vectors.  The \textbf{natural
  coframe}\index{Coframe!Natural} is the set of differentials \(\d
x^{i}\) of the coordinate functions \( x \mapsto x^{i}.\) A frame (sets
of \(n\) vectors \(e_{(i)}, i=1,...,n\)) and coframe (sets of \(n\)
1-forms \(\theta^{(j)}, j=1,...,n\)) are \textbf{dual
  frames}\index{Frame!Dual} if 
\[\theta^{(j)} e_{(i)} = \delta_{i}^{j}, \quad \text{the Kronecker
  symbol.}\]
The differential of a differentiable function \(f\) is a covariant
vector field denoted by \(\d f,\) and is called an exact 1-form.

Like vectors, forms can also be Lie dragged and therefore Lie derived.

\subsection{Tensors}
Now that we have introduced both contravariant vectors and covariant
ones (1-forms,) we can define more general tensors.  However before
doing so, we will introduce two conventions that we will be using as
from now on.  The first is called the \textbf{abstract index
  notation,}\index{Abstract Index} and was introduced by
Penrose~\cite{Pen87} to write formulae using representatives of vector
and tensor fields in arbitrary frames, instead of the geometric
objects themselves.  These formulae are equivalent to the geometric
ones, and tell us how the equivalence classes of representatives
behave.  As a result, geometric objects, when we will look at them
will not have indices, but formulae involving these geometric objects
will have indices to make calculations easier.

The second convention is called the \textbf{Einstein summation
  convention}.  Stated simply, whenever the same letter index appears
both upstairs and downstairs in formulae, an implied sum over that
index is assumed.  This is done to de-clutter our formulae from the
numerous summation signs that would otherwise be needed to denote
contraction: a tensor operation.

With these two conventions, we define \textbf{tensors} through their
transformation properties as follows: A covariant
\(p\)-tensor\index{Tensor!Covariant} at a point \(x \in X\) is a \(p\)
multilinear form on the \(p\) direct product of the tangent space
\(T_{x}X.\) Similarly contravariant
tensors\index{Tensor!Contravariant} are multilinear forms on the
direct products of the cotangent space \(T^{*}_{x}X.\) As an example a
covariant 2-tensor \(T\) at \(x \in X\) is an equivalence class of
triplets \(\left( U_{I}, \varphi_{I}, T_{\varphi I, ij} \right),
i,j=1,...,n,\) with the equivalence relation allowing for the
components to change from chart \((U, \varphi)\) to \((U, \varphi')\)
through
\begin{equation}
  \label{eq:Prel.CoTensor}
  T_{i'j'} = \pderiv{x^{h}}{x^{i'}} \pderiv{x^{k}}{x^{j'}} T_{hk}.
\end{equation}
A \textbf{contravariant tensor} would then transform analogously
through
\begin{equation}
  \label{eq:Prel.ConTensor}
  T^{i'j'} = \pderiv{x^{i'}}{x^{h}} \pderiv{x^{j'}}{x^{k}} T^{hk}.
\end{equation}
The \textbf{space} of covariant[contravariant] 2-tensors at \(x\) is
denoted by \(T_{x}^{*} \otimes T_{x}^{*}\) [respectively \(T_{x}
\otimes T_{x}\).]  The natural basis of this space associated to the
chart with local coordinates \(x^{i}\) is denoted \(\d x^{i} \otimes
\d x^{j}\) [respectively \(e_{(i)} \otimes e_{(j)}\)].  Therefore the
covariant 2-tensor \(T\) is given by \[T = T_{ij} \d x^{i} \otimes \d
x^{j},\] where \(\d x^{i} \otimes \d x^{j}\) is the covariant 2-tensor,
bilinear form on \(T_{x}X \otimes T_{x}X,\) such that for any pair of
vectors \(u\) and \(v\) with natural frame components \(u^{i}\) and
\(v^{i}\) respectively, it holds that \((\d x^{i} \otimes
\d x^{j})(u,v) = u^{i}v^{j}.\)

The \textbf{tensor direct product}\index{Tensor!Direct Product} \(S \otimes
T\) of a \(p \)--tensor \(S\) and a \(q \)--tensor \(T\) is a \(p+q\)
tensor with components defined by products of components.  Although
products of components are commutative, tensor products are
non-commutative, and \(S \otimes T\) and \(T \otimes S\) are different
objects belonging to different spaces.  As an example, if \(T\) is a
contravariant 2-tensor, and \(S\) a covariant vector, the mixed tensor
product \(S \otimes T\) is written as \[(S \otimes T)_{i}{}^{jk} =
S_{i}T^{jk}.\]

The \textbf{contracted product}\index{Tensor!Contracted Product} or a
\(p\) contravariant tensor and a \(q\) covariant tensor is a tensor of
order \(p+q-2\) whose components are obtained by summing over a
repeated index appearing once upstairs and once downstairs.  For
example, we can contract the tensors \(S\) and \(T\) above in these
different ways: \(W^{j} = S_{i}T^{ij} = \sum_{i} T^{ij}S_{i},\) or
\(V^{a} = S_{j}T^{aj} = \sum_{j} S_{j}T^{aj},\) where we have eschewed
Einstein's convention for clarity.

From these definitions, it can be proved that certain properties of
tensors are intrinsic, and this independent of coordinates or charts.
The symmetry and antisymmetry properties of similarly places indices
(i.e.\ all indices being considered being either all upstairs, or
downstairs) is intrinsic, as is that of a tensor vanishing.

Like vectors and forms, we can define \textbf{tensor
  fields}\index{Tensor!Field} as tensors at \(x\) for each point
\(x \in X.\) Differentiability is defined again on the charts, and the
notion of a \(C^{k}\) tensor is chart independent if the underlying
manifold is at least of class \(C^{k+1}.\) 

The tensor fields are defined on structures called fibre bundles.  The
basic notion to define a fibre bundle is that of a fibre.

A \textbf{bundle}\index{Bundle} is a triple \((E,B, \pi)\) consisting
of two topological spaces \(E\) and \(B\) and a continuous surjective
mapping \(\pi: E \to B.\) The space \(B\) is called the base.  We will
restrict ourselves to situations in which the topological spaces
\(\pi^{-1}(x),\) for all \(x \in B\) are homeomorphic to a space
\(F.\) Then \(\pi^{-1} (x)\) is called the \textbf{fibre}\index{Fibre}
at \(x\) and is denoted \(F_{x}.\) The space \(F\) itself is called
the \textbf{typical fibre}\index{Fibre!Typical}.  If the bundle also
has certain additional structure involving a group of homeomorphisms
of \(F\) and a covering of \(B\) involving open sets, it is called a
fibre bundle.  If \(F\) is a vector space and the group is the linear
group, the fibre bundle is called a \textbf{vector
  bundle}\index{Bundle!Vector}.

A formal definition of a fibre bundle will take us too far into
category theory, and the simple notion above will suffice for our
purpose. As a example we state the relevant spaces for
\begin{enumerate}
\item a \textbf{tangent bundle}\index{Bundle!Tangent}.  Let \(T(X^n)\)
  be the space of pairs \((x,v_{x})\) for all \(x\) in the
  differential manifold \(X^{n}\) and all \(v_{x} \in T_{x}(X^{n}),\)
  the tangent space of \(x.\) This space of pairs can be given the following fibre
  bundle structure \((T(X^{n}), X^{n}, \pi, \mathrm{GL}(n,\R)) \) :
  \begin{itemize}
  \item the fibre \(F_{x}\) at \(x\) is \(T_{x}(X^{n}),\)
  \item the typical fibre F is \(\R^n,\)
  \item the projection \(\pi: (x,v_{x}) \mapsto x, \)
  \item the covering of \(X^n\) is \(\{U_{j}; \{U_{j}, \psi_{j}\}\} \text{is an atlas of} X^{n},\)
  \item the coordinates of a point \(p = (x,v_{x}) \in \pi^{-1}(U_{j}) \subset T(X^{n})\) are 
    \[(x^{1},\dots x^{n}, v^{1}_{x}, \dots v^{n}_{x}).\]
  \item the structural group \(G\) is \(\mathrm{GL}(n,\R),\) the group
    of linear automorphisms of \(\R^{n}\) whose matrix representations
    is the set of \(n \times n\) matrices with non-vanishing determinant.
  \end{itemize}

\item a \textbf{tensor bundle}\index{Bundle!Tensor} of order
  \(s = (p+q).\) Let \(T(X)\) be the space of pairs \((x,R_{x})\)
  for all \(x\) in the differential manifold \(X\) and all
  \(R_{x} \in ,\bigotimes^{p}_{i=1} (T_{x}X)_{i} \bigotimes^{q}_{j=1}
  (T^{*}_{x}X)_{j}\) the set of vector spaces of \(x.\) This space
  of pairs can be given the following fibre bundle structure
  \((T(X), X, \pi, \mathrm{GL}(n,\R)) \):
  \begin{itemize}
  \item the fibre \(F_{x}\) at \(x\) is \(T(X),\) i.e.\ each the
    \(n-\)components representation of elements of the \(p\) tangent and \(q\) cotangent spaces.
  \item the typical fibre F is \(\otimes^{s}_{i=1} \R^n,\)
  \item the projection \(\pi: (x,R_{x}) \mapsto x,\) mapping the tensor to its point,  
  \item the covering of \(X\) is \(\{U_{j}; \{U_{j}, \psi_{j}\}\} \text{is an atlas of} X,\)
  \item the coordinates of a point \(p = (x,R_{x}) \in \pi^{-1}(U_{j}) \subset T(X)\) are 
    \[(x^{1},\dots, x^{n}, R^{1}_{x}, \dots,  R^{n^{s}}_{x}),\]
    the \(n^{s}\) real components of the tensor associated with each point.
  \item the structural group \(G\) is \(\mathrm{GL}(n,\R),\) the group
    of linear automorphisms of \(\R^{n}\) whose representations are
    multidimensional objects.
  \end{itemize}

\end{enumerate}
We now have notions of vectors, forms, tensors and Lie dragging on
differential manifolds.  What we still lack to characterise the
geometry completely enough to do physics are the notions of distance
and parallelism.  These two concepts will allow us to eventually talk
about curvature, and from there lead us to Einsteinian relativity.  We
start with the notion of distance through the definition of a metric,
and introduce the two different types of derivatives we use in this
thesis: the Lie, and the covariant derivative.

\subsection{Derivative of tensors}
In this Section we discuss the various ways calculus can be done on
tensor fields.  Calculus is difficult in general manifolds because
each point has its own tangent fibre on which the tensors are defined,
and since calculus is involved with the comparing of objects at
different points, a method of mapping fibres in the fibre bundles is
needed.  There are various ways to achieve this.  Here, we use a
completely operational approach involving tensor components instead of
the abstract geometrical objects.  However the 2 approaches are
equivalent.  We follow mostly the Refs.~\cite{Inv92,Cho08},
and~\cite{sch09}

Lie dragging introduces the concept of the Lie derivative, and
parallel transport introduces the covariant derivative.  Before going
into the details however, the notation used throughout this thesis
concerning derivatives is as follows:
\begin{enumerate}
\item The ordinary partial derivative will be denoted with a comma in
  the index.  Thus for a tensor \(T_{ab}\) for example,
  \[ T_{ab,c} = \pderiv{T_{ab}}{x^{c}}.\]
\item An equivalent notation that will be used for partial derivatives
  when it is convenient is the \(\partial\) notation.  In this
  notation, derivatives w.r.t the coordinate \(x^{c}\) is written as
  \(\partial_{c},\) and for example the above equation is written
\[\pderiv{T_{ab}}{x^{c}} = \partial_{c} T_{ab}.\]
\item The Lie derivative induced by a vector field \(u\) with
  components \(u^{i}\) will be given by
  \[ (\LieD{u}T)_{ab} = u^{i} \partial_{i} T_{ab} +
    T_{ib}\partial_{a}u^{i} + T_{ai} \partial_{b}u^{i} = u^{i}T_{ab,i} + T_{ib}u^{i}_{,a} + T_{ai} u^{i}_{,b}.\]
\item The covariant derivative of tensor \(T_{ab}\) induced by a
  connection whose components are given as \(\Gamma^{a}_{bc}\) will be
  written either with the \(\nabla\) symbol or with a semicolon (;) thus,
  \[\nabla_{c}T_{ab} = T_{ab;c} = T_{ab,c} - \Gamma^{i}_{ac} T_{ib}
    -\Gamma^{i}_{bc} T_{ai} = \partial_{c}T_{ab} - \Gamma^{i}_{ac}
    T_{ib} -\Gamma^{i}_{bc} T_{ai}\]
\end{enumerate}

Now we look into Lie derivatives.

\subsection{The Lie derivative}\index{Derivative!Lie}
The basic idea behind the notion of a Lie derivative is the following:
Consider a vector field \(v\) in the same neighbourhood \(N,\) in a
manifold \(M,\) and two points \(p, q \in N.\) Then the vector field
assigns a vector to each of \(p\) and \(q,\) named \(v(p),\) and
\(v(q),\) respectively. Choose a congruence of curves defined by
another vector field \(u\) which has one curve through \(p\) that also
goes through \(q.\) Then \(v(p)\) can be Lie dragged along the curve
defined by \(u\) up to \(q.\) If the one parameter group of
diffeomorphism generated by \(Y,\) is denoted by \(f_{t},\) the
``dragged along'' point \(p\) is \(q = f_{t}(p),\) and therefore the
vector at point \(q\) can also be denoted as \(v(q) = v(f_{t}(p)).\)
Then the \(v(q)\) Lie dragged vector at \(p\) is
\(f^{-1}_{t} (v(f_{t}(p))). \)  Both of \(v(p)\) and
\(f^{-1}_{t} (v(f_{t}(p)))\) are now vectors at point \(p,\) and hence in
the same tangent space, so that vector operations can be performed on
them.  Indeed, by subtracting the two vectors above, and the addition
of a limit process on the parameter \(t,\) the Lie derivative of
vector \(v\) may be defined through Lie dragging along the curve defined
through vector \(u\) by
\begin{equation}
  \label{A.eq:LieV}
  \LieD{u}v(p) \eqqcolon \lim_{t \to 0}\left[ (f^{-1}_{t} (v(f_{t}(p)))) - v(p) \right].
\end{equation}
This derivative produces another vector \(w,\) which in component form
is given by
\[ w^{j} = (\LieD{u} v )^{j} = u^{i} \pderiv{v^{j}}{x^{i}} - v^{i}
  \pderiv{u^{j}}{x^{i}}.\] 

The same procedure can be applied to forms, and tensors, both of which
can be Lie dragged and compared along the same congruence of curves.
We will here just provide the general expression that can be used to
compute this derivative.  For a 1-forms \(\alpha\), the derivative
produces another 1-form \(\beta,\)
\[ \beta_{j} = (\LieD{u} \alpha )_{j} = u^{i} \pderiv{\alpha_{j}}{x^{i}} + \alpha_{i}
  \pderiv{u^{i}}{x^{j}}.\] 

For the general \(p\) contravariant and \(q\) covariant tensor, the
mathematical expression is complicated, but if the tensor is given in
component form, following the above prescription for vectors and
1-forms, the first derivative is the partial derivative of the tensor
components contracted with the vector. We then get an additional term
for each upstairs index, of the form
\(T^{a_{1},\dots, i, \dots, a_{p}} u^{a_{i}}_{,i}\) and which carries a
positive sign.  Another additional term of the form
\(T^{a_{1},\dots,a_{p}}_{b_{1},\dots,i,\dots,b_{q}} \partial_{b_{i}}u^{i}
\)is then added with a negative sign for each downstairs index.

\(T^{a_{1},\dots,a_{p}} _{b_{1},\dots,b_{q}}\) as
\begin{align*}
S^{a_{1},\dots,a_{p}}_{b_{1},\dots,b_{q}} = 
(\LieD{u} T)^{a_{1},\dots,a_{p}}_{b_{1},\dots,b_{q}} = 
  & u^{i} \partial_{i} T^{a_{1},\dots,a_{p}}_{b_{1},\dots,b_{q}}\\
  &-\sum_{i=1}^{p} T^{a_{1}, \dots,i,\dots,a_{p}}_{b_{1},\dots,b_{q}} \partial_{i}u^{a_{i}}\\
  &+\sum_{i=1}^{q} T^{a_{1},\dots,a_{p}}_{b_{1},\dots,i,\dots,b_{q}} \partial_{b_{i}}u^{i}.
\end{align*}
We now have a prescription for the computation of the Lie
derivatives.  We state some important properties of the Lie derivative.
Note that a vector field is needed for its computation and
construction. 
\begin{enumerate}
\item It is \textbf{type-preserving}\index{Type Preserving} in that
  the Lie derivative of a \((p+q)\) tensor is also a \((p+q)\) tensor.
\item As can de derived from its construction, it is
linear:
\[\LieD{u}(\lambda v^{a} + \mu w^{a}) = \lambda \LieD{u}v^{a} + \mu \LieD{u} w^{a},\]
for \(\lambda\) and \(\mu\) constants.
\item It follows the Leibniz law, just like the ordinary derivative so
  that
\[ \LieD{u} Y^{a}Z_{bc} = Y^{a}(\LieD{u}Z_{bc}) +
  (\LieD{u}Y^{a})Z_{bc}.\] 
\item It commutes with contraction, so that for example, if the
  contraction \(S_{a}T^{a}_{b} = Q_{b}\) 
\[ S_{a}(\LieD{u}T)^{a}_{b} = \LieD{u} (S_{a}T^{a}_{b}) = \LieD{u}Q_{b}\]
\item The Lie derivative of a scalar field \(\phi\) is given by
\[\LieD{u} \phi = u^{a}\partial_{a}\phi,\]
i.e.\ the contraction of the vector components with the gradient of
the scalar field : the standard directional derivative in the
direction of the vector field.
\end{enumerate}

Lie derivatives are useful because they allow the definition of
\textbf{isometries}\index{Isometry} and
\textbf{symmetries}\index{Symmetry} of tensor, and in this thesis of
the metric.  We shall make use of these notions in the appropriate
section below.
  
\subsection{The covariant derivative}\label{A.ssec:CovD}
We now introduce another derivative that is used in this thesis, the
covariant derivative.  We will do so in an operational manner in
component form, neglecting the deeper mathematical underpinnings of
parallel transport which would take longer than we have to introduce.
Consider a contravariant vector field \(X^{a}(x)\) evaluated at a
point \(Q\) with coordinates \(x^{a} +\delta x^{a},\)
near\footnote{Note here that we are expressly working with a vector
  field, and so defined at every point of the manifold. The concept of
  nearness has not been defined yet, but the more rigorous method
  that takes this into account is lengthy.} a point \(P,\) with
coordinates \(x^{a}.\) Then by using Taylor's theorem, the vector
field's components at \(Q\) can be expressed as
\(X^{a}(x+\delta x) = X^{a}(x) + \delta x^{b} \partial_{b} X^{a}\) to
first order.  Denoting the second term by
\begin{equation}
  \label{A.eq:Tayl}
\delta X^{a}(x) = \delta x^{b} \partial_{b} X^{a} = X^{a}(x+ \delta
x) - X^{a}(x),  
\end{equation}
we find that this quantity is not tensorial: It involves the
subtraction of vectors at different points of the manifold, and hence
belonging to different fibres.  We proceed to define a tensorial
derivative by introducing a vector at \(Q\) which is
parallel\footnote{This is what the rigorous notion of a linear
  connection establishes.} to \(X^{a}\) at \(P.\) Since
\(x^{a} + \delta x^{a}\) is close to \(x^{a},\) we assume that the
parallel vector differs from \(X^a\) by a small amount denoted by
\(\bar{\delta}X^{a}(x).\) \(\bar{\delta}X^{a}\) is not tensorial since
it also involves the difference of vectors a 2 different points.
However we can construct another quantity, the difference between the
first~\eqref{A.eq:Tayl} and the parallel vector's deviation through
\[\delta X^{a}(x) - \bar{\delta} X^{a}(x) = X^{a}(x) + \delta X^{a}
  (x) - [X^{a}(x) + \bar{\delta} X^{a} (x)],\] is tensorial.  To do
so, consider that \(\bar{\delta X^{a}(x)}\) would be zero if either
\(\delta x,\) or \(X^{a}(x)\) vanishes.  As a result assuming that
\(\bar{\delta X^a}\) be linear in both is the first step.  This
implies that there are objects (multiplicative factors) which we will
call \(\Gamma^{a}_{bc}\) such that
\begin{equation}
  \label{A.eq:ConnDelta}
  \bar{\delta}X^{a}(x) = -\Gamma^{a}_{bc}X^{b}(x) \delta x^{c}
\end{equation}
where the negative sign is introduced to agree with convention.  We
have therefore introduced \(n^{3}\) functions \(\Gamma^{a}_{bc}\) on
the manifold.  The transformation properties of these objects is what
the rest of this section looks into.  However once these have been
introduced, a \textbf{covariant
  derivative}\index{Derivative!Covariant} of \(X^{a}\) can be defined
through the limiting process
\begin{equation}
  \label{A.eq:CovDef}
  X^{a}{}_{;c} = \lim_{\delta x^{c} \to 0} \f{X^{a}(x+\delta x) - [X^{a}(x) + \bar{\delta} X^{a} (x)]}{\delta x^{c}}.
\end{equation}
This is the difference between the vector \(X^{a}\) at \(Q\) and the
vector parallel to \(X^{a}\) at \(P\) parallel transported to \(Q.\)
This notion of parallel transport can be made rigorous.  This
limiting process then gives an expression for computing the covariant
derivative of a vector as
\begin{equation}
  \label{A.eq:CovVec}
  X^{a}{}_{;c} = \partial_{c}X^{a} + \Gamma^{a}_{bc}X^{b}.
\end{equation}

The requirement that this object \(X^{a}{}_{;c}\) be a (1+1) rank tensor
through the use of equation~\eqref{eq:Prel.CoTensor}
and~\eqref{eq:Prel.ConTensor} then give the transformation properties
of the objects \(\Gamma^{a}_{bc}\) which are not tensors, but are
called \textbf{affine connections}\index{Connection!Affine}.  The
transformation properties on changing coordinate systems
\((x^{a} \to x^{a'})\) is
\begin{equation}
  \label{A.eq:ConnTrans}
  \Gamma'^{a}_{bc} =
  \pderiv{x^{a'}}{x^{d}}\pderiv{x^{e}}{x^{b'}}\pderiv{x^{f}}{x^{c'}}\Gamma^{d}_{ef}
  - \pderiv{x^{d}}{x^{b'}}\pderiv{x^{e}}{x^{c'}} \f{\partial^{2}
    x^{a'}}{\partial x^{d} \partial x^{e}}.
\end{equation}
 
A manifold that has a connection defined on it is called an
\textbf{affine} manifold\index{Manifold!Affine}.

We also define the covariant derivative of a scalar field \(\phi\) to
be the same as the ordinary derivative through
\( \nabla_{c} \phi = \phi_{;c} = \partial_{c}\phi.\) As a result, the covariant
derivatives of forms can be defined as well since, vectors and forms
contract to give rise to scalars, and demanding that the covariant
derivative obey the Leibniz law yields
\[
X_{a;c} = \partial_{c}X_{a} - \Gamma^{b}_{ac}X_{b}.
\]
The name covariant derivative comes from the fact that one additional
covariant index get attached to the object being derived.  Indeed the
covariant derivative is not type-preserving like the Lie derivative,
since a \((p+q)\) tensor's covariant derivative is the \((p+q+1)\)
tensor given by 
\begin{align*}
S^{a_{1},\dots,a_{p}}_{b_{1},\dots,b_{q}, b_{q+1}} = 
 T^{a_{1},\dots,a_{p}}_{b_{1},\dots,b_{q}, b_{q};c} = 
  & \partial_{c} T^{a_{1},\dots,a_{p}}_{b_{1},\dots,b_{q}}\\
  &+\sum_{i=1}^{p} \Gamma^{a_{i}}_{dc} T^{a_{1}, \dots,d,\dots,a_{p}}_{b_{1},\dots,b_{q}}\\
  &-\sum_{i=1}^{q} \Gamma^{d}_{b_{i}c} T^{a_{1},\dots,a_{p}}_{b_{1},\dots,d,\dots,b_{q}} .
\end{align*}
From the transformation properties of the connection
\(\Gamma^{a}_{bc},\) it can be deduced that the difference between the
connections with covariant index swapped, i.e.\ \(\Gamma^{a}_{bc}\)
and \(\Gamma^{a}_{cb}\) gives a tensor.  This is because the last term
of equation~\eqref{A.eq:ConnTrans} vanishes upon subtraction, and the
resulting tensor is called the \textbf{torsion tensor}
\index{Tensor!Torsion} and is given
by\[T^{a}{}_{bc} = \Gamma^{a}_{[bc]} = \Gamma^{a}_{bc} -
  \Gamma^{a}_{cb}.\] The covariant derivative is the most useful
derivative from a physical point of view, since all the conservation
laws can be most succinctly expressed in terms of this derivative.  A
usual rule of thumb given to students when going from special to
general relativity is that most differential equations in SR involving
partial derivatives get ``promoted'' to covariant derivatives,
everything else being unchanged.  We make use of this derivative in
the geometrical aspects in the next section, and in computing
conservation laws in a later Section.

We now have two notions of derivatives on the manifold. We proceed by
defining a measure of distance and angle through the metric.

\subsection{The metric}
A metric will be an object that associates a notion of distance
between two points.  There are many equivalent ways of doing this, and
a good place to start is with Pythagoras' theorem which allow us to
calculate rectilinear distances between points in Euclidean space.
Here because we only have a general, not necessarily euclidean,
manifold, we generalise this idea of rectilinear distance to apply
only infinitesimally.

A \textbf{Riemannian manifold}\index{Manifold!Riemannian} is a smooth
manifold \(X\) together with a continuous 2-covariant tensor field
\(g,\) called the \textbf{metric tensor}\index{Tensor!Metric}, such that
\begin{enumerate}
\item \(g\) is symmetric.  We expect this for the usual distance
  function, and require it here too, since the distance between two
  points does not depend on which point we consider first.
\item for each \(x \in X,\) the bilinear form \(g_{x}\) is
  non-degenerate.  This means that \(g_{x}(v,w) = 0\) for all \(v \in
  T_{x}(X)\) if and only if \(w = 0.\) This requirement ensures that
  the metric is invertible.
\end{enumerate}
Such a manifold is said to possess a Riemannian structure.  It is a
\textbf{proper} Riemannian manifold if we have further that
\(g_{x}(v,v) > 0\) for all possible \(v \in T_{x}(X)\) such that \(v
\neq 0.\) If the manifold is not proper, we call it a
\textbf{pseudo-Riemannian}\index{Manifold!Pseudo-Rimannian} manifold
endowed with an \textbf{indefinite metric.}\index{Metric!Indefinite}

The condition~2 above is necessary if we want the metric to have an
inverse, and can be expressed more conveniently in component form.  In
a local coordinate chart, \(g_{x}(v,w)\) is written as \(g_{ij} v^{i}
w^{j},\) and the non-degeneracy requirement above implies that the
determinant of \(g,\) with elements \(g_{ij}\) does not vanish in any
chart.  This property is not dependant on the choice of the charts
since local coordinate changes, \((x^{i'}) \to (x^{i})\) result in
\(g\) transforming through \[g_{i'j'} = \pderiv{x^{h}}{x^{i'}}
\pderiv{x^{k}}{x^{j'}} g_{hk} = [\Lambda^{k}{}_{j'}]
[\Lambda^{h}{}_{i'} ] g_{hk}.\] If initially we require \(\Det(g')
\neq 0,\) then since \[\Det(g) = \Det(g') \Det(\Lambda^{k}{}_{j'})
\Det (\Lambda^{h}{}_{i'}),\] the determinant of \(g\) is never zero
for local coordinate transformations.

The inverse of the matrix \((g_{ij})\) is denoted \((g^{ij})\) and
defines the components of a contravariant symmetric 2-tensor.  Both
the metric and its inverse can be used to respectively ``lower'' and
``raise'' indices on other tensors.  This works by contraction with
the metric of the covariant or contravariant components of the
geometrical object involved.  As an example, the vector \(v\) having
contravariant components \((v^{i}),\) can be expressed as to the
object \(v\) having covariant components \((v_{i}),\) related to the
contravariant components through \(v_{i} = g_{ij}v^{j},\) and
vice-versa by \(v^{i} = g^{ij}v_{j}.\) As mentioned previously, the
kernel \(v\) represents the geometrical object that has covariant and
contravariant components depending on the context, as required in the
abstract index notation.  Similarly, we can build
\textbf{mixed}\index{Tensor!Mixed} tensors out of purely covariant or
contravariant ones, e.g.\ \(T^{i}{}_{j} = g_{aj}T^{ia}.\) Since we
defined the metric \(g_{ij}\) and its inverse \(g^{ij}\) as matrix
inverses, it holds trivially that \(g_{ij}g^{ik} = \delta_{j}{}^{k},\)
where \(\delta_{j}{}^{k}\) is the Kronecker symbol.  The
\textbf{norm}\index{Vector!Norm} of a vector \(v \in T_{x}(X)\) is
only defined if we have a metric, and is given by
  \begin{equation}
    \label{A.eq:VecNorm}
|v|^{2} =
g_{x}(v,v) = g_{ij}v^{i}v^{j}.    
  \end{equation}
  If the our manifold is Riemannian proper, then this norm will always
  be positive, and if \(|v| = 0,\) we call \(v\) a
  \textbf{null}\index{Vector!Null} vector.  We will take a null vector
  to be orthogonal to itself.  At each point \(x \in X\) the null
  vectors can be imagined to form a cone in \(T_{x}(X)\) called the
  \textbf{null cone.}\index{Null Cone}

\subsection{Metric signature and orthogonality of vectors}
Before we can turn to physics, we investigate the types of relations
that the metric give us, once it is defined.  The norm of a vector
\(v \in T_{x}(X)\), and equivalently the scalar from the quadratic
form \(g_{ij}v^{i}v^{j}\) in some chosen basis can be expressed as a
sum of \(k\) positive squares and \(n-k\) negative ones, where \(n\)
is the dimension of the manifold \(X\)
\[g_{ij} v^{i} v^{j} = \sum_{i=1}^{k} (v^{i})^{2} - \sum_{i = k+1}^{n}
  (v^{i})^{2}.\] The number \(k\) is then called the
\textbf{index}\index{Index} of the quadratic form, and is independent
of the basis.  Then then number \(k - (n-k)\) is called the
\textbf{metric signature.}\index{Metric!Signature} Since we have
defined the metric \(g\) to be continuous, the index and thus the
signature of the metric is the same at each point \(x \in X,\) and one
can speak of the signature of the whole manifold.  The index of a
proper Riemannian manifold is the same as its dimension \(n.\) This
follows simply from our two metric axioms.  A pseudo-Riemannian
metric \(g\) is called a \textbf{Lorentzian
  metric}\index{Metric!Lorentzian} if the signature of the quadratic
form is \((+,-,-,\dots,-).\) In the case of a Lorentzian manifold we
denote its dimensions by \(n+1\) and we use Latin letters
\((a, b,\dots, = 0,1,\dots, n)\) for labelling local coordinates and
tensor components, and use Greek letter
\((\alpha, \beta, \dots = 1,\dots,n)\) for the spatial components.

In this thesis, because of the assumptions of classical general
relativity, we restrict the dimension of the Lorentzian manifold we
look at to 4, the index to 1, for a signature for -2.  We will choose
to have the time component of our metric be positive, and the spatial
ones to be negative.

A vector \(v \in T_{x}(X)\) such that \(g_{ab}v^{a}v^{b} <0,\) that is
one outside the null cone is called
\textbf{spacelike}\index{Vector!Spacelike}.  A vector
\(u \in T_{x}(X),\) such that \(g_{ab}v^{a}v^{b} > 0,\) that is inside
the null cone is called \textbf{timelike}\index{Vector!Timelike}.  The
null cone \(C_{x}\) is made up of two half-cones.  If one of the
half-cone is chosen and called the future half-cone \(C^{+}_{x},\)
then the tangent space \(T_{x}(X)\) is said to be \textbf{time
  oriented}\index{Time oriented}. A timelike vector in \(C^{+}_{x}\)
is said to be \textbf{future-directed}\index{Future directed}; a timelike vector in \(C^{-}_{x}\)
is said to be \textbf{past-directed}\index{Past directed}.

 The metric itself is usually written in the natural
coordinate frame, and as a tensor usually expressed
through
\begin{equation}
  \label{A.eq:metric}
g = g_{ab} \d x^{a} \d x^{b}.  
\end{equation}

As mentioned earlier, the metric is used to define length, surface and
volume measures on a manifold.  The \textbf{length}\index{length} of a
parametrized curve \(\tau \mapsto x(\tau)\) joining two point of
manifold \(M\) with parameters \(\tau_{1}\) and \(\tau_{2}\) is
\begin{equation}
  \label{A.eq:length}
l \coloneqq \int_{\tau_{1}}^{\tau_{2}} \left|
    g_{ab}(x(\tau))\deriv{x^{a}}{\tau}\deriv{x^{b}}{\tau} \right|\d
  \tau   
\end{equation}
The curve \(s \mapsto x(s)\) is said to be parametrized by arc length
if
\begin{equation}
  \label{A.eq:arclengthparam}
  \left| g_{ab}(x(\tau))\deriv{x^{a}}{\tau}\deriv{x^{b}}{\tau} \right|
  = 1.
\end{equation}
Since we now have a metric tensor through
equation~\eqref{A.eq:metric}, we use the component form of vectors and
tensors to perform calculations.  The norm~\eqref{A.eq:VecNorm} of a
vector having components \(X^{a}\) is thus
\[|X|^{2} = g_{ab}X^{a}X^{b}.\] For two vectors \(X^{a}\) and
\(Y^{a}\), neither null, the angle between then is defined through the
cosine of the angle between then.  This is given by
\[\cos(X,Y) = \f{g_{ab}X^{a}Y^{b}}{\sqrt{g_{cd}X^{c}X^{d}}
    \sqrt{g_{ef}Y^{e}Y^{f}}}.\]

The next step in getting to GR is the concept of geodesics.  We define
a \textbf{timelike metric geodesic}\index{Geodesic!Metric} between
points \(p,q \in X,\) as the privileged curve joining the two points
whose length is \emph{stationary} under small variations that vanish
at the endpoints.  This length may this be a maximum, a minimum, or a
saddle point.  To implement this condition, we require the calculus of
variations, and the Euler--Lagrange equation, which we are going to
assume without giving details.  We refer the interested reader
to~\cite{Inv92, Cho82} for details.  The Euler--Lagrange (E--L)
equation here has to be applied to the length which behaves as the
action in equation~\eqref{A.eq:length}; the metric is the generalised
coordinate; and the curve parameter \(\tau,\) the independent variable.
The application of the E--L equation results in the equation of motion
which is this case is the equation of a geodesic in a general
Lorentzian manifold:
\begin{equation}
  \label{A.eq:geodesic1}
g_{ab} \sderiv{x^{b}}{\tau} + \{bc,a\} \deriv{x^{b}}{\tau}
  \deriv{x^{c}}{\tau} = \left( \sderiv{l}{\tau} \Bigg{/}
    \deriv{s}{u}\right) g_{ab} \deriv{x_{b}}{\tau}.  
\end{equation}
 The quantities
denoted by \(\{ab,c\}\) are known as the \textbf{Christoffel symbols
  of the first kind}\index{Christoffel symbols!First kind}, and are given by
\begin{equation}
  \label{A.eq:1christof}
  \{ab,c\} = \f{1}{2}\left(g_{ac,b} + g_{bc,a} -g_{ab,c} \right).
\end{equation}
The equation~\eqref{A.eq:geodesic1} can be simplified by using
arc-length parametrisation~\eqref{A.eq:arclengthparam}, and finding an
expression for the derivatives on the RHS in~\eqref{A.eq:geodesic1}, which can
easily be done with the definitions we have.  This results in the
geodesic equation simplifying to
\begin{equation}
  \label{A.eq:geodesic2}
\sderiv{x^{a}}{l} + \Christoffel{a}{b}{c} \deriv{x^{b}}{\tau}
  \deriv{x^{c}}{\tau} = 0.  
\end{equation}
with the \(\Christoffel{a}{b}{c}\) being the \textbf{Christoffel
  symbols of the second kind}\index{Christoffel symbols!Second kind}
given by
\begin{equation}
  \label{A.eq:2christof}
\Christoffel{a}{b}{c} = g^{ad} \{bc,d\} = \f{1}{2} g^{ad} \left(g_{bd,c} + g_{cd,b} -g_{bc,d} \right).
\end{equation}

We can now define the \textbf{Einstein metric}, which is the special
choice of the metric so that the connection is the same as the
Christoffel symbol of the second kind.  If this choice is made (the
two objects transform similarly), the connection is called a
\textbf{metric connection}\index{Connection!Metric}.  With this
particular choice we have
\begin{equation}
  \label{A.eq:metricConn}
  \Gamma^{a}_{bc} = \Christoffel{a}{b}{c} =  \f{1}{2} g^{ad} \left(g_{bd,c} + g_{cd,b} -g_{bc,d} \right).
\end{equation}
As a result, \(\Gamma^{a}_{bc}\) is automatically symmetric, with
\(\Gamma^{a}_{bc} = \Gamma^{a}_{cb}\) and the torsion \(T^{a}{}_{bc}\)
vanishes.

The consequence of this identification, of the geodesic coefficients
\(\Christoffel{a}{b}{c}\) on the one hand and the connection
\(\Gamma^{a}_{bc}\) on the other, means that the covariant derivative
of the metric is zero\[\nabla_{c} g_{ab} = g_{ab;c} = 0,\] as can be
readily computed.  We now have chosen the manifold, its dimension, the
metric, the metric connection to be the ingredients of the theory of
classical GR.

At each point of a curved Lorentzian manifold, we can choose a
coordinate system such that at that point the metric is locally
Minkowski.  This is know as the local flatness theorem, and we state
it here without proof.  The interested reader is referred to
Refs.~\cite{sch09,Poi04} where the theorem is proved through a first
order Taylor expansion of the metric coefficients around point \(P\)
and comparison with the curved metric transformation laws.
\begin{myTheorem}
  For a given point \(P\) in space-time it is always possible to find
  a coordinate system \(x^{a'}\) such
  that\[g_{a'b'}(P) = \eta_{a'b'}, \quad\text{and}\quad
    \Gamma^{a'}_{b'c'}(P) = 0. \] where
  \(\eta_{a'b'} = \diag{(1,-1,-1,-1)}\) is the \textbf{Minkowski
    metric}\index{Metric!Minkowski} of flat space.  Such a coordinate
  system is called a \textbf{Lorentz frame}\index{Frame!Lorentz} at P.
  The physical interpretation of this theorem leads directly to
  Einstein's equivalence principle which states that free-falling
  observers do not see any gravitational effects in their immediate
  vicinity.  Note however, that the derivatives of the connection
  coefficients \(\partial_{d}\Gamma^{a'}_{b'c'}\) and therefore the
  second derivatives of the metric \(g_{a'b',c'd'}\) are \emph{not}
  zero.
  \label{A.th:LocFlat}
\end{myTheorem}
This theorem is used in the following sections, and sometimes even
assumed without explicit warning.

 We have a metric algebraically symmetric in
its indices which reduces the number of components of \(g_{ab}\) from
\(n^2 = 16\) to \(\f{1}{2}n(n+1) = 10.\) However we need to be able to
enforce physical symmetries on this metric, and other tensors.  We
will now use Lie derivatives to show how symmetries are enforced.

\subsection{Symmetry}
An \textbf{isometry} of a Lorentzian manifold \((V,g)\) is a
diffeomorphism \(f\) which leaves the metric \(g\) invariant; that is
\(f^{*} g = g.\) A metric is invariant by a 1-parameter group of
isometries generated by a vector field \(X\) if its Lie derivative
with respect to \(X\) vanishes.  This is expressed as
\begin{equation}
  \label{A.eq:Killing}
 \LieD{X} g_{ab} = X^{i}g_{ab,i} + g_{ib}X^{i}_{,a} + g_{ai} X^{i}_{,b} = 0, 
\end{equation}
if \(X\) is the vector field that generates the symmetries we are
concerned with. Equation~\eqref{A.eq:Killing} is called
\textbf{Killing's equation}\index{Killing's equation}, and the
associated vector field a Killing field\index{Field!Killing}.

This vector field that generates the symmetries can be any of the
numerous ones we usually see in physics: for example there exists
suitable vectors encoding time invariance, Lorentz invariance,
spherical symmetry, etc.\ In this thesis we will be concerned with
spherical symmetry and staticity and therefore briefly define what
this entails.

A family of \textbf{hypersurfaces}\index{Hypersurface} is given by the
equation \(f(x^{a}) = \mu,\) where the different members of the family
have different values of \(\mu.\) This is similar to the usual concept
of surfaces in Cartesian 3-space, but here generalised to manifolds.
We can define a covariant vector field \(N_{a}\) of vectors
\textbf{normal}\index{Vector!Normal} to the hypersurface by
\(N_{a} = \pderiv{f}{x^{a}}.\) In the same analogy, these are the
gradient of the surfaces in 3-space.  This normal vector can be made
to be of unit length by the usual process of dividing by its norm if
the normal vector is nowhere null. This is defined
through \(n_{a} = \f{N_{a}}{|N_{a}N^{a}|}.\) Then depending on the type
of hypersurface, the unit normal vector \(n_{a}\) can be classified
through
  \begin{equation}
    \label{A.eq:HyperNormal}
n_{a}n^{a} = \epsilon \equiv \begin{cases} -1 \qquad \text{if the hypersurface is timelike},\\
    + 1 \qquad \text{if the hypersurface is spacelike} \end{cases}    
  \end{equation}
  A vector field \(X^{a}\) is said to be
  \textbf{hypersurface-orthogonal}\index{Hypersurface-orthogonal} if
  it is everywhere orthogonal to the family of hypersurfaces, and
  proportional to the normal vector \(n_{a}\) everywhere, so that
  \(X_{a} = \lambda(x)n_{a}.\)

A space-time is said to be
\textbf{stationary}\index{Space-time!Stationary}\index{Symmetry!Stationary}
if and only if it admits a timelike Killing vector field.  If the
vector field is additionally hypersurface-orthogonal, the space-time
is called
\textbf{static}\index{Symmetry!Static}\index{Space-time!Static} . In a
static space-time, there exists a coordinate system \textbf{adapted}
to the timelike Killing field above in which the metric is
time-independent and has no cross-terms in the line element involving
the time.

A space-time is said to be \textbf{spherically
  symmetric}\index{Symmetry!Spherical} if and only if it admits three
linearly independent spacelike Killing vector fields \(X^{a}, Y^{a}\)
and \(Z^{a}\) whose orbits are closed, and which obey the following relations
\[[X^{a},Y^{a}] = Z^{a}, \qquad [Y^{a},Z^{a}] = X^{a}, \qquad
  [Z^{a},X^{a}] = Y^{a}.\] The Lie brackets \([X,Y]\) being defined
through \([X,Y] = \LieD{X}Y - \LieD{Y}X.\) These vectors are usually
picked in a Cartesian frame so that
\begin{align}
  \label{A.eq:AngMomentum}
X &= x^{2}\pderiv{}{x^{3}} - x^{3}\pderiv{}{x^{2}}\\
Y &= x^{3}\pderiv{}{x^{1}} - x^{1}\pderiv{}{x^{3}}\\
Z &= x^{1}\pderiv{}{x^{2}} - x^{2}\pderiv{}{x^{1}}
\end{align}
The adapted coordinates to these vectors are the usual spherical
coordinates.  These vectors together also generate the \(SO(3)\)
symmetry Lie group.

\subsection{Curvature and the Riemann tensor}
Curvature is going to play an important part in this thesis.
Geometrically \textbf{curvature}\index{Curvature} is signalled by the
non-commutativity of the covariant derivatives.  This essentially
means that a vector that has been parallel transported along a
closed loop up to its starting point in no longer the same.  This
deviation of the 2 vectors is an effect of curvature.

The covariant derivative, unlike the partial derivative is not
commutative.  We define the \textbf{commutator} of a tensor
\(T^{a\dots}_{b\dots}\) to be
 \[\nabla_c \nabla_d T^{a\dots}_{b\dots} - \nabla_{d} \nabla_{c} T^{a\dots}_{b\dots}.\]
 To compute the curvature we use the definition above and calculate
 the commutator of a vector directly.  After a lengthy process, we
 obtain
 \begin{align}
   \label{A.eq:DefRiemGen}
\nonumber
\nabla_{c}\nabla_{d} X^{a} - \nabla_{d}\nabla_{c} X^{a} &= (\partial_{c} \Gamma^{a}_{bd} - \partial_{d}\Gamma^{a}_{bc} + \Gamma^{e}_{bd} \Gamma^{a}_{ec} - \Gamma^{e}_{bc} \Gamma^{a}_{ed})X^{b} 
+ (\Gamma^{e}_{cd} - \Gamma^{e}_{dc}) \nabla_{e} X^{a}, \\
&= R^{a}{}_{bcd} X^{b} + T^{e}{}_{cd}\nabla_{e} X^{a}.
 \end{align}
 The last equation uses the definition of the torsion, and also
 defines the \textbf{Riemann tensor}\index{Tensor!Riemann} as the
 (1+3) tensor \(R^{a}{}_{bcd}\) that measures the curvature as a vector
 moves around a loop.  If additionally we choose a metric
 connection~\eqref{A.eq:metricConn}, the torsion vanishes and the
 expression of the Riemann tensor~\eqref{A.eq:DefRiemGen} simplifies to
 \begin{equation}
   \label{A.eq:DefRiem}
   R^{a}{}_{bcd} = \partial_{c} \Gamma^{a}_{bd} - \partial_{d}\Gamma^{a}_{bc} + \Gamma^{e}_{bd} \Gamma^{a}_{ec} - \Gamma^{e}_{bc} \Gamma^{a}_{ed}
 \end{equation}
 Since the metric connection depends on the metric's first derivatives
 as given in equation~\eqref{A.eq:metricConn}, The Riemann tensor
 depends on the first \emph{and} second derivatives of the metric.
 This becomes important when the boundary conditions have to be
 applied on Einstein's equations which depend on the Riemann tensor,
 and hence up to the second derivative of the metric.

 The Riemann tensor has a number of algebraic symmetries inherent in
 it.  We state a few here, as these can easily be proved from the
 definitions we have given so far.
 \begin{enumerate}[label=(\roman*)]
 \item \label{A.en.R1} It is antisymmetric in its last two indices so that  \(R^{a}{}_{bcd} = -R^{a}{}_{bdc}.\)
 \item \label{A.en.R2}A symmetric connection, and zero torsion leads
   to the following identity
   \(R^{a}{}_{[bcd]} = R^{a}{}_{bcd} + R^{a}{}_{dbc} +R^{a}{}_{cdb} =
   0\)
 \item \label{A.en.R3}When all the indices of the Riemann tensor are lowered
   \(R_{abcd} = g_{ad}R^{d}{}_{bcd},\) the interchange of the first
   and last pair of indices do not change the tensor, that
   is \(R_{abcd} = R_{cdab}.\)
 \item \label{A.en.R4}The only way the above identity can work is if the Riemann
   tensor is antisymmetric in its first 2 indices, and indeed it is, \(R_{abcd} = -R_{bacd}.\)
 \end{enumerate}
 All these symmetries reduce the number of components of the Riemann
 tensor from the naive \(n^{4} = 256\) to
 \(\f{1}{12}n^{2}(n^{2}-1) = 20\) independent components.  The
 additional symmetries we will have on the metric tensor in this
 thesis reduce these even further.

 The \textbf{Bianchi identities}\index{Bianchi identity} can be stated
 in terms of the Riemann tensor directly.  The contracted form of
 these identities are used in this thesis to simplify some of the
 differential equations, and in the derivation of the TOV equation.
 The identities read
 \begin{equation}
   \label{A.eq:bianchi}
   R_{de[bc;a]} = \nabla_{a}R_{debc} + \nabla_{c}R_{deab} +\nabla_{b}R_{deca} = 0.
 \end{equation}

 Once the Riemann tensor has been defined on the manifold, we build
 the Einstein tensor from different contractions of the tensor.

\subsection{The Ricci tensor}
The \textbf{Ricci tensor}\index{Tensor!Ricci} is obtained by
contracting the Riemann tensor in its first and third indices. The
resulting tensor is a rank 2 tensor given by
\begin{equation}
  \label{A.eq:RicciT}
  R_{ab} = g^{cd} R_{dacb} = R^{c}{}_{acb}.
\end{equation}
This is a symmetric tensor since \(R_{ab} = R_{ba},\) as can be
confirmed from the symmetry of \(R_{abcd}\) in its interchange of the
first and last pair of indices.

\subsection{The Ricci scalar}
The \textbf{Ricci scalar}\index{Scalar!Ricci} is generated by a further contraction of the Ricci
tensor~\eqref{A.eq:RicciT}. It is given by
\begin{equation}
  \label{A.eq:RicciS}
  R = G^{ab} R_{ab} = R^{a}{}_{a}.
\end{equation}
It is the trace of the Ricci tensor, and can be considered as the
average curvature in a certain sense.  It is used to define Einstein's
tensor which we do next.

\subsection{The Einstein tensor}
Finally the \textbf{Einstein tensor}\index{Tensor!Einstein} is defined
as
\begin{equation}
  \label{A.eq:EinT}
  G_{ab} = R_{ab} - \f{1}{2}g_{ab}R.
\end{equation}
in terms of the Ricci tensor and the Ricci scalar.  This is the tensor
that the Einstein field equations are written with, and encode all the
geometrical aspects that are directly influenced by matter and fields.
The curvature, and metric are all affected through climbing up the
contraction ``ladder'' just presented.

The Einstein tensor obeys the \textbf{contracted Bianchi
  identity}\index{Bianchi identity!Contracted}, given by
\[\nabla_{b} G^{b}{}_{a} = G^{b}{}_{a;b} = 0.\]

We now turn to the other part of the Einstein field equations, having
completed the geometrical aspects of it. The next part concerns the
source of this geometrical curvature, matter and fields.

\subsection{The Weyl tensor}
The \textbf{Weyl tensor}\index{Tensor!Weyl} \(C_{abcd}\) is defined
through
\begin{equation}
  \label{A.eq:WeyT}
  C_{abcd} = R_{abcd} - g_{a[c}R_{d]b} + g_{b[c}R_{d]a} + \f{R}{3} g_{a[c}g_{d]b},
\end{equation}
and satisfies the symmetries~\ref{A.en.R1},~\ref{A.en.R2},
and~\ref{A.en.R4} of the Riemann tensor.  Additionally it is
trace-free in all of its indices, and is hence also known as the
``trace-free'' part of the Riemann tensor.  Under conformal
transformation of the metric, \(\bar{g}_{ab} \to \Omega^{2} g_{ab},\)
the Weyl tensor remains invariant, and sometimes this tensor is also
called the \textbf{conformal tensor}\index{Tensor!Conformal}.

\section{Matter}
\label{S:Matter}
To our best knowledge, matter is discontinuous at all scales, and
quantum mechanics confirms this.  However most of the time we are not
concerned with this discontinuity and the advantages obtained by
``smoothing'' out these discontinuities are so many, that it is usual
to describe matter in physics through a fluid.  This approximation is
usually valid only when we want to look at the behaviour of volumes
big enough that quantum effects do not come into play, and small
enough that arguments based on the infinitesimal are valid.  Therefore
if we can define measurable quantities associated with some finite
volume (at whatever scale) at space-time events, we deem our model for
matter to be continuous.  This has to be taken with the grain of salt
that the model breaks down when quantum effects come into play, but
this breaking down only occurs at scales we are unconcerned with, and
the macroscopic picture (with quantities corresponding to averages
over microscopic ones) remains more or less faithful to reality.  In
this thesis we will model matter as a \textbf{fluid.}\index{Fluid}

A fluid is a model of matter where the only interaction possible
between fluid elements occur at the interface between the elements, if
no external forces are acting~\cite{Cho08}.  These interactions might
be of any type, including slipping, compression, pushing, etc.\ In
this thesis we will be looking at three specific cases of fluids:
dust\index{Matter!Dust}, perfect fluids\index{Matter!Perfect fluid}
and fluids with anisotropic pressures\index{Matter!Anisotropic fluid}.
Dust is the simplest with no possible interaction between fluid
elements.  The next are perfect fluids which have minimal interactions
between fluid elements, and the most complicated of the three is a
simple generalization of perfect fluids.

To be able to use matter as a source of gravitation (an idea we wish
to preserve from Newtonian physics), we need an object, preferably
tensorial, that encodes matter and its aspects (energy, momentum,
temperature, enthalpy).  This will then allow us to specify a source
for the Einstein equations.  We proceed in the footsteps of
Tolman~\cite{Tol66} who derives a tensor encoding aspects of matter
from very general considerations of a continuum.  The reason we do not
start with the relativistic dynamics of particles is that it is not
possible to derive the equations of the continuum, in either
Newtonian or relativistic physics from particle dynamics without a
fair bit of quantum mechanics~\cite{Tol66, Ber76, Pau58}.  A more
modern approach to this problem is given in~\cite{sch09} and while
many details are eschewed, the concept of a momentarily comoving frame
of reference (MCFR), which we use extensively is explained in detail.

\subsection{Newtonian Analysis}
\begin{figure}[h!]
  \centering 
  \includegraphics[width=12cm]{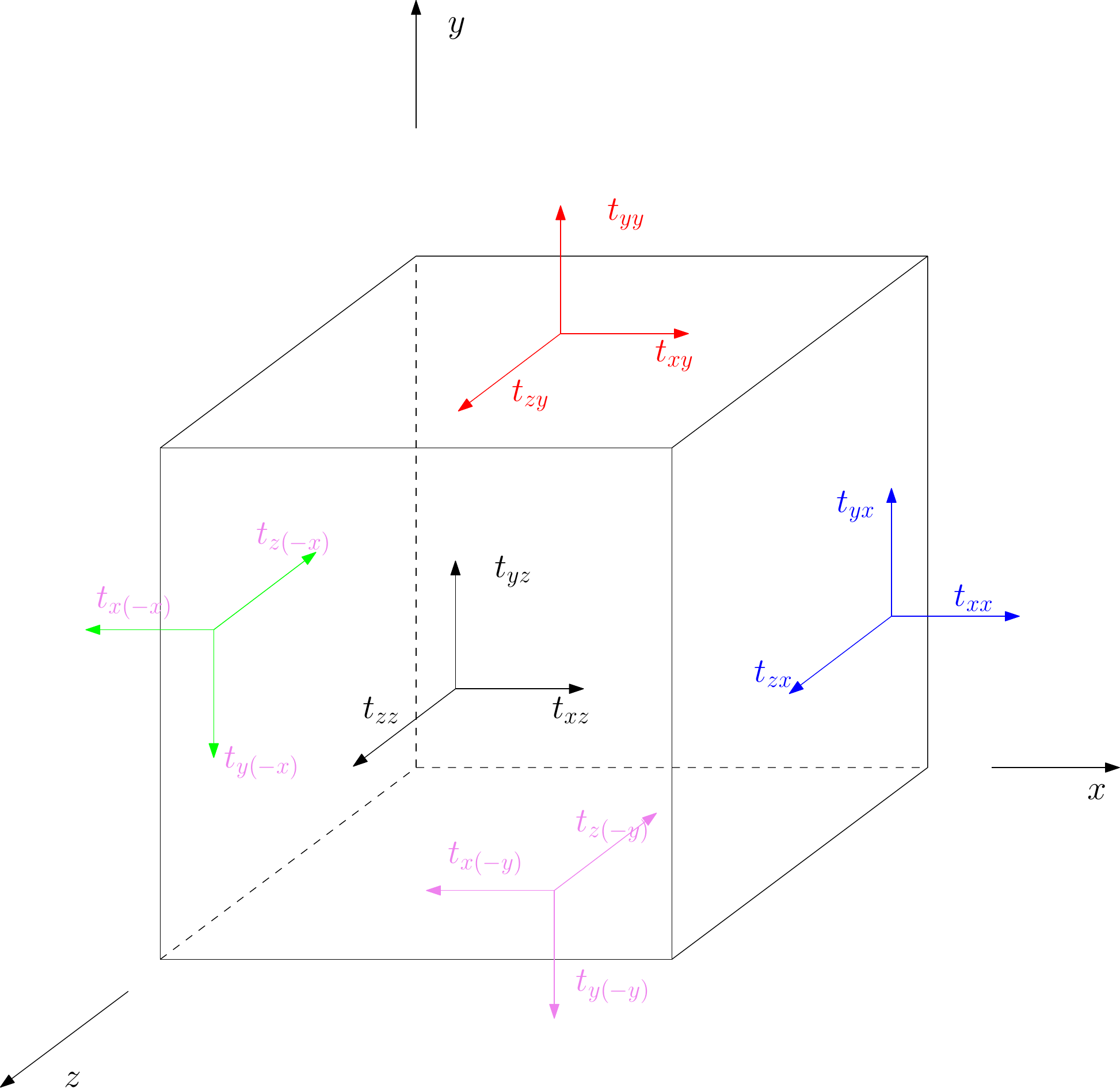}
  \caption[Stress direction convention]{The convention for specifying
    directions of the components of the stress tensor in Cartesian
    coordinates}
  \label{fig:Prel.Stress}
\end{figure}

The first aspect of matter we wish to capture is momentum and its
conservation.  In a continuum, at any point we can define nine
quantities, usually called the stress matrix\index{Tensor!Stress} that
give both the tangential and perpendicular components of the force
acting on an imaginary surface at that point.  We will label these
quantities with two indices (as with a matrix).  The first index will
correspond to the direction in which the component of the force is
acting, and the second index will refer to the direction normal to the
surface to which the force component is acting.  In a Cartesian
coordinate system this will correspond to
\begin{equation}
\label{Prel.eq:stress}
t_{ij} =
\begin{pmatrix}
  t_{xx} &t_{xy} &t_{xz}\\
  t_{yx} &t_{yy} &t_{yz}\\
  t_{zx} &t_{zy} &t_{zz}
\end{pmatrix},  
\end{equation}
where e.g.\ \(t_{zy}\) refers to the z-component of the force acting
on a surface oriented in the y-direction: the imaginary surface that
spans part of the \(xy\)-plane at the point.  This force is caused by
the material around the point in question and is thought of in the
above example to be due to matter present at lower y-coordinate
values.  We show this in detail in figure~\ref{fig:Prel.Stress}.  The
major source of confusion in this set up is the direction of the
normal for the imaginary surfaces.  These normals are degenerate and
could be in two opposing directions, however since we have not imposed
any external forces on our continuum thus far, we expect Newton's
third law to hold, and we impose this by requiring that the force at
the same point, but on the plane with the normal opposite to the one
from the first surface be the same.  Thus in our diagram, parallel
surfaces of the cube would have opposite forces in the absence of
volumetric external forces, as shown by both the complementary colour
and notation: \(t_{i(-j)}\) refers to the force in the \(i\) direction
with respect to the surface normal to the \((-j)\) direction.

In the presence of external volumetric forces (electromagnetic,
gravitational, etc.\ ) the forces on parallel surfaces will not cancel
each other as above.  Instead the difference between these forces will
tell us about how the external forces are acting on the volume under
consideration.  Hence instead of \(t_{zy}\) on the upper surface and
\(t_{z(-y)} = -t_{zy}\) on the lower one, we will have \(-(t_{zy} +
\pderiv{t_{zy}}{y})\) on the bottom surface.  The partial derivative
will give us a measure of how the external volumetric forces are acting,
and by summing all the contributions in say the \(z\) direction in
particular, we get\[f_{z} = - \pderiv{t_{zx}}{x} - \pderiv{t_{zy}}{y}
- \pderiv{t_{zz}}{z},\] the total external force in the \(z\)
direction, per unit volume.  This argument applies to the other
directions, and we can immediately conclude that 
\begin{equation}
  \label{eq:Prel.Force}
f_{i} =-\pderiv{t_{ij}}{x_{j}} = t_{ij}{}^{,j} \quad,  
\end{equation} where we are using Einstein's
notation again.  We wish to relate this force to the change in
momentum of the volume under consideration in an attempt to get back a
dynamical rule like Newton's second law, and to do so we will
introduce, following Tolman~\cite{Tol66, Ber76} a \textbf{momentum
  volume density, }\index{Momentum!Volume density}whose component in
the \(i\) direction is denoted by \(g_{i}.\) The rate of change of
this momentum density will be related to the force \(f_{i}\) through
the usual relation,
\begin{equation}
  \label{eq:Prel.N2}
f_{i} \delta V = \deriv{}{t}(g_{i} \delta V),  
\end{equation}
where \(\delta V\) is the volume of the cube we are considering.
Combining the two relations~\eqref{eq:Prel.N2} and
\eqref{eq:Prel.Force} we obtain 
\begin{equation}
  \begin{aligned}
- \pderiv{t_{ij}}{x_{j}} \delta V &= \deriv{}{t}(g_{i} \delta V) \\
  &=\deriv{g_{i}}{t} \delta V + g_{i} \deriv{(\delta V)}{t},
  \end{aligned}
  \label{eq:Prel.N2simp1}
\end{equation}
which can immediately be simplified since the momentum density can
change either instantaneously \emph{at one point} or by the movement
of the fluid element.  This is usually expressed though a material
derivative, but we will not pursue this matter further than to
simplify the first term of the right hand side in terms of the
velocities of the fluid element defined though \(u_{i} = \partial
x_{i} / \partial t.\)  Therefore we have
\begin{equation}
  \label{eq:Prel.Momentum}
  \deriv{g_{i}}{t} = \pderiv{g_{i}}{t} + \pderiv{g_{i}}{x_{j}} \pderiv{x_{j}}{t} = 
\pderiv{g_{i}}{t} + u_{j} \pderiv{g_{i}}{x_{j}}. 
\end{equation}
Similarly, the volume element itself changes as it moves and the way
each surface of our initial cube moves with the velocities defined
above allows us to determine the quantitative change through
\begin{equation}
  \label{eq:Prel.Volume}
  \deriv{}{t} (\delta V) = \left( \pderiv{u_{x}}{x} + \pderiv{u_{y}}{y} + \pderiv{u_{z}}{z}\right) \delta V
  = \pderiv{u_{j}}{x_{j}} \delta V.
\end{equation}
The derivative of the velocities appear instead of the velocities only
since both parallel surfaces in our initial cube move with different
velocities.  Substituting the above results~\eqref{eq:Prel.Volume}
and~\eqref{eq:Prel.Momentum} in~\eqref{eq:Prel.N2simp1}, we get the
final dynamical equation of motion of our fluid element in terms of
the momentum density
\begin{equation}
  \label{eq:Prel.N2fin}
  \begin{aligned}
    - \pderiv{t_{ij}}{x_{j}} &= \pderiv{g_{i}}{t} + u_{j} \pderiv{g_{i}}{x_{j}} + g_{i} \pderiv{u_{j}}{x_{j}} \\
    &= \pderiv{g_{i}}{t} + \pderiv{}{x_{j}} (g_{i}u_{j}).
  \end{aligned}
\end{equation}
This becomes our equation of momentum conservation.  Additionally we
require an equation of mass conservation, which in this framework we
will take as the conservation of mass density, \(\rho\).  Typically,
we expect mass to exit or enter our cube through a momentum flow in
and out of the surfaces of the cube.  This means that
\[-\pderiv{M}{t} = -\pderiv{}{t} \int \rho \d V = \oint_{\partial V}
g_{i} n^{i} \d A. = \int g_{i}{}^{,i} \d V\] The negative sign comes from
the fact that the momentum density is taken to be pointing outward
from the cube in our definition, and thus results in a reduction in
mass of our initial volume; \(n_{i}\) is the normal vector in the
\(i\) direction, pointing outwards too.  Straightforward application
of Gauss's theorem on the surface integral then results in the
following conservation law for the mass density:
\begin{equation}
  \label{eq:Prel.Density}
  -\pderiv{\rho}{t} = \pderiv{g_{i}}{x_{i}}
\end{equation}
All the above discussion is valid in Newtonian physics. We now wish to
generalise it to a relativistic framework, and to do so we require the
transformation properties of our dynamical variables under Lorentz
transformations, since according to the first principle of relativity,
the form of the above laws are invariant in all frames under uniform
motion.  To do so we refer to the transformation laws involving mass,
force, momentum and area, and deduce the transformation laws for the
stress tensor.

\subsection{Special relativistic generalisation}
In order to set up the problem, we will require two coordinate
systems, related to each other through a proper motion given by a
velocity vector \(\vec{V}\).
The first system \(S,\)
will be assumed to be oriented (with no loss of generality) so that
the fluid is moving along one particular direction, say the
\(x-\)direction,
with no components in the other spatial directions.  We call the
velocity of the fluid element with respect to the \(S-\)system,
\(\vec{u}.\)
In order to simplify the derivation of the transformation properties
we shall also assume that the second coordinate system \(S^{0}\)
is also moving in the \(x-\)direction
with respect to \(S\)
with the same velocity, so that in \(S^{0}\)
the components of the fluid element is given by
\(\vec{u} = (u^{0}_{x},u^{0}_{y}, u^{0}_{z})^{\top} = \vec{0}.\)
This frame \(S^{0}\)
is what is referred to in~\cite{sch09} as the MCRF.

Since we have expressions for the stress components \(t^{0}_{ij}\)
in a rest frame: which we will identify with \(S^{0},\)
we can use the Lorentz transformations to generate the expressions of
these quantities in the frame \(S.\)
To do this we first transform all the forces~\eqref{eq:Prel.Force} to
the moving frame \(S,\)
with the velocity \(\vec{u}\) relating the proper motions between the two frames, to obtain
\begin{equation}
  \label{Prel.eq:ForceTransformation}
  f_{x} = f_{x}^{0}, \qquad f_{y} = f_{y}^{0} \sqrt{1-(u/c)^{2}}, \qquad f_{z} = f_{z}^{0} \sqrt{1-(u/c)^{2}},
\end{equation}
since the proper motion is in the \(x-\)direction.
Similarly, the surface areas of the cube faces that are normal to the
\(y-\)
and \(z-\)
axes will be contracted, with the areas normal to the direction of
motion remaining the same, so that
\begin{equation}
  \label{Prel.eq:AreaTransformation}
A_{x} = A_{x}^{0}, \qquad A_{y} = A_{y}^{0} \sqrt{1-(u/c)^{2}}, \qquad A_{z} = A_{z}^{0} \sqrt{1-(u/c)^{2}}.
\end{equation}
From the definition of the stress tensor we used
previously~\eqref{Prel.eq:stress} of force per unit area, and
correspondence we just derived from the Lorentz
transformations~\eqref{Prel.eq:ForceTransformation},
and~\eqref{Prel.eq:AreaTransformation}, we infer that the stress
components in the \(S\) frame become:
\begin{equation}
\label{Prel.eq:stressTransUn}
t_{ij} =
\begin{pmatrix}
  t^{0}_{xx} &\f{t^{0}_{xy}}{\sqrt{1-(u/c)^{2}}} &\f{t^{0}_{xz}}{\sqrt{1-(u/c)^{2}}}\\
  t^{0}_{yx}\sqrt{1-(u/c)^{2}} &t^{0}_{yy} &t^{0}_{yz}\\
  t^{0}_{zx}\sqrt{1-(u/c)^{2}} &t^{0}_{zy} &t^{0}_{zz}
\end{pmatrix}.  
\end{equation}
The above equation is surprising in a two ways.  First, we notice that
since the velocity relating the frames is only in the \(x-\)direction,
equation~\eqref{Prel.eq:stressTransUn} is specialized for that
particular case.  Secondly it is quite surprising that while the
stress tensor is a symmetrical array in the rest frame, that is
\(t^{0}_{ij} = t^{0}_{ji},\)
it is quite clear that even more so in the case of general velocities
between the frames, the transformed stress array will not be
symmetrical, so that in general relativistic fluids,
\(t_{ij} \neq t_{ji}.\)
However now we have a way to relate the stresses measured by an
observer at rest with respect to the fluid, with stresses measured in
an arbitrary frame.

To get a complete tensorial description of matter, we should also
include the transformations of energy-momentum\index{Energy-momentum}
\(g_{i}\)
and density \(\rho\)
between frames.  From the expressions we have obtained before, we
should then, following \citeauthor{Tol66} find an expression for the
momentum, calculate the force acting on the stressed fluid, and thus
calculate the work done, and energy change on our moving cube in terms
of its mass, energy, velocity and stresses.  The complication that
arises however is that the change in momentum is not only due to the
motion of the cube, but also due to the work done by the stress on the
moving faces, and hence the volume of the cube.  We shall initially
assume as we did previously that the velocity is only in the
\(x-\)direction, so that the momentum density is given by
\begin{equation}
  \label{Prel.eq:pDensity}
 \vec{g} =  \begin{pmatrix}
    g_{x}\\g_{y}\\g_{z}
  \end{pmatrix}
 =
 \begin{pmatrix}
   \rho u + \f{t_{xx} u }{c^{2}} \\
   \f{t_{xy} u}{c^{2}}\\
   \f{t_{xz} u }{c^{2}}
 \end{pmatrix},
\end{equation}
where only the stress components \(t_{xj},\) with \(j=x,y,z\)
are chosen since the cube is moving in the \(x-\)direction
only. The products \(t_{xj} u\)
thus give the energy density flow in the \(j\)
direction due to the stresses and we divide by \(c^{2}\)
to make the units match.  As is evident and expected, the momentum
density in the \(i\)
direction is also affected by stresses \emph{perpendicular} to that
direction in relativity.

Since the cube we are considering is infinitesimal, integrating the
above~\eqref{Prel.eq:pDensity} gives the total momentum change in some
volume V as
\[
 \vec{G} =  \begin{pmatrix}
    G_{x}\\G_{y}\\G_{z}
  \end{pmatrix}
 =
 \begin{pmatrix}
   \f{E + V t_{xx} }{c^{2}} u \\
   \f{t_{xy} V}{c^{2}} u\\
   \f{t_{xz} V }{c^{2}} u
 \end{pmatrix},
\]
which allows us to use Newton's second law to find the force exerted
on the volume \(V\) to change its velocity \(u\) in the \(x-\)direction as
\begin{equation}
  \label{Prel.eq:fTotal}
 \vec{F} = \deriv{}{t} \begin{pmatrix} G_{x}\\G_{y}\\G_{z} \end{pmatrix}
= \deriv{}{t}
 \begin{pmatrix}
   \f{E + V t_{xx} }{c^{2}} u \\
   \f{t_{xy} V}{c^{2}} u\\
   \f{t_{xz} V }{c^{2}} u
 \end{pmatrix}.
\end{equation}

To calculate the work done on the stressed volume, we start with the
initial volume in the observer's rest frame, characterised by
\((V^{0}, t^{0}_{ij},\)
and \(E^{0}),\)
and bring it from rest up to some velocity by an adiabatic acceleration
(boost), so that the observer is also moving with the accelerated
material.  The volume gets Lorentz contracted in the direction of
motion through
\begin{equation}\label{Prel.eq:volumeTransformation}
V = V^{0} \sqrt{1- (u/c)^{2}},  
\end{equation}
and throughout the boost, according to~\eqref{Prel.eq:stressTransUn},
the stresses transform through \(t_{xx} = t^{0}_{xx}.\)
Therefore the change in energy of the volume, which comes from both
the force accelerating the volume, and the work done by the stresses
to contract the volume in the \(x-\)direction is
\begin{equation}
  \label{Prel.eq:energy1}
  \deriv{E}{t} = F_{x}\deriv{x}{t} - t_{xx}\deriv{V}{t}.
\end{equation}
This can be expanded with the expression for the force
from~\eqref{Prel.eq:fTotal}, and by holding \(t_{xx}\)
constant as per the transformation law into
\begin{equation}
  \label{Prel.eq:energy2}
  \deriv{E}{t} = \deriv{E}{t} \left(\f{u}{c} \right)^{2} + E \f{u}{c^{2}} \deriv{u}{t} + t_{xx} \left(\f{u}{c} \right)^{2} \deriv{V}{t} + t_{xx} \f{u}{c^{2}} \deriv{u}{t} V - t_{xx}\deriv{V}{t}, 
\end{equation}
after which it can be factorised into
\[ \left( 1 - \f{u^{2}}{c^{2}} \right) \deriv{}{t} (E + t_{xx} V) = (E
+ t_{xx}V) \f{u}{c^{2}} \deriv{u}{t}.\]
To get the energy change in the moving frame, we have to integrate
from zero velocity at time \(t=0\)
to a velocity of \(u'= u,\)
at \(t = t',\) after rearrangement into logarithmic integrals:
\[ 
\f{1}{E + t_{xx}V} \deriv{}{t} (E + t_{xx} V) = -\f{1}{2} \f{-\f{2u'}{c^{2}} \deriv{u'}{t}} {\left(1 - \left( \f{u'}{c} \right)^{2} \right)},
\]
giving rise to
\[
\log{\left( E + t_{xx} V \right)}\Big|_{t=0, u'=0}^{t=t', u'=u} = -\f{1}{2} \int_{t=0, u'=0}^{t=t', u'=u} \f{-\f{2u'}{c^{2}} \deriv{u'}{t}} {\left(1 - \left( \f{u'}{c} \right)^{2} \right)} \d t = \log{\left[1 - \left( \f{u}{c} \right)^{2} \right]}^{-(1/2)},
\]
which is then simplified into the simple
\begin{equation}
  \label{Prel.eq:energyTransformation}
E + t_{xx} V = \f{E^{0} + t^{0}_{xx} V^{0}}{\sqrt{\left(1 -(u/c)^{2} \right)}}.
\end{equation}
The zero-superscripted variables being evaluated in the rest observer
frame prior to the boost.  With
equation~\eqref{Prel.eq:energyTransformation}, we can now deduce the
transformation of energy densities in a continuous fluid due to both
relativistic motion, and stresses.  To do so we first convert the
energy \(E\)
to an energy density by dividing by \(c^{2}V,\)
followed by a conversion of all the volumes \(V\)
into \(V^{0}\)
through~\eqref{Prel.eq:volumeTransformation}, such that
equation~\eqref{Prel.eq:energyTransformation} is converted into the
equivalent
\begin{equation}
  \label{Prel.eq:densityTransformation}
\rho = \f{\rho_{00} + t^{0}_{xx}u^{2}/c^{4}}{1 - (u/c)^{2}}.
\end{equation}
In the above~\eqref{Prel.eq:densityTransformation}, \(\rho_{00}\)
is the proper energy density in the cube at rest (i.e.\ in the
observer's frame, such that \(\rho_{00} = E^{0}/V^{0}\).)

We can now write down the momentum densities in the moving frame by
combining equations~\eqref{Prel.eq:densityTransformation},
\eqref{Prel.eq:stressTrans}, and~\eqref{Prel.eq:pDensity} to get
\begin{equation}
\label{Prel.eq:pDensityTransformed}
 \vec{g} =  \begin{pmatrix}
    g_{x}\\g_{y}\\g_{z}
  \end{pmatrix}
 =
 \begin{pmatrix}
   \f{c^2 \rho_{00} + t^0_{xx}}{1 - (u/c)^2} \f{u} {c^{2}} \\
   \f{t^0_{xy}}{\sqrt{1 - (u/c)^2}} \f{u}{c^{2}}\\
   \f{t^0_{xz}}{\sqrt{1 - (u/c)^2}} \f{u}{c^{2}}
 \end{pmatrix},
\end{equation}
This equation is important because it will allow us to compute the
mass density \(\rho\)
and momentum density \(g_{i},\)
at some point in a medium moving with a velocity \(u\)
in terms of this velocity, and the values of the densities
\(\rho_{00}\)
and stresses \(t^{0}_{ij}\)
measured by an observer moving with the fluid element, if the velocity
\(u\)
is oriented along the \(x-\)direction.
From the form of the equations, extension to other more general
directions immediately becomes possible, and can even be done by
inspection, but we will not give those expressions here.  Before
continuing onto the transformations and definitions of pressures in
this fluid, we remember that no quantum mechanical considerations has
gone into this derivation: only basic Newtonian thermodynamics and
relativity principles have been used.

\subsection{Pressures}
We mentioned before that surprisingly the relativistic stress matrix
is no longer symmetrical.  However it is a known feature of continuum
mechanics that the symmetry of the stress tensor helps in the
interpretation, and indeed with many calculations in Newtonian
mechanics.  We now look for a convenient relativistic framework that
will achieve the same end.  The formalism as presented so far is
complete inasmuch it allows the calculation of all stresses and
energies in frames moving with respect to each other.  To retrieve
symmetrical matrices and then tensors, we first define ``absolute
stresses,'' or pressures in this continuum.  We consider the array of
quantities given by
\begin{equation}
\label{Prel.eq:pDef}
p_{ij} = t_{ij} + g_{i}u_{j},
\end{equation}
where \(t_{ij}\)
are the stresses defined previously in~\eqref{Prel.eq:stress},
\(g_{i}\)
are the momentum densities defined in~\eqref{Prel.eq:pDensity}, and
\(u_{i}\)
are the velocity components of the continuum at that point.  If we
place ourselves in the observer frame moving with the fluid, according
to the symmetry of the stresses \(t_{ij}\)
in the rest frame, we automatically have
\[p^{0}_{ij} = p^{0}_{ji} = t^{0}_{ij} = t^{0}_{ji},\]
since the \(u_{i} = 0.\)
In the moving frame however, here again assumed to be moving in the
\(x-\)direction, we have the \(p_{ij}\) transform to 
\begin{equation}
\label{Prel.eq:stressTrans}
p_{ij} =
\begin{pmatrix}
  \f{p^{0}_{xx} + \rho_{00} u^{2}}{1-(u/c)^{2}} &\f{p^{0}_{xy}}{\sqrt{1-(u/c)^{2}}} &\f{p^{0}_{xz}}{\sqrt{1-(u/c)^{2}}}\\
  \f{p^{0}_{yx}}{\sqrt{1-(u/c)^{2}}} &p^{0}_{yy} &p^{0}_{yz}\\
  \f{p^{0}_{zx}}{\sqrt{1-(u/c)^{2}}} &p^{0}_{zy} &p^{0}_{zz}
\end{pmatrix},  
\end{equation}
as can be computed directly through equation~\eqref{Prel.eq:pDef}.  As
we can see this array \emph{is} symmetrical, as we set out to do.  If
we now go back to equations~\eqref{Prel.eq:densityTransformation}
and~\eqref{Prel.eq:pDensityTransformed}, we can re-express them in
terms of the pressures instead of the stresses to get,
\begin{equation}
  \label{Prel.eq:pDen,Den,WrtPressure}
\begin{pmatrix}
    g_{x}\\g_{y}\\g_{z}
  \end{pmatrix}
 =
 \begin{pmatrix}
   \f{c^2 \rho_{00} + p^0_{xx}}{1 - (u/c)^2} \f{u} {c^{2}} \\
   \f{p^0_{xy}}{\sqrt{1 - (u/c)^2}} \f{u}{c^{2}}\\
   \f{p^0_{xz}}{\sqrt{1 - (u/c)^2}} \f{u}{c^{2}}
 \end{pmatrix},
\qquad \text{and,} \qquad
 \rho = \f{\rho_{00} + p^{0}_{xx}u^{2}/c^{4}}{1 - (u/c)^{2}}.
\end{equation}
As a result of re-expressing everything in terms of the pressures
instead, the equation of motion~\eqref{eq:Prel.N2fin} can be put in a
similar form as the continuity equation~\eqref{eq:Prel.Density}, the simple
\begin{equation}
  \label{Prel.eq:3continuity}
  \pderiv{p_{ij}}{x_{j}} + \pderiv{g_{i}}{t} = 0.
\end{equation}
This simple form of both of these equations is the clue that leads to
the generalisation to 4-quantities required for a relativistic
treatment.  The only difference in the general case where the fluid is
not assumed to be moving only along the \(x-\)direction is that the
transformations seen above get more complicated, the general idea of
how these work having been made clear.

However before an extension to four dimensional
quantities, a note on terminology.  As mentioned previously, the
quantities \(t_{ij}\)
are called stresses, and corresponds to the forces one side of the
imaginary cube pictured previously exerts on another potion of the
fluid.  This aspect is sometimes used to call the \(t_{ij}\)
the ``relative stresses,'' particularly in continuum mechanics of
solids.  By contrast since the \(p_{ij}\)
take into account the total momentum densities at every point of the
fluid, independently of the surroundings, in one particular coordinate
system, they correspond to ``absolute stresses'' in the same
terminology.  Now we generalize these notions to four dimensional
space.

\subsection{Four dimensional quantities}
The relativistic treatment we started in the previous section suggests
that both the pressures \(p_{ij}\)
and the density \(\rho_{00},\)
by transforming according to the Lorentz transformations of spatial
and time like variables respectively can be combined in one
four-tensor. To do so we consider proper coordinates that move along
with the fluid as previously such that the fluid has zero spatial
velocity in these coordinates.  If we have
\((t,x,y,z) \equiv (x_{0}^{0}, x_{0}^{1},x_{0}^{2},x_{0}^{3}),\)
as coordinates, the spatial velocities are expressed as
\[ \deriv{x_{0}^{1}}{s} = \deriv{x_{0}^{2}}{s} = \deriv{x_{0}^{3}}{s}
= 0,\]
where \(s\)
is the proper distance along the fluid world line.  In such a system
we can \emph{define} an
\textbf{energy-momentum}\index{Tensor!Energy-Momentum} tensor through
the components of the pressures and the density such that this tensor
transforms according to the general diffeomorphic invariance we expect
of tensors,
\begin{equation}
  \label{Prel.eq:energy-momentum}
T_{0}^{ab} = \begin{pmatrix}
c^{2} \rho_{00} & 0 & 0 & 0\\
            0 & p^{0}_{xx} & p^{0}_{xy} & p^{0}_{xx}\\
            0 & p^{0}_{yx} & p^{0}_{yy} & p^{0}_{yz}\\
            0 & p^{0}_{zx} & p^{0}_{zy} & p^{0}_{zz}\\
\end{pmatrix}, \qquad \text{with,}\qquad T^{ab} = \pderiv{x^{a}}{x^{c}_{0}} \pderiv{x^{b}}{x^{d}_{0}} T^{cd}_{0},
\end{equation}
the second equation permitting the transformation of the quantities in
the observer frame to any other frame at any other point.  Since we
just stated a definition without really proving that the new tensor
\(T\)
really transforms like one, we can check that this is indeed the case.
To do so, we start with an observer with coordinates
\((x_{0}^{0}, x_{0}^{1},x_{0}^{2},x_{0}^{3}),\)
who perceives the fluid to be at rest, and hence has an energy
momentum tensor given by~\eqref{Prel.eq:energy-momentum} and transform
to a frame where the fluid is moving parallel to the \(x-\)axis
with velocity \(u,\)
as we did previously, but at the \emph{same point,} the coordinates
transform according the Lorentz transformation with respect to a
velocity of \(-u\) through
\begin{equation}
  \begin{pmatrix}
    x^{0} \\ x^{1} \\ x^{2} \\ x^{3}
  \end{pmatrix}
  = 
\begin{pmatrix}
    \f{x^{0}_{0} + u x_{0}^{1}/c}{\sqrt{1 - (u/c)^{2}}} \\ \f{x^{1}_{0} + u x_{0}^{0}/c}{\sqrt{1 - (u/c)^{2}}} \\ x^{2}_{0} \\ x^{3}_{0}
  \end{pmatrix}.
\end{equation} 
This then allows the computation of the derivatives that we had in the
second equation of~\eqref{Prel.eq:energy-momentum}, the symmetric matrix:
\begin{equation}
  \label{Prel.eq:TransDerivs}
\pderiv{x^{a}}{x^{c}_{0}} =
\begin{pmatrix}
  \f{1}{\sqrt{1-(u/c)^{2}}}  & \f{u/c}{\sqrt{1-(u/c)^{2}}} & 0 & 0 \\
  \f{u/c}{\sqrt{1-(u/c)^{2}}}& \f{1}{\sqrt{1-(u/c)^{2}}}   & 0 &0 \\
  0                         & 0                          & 1 & 0 \\
  0                         & 0                          & 0 & 1
\end{pmatrix}
\end{equation}
which in tern allows the computation of the energy-momentum tensor in
the moving frame with equation~\eqref{Prel.eq:energy-momentum}, since
we now possess all the pieces of the latter.  The computation gives
\begin{equation}
  \label{Prel.eq:TransEnergy-momentum}
T^{ab} =
\begin{pmatrix}
\f{c^{2}\rho_{00} + p^{0}_{xx}u^{2}/c^{2}}{1 - (u/c)^{2}} & \f{c^2 \rho_{00} + p^0_{xx}}{1 - (u/c)^2} \f{u}{c} & \f{p^0_{xy}}{\sqrt{1 - (u/c)^2}} \f{u}{c} & \f{p^0_{xz}}{\sqrt{1 - (u/c)^2}} \f{u}{c}   \\
\f{c^2 \rho_{00} + p^0_{xx}}{1 - (u/c)^2} \f{u}{c}     & \f{p^{0}_{xx} + \rho_{00} u^{2}}{1-(u/c)^{2}}       & \f{p^{0}_{xy}}{\sqrt{1-(u/c)^{2}}}          &\f{p^{0}_{xz}}{\sqrt{1-(u/c)^{2}}} \\
\f{p^0_{xy}}{\sqrt{1 - (u/c)^2}} \f{u}{c}            & \f{p^{0}_{yx}}{\sqrt{1-(u/c)^{2}}}                &p^{0}_{yy}                                  &p^{0}_{yz}\\
\f{p^0_{xz}}{\sqrt{1 - (u/c)^2}} \f{u}{c}            & \f{p^{0}_{zx}}{\sqrt{1-(u/c)^{2}}}                &p^{0}_{zy}                                  &p^{0}_{zz}
\end{pmatrix},
\end{equation}
directly from the tensor equations, without any notions of momentum
gain or loss through the cubes faces being evident.  Upon comparing
with the actual momentum flow equations~\eqref{Prel.eq:stressTrans},
and~\eqref{Prel.eq:pDen,Den,WrtPressure}, we can easily see by
inspection that the above equation can be expressed very simply in the
form
\begin{equation}
  \label{Prel.eq:Energy-Momentum}
T^{ab} =
\begin{pmatrix}
  c^{2}\rho & cg_{x} & cg_{y} & cg_{z} \\
  cg_{x} & p_{xx}& p_{xy} & p_{xy} \\
cg_{y} & p_{yx}& p_{yy} & p_{yz} \\
cg_{z} & p_{zx}& p_{zy} & p_{zz} \\
\end{pmatrix},
\end{equation}
just as we wanted the forces, densities and momenta to transform in
any frame: this \emph{proves} that indeed the object \(T^{ab}\)
is a well defined tensor, and we have also shown that it summarizes
all the mechanical fluid properties of interest in any Lorentzian
frame within the scope of special relativity.  This will serve as a
starting point for the general relativistic notion of a stress-energy,
or energy-momentum tensor.

As a summary, the components of the energy-momentum tensor can be
interpreted as
\begin{enumerate}
\item \(T^{00}\) is the energy density of the fluid.
\item \(T^{0 \alpha}\) is the energy flux across the surface of the cube associated with the observer.
\item \(T^{\alpha 0}\)
  is the momentum density across the surface of the same cube.  In all
  our considerations, and through construction \(T^{0\alpha} = T^{\alpha 0}.\)
\item \(T^{\alpha \beta}\) are the pressures, or stresses compensated by momentum flux.
\end{enumerate}
All of the above only holds in the rest frame where the observer moves
with the fluid. In moving frames, these quantities get
Lorentz-transformed and their straight forward interpretation become
problematic, in the same way as the concepts of space and time
separately become problematic.

Before we extend this formalism to general relativity however, we
should conveniently notice that the conservation
equations~\eqref{Prel.eq:3continuity}, and~\eqref{eq:Prel.Density} can
be expressed in the short form of
\begin{equation}
  \label{Prel.eq:continuity}
  \pderiv{}{x^{a}} T^{ab} = 0 \implies T^{ab}{}_{,a} = 0,
\end{equation}
an equation valid only in Lorentzian frames.  As a result it is a
tensorial equation that was in terms of partial derivatives only
because of the procedure of our derivation in the special relativistic
case.

However, through the correspondence referred to previously, partial
derivatives being ``promoted'' to covariant ones in general
relativity, where the coordinate transformations are more general, we
would obtain the equivalent general relativistic postulate
encapsulating all of the matter equations through
\begin{equation}
  \label{Prel.eq:continuityGR}
  T^{ab}{}_{,a} \longrightarrow T^{ab}{}_{;a} = \pderiv{}{x^{a}} T^{ab} + \Gamma^{a}_{ad}T^{db} + \Gamma^{b}_{ad}T^{ad},
\end{equation}
this being the beginning of the generalization to general relativity of
the continuum.

\subsection{The energy momentum tensor for specific fluids}
We now look at different types of fluids that will be used in this
thesis.  We only look at simplified fluids, mostly because of the
symmetry requirements we have.  More general fluids can be reduced to
these simple cases in the symmetries we consider, and the proof of
this follows the same proof that allows the reduction of the
components of the metric tensor given in the Section~

\subsubsection{Dust}
\textbf{Dust}\index{Dust} is also called \textbf{incoherent matter}
because neighbouring fluid elements (components of the dust) do not
exert any force whatsoever on each other.  As a result, all the dust
elements exert no stress, or pressure on each other, so that they can
be characterised by the energy density of the fluid only.  The
energy-momentum tensor can then be expressed in the very simple
coordinate independent, and coordinate dependant forms respectively:
\begin{equation}
  \label{Prel.eq:Dust}
T = \rho \vec{u} \otimes \vec{u} \implies T^{ab} = \rho u^{a}u^{b},
\end{equation}
where \(u^{a} = \deriv{x^{a}}{s},\) the proper velocity of the fluid.

\subsubsection{Perfect fluids}
A \textbf{perfect fluid}\index{Perfect Fluid} is a mechanically
continuous medium that is incapable of exerting any transverse
stresses on other fluid elements.  This is equivalent to demanding
that the fluid admits no viscosity, since viscous forces are always
tangential to the surfaces of the cubes in the fluid element picture
we had previously.  Therefore non-diagonal elements of the stress
energy array have to vanish.  Furthermore, the no viscosity criteria
being global to the fluid, this has to hold at every single point.  It
is easy to see that the only array that admits this even through
coordinate transformations is one that is diagonal in all frames, and
linear algebra ensures that only a tensor that is a multiple of the
identity array holds this property~\cite{sch09}.

Therefore in the MCFR, together with having no transverse momenta, no
non-diagonal pressures are present either, so that
\begin{equation}
  \label{Prel.eq:PerfectFluid}
T_{0}^{ab} = 
\begin{pmatrix}
  \rho_{00} & 0 & 0 & 0\\
  0 & p_{0} & 0 & 0 \\
  0 &  0 &p_{0} & 0 \\
  0&0&0&p_{0}
\end{pmatrix}.
\end{equation}

\subsubsection{Non-perfect fluids}
\textbf{Non-perfect fluids}\index{Non-perfect fluid} are the most
general case of continuous media one can look into, however this is
only possible in reduced symmetry systems.  Many classes of these
fluids exist: fluids admitting heat fluxes and heat conduction would
have non-zero momenta components in the MCFR:
\(T_{0}^{0\alpha} =T_{0}^{\alpha 0} \neq 0,\)
for example.  Other examples like ones admitting \textbf{anisotropic pressures}
will be of prime interest to us, and can be given simply in the MCFR as
\begin{equation}
  \label{Prel.eq:AnisotropicFluid}
T_{0}^{ab} = 
\begin{pmatrix}
  \rho_{00} & 0 & 0 & 0\\
  0 & p_r & 0 & 0 \\
  0 &  0 &p_t & 0 \\
  0&0&0&p_t
\end{pmatrix},
\end{equation}
where the only difference from a perfect fluid is the unequal
pressures in the diagonal entries.  How this particular characteristic
comes about is complicated, but we will investigate them and their
structure only after introducing a general relativistic formulation.

\subsection{The matter continuum in general relativity}
The previous section's analysis of the continuum can be extended to
the general relativistic cases through the canonical ``promotion'' of
the derivative operators.  However many interpretation of the tensor
components can be carried out through the notions given previously in
the special relativistic case.  Notions like angular momentum: which
we shall not be concerned with, and mass: which we will have to
consider, being a few examples.  However since we already have a local
conservation equation of energy density and of momentum through the
energy momentum tensor in the MCFR, we shall continue and generalize
all these concepts to the general relativistic case.

In addition to the promotion of the derivative operators, the other
canonical transformation that can be carried out is the promotion of
the Lorentzian metric \(\eta^{ab}\)
to the general metric \(g^{ab}.\)

We start with the a special relativistic perfect fluid and investigate
how its general expression changes in moving frames.  Since the MCFR
is a Lorentzian frame, locally the metric in this frame is
\(\eta^{ab},\)
and the energy-momentum tensor for a perfect fluid is still
\(T_{0}^{ab}\)
given by equation~\eqref{Prel.eq:PerfectFluid}, while the velocity of
the fluid in the MCFR is \(u^{a}_{0} = (1,0,0,0)^{\top}.\)

To get the general form of the energy-momentum in arbitrary frames, we
proceed by the general tensor transformation rule,
\[T^{ab} = \pderiv{x^{a}}{x^{c}_{0}} \pderiv{x^{b}}{x^{d}_{0}}
T^{cd}_{0},\]
which simplifies due to \(T^{ab}_{0}\) being diagonal into
\begin{equation}
\label{Prel.eq:EMperf}
T^{ab} = \pderiv{x^{a}}{x_{0}^{0}} \pderiv{x^{b}}{x_{0}^{0}}
\rho_{00} + \pderiv{x^{a}}{x_{0}^{1}} \pderiv{x^{b}}{x_{0}^{1}} p_{0}
+ \pderiv{x^{a}}{x_{0}^{2}} \pderiv{x^{b}}{x_{0}^{2}} p_{0}
+\pderiv{x^{a}}{x_{0}^{3}} \pderiv{x^{b}}{x_{0}^{3}} p_{0}.  
\end{equation}
The velocity vector transforms according to
\[u^{a} = \pderiv{x^{a}}{x_{0}^{c}} u^{c}_{0} \implies \deriv{x^{a}}{s}  = \pderiv{x^{a}}{x_{0}^{0}},\]
while the metric in a general frame is given by
\[g^{ab} = \pderiv{x^{a}}{x_{0}^{c}} \pderiv{x^{b}}{x_{0}^{d}}
\eta^{cd} =
\begin{pmatrix}
\pderiv{x^{a}}{x_{0}^{0}} \pderiv{x^{b}}{x_{0}^{0}} & 0 & 0 & 0\\
0 & -\pderiv{x^{a}}{x_{0}^{1}} \pderiv{x^{b}}{x_{0}^{1}} & 0 & 0 \\
0 & 0 & -\pderiv{x^{a}}{x_{0}^{2}} \pderiv{x^{b}}{x_{0}^{2}} &0 \\
0& 0& 0&-\pderiv{x^{a}}{x_{0}^{3}} \pderiv{x^{b}}{x_{0}^{3}}
\end{pmatrix}
\]
These two relations when substituted in~\eqref{Prel.eq:EMperf} yield
the coordinate invariant general energy momentum tensor in arbitrary frames:
\begin{equation}
  \label{Prel.eq:PerfectFluidCordInv}
T^{ab} = (\rho_{00} + p_{0}) u^{a} u^{b} - g^{ab} p_{0} \implies T =  (\rho + p) \vec{u} \otimes \vec{u} - p g^{-1}  ,
\end{equation}
in terms of the fluid velocity in those frames, and the metric at
those point.  This will be the stating point for most of the work in
this thesis.

\subsubsection{Non-perfect fluids}
We will also be concerned with non-perfect fluids in this thesis.  The
reasons leading to their consideration are many, but stem from the
fact that assuming local pressure isotropy in the energy-momentum tensor
is an oversimplification due to a perfect fluid assumption, and does
not follow from spherical symmetry.  Local anisotropy is more
interesting since~\cite{HerSan97}
  \begin{enumerate*}[label=(\roman{*}), ref=(\roman{*})]
  \item Anisotropic models can naturally incorporate charged
    distributions, with the anisotropy proportional to some static
    charge distribution~\cite{BohMus10}.
  \item Any solution requiring more than one perfect fluid matter
    component with minimal interaction has to be modelled with an
    anisotropic stress tensor~\cite{Let80, Bay82, Bay84}.  Anisotropy would
    allow the possibility of different species interacting with each
    other, instead of just one homogeneous fluid species.  Cases
    like quark stars and neutron stars above would require this for
    example, since these require multiple types of
    particles for any form of quantum stability.
  \item An isotropic energy-momentum tensor means that all interaction
    of the fluid with itself has to be modelled with one barotropic
    equation of state, \(p(\rho)\).  This might not provide enough
    degrees of freedom if complex interactions is to take place in the
    fluid.
\end{enumerate*}
All of these reasons suggest that more complicated fluid profiles that
perfect fluids with one barotropic equation of state would be useful
in modelling actual physical objects.  Such generalizations could
occur in many directions: Some include conductivity terms in the
energy momentum tensor in an attempt to model conductive
fluids. Others embed the interior solutions in an external magnetic
field, and generate a different energy momentum tensor through this
external field.  

We will instead follow~\citeauthor{Let80} and more
recently~\citeauthor{BooNgaVis15} who couple multiple fluids minimally
to generate anisotropy.  \citeauthor{Let80}'s method is more involved
mathematically, but is more transparent.  Starting with two perfect
fluids labelled with their proper future oriented velocities \(u,\)
and \(v,\)
prescribed by our coordinate independent model of
fluids~\eqref{Prel.eq:PerfectFluidCordInv}, we use the additive
property\footnote{The total energy momentum tensor of multiple fluids
  is just the addition of the individual energy momentum tensors of
  the separate fluids.  This simple but counter-intuitive canonical
  relation comes from the alternative definition of the tensor in the
  Lagrangian formulation of general relativity, where the energy
  momentum tensor behaves very much like a Lagrange density, and is
  thus additive. This is investigated in Section~\ref{A.sec:Lagrangian}}
\begin{equation}
  \label{Prel.eq:addEMTensor}
T^{ab}(u+v) = T^{ab} (u) + T^{ab} (v),
\end{equation}
of the energy momentum tensor to build the total
energy momentum tensor of the system:
\begin{subequations}
\label{Prel.eq:perf1+2}
\begin{equation}
\label{Prel.eq:perf1}  
T^{ab}(u) = (\rho + p) u^{a} u^{b} - p g^{ab},
\end{equation}
\begin{equation}
  \label{Prel.eq:perf2}
T^{ab}(v) = (\tau + q) v^{a} v^{b} - q g^{ab},
\end{equation}
\end{subequations}
where \(\rho\) and \(\tau\) are the rest energy densities, while \(p\)
and \(q\) are the pressures of the fluids \(u\) and \(v\)
respectively. Therefore \(T^{ab} (u+v)\) is the energy momentum tensor
of the combination of the two fluids.  Additionally we have the usual
normalization conditions of the fluid velocities \(u_{a}u^{a} = 1,\)
and \(v^{a}v_{a}=1,\) with \(v_{a} \neq u_{a}.\)

The consistency conditions for energy momentum
tensors~\eqref{Prel.eq:addEMTensor} is
\begin{equation}
\label{Prel.eq:Conservation}
G^{ab}{}_{;a} = 0 \implies T^{ab}{}_{;a} (u+v) = 0,  
\end{equation}
for the final energy momentum \(u+v.\) These however are not enough to
completely determine all the unknowns of the
system~\eqref{Prel.eq:addEMTensor}, and we need additional conditions
to close the system.  To see why this is, we consider the fact
that~\eqref{Prel.eq:addEMTensor} has as unknowns \(g^{ab}\) with 10
independent components, \(u^{a}\) and \(v^{a}\) each having 3
independent components (since the normalisation conditions reduce the
4 components by one each,) and 4 from the matter variables
\(\rho, \tau, p\) and \(q,\) for a total of 20 unknowns.  As
constraints, we have the Einstein equations~\eqref{pr.eq:EFE} which
provide a total of 10 constraint equations, and the Bianchi
identities~\eqref{Prel.eq:Conservation} which provide 4.  To close the
system we need to provide 6 more equations.
\begin{enumerate*}[label=(\roman{*}), ref=(\roman{*})]
\item First we
assume a form of minimal coupling for the individual energy momentum
equations so that
\end{enumerate*}
\begin{equation}
  \label{Prel.eq:consIndEM}
T^{ab}{}_{;a}(u) = 0, \qquad\text{and,}\qquad T^{ab}{}_{;a}(v) = 0.
\end{equation}
Since we already have constraints on the metric, these two equations
only provide constraints on the matter variables individually, for 4
equations in total from~\eqref{Prel.eq:consIndEM}.  The form of the
above equations also ensures that~\eqref{Prel.eq:Conservation} is
automatically satisfied.
\begin{enumerate*}[label=(\roman{*}), ref=(\roman{*})]
\setcounter{enumi}{1}
\item Additionally we will also assume that there exist an equation of
  state relating the state variables of each fluids in the form of
\end{enumerate*}
\begin{equation}
  \label{Prel.eq:EOSInd}
h_{1}(p, \rho) = 0 , \qquad\text{and,}\qquad h_{2}(q, \tau) = 0
\end{equation}
for another two constraints, closing the system: so that we have 20
unknowns, and 20 constraints.  With these two assumptions: minimal
coupling, and the existence of an equation of state, we now try to
find a simple form of the energy momentum for a combination of two
perfect fluids.  What we want is for \(T^{ab}(u+v)\) to be expressed as
\begin{equation}
  \label{Prel.eq:Tcombined}
  T^{ab}(u+v) = D U^{a} U^{b} + Q^{ab},
\end{equation}
where \(U^{a}\)
is normalised such that \(U^{a}U_{a} = 1,\)
like the previous 4-velocities, in some coordinate system, and
\(Q^{ab}\)
is a stress tensor as we defined previously as \(p^{ij}\)
in~\eqref{Prel.eq:pDef}.  It obeys the normalisation condition
\(Q^{ab}U_{a} = 0,\)
as expected from the stress tensor.  We also want \(D >0,\)
since we want it to correspond to a rest energy density.

Considering~\eqref{Prel.eq:addEMTensor}, and
using~\eqref{Prel.eq:perf1} and~\eqref{Prel.eq:perf2} to expand it, we get
\[
T^{ab}(u+v) = (\rho+p)u^{a}u^{b} + (\tau + q) v^{a}v^{b} -(p+q)g^{ab}.
\]
Now, following the insight of~\citeauthor{Let80}, and using the
diffeomorphic invariance of general relativity, we transform to a
different coordinate system, in such a way that the four velocities
transform as 
\begin{subequations}
  \begin{equation}
    \label{Prel.eq:uTrans}
    u^{a} \to \tilde{u}^{a} = u^a \cos{\alpha} + v^a \sqrt{\left(\f{\tau + q}{\rho+p}\right)} \sin{\alpha},
  \end{equation}
  \begin{equation}
    \label{Prel.eq:vTrans}
    v^{a} \to \tilde{v}^{a} = -u^{a} \sqrt{\left( \f{\rho + p}{\tau+q}\right)} \sin{\alpha} + v^{a} \cos{\alpha}, 
  \end{equation}
\label{Prel.eq:cTrans}
\end{subequations}
where \(\alpha\)
is undetermined as of yet.  By substituting the expressions of the
transformed velocities in the energy-momentum tensor, we find that the
rotation indeed preserves the tensor, that is,
\begin{equation}
  \label{Prel.eq:coodIndep}
T^{ab}(u,v) = T^{ab}(\tilde{u},\tilde{v})
\end{equation}

We now pick the new coordinate system so that it allows us to easily
interpret the new velocities.  We choose \(\tilde{u}^{a}\) to be
a timelike vector\index{Vector!Timelike}, and \(\tilde{v}^{a}\) to be
a spacelike vector\index{Vector!Spacelike} respectively.  As a result
\(\tilde{u}\) and \(\tilde{v}\) are orthogonal, so that
\(\tilde{u}^{a} \tilde{v}_{a} = 0.\) Demanding that this condition be
satisfied uniquely determines the ``angle'' \(\alpha\) in the
coordinate transformation~\eqref{Prel.eq:cTrans} to be
\[\tan{(2\alpha)} = 2 u^{a}v_{a} \f{\sqrt{(\rho + p)(\tau + q)}}{\rho + p-\tau - q }.\]
In addition to the spacelike and timelike nature of the new
velocities, the fact that all 4-velocities must be future oriented
implies additionally that \(\tilde{u}^{a}\tilde{u}_{a} > 0\)
and \(\tilde{v}^{a}\tilde{v}_{a} < 0\)
respectively.  With this transformation completely specified, we can
now express the tensors in~\eqref{Prel.eq:Tcombined} in terms of the
rotated vectors.  A straight forward calculations then gives that
\begin{subequations}
  \begin{align}
    \label{Prel.eq:TcombU} U^{a} &\coloneqq \f{\tilde{u}^{a}}{\sqrt{\tilde{u}^{a} \tilde{u}_{a}}}, \\
    \label{Prel.eq:TcombV} V^{a} &\coloneqq \f{\tilde{v}^{a}}{\sqrt{-\tilde{v}^{a} \tilde{v}_{a}}}, \\
    \label{Prel.eq:TcombD} D    &\coloneqq T^{ab}U_{a}U_{b} = (\rho + p) \tilde{u}^{a} \tilde{u}_{a} - (p+q), \\
    \label{Prel.eq:TcombS} S    &\coloneqq T^{ab}V_{a}V_{b} = (p + q) - (\tau + q) \tilde{v}^{a} \tilde{v}_{a},\\
    \label{Prel.eq:TcombT} T    &\coloneqq p+q,
  \end{align}
\label{Prel.eq:TcombVariables}
\end{subequations}
To understand what \(D\)
and \(S\)
mean, consider that by expanding~\eqref{Prel.eq:TcombD}
and~\eqref{Prel.eq:TcombS} with the expressions of \(\tilde{u},\)
and \(\tilde{v},\) we explicitly get
\begin{subequations}
\label{Prel.eq:AnisPs}
  \begin{align}
  D    &= \f{1}{2}(\rho - p + \tau - q) + \f{1}{2} \sqrt{\left\{ (p + \rho + \tau + q)^{2} + 4(p+\rho)(q+\tau)\left[(u^{a}v_{a})^{2} - 1 \right] \right\} } ,\\  
  S    &= -\f{1}{2}(\rho - p + \tau - q) + \f{1}{2} \sqrt{\left[ (p + \rho - \tau - q)^{2} + 4(p+\rho)(q+\tau)(u_{a}v^{a})^{2} \right] }
  \end{align}
\end{subequations}
which we notice to be both positive since the terms in the square root
is always larger than the first term.  Therefore the interpretation of
\(D\)
and \(S\) as a density and a stress respectively is not far-fetched. 
With this is mind, we re-express the energy momentum tensor given in~\eqref{Prel.eq:coodIndep} through
\begin{equation}
\label{Prel.eq:Qdef}
  T^{ab} = (D+A)U^{a}U^{b} + (S-A)V^{a}V^{b} -Ag^{ab} \eqqcolon D U^{a}U^{b} + Q^{ab}
\end{equation}
in the rotated velocities, and extract the quantity \(Q^{ab}\) given by
\begin{equation}
\label{Prel.eq:Qexplicit}
  Q^{ab} = S V^{a}V^{b} + A \left( U^{a}U^{b} - V^{a}V^{b} - g^{ab} \right).
\end{equation}
Another rotation of the coordinate system to the canonical tangent
space of the metric, where \(g^{ab} = \eta^{ab},\) the Minkowski
metric, which is always possible at a point according to
theorem~\eqref{A.th:LocFlat}, then
diagonalizes~\eqref{Prel.eq:Qexplicit} into
\begin{equation}
\label{Prel.eq:Qdiag}
Q^{ab} =
\begin{pmatrix}
  0 & 0 & 0 & 0\\
  0 & S & 0 & 0\\
  0 & 0 & A & 0\\
  0 & 0 & 0 & A
\end{pmatrix},
\end{equation}
since in the new coordinates, we have \(U^{a} = \delta^{a}_{0},\)
and \(V^{a} = \delta^{a}_{1}.\)
This is the result we are after, since in this rotated system, we have
just expressed the mixture of two ideal fluid into an anisotropic
energy-momentum tensor, by just exploiting the coordinate
transformations allowed by general relativity. Thus \[T^{ab}(u+v) = \begin{pmatrix}
  D & 0 & 0 & 0\\
  0 & S & 0 & 0\\
  0 & 0 & A & 0\\
  0 & 0 & 0 & A
\end{pmatrix},\]
with \(D > 0,\)
the rest energy density, \(S = p_{r} > 0\)
the pressure in the \(\delta^{a}_{1}\)
direction: corresponding to the radial direction in spherical
coordinates, and another pressure \(A = \ppen < S,\)
from~\eqref{Prel.eq:AnisPs}. This last pressure has to be
perpendicular to \(\delta^{a}_{1}\)
direction and we identify it with the other non-radial direction in
spherical symmetry.  We also note that since \(u\)
and \(v\)
both have to be zero at \(r=0,\)
the centre by symmetry, \(A = S\)
there too, in accordance to anisotropic but spherically symmetric
pressures.

A similar derivation, which we will not give since it is simpler and
follows this exact same procedure for the combination of a perfect
fluid (interacting matter) and a null fluid (dust).  This alternative
derivation produces different expressions for \(D, S\)
and \(A,\)
but the diagonalization proceeds through exactly in the same way, and
we can say that any combination of more that one fluid, perfect or
null can be transformed into \emph{one anisotropic} fluid with
calculable properties.

This ends this section on fluids, and we have justified the physical
relevance of anisotropic fluids for the modelling of physical stars,
if the latter can be thought as to be a combination of perfect fluid
species interacting minimally with each other.

\section{Electromagnetic fields}
Electromagnetic fields can be added to general relativity in a similar
manner as matter was.  Matter was introduced in section~\ref{S:Matter}
through an energy-momentum tensor, \(T_{ab}.\) In classical Maxwell
theory, an energy-momentum tensor can be constructed in terms of the
known electric and magnetic fields.  The simplest way to achieve this
is to use the potential formulation of the electromagnetic field.  In
this formulation the electromagnetic field is encoded in a 4-vector
\(A,\) whose components are given as \(A_{a} = (\phi/c, -\vec{A}),\)
where \(\phi\) is the electric potential and \(\vec{A}\) the magnetic
vector potential. From this 4-potential, we can then define the
\textbf{Faraday tensor}\index{Faraday Tensor}\index{Tensor!Faraday}
\begin{equation}
  \label{A.eq:Faraday}
  F_{ab} = \partial_{[a}A_{b]} = \partial_{a}A_{b} - \partial_{b}A_{a}.
\end{equation}

Then the addition of Maxwell's equations to those of the Einstein's
field equations (EFE) can be done canonically through a transformation
involving the substitution of the flat Lorentz metric \(\eta_{ab},\)
which is a natural component of Maxwell's theory into a general metric
tensor \(g_{ab}.\) Following this substitution, the Maxwell's
equations which in standard tensorial notation in flat space-time can
be expressed as
\begin{subequations}
  \begin{equation}
    \label{pr.eq:FlatMaxwell1}
    \partial_{[a}F_{bc]} = 0,    
  \end{equation}
  \begin{equation}
    \label{pr.eq:FlatMaxwell2}
    \partial_{a} \left( \eta^{ac}\eta^{bd}F_{cd} \right)  = \partial_{a} F^{ab} = \mu_{0}j^{b},
  \end{equation}
\end{subequations}
get transformed into the Maxwell's equation for curved space
\begin{subequations}
\begin{equation}
  \label{A.eq:Maxwell1}
  \nabla_{b} F^{ab} = \f{1}{\sqrt{-g}} \partial_{a} \left( \sqrt{-g} F^{ab} \right) = j^{a}, \qquad \text{subject to} \qquad \nabla_{a}j^{a} = 0,
\end{equation}
\begin{equation}
  \label{A.eq:Maxwell2}
  \partial_{[a}F_{bc]} = \partial_{a}F_{bc} + \partial_{b}F_{ca} + \partial_{c}F_{ab}  = \nabla_{[a}F_{bc]} = 0,
\end{equation}
\end{subequations}
where \(j^{b}\) is the current density 4-vector.  We see that the
partial derivatives are canonically promoted to covariant derivatives
one as is discussed in Section~\ref{A.ssec:CovD}.

The coupling of matter with the electromagnetic field is most easily
done in a Lagrangian formulation, and we investigate this in
Section~\ref{A.sec:Lagrangian}.

\section{Energy conditions and conservation laws}\label{A.sec:Energy}
\index{Energy Conditions} We now have all the structure and matter
variables to talk about energy, and conservation laws, and how these
are implemented.  In general relativity total energy is not well
defined in a coordinate independent manner.  For this reason, the
energy conservation conditions are expressed through the
energy-momentum tensor and timelike vectors.  We state the three main
energy conditions usually used in the literature, and in this thesis:
\begin{enumerate}
\item The \textbf{weak energy condition}\index{Energy Conditions!Weak}
  is satisfied if the energy-momentum tensor \(T_{ab}\) satisfies
\[T_{ab}X^{a}X^{b} \geq 0, \]
for all timelike vectors \(X.\)
\item The \textbf{strong energy condition}\index{Energy
    Conditions!Strong} is satisfied if instead
  \[\left(T_{ab} - \f{1}{2} g_{ab} T^{c}{}_{c}\right) X^{a} X^{b} \geq 0.\]
  This also called the Ricci positivity condition, because the tensor
  that is contracted with the timelike vectors is the Ricci tensor if
  Einstein equations hold.
\item The \textbf{dominant energy condition}\index{Energy
    Conditions!Dominant} is satisfied if the energy-momentum tensor is
  such that the vector \(-T^{a}{}_{b} X^{b}\) is timelike and future
  directed for all timelike and future directed vectors \(X.\)
\end{enumerate}

We now state how conservation laws are expressed in curved space-time,
and provide a few formulae which are useful for their computations.
As stated in Section~\ref{A.ssec:CovD} we use covariant derivatives in
GR where we would have used partial derivatives in SR.

A scalar \(\phi\) is locally conserved if
\(\nabla_{a} \phi = \phi_{;a}= 0.\) Since covariant derivatives of
scalars are defined to be the same as partial derivatives, the latter
equation reduces to \(\partial_{a} \phi = \phi_{,a}= 0.\)

A vector \(X^{a}\) is locally conserved if
\begin{equation}
  \label{A.eq:VecCons}
\nabla_{a} X^{a} = X^{a}{}_{;a} = X^{a}{}_{,a} +\Gamma^{a}_{ab}X^{b}= 0. 
\end{equation}
This is also called the covariant divergence of the vector.

A tensor \(T^{ab}\) is locally conserved if
\(\nabla_{a}T^{ab} = T^{ab}{}_{,a} + \Gamma^{a}_{ac} T^{cb} +
\Gamma^{a}_{ac} T^{bc} = 0.\)

A common theme in the above equations is the appearance of the
contracted connection coefficients \(\Gamma^{a}_{ab}.\) A convenient
method to compute these is now given. From the metric compatibility of
\(\Gamma^{a}_{bc},\) we can use the definition of the cnnection in
equation~\eqref{A.eq:metricConn}.  Contracting this equation, we have
\begin{align}
 \nonumber \Gamma^{a}_{ab} = \Gamma^{a}_{ba} &= \f{1}{2}g^{am}\pderiv{g_{am}}{x^{b}} \\
 \nonumber                                   &= \f{1}{2g} \pderiv{g}{x^{b}} \\
                                             &= \pderiv{\log{\sqrt{-g}}}{x^{k}} 
\label{A.eq:contractedGamma}
\end{align}
The second equation is due to the symmetry of the connection (there is
no torsion in GR), the third from a simple contraction of
equation~\eqref{A.eq:metricConn}. The next equality results from the
definition of the metric determinant \(g = \det{(g)}.\) From this
definition, we can easily calculate
\(\partial_{c} g = g g^{ab} \partial_{c} g_{ab},\) from which the
fourth equation results.  The final equation, the one we use in
chapter~\ref{C:Stability} can be seen to be true from a simple
differentiation identity.  In this form, equation
~\eqref{A.eq:contractedGamma} is very useful for the computation of
conserved quantities and divergences.

\section{Lagrangian approach}\label{A.sec:Lagrangian}
Another way to approach general relativity is through a Lagrangian
approach, and we introduce this here because it makes transparent how
different fields, in particular the electromagnetic field can be
included into the EFE without a complicated process.

The \textbf{action}\index{Action} of a gravitational field is
geometrical in nature and is encoded in the \textbf{Einstein
  Lagrangian}\index{Lagrangian!Einstein} given by
\(\La_{G} = \sqrt{-g}R,\) where \(R\) is the Ricci curvature scalar
defined in equation~\eqref{A.eq:RicciS}.  Since we need both the
metric and the Ricci scalar, all the notions of manifolds,
connections, and covariant derivatives are also needed.  The Einstein
Lagrangian is a functional dependant on the metric and its first and
second derivatives, since \(R\) which is defined in terms of the
Riemann tensor which in turn depends on the second derivatives of
\(g_{ab}.\) As a result of this dependence, the Euler-Lagrange
equations of the action
\begin{equation}
  \label{A.eq:EinAct}
  I = \int_{\Omega} \La_{G}(g_{ab},g_{ab,c},g_{ab,cd}) \d \Omega,
\end{equation}
with respect to the metric is
\begin{equation}
  \label{A.eq:E-LVac}
  \f{\delta \La_{G}}{\delta g_{ab}} = \pderiv{\La_{G}}{g_{ab}} 
  - \left( \pderiv{\La_{g}}{g_{ab,c}}\right)_{,c} 
  + \left( \pderiv{\La_{G}}{g_{ab,cd}}\right)_{,cd} = 0.
\end{equation}
The calculation of all the terms in equation~\eqref{A.eq:E-LVac} is
very lengthy and cumbersome, and we only state that the final answer
is indeed the Einstein tensor density
\begin{equation}
  \label{A.eq:VarEin}
  \La^{ab}_{G} = \f{\delta \La_{G}}{\delta g_{ab}} = -\sqrt{-g} G^{ab}, 
\end{equation}
with \(G^{ab}\) the Einstein tensor defined in equation~\eqref{A.eq:EinT}

In the presence of matter and electromagnetic fields, since we have a
Lagrangian theory now, we only need modify the Lagrangian to include
the different fields, with some coupling.  For matter we include a
matter Lagrangian of the form \(\La_{M},\) which couples to the
gravitational Lagrangian through the coupling constant \(\kappa\)
manifestly in the action:
\begin{equation}
  \label{A.eq:EinActM}
  I = \int_{\Omega} (\La_{G} + \kappa \La_{m})\d \Omega.
\end{equation}
The variation of each Lagrangian then gives the full Einstein
equations if we define the energy-momentum tensor to be
\begin{equation}
  \label{A.eq:LagragianEM}
  \f{\delta \La_{M}}{\delta g_{ab}} = \sqrt{-g}T^{ab},
\end{equation}
since from~\eqref{A.eq:VarEin}, we get the corresponding geometrical
part.  Together the variation of the whole action\eqref{A.eq:EinActM}
gives the full Einstein equations \(G^{ab} = \kappa T^{ab}.\) We
already derived the full matter \(T^{ab}\) in the
Section~\ref{pr.sec:Matter}, and we use this as the \(T^{ab}\) to
generate the matter Lagrangian \(\mathcal{L}_{M}.\)

To couple with an electromagnetic field, we only need to include the
Lagrangian \(\mathcal{L}_{E}\) for the electromagnetic field.
Classical Maxwell field theory already has an answer ready for this
Lagrangian, and it is in terms of the metric \(g_{ab}\) and the
Faraday tensor \(F_{ab}\). It is given by
\begin{equation}
  \label{A.eq:EMLagrangian}
\mathcal{L}_{E}(A_{a}, F_{ab}) = \f{\sqrt{-g}}{8\pi} g^{ac} g^{bd} F_{ab} F_{cd},
\end{equation}
where \(g\) is the metric determinant.
Then the equation of motion resulting from the Euler-Lagrange equation
and the action \[I = \int_{\Omega}\mathcal{L}_{E} \d \Omega,\] give
the Maxwell equations, and the definition for the Faraday tensor in
terms of the 4-potentials.
In the above
\begin{equation}
  \label{A.eq:EOMMaxwell}
  \f{\delta \mathcal{L}_{E}}{\delta g_{ab}} = \pderiv{\mathcal{L}_{E}}{g_{ab}} = 
 -\f{\sqrt{-g}}{4\pi} \left(-g^{cd}F^{a}{}_{c}F^{b}{}_{d} + \f{1}{4}F_{cd}F^{cd}g^{ab}\right).
\end{equation}
The electromagnetic stress-energy tensor can then be written as
\begin{equation}
  \label{A.eq:EMS-E}
  T^{ab} = \f{1}{4\pi}\left(-g^{cd}F^{a}{}_{c}F^{b}{}_{d} + \f{1}{4}F_{cd}F^{cd}g^{ab}\right),
\end{equation}
following the identification of
\[ \f{\delta \mathcal{L}_{E}}{\delta g_{ab}} = -\sqrt{-g} T^{\text{Electromagnetic}}_{ab}\] in
line with equation~\eqref{A.eq:LagragianEM} for the matter Lagrangian
\(\mathcal{L}_{M}.\)

Once the stress-energy tensor of the electromagnetic field has been
specified, coupling the stress-energy of matter and of the
electromagnetic field through minimal coupling is achieved through
\begin{equation}
  \label{A.eq:CoupledS-E}
  T^{\text{Total}}_{ab} = T^{\text{Electromagnetic}}_{ab} + T^{\text{Matter}}_{ab}.
\end{equation}
This is possible since we write the complete action of the total
system: curvature, matter and electromagnetic field as
\begin{equation}
  \label{A.eq:EinActComplete}
  I = \int_{\Omega} (\La_{G} + \kappa \La_{M} + \La_{E})\d \Omega,
\end{equation}
then variation of the whole action gives the full Einstein equations
with the total stress energy tensor as in~\eqref{A.eq:CoupledS-E}.

Because of the contracted Bianchi identity requiring that
\(\nabla_{a}G^{a}{}_{b} = 0,\) from Einstein's equation we must have
\(\nabla_{a}T^{a}{}_{b} =0.\) We constructed the matter \(T_{ab}\) to
behave this way explicitly, and the electromagnetic one obeys this
equation because of the form of the Maxwell's equation.  Indeed, we
have \(T^{ab}{}_{,b} = 0 \implies T^{ab}{}_{;b} = 0\) for the
\(T^{ab}_{\text{Electromagnetic}},\) and therefore as is required from
the EFE,
\[\nabla_{b} T^{ab}_{\text{Total}} = 0.\]

This completes the introduction to Einstein's theory of gravitation as
used in this thesis.  Most of the aspects introduced here is used in
one part or another of the main text, and an index provides the
relevant section where definitions and theorems may be found.

\section{Differential equations}
\label{S:Sturm-Liouville}
All Einstein equations are partial differential
equations (PDEs). However, our symmetry requirements (spherical
symmetry and staticity) force the equations to simplify into ordinary
differential equations (ODEs), except when we have to consider a
relaxation of the static condition with pulsations for the stability
analysis in Chapter~\ref{C:Stability}.  In Subsection~\ref{A.ssec:I-D}
we consider how to find and apply boundary conditions on the EFE we
have to solve.  In Chapter~\ref{C:Stability}, we use Sturm-Liouville
theory to determine linear stability of the EFE for our solutions, and
in Subsection~\ref{A.ssec:S-L} we sketch the main results that will be
useful for this conclusion.

\subsection{Israel-Darmois junction conditions}\label{A.ssec:I-D}
This Section is heavily influenced from~\cite{Poi04,MisSha64,Isr66}
where these conditions are extensively treated.  The EFE form a set of
PDEs.  Taking the cue from classical mechanics, where a complete
solutions to the equations of motion can be obtained uniquely from the
initial values on the positions and velocities, we expect that in
equations involving the metric \(g_{ab},\) initial conditions on
\(g_{ab}\) and \(g_{ab,t}\) should be sufficient to find unique
solutions.  This is essentially correct, but not very helpful in
practice where no such information is usually available.

The initial value problem of GR begins with the selection of a
spacelike hypersurface\index{Hypersurface} \(\Sigma\) which represents
an `instant of time'. This hypersurface can be chosen freely, and we
place an arbitrary system of coordinates \(y^{\alpha}\) on it. (Note,
we are using Greek indices spanning 1,2,3 only here, because we are on
a hypersurface of 3-dimensions)

We can define a normal vector to the hypersurface
through~\eqref{A.eq:HyperNormal}.  Additionally we can define a metric
intrinsic\index{Metric!Induced} to the hypersurface \(\Sigma\) by
restricting the line element to displacements confined to the
hypersurface.  If we define curves \(x^{a} = x^{a}(y^{\alpha})\) on
the hypersurface, we have that the vectors
\[ e^{a}_{\alpha} = \pderiv{x^{a}}{y^{\alpha}},\] are tangent to the
curves contained in \(\Sigma.\) As a result we have trivially that
\(e^{a}_{\alpha} n_{a} = 0,\) that is the tangent vectors on the
hypersurface are normal to the hypersurface orthogonal vector
\(n_{a}.\)  For displacements within \(\Sigma\) we have
\begin{align*}
  \d s^{2}_{\Sigma} & = g_{ab} \d x^{a} \d x^{b}\\
                  & = g_{ab}\left( \pderiv{x^{a}}{y^{\alpha}} \d y^{\alpha}\right) \left( \pderiv{x^{b}}{y^{\beta}} \d y^{\beta} \right)\\
                    & =h_{\alpha \beta} \d y^{\alpha} \d y^{\beta},
\end{align*}
where
\begin{equation}
    \label{A.eq:InducedMetric}
  h_{\alpha \beta} \coloneqq  g_{ab} e^{a}_{\alpha} e^{b}_{\beta}
\end{equation}
is the \textbf{induced metric}, also called the \textbf{first
  fundamental form}\index{First fundamental form}, of the
hypersurface.  The inverse metric can also be expressed in terms of
the induced metric and the normal vector through
\[g^{ab} = \epsilon n^{a}n^{b} + h^{\alpha \beta} e^{a}_{\alpha}
  e^{b}_{\beta},\] where \(h^{\alpha \beta}\) is the metric on the
hypersurface.

Having defined an intrinsic metric on the hypersurface \(\Sigma\), it
should come to no surprise that both a metric connection, and a
covariant derivative on the hypersurface is possible.  We just state
this, and refer the reader to the beginning of the appendix which
treats all this in a dimension independent fashion. The extension to
lower dimensional spaces should not be difficult.  

The next object we define is the \textbf{extrinsic
  curvature}\index{Extrinsic Curvature} of the hypersurface.  This
three-tensor (on the hypersurface) \(K_{\alpha \beta}\) is defined in
terms of vectors not all present on the hypersurface, hence the name.
It characterises how the hypersurface is embedded into the higher
dimensional space it is a surface in.  We have
\[K_{\alpha\beta} \coloneqq n_{a;b} e^{a}_{\alpha} e^{b}_{\beta}.\]
\(K_{\alpha \beta}\) is also called the \textbf{second fundamental
  form}\index{Second fundamental form} of the hypersurface.  From the
definition, we can show that the extrinsic curvature is symmetric,
\(K_{\alpha \beta} = K_{\beta \alpha}.\)

In terms of the first and second fundamental forms of the hypersurface
\(\Sigma,\) we have a complete characterisation in terms of intrinsic
properties, and embedding properties.  We can now continue
investigating the initial value problem.  The space-time metric
\(g_{ab}\) when evaluated on \(\Sigma\) has some components that
characterise displacements outside of the hypersurface (e.g.\
\(g_{tt}\) if \(\Sigma\) as in our case is a surface of constant
\(t\)). The initial values of these components cannot be given from the
intrinsic geometrical properties of \(\Sigma\) alone.

The initial data for the \(g_{ab}\) corresponding to the `positions'
in our analogy have to come from the first fundamental form
\(h_{\alpha\beta}\) of the chosen hypersurface (for a total of 6
components).  The data for the remaining four components of \(g_{ab}\)
are expressed in the choice of the hypersurface and its arbitrary
coordinate system.

Similarly the initial data for \(g_{ab,t}\) corresponding to the
`velocities' in our analogy, come from the second fundamental form
\(K_{\alpha\beta},\) of \(\Sigma.\) Together these provide the initial
data for the initial value problem of GR.  In the complete space-time
these cannot be arbitrarily specified and have to obey the EFE.

We can now state the form of the boundary conditions for the junction
of two space-times and their corresponding metrics.  The question one
asks when solving these equations is the following: A hypersurface
\(\Sigma\) partitions spacetime into two regions \(\mathcal{V}^{+}\)
and \(\mathcal{V}^{-}.\) In \(\mathcal{V}^{+}\) the metric is
\(g^{+}_{ab}\) expressed in coordinates \(x_{+}^{a}\) and in
\(\mathcal{V}^{-},\) it is \(g^{-}_{ab}\) expressed in coordinates
\(x_{+}^{a}.\) How do we get a consistent solution for the whole
space-time, and more precisely, what conditions must be put on the
metrics \(g^{+}_{ab}\) and \(g^{-}_{ab},\) to ensure that
\(\mathcal{V}^{+}\) and \(\mathcal{V}^{-}\) are joined smoothly at
\(\Sigma.\)

We will state the answer without proof, with the explanation hinging
on the fundamental forms defined above, since details will not be used
in this thesis.  Before however we introduce the notation\( [A] \) for
any object \(A\) to
mean\[[A] \left. \coloneqq A(\mathcal{V}^{+})\right|_{\Sigma} -
  \left. A(\mathcal{V}^{-})\right|_{\Sigma},\] i.e.\ the difference in the value of
\(A,\) an object defined on both sides of the hypersurface.  \([A]\)
can be considered as the ``jump'' in the value of \(A\) as one crosses
the hypersurface.

To answer the question, we state
\begin{myTheorem}
  The \textbf{Israel-Darmois junction condition}\index{Junction
    Condition!Israel-Damois} states that the two space times
  \(\mathcal{V}^{+}\) and \(\mathcal{V}^{-}\) are joined smoothly at
  \(\Sigma,\) in their metric structure, and the full space-time obeys
  the EFE if
  \begin{enumerate}
  \item the intrinsic curvature induced by both metrics on \(\Sigma,\) \[[h_{\alpha\beta}] = 0, \]
  \item the extrinsic curvature induced by both metrics on \(\Sigma,\) \[[K_{\alpha\beta}] = 0. \]
  \end{enumerate}
  If the extrinsic curvature is not the same on both sides, additional
  details about the surface must be considered.
\end{myTheorem}

In our case, we shall match an interior solution we find to the
Schwarzschild exterior solution. Since we are in a static spacetime,
\(n_{a}\) is a timelike killing vector that is hypersurface
orthogonal.  This simplifies the calculation of both
\(h_{\alpha\beta}\) and \(K_{\alpha\beta},\) for both the
Schwarzschild exterior metric in \(\mathcal{V}^{-},\)
\[\d s^{2}_{-} = g^{-}_{ab} \d x^{a} \d x^{b} = \left( 1 - \f{2M}{r} \right) \d t^{2} - \left( 1 - \f{2M}{r} \right)^{-1} \d r^{2} - r^{2}\d \Omega^{2},\]
and the
interior solution's metric in \(\mathcal{V}^{+},\)
\[ \d s^{2} = g^{+}_{ab} \d x^{a} \d x^{b} = \e^{\nu(r)} \d t^{2} -\e^{\lambda(r)} \d r^{2} - r^{2}
  \d \Omega^{2}.\] The computation of the two conditions is still
lengthy, and results in
\begin{enumerate}
\item the condition on \([h_{\alpha\beta}]\) reducing to matching the
  metric function on both sides and specifying that
  \[m(r=0) = \left.\deriv{m}{r}\right|_{(r=0)}=0,\] leading to a the
    definition of \(m(r)\) given through
  \[\ e^{-\lambda(r_{b})} = \e^{\nu(r_{b})} = 1 - \f{2M}{r_{b}}, \quad
    \text{with} \quad M = m(r_{b}) = 4\pi \int_{0}^{r_{b}} \rho(\bar{r})
    \bar{r}^{2 }\rmd \bar{r}. \]
\item The computation on \([K_{\alpha\beta}]\) is lengthy and results in 
  \[p(r_{b}) = 0,\] after simplifications, as was shown in~\cite{MisSha64}
\end{enumerate}

We now have the two boundary condition on the system of ODE for
the solution of the complete space-time.

\subsection{Sturm-Liouville theory}\label{A.ssec:S-L}
A \textbf{Sturm--Liouville}\index{Sturm-Liouville} problem consists in
finding eigenvalues \(\sigma^{2}\) and eigenfunctions \(f(x)\) for the
differential equation
\begin{equation}
  \label{eq:SLDiff}
\deriv{}{x} \left[ P(x) \deriv{f}{x} \right] - Q(x)f(x) + \sigma^{2} W(x) f(x) = 0, \qquad (a < x < b),
\end{equation}
satisfying the boundary conditions at \(a\) and \(b\) given through
\begin{subequations}
  \begin{equation}
    \label{eq:SLBC1}
    \alpha_{1} f(a) + \alpha_{2} \left. \deriv{f}{x} \right|_{a} = 0, \qquad \alpha_{1}^{2} + \alpha_{2}^{2} >0,
  \end{equation}
  \begin{equation}
    \label{eq:SLBC2}
    \beta_{1} f(b) + \beta_{2} \left. \deriv{f}{x} \right|_{b} = 0, \qquad \beta_{1}^{2} + \beta_{2}^{2} > 0,
  \end{equation}
\label{eq:SLBC}
\end{subequations}
In our work we also use the generalised S--L equation, given instead by
\begin{equation}
  \label{eq:gSLDiff}
\deriv{}{x} \left[ P(x) \deriv{f}{x} \right] - Q(x)f(x) + \sigma^{2} W(x) f(x) = R(x,f), \qquad (a < x < b).
\end{equation}
Both of these equations have been investigated in the past, and for
our purposes, the reduction of~\eqref{eq:gSLDiff}
into~\eqref{eq:SLDiff}, through the absorption of \(R(x,f)\)
into \(Q(x)\)
is what we strive for. Once we have the equation in the form
of~\eqref{eq:SLDiff}, we express the latter in a variational form, in
terms of functionals, so that an Euler-Lagrange technique can be
applied to it.  As can be checked by expansion and simplification, if
the functional
\[\mathcal{I}[f(x)] = \int_{a}^{b} \left[p(x) \left(
    \deriv{f}{x}\right)^{2} + Q(x) f^{2} \right] \d x,\]
is minimized through the Euler-Lagrange method, subject to the
condition that the functional
\begin{equation}
\label{eq:SLCons}
 \mathcal{J}[f(x)] = \int_{a}^{b} W(x)f^{2} \d x = \text{constant}, 
\end{equation}
is a constant, the eigenvalue \(\sigma^{2}\)
appears as a Lagrange multiplier.  Since~\eqref{eq:SLCons} is the
normalisation condition for \(f(x)\)
under a weight function \(W(x),\)
the variation technique is equivalent to minimizing the
functional
\[\mathcal{K}[f(x)] = \f{\mathcal{I}[f(x)]}{\mathcal{J}[f(x)]}.\]
Once this functional has been defined, we can use the main theorem of
the theory given below,
\begin{myTheorem}
  A Sturm--Liouville problem~\eqref{eq:SLDiff} is
  \textbf{regular}\index{Sturm-Liouville Problem!Regular} if
  \(P(x) > 0, W(x) > 0,\) and
  
  \(P(x), P'(x), Q(x), W(x) \) are all continuous functions over the
  finite interval \([a,b],\) and additionally satisfy the boundary
  conditions given through~\eqref{eq:SLBC}.  If the Sturm--Liouville
  problem is regular, then the eigenvalues \(\sigma_{i}^{2}\)
  of~\eqref{eq:SLDiff} are real and can be ordered such that
\[
\sigma_{1}^{2} < \sigma_{2}^{2} < \sigma_{3}^{2}< \cdots < \sigma_{n}^{2} < \cdots \to \infty.
\]
To each eigenvalue \(\sigma^{2}_{i},\) there corresponds a unique
eigenfunction \(f_{i}(r),\) which is called the
\(\mathbf{i^{\text{th}}}\) \textbf{fundamental solution}\index{Sturm-Liouville Problem!Fundamental
solution} satisfying the regular Sturm-Liouville problem.
Furthermore, the normalized eigenfunctions form an orthonormal basis
in the Hilbert space \(L^{2}([a,b])\) with weight \(W(x)\) and norm
\[
\int^{a}_{b} f_{m}(x) f_{n}(x) W(x) \d x = \delta_{nm}.
\]
\label{A.th:S-L}
\end{myTheorem}



\chapter{}
\label{C:AppendixB}
We provide a complete reference of all the new solutions mentioned in
the thesis in the form of the metric functions associated with each,
and the different classes of parameters each solution is valid for,
with simplifications.  This is meant as complete reference of the
solutions and a useful ``cheat-sheet'' for looking up values of
parameters mid-text.
\section{The Tolman VII solution, with anisotropic pressure and no charge}
This section gives the expressions for the uncharged case with \(k=0\)
but with anisotropy, \(\beta \neq 0.\)
\subsection{The $\phi^{2} = 0$ case}
{\tabulinesep=1.2mm
\begin{tabu}[c]{| >{$\displaystyle}l<{$} | >{$\displaystyle}c<{$} |}
  \hline
  \phi^2 &  0\\
  \hline
  \mu &  0 < \mu \leq 1\\
  \hline
  Z & 1 - \left( \f{\kappa \rho_c}{3} \right) r^2  + \left( \f{\kappa\mu\rho_c}{5r_b^2}\right) r^4\\
\hline  
  Y & \gamma + \f{2\alpha r_{b}}{\sqrt{\kappa \rho_{c}\mu /5}}\left[ 
      \acoth{\left(\f{1-\sqrt{Z(r)}}{r^{2}\sqrt{\f{\kappa \rho_{c} \mu} {5r_{b}^{2}}}}\right) }  -\acoth{\left( \f{1-\gamma}{r_{b}\sqrt{\kappa \rho_{c} \mu /5}}\right)}  \right] \\
\hline
  \rho & \rho_{c} \left[ 1 - \mu \left( \f{r}{r_{b}}\right)^{2}\right]  \\
\hline
  a & \left( \f{\kappa\mu\rho_c}{5r_b^2}\right) \\\hline
  b & \f{\kappa \rho_c}{3} \\\hline
  \xi & \f{2}{\sqrt{a}} \acoth \left( \f{1+\sqrt{1 - br^2 + a r^4}}{\sqrt{a} r^2} \right)\\\hline
  c_1 & \gamma - \f{2 \alpha}{\sqrt{b}} \\\hline
  c_2 & \alpha \\\hline
  \alpha & \f{\kappa \rho_{c}(5-3\mu)}{60} \\\hline
  \beta & -a\\\hline
  \gamma & \sqrt{1+ \f{\kappa \rho_{c}r_{b}^{2}(3\mu-5)}{15} }\\\hline
  p_r & \begin{aligned} &\f{2\kappa\rho_{c}}{3} - \f{4\kappa\rho_{c}\mu r^{2}}{5r_{b}^{2}} -\kappa\rho_{c} \left[ 1 - \mu \left( \f{r}{r_{b}}\right)^{2}\right] + \\
&+ \left( \f{\kappa\rho_{c}}{3} - \f{\kappa \rho_{c} \mu}{5}\right) \f{\sqrt{1-\f{\kappa\rho_{c}}{3}r^{2} + \f{\kappa\mu\rho_{c}}{5r_{b}^{2}} r^{4}}}{\gamma + \f{2\alpha r_{b}}{\sqrt{\kappa \rho_{c}\mu /5}}\left[ 
  \acoth{\left(\f{1-\sqrt{Z(r)}}{r^{2}\sqrt{\f{\kappa \rho_{c} \mu} {5r_{b}^{2}}}}\right) }  -\acoth{\left( \f{1-\gamma}{r_{b}\sqrt{\kappa \rho_{c} \mu /5}}\right)}  \right]} \end{aligned} \\\hline
  \ppen & p_{r} + \f{\kappa\rho_{c}\mu}{5r_{b}^{2}} r^{2}\\\hline
\end{tabu}}
\subsection{The $\phi^{2} < 0$ case}
{\tabulinesep=1.2mm
\begin{tabu}[c]{| >{$\displaystyle}l<{$} | >{$\displaystyle}c<{$} |}
  \hline
  \phi^2 &  a + \beta  < 0\\
  \hline
  \mu &  0 < \mu \leq 1\\
  \hline
  Z & 1 - \left( \f{\kappa \rho_c}{3} \right) r^2  + \left( \f{\kappa\mu\rho_c}{5r_b^2}\right) r^4\\
\hline  
  Y &\begin{aligned}
     &\left[ \gamma \cosh{(\phi\xi_{b})} - \f{\alpha}{\phi} \sinh{(\phi\xi_{b})} \right] 
    \cosh{\left[ \f{2\phi}{\sqrt{a}} \coth^{-1} \left( \f{1+\sqrt{1-br^2+ar^4}}{r^2\sqrt{a}}\right)  \right]} +\\
    &+ \left[ \f{\alpha}{\phi} \cosh{(\phi\xi_{b})} - \gamma \sinh{(\phi\xi_{b})} \right] 
    \sinh{\left[\f{2\phi}{\sqrt{a}} \coth^{-1} \left( \f{1+\sqrt{1-br^2+ar^4}}{r^2\sqrt{a}}\right) \right]},   
      \end{aligned} \\\hline
  \rho & \rho_{c} \left[ 1 - \mu \left( \f{r}{r_{b}}\right)^{2}\right]  \\
\hline
  a & \left( \f{\kappa\mu\rho_c}{5r_b^2}\right) \\\hline
  b & \f{\kappa \rho_c}{3} \\\hline
  \xi & \f{2}{\sqrt{a}} \acoth \left( \f{1+\sqrt{1 - br^2 + a r^4}}{\sqrt{a} r^2} \right)\\\hline
  c_1 & \f{\alpha}{\phi} \cosh{(\phi\xi_{b})} - \gamma \sinh{(\phi\xi_{b})} \\
\hline
  c_2 & \gamma \cosh{(\phi\xi_{b})} - \f{\alpha}{\phi} \sinh{(\phi\xi_{b})} \\\hline
  \alpha & \f{\kappa \rho_{c}(5-3\mu)}{60} \\\hline
  \beta & < -a \implies < -\f{\kappa\mu\rho_c}{5r_b^2} \\\hline
  \gamma & \sqrt{1+ \f{\kappa \rho_{c}r_{b}^{2}(3\mu-5)}{15} }\\\hline
  p_r & \begin{aligned} &\f{1}{\kappa}\left\{ \f{2\kappa\rho_{c}}{3} - \f{4\kappa\rho_{c}\mu r^{2}}{5r_{b}^{2}} -\kappa\rho_{c} \left[ 1 - \mu \left( \f{r}{r_{b}}\right)^{2}\right] + 4 \phi\sqrt{Z} \times \right. \\
&\left. \times \f{ \left[ \f{\alpha}{\phi} \cosh{(\phi\xi_{b})} -\gamma \sinh{(\phi\xi_{b})} \right] \cosh{(\phi \xi)} + \left[ \gamma \cosh{(\phi\xi_{b})} - \f{\alpha}{\phi} \sinh{(\phi\xi_{b})} \right] \sinh{(\phi \xi)} }{\left[ \gamma \cosh{(\phi\xi_{b})} - \f{\alpha}{\phi} \sinh{(\phi\xi_{b})}\right] \cosh{(\phi \xi)} + \left[\gamma \cosh{(\phi\xi_{b})} - \f{\alpha}{\phi} \sinh{(\phi\xi_{b})} \right]\sinh{(\phi \xi)}} \right\} \end{aligned} \\\hline
  \ppen & p_{r} - \beta r^{2}\\\hline
\end{tabu}}

\subsection{The $\phi^{2} > 0$ case}
{\tabulinesep=1.2mm
\begin{tabu}[c]{| >{$\displaystyle}l<{$} | >{$\displaystyle}c<{$} |}
  \hline
  \phi^2 &  a + \beta  > 0\\
  \hline
  \mu &  0 < \mu \leq 1\\
  \hline
  Z & 1 - \left( \f{\kappa \rho_c}{3} \right) r^2  + \left( \f{\kappa\mu\rho_c}{5r_b^2}\right) r^4\\
\hline  
  Y &\begin{aligned}
     &\left[ \gamma \cos{(\phi\xi_{b})} - \f{\alpha}{\phi} \sin{(\phi\xi_{b})} \right] 
    \cos{\left[ \f{2\phi}{\sqrt{a}} \coth^{-1} \left( \f{1+\sqrt{1-br^2+ar^4}}{r^2\sqrt{a}}\right)  \right]} +\\
    &+ \left[ \f{\alpha}{\phi} \cos{(\phi\xi_{b})} + \gamma \sin{(\phi\xi_{b})} \right] 
    \sin{\left[\f{2\phi}{\sqrt{a}} \coth^{-1} \left( \f{1+\sqrt{1-br^2+ar^4}}{r^2\sqrt{a}}\right) \right]},   
      \end{aligned} \\\hline
  \rho & \rho_{c} \left[ 1 - \mu \left( \f{r}{r_{b}}\right)^{2}\right]  \\
\hline
  a & \left( \f{\kappa\mu\rho_c}{5r_b^2}\right) \\\hline
  b & \f{\kappa \rho_c}{3} \\\hline
  \xi & \f{2}{\sqrt{a}} \acoth \left( \f{1+\sqrt{1 - br^2 + a r^4}}{\sqrt{a} r^2} \right)\\\hline
  c_1 & \f{\alpha}{\phi} \cos{(\phi\xi_{b})} + \gamma \sin{(\phi\xi_{b})} \\
\hline
  c_2 & \gamma \cos{(\phi\xi_{b})} - \f{\alpha}{\phi} \sin{(\phi\xi_{b})} \\\hline
  \alpha & \f{\kappa \rho_{c}(5-3\mu)}{60} \\\hline
  \beta & > -a \implies > -\f{\kappa\mu\rho_c}{5r_b^2} \\\hline
  \gamma & \sqrt{1+ \f{\kappa \rho_{c}r_{b}^{2}(3\mu-5)}{15} }\\\hline
  p_r & \begin{aligned} &\f{1}{\kappa}\left\{ \f{2\kappa\rho_{c}}{3} - \f{4\kappa\rho_{c}\mu r^{2}}{5r_{b}^{2}} -\kappa\rho_{c} \left[ 1 - \mu \left( \f{r}{r_{b}}\right)^{2}\right] + 4 \phi\sqrt{Z} \times \right. \\
&\left. \times \f{\left[ \gamma \sin{(\phi\xi_{b})} + \f{\alpha}{\phi} \cos{(\phi\xi_{b})} \right] \cos{(\phi \xi)} - \left[ \gamma \cos{(\phi\xi_{b})} - \f{\alpha}{\phi} \sin{(\phi\xi_{b})} \right] \sin{(\phi \xi)} }{\left[ \gamma \sin{(\phi\xi_{b})} + \f{\alpha}{\phi} \cos{(\phi\xi_{b})}\right] \sin{(\phi \xi)} + \left[\gamma \cos{(\phi\xi_{b})} - \f{\alpha}{\phi} \sin{(\phi\xi_{b})} \right]\cos{(\phi \xi)}} \right\}, \end{aligned}\\\hline

  \ppen & p_{r} - \beta r^{2}\\\hline
  
\end{tabu}}

\section{The Tolman VII solution, with anisotropic pressure and charge}
\subsection{Anisotropised charge, with charge matching anisotropy, but $\Phi^{2} \neq 0$}
{\tabulinesep=1.2mm
\begin{tabu}[c]{| >{$\displaystyle}l<{$} | >{$\displaystyle}c<{$} |}
  \hline
  \phi^2 &  \Phi^{2} = \f{1}{20}\left(\f{\kappa\rho_{c}\mu}{r_{b}^{2}} - k^{2} \right),\\\hline
  \mu &  0 < \mu \leq 1\\\hline
  Z & 1 - \left(\f{\kappa \rho_{c}}{3} \right) r^{2} + \f{1}{5}\left( \f{\kappa \rho_{c}\mu}{r_{b}^{2}} - k^{2}\right) r^{4}\\\hline  
  Y & \begin{aligned} &\left( \gamma \cos{(\Phi\xi_{b})} - \f{\alpha}{\Phi} \sin{(\Phi\xi_{b})} \right) 
    \cos{\left( \f{2\Phi}{\sqrt{a}} \coth^{-1} \left( \f{1+\sqrt{1-br^2+ar^4}}{r^2\sqrt{a}}\right)  \right)} +\\
    &+ \left(\gamma \sin{(\Phi\xi_{b})} + \f{\alpha}{\Phi} \cos{(\Phi\xi_{b})} \right) 
    \sin{\left(\f{2\Phi}{\sqrt{a}} \coth^{-1} \left( \f{1+\sqrt{1-br^2+ar^4}}{r^2\sqrt{a}}\right) \right)},\end{aligned}\\\hline
  \rho & \rho_{c} \left[ 1 - \mu \left( \f{r}{r_{b}}\right)^{2}\right]  \\\hline
  a & \f{1}{5}\left( \f{\kappa \rho_{c}\mu}{r_{b}^{2}} - k^{2}\right) \\\hline
  b & \f{\kappa \rho_c}{3} \\\hline
  \xi & \f{2}{\sqrt{a}} \acoth \left( \f{1+\sqrt{1 - br^2 + a r^4}}{\sqrt{a} r^2} \right)\\\hline
  c_1 & \gamma \cos{(\Phi\xi_{b})} - \f{\alpha}{\Phi} \sin{(\Phi\xi_{b})} \\\hline
  c_2 & \gamma \sin{(\Phi\xi_{b})} + \f{\alpha}{\Phi} \cos{(\Phi\xi_{b})} \\\hline
  \alpha & \f{\left( \kappa\rho_{c}(5 -3\mu) - 12k^{2}r_{b}^{2} \right)}{60}\\\hline
  \Delta & 2k^{2} r^{2} = \f{qr}{2k} \\\hline
  \gamma & \sqrt{1+ \f{\kappa \rho_{c}r_{b}^{2}(3\mu-5)}{15} -\f{k^{2}r_{b}^{2}}{5}} \\\hline
  p_r & \begin{aligned} & \f{2\kappa\rho_{c}}{3} - \f{4}{5}\left( \f{\kappa \rho_{c}\mu}{r_{b}^{2}} - k^{2}\right)r^2 -\kappa\rho_{c} \left[ 1 - \mu \left( \f{r}{r_{b}}\right)^{2}\right] + 4 \Phi\sqrt{Z} \times \\
&\times \left\{ \f{ \left[ \gamma \sin{(\Phi\xi_{b})} + \f{\alpha}{\Phi} \cos{(\Phi\xi_{b})} \right] \cos{(\Phi \xi)} - \left[ \gamma \cos{(\Phi\xi_{b})} - \f{\alpha}{\Phi} \sin{(\Phi\xi_{b})} \right] \sin{(\Phi \xi)} }{\left[ \gamma \sin{(\Phi\xi_{b})} + \f{\alpha}{\Phi} \cos{(\Phi\xi_{b})}\right] \sin{(\Phi \xi)} + \left[\gamma \cos{(\Phi\xi_{b})} - \f{\alpha}{\Phi} \sin{(\Phi\xi_{b})} \right]\cos{(\Phi \xi)}} \right\} \end{aligned} \\\hline
  \ppen & p_{r} - 2k^{2}r^{2}\\\hline
  
\end{tabu}}

\subsection{The $\Phi^{2} = 0$ case}
{\tabulinesep=1.2mm
\begin{tabu}[c]{| >{$\displaystyle}l<{$} | >{$\displaystyle}c<{$} |} 
  \hline
  \Phi^2 &  0\\\hline
  \mu &  0 < \mu \leq 1\\\hline
  Z & 1 - \left( \f{\kappa \rho_c}{3} \right) r^2  + \f{2}{11}\left( \f{\kappa\mu\rho_c}{r_b^2} - \f{\beta}{2}\right) r^4\\
\hline  
  Y & c_1 + c_2\xi \\\hline
  \rho & \rho_{c} \left[ 1 - \mu \left( \f{r}{r_{b}}\right)^{2}\right]  \\
\hline
  a & \f{2}{11} \left( \f{\kappa\mu\rho_c}{r_b^2} - \f{\beta}{2}\right) \\\hline
  b & \f{\kappa \rho_c}{3} \\\hline
  \xi & \f{2}{\sqrt{a}} \acoth \left( \f{1+\sqrt{1 - br^2 + a r^4}}{\sqrt{a} r^2} \right)\\\hline
  c_1 &  \gamma - \alpha \xi_b \\\hline
  c_2 & \alpha \\\hline
  k   & \sqrt{(a + \beta)/2}\\\hline
  \alpha & \f{1}{4}\left(\f{\kappa \rho_c}{3} - \f{3 \kappa \rho_c \mu}{11} - \f{4r_b^2\beta}{11}\right) \\\hline
  \beta & 2k^2 - a \\\hline
  \gamma & \sqrt{1+ r_b^2 \kappa \rho_b \left(\f{2\mu}{11}-\f{1}{3}\right) -\f{\beta r_b^4}{11}} \\\hline
  p_r & \f{1}{\kappa} \left[ \f{4c_{2} \sqrt{1 -br^{2} + ar^{4}}}{c_{1}+c_{2}\xi} +2b -4ar^{2} \right] - \rho(r) \\\hline
  \ppen & p_{r} - \f{\beta r^{2}}{\kappa} \\\hline
  
\end{tabu}}

\subsection{The $\Phi^{2} < 0$ case}
{\tabulinesep=1.2mm
\begin{tabu}[c]{| >{$\displaystyle}l<{$} | >{$\displaystyle}c<{$} |}
  \hline
  \Phi^2 &  a + \beta -2k^2  < 0\\
  \hline
  \mu &  0 < \mu \leq 1\\
  \hline
  Z & 1 - \left( \f{\kappa \rho_c}{3} \right) r^2  + \f{1}{5}\left( \f{\kappa \rho_{c}\mu}{r_{b}^{2}} - k^{2}\right) r^4\\
\hline  
  Y &\begin{aligned}
     &\left[ \gamma \cosh{(\Phi\xi_{b})} - \f{\alpha}{\Phi} \sinh{(\Phi\xi_{b})} \right] 
    \cosh{\left[ \f{2\Phi}{\sqrt{a}} \coth^{-1} \left( \f{1+\sqrt{1-br^2+ar^4}}{r^2\sqrt{a}}\right)  \right]} +\\
    &+ \left[ \f{\alpha}{\Phi} \cosh{(\Phi\xi_{b})} - \gamma \sinh{(\Phi\xi_{b})} \right] 
    \sinh{\left[\f{2\Phi}{\sqrt{a}} \coth^{-1} \left( \f{1+\sqrt{1-br^2+ar^4}}{r^2\sqrt{a}}\right) \right]},   
      \end{aligned} \\\hline
  \rho & \rho_{c} \left[ 1 - \mu \left( \f{r}{r_{b}}\right)^{2}\right]  \\
\hline
  a & \f{1}{5}\left( \f{\kappa\mu\rho_c}{r_b^2} - k^2 \right) \\\hline
  b & \f{\kappa \rho_c}{3} \\\hline
  \xi & \f{2}{\sqrt{a}} \acoth \left( \f{1+\sqrt{1 - br^2 + a r^4}}{\sqrt{a} r^2} \right)\\\hline
  c_1 & \f{\alpha}{\Phi} \cosh{(\Phi\xi_{b})} - \gamma \sinh{(\Phi\xi_{b})} \\
\hline
  c_2 & \gamma \cosh{(\Phi\xi_{b})} - \f{\alpha}{\Phi} \sinh{(\Phi\xi_{b})} \\\hline
  \alpha & \f{1}{4} \left( \f{\kappa\rho_c}{3} - \f{\kappa \rho_c \mu}{5} - \f{4 k^2 r_b^2}{5}\right) \\\hline
  \beta & < 2k^2 -a \implies < \f{1}{5} \left( 11k^2  - \f{\kappa \mu \rho_c}{r_b^2}\right)\\\hline
  \gamma & \sqrt{1+ \kappa \rho_c r_b^2\left( \f{\mu}{5} - \f{1}{3}\right) -\f{k^2 r_b^4}{5} }\\\hline
  p_r & \begin{aligned} &\f{1}{\kappa}\left\{ \f{2\kappa\rho_{c}}{3} - \f{4\kappa\rho_{c}\mu r^{2}}{5r_{b}^{2}} -\kappa\rho_{c} \left[ 1 - \mu \left( \f{r}{r_{b}}\right)^{2}\right] + 4 \Phi\sqrt{Z} \times \right. \\
&\left. \f{ \left[ \f{\alpha}{\Phi} \cosh{(\Phi\xi_{b})} -\gamma \sinh{(\Phi\xi_{b})} \right] \cosh{(\Phi \xi)} + \left[ \gamma \cosh{(\Phi\xi_{b})} - \f{\alpha}{\Phi} \sinh{(\Phi\xi_{b})} \right] \sinh{(\Phi \xi)} }{\left[ \gamma \cosh{(\Phi\xi_{b})} - \f{\alpha}{\Phi} \sinh{(\Phi\xi_{b})}\right] \cosh{(\Phi \xi)} + \left[\gamma \cosh{(\Phi\xi_{b})} - \f{\alpha}{\Phi} \sinh{(\Phi\xi_{b})} \right]\sinh{(\Phi \xi)}} \right\} \end{aligned} \\\hline
  \ppen & p_{r} - \beta r^{2}\\\hline  
\end{tabu}}

\subsection{The $\Phi^{2} > 0$ case}
{\tabulinesep=1.2mm
\begin{tabu}[c]{| >{$\displaystyle}l<{$} | >{$\displaystyle}c<{$} |}
  \hline
  \Phi^2 &  a + \beta -2k^2 > 0\\
  \hline
  \mu &  0 < \mu \leq 1\\
  \hline
  Z & 1 - \left( \f{\kappa \rho_c}{3} \right) r^2  + \f{1}{5}\left( \f{\kappa \rho_{c}\mu}{r_{b}^{2}} - k^{2}\right)r^4\\
\hline  
  Y &\begin{aligned}
     &\left[ \gamma \cos{(\Phi\xi_{b})} - \f{\alpha}{\Phi} \sin{(\Phi\xi_{b})} \right] 
    \cos{\left[ \f{2\Phi}{\sqrt{a}} \coth^{-1} \left( \f{1+\sqrt{1-br^2+ar^4}}{r^2\sqrt{a}}\right)  \right]} +\\
    &+ \left[ \f{\alpha}{\Phi} \cos{(\Phi\xi_{b})} + \gamma \sin{(\Phi\xi_{b})} \right] 
    \sin{\left[\f{2\Phi}{\sqrt{a}} \coth^{-1} \left( \f{1+\sqrt{1-br^2+ar^4}}{r^2\sqrt{a}}\right) \right]},   
      \end{aligned} \\\hline
  \rho & \rho_{c} \left[ 1 - \mu \left( \f{r}{r_{b}}\right)^{2}\right]  \\
\hline
  a & \f{1}{5}\left( \f{\kappa\mu\rho_c}{r_b^2} - k^2 \right) \\\hline
  b & \f{\kappa \rho_c}{3} \\\hline
  \xi & \f{2}{\sqrt{a}} \acoth \left( \f{1+\sqrt{1 - br^2 + a r^4}}{\sqrt{a} r^2} \right)\\\hline
  c_1 & \f{\alpha}{\Phi} \cos{(\Phi\xi_{b})} + \gamma \sin{(\Phi\xi_{b})} \\
\hline
  c_2 & \gamma \cos{(\Phi\xi_{b})} - \f{\alpha}{\Phi} \sin{(\Phi\xi_{b})} \\\hline
  \alpha & \f{1}{4} \left( \f{\kappa\rho_c}{3} - \f{\kappa \rho_c \mu}{5} - \f{4 k^2 r_b^2}{5}\right) \\\hline
  \beta & > 2k^2 - a \implies >  \f{1}{5} \left( 11k^2  - \f{\kappa \mu \rho_c}{r_b^2}\right) \\\hline
  \gamma & \sqrt{1+ \kappa \rho_c r_b^2\left( \f{\mu}{5} - \f{1}{3}\right) -\f{k^2 r_b^4}{5} }\\\hline
  p_r & \begin{aligned} &\f{1}{\kappa}\left\{ \f{2\kappa\rho_{c}}{3} - \f{4\kappa\rho_{c}\mu r^{2}}{5r_{b}^{2}} -\kappa\rho_{c} \left[ 1 - \mu \left( \f{r}{r_{b}}\right)^{2}\right] + 4 \Phi\sqrt{Z} \times \right. \\
&\left. \times \f{\left[ \gamma \sin{(\Phi\xi_{b})} + \f{\alpha}{\Phi} \cos{(\Phi\xi_{b})} \right] \cos{(\Phi \xi)} - \left[ \gamma \cos{(\Phi\xi_{b})} - \f{\alpha}{\Phi} \sin{(\Phi\xi_{b})} \right] \sin{(\Phi \xi)} }{\left[ \gamma \sin{(\Phi\xi_{b})} + \f{\alpha}{\Phi} \cos{(\Phi\xi_{b})}\right] \sin{(\Phi \xi)} + \left[\gamma \cos{(\Phi\xi_{b})} - \f{\alpha}{\Phi} \sin{(\Phi\xi_{b})} \right]\cos{(\Phi \xi)}} \right\}, \end{aligned}\\\hline

  \ppen & p_{r} - \beta r^{2}\\\hline
  
\end{tabu}}


\chapter{}
\label{C:AppendixC}
\lstset{numbers=left, breaklines=true, basicstyle=\scriptsize, frame=single}
We provide the source code for some selected MAXIMA\cite{maxima} functions,
  definitions and procedures that were used in the writing of this thesis.
\section{Stability routines}
The type of input file MAXIMA accept resembles the following, and for
this program, the output consists of the integrals, with their
uncertainties, and we also provide those below
\begin{lstlisting}[caption=Stability routines ]
/* [wxMaxima batch file version 1] [ DO NOT EDIT BY HAND! ]*/
/* [ Created with wxMaxima version 13.04.2 ] */

/* [wxMaxima: input   start ] */
/* reset all variables */
kill(all);  

/* for numerical methods, newton_raphson, etc. 
   Used for finding critical densities          */
load(newton1)$ 

/* Set up constant boundaries, and ranges; 
    allows certain simplifications to occur     */
assume(rho_c>0, r_b>0, mu>=0 and mu<=1, r>=0, a>0, b>0, d>0)$

/* coordinate transformations */
x(r):= r^2$
xi(r):= -1/sqrt(a) * log((b-2*a*x(r)+2*sqrt(a)*sqrt(1-b*x(r)+a*x(r)^2))/(b+2*sqrt(a)))$

/*  metric, matter variables and their derivatives for 
    TVII if beta=k=0, and more general ones otherwise  */ 
Z(r):=      1-b*x(r)+a*x(r)^2$
Y(r) :=     c2*sin(phi*xi(r))+c1*cos(phi*xi(r))$
lamb(r):=   -log(Z(r))$
nu(r):=     2*log(Y(r))$

rho(r):=rho_c*(1-mu*(r/r_b)^2)$
p_r(r) :=   ''(2*Z(r)/(r*Y(r))*diff(Y(r),r) + -1/r*diff(Z(r),r) - kappa*rho(r))$
r(rho):=    r_b*sqrt((1-rho/rho_c)/mu)$

/*polytropic index*/
dp_rdrho(r):=   ''(subst(rho = rho(r),diff(p_r(r(rho)),rho)))$
Gamma(r) :=     (rho(r)+p_r(r))/p_r(r) * dp_rdrho(r)$

/* Constant values. Needed for numeric computations */
cmu:        mu=     1$
crho_c:     rho_c=  1.0e18*7.42591549e-28 $
cr_b:       r_b=    1e4$
ck:         k=      1e-10$
cbeta:      beta=   0$
ckappa:     kappa=  8*%pi$


/* TVIIca substitutions */
ca:     a = (kappa*rho_c*mu/r_b^2 - k^2)/5$
cb:     b = kappa*rho_c/3$
calpha: alpha = (kappa*rho_c/3 - kappa*rho_c*mu/5 - 4*k^2*r_b^2/5)/4$
cphi:   phi = sqrt(a+beta-2*k^2)/2$
cZb:    Zb = Z(r_b)$
cxib:   xib = xi(r_b)$
cxb:    xb = x(r_b)$
cc1:    c1 = sqrt(Zb)*cos(phi*xib) - alpha/phi*sin(phi*xib)$
cc2:    c2 = alpha/phi*cos(phi*xib) + sqrt(Zb)*sin(phi*xib)$
cxi:    xi = xi(r)$
cx:     x = x(r)$
cY :    Y = Y(r)$
cZ:     Z = Z(x)$
crho:   rho = rho(r)$
cr:     r = r(rho)$
clambi:  lamb = lamb(r)$
cnu:    nu = nu(r)$

/* Substitution list applied in this specific order to generate graphs */
    /* rho_c, mu, and r_b free */
subls:      [cY, cZ, crho]$
tovalls:    [cc1,cc2,cZb,cphi,calpha,cxib,cxi,cx,cxb,ca,cb]$
constls:    [kappa=8*%pi*G/c^2, G=1,c=1]$
ultls:      [rho_c=3/(16*%pi), r_b=1.]$
cnoParams:  append(subls, tovalls, constls)$

/*  Definitions of integrands, and integrating factors used to build 
    Sturm-Liouville (SL) coefficients */

Ip_r(r):=        subst(cnoParams,p_r(r))$
Ip_rprime(r):=   subst(cnoParams,dp_rdrho(r))$
Igamma(r):=      subst(cnoParams,Gamma(r))$
Ifacd(r):=       2*r/(Igamma(r)*Ip_r(r)) *(Ip_rprime(r) - 1)$

IfacdN(r):=      subst([ckappa,cmu,crho_c,ck,cr_b,cbeta],Ifacd(r))$

Ifac(r,params):= 
block([llim:1,ulim:r,const:params],
    prefac: subst(const, beta - 6*k^2/kappa),
    exp(prefac * quad_qags(IfacdN(t),t,llim,ulim)[1])
)$

SL_sqBraPrime(r) := diff((exp((lamb(r)+3*nu(r))/2)*(Ip_rprime(r)-1)/r),r)$

SL_P(r) :=  Gamma(r)*p_r(r)*exp((lamb(r)+3*nu(r))/2)/r^2*(
            Ifac(r,[ckappa,cmu,crho_c,ck,cr_b,cbeta])
)$

SL_Q(r) :=  Ifac(r,[ckappa,cmu,crho_c,ck,cr_b,cbeta])*(
            -4*exp((3*nu(r)+lamb(r))/2)/r^3*diff(p_r(r),r) +
            -8*%pi*exp(3*(lamb(r)+nu(r))/2)/r^2*(p_r(r)+ (beta+k^2/kappa)*r^2)*(p_r(r)+rho(r)) +
            -24*exp((3*nu(r)+lamb(r))/2)/r^2*(k^2/kappa - beta/2) +
            4*exp((3*nu(r)+lamb(r))/2)/(p_r(r)+rho(r))*(3*k^2/kappa-beta)*(9*k^2/kappa-beta) +
            -4*diff(p_r(r),r)*exp((3*nu(r)+lamb(r))/2)/(r*(p_r(r)+rho(r)))*(6*k^2/kappa-beta) +
            2*(beta - 6*k^2/kappa)*SL_sqBraPrime(r) +
            (diff(p_r(r),r))^2*exp((3*nu(r)+lamb(r))/2)/(r^2*(p_r(r)+rho(r)))
)$

SL_W(r) :=  Ifac(r,[ckappa,cmu,crho_c,ck,cr_b,cbeta])*(
            ((p_r(r)+rho(r))*exp((3*lamb(r)+nu(r))/2)/r^2)
)$
/* [wxMaxima: input   end   ] */

/* [wxMaxima: input   start ] */
/* testing plots, and functions */

appenList:  append(subls, tovalls, constls, [cmu,crho_c,ck,cr_b,cbeta])$

PSL_P :     subst(appenList,SL_P(r))$
PSL_Q :     subst(appenList,SL_Q(r))$
PSL_W :     subst(appenList,SL_W(r))$

plot2d(PSL_Q,[r,1,10000]);
/* [wxMaxima: input   end   ] */

/* [wxMaxima: input   start ] */
/* Evaluate the integrals with different function f's*/

appenList:  append(subls, tovalls, constls, [cmu,crho_c,ck,cr_b,cbeta])$

TestF(r):=     r^3$ 
TestdiffF(r):= diff(TestF(r),r)$

ulim:   10000$
llim:   0$

firstnIntegrand(r) :=   subst(appenList, SL_P(r)*(TestdiffF(r)^2))$
secondnIntegrand(r) :=  subst(appenList, SL_Q(r)*(TestF(r)^2))$
denomIntegrand(r) :=    subst(appenList, SL_W(r)*(TestF(r)^2))$

fInt:   quad_qags(firstnIntegrand(r),r,llim,ulim);
sInt:   quad_qags(secondnIntegrand(r),r,llim,ulim);
tInt:   quad_qags(denomIntegrand(r),r,llim,ulim);

sqfrequency0: (fInt[1]-sInt[1])/tInt[1];
/* [wxMaxima: input   end   ] */


/* Maxima can't load/batch files which end with a comment! */
"Created with wxMaxima"$
\end{lstlisting}

The output of this routine then gives the frequencies associated with each solution.
\begin{lstlisting}[caption=Stability output]
  (%o72) [2727.549758809813,4.210599264169478*10^-7,21,0]
  (%o73) [1159.023634396673,5.878771063296232*10^-7,21,0]
  (%o74) [8.581280570195294*10^9,4.979393690689304,21,0]
  (%o75) 1.827846218967577*10^-7
\end{lstlisting}
As mentioned previously, the integration routine used is from the
QUADPACK~\cite{PieDeDUeb83} function \texttt{quag\_qags}, which
provides the output shown. The last line of the output is the value of
the fundamental frequency squared, and its value for different
parameter values are shown in the main text of this thesis.

\section{Tensor routines}
These routines were used to calculate the components of tensors,
particularly the Einstein tensor. We show a sample here, working with
a static spherically symmetric metric as given in
equation~\eqref{pr.eq:SymMetric}.
\begin{lstlisting}[caption=Tensor routines]

NIL
(%o3)                                done
(%i4) 
(%o0)                                done

(%i1) load(ctensor);

(%o1)          /usr/share/maxima/5.37.2/share/tensor/ctensor.mac

(%i2) csetup();

(%o2)                                done

(%i3) writefile("x.txt");

(%o3)                                done
(%o4)                                done
(%i5)                                            n - l
                                         %e      n
                                                  r
(%t5)                       mcs        = ----------
                               1, 1, 2       2

                                             n
                                              r
(%t6)                           mcs        = --
                                   1, 2, 1   2

                                             l
                                              r
(%t7)                           mcs        = --
                                   2, 2, 2   2

                                             1
(%t8)                           mcs        = -
                                   2, 3, 3   r

                                             1
(%t9)                           mcs        = -
                                   2, 4, 4   r

                                             - l
(%t10)                      mcs        = - %e    r
                               3, 3, 2

                                         cos(theta)
(%t11)                      mcs        = ----------
                               3, 4, 4   sin(theta)

                                       - l      2
(%t12)                mcs        = - %e    r sin (theta)
                         4, 4, 2

(%t13)               mcs        = - cos(theta) sin(theta)
                        4, 4, 3

(%o13)                               done
(%i14)                                [       n - l          ]
                               [     %e      n        ]
                               [              r       ]
                               [ 0   ----------  0  0 ]
                               [         2            ]
                               [                      ]
                        mcs  = [ n                    ]
                           1   [  r                   ]
                               [ --      0       0  0 ]
                               [ 2                    ]
                               [                      ]
                               [ 0       0       0  0 ]
                               [                      ]
                               [ 0       0       0  0 ]

                                   [ n            ]
                                   [  r           ]
                                   [ --  0   0  0 ]
                                   [ 2            ]
                                   [              ]
                                   [     l        ]
                                   [      r       ]
                                   [ 0   --  0  0 ]
                            mcs  = [     2        ]
                               2   [              ]
                                   [         1    ]
                                   [ 0   0   -  0 ]
                                   [         r    ]
                                   [              ]
                                   [            1 ]
                                   [ 0   0   0  - ]
                                   [            r ]

                           [ 0      0      0      0      ]
                           [                             ]
                           [               1             ]
                           [ 0      0      -      0      ]
                           [               r             ]
                    mcs  = [                             ]
                       3   [        - l                  ]
                           [ 0  - %e    r  0      0      ]
                           [                             ]
                           [                  cos(theta) ]
                           [ 0      0      0  ---------- ]
                           [                  sin(theta) ]

          [ 0            0                       0                 0      ]
          [                                                               ]
          [                                                        1      ]
          [ 0            0                       0                 -      ]
          [                                                        r      ]
   mcs  = [                                                               ]
      4   [                                                    cos(theta) ]
          [ 0            0                       0             ---------- ]
          [                                                    sin(theta) ]
          [                                                               ]
          [        - l      2                                             ]
          [ 0  - %e    r sin (theta)  - cos(theta) sin(theta)      0      ]

(%o14)                               done
(%i15) (%o15)                               done
(%i16)                               einstein = einstein

(%o16)                               done
(%i17)                 - l            l
              %e    ((- 1) + %e  + l  r)
                                    r
ein = matrix([--------------------------, 0, 0, 0], 
                           2
                          r
        - l        l
      %e    (1 - %e  + n  r)
                        r
[0, - ----------------------, 0, 0], [0, 0, 
                 2
                r
    - l                                  2
  %e    ((- 2 l ) + 2 n  + (2 n    + (n )  - l  n ) r)
               r       r       r r     r      r  r
- ----------------------------------------------------, 0], 
                          4 r
              - l                                  2
            %e    ((- 2 l ) + 2 n  + (2 n    + (n )  - l  n ) r)
                         r       r       r r     r      r  r
[0, 0, 0, - ----------------------------------------------------])
                                    4 r

(%o17)                               done
(%i18)                              n - l           n - l     2     n - l
                           %e      n       %e      (n )    %e      n  (n  - l )
                                    r r              r              r   r    r
(%t18) riem           = (- ------------) + ------------- - --------------------
           1, 2, 1, 2           2                4                  2
                                                                       n - l
                                                                  l  %e      n
                                                                   r          r
                                                                - -------------
                                                                        4

                                              n - l
                                            %e      n
                                                     r
(%t19)                   riem           = - ----------
                             1, 3, 1, 3        2 r

                                              n - l
                                            %e      n
                                                     r
(%t20)                   riem           = - ----------
                             1, 4, 1, 4        2 r

                                                    2
                                    (- l  n ) + (n )  + 2 n
                                        r  r      r        r r
(%t21)           riem           = - --------------------------
                     2, 2, 1, 1                 4

                                               l
                                                r
(%t22)                      riem           = - ---
                                2, 3, 2, 3     2 r

                                               l
                                                r
(%t23)                      riem           = - ---
                                2, 4, 2, 4     2 r

                                              - l
                                            %e    n  r
                                                   r
(%t24)                   riem           = - ----------
                             3, 3, 1, 1         2

                                             - l
                                           %e    l  r
                                                  r
(%t25)                    riem           = ----------
                              3, 3, 2, 2       2

                                             - l
(%t26)                    riem           = %e    - 1
                              3, 4, 3, 4

                                        - l         2
                                      %e    n  r sin (theta)
                                             r
(%t27)             riem           = - ----------------------
                       4, 4, 1, 1               2

                                       - l         2
                                     %e    l  r sin (theta)
                                            r
(%t28)              riem           = ----------------------
                        4, 4, 2, 2             2

                                    - l    l         2
(%t29)           riem           = %e    (%e  - 1) sin (theta)
                     4, 4, 3, 3

(%o29)                               done
(%i30) 
\end{lstlisting}
\section{Plotting routines}
These routines were used to generate data for the plot we produced in
this thesis. We only show a sample of the files used.
\begin{lstlisting}[caption=Data generating routines]
/* [wxMaxima batch file version 1] [ DO NOT EDIT BY HAND! ]*/
/* [ Created with wxMaxima version 13.04.2 ] */

/* [wxMaxima: input   start ] */
kill(all);
assume(a>0, b>0, mu<1 and mu>0, x_b>0, rho_c>0,r>0)$
x(r)        := r^2$
xi(x)       := 2/sqrt(a) * acoth( (1+sqrt(Z(x)))/(sqrt(a)*x) )$
Z(x)        := 1 - b*x + a*x^2$
Z_x(x)      := diff(Z(x),x)$
M(r)        := 4*%pi*r_b^3*(1/3-mu/5)*rho_c+ k^2*r_b^5/10$
Q(r)        := k*r_b^3$
Y(xi)       := c1*cosh(phi*xi) + c2*sinh(phi*xi)$
Y_xi(xi)    := diff(Y(xi),xi)$
Y_r(r)      := diff(Y(xi(x(r))),r)$
Y_rr(r)     := diff(Y_r(r), r)$
Y_xir(r)    := diff(Y_xi(xi(x(r))), r)$
rho(r)      := rho_c*(1-mu*(r/r_b)^2)$
r(rho)      := r_b*sqrt((1-rho/rho_c)/mu)$
lamb(x)     := -log(Z(x))$
nu(xi)      := 2*log(Y(xi))$
nu_r(r)     := diff(nu(xi(x(r))),r)$
Zsurf(mu)   := (1 - 2*M(r_b)/r_b)**(-1/2) - 1 $


elambi(x)   :=1/Z(x)$
enui(xi)    :=(Y(xi))^2$
elambo(r)   :=1/(1-2*G*M(r_b)/(r*c^2))$
enuo(r)     :=(1-2*G*M(r_b)/(r*c^2))$

ca          :a = (kappa*rho_c*mu/r_b^2 - k^2)/5$
cb          :b = kappa*rho_c/3$

cphi:       phi = sqrt(-(a+beta-2*k^2))/2$
calpha:     alpha = (kappa*rho_c/3 - kappa*rho_c*mu/5 - 4*k^2*r_b^2/5)/4$
cgamma:     gamma = sqrt(Zb)$

cZb:        Zb = Z(xb)$
cxib:       xib = xi(xb)$
cxb:        xb = x(r_b)$
cc1:        c1 = gamma*cosh(phi*xib) - alpha/phi*sinh(phi*xib)$
cc2:        c2 = alpha/phi*cosh(phi*xib) - gamma*sinh(phi*xib)$
cxi:        xi = xi(x)$
cx:         x = x(r)$
cY :        Y = Y(xi)$
cY_xi:      Y_xi = Y_xi(xi)$
cZ:         Z = Z(x)$
cY_r:       Y_r = Y_r(r)$
cZ_x:       Z_x = Z_x(x)$
crho:       rho = rho(r)$
cr:         r = r(rho)$
csigma:     rho = exp(sigma)$
clambi:     lamb = lamb(x)$
cnu:        nu = nu(xi)$
cnu_r:      nu_r = nu_r(r)$

cElambi:    elambi = elambi(x)$
cEnui:      enui = enui(xi)$
cElambo:    elambo = elambo(r)$
cEnuo:      enuo = enuo(r)$

v2:         v2 = c^2*Y_r/Y*(2*sqrt(Z)*Y_xi/Y-Z_x)*r_b^2/(rho_c*mu*r*kappa)$
pr:         pr = c^2*(4*sqrt(Z)*Y_xi/Y - 2*Z_x - kappa*rho)/kappa$
pt:         pt = rhs(pr) - c^2*beta*x/kappa$ 
Bul:        K = rho*rhs(v2)$
Zsurf:      Zsurf = Zsurf(mu)$

myterminal: "pdfcairo enhanced transparent dashed linewidth 3"$

subls:[cY, cY_r, cY_xi, cZ, cZ_x, crho]$
tovalls:[cc1,cc2,cphi,calpha,cgamma,cZb,cxib,cxi,cx,cxb,ca,cb]$
tovallsEOS:[cc1,cc2,cZb,cd,cbeta,cxib,cxi,cx,cxb,ca,cb, cr]$
constls:[kappa=8*%pi*G/c^2, G=6.673e-11,c=2.99792458e8]$
ultls:[rho_c=1e18, r_b=10000]$
ultlsmuHalf:[rho_c=1e18, mu=0.9, r_b=10000]$
noMu:[rho_c=1e18, r_b=10000]$
/* [wxMaxima: input   end   ] */

/* [wxMaxima: input   start ] */
/*Boundary metric compatibility plot*/
comp_post: "set key center left;set xlabel 'r (m)';set ylabel 'metric coefficients';set grid;"$
my_file: "/home/Ambrish/Documents/Phd/Text/NewSolutions/Figures/chargedAnisotropicMetricBoundaryPhiN.data"$

tovalls:[cc1,cc2,cphi,calpha,cgamma,cZb,cxib,cxi,cx,cxb,ca,cb]$
constb:[beta=-1e-16,ca,k=8e-9,r_b=1e4, rho_c=7.42564845e-28*1e18, mu=1,kappa=8*%pi]$
const2b:[beta=-5e-17,ca,k=6e-9,r_b=1e4, rho_c=7.42564845e-28*1e18, mu=1,kappa=8*%pi]$
pGrrb:subst(append(tovalls,constb), Z(x(r)))$
pGttb:subst(append(tovalls,constb), Y(xi(x(r)))**2)$
pGrr2b:subst(append(tovalls,const2b), Z(x(r)))$
pGtt2b:subst(append(tovalls,const2b),Y(xi(x(r)))**2)$

plot2d([pGrrb,pGttb, pGrr2b,pGtt2b],[r,0,1e4],
[xlabel,"r"],
[ylabel,"Metric coefficients 1/g_11 and g_00"],
[legend, "1/g_{rr}, {/Symbol b}=-1e-16,k=8e-9","g_{tt}, {/Symbol b}=-1e-16,k=8e-9","1/g_{rr}, {/Symbol b}=-5e-17,k=6e-9","g_{tt}, {/Symbol b}=-5e-17,k=6e-9"],
[plot_format, gnuplot_pipes],
[gnuplot_out_file, my_file],[gnuplot_postamble,comp_post]);
/* [wxMaxima: input   end   ] */

/* [wxMaxima: input   start ] */
/* Density plot */
density_post: "set key bottom left;set xlabel 'r (m)';set ylabel 'density (kgm^{-3})';set grid;"$
my_file: "/home/Ambrish/Documents/Phd/Text/Analysis/Figures/chargedAnisotropicPhiN/density.data"$

pD60:   subst(append(constls, tovalls,ultls,[mu=0.6]), rhs(crho))$
pD70:   subst(append(constls, tovalls,ultls,[mu=0.7]), rhs(crho))$
pD80:   subst(append(constls, tovalls,ultls,[mu=0.8]), rhs(crho))$
pD90:   subst(append(constls, tovalls,ultls,[mu=0.9]), rhs(crho))$
pD100:  subst(append(constls, tovalls,ultls,[mu=1]), rhs(crho))$

plot2d([pD60,pD70,pD80,pD90,pD100],[r,0,10000],
[legend,"{/Symbol m} = 0.6","{/Symbol m} = 0.7", "{/Symbol m} = 0.8", "{/Symbol m} = 0.9", "{/Symbol m} = 1.0"],
[plot_format, gnuplot_pipes],
[gnuplot_out_file,my_file],[gnuplot_postamble,density_post])$
/* [wxMaxima: input   end   ] */

/* [wxMaxima: input   start ] */
/* Radial Pressure plot, beta=-2e-16, k=1e-9 */
pressure_post: "set key top left;set xlabel 'r (m)';set ylabel 'pressure (Pa)';set grid;"$
my_file: "/home/Ambrish/Documents/Phd/Text/Analysis/Figures/chargedAnisotropicPhiN/pressureRbetaAK9.data"$

cbeta:  beta = -2e-16*mu$
ck :    k = 1e-9$
pP60:   subst(append(subls,tovalls,constls,ultls,[cbeta,ck,mu=0.6]), rhs(pr))$
pP70:   subst(append(subls,tovalls,constls,ultls,[cbeta,ck,mu=0.7]), rhs(pr))$
pP80:   subst(append(subls,tovalls,constls,ultls,[cbeta,ck,mu=0.8]), rhs(pr))$
pP90:   subst(append(subls,tovalls,constls,ultls,[cbeta,ck,mu=0.9]), rhs(pr))$
pP100:  subst(append(subls,tovalls,constls,ultls,[cbeta,ck,mu=1]), rhs(pr))$

plot2d([pP60,pP70,pP80,pP90,pP100],[r,0,10000],
[legend,"{/Symbol m} = 0.6","{/Symbol m} = 0.7", "{/Symbol m} = 0.8", "{/Symbol m} = 0.9", "{/Symbol m} = 1.0"],
[plot_format, gnuplot_pipes],
[gnuplot_out_file,my_file],[gnuplot_postamble,pressure_post])$
/* [wxMaxima: input   end   ] */

/* [wxMaxima: input   start ] */
/* Radial Pressure plot, mu=1, beta=-2e-16 */
pressure_post: "set key top left;set xlabel 'r (m)';set ylabel 'pressure (Pa)';set grid;"$
my_file: "/home/Ambrish/Documents/Phd/Text/Analysis/Figures/chargedAnisotropicPhiN/pressureRbetaMu1.data"$

cbeta:  beta = -2e-16*mu$
cmu:    mu=1$
pP60:   subst(append(subls,tovalls,constls,ultls,[cbeta,k=1e-9,cmu]), rhs(pr))$
pP70:   subst(append(subls,tovalls,constls,ultls,[cbeta,k=2e-9,cmu]), rhs(pr))$
pP80:   subst(append(subls,tovalls,constls,ultls,[cbeta,k=3e-9,cmu]), rhs(pr))$
pP90:   subst(append(subls,tovalls,constls,ultls,[cbeta,k=4e-9,cmu]), rhs(pr))$
pP100:  subst(append(subls,tovalls,constls,ultls,[cbeta,k=5e-9,cmu]), rhs(pr))$

plot2d([pP60,pP70,pP80,pP90,pP100],[r,0,10000],
[legend,"k = 1 x 10^{-9}","k = 2 x 10^{-9}","k = 3 x 10^{-9}","k = 4 x 10^{-9}","k = 5 x 10^{-9}"],
[plot_format, gnuplot_pipes],
[gnuplot_out_file,my_file],[gnuplot_postamble,pressure_post])$
/* [wxMaxima: input   end   ] */

/* [wxMaxima: input   start ] */
/* Radial Pressure plot, mu=1, beta=-2e-16 */
pressure_post: "set key top left;set xlabel 'r (m)';set ylabel 'pressure (Pa)';set grid;"$
my_file: "/home/Ambrish/Documents/Phd/Text/Analysis/Figures/chargedAnisotropicPhiN/pressureRk9Mu1.data"$

cmu:    mu = 1$
ck :    k = 1e-9$
pP60:   subst(append(subls,tovalls,constls,ultls,[beta = -2e-16,ck,cmu]), rhs(pr))$
pP70:   subst(append(subls,tovalls,constls,ultls,[beta = -3e-16,ck,cmu]), rhs(pr))$
pP80:   subst(append(subls,tovalls,constls,ultls,[beta = -4e-16,ck,cmu]), rhs(pr))$
pP90:   subst(append(subls,tovalls,constls,ultls,[beta = -5e-16,ck,cmu]), rhs(pr))$
pP100:  subst(append(subls,tovalls,constls,ultls,[beta = -6e-16,ck,cmu]), rhs(pr))$

plot2d([pP60,pP70,pP80,pP90,pP100],[r,0,10000],
[legend,"{/Symbol b} = -2 x 10^{-16}", "{/Symbol b} = -3 x 10^{-16}", "{/Symbol b} = -4 x 10^{-16}", "{/Symbol b} = -5 x 10^{-16}", "{/Symbol b} = -6 x 10^{-16}"],
[plot_format, gnuplot_pipes],
[gnuplot_out_file,my_file], [gnuplot_postamble,pressure_post])$
/* [wxMaxima: input   end   ] */

/* [wxMaxima: input   start ] */
/* Tangential Pressure plot */
pressure_post: "set key top left;set xlabel 'r (m)';set ylabel 'pressure (Pa)';set grid;"$
my_file: "/home/Ambrish/Documents/Phd/Text/Analysis/Figures/chargedAnisotropicPhiN/pressureT.data"$

pP60:   subst(append(subls,tovalls,constls,ultls,[mu=0.6]), rhs(pt))$
pP70:   subst(append(subls,tovalls,constls,ultls,[mu=0.7]), rhs(pt))$
pP80:   subst(append(subls,tovalls,constls,ultls,[mu=0.8]), rhs(pt))$
pP90:   subst(append(subls,tovalls,constls,ultls,[mu=0.9]), rhs(pt))$
pP100:  subst(append(subls,tovalls,constls,ultls,[mu=1]), rhs(pt))$

plot2d([pP60,pP70,pP80,pP90,pP100],[r,0,10000],
[legend,"{/Symbol m} = 0.6","{/Symbol m} = 0.7", "{/Symbol m} = 0.8", "{/Symbol m} = 0.9", "{/Symbol m} = 1.0"],
[plot_format, gnuplot_pipes],
[gnuplot_out_file,my_file],[gnuplot_postamble,pressure_post])$
/* [wxMaxima: input   end   ] */

/* [wxMaxima: input   start ] */
/* Y metric plot */
Y_post: "set key top right;set xlabel 'r (m)';set ylabel 'metric function, Y(r)';set grid;"$
my_file: "/home/Ambrish/Documents/Phd/Text/Analysis/Figures/chargedAnisotropicPhiN/Ymetric.data"$

cbeta:  beta = -2e-16*mu$
ck:     k=1e-9$

pY60:   subst(append(subls,tovalls,constls,ultls,[cbeta,ck,mu=0.6]), rhs(cY))$
pY70:   subst(append(subls,tovalls,constls,ultls,[cbeta,ck,mu=0.7]), rhs(cY))$
pY80:   subst(append(subls,tovalls,constls,ultls,[cbeta,ck,mu=0.8]), rhs(cY))$
pY90:   subst(append(subls,tovalls,constls,ultls,[cbeta,ck,mu=0.9]), rhs(cY))$
pY100:  subst(append(subls,tovalls,constls,ultls,[cbeta,ck,mu=1]), rhs(cY))$

plot2d([pY60,pY70,pY80,pY90,pY100],[r,0,10000],
[legend,"{/Symbol m} = 0.6","{/Symbol m} = 0.7", "{/Symbol m} = 0.8", "{/Symbol m} = 0.9", "{/Symbol m} = 1.0"],
[plot_format, gnuplot_pipes],
[gnuplot_out_file,my_file],[gnuplot_postamble,Y_post])$
/* [wxMaxima: input   end   ] */

/* [wxMaxima: input   start ] */
/* Z metric plot */
Z_post: "set key top right;set xlabel 'r (m)';set ylabel 'metric function, Z(r)';set grid;"$
my_file: "/home/Ambrish/Documents/Phd/Text/Analysis/Figures/chargedAnisotropicPhiN/Zmetric.data"$

cbeta:  beta = -2e-16*mu$
ck:     k=1e-9$

pZ60:   subst(append(subls,tovalls,constls,ultls,[cbeta,ck,mu=0.6]), rhs(cZ))$
pZ70:   subst(append(subls,tovalls,constls,ultls,[cbeta,ck,mu=0.7]), rhs(cZ))$
pZ80:   subst(append(subls,tovalls,constls,ultls,[cbeta,ck,mu=0.8]), rhs(cZ))$
pZ90:   subst(append(subls,tovalls,constls,ultls,[cbeta,ck,mu=0.9]), rhs(cZ))$
pZ100:  subst(append(subls,tovalls,constls,ultls,[cbeta,ck,mu=1]), rhs(cZ))$

plot2d([pZ60,pZ70,pZ80,pZ90,pZ100],[r,0,10000],
[legend,"{/Symbol m} = 0.6","{/Symbol m} = 0.7", "{/Symbol m} = 0.8", "{/Symbol m} = 0.9", "{/Symbol m} = 1.0"],
[plot_format, gnuplot_pipes],
[gnuplot_out_file,my_file],[gnuplot_postamble,Z_post])$
/* [wxMaxima: input   end   ] */

/* [wxMaxima: input   start ] */
/* lambda metric plot */
Lamb_post: "set key top left;set xlabel 'r (m)';set ylabel 'metric function, {/Symbol l}(r)';set grid;"$
my_file: "/home/Ambrish/Documents/Phd/Text/Analysis/Figures/chargedAnisotropicPhiN/Lmetric.data"$

cbeta:  beta = -2e-16*mu$
ck:     k=1e-9$

plamb60:   subst(append(subls,tovalls,constls,ultls,[cbeta,ck,mu=0.6]), rhs(clambi))$
plamb70:   subst(append(subls,tovalls,constls,ultls,[cbeta,ck,mu=0.7]), rhs(clambi))$
plamb80:   subst(append(subls,tovalls,constls,ultls,[cbeta,ck,mu=0.8]), rhs(clambi))$
plamb90:   subst(append(subls,tovalls,constls,ultls,[cbeta,ck,mu=0.9]), rhs(clambi))$
plamb100:  subst(append(subls,tovalls,constls,ultls,[cbeta,ck,mu=1]), rhs(clambi))$

plot2d([plamb60,plamb70,plamb80,plamb90,plamb100],[r,0,10000],
[legend,"{/Symbol m} = 0.6","{/Symbol m} = 0.7", "{/Symbol m} = 0.8", "{/Symbol m} = 0.9", "{/Symbol m} = 1.0"],
[plot_format, gnuplot_pipes],
[gnuplot_out_file,my_file],[gnuplot_postamble,Lamb_post])$
/* [wxMaxima: input   end   ] */

/* [wxMaxima: input   start ] */
/* exp(lambda) metric plot IN and OUT */
mIO_post: "set key top left;set xlabel 'r (m)';set ylabel 'metric function, exp({/Symbol l})';set grid;"$
my_file: "/home/Ambrish/Documents/Phd/Text/NewSolutions/Figures/LIOmetricPhiN.data"$

plamb60i:   subst(append(subls,tovalls,constls,ultls,[mu=0.6]), unit_step(r_b-r)*rhs(cElambi))$
plamb70i:   subst(append(subls,tovalls,constls,ultls,[mu=0.7]), unit_step(r_b-r)*rhs(cElambi))$
plamb80i:   subst(append(subls,tovalls,constls,ultls,[mu=0.8]), unit_step(r_b-r)*rhs(cElambi))$
plamb90i:   subst(append(subls,tovalls,constls,ultls,[mu=0.9]), unit_step(r_b-r)*rhs(cElambi))$
plamb100i:  subst(append(subls,tovalls,constls,ultls,[mu=1]),   unit_step(r_b-r)*rhs(cElambi))$

plamb60o:   subst(append(subls,tovalls,constls,ultls,[mu=0.6]), unit_step(r-r_b)*rhs(cElambo))$
plamb70o:   subst(append(subls,tovalls,constls,ultls,[mu=0.7]), unit_step(r-r_b)*rhs(cElambo))$
plamb80o:   subst(append(subls,tovalls,constls,ultls,[mu=0.8]), unit_step(r-r_b)*rhs(cElambo))$
plamb90o:   subst(append(subls,tovalls,constls,ultls,[mu=0.9]), unit_step(r-r_b)*rhs(cElambo))$
plamb100o:  subst(append(subls,tovalls,constls,ultls,[mu=1]),   unit_step(r-r_b)*rhs(cElambo))$

plamb60:    plamb60i+plamb60o$
plamb70:    plamb70i+plamb70o$
plamb80:    plamb80i+plamb80o$
plamb90:    plamb90i+plamb90o$
plamb100:   plamb100i+plamb100o$

plot2d([plamb60,plamb70,plamb80,plamb90,plamb100,1.8*unit_step(r-10000)],[r,0,20000],
[legend, "{/Symbol m} = 0.6","{/Symbol m} = 0.7", "{/Symbol m} = 0.8", "{/Symbol m} = 0.9", "{/Symbol m} = 1.0", "Boundary"],
[plot_format, gnuplot_pipes],
[gnuplot_out_file,my_file],[gnuplot_postamble,mIO_post])$
/* [wxMaxima: input   end   ] */

\end{lstlisting}.  

With the data generated, I then used gnuplot, a plotting program to
plot the data.  A sample of the gnuplot file is given below
\begin{lstlisting}[language=Gnuplot, caption=Gnuplot routines]
set term pdfcairo enhanced transparent dashed
set output '/home/Ambrish/Documents/Phd/Text/Analysis/Figures/chargedAnisotropicPhiP/EOSbeta1k9.pdf

set style line 1 lc rgb '#0000FF' lt 1 dt 1
set style line 2 lc rgb '#3300CC' lt 2 dt 2
set style line 3 lc rgb '#5500AA' lt 3 dt 3
set style line 4 lc rgb '#770088' lt 4 dt 4
set style line 5 lc rgb '#990066' lt 5 dt 5
set style line 6 lc rgb '#BB0044' lt 6 dt 6
set style line 7 lc rgb '#AA0022' lt 7 dt 7
set style line 8 lc rgb '#FF0000' lt 8 dt 8

set multiplot

set key bottom right
set xlabel '{/Symbol r} ( {/Symbol \264} 10^{18} kg{/Symbol \267}m^{-3})'
set ylabel 'pressure ({/Symbol \264} 10^{34} Pa)'
set size ratio 0.618

set yrange [0:]
set xrange [0:]

plot 'EOSbeta1k9.data' using ($1/1e18):($2/1e34) index 0 with lines ls 1 title "{/Symbol m} = 1.0", \
     ''              using ($1/1e18):($2/1e34) index 1 with lines ls 2 title "{/Symbol m} = 0.9" , \
     ''              using ($1/1e18):($2/1e34) index 2 with lines ls 3 title "{/Symbol m} = 0.8",  \
     ''              using ($1/1e18):($2/1e34) index 3 with lines ls 4 title "{/Symbol m} = 0.7", \
     ''              using ($1/1e18):($2/1e34) index 4 with lines ls 5 title "{/Symbol m} = 0.6 "
set size 0.5
set origin 0.15,0.45
set xrange[0.8:0.9]
set yrange[4.5:]
unset key
set xtics 0.02
set xlabel ""
set ylabel ""
set grid x,y

plot 'EOSbeta1k9.data' using ($1/1e18):($2/1e34) index 0 with lines ls 1 title "{/Symbol m} = 1.0", \
     ''              using ($1/1e18):($2/1e34) index 1 with lines ls 2 title "{/Symbol m} = 0.9" , \
     ''              using ($1/1e18):($2/1e34) index 2 with lines ls 3 title "{/Symbol m} = 0.8",  \
     ''              using ($1/1e18):($2/1e34) index 3 with lines ls 4 title "{/Symbol m} = 0.7", \
     ''              using ($1/1e18):($2/1e34) index 4 with lines ls 5 title "{/Symbol m} = 0.6"

unset multiplot
\end{lstlisting}.  



\end{appendices}
\printindex
\end{document}